# Eigenvalue equations from the field of theoretical chemistry and correlation calculations

Sajátérték egyenletek az elméleti kémia és korreláció számítás területéről

SÁNDOR KRISTYÁN

KRISTYÁN SÁNDOR

Doctor of the Hung. Acad. Sci. dissertation

MTA doktori értekezés

Institute of Materials and Environmental Chemistry,
Research Centre for Natural Sciences,
Hungarian Academy of Sciences

Anyag- és Környezetkémiai Intézet
MTA Természettudományi Kutatóközpont

Budapest, 2017

Mottó: $a^2+b^2=c^2$, $F=ma$, $E=mc^2$, $H\Psi=E\Psi$, stb.

A matematika mondja meg a fizikának mit kell tenni. (Math tells the phys. what to do.)

Az ördög a részletekben rejlik. (ősi bölcsesség)

Az a baj, hogy azt hiszed, van időd. (Buddha)

Nostradamus vers a tudós intuíciójáról:
Titokban az éjszakában, néma csendben
Egyedül ül a tudós gyertya vele szemben
Magányban ül, a kicsiny lángnyelv lobban
Mit más nem láthat, arra fényt vet nyomban
Szörnyű hang riasztja, reszket köpenyében
Túlvilági hírnök, Isten ül a közelében.

Hekler Antal (Bp, 1882-1940) művészettörténész, régész, egyetemi tanár, az MTA tagja:
Az alkotó munka, a teremtés az, ami az embert valódi istenközelségbe emeli.
Ez a legegyenesebb út Istenhez. Alkotó ember éppen ezért sohasem lehet istentelen.

A részletek részletes analízise. . .(K.S.)

A kémia a fizika királynője. . .(K.S.)

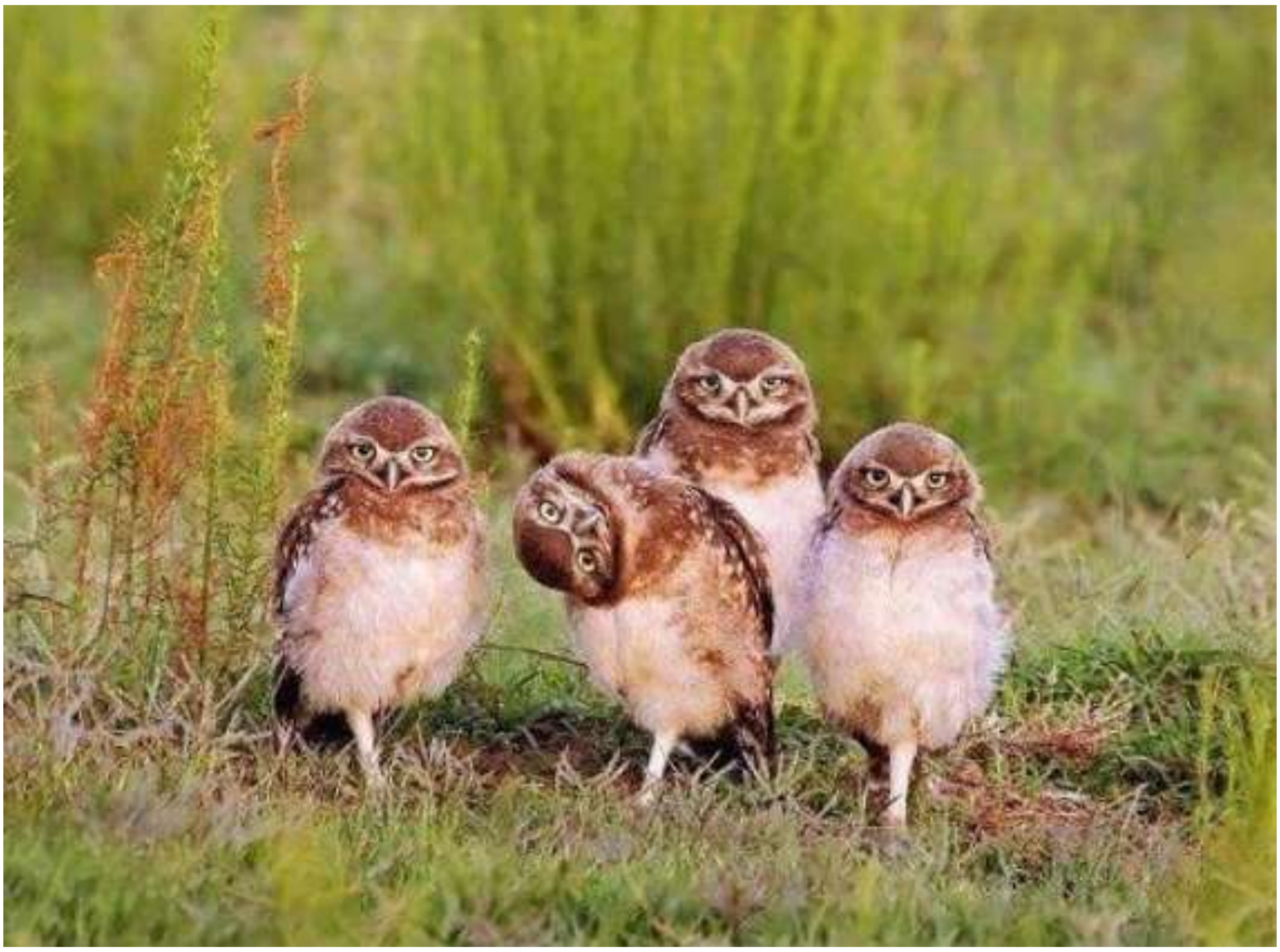

(Nézőpont kérdése…)

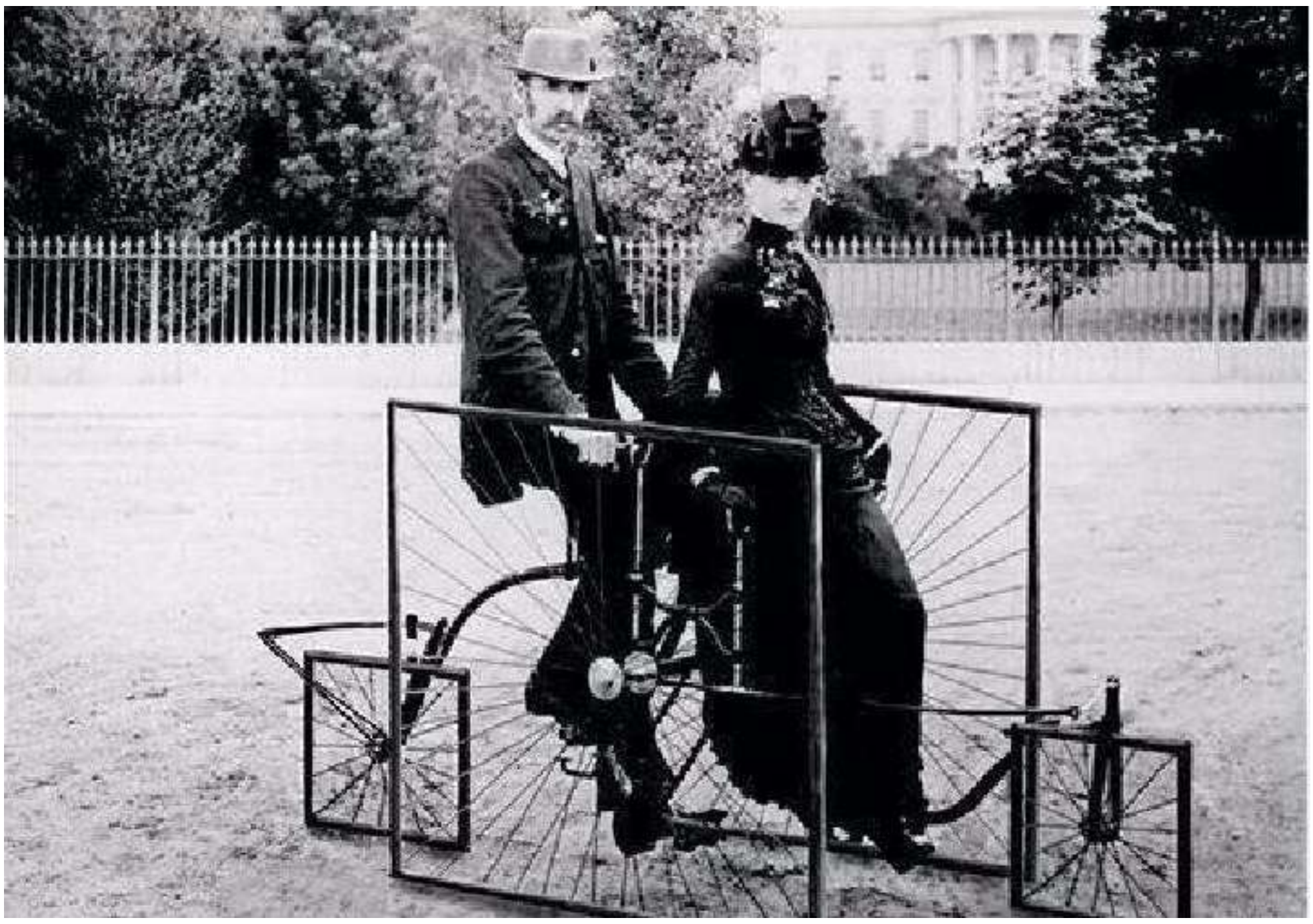

(This device does not work smoothly, but “DFT” [does function theoretically]…)

# Tartalom/Content

# 1. Előszó: MIÉRT ÍGY ÍRTAM MEG?

Disszertációm technikai olvasásához megjegyzem, hogy a Doktori Ügyrend VII. Osztály szabályzata az érdemi részt maximum 150 oldal terjedelemben kéri, mint „minden tekintetben komplett és önmagában értelmezhető és értékelhető munka". Ezen disszertációban a második (2.x) fő fejezet kb. 100 oldal terjedelemben írja le tudományos tevékenységem ide vonatkozó részét, a négy alfejezet (2.1.x, 2.2.x, 2.3.x, 2.4.x) önálló – természetesen egymástól nem független – gondolatmeneteket ír le. A harmadik (3.x) fő fejezet tézisszerűen tárgyal eredményeket vázlatszerű 2-4 oldal terjedelmekben, ahol a tézisek írnak le önálló gondolatokat összesen kb. 150 oldal terjedelemben, (kb. ötször annyi témát de kb. tízszer rövidebben, mint a 2.x fejezet). Természetesen örülnék ha az olvasó mindent kritikusan elolvas, de ha kutatásaimat valaki rövidebb olvasatban akarja megismerni, akkor az is elég, ha csak az egyiket olvassa a 2.x és 3.x fő fejezetek közül; ha csak ízelítőként egy-egy szemelvényre kíváncsi, akkor választhat a 2.1.x - 2.4.x alfejezetek illetve a 3.x tézisek közül.

Disszertációm emocionális olvasásához megjegyzem, hogy a Velencei tó partján álló Pákozdi Emlékmű melletti hangulatos halászcsárdában, és máshol is persze, olvasható a borlapon, hogy ha megiszol egy pohár magyar bort, mindig gondolj arra, mennyi kemény munka van addig, amíg az a pohár bor eljut az asztalodig. Aki saját maga izzadt ki tudományos publikációkat, akár egyet is, különösen egy szerzőset, vagy több szerzőset, de amiben saját maga volt a fő szerző és irányító, az tudja mennyi szenvedés és elszántság van abban, amíg egy téma megismerése, irodalom követése, ötlet fejlesztése (esszencia), megvalósítása és kidolgozása, összefoglalása, megírása és végül publikálása megvalósul, elkészül. Ha valaki meg akarja ismerni a munkáimat, eredményeimet, annak mindenekelőtt (az ő érdeklődési körébe tartozó) cikkeim elolvasását javaslom. A publikációnak mindig egy kis egésznek kell lennie, ellentmondásmentesnek, megalapozottnak, érthetőnek, továbbá jelezve, hogy a rövidségre és tömörségre való állandó igény miatt hol találhatók meg a szükséges, de a cikkben nem közölhető részletek, valamit a nyitva maradó kérdéseket is jelezni illik. (Mint ahogy Michelangelo mondta: Minden macska kölyök egy tökéletes alkotás, – hát valahogy így kell lenni a tudományos publikációnak is eredményét és megírását tekintve…) Publikációim nagy részében fő szerzőként vettem részt, ezt a levelező szerző jelzések is mutatják, illetve nyilvánvaló a számos egy szerzős (18, azaz tizennyolc) munkáimból. Mivel összefoglaló (review) jellegű cikkeim is vannak, nem látom értelmét, hogy a verejtékes munkával írt és hosszasan csiszolt cikkeimből összeollózzak, összesimítsak egy egységes szöveget, lefordítsam magyarra, átszámozzam az egyenleteket. Inkább fő fejezetekként gyűjtöm össze a review-jellegű munkákat, és kivonatosan, vázlatosan írom le a fő eredményeimet, de azért azt szem előtt tartva, hogy viszonylag szilárd alapon álljanak a bemutatott eredmények. (A szilárd alap elvénél az vezérelt, miszerint néhány évvel ezelőtt egy vitatott Nobel díj kapcsán valaki megjegyezte – tartalmilag idézem: És mit mondanátok, ha édesanyátoknak kellene elmondani, hogy mire kaptátok a Nobel díjat?) Néhány eredmény nemzetközileg ismert és magasan jegyzett kutatokkal, pl. Prof. Pulay Péter (University of Arkansas at Fayetteville) ill. Prof. Aron Kuppermann+ (California Institute of Technology) született. (Bár szívesen dolgoztam együtt pl. ezekkel a neves kutatókkal, de a „nagy fák alatt nem nő virág” elve alapján csak rövid ideig szándékoztam mellettük dolgozni, hogy saját kutatási irányomba fektethessem energiámat.) A kompaktabb téziseket részletesebb, vázlatszerű eredmény bemutatása egészíti ki.

Disszertációm tudományos olvasásához megjegyzem, hogy első sorban a molekulák elektron szerkezete érdekel, ill. azon belül a sűrűség funkcionál elmélet (density functional theory [DFT]). Sherlock Holmes szerint először a tények, utána a magyarázat. Elméleti kutatásban először a kísérlet, utána az elmélet vagy modellezés, majd az elmélet, ill. modell tesztelése. Első sorban a DFT fejlesztése érdekel, és bár a tesztelés alapvető, figyelmem leginkább nem a különböző kémiai rendszerek leírására irányul (főleg már jól bevált DFT módszerekkel), talán ez is magyarázza publikációim nem nagy számát. A DFT mellett a termodinamika, reakció kinetika és statisztikus mechanika is érdekel, e területekről is bemutatok saját eredményeket (nem a fő címbeli kutatási területem helyett, hanem mellette…).

Miután 2014-ben túlléptem az 1000-es hivatkozási indexet és az 50-dik publikációmat, ezért úgy döntöttem, hogy beadom az MTA doktori értekezésemet. Csak azon tudományos eredményeimet bocsátom szemlére itt, amelyek a nemzetközi tudományos folyóiratok őrlőfogai közt lecsiszolódtak, megmérettettek, és a „tíz körmömmel kapartam ki”, ezek téma szerinti (de nem szigorú) fontossági sorrendben a következő folyóiratok:
International Journal of Quantum Chemistry,
Journal of Theoretical and Applied Physics ,
Computational and Theoretical Chemistry,
Journal Computational Chemistry,
Journal of Molecular Structure: THEOCHEM ,
Theoretical Chemistry Accounts,
Chemical Physics Letters,
Chemical Physics,
Journal of Chemical Physics,
Journal of Physical Chemistry A and C,
Physical Review A,
Computers in Physics,
Journal of the Chemical Society - Faraday Transactions I,
Journal of Colloid and Interface Science ,
Biochimica et Biophysica Acta - Biomembranes,
Journal of Catalysis,
Tetrahedron: Asymmetry,
Catalysis Today,
Langmuir,
Canadian Journal of Chemical Engineering,
Surface Science Letters,
Surface Science,
Il Nuovo Cimento D,
Reaction Kinetics and Catalysis Letters,
Periodica Polytechnica Chemical Engineering and Electrical Engineering,

Acta Chimica Hungarica,
Acta Physica Hungarica,
Fizikai Szemle.

Végezetül, 1.: Görgey Artúr (1818-1916) az 1848-49-es forradalom és szabadságharc magyar tábornokától idézek, akinek eredeti szakmája kémikus volt: Mert a tisztán elméleti vegytan, hasonló a puszta parlaghoz, melyen a tévtanok gyomjai tenyésznek, ha nem vetjük be azt a lelkiismeretes kísérletek vetőmagjával, hogy rajta igazságokat arassunk. 2.: Amilyen közel áll az $mv^2/2$ képlet (egy m tömeggel rendelkező, egyenes vonalban, tömeg-középpontjának egyenletes v sebességgel mozgó test kinetikus energiája a klasszikus Newtoni mechanika szerint) algebrailag az $mc^2$ (tömeg-energia ekvivalencia Einsteini relativitáselmélet szerinti nyugalmi energia) képletéhez, ahol c jelöli a fénysebességet, ugyan olyan távol is áll a szemlélet ami megettük van, mégis a kettő ugyananak a dolognak a része: az Anyatermészet (legalábbis egy részének) leírása. A világ egyik legfontosabb sajátérték egyenletének (Schrödinger) megoldásához közvetlenül, vagy akár távolabbi értelemben vett közvetett megoldásához, (mert némelyik más területre csatangolt át) legjobb tudásommal kidolgozott saját egyenleteim felsorolnám itt, az Előszóban, és hogy ne tűnjék „fekete mágiának” tisztelettel veszem mindenkitől, aki elolvassa a kb. 250 oldalas részletezést is hozzá ezen oldalak folytatásában, hátha olyan varázsnak tűnnek, amelyeket a kritikus rendelkezésre álló szavak nem tudnak megközelíteni, és hátha olyanok, mint egy mély, sötét kút, minél tovább nézünk bele, annál fényesebben ragyognak belőle vissza a csillagok:

Charge-Exchange between atoms and metal surfaces during collision:

$$P_1^{fm}= \sin^2[2\alpha/( E_{kin}^{init})^{0.5}] \quad \text{for} \quad 2\alpha/( E_{kin}^{init})^{0.5}\leq\pi$$

$$\alpha\equiv (m/2)^{0.5} \int_{[0,\propto]}V_{12}^{d}(R)dR$$

$$P_1^{fm}= 1-\sin^2[\sin(\beta^*)\alpha/( E_{kin}^{init})^{0.5}]$$

$$P_1^{fm}= 1/(1 + \alpha^2/E_{kin}^{init})$$

$$P_1^{fm}= 2\alpha^2E_{kin}^{init}/(\alpha^2 + E_{kin}^{init})^2$$

$$E_{kin}^{init}(max)= \alpha^2 = (½)m(\int_{[0,\propto]}V_{12}^{d}(R)dR)^2$$

$$P_1^{fm}= 1/[1 + \alpha^2(E_{kin}^{init})^{-1} + (\alpha^*)^4(E_{kin}^{init})^{-2}]$$

$$\alpha^*\equiv (m/2)^{0.5} \int_{[0,\propto]}\{ (V_{12}^{d})^2[ 8(V_{12}^{d})^2-( V_{11}^{d}-V_{22}^{d})^2 ]/12 \}^{0.25}dR$$

# # # # # #

Semilocal DFT vs. highly ionized atoms:
The quality of DFT methods in estimating $E_{total\ electr,0}(N,Z)$: B-LYP > B > D-S.
In contrast, for the dispersion forces (in weakly interacting systems) between noble atoms ($He_2$, $Ne_2$, and $Ar_2$) the simplest D-S method gives the best, though not satisfactory, results.

# # # # # #

Basis set choice in DFT:
The MP2 with large basis set accounts for 30-45% (strong under shot) of $E_{corr}$ for closed shell ground state atoms with $2\leq N\leq Z\leq 18$,
this is 95-110% (slight under and over shot) in case of LYP with large and small basis sets.

Although the LYP approximation for correlation energy, $E_{corr}(LYP)$, is not variational, a 3-21G basis set can produce almost the same value as the TZ or TZP basis set for small molecules within the chemical accuracy.

######

Calculating $E_{corr}$ with RECEP:

$E_{corr}$ in molecular systems mainly and quasi-linearly depends on electron content (N) as

$$-0.035(N-1) > E_{corr}[hartree] > -0.045(N-1),$$

for large molecules, macroscopic media, crystals and metals it reads as

$$-(\partial E_{corr}/\partial N)_{\{ZA,RA\}M} \approx 0.035 \div 0.045.$$

$$E_{corr}(ESP\text{-}i) = \Sigma_{A=1\ldots M} E_{corr}(N_A, Z_A)$$

$$E_{corr}(N_A, Z_A) \equiv [(N_A^*+1)-N_A]E_{corr}(N_A^*,Z_A) + (N_A-N_A^*)E_{corr}(N_A^*+1,Z_A)$$

$E_{corr}(N_A^*,Z_A)$= HF-SCF/basis atomic correlation energy, which is transferable to molecular systems, i.e. a nuclear frame quasi-independent property.

######

Development in RECEP:

$$E_{corr}(RECEP) = \Sigma_{A=1\ldots M} E_{corr}(N_A, Z_A)$$

$E_{corr}$(N,Z,CI, low spin)≡$E_0$(N,Z,CI, high spin) - $E_0$(N,Z,HF-SCF, high spin) + + "high-to-low-spin-correction"

$E_{corr}$(N,Z,B3LYP, low spin)≡ $E_0$(N,Z,B3LYP, low spin) - $E_0$(N,Z,HF-SCF, low spin)

Hydrogen atoms require special attention because their partial charges fall frequently in $0<N_A<1$: In RECEP method the algorithm for $E_{corr}(N_A, Z_A)$ needs the wider $0<N_A<2$.

######

Atomic correlation parameters in RECEP:

Comparing "atomic correlation energies in free space" to "atomic correlation energies in molecular environment" based on different partial charges reveal that these values are very close to each other at any (N,Z) value.

The abacus or recipe of RECEP-fit method (approximating the prestigious G2 results with mean absolute difference (MAD) < 2 kcal/mol and even the G3 as well) is represented via the case of methyl-nitrite ($CH_3$-O-N=O) using fitted RECEP atomic correlation parameters:

| $Z_A$ | NPA | $N_1$ | $N_A$ | $N_2$ | $E_{corr}(N_1,Z_A)$ | $E_{corr}(N_2,Z_A)$ | $E_{corr}(N_A,Z_A)$ |
|---|---|---|---|---|---|---|---|
| 6 | -0.133 | 6 | 6.133 | 7 | -0.1659 | -0.1909 | -0.1692 |
| 8 | -0.490 | 8 | 8.490 | 9 | -0.2703 | -0.2790 | -0.2746 |
| 1 | 0.171 | 0 | 0.829 | 2 | 0.0 | -0.0376 | -0.0156 |
| 1 | 0.165 | 0 | 0.835 | 2 | 0.0 | -0.0376 | -0.0157 |
| 1 | 0.165 | 0 | 0.835 | 2 | 0.0 | -0.0376 | -0.0157 |
| 7 | 0.504 | 6 | 6.496 | 7 | -0.2227 | -0.2259 | -0.2243 |
| 8 | -0.382 | 8 | 8.382 | 9 | -0.2703 | -0.2790 | -0.2736 |

######

HF-SCF limit (based on G3 calculations):

$$E_{total\ electr,0}(\text{HF-SCF limit}) = 1.000\ 178\ E_{total\ electr,0}\ (\text{HF/6-311+G(2d,p)})$$

######

Nearly experimental quality $E_{total\ electr,0}$ with small basis set RECEP/6-31G(d):

RECEP enthalpy of formation: $\Delta H^0_f$(molecule, RECEP, expt., charge def.)=

$E_{total\ electr,0}$(molecule, RECEP, expt., charge def.) +

+ $E_{ZPE}$(molecule, G3) +

+ $E_{therm}$(molecule, G3) +

+ $\Sigma_{Atom=1\ldots M}$ [$\Delta H^0_f$(Atom, expt.) - $E_{total\ electr,0}$(Atom, G3) - $E_{therm}$(Atom, G3)]

######

Calculating $E_{corr}$ with REBECEP with Mulliken matrix:

The REBECEP (rapid estimation of basis set error and correlation energy from partial charges) formula to calculate $E_{corr}$ and basis set error for ground state covalent neutral molecules in the vicinity of stationary points is

$$E_{corr}(\text{REBECEP, method, charge def., basis set}) \equiv$$
$$\Sigma_{A=1,\dots,M} E_{corr}(N_A, Z_A, \text{method, charge def., basis set}),$$
$$E_{corr}(N_A, Z_A, \text{method, charge def., basis set}) =$$
$$(N_A - N_1)E_{fitpar}(N_2, Z_A, \text{method, charge def.,basis set})$$
$$+(N_2 - N_A)E_{fitpar}(N_1, Z_A, \text{method, charge def., basis set})$$

where $N_1$ and $N_2$ are integer numbers of electrons, with $N_1 \leq N_A \leq N_2 = N_1 + 1$, and $N_A$ is the electron content around atom A based on the chosen partial charge.

```
> # hf/6-31G*
> Formaldehyde (H2C=O)
> SCF Done: E(RHF) = -113.863712881 A.U. after 6 cycles
> Convg = 0.8513D-04 -V/T = 2.0037
> S**2 = 0.0000
> Condensed to atoms (all electrons):
>      1        2        3        4
> 1 O  8.013325 0.522645 -0.049815 -0.049815   ← 4x4 Mulliken matrix
> 2 C  0.522645 4.600151  0.371281  0.371281
> 3 H -0.049815 0.371281  0.596456 -0.068770
> 4 H -0.049815 0.371281 -0.068770  0.596456
> Total atomic charges:
> 1
> 1 O -0.436341   ← Mulliken partial charge
> 2 C  0.134643
> 3 H  0.150849
> 4 H  0.150849
> Sum of Mulliken charges= 0.00000
```

######

Calculating zero point energy (ZPE) with REZEP:

For closed- or open shell ground state covalent molecules in the vicinity of equilibrium geometry, the ZPE estimation by an inexpensive atom by atom method:

$$\text{ZPE(REZEP)} \equiv \sum E_{ZPE}(N_A, Z_A)$$
$$E_{ZPE}(N_A, Z_A) = (N_A - N_1)\, E_{ZPEpar}(N_2, Z_A) + (N_2 - N_A)\, E_{ZPEpar}(N_1, Z_A)$$
$$E_{ZPEpar}(N_A,Z_A{=}1) = N_A\, E_{ZPEpar}(N_2{=}2,Z_A{=}1)/2 \qquad \text{for } 0 \leq N_A \leq 2$$

Empirical estimation:

$$\text{ZPE (modified Politzer, kcal/mol)} = 6.99n_H + 3.74n_C + 3.98n_N + 3.45n_O + 2.79n_F - 4.63$$

######

Dependence of $E_{corr}$ and ZPE on the nuclear frame and number of electrons:

From quantum chemical calculations:

$$0 < \text{ZPE} \leq b(N-1) \qquad \text{with } b \approx 0.0036 \text{ hartree}$$
$$0 > E_{corr} \approx a(N-1) \qquad \text{with } a \approx -0.04 \text{ hartree}$$

From classical mechanics modeling:

$$\text{ZPE} \approx E_{kin} = \Sigma\, h(2\pi)^{-1}(k_A/m_A)^{1/2}$$
$$\text{ZPE[hartree]} \approx (-0.00304 \pm 0.000555) + (0.01193 \pm 0.000083)\Sigma m_A[\text{a.u.}]^{-1//2} \approx$$
$$\approx c\, \Sigma m_A[\text{a.u.}]^{-1//2} \qquad \text{with } c=0.012$$
$$\text{ZPE[hartree]} = (h/2)\Sigma\nu_i \approx 0.012\Sigma m_A[\text{a.u.}]^{-1//2} \leq 0.0036(N-1)$$

######

Calculating $E_{corr}$ and basis set error with scaling operators T and V in HF-SCF:

$(1 + k_c)<S|H_\nabla|S> + <S|H_{Rr}|S> + (1 + k_{ee})<S|H_{rr}|S>$

######

Variational calculation with a scaling correct moment functional:

Truncation n=1: $(5/3)A_1\rho_0^{2/3} + (4/3)B_1\rho_0^{1/3} + v(\mathbf{r}_1) \approx \lambda \equiv E_{electr,0} / N$

Truncation n=2: $(5/3)A_1u^4 + (8/3)A_2[\int u^8 d\mathbf{r}_1]u^2 + (4/3)B_1u^2 + (7/3)B_2[\int u^7 d\mathbf{r}_1]u + v(\mathbf{r}_1) \approx \lambda$

$\lambda = \partial E_{electr,0}/\partial N$= Lagrange multiplier or chem. pot., $\rho_0(\mathbf{r}_1)$= one-electron density, $u(\mathbf{r}_1) \equiv \rho_0^{1/6}$.

######

Variational calculation with general density functional for ground state, recipe for SCF:

Key property for the algorithm is that the absolute values of the three main energy terms are in about the same magnitude in molecular systems, (recall e.g. the virial theorem for stationary systems as $(V_{ee}+ V_{ne}+ V_{nn})/T = -2$, wherein the 2 is exact), for example:

H(Z=N=1) atom: $E_{electr,0}= T+ V_{ee}+ V_{ne}= 0.5 +0 -1 = -0.5$ hartree $\Rightarrow$ 0.5 : 0 : 1,

while the much larger Ar(Z=N=18) atom has similar ratio among the three terms:

$E_{electr,0}= (0.527544 +0.264456 -1.319544)\text{x}10^3= -527.544$ hartree $\Rightarrow$ 0.53 : 0.26 : 1.32,

as well as the abs($V_{ne}$) term is the largest among the three, furthermore:

$$|\partial T(N)/\partial\rho_0|,\ |\partial V_{ee}(N)/\partial\rho_0| < |\int(v(\mathbf{r}_1)-\lambda)\, \rho_{0i} d\mathbf{r}_1|.$$

The "Lagrange's method of undetermined multiplier" for the 2nd Hohenberg-Kohn theorem minimizes the Lagrangian, yielding the quasi-linear equation system to solve iteratively (i=1,…,L, and the left hand side is from a particular DFT functional used):

$$\Sigma_{k=1\dots L}\, d_k^{\text{iter } m+1}\int(v(\mathbf{r}_1)-\lambda^{\text{iter } m})b_i(\mathbf{r}_1)b_k(\mathbf{r}_1)d\mathbf{r}_1 = -(1/2)\partial T[\rho_0^{\text{iter } m}(\mathbf{r}_1)]/\partial d_i - (1/2)\partial V_{ee}[\rho_0^{\text{iter } m}(\mathbf{r}_1)]/\partial d_i$$

######

Compact one-electron DFT energy functional approximation for $E_{electr,0}$ with $T_{Thomas–Fermi}$ and Parr $V_{ee}$:

Crudely

$$c_{10}c_1\rho^{5/3} + P(\mathbf{r}_1)\rho + c_{20}c_2\rho^{4/3} \approx \rho E_{electr}$$

where $c_1 \equiv Nc_F$, $P(\mathbf{r}_1) \equiv -N\Sigma_{A=1,\dots,M}Z_AR_{Ai}^{-1}$ and $c_2 \equiv N2^{-1/3}(N-1)^{2/3}$, with $z \equiv \rho_0^{1/3} \geq 0$ it reduces to a 2nd order algebraic equation:

$$c_{10}c_1z^2 + c_{20}c_2z + (P(\mathbf{r}_1)-E_{electr,0}) \approx 0.$$

With $A \equiv c_{10}c_1$, $B \equiv c_{20}c_2$, $P(\mathbf{r}_1)$, $\text{discr}(\mathbf{r}_1) \equiv B^2-4A(P(\mathbf{r}_1)-E_k)$, its solution is

$$\rho_{0,k}(\mathbf{r}_1)=C[(+(\text{discr}(\mathbf{r}_1))^{1/2}-B)/(2A)]^3 \quad \text{with } C \equiv N/[\int[(+(\text{discr}(\mathbf{q}_1))^{1/2}-B)/(2A)]^3 d\mathbf{q}_1]$$

for any nuclear configuration (index k refers to the kth approximation). The minimum with respect to $E_k$ for the approximate functional:

$$E_{electr,0}(\text{approx.},E_k)= (1/N)\int(c_{10}c_1\rho_{0,k}^{5/3} + P(\mathbf{r}_1)\rho_{0,k} + c_{20}c_2\rho_{0,k}^{4/3})d\mathbf{r}_1.$$

######

Multi-electron densities from Hohenberg–Koh theorems to variational principle:

<u>Statement</u>: On N+2 level, the kinetic functional is $T= F_{kin}[b_{N+2} \equiv \Psi]= <\Psi|H_\nabla\Psi>$, but serious difficulty starts even from the next lower level, N+1, for $T= F_{kin}[b_{N+1} \equiv |\Psi|^2]$, i.e. finding the analytical form of $F_{kin}$ as a functional of $b_{N+1}$, generally for cases i=1,2,…,N+1. For i=1, the weak Thomas –Fermi approximation, $t_1[b_1 \equiv \rho] \sim \rho^{5/3}$, holds, while for i=N+2 the exact $t_{N+2}[b_{N+2}] = -(1/2)b_{N+2}^*\nabla_1^2 b_{N+2}$ holds (*= complex conjugate).

Statement: In contrary to kinetic operators, all cases i=1,2,…, N+1,N+2 yield easy and simple nuclear-electron attraction functional: $v_{en}[b_i] = -\Sigma_{A=1,\dots,M}Z_A\, b_i\, R_{A1}^{-1} = b_i\, v(\mathbf{r}_1)$.

Statement: For cases i=2,…,N+1,N+2, the electron-electron repulsion func. is also simple: $v_{ee}[b_i] = ((N-1)/2)\, b_i\, r_{12}^{-1}$, but problematic for i=1. Technically, $v_{ee}[b_{N+2}] = ((N-1)/2)\, b_{N+1}\, r_{12}^{-1}$.

Generalization of 1st HK theorem: The external potential $v(\mathbf{r}_i) \equiv -\Sigma_{A=1,\dots,M}Z_A\, R_{Ai}^{-1}$ is determined, within a trivial additive constant, by the multi-electron density $b_i$ for i=1,2,3,…,N+2. (The 1st HK theorem states it for i=1.)

Generalization of 2nd HK theorem: VP holds for all i=1,2,3,…,N+1 as $E_{electr,0} \leq E_{v,i}[b_{i,trial}]$, if the trial i-electron density $b_{i,trial}$, is $b_{i,trial} \geq 0$ at any $(\mathbf{r}_1,\mathbf{r}_2, \mathbf{r}_3\dots,\mathbf{r}_i)$ 3i-dimensional real space point and normalized to N ($\int b_{i,trial}(\mathbf{r}_1,\mathbf{r}_2, \mathbf{r}_3\dots,\mathbf{r}_i)d\mathbf{r}_1 d\mathbf{r}_2 d\mathbf{r}_3\dots d\mathbf{r}_i = N$) and $E_{v,i}[b_{i,trial}]$ is the energy functional. (The 2nd HK theorem states it for i=1, and the VP states it for i=N+2.)

Integro-differential forms: $D[\rho] \equiv D_\nabla[\rho] + D_{Rr}[\rho] + D_{rr}[\rho] = \rho E_{electr}$

$$D_\nabla[\rho] = -(1/2)\Sigma_{i=1,\dots,N}\int\Psi^*\nabla_i^2\Psi ds_1 d\mathbf{x}_2\dots d\mathbf{x}_N =$$

$$= -(1/2)\int\Psi^*\nabla_1^2\Psi ds_1 d\mathbf{x}_2\dots d\mathbf{x}_N -((N-1)/2)\int\Psi^*\nabla_2^2\Psi ds_1 d\mathbf{x}_2\dots d\mathbf{x}_N$$

$$\int D_\nabla[\rho]d\mathbf{r}_1 = -(N/2)\int\Psi^*\nabla_1^2\Psi d\mathbf{x}_1 d\mathbf{x}_2\dots d\mathbf{x}_N$$

$$D_{Rr}[\rho] = \rho(\mathbf{r}_1)v(\mathbf{r}_1) + (N-1)\int b_2(\mathbf{r}_1,\mathbf{r}_2)v(\mathbf{r}_2)d\mathbf{r}_2$$

$$\int D_{Rr}[\rho]d\mathbf{r}_1 = N\int\rho(\mathbf{r}_1)v(\mathbf{r}_1)d\mathbf{r}_1$$

$$D_{rr}[\rho] = (N-1)\int b_2(\mathbf{r}_1,\mathbf{r}_2)r_{12}^{-1}d\mathbf{r}_2 + [N(N-1)/2 - (N-1)]\int b_3(\mathbf{r}_1,\mathbf{r}_2,\mathbf{r}_3)r_{23}^{-1}d\mathbf{r}_2 d\mathbf{r}_3$$

$$\int D_{rr}[\rho]d\mathbf{r}_1 = (N(N-1)/2)\int\Psi^*\Psi r_{12}^{-1}d\mathbf{x}_1 d\mathbf{x}_2\dots d\mathbf{x}_N = (N(N-1)/2)\int b_2(\mathbf{r}_1,\mathbf{r}_2)r_{12}^{-1}d\mathbf{r}_1 d\mathbf{r}_2.$$

For H-like atoms and one-electron molecules for ground and excited states the exact

$$D[N=1,\rho(\mathbf{r}_1)] \equiv -(1/4)\nabla_1^2\rho(\mathbf{r}_1) + (1/8)\rho(\mathbf{r}_1)^{-1}|\nabla_1\rho(\mathbf{r}_1)|^2 + \rho(\mathbf{r}_1)v(\mathbf{r}_1) = E_{electr}\,\rho(\mathbf{r}_1)$$

holds, while for ground state the exact

$$E_{electr,0} \leq \int [(1/8)\rho_{0,trial}(\mathbf{r}_1)^{-1}|\nabla_1\rho_{0,trial}(\mathbf{r}_1)|^2 + \rho_{0,trial}(\mathbf{r}_1)v(\mathbf{r}_1)]d\mathbf{r}_1$$

holds for the HK energy functional.

For ground- and excited states as well as HF-SCF ground state one-electron density

$$\int\nabla_1^2\rho(\mathbf{r}_1)d\mathbf{r}_1 = 0$$

holds, useful e.g. in correlation calculation.

# # # # # #

Electron-electron repulsion energy participation in
non-relativistic electronic Schrödinger equation
via the coupling strength parameter (a) in Hartree-Fock theory:

The electronic Hamiltonian can be extended with coupling strength parameter (a) as

$$H(a)y_k(a) \equiv (H_\nabla + H_{ne} + aH_{ee})y_k(a) = enrg_{electr,k}(a)y_k(a)$$

of which only a=1 makes physical sense (for which the familiar notation is $\{enrg_{electr,k},y_k\} = \{E_{electr,k},\Psi_k\}$), and the simplest mathematical case a=0 is

$$(H_\nabla + H_{ne})Y_k = e_{electr,k}Y_k\ ,$$

in which no electron-electron interaction at all, the totally non-interacting reference system (TNRS); the $S_0$ is the HF-SCF/basis/a=1 approximation of $\Psi_0(a=1)$.

The emblematic Hund's rule and virial/ Møller-Plesset/ Hohenberg-Kohn/ Koopmans/ Brillouin theorems as well as the configuration interactions formalism in relation to coupling strength parameter (a) have been generalized.

Relations: $E_{electr,k} = e_{electr,k'} + (N(N-1)/2)\langle\Psi_k|r_{12}^{-1}|Y_{k'}\rangle/\langle\Psi_k|Y_{k'}\rangle$

$$E_{electr,0}= e_{electr,0} + <\Psi_0|H_{ee}|Y_0>/<\Psi_0|Y_0>$$
$$e_{electr,0} << (e_{electr,0}+<\Psi_0|H_{ee}|\Psi_0>) \le E_{electr,0}= (e_{electr,0}+<\Psi_0|H_{ee}|Y_0>/<\Psi_0|Y_0>) \le$$
$$\le (e_{electr,0}+<Y_0|H_{ee}|Y_0>)$$
$$<\Psi_0|H_{ee}|\Psi_0>) \le <\Psi_0|H_{ee}|Y_0>/<\Psi_0|Y_0> \le <Y_0|H_{ee}|Y_0>$$
$$\partial enrg_{electr,0}(a)/\partial a = (N(N-1)/2) <y_0(a)|r_{12}^{-1}|y_0(a)>$$
$$\partial^2 enrg_{electr,0}(a)/\partial a^2 = N(N-1) <y_0(a)|r_{12}^{-1}| \partial y_0(a)/\partial a>$$
$\partial enrg_{electr,0}(a)/\partial a$ is nearly constant

Extension of 1st Hohenberg-Kohn theorem in relation to „a”:

$$\Psi_0(a=1) \Leftrightarrow H_{ne} \Leftrightarrow Y_0(a=0).$$

Generalization of the 1st Hohenberg-Kohn theorem in relation to „a”:

$\rho_0(\mathbf{r}_1,a_1)$ or $y_0(a_1)$ $\Leftrightarrow$ $\rho_0(\mathbf{r}_1,a_2)$ or $y_0(a_2)$,

$\rho_0(\mathbf{r}_1,a=0)$ from $H_\nabla+H_{ne}$ $\Leftrightarrow$ $E_{electr,0}$ from $H_\nabla+H_{ne}+H_{ee}$.

If an r-symmetric w is good enough, $wY_0$ may approach $\Psi_0$ more efficiently than $S_0$ and
$$E_{electr,0}(a)= e_{electr,0} -(N/2)<wY_0|Y_0\nabla_1{}^2 w> -N<wY_0|\nabla_1 Y_0\nabla_1 w> + <wY_0|H_{ee}|wY_0>,$$
solve it for $w(\mathbf{r}_1,...,\mathbf{r}_N)$ or it is the variation equation with a pre-calculated $(Y_0,e_{electr,0})$.

For ground (k=0) and excited (k>0) states, using the nuclear frame generated orto-normalized Slater determinant basis set $\{Y_k\}$ from TNRS (a=0) for different levels of CI calculation, the diagonal elements (k'=k):
$$<Y_k|H_\nabla+H_{ne}+aH_{ee}|Y_k>= e_{electr,k} + a(N(N-1)/2)<Y_k|r_{12}^{-1}|Y_k>,$$
making the link between case a=0 and 1 for ground (k=0) and excited (k>0) states as:
$$E_{electr,k} \approx E_{electr,k}(TNRS) \equiv e_{electr,k} + (N(N-1)/2)<Y_k|r_{12}^{-1}|Y_k>.$$
The off-diagonal elements (k'≠k):
$$<Y_{k'}|H_\nabla+H_{ne}+aH_{ee}|Y_k> = a(N(N-1)/2)<Y_{k'}|r_{12}^{-1}|Y_k> \quad \text{if } k'\neq k,$$
i.e. the off diagonal elements contain the electron-electron interaction only. (If the known/regular HF-SCF/basis/a=1 determinant basis set $\{S_k\}$ is used (generally $\{s_k(a)\}$), the molecular orbital energies ($\varepsilon_i$) show up in the off-diagonal elements.)

TNRS-CI matrix for ground and excited (k≥0) states: diagonal elements are the crude approximations at the same time:
$$E_{electr,k} \approx E_{electr,k}(TNRS) \equiv e_{electr,k}+ (N(N-1)/2)<Y_k|r_{12}^{-1}|Y_k>,$$
the off-diagonal elements
$$<Y_{k'}|H_{ee}|Y_k>.$$

Extension of Brillouin's theorem w/r to coupling strength parameter for HF-SCF/basis/a (which approximates $y_0(a)$ by single determinant $s_0(a)$) is formally the same for a=1 vs. a≠1:
$$<s_0(a)|H_\nabla+H_{ne}+aH_{ee}|s_{0,b}{}^r(a)>= 0.$$
An important consequence of this is that, the $\{Y_0, \{Y_{0,b}{}^r\}\}$ truncated basis set from a=0 (using the minimal, singly-excited ones) can already be used as a basis set to estimate $\Psi_0(a=1)$ better than the (1,1) diagonal element ($E_{electr,0}(TNRS)$), even to estimate $\Psi_1$ also by the eigenvectors of the Hamiltonian matrix (TNRS-CI). This means that, it can provide the large part of correlation energy, and the doubly excited determinants do not have to be used to save computer time and disc space unless one needs more accurate results or higher excited states.

######

Analytic evaluation of Coulomb integrals for one, two and three-electron operators,

$R_{C1}^{-n}R_{D1}^{-m}$, $R_{C1}^{-n}r_{12}^{-m}$ and $r_{12}^{-n}r_{13}^{-m}$ with n, m=0,1,2:

The $\int_{(R3)} \exp(-pR_{P1}^2)R_{C1}^{-n}\, d\mathbf{r}_1 = (2\pi/p)F_0(v)$ if n=1 & $(2\pi^{3/2}/p^{1/2})e^{-v}F_0(-v)$ if n=2, with $v \equiv pR_{CP}^2$ and Boys function ($F_0$), generate <u>analytic expressions for Coulomb integrals</u> with higher distance moment for n, m=0, 1, 2, i.e. for

$$\int\rho(1)R_{C1}^{-n}R_{D1}^{-m}d\mathbf{r}_1,$$
$$\int\rho(1)\rho(2)R_{C1}^{-n}r_{12}^{-m}d\mathbf{r}_1d\mathbf{r}_2,$$
$$\int\rho(1)\rho(2)\rho(3)r_{12}^{-n}r_{13}^{-m}d\mathbf{r}_1d\mathbf{r}_2d\mathbf{r}_3.$$
$$R_{C1}^{-2} = \int_{(-\infty,0)} \exp(R_{C1}^2t)dt = \int_{(0,\infty)} \exp(-R_{C1}^2t)dt \text{ (new trick).}$$

$$V_{P,C}^{(n)} \equiv \int_{(R3)} \exp(-p\, R_{P1}^2)\, R_{C1}^{-n}\, d\mathbf{r}_1$$
$$V_{P,C}^{(2)} = (2\pi^{3/2}/p^{1/2}) \int_{(0,1)} \exp(p\, R_{CP}^2\, (w^2-1))dw = (2\pi^{3/2}/p^{1/2})e^{-v}F_0(-v)$$
$$0 < \exp(-v) < [p^{1/2}/(2\pi^{3/2})]\, V_{P,C}^{(2)} < 1$$

$$V_{P,C}^{(2)}(R_{CP}=0)/\, V_{P,C}^{(1)}(R_{CP}=0)= (2\pi^{3/2}/p^{1/2})/(2\pi/p)= (\pi p)^{1/2}$$

$${}^{full}V_{P,C}^{(2)}= 2\Sigma_1\Gamma_1 D\, p^{-(m1+1)/2} \int_{(0,1)} (w^2-1)^{n1-m1}\, w^{m1}\, \exp(p\, R_{CP}^2(w^2-1))\, dw$$

$${}^{full}V_{P,C}^{(1)}= 2p^{-1}\pi^{-1/2}\Sigma_1\Gamma_1 D\, p^{-m1/2}\int_{(0,1)}(-w^2)^{n1-m1}\, (1-w^2)^{m1/2}\, \exp(-p\, R_{CP}^2\, w^2)\, dw.$$

$$V_{P,CD}^{(n,m)} \equiv \int_{(R3)}\exp(-pR_{P1}^2)R_{C1}^{-n}R_{D1}^{-m}d\mathbf{r}_1$$
$$V_{P,CD}^{(1,2)}= \pi\int_{t=(-\infty,\infty)}\int_{u=(0,\infty)}g^{-3/2}\exp(-f/g)dudt$$
$$g \equiv p + t^2 +u,\ f \equiv p\, t^2\, R_{PC}^2 +p\, u\, R_{PD}^2 +u\, t^2\, R_{CD}^2.$$

$$V_{PQ}^{(n)} \equiv \int_{(R6)} \exp(-p\, R_{P1}^2)\, \exp(-q\, R_{Q2}^2)\, r_{12}^{-n}\, d\mathbf{r}_1d\mathbf{r}_2$$
$$V_{P,C}^{(2)} = \int_{(R3)} \exp(-p\, R_{P1}^2)r_{12}^{-2}\, d\mathbf{r}_1 = (2\pi^{3/2}/p^{1/2}) \int_{(0,1)} \exp(p\, R_{P2}^2\, (w^2-1))dw$$
$$V_{PQ}^{(2)} = 2\pi^3(pq)^{-1/2}(p+q)^{-1}\int_{(0,1)} \exp(v(w^2-1))dw = (2\pi^3(pq)^{-1/2}(p+q)^{-1})e^{-v}F_0(-v)$$
$$0 < \exp(-v) < [(pq)^{1/2}(p+q)/(2\pi^3)]V_{PQ}^{(2)} < 1$$

$$V_{PQ}^{(2)}(R_{PQ}=0)/\, V_{PQ}^{(1)}(R_{PQ}=0)= (2\pi^3(pq)^{-1/2}(p+q)^{-1}/(2c\pi^{5/2}/(pq))= (\pi pq/(p+q))^{1/2}.$$

$$\int_{(R6)}\exp(-pR_{P1}^2)\exp(-qR_{Q2}^2)R_{C1}^{-1}r_{12}^{-1}d\mathbf{r}_1d\mathbf{r}_2=(2\pi^2/q)\int_{u=(0,1)}\int_{t=(-\infty,\infty)} g^{-3/2}\exp(-f/g)\, dtdu$$
$$f \equiv pqR_{PQ}^2u^2+pR_{PC}^2t^2+qR_{QC}^2u^2t^2\ ,\ g \equiv p+qu^2+ t^2, \text{ or}$$
$$\int_{(R6)}\exp(-pR_{P1}^2)\exp(-qR_{Q2}^2)R_{C1}^{-1}r_{12}^{-1}d\mathbf{r}_1d\mathbf{r}_2=(4\pi^2/q)\int_{(0,1)}F_0(gR_{WC}^2)g^{-1}\exp(-f/g)du$$
$$f \equiv pqR_{PQ}^2u^2,\ g \equiv p+qu^2\ .$$

$$V_{PQS}^{(n,m)} \equiv \int_{(R9)} \exp(-p\, R_{P1}^2)\, \exp(-q\, R_{Q2}^2)\, \exp(-s\, R_{S3}^2)\, r_{12}^{-n}r_{13}^{-m}\, d\mathbf{r}_1d\mathbf{r}_2d\mathbf{r}_3$$
$$V_{PQS}^{(1,1)}= (4\pi^{7/2}/(qs))\int_{(0,1)}\int_{(0,1)} g^{-3/2}\exp(-f/g)dudt$$
$$f \equiv pqR_{PQ}^2u^2+psR_{PS}^2t^2+qsR_{QS}^2u^2t^2,\ g \equiv p+qu^2+st^2. \text{ Or}$$
$$V_{PQS}^{(1,1)}= (4\pi^2/(qs))\int_{(R3)} F_0(qR_{Q1}^2)\, F_0(sR_{S1}^2)\, \exp(-pR_{P1}^2)d\mathbf{r}_1, \text{ or}$$
$$V_{PQS}^{(1,1)}= (4\pi^{7/2}/(qs)) \int_{(0,1)}\, h(u)\, g^{-1}\exp(-f/g)\, du$$
$$h(u) \equiv \int_{(0,c)}\exp(-g\, s\, R_{VS}^2\, w^2)dw,\ c \equiv (g+s)^{-1/2},\ f \equiv pqR_{PQ}^2u^2,\ g \equiv p+qu^2.$$

For product of 3 or more Gaussians: $\Sigma_J\, p_J\, R_{J1}^2 = (\Sigma_J\, p_J)\, R_{W1}^2 + (\Sigma_J\Sigma_K\, p_J\, p_K\, R_{JK}^2)/(2\Sigma_J\, p_J)$

$$\mathbf{R}_W \equiv (\Sigma_J\, p_J\, \mathbf{R}_J)/(\Sigma_J\, p_J)$$

$$\text{POLY}(x,P,S,n) \equiv (x-x_P)^n= \Sigma_{i=0 \text{ to } n}\, \binom{n}{i}(x_S-x_P)^{n-i}\, (x-x_S)^i$$

Recursive formula for Boys function: $2vF_{L+1}(v)= (2L+1)F_L(v) - \exp(-v)$.

######

Reformulation of Gaussian error propagation:

$$df = \Sigma_{(i=1\ldots n)}(\partial f/\partial x_i)dx_i + (\partial f/\partial z)dz = \Sigma_{(i=1\ldots n)}(\partial f/\partial x_i)dx_i + (\partial f/\partial z)(\Sigma_{(i=1\ldots n)}(\partial z/\partial x_i)dx_i) =$$

$$= \Sigma_{(i=1\ldots n)}[(\partial f/\partial x_i) + (\partial f/\partial z)(\partial z/\partial x_i)]dx_i$$

$$(\Delta f)^2 = \Sigma_{(i=1\ldots n)}[(\partial f/\partial x_i) + (\partial f/\partial z)(\partial z/\partial x_i)]^2(\Delta x_i)^2$$

$$(\Delta f)^2 = \Sigma_{(i=1\ldots n)}[(\partial f/\partial x_i) + \Sigma_{(j=1\ldots m)}(\partial f/\partial z_j)(\partial z_j/\partial x_i)]^2(\Delta x_i)^2$$

######

Role of the surface free enthalpy excess of solid chemical elements in their melting and critical temperature:

"Free enthalpy excess of the surface" of solid chemical elements: $\gamma$[J/mol]=$\alpha$ $\Delta H'$, $\Delta H'$= internal enthalpy (heat) of atomization, <u>$\alpha$ changes very slightly with temperature, but its change is fundamental in phase transitions</u>, $\alpha \approx (z- n_{avrg})/z$, z= bulk effective coordination number, mainly the 1st, or by any chance 2nd, 3rd, etc. nearest neighbors, as well as $0.2<\alpha<0.3$:

$$\partial\alpha/\partial T = -R \ln(m)/\Delta H'$$

$$\partial\alpha/\partial T \approx -5x10^{-5} K^{-1} \text{ for metals}$$

$$-0.1 < \Delta\alpha_{cr} < -0.055 \text{ for metals}$$

$$T_m = |\Delta\alpha_{cr}|\Delta H'/(R \ln(m)) = \Delta n_{cr}\Delta H'/(z R \ln(m))$$

$$-\Delta\alpha_{cr} \approx 1 - (\rho_{liq}(T_m)/\rho_{sol})$$

$$\Delta H_m= T_m R \ln(m) \approx -T_m (\partial\gamma/\partial T)_{avrg}$$

$$\Delta S_m= \Delta H_{m,calc}/T_m = R \ln(m) \approx -(\partial\gamma/\partial T)_{avrg}$$

$$T_c= \alpha(T=0K) \Delta H'/R \ln(m) = \gamma/R \ln(m) \approx -\gamma/(\partial\gamma/\partial T)_{avrg}$$

$$T_{c1}/T_{c2} \approx \Delta H'_1/ \Delta H'_2 \approx T_{m1}/T_{m2}$$

$$T_c/T_m \approx \alpha(T=0K)/|\Delta\alpha_{cr}| \approx 3.6$$

$$T_c/T_m \approx \gamma/\Delta H_m$$

$$\alpha(T=0K) \approx RT_c \ln(5)/\lambda ,$$

$$T_c/T_m \approx 2.5 \text{ (critical and melting T of metals)}$$

$$\lambda \approx 8 a \ln(m)/(27 b\, \alpha(T=0K)) \approx 2a/b \text{ (gas law parameters)}$$

######

Quaternionic Treatment of the Electromagnetic Wave Equation:

$$\nabla * E = -\rho/\varepsilon - \partial B/\partial t,$$

$$\nabla * B = \mu J + \mu\varepsilon\partial E/\partial t,$$

$$w= -(1/2)\varepsilon E*E \text{ and } -(1/2)\mu^{-1}B*B$$

######

Generalization of Savitzky-Golay parameters to least-square smoothing and differentiation of two-dimensional data:

The two dimension version has been worked out (extended), for example, the 2nd degree 2 dimension smoothing parameter sets for polynomial $f(x,y)= a_0+a_1x+a_2y+a_3x^2+a_4xy+a_5y^2$ are

```
to smooth:        -13   2   7   2 -13
                    2  17  22  17   2
                    7  22  27  22   7      N= 175
                    2  17  22  17   2
                  -13   2   7   2 -13,
to calc. ∂²f/∂x∂y: -4  -2   0   2   4
                   -2  -1   0   1   2
                    0   0   0   0   0      N= 100
                    2   1   0  -1  -2
                    4   2   0  -2  -4, etc.
```

# # # # # #

Kinetics of heterogeneous catalytic hydrogenolysis of ethane:

$C_2H_6$ + (7-m)* $\rightarrow$ $K_E$ (fast )$\leftarrow$ $C_2H_m$* + (6-m)H*      with $K_E = \theta_m \theta_H^{6-m}/(p_E \theta_0^{7-m})$

$H_2$ +2* $\rightarrow$ $K_H$ (fast) $\leftarrow$ 2H*      with $K_H = \theta_H^2/(p_H \theta_0^2)$

$C_2H_m$* + B –k (irrev., slow)$\rightarrow$ $CH_u$*+$CH_v$* -(H* or $H_2$, fast)$\rightarrow$ $CH_4$(g)

rate = k $\theta_m$ B and B= * or H* or $H_2$(g).

For $CH_{u \text{ or } v}$* the $\theta_i \approx 0 \Rightarrow \theta_0+\theta_H+\theta_m \approx 1$, using $y \equiv K_E\, p_E$, $x \equiv (K_H\, p_H)^{1/2}$, $G_6 \equiv x^{6-m}$, $G_7 \equiv x^{7-m}$ and $D \equiv y+G_6+G_7 \Rightarrow \theta_0 = G_6/D$, $\theta_m = y/D$ and $\theta_H = G_7/D \Rightarrow$ the mathematically possible rates are

rate = k y $G_6/D^2$ if B= $\theta_0$

rate = k y $G_7/D^2$ if B= $\theta_H$

rate = k y $p_H/D$ if B= $p_H$.

Experiment vs. theory $\Rightarrow$ $H_2$(g) is responsible for the C-C rupture on Ni and Pd catalysts.

# # # # # #

A hobby theme: On the statistical distribution of prime numbers,
a view from where the distribution of prime numbers is not erratic:
The m= 4ab+2(a+b)+1 jumps over all odd primes (except 2) and makes multiple hits on all odd composite (non-prime) numbers for a,b=1,2,…,
while if zero is also allowed for a and b, it generates all odd numbers.

# 2. Examination of the form of Schrödinger equation (eigenvalue equations in focus)

### 2.1. Multi-electron densities "between" the Hohenberg-Kohn theorems and variational principle

#### 2.1. Preliminary

The properties of multi-electron densities are analyzed with respect to the two Hohenberg –Kohn theorems, and the fundamental extensions are established of the latter. This analysis is continued with the form of density functionals and density differential operators on different levels of multi-electron densities and the connection between the variational principle and Hohenberg-Kohn theorems. The trend in ionization potentials is commented upon. The exact density functional operator of H-like atoms and 1-electron systems is also discussed with the 2-electron systems, the latter only on a Slater determinant level; these are the prototypes of the two-electron density based solutions or approximations. A nice perspective on the field is tried to be given, which contains a rather comprehensive and lucid review of the literature.

#### 2.1.1. Introduction

The non-relativistic Schrödinger equation is capable of describing the nuclear and electronic motion of molecular systems. This equation is one of the most important known fundamental equations governing our material world. Two shorthand notations are used next: VP for variational principle and $1^{st}$ and $2^{nd}$ HK for the Hohenberg-Kohn theorems. To introduce the properties indicated in the title, based on mathematical physics, a few textbook definitions are included for a complete description. The Born – Oppenheimer approximation separates [2.1.1, 2] the two types of motion based on the huge mass difference between electrons and nuclei. After separation, the nuclei can be considered as "moving on the potential electronic surface" defined by the stationary motion of electrons on the "fixed" nuclear frame at all possible nuclear configurations. For physically or chemically important systems one must solve this equation to predict its properties (e.g. chemical rates and equilibriums or energetics). Before studying the nuclear motion, the electronic potential energy surface, $E_{total\ electr}$, (mostly for ground state) must be mapped as a function of nuclear coordinates. Most of the chemical properties can be deduced by calculating $E_{total\ electr}$. However, the gradient vector $\{\partial E_{total\ electr}/\partial R_{Au}\}$ and Hessian matrix $\{\partial^2 E_{total\ electr}/(\partial R_{Au}\partial R_{Bv})\}$ are both fundamental in finding stationary points (i.e. minimums or equilibrium geometries and transition states) on electronic potential energy surfaces as well as calculating vibration frequencies. Below, $R_{Au}$ notates the u = x, y or z Cartesian coordinate of the $A^{th}$ nucleus with nuclear charges $Z_A$ in the molecule or system (A = 1,2,…,M) containing N electrons and M atoms. The non-stationary regions (the range of van der Waals distances or London dispersion forces) are also important in many areas, such as chemical kinetics, liquid structures and, adhesive forces (e.g. in biochemistry, or heterogeneous catalysis, etc.) Studying crystals, solid states or heterogeneous catalytic problems, metal atoms must be included. In the latter case, problems of convergence, basis set, size and inclusion of relativistic effects, etc. appear currently. Below, the $E_{total\ electr}$ notates the non - relativistic total electronic energy of the molecule at a fixed nuclear geometry. This includes the nuclear repulsion terms although the $E_{electr}$ notes it without nuclear repulsion. For the N electronic spin-orbit variables or coordinates we use $\mathbf{x}_i = (\mathbf{r}_i,s_i) = (x_i, y_i, z_i, s_i)$. Having the potential surface, $E_{total\ electr}(\mathbf{R}_1, \mathbf{R}_2,...,\mathbf{R}_M, Z_1,Z_2,...,Z_M,N)$, available wherein the boundary conditions are periodic or (at infinitely far away) zero [2.1.3], one can calculate elementary rates solving the nuclear Schrödinger equation [2.1.4]. The latter is more difficult, seeing that, mathematically, 1.: the explicit Coulomb potentials are replaced by the generally not explicit, or approximate analytical potential energy surface, 2.: boundary conditions are the more difficult nuclear vibronic eigenfunctions.

The known form of non - relativistic, spinless, fixed nuclear coordinate (Born - Oppenheimer approximation) electronic Schrödinger equation for a molecular system containing M atoms and N electrons with nuclear configuration $\{\mathbf{R}_A, Z_A\}_{A=1,...,M}$ in free space is

$$(-(1/2)\Sigma_{i=1,...,N}\nabla_i^2 - \Sigma_{i=1,...,N}\Sigma_{A=1,...,M}Z_A R_{Ai}^{-1} + \Sigma_{i=1,...,N} \Sigma_{j=i+1,...,N} r_{ij}^{-1})\Psi = E_{electr}\Psi \quad \text{(Eq.2.1.1)}$$

where $\Psi$ and $E_{electr}$ are the $i^{th}$ excited state (i=0,1,2,...) anti-symmetric wavefunction (with respect to all spin-orbit electronic coordinates $\mathbf{x}_i$) and electronic energy respectively. (The notation $R_{Ai} \equiv |\mathbf{R}_A-\mathbf{r}_i|$ and $r_{ij} \equiv |\mathbf{r}_i-\mathbf{r}_j|$ are used.) Below, we will discuss in particular the ground state. To call attention to this, we have used the notation $\Psi_0$ and $E_{electr,0}$ as, otherwise these could mean any of the $i^{th}$ excited states, including the ground state. The electronic energy, $E_{electr}$, parametrically depends on all $\mathbf{R}_A$ and $Z_A$, while $\Psi$ depends on these and on all $\mathbf{x}_i$. (The Hamilton operator in Eq.2.1.1 has no spin variable, as well as symmetric with respect to the exchange of any $\mathbf{r}_i$ and $\mathbf{r}_j$.) The total electronic energy is

$$E_{total\ electr} = E_{electr} + \Sigma_{A=1,...,M} \Sigma_{B=A+1,...,M} Z_A Z_B/|\mathbf{R}_A-\mathbf{R}_B| \quad \text{(Eq.2.1.2)}$$

where the double sum is the nuclear-nuclear repulsion (notation $V_{nn}$ is used below for it). The $V_{nn}$ can be calculated after solving Eq.2.1.1 since it does not contain electron spatial coordinates $\mathbf{r}_i = (x_i,y_i,z_i)$ for the i=1,2,3,…N electrons. The electronic Hamiltonian operator in free space on the left-hand side of Eq.2.1.1 contains the three known operators. In atomic units these are the kinetic energy $H_\nabla = -(1/2)\Sigma_{i=1,...,N}\nabla_i^2$, the nuclear – electron attraction $H_{Rr} = -\Sigma_{i=1,...,N}\Sigma_{A=1,...,M}Z_A R_{Ai}^{-1}$, and the electron – electron repulsion $H_{rr} = \Sigma_{i=1,...,N} \Sigma_{j=i+1,...,N} r_{ij}^{-1}$ operators. These are spinless operators; spin coordinates are introduced in $\Psi$ only. (The use of spin in this way is a handicap being Eq.2.1.1 non-relativistic. If atoms with $Z_A > 16$ appear in the molecule, additional relativistic correction is absolutely required for adequate accuracy.) If external forces apply, its operator may contain spin coordinates, however we have not consider those cases in this work, although in many cases the extension is straightforward.

The "well behaving" property (that is $\Psi$(at least one component of an $\mathbf{r}_i \to \pm\infty$)=0 and the integral of $\Psi$ over the 3N dimensional real space is finite) is required as boundary condition. $\Psi$ is normalized in cases for convenience.

### 2.1.2. Properties of the multi-electron densities "between" the two Hohenberg-Kohn theorems and variational principle

Similar to the diagonal elements of "spinless density matrices" [2.1.2] (the difference here relates to a constant factor), let us define the multi-electron densities $b_1,…,b_N$ (for any $i^{th}$ excited state i=0,1,2,…) as

$$|\Psi(\mathbf{x}_1,\mathbf{x}_2,\mathbf{x}_3,...,\mathbf{x}_N)|^2 = \Psi^* \Psi \quad \text{(Eq.2.1.3)}$$

$$b_N(\mathbf{r}_1,\mathbf{r}_2, \mathbf{r}_3…, \mathbf{r}_{N-1},\mathbf{r}_N) = \int\Psi^* \Psi\, ds_1ds_2\, ds_3… ds_{N-1}ds_N \quad \text{(Eq.2.1.4)}$$

$$b_{N-1}(\mathbf{r}_1, \mathbf{r}_2, \mathbf{r}_3…, \mathbf{r}_{N-1}) = \int\Psi^* \Psi\, ds_1ds_2ds_3… ds_{N-1}d\mathbf{x}_N \quad \text{(Eq.2.1.5)}$$

$$b_2(\mathbf{r}_1, \mathbf{r}_2) = \int\Psi^* \Psi\, ds_1ds_2d\mathbf{x}_3… d\mathbf{x}_{N-1}d\mathbf{x}_N \quad \text{(Eq.2.1.6)}$$

$$\rho(\mathbf{r}_1) \equiv b_1(\mathbf{r}_1) = \int\Psi^* \Psi\, ds_1d\mathbf{x}_2d\mathbf{x}_3… d\mathbf{x}_{N-1}d\mathbf{x}_N \quad \text{(Eq.2.1.7)}$$

where the * marks the complex conjugate. Quantity in Eq.2.1.3 is the known probability distribution associated with a solution of Schrödinger equation in Eq.2.1.1, Eq.2.1.6 is for the geminals (two particle orbitals), and Eq.2.1.7 defines the known one-electron density. For example, the second-order reduced density matrix is defined in the literature as $n_2(\mathbf{x}_1,\mathbf{x}_2;\mathbf{x'}_1,\mathbf{x'}_2) = (N(N-1)/2) \int\Psi^*(\mathbf{x'}_1,\mathbf{x'}_2,\mathbf{x}_3,...,\mathbf{x}_N)\Psi(\mathbf{x}_1,\mathbf{x}_2,\mathbf{x}_3,...,\mathbf{x}_N)d\mathbf{x}_3...d\mathbf{x}_N$ and from this, the diagonal of the spin-independent second-order density matrix is $n_2(\mathbf{r}_1,\mathbf{r}_2) = \int n_2(\mathbf{x}_1,\mathbf{x}_2;\mathbf{x}_1,\mathbf{x}_2)ds_1ds_2 = (N(N-1)/2)\, b_2(\mathbf{r}_1, \mathbf{r}_2)$; where the last part of the latter expression shows the simple connection to the definitions in Eqs.2.1.3-7. The insights into many-electron densities are not new, it starts with Ziesche's work in 1994 [2.1.5, 6], and there has been a wealth of work on this topic. The early work focused on the "Hohenberg-Kohn theorems" [2.1.5, 6] and later work concentrated on kinetic energy functionals and variational procedures [2.1.7-15]. Now we have the finite series of N+2 functions:

$$\{b_{N+2}\equiv\Psi(\mathbf{x}_1,\mathbf{x}_2,\mathbf{x}_3,...,\mathbf{x}_N),\ b_{N+1}\equiv|\Psi(\mathbf{x}_1,\mathbf{x}_2,\mathbf{x}_3,...,\mathbf{x}_N)|^2,\ b_N(\mathbf{r}_1,\mathbf{r}_2, \mathbf{r}_3..., \mathbf{r}_{N-1},\mathbf{r}_N),\ b_{N-1},\ b_{N-2}, ..., b_3,\ b_2(\mathbf{r}_1, \mathbf{r}_2),\ \rho(\mathbf{r}_1) \equiv b_1(\mathbf{r}_1)\} \quad \text{(Eq.2.1.8)}$$

Let us call "x-anti-symmetric" a function if it is anti-symmetric for the exchange of any $\mathbf{x}_i$ and $\mathbf{x}_j$ and "r-anti-symmetric" if it is anti-symmetric for any $\mathbf{r}_i$ and $\mathbf{r}_j$. Similar definitions are used for "x-symmetric" and "r-symmetric". The series in Eq.2.1.8 has the following trivial properties: The first element (the wave function itself) is x-anti-symmetric, the second (probability distribution function) is x-symmetric and the rest (the multi-electron densities down to $b_2$) are r-symmetric. The first element can have an alternate sign (e.g. the $p_x$ orbital of H-like atoms), while the rest are always greater or equal to zero in the entire 3(N-i) dimensional real space (i= -1,0,1,2,…,N-1). The first element ($\Psi$) is, of course, the "generator element" of the series. The first element ($\Psi$) is used in the VP, while the last one ($\rho(\mathbf{r}_1)$) is used in the two HK theorems. The property is such that

$$\int b_{i+1}d\mathbf{r}_{i+1} = b_i \quad i=1,2,3,...,N-1 \quad \text{(Eq.2.1.9)}$$

can be seen immediately.

In density functional theory (DFT) the known normalization condition is

$$\int\rho(\mathbf{r}_1)d\mathbf{r}_1 = N \quad \text{(Eq.2.1.10)}$$

where the domain of the integral is the entire three dimensional Cartesian space (keeping the index 1 on the electron spatial coordinate). Eq.2.1.10 can be enforced on $\rho$ because if $\Psi$ is a solution, then const.$\Psi$ is also a solution, and $\Psi$ is "well behaving". Ground state will be marked with 0 in the index as $\rho_0$ if necessary. For normalization of $\Psi$, generally the unity is chosen. Multiplying Eq.2.1.1 by $\Psi^*$ from the left, and integrating both sides over the N spin-orbit coordinates ($\int...d\mathbf{x}_1d\mathbf{x}_2d\mathbf{x}_3...d\mathbf{x}_{N-1}d\mathbf{x}_N$) and using the bra-ket notation:

$$E_{electr} = \langle\Psi|(H_\nabla+H_{Rr}+H_{rr})\Psi\rangle / \langle\Psi|\Psi\rangle = (1/N)\langle\Psi|(H_\nabla+H_{Rr}+H_{rr})\Psi\rangle \quad \text{(Eq.2.1.11)}$$

where in the far right side the N normalization of $\Psi$ is supposed. Eq.2.1.11 is the known equation for the VP [2.1.1]: For any well behaving anti-symmetric N-electron $\Psi_{trial}$, which is normalized to N

$$E_{electr,0} \leq (1/N)\langle\Psi_{trial}|(H_\nabla+H_{Rr}+H_{rr})\Psi_{trial}\rangle \equiv E_{electr}[\Psi_{trial}] \quad \text{(Eq.2.1.12)}$$

is hold, and the equality is hold for the ground state $\Psi_0$. In other words, the functional $E_{electr}[\Psi_{trial}]$ must be fully minimized with respect to all allowed N-electron wave functions $\Psi_{trial}$ to obtain the true ground state wave function $\Psi_0$ and ground state energy $E_{electr,0} = E_{electr}[\Psi_0]$. In the literature, Eq.2.1.12 is stated in the form wherein $\Psi$ is normalized to 1, and in that case there is no 1/N factor, however, it is only a technical manner. The reason it must be mentioned is that we will consider next the related properties for the series in Eq.2.1.8, and must be careful with the normalization. It must also be mentioned that the variational property in Eq.2.1.12 comes from a similar general property of general linear partial differential equations (more generally speaking: linear operators) [2.1.1], not a particular property of the Schrödinger equation.

Multiplying Eq.2.1.1 by $\Psi^*$ from the left, using the definition of $\rho(\mathbf{r}_1)$ and the anti-symmetric property of $\Psi$, and integrating over all N spin-orbit coordinates on the entire 3N dimensional real and N dimensional spin space:

$$-(N/2)\int(\int\Psi^*\nabla_1^2\Psi ds_1 d\mathbf{x}_2\ldots d\mathbf{x}_N)d\mathbf{r}_1 - N\Sigma_{A=1,\ldots,M}Z_A\int\rho(\mathbf{r}_1)R_{A1}^{-1}d\mathbf{r}_1 + $$
$$+(N(N-1)/2)\int(\int\Psi^*\Psi r_{12}^{-1} ds_1 ds_2 d\mathbf{x}_3\ldots d\mathbf{x}_N)d\mathbf{r}_1 d\mathbf{r}_2) = E_{electr}\int\rho(\mathbf{r}_1)d\mathbf{r}_1 \quad \text{(Eq.2.1.13)}$$

With the normalization in Eq.2.1.10 one yields the energy functional (for later discussion, not using the cancellation for N yet):

$$E_{electr}[\rho] = (1/N)[-(N/2)\int(\int\Psi^*\nabla_1^2\Psi ds_1 d\mathbf{x}_2\ldots d\mathbf{x}_N)d\mathbf{r}_1 - N\Sigma_{A=1,\ldots,M}Z_A\int\rho(\mathbf{r}_1)R_{A1}^{-1}d\mathbf{r}_1 + $$
$$+(N(N-1)/2)\int(\int\Psi^*\Psi r_{12}^{-1} ds_1 ds_2 d\mathbf{x}_3\ldots d\mathbf{x}_N)d\mathbf{r}_1 d\mathbf{r}_2] \quad \text{(Eq.2.1.14)}$$

The second nuclear-electron attraction term in Eq.2.1.14 is a functional which is expressed compactly, shortly and totally accurately with $\rho(\mathbf{r}_1)$. The first kinetic-, and third electron-electron repulsion terms are not expressed so easily with $\rho(\mathbf{r}_1)$. However, for example, the gradient, $\partial E_{electr}/\partial R_{au}$, has also a simple exact DFT functional [2.1.2, 16] with respect to $\rho$, stated in the Hellmann-Feynmann theorem. The latter functional for gradient is very sensitive if e.g. an approximate density (e.g. $\rho_{0,HF\text{-}SCF}$) is used instead of the accurate one. Trick must be employed [2.1.17] to get the required accuracy for predicting molecular configuration. (For example, in the case of chemical shift calculation [2.1.18], one needs very accurate molecular geometry.) The short-hand notation of Eq.2.1.14 is

$$E_{electr}[\rho] = E_{v,1}[\rho(\mathbf{r}_1)] \quad \text{or} \quad E_{v,1}[b_1(\mathbf{r}_1)] \quad \text{(Eq.2.1.15)}$$

where the $E_{v,1}$ is the universal DFT energy functional used in the 2nd HK theorem [2.1.2, 19]. The index v in $E_{v,1}$ indicates [2.1.2] that in Eq.2.1.1 the Hamiltonian is characterized by the nuclear frame that defines the molecular system by knowing how many electrons it has. The index 1 in $E_{v,1}$ indicates that the function variable is on $b_1$ level in that functional. The origin of the two expressions on the right of Eq.2.1.15 simply comes from the two notations in Eq.2.1.7. Eq.2.1.14 shows that $E_{v,1}$ in Eq.2.1.15 is the sum of three terms (kinetic-, nuclear-electron attraction-, and electron-electron repulsion terms) similar to Eq.2.1.12.

The 2nd HK theorem states [2.1.2, 19] the energy VP for a trial one-electron density $\rho_{trial}(\mathbf{r}_1)$: For any $\rho_{trial}(\mathbf{r}_1)$ which $\rho_{trial}(\mathbf{r}_1) \geq 0$ at any $\mathbf{r}_1$ real space point and normalized to N (as in Eq.2.1.10) the

$$E_{electr,0} \leq E_{v,1}[\rho_{trial}(\mathbf{r}_1)] \quad \text{or} \quad E_{v,1}[b_{1,trial}] \quad \text{(Eq.2.1.16)}$$

is hold and the equality is hold for the ground state $\rho_0 \equiv b_{1,0}$. In other words, the functional $E_{v,1}[\rho_{trial}]$ must be fully minimized with respect to all allowed one-electron densities $\rho$ to obtain the true ground state one-electron density $\rho_0$ and ground state energy $E_{electr,0} = E_{v,1}[\rho_0]$. The $E_{v,1}$ functional is unique [2.1.2] and determined by Eq.2.1.14. It must be derived, a difficult task and see the notes in the next two chapters about this. It is analogous to the VP for wave functions in Eq.2.1.12. Similarly to Eq.2.1.15, the Eq.2.1.12 can be re-notated as

$$E_{electr,0} \leq E_{v,N+2}[b_{N+2,trial}] \quad \text{(Eq.2.1.17)}$$

using the definitions in Eq.2.1.8. For Eq.2.1.17, the unique functional is in the middle part of Eq.2.1.12, containing the Hamiltonian in Eq.2.1.1 (or Eq.2.1.14 with 7). Now the functional form is simply expressed via Eq.2.1.1.

Going back to index v (indication of external potential), the literature [2.1.2] uses the notation

$$v(\mathbf{r}_i) \equiv -\Sigma_{A=1,\ldots,M}Z_A R_{Ai}^{-1} \quad \text{(Eq.2.1.18)}$$

The Hamiltonian operator contains it in the $H_{Rr} = -\Sigma_{i=1,\dots,N}\Sigma_{A=1,\dots,M}Z_A\ R_{Ai}^{-1} = \Sigma_{i=1,\dots,N}v(\mathbf{r}_i)$ in Eq.2.1.1. The 1st HK theorem states [2.1.2, 19] that the external potential $v(\mathbf{r}_1)$ is determined, within a trivial additive constant, by the one-electron density $\rho(\mathbf{r}_1)$. Furthermore, $\rho$ determines N via Eq.2.1.10 as well => $\rho_0$ determines $\Psi_0$, and all other electronic properties of the system. The $v(\mathbf{r}_i)$ is not restricted to Coulomb potentials. It should be noticed that the simpler $v(\mathbf{r}_1)$ is mentioned in the 1st HK theorem, not the $H_{Rr} = \Sigma_{i=1,\dots,N}v(\mathbf{r}_i)$ appearing in Eq.2.1.1.

Of course, all $E_{v,i}$ functionals contain 3 groups of terms for i=1,2,…,N+2, as in Eq.2.1.12 and Eq.2.1.14. Continuing our discussion, the i=N+1 case ($b_{N+1}\equiv|\Psi|^2$) is a bit different from i=1,2,3,…,N cases in Eqs.2.1.3-8, as the former contains spin coordinates while the latter cases do not. Let us use the common notation t, $v_{en}$, and $v_{ee}$ for $<\Psi|\Psi>=N$ normalized wave functions as

$$E_{v,i}[b_i] = <\Psi|H_\nabla\Psi>+<\Psi|H_{Rr}\Psi>+<\Psi|H_{rr}\Psi> \equiv$$
$$\int t[b_i]+\int v_{en}[b_i]+\int v_{ee}[b_i] \equiv T+V_{en}+V_{ee} \qquad \text{(Eq.2.1.19)}$$

where the three terms are obviously the kinetic-, electron-nuclear attraction-, and electron-electron repulsion energy terms respectively. The integration in Eq.2.1.19 is as required by sense via Eqs.2.1.3-7, i.e. over $d\mathbf{r}_1 d\mathbf{r}_2\dots d\mathbf{r}_i$ for i=1,2,…,N and for all spin-orbit coordinates ($d\mathbf{x}_1 d\mathbf{x}_2\dots d\mathbf{x}_N$) integration for i=N+1 and N+2. The normalization $\int b_i d\mathbf{r}_1 d\mathbf{r}_2\dots d\mathbf{r}_i= N$ is also necessary for i=1,2,…,N and $\int b_{N+1}d\mathbf{x}_1 d\mathbf{x}_2\dots d\mathbf{x}_N= N$ for i=N+1 and N+2. For the latter, notice that in Eq.2.1.8 i=N+2 contains only $\Psi$, while in all other terms the core contains $\Psi^*\Psi$. This stems from the fact that on the N+2 level, the kinetic functional is $T= F_{kin}[b_{N+2}\equiv\Psi]= <\Psi|H_\nabla\Psi>$. Even this can not be easily reduced to an (N+1)th level $T= F_{kin}[b_{N+1}\equiv|\Psi|^2]$, i.e. finding the analytical form of $F_{kin}$ on this (N+1)th level which is expressed as a function of $b_{N+1}$. An exemption for this difficult task is described for H-like atoms and one-electron systems below.

In the case of i=N+2 the three energy functionals are expressed with $b_{N+2}\equiv\Psi$ as in Eq.2.1.12 (see explicitly with the help of Eq.2.1.1). The advantage of this form is the linearity (nonlinear differential equations have more complexities). Beside the many numerical complexities to solve the electronic structure and nuclear motion problem documented in the literature, generally speaking the anti-symmetric property (what i = N+2 case has) makes more difficulties than a symmetric property. In the case of i=1, the 4N dimensional wave function ($\Psi$) was reduced to a three-dimensional spatial variable one-electron density ($\rho(\mathbf{r}_1)$) and the anti-symmetric property need not be included any more (that has been absorbed in the form of energy functionals). However, the heavy price for this reduction is the lost of linearity in the full energy functional, as well as the difficult and not thoroughly adequate approximations of t and $v_{ee}$ functionals known now.

For cases i=1,2,…,N+1, the integrand of kinetic functional, $t[b_i]$, is difficult to obtain, and all are nonlinear functions containing differential operators, which is disadvantageous, however, these $b_i$ are x-symmetric for i=N+1 and r-symmetric for i=2,…,N. For i=1 ($\rho$) the algebraic r-symmetry property is meaningless; (its approximation for T is fundamental in the practice). However, the geometric symmetry of it in three-dimensional space, of course, exists and is the subject of chemical group theory [2.1.20]. For i=2 see ref. [2.1.8, 15]; it is very easy to construct a kinetic energy functional T for i = N+1 and also for i = N. It is trivial from integration by parts to show that $T(i=N) = (1/2)\Sigma_{k=1toN}<\Psi\nabla_k^2\Psi> = N<\Psi\nabla_1^2\Psi/2> = N<|\nabla_1(\Psi^2)|^2/(8\Psi^2)>$. This is true for any real wavefunction, and so it is true for any stationary state of a molecule (or any other system) as long as $v(\mathbf{r}_1)$ is real (because then the wavefunction can be chosen to be real). This is true, in particular, for Eq.2.1.18, but in practice it seems true for any electronic system in the absence of magnetic fields.

In contrary to kinetic operators, the cases i=1,2,…, N+1,N+2 yield easy and simple nuclear-electron attraction functional, as the x-anti-symmetric property of generator function $\Psi$ and the one-electron variable property in Eq.2.1.18 particularly allow expressing these as

$$v_{en}[b_i] = -\Sigma_{A=1,\dots,M} Z_A\, b_i\, R_{A1}^{-1} = b_i\, v(\mathbf{r}_1) \qquad \text{(Eq.2.1.20)}$$

as was shown already above in Eq.2.1.14 for i=1, but it can be recognized in Eq.2.1.12 as well for i=N+2. (Again, for i=N+2, not $b_{N+2} = \Psi$, but $\Psi^*\Psi$ comes up ($v_{en}[b_{N+2}] = b_{N+1}\, v(\mathbf{r}_1)$) – see note above for series in Eq.2.1.8.)

For cases i=2,…,N+1,N+2, the electron-electron repulsion functional is also simple, because the x-anti-symmetric $\Psi$ again allows the simple form

$$v_{ee}[b_i] = ((N-1)/2)\, b_i\, r_{12}^{-1} \qquad \text{(Eq.2.1.21)}$$

as was shown already above in Eq.2.1.14 for i=2. It can be recognized in Eq.2.1.12 as well for i=N+2. (Technically, again, $v_{ee}[b_{N+2}] = ((N-1)/2)\, b_{N+1}\, r_{12}^{-1}$.) One can realize that the simple forms in Eqs.2.1.20-21 come from 1a: $V_{en}$ depends on one-electron operator ($R_{Ai}$), 1b: $V_{ee}$ depends on two-electron operator ($r_{ij}$), (and the latter is the reason Eq.2.1.21 does not hold for i=1), as well as 2: the $\Psi$ is x-anti-symmetric. (An x-symmetric property would allow the same mathematical simplification as well. Now it is discussed for anti-symmetric (fermions, like electron) property $\Psi$, but true for hypothetic symmetric (boson) property $\Psi$ functions as well.) To prove Eqs.2.1.20-21 generally is as easy as the case of Eq.2.1.12 (i=N+2) and Eq.2.1.14 (i=1 and 2 together), the two far side terms ($\Psi$ and $\rho$ or $b_2$ respectively) of series in Eq.2.1.8: It is just the matter of where the right bracket is placed in the integrand of the last two terms in the right hand side of Eq.2.1.14. An interesting property is that $t[b_i]$ and $v_{ee}[b_i]$ can be expressed with $b_i$ only and $v(\mathbf{r}_1)$ is not necessary in the algebraic expression but, of course, $b_i$ and $v(\mathbf{r}_1)$ mutually determine each other (see theorem below). Eq.2.1.21 shows this, and for case i=1 ($\rho$) see Eq.2.1.26 and Eq.2.1.29 below for ground states as examples for approximations. It is known for the $\rho(\mathbf{r}_1)$ case (i=1), but in an indirect way it can be proven with the help of Eq.2.1.9. For $v_{en}[b_i]$ both, $b_i$ and $v(\mathbf{r}_1)$ are necessary explicitly in the expression as shown in Eq.2.1.20. Physically this is due to the fact that $t[b_i]$ and $v_{ee}[b_i]$ are the motion of the electron itself and the Coulomb interaction between two electrons respectively, while $v_{en}[b_i]$ is a Coulomb interaction between electron and nuclei. Mathematically, it comes easily from Eq.2.1.19, such as among the three integrands, only the $H_{Rr}$ contains $v(\mathbf{r}_i)$. Another important thing is that in Eq.2.1.19 the $t[b_i]$ is symbolic only: Unlike Eqs.2.1.20-21 wherein $b_i$ is in the argument, in Eq.2.1.19 the form of t changes with indices i. For i=1 the weak Thomas –Fermi approximation, $t_1[b_1 \equiv \rho] \sim \rho^{5/3}$, holds (see Eq.2.1.26 below), while for i=N+2 the exact $t_{N+2}[b_{N+2}] = -(1/2)b_{N+2}^*\nabla_1^2 b_{N+2}$ holds.

Ayers and Levy [2.1.21, 22] have pointed out that as for one-electron density ($\rho$ or $b_1$) in the HK theorem the multi electron densities ($b_i$, more, reduced density matrices) have similar properties. They also discuss [2.1.21-23] the N-representability (and v-representability) problem: The difficulty of the N-representability problem for a general $b_i$ (for $1 < i < N+2$) can be "converted" into a functional approximation problem, similarly to the great success in density-functional theory (using the level of $\rho \equiv b_1$), where there is no N-representability problem but, instead, one needs to approximate the exact energy functionals. (Since the variational principle for the wave function or the N-electron density matrix is restricted to antisymmetric and normalized wave functions, the variational principle for $b_i$ must be restricted to those densities that can be expressed in the form of Eqs.2.1.3-7. Such $b_i$ are said to be N-representable. Notice that Eq.2.1.9 always makes a connection between levels i and i+1 if an approximation is derived for level i.) Following theorems are the generalization of the two HK theorems on the ground of functional analysis above for $b_i$, inspired from Eqs.2.1.16-17. It states: With the definition of elements of series in Eq.2.1.8

1: Generalization of 1st HK theorem: The external potential $v(\mathbf{r}_1)$ is determined, within a trivial additive constant, by the multi-electron density $b_i$ for i=1,2,3,…,N+2. (The 1st HK theorem states it for i=1.)

2: Generalization of 2nd HK theorem: Similar VP holds as in Eqs.2.1.16-17, namely

$$E_{electr,0} \le E_{v,i}[b_{i,trial}] \quad \text{for } i=1,2,3,\ldots,N+1 \qquad \text{(Eq.2.1.22)}$$

if the trial i-electron density $b_{i,trial}$, is $b_{i,trial} \ge 0$ at any $(\mathbf{r}_1,\mathbf{r}_2, \mathbf{r}_3\ldots,\mathbf{r}_i)$ 3i-dimensional real space point and normalized to N (analogously to Eq.2.1.10 it is: $\int b_{i,trial}(\mathbf{r}_1,\mathbf{r}_2, \mathbf{r}_3\ldots,\mathbf{r}_i)d\mathbf{r}_1 d\mathbf{r}_2 d\mathbf{r}_3\ldots d\mathbf{r}_i = N$) and $E_{v,i}[b_{i,trial}]$ is the energy functional (systematically obtained as Eqs.2.1.12, 14 or 19 above). Notice that for the last i=N+1 case the (N+1)th variables are the spin variables. (The 2nd HK theorem states it for i=1, and the VP states it for i=N+2. In the latter (VP) there is no restriction to the sign of $\Psi$ and if the form is written like the middle part of Eq.2.1.11, even the normalization is not necessary.) The equality in Eq.2.1.22 is hold for the ground state $b_{i,0}$. In other words, the functional $E_{v,i}[b_i]$ must be fully minimized with respect to all allowed i-electron densities $b_i$ to obtain the true ground state i-electron density $b_{i,0}$ and ground state energy $E_{electr,0} = E_{v,i}[b_{i,0}]$. These $E_{v,i}$ functionals are also unique functionals like $E_{v,1}$ or $E_{v,N+2}$, the latter is expressed with the Hamiltonian operator (see Eq.2.1.1) in Eq.2.1.12. The variational principle is also restricted by using the N-representability conditions, alternative the formulation by Levy could be used [2.1.91], but then the functional is not the same as the one systematically obtained as Eqs.2.1.12, 14 or 19 above.

The means to prove the first is similar to case i=1 [2.1.2]: Let us use VP for ground state and Eq.2.1.20 which states that its form holds for i ≥ 1. Actually, the main point is that $v(\mathbf{r}_1)$ is only a one-electron operator, so $\Psi$ can be "degraded" algebraically to $b_1(\mathbf{r}_1)$ in the statement. Consider the multi-electron density $b_i$ for the non-degenerate ground state of some N-electron system. It determines N by its normalization analogous to Eq.2.1.10. If there were two external potentials v and v' differing by more than a constant, each giving the same $b_i$ for its ground state, we would have two Hamiltonians H and H' whose ground state $b_i$ were the same, though the normalized wave functions $\Psi$ and $\Psi'$ would be different. Take $\Psi'$ as a trial for the H problem: $E_{electr,0} < \langle\Psi'|H\Psi'\rangle = \langle\Psi'|H'\Psi'\rangle + \langle\Psi'|H-H'|\Psi'\rangle = E'_{electr,0} + \int b_i[v(\mathbf{r}_1) - v'(\mathbf{r}_1)]$ where $E_{electr,0}$ and $E'_{electr,0}$ are the ground state energies for H and H', respectively. Similarly, taking $\Psi$ as a trial for the H' problem: $E'_{electr,0} < E_{electr,0} - \int b_i[v(\mathbf{r}_1) - v'(\mathbf{r}_1)]$. Adding these two inequalities yields $E_{electr,0} + E'_{electr,0} < E_{electr,0} + E'_{electr,0}$, a contradiction. Meaning there cannot be two different v external potentials that give the same $b_i$ for their ground states. Thus, $b_i$ determines N and v and hence, all properties of the ground state, e.g. energies T, $V_{en}$, $V_{ee}$ and total ground state electronic energy, $E_{total\ electr,0}$.

The proof is also easy for the second. Simply put, one would end up with a contradiction to VP if the opposite relationship is supposed. Another way is to pick an i > 1 value indicated in Eq.2.1.22 and an allowed i-electron density $b_{i,trial}$. If for this $b_{i,trial}$ the $E_{v,i}[b_{i,trial}] < E_{electr,0}$ would hold, a successive integration such as Eqs.2.1.3-7 and Eq.2.1.9 would end up Eq.2.1.16 with opposite relation, and one would encounter a contradiction with the 2nd HK. (The original 2nd HK (for i=1) is also proved with the help of VP.) A third way to prove this is to use the known "policeman rule" from mathematical analysis. It is generally stated for a series of real numbers: If for all n in three series $a_n \le b_n \le c_n$ holds with $\lim a_n = \lim c_n = d$ then $\lim b_n = d$.

What about continuing the integration in Eqs.2.1.3-7? The one-electron property of $v(\mathbf{r}_1)$ in Eq.2.1.18 or Eq.2.1.20 is the origin of the possibility to use the one-electron density as a variable, $\rho(\mathbf{r}_1)$, in the two HK theorems – i.e. the "lowest" level in dimensionality is three in its argument, namely the $\mathbf{r}_1$. However, continuing the integration from Eq.2.1.7 to expanding the series in Eq.2.1.8, one can define, the -1,-2,-3$^{rd}$ elements as $b_{-1z}(x_1,y_1) = \int\rho(\mathbf{r}_1)dz_1$ and similarly for the different but similar kind of $b_{-1y}(x_1,z_1)$ and $b_{-1x}(y_1,z_1)$ functions. Furthermore, $b_{-2x}(x_1) = \int\rho(\mathbf{r}_1)dz_1dy_1$ and similarly the $b_{-2y}(y_1)$ and $b_{-2z}(z_1)$, and $b_{-3} = \int\rho(\mathbf{r}_1)dz_1dy_1dx_1 = N$ where $\mathbf{r}_1 \equiv (x_1,y_1,z_1)$. The latter is not a function, it is degraded to a constant (or constant function). These $b_{-1}$, $b_{-2}$, $b_{-3}$ functions are not necessarily symmetric or anti-symmetric with respect to their variables because of the known geometrical polarization property of a general one-electron density of a molecular system. The form Eq.2.1.20 for Eq.2.1.19 does not hold for $b_{-1}$, $b_{-2}$, $b_{-3}$, because there is a common variable ($x_1$, $y_1$ or $z_1$) between $b_{-1}$, $b_{-2}$, $b_{-3}$ and $v(\mathbf{r}_1)$. However, $\rho(\mathbf{r}_1)$ is determined mutually by $v(\mathbf{r}_1)$ (1$^{st}$ HK theorem), and the functions $b_{-1}$, $b_{-2}$ and $b_{-3}$ are also determined. For $b_{-3}$ with respect to 1$^{st}$ HK one can say that, because the molecular frame ($v(\mathbf{r}_1)$ in Eq.2.1.1) determines the energy (ground and excited states) with the number of electrons, indeed $b_{-3}=N$ – and we are at the beginning of the problem... The 2$^{nd}$ HK theorem in the theorem above degrades to $E_{electr,0} \leq E_{electr}(\mathbf{R}_1,\mathbf{R}_2,\ldots,\mathbf{R}_M, Z_1, Z_2,\ldots,Z_M, N)$ if in Eq.2.1.22 one continues (or evaluates) the integration appearing in Eq.2.1.13, which tells trivial things: the ground state is the lowest, as well as the energy of ground and excited states depend on the nuclear frame and number of electrons. (If other parameters also show up in the external potential, other than the nuclear frame, like external electric or magnetic fields, those can also be included.) The right hand side of this relationship has been reduced to a function from functional. (A known form of this right hand side for ground state H-like atoms (N=M=1) is $-Z_1^2/2$ hartree placed anywhere in the three-dimensional real geometrical space.) Furthermore, this relationship can be connected to the known fact as "successive ionization potentials are not decreasing (for fixed external potential $v(\mathbf{r}_1)$)" [2.1.2]. Meaning that for a fixed $v(\mathbf{r}_1)$ molecular frame, the addition of $N=1,2,3,\ldots,1+\Sigma_{A=1,\ldots,M}Z_A$ electrons, the $E_{electr,0}$ of the system is a decreasing function with respect to N (now we end the discussion for some more precise facts such that e.g. $Ne^{-1}$, or double anions do not exist in gas phase, etc.). Finally, it is not surprising that the theorem above is also true for "intermediate" terms to series in Eq.2.1.8 as for $\int b_i du_i$ or $\int b_i du_i dw_i$ for u,w=x,y,z and i=2,3,…,N as well as these kinds of terms "between" $b_N$ and $b_{N+1}$ (partial integration for some but not all spin variables). In these cases Eqs.2.1.20-21 hold too for indices specified there, and the proof is the same. Start with the Schrödinger equation (Eq.2.1.1) to get a similar equation to Eq.2.1.14 as $E_{electr}[b_i] = (-½)\langle\Psi^*|\nabla_1^2\Psi\rangle - \Sigma_{A=1,\ldots,M}Z_A\langle\Psi^*|R_{A1}^{-1}\Psi\rangle + ((N-1)/2)\langle\Psi^*|r_{12}^{-1}\Psi\rangle$ and the full bra-ket integration for $\mathbf{x}_1,\mathbf{x}_2,\ldots,\mathbf{x}_N$ is separated to the proper level of integration in the integrand according to $b_i$ or terms between $b_i$ and $b_{i+1}$ (as it was separated in Eq.2.1.14 for i=1,1 and 2 respectively for the three kinds of energy terms). The same proof holds for these terms as well in the theorem above.

Recently, a lot of work has been done on the *N*-representability problem (restricting the VP to $b_i$ that correspond to N-fermion systems) for many-electron distribution functions [2.1.15, 21-28]. Even if the exact kinetic energy functional, for example for $b_2(\mathbf{r_1},\mathbf{r_2})$ was known, the range of variation has to be restricted to $b_2(\mathbf{r_1},\mathbf{r_2})$ that correspond to some wave function. This is by no means trivial. Davidson [2.1.27] has pointed out that the scaling in electron pair density ($b_2$) is not as easy as in case of one-electron density ($\rho$) in respect to its *N*-representability property.

### 2.1.3. Approximations for the functionals

In practice, the accuracy of $E_{electr}$, $\rho(\mathbf{r}_1)$ and $\Psi$ are not the same in different levels of approximate solutions [2.1.1, 2]. Generally the energy is targeted to be the most accurate. The most popular calculations to solve Eq.2.1.1 are the expensive configuration interactions (CI) method [2.1.1] for ground and excited states and the less accurate but faster and less memory taxing Hartree – Fock self consistent field (HF-SCF) method for ground state with or without correlation corrections [2.1.1, 29]. The former is for any nuclear geometry while the latter is only for the vicinity of stationary points. These two groups of calculations have been the focus of theoretical chemistry for a long time and have vast literature. Beside many other numerical tools, for both calculations the VP provides the ground state. The term "functional" is used mostly in DFT rhetoric however very generally speaking, the CI and HF-SCF methods work with functionals as well. The path to the solution is a minimization of functions in the integrals. In the HF-SCF approximation $\Psi_0$ is approximated by a single Slater determinant (S) which is a plausible approximation in the vicinity of stationary points of the ground state electronic potential surface where the assumption of spin pairing effect is plausible (open shell cases have more complexity in comparison to closed shells). In the HF-SCF method for N electron systems, N one-electron molecular orbitals (MO) are used ($f_k(\mathbf{r}_1)$), and because of the enforced ortonormalized property upon them, the one-electron density is in the form of $\rho_{0,HF\text{-}SCF}(\mathbf{r}_1) = \Sigma_{k=1,\dots,N} f_k^2(\mathbf{r}_1)$. The $f_k(\mathbf{r}_1)$ functions are a linear combination of (contracted or not contracted) Gaussian type atomic orbital (GTO) or Slaterian type atomic orbital (STO) functions. Generally GTO's are used for a technical purpose: to employ analytical formulas for the integrals in the calculation. By spin pairing, $f_k$'s are pair-wise the same in the vicinity of stationary points. Of course, one must use a reasonable basis set [2.1.30]. In most cases the HF-SCF method is chosen as a starting point for the full or "reduced" CI calculations. There, virtual or unoccupied $f_k$'s are used for correction.

In the recent past, when even the HF-SCF calculations had problems with early computer capabilities, some of its approximations were based on the fact that near zero integrals were neglected in the algorithm to speed up the huge computational demand. Currently, with the vast increase in computer speed and disc space capabilities, these approximations are rarely used and only for very large molecular systems. Another well known but "non - *ab initio*" method is the family of molecular mechanics calculations (e.g. ref.[2.1.31]) based on the classical valence bond stretch, angle bend, dihedral angle rotation, etc. motions. This method is much faster, requires less memory and can treat much larger systems, up to thousands of atoms. In its force field it uses thousands of parameters. However, it is most trustworthy in carbon chemistry (molecules, containing mostly H, C, N, O, S atoms) where the bond is relatively well characterized. For example, if metal atoms show up in the molecular system, or the molecule is charged or radical, this method faces major difficulties with its reliability. Also, it can not account for bond formation and breaking. However, in many cases it is extremely useful for pre-calculating equilibrium molecular geometry and often, chiefly in carbon chemistry, the energy values provided reach the chemical accuracy for energy differences.

In case of basis set limit solution, the correlation energy is defined as $E_{electr,0}$(HF-SCF, basis set limit) + $E_{corr}$ = $E_{electr,0}$. This error is due to the fact that wave function $\Psi$ is approximated by a single Slater determinant. The VP provides that $E_{corr} < 0$. Because HF-SCF does not reach the chemical accuracy (1 kcal/mol) sometimes even for energy differences, the estimation of correlation energy is necessary [2.1.32-36], which has a vast literature. Popular correlation calculations are the Møller-Plesset (MP2, MP3,...) [2.1.1, 37-38], or the coupled cluster [2.1.1, 39], etc.. An extrapolated or composite method for including correlation effect is the Gaussian-2 (G2) [2.1.40] and Gaussian-3 (G3) methods [2.1.41] tested on the named G2 and G3 molecule set. These methods require great disc space and CPU time, sometimes more than the HF-SCF routine itself. The maximum size of the molecule feasible with these methods with average capacity computers currently is about the size of naphthalene with $Z_A < 18$. Otherwise the calculation is lengthy or faces a convergence problem and requires relativistic correction. Along with other methods, these corrections are built in commercial packages like Gaussian package [2.1.29]. The magnitude of correlation energy [2.1.42-44] is $E_{corr} \approx -0.039(N-1)$ hartree for an N-electron system, but it non-negligibly depends on nuclear geometry with respect to chemical accuracy. We mention here an almost instantaneous and accurate new calculation (rapid estimation of basis set error and correlation energy from partial charges, abbreviated as REBECEP) [2.1.42, 45-50] for the vicinity of stationary points, worked out for neutral, closed shell molecules. It can be easily extended to open shell and/or charged molecules. This states that $E_{corr}$ can be estimated as a linear combination of atomic correlation parameters weighted by the partial charges on atoms. (The atomic correlation parameters have the same magnitude as the atomic correlation energies in free space and one can associate similar physical meanings to them in molecular environment). Moreover, these atomic correlation parameters can absorb the basis set error even for a smaller basis; one does not have to reach the HF-SCF limit, which is a great advantage. This latter method for $E_{corr}$ is a rather non-conventional DFT method, contrary to the aforementioned methods, which are based on perturbation methods or method with series expansion of the wave function, etc..

In the last decade, the density functional theory (DFT) methods [2.1.2] have become very popular. (Sometimes the name *ab initio* is avoided because it contains few empirical parameters to fit CI quality results on smaller systems.) These newer DFT methods [2.1.44] leave the calculation at about the HF-SCF level with respect to CPU time and disc space but with increasing accuracy. However, its accuracy and convergence are not guaranteed for all systems. In the case when certain semi-empirical corrections are built in the routine, the estimated total ground state energy is not necessarily variational. A drawback of DFT is that the exact analytical forms of functionals are not known for kinetic energy (for $b_1,\ldots,b_{N+1}$ terms in Eq.2.1.8) and electron – electron repulsion energy (for $\rho(\mathbf{r}_1) \equiv b_1(\mathbf{r}_1)$ term in Eq.2.1.8), only approximate formulas exist. Not only the approximate, but very likely the exact forms of these functionals are non – linear functions of $\rho$, which makes for complexities. Besides their inaccuracy, a consequence of non-linearity (e.g. in $\rho(\mathbf{r}_1)$) is the need for numerical integration, which is another source of error. Generally, the evaluation of DFT functionals for electronic structures [2.1.2, 43, 44, 51] can not be done analytically. The shapes of DFT integrands for electronic structures (see e.g. Eq.2.1.20 for i=1) frequently follow the shape of one-electron density, i.e. sharp high peaks on nuclei and lower value "bridges" along the chemical bonds. For this case, specially designed numerical integral methods are available. It starts with a peak partitioning procedure (Voronoi polygons) which divides the molecule into atoms by perpendicular plains between atom-atom neighbors [2.1.52-54]. This is followed by Chebyshev numerical radial integration [2.1.55], where the weights are defined by easy analytical equations, and Lebedev numerical spherical integration [2.1.56-58], which integrates the spherical harmonics accurately up to high degree. In practice, only the $\{\mathbf{R}_A\}_{A=1,2\ldots,M}$ coordinates of nuclear frame are used for polygon centers when applying the partitioning method in refs.[2.1.52-54] with sigmoid smoothing between the parts. Certain scaling is described in ref.[2.1.59] as an improving device. (In certain circumstances the partitioning only by the peaks on atoms may cause inaccuracies, and inter-atomic centers must be considered for improvement. It is often neglected in the literature when numerical integration is used in DFT correlation calculations for electronic structures.)

Because exact analytical form of the kinetic DFT operator is problematic for $1 \leq i \leq N+1$ in the series in Eq.2.1.8, the Kohn – Sham (KS) [2.1.2] formalism uses the kinetic term as $\int t[b_1] = \int F_{\nabla}(\rho(\mathbf{r}_1))d\mathbf{r}_1 = (-1/2)\Sigma_{k=1,\dots,N}<\phi_k|\nabla_1^2|\phi_k> + \int F_{corr}(\rho(\mathbf{r}_1))d\mathbf{r}_1$ for the i=1 case, not only because the exact analytical expression for $F_{\nabla}$ is not known, but its known approximate forms, containing non-linear terms like $\rho(\mathbf{r}_1)^a$ and $|\nabla_1\rho(\mathbf{r}_1)|^b/\rho(\mathbf{r}_1)^c$, etc., is very difficult to treat. In fact, this kinetic formalism is very similar to the i=N+2 level in the series of Eq.2.1.8, or to the kinetic operator itself in Eq.2.1.1 (however, not $\Psi$ but the one-electron orbitals are in the argument). In this way that $F_{corr}$ above is only a small correction but necessary for chemical accuracy. The one-electron KS orbitals, $\phi_k(\mathbf{r}_1)$, are analogous of the Slaterian MO's but not the same and importantly, the $<\phi_k|\nabla_1^2|\phi_k>$ terms can be evaluated analytically also. However, $F_{corr}$ contains nonlinear terms and differential operators, and can be evaluated only numerically. Its exact form is not known, only approximate formulas. The electron-electron repulsion energy $\int v_{ee}[b_i]$ is a slightly easier case, because for $2 \leq i \leq N+2$, the exact and simple form in Eq.2.1.21 is known. However, the i=1 case is preferred in the practice via the one-electron density ($\rho(\mathbf{r}_1)$), for which MO and KS orbitals are functions of variable $\mathbf{r}_1$. For i=1 the $\int v_{ee}[b_1] = \int F_{rr}(\rho(\mathbf{r}_1))d\mathbf{r}_1 = \int\rho(\mathbf{r}_1)\rho(\mathbf{r}_2)r_{12}^{-1}d\mathbf{r}_1d\mathbf{r}_2 + \int F_{exchange}(\rho(\mathbf{r}_1))d\mathbf{r}_1$ formalism is a theoretically established one. It originates from classical electrodynamics. It is a relatively accurate form, meaning that the second part is a small correction only [2.1.2] (like in case of the kinetic operator) and, for this reason, it is considered in great detail in the literature. Similar to the kinetic case the exact form of $F_{exchange}$ is also not known currently. However, the first integral can be evaluated analytically in HF-SCF or KS formalism with GTO's, similarl to the kinetic one. We do not address here, as the literature calls, the local and non-local properties of these operators as well as the inclusion of spin dependence of these $F_{corr}$ (correlation) and $F_{exchange}$ (exchange) functionals. The integral of the sum of these two yields the exchange-correlation energy ($E_{xc}$) energy in DFT which corresponds to the $E_{corr}$ correlation energy in HF-SCF method [2.1.44]. For the exact definition of $E_{xc}$ one must go back to KS equations. It is defined in the context of one-electron differential equations – the KS equations are an extension of Hartree equations with the additive $E_{xc}$ in every line. $E_{xc} = E_x + E_c$, and $E_c$ (correlation) is less than 10 % of $E_x$ (exchange) generally [2.1.44, 60]. As a starting point the local density approximation (LDA) is in the form of $E_{xc}^{LDA}[\rho] = \int\varepsilon_{xc}(\rho)\rho d\mathbf{r}_1$, which is improved by the local spin-density approximation (LSDA). More adequate is the generalized gradient approximation (GGA), which is in the form of $E_{xc}^{GGA}[\rho] = \int f(\rho,|\nabla_1\rho|)d\mathbf{r}_1$, but it turned out that non-local functionals must be used for the final chemical accuracy (1 kcal/mol) in general cases. An overview summary is documented e.g. in refs. [2.1.2, 60], here its algebraic origin is outlined only. The $E_{corr}$ and $E_{xc}$ have about the same magnitude (at least on the similar basis set levels). Worth to mention that Levy has extended the Kohn-Sham procedure to pair densities ($b_2$) [2.1.91].

There are many proposed forms in the literature for these two parts of $E_{xc}$. For example the so called popular B3LYP has been built into packages e.g. in ref.[2.1.29]. It approximates $E_{xc}$ with a three-parameter hybrid functional (abbreviated as B3) accounting for the $E_x$ part [2.1.61] and is combined with a correlation functional (abbreviated as LYP) accounting for the $E_c$ part [2.1.62-64]. Another version, the X3LYP, has been appeared recently [2.1.65].

Also, high quality and empirical parameter free approximation [2.1.66] for $E_x$ is worked out using the ideas in refs.[2.1.67, 68] and for $E_c$ in refs.[2.1.69-71]. There is a vast amount of literature on these functionals, and an enormous number of tests on different systems. The KS method is also supposed to be used in the vicinity of stationary points (at least in current available approximation) as the Slater determinant in HF-SCF method, although in principle it is not restricted. Technically, the advantage of the Kohn-Sham method with respect to programming is that existing HF-SCF routines can be easily modified with building in some correction functionals (the exchange-correlation). Generally it gives a better approximation for $E_{electr,0}$ than the HF-SCF method, because in the vicinity of stationary points, this DFT calculation (KS) is supposed to include the correlation (more exactly the exchange correlation $E_{xc}$) energy for ground state $E_{electr,0}$ what HF-SCF itself does not possess (i.e. $E_{corr}$ must be calculated thereafter). It should be remembered that in electronic energy calculations (i.e. solving Eq.2.1.1) the energy differences between two geometries (configurations) on the potential surface is important chiefly in chemistry, not the absolute value of $E_{electr,0}$, so cancellation of errors ($E_{corr}$) can fortunately occur sometimes. DFT calculations for $E_{electr,0}$ is supposed to provide better estimation for ground state, but still today it has problems with accuracy and sometimes with convergence [2.1.2, 51]. To develop this KS formalism for improving $E_{xc}$ seems a daunting task. We also note that for example, the equilibrium geometry estimation can be accomplished better with the KS method than with the HF-SCF with correlation calculations. For example, DFT methods are able to give better results for energy differences than the HF-SCF method, or HF-SCF improved with MP2. However, sometimes the MP2 corrections provide a better result for estimating e.g. $\pi-\pi$ interaction among molecules. The DFT methods have problem accounting for London dispersion forces [2.1.51, 72-73]. Interestingly, a HF-SCF level calculation alone gives surprisingly good results for description of oligosacharide adducts with proteins [2.1.74, 75], better still than using some correlation methods. Generally speaking, there is as yet no superior method in this class that could provide the best quality result in any circumstances.

A final note on the form of the kinetic operator on the $b_1$ level is that if the moment vector operator ($\sim\nabla_i$), not the kinetic energy operator ($\sim\nabla_i^2$) is considered (they relate as $E_{kin} = \mathbf{p}^2/2m$), the functional is surprisingly simple, at least on the HF-SCF level. In relation to the HF-SCF density mentioned above, a Slater determinant gives $<S|\Sigma_{i=1,\dots,N}\nabla_i|S> = \Sigma_{k=1,\dots,N}\int f_k\nabla_1 f_k d\mathbf{r}_1$ with the ortonormalized MO's. Furthermore, $\nabla_1\rho_{0,HF\text{-}SCF}(\mathbf{r}_1) = 2\Sigma_{k=1,\dots,N} f_k(\mathbf{r}_1)\nabla_1 f_k(\mathbf{r}_1)$ holds, so the expectation value is $<S|\Sigma_{i=1,\dots,N}\nabla_i|S> = (1/2)\int\nabla_1\rho_{0,HF\text{-}SCF}(\mathbf{r}_1)d\mathbf{r}_1$ for the vector. For the $|\mathbf{p}|^2/2$, an associated kinetic energy quantity is $(1/8)|\int\nabla_1\rho_{0,HF\text{-}SCF}(\mathbf{r}_1)d\mathbf{r}_1|^2 = (1/8)\Sigma_{u=x,y,z}(\int(\partial\rho_{0,HF\text{-}SCF}(\mathbf{r}_1)/\partial u_1)d\mathbf{r}_1)^2$ which is a functional of $\rho_{0,HF\text{-}SCF}$ and analytical integral evaluation is available. For this second form the Weizsacker term must be recalled. The Weizsacker functional for Hartree-Fock densities has already been done for atoms including even the "problematic" $4^{th}$ order term [2.1.83, 92].

Integrating both sides of Eq.2.1.1 for all spin-orbit variables $\mathbf{x}_i = (\mathbf{r}_i,s_i)$ except $\mathbf{r}_1$ after multiplying by the complex conjugate of the same $i^{th}$ exited state wave function from left, one yields for $\rho(\mathbf{r}_1)$ that

$$D[\rho] \equiv D_\nabla[\rho] + D_{Rr}[\rho] + D_{rr}[\rho] = \rho E_{electr} \qquad \text{(Eq.2.1.23)}$$

Here D is the density operator corresponding to the Hamiltonian in Eq.2.1.1 for ground or excited states. The $\rho$ and $E_{electr}$ depend parametrically on nuclear variables $\{\mathbf{R}_A,Z_A\}_{A=1,\dots,M}$, and $\rho$ depends on one-electron coordinate $\mathbf{r}_1$ too. The indices of D operators in Eq.2.1.23 refer to the same as the indices in Eq.2.1.1. Eq.2.1.23 is an analogue of Eq.2.1.1 with the advantage that $\rho$ depends only on the three spatial coordinates, $\mathbf{r}_1$, while $\Psi$ depends on all $\mathbf{x}_i$, having 4N spin-orbit coordinates all together; both of course depend parametrically on nuclear coordinates. The disadvantage of D against Eq.2.1.1 is that Eq.2.1.1 is a linear partial differential eigenvalue equation while Eq.2.1.23 is a non-linear one. The $D_k$ operators ($k=\nabla$, Rr, rr) in Eq.2.1.23 should be called N-electron DFT operators as a counterpart to the N-electron DFT functionals e.g. in Eq.2.1.19. Integration of Eq.2.1.23 with using Eq.2.1.10:

$$(1/N)\int D[\rho]d\mathbf{r}_1 = E_{electr} \qquad \text{(Eq.2.1.24)}$$

proves that the relationship is a multiple factor of N only with respect to the conventional writing of DFT functional forms in Eqs.2.1.19, 22. There are many expressions for $\int D_k[\rho]d\mathbf{r}_1$ known in the DFT theory [2.1.2] for electronic structure of ground states which are separate form the exchange-correlation formalism above. As in Eq.2.1.19, the level of i can also be developed, with respect to the elements in Eq.2.1.8: For example Eq.2.1.23 is a level of $b_1$ equation, and the $b_2$ (not integrating with respect to $\mathbf{r}_1$ and $\mathbf{r}_2$) and $b_3$, etc. levels are analogous.

Similarly to Eq.2.1.19, use the anti-symmetric property of $\Psi$ for the kinetic term to get

$$D_\nabla[\rho] = -(1/2)\Sigma_{i=1,\ldots,N}\int\Psi^*\nabla_i^2\Psi ds_1 d\mathbf{x}_2\ldots d\mathbf{x}_N =$$

$$= -(1/2)\int\Psi^*\nabla_1^2\Psi ds_1 d\mathbf{x}_2\ldots d\mathbf{x}_N -((N-1)/2)\int\Psi^*\nabla_2^2\Psi ds_1 d\mathbf{x}_2\ldots d\mathbf{x}_N \qquad \text{(Eq.2.1.25)}$$

with $\int D_\nabla[\rho]d\mathbf{r}_1 = -(N/2)\int\Psi^*\nabla_1^2\Psi d\mathbf{x}_1 d\mathbf{x}_2\ldots d\mathbf{x}_N$. An approximation [2.1.2] for its measure (or integral over the three dimensional real space as a function of $\rho$) is

$$T\equiv (1/N)\int D_\nabla[\rho_0]d\mathbf{r}_1 \approx \int[c_F\,\rho_0^{5/3} + (\lambda/8)|\nabla_1\rho_0|^2/\rho_0 + (4^{th}\text{ order term})+(6^{th}\text{ order term})+\ldots]d\mathbf{r}_1 \qquad \text{(Eq.2.1.26)}$$

for ground state. The higher order (4, 6,…) terms in Eq.2.1.26 are necessary for adequate precision but unfortunately they are too difficult to use in programming in an algorithm. Most importantly, the gradient expansion for the kinetic energy in Eq.2.1.26 is known to have divergence problem for atoms and molecules [2.1.76], especially from the $4^{th}$ term. This means, that this expression may not useful for chemistry as a direct starting point. The kinetic energy functional in the KS formalism mentioned above (having N terms, one for each electron or KS orbital) owns its success by overcoming this difficulties, however DFT is a better representation of the kinetic energy solely in terms of the density. If this is true, KS orbitals will be completely eliminated from DFT formulation [2.1.77], and the density can be solved directly from the functionals originating from an adequate development of Eq.2.1.26. There has been a lot of work on kinetic-energy functionals [2.1.77-80] in literature. For example, the promising properties and development of the first two terms in Eq.2.1.26 is reported in ref.[2.1.78] by Handy et al., while a general perspective is documented in ref.[2.1.77] by Wang and Carter. Beside the approximate form for $T[b_1]$ in Eq.2.1.26, important contribution by Levy and Ziesche [2.1.7] is the approximation for $T[b_2]$. Eq.2.1.26 is considered in literature [2.1.2] and based on the anti-symmetric property of $\Psi$ with respect to $\mathbf{x}_i$ and $\mathbf{x}_j$, etc.. One must observe (see e.g. Eq.2.1.26) that Eq.2.1.23 is a non-linear partial differential equation for $\rho$, while in DFT the equations considered (as functionals of $\rho$) are subjected for full integration, i.e. for over $\mathbf{r}_1$ as well. Generally, if $\int f=\int g$ on a domain, it does not mean that f=g, but under certain circumstances it could be. (On the other hand the reverse is true.) The $D_\nabla$ term in Eq.2.1.23 or 25 can be approximated with the integrand in Eq.2.1.26 (for ground state) although the integral form (i.e. Eq.2.1.26) was derived/ considered originally in the literature. The most transparent example is the case of $D_{Rr}[\rho]$ term (see Eq.2.1.20 above or Eq.2.1.27 below) in which the integrand practically survives in its simple and totally accurate form. The $2^{nd}$ order correction part of the kinetic term in Eq.2.1.26, the $|\nabla_1\rho|^2/\rho$ in $D_\nabla[\rho]$, will explicitly show up below for N=1 and 2 electron systems. This second term in the integral in Eq.2.1.26 is called the Weizsacker correction for kinetic energy [2.1.2] for an N electron system. All these will manifest in Eqs.2.1.31 and 34 below. In Eq.2.1.26, the $c_F = (3/10)(3\pi^2)^{2/3} = 2.871234$ is the famous Thomas – Fermi constant, alone it accounts for atoms only, and $\lambda \approx 1/9$; for more details see ref. [2.1.2] on p.49 and pp.127-138. More detailed investigations [2.1.2] have yielded that 1/5 may be more accurate for $\lambda$, as well as the higher order terms (4, 6, …) have a problem with convergence beside their computationally almost non-tractable algebraic form. Eq.2.1.26 is supposed to work for any nuclear configuration for the price of losing the linearity.

The $D_{Rr}$ nuclear-electron attraction term is easier. Like Eq.2.1.20 for the i=1 case, this term in Eq.2.1.23 is $D_{Rr}[\rho] = -\Sigma_{i=1,\dots,N}\Sigma_{A=1,\dots,M} Z_A \int\Psi^*\Psi R_{Ai}^{-1} ds_1 d\mathbf{x}_2 \dots d\mathbf{x}_N = \Sigma_{i=1,\dots,N} \int\Psi^*\Psi\, v(\mathbf{r}_i) ds_1 d\mathbf{x}_2 \dots d\mathbf{x}_N = \int\Psi^*\Psi v(\mathbf{r}_1) ds_1 d\mathbf{x}_2 \dots d\mathbf{x}_N + (N-1)\int\Psi^*\Psi v(\mathbf{r}_2) ds_1 d\mathbf{x}_2 \dots d\mathbf{x}_N$ which reduces to

$$D_{Rr}[\rho] = \rho(\mathbf{r}_1)v(\mathbf{r}_1) + (N-1)\int b_2(\mathbf{r}_1,\mathbf{r}_2)v(\mathbf{r}_2)d\mathbf{r}_2 \qquad \text{(Eq.2.1.27)}$$

with $\int D_{Rr}[\rho]d\mathbf{r}_1 = N\int\rho(\mathbf{r}_1)v(\mathbf{r}_1)d\mathbf{r}_1$, for ground and excited states coming from the anti-symmetric property of $\Psi$. (This latter integral form is a simple and totally accurate expression of $\rho$. Of course, $V_{en} \equiv (1/N)\int D_{Rr}[\rho]d\mathbf{r}_1 = \int\rho(\mathbf{r}_1)v(\mathbf{r}_1)d\mathbf{r}_1$. If e.g. $\rho_0$ is a linear combination of GTO's, what is possible in HF-SCF method, analytical evaluation is available for integrating Eq.2.1.27 with respect to $\mathbf{r}_1$.) It is important how the corresponding integrand in the functional Eq.2.1.19 (with Eq.2.1.20) survives as differential form in Eq.2.1.23 (see Eq.2.1.27) up to a multiplier N, see below for particular examples.

The $D_{rr}$ electron-electron repulsion term is not easy again for $\rho$ ($\equiv b_1$) but simpler on $b_2$ - $b_3$ level and up. This term in Eq.2.1.23 (similar to Eq.2.1.21) is $D_{rr}[\rho] = \Sigma_{i=1,\dots,N}\Sigma_{j=i+1,\dots,N}\int\Psi^*\Psi\, r_{ij}^{-1} ds_1 d\mathbf{x}_2 \dots d\mathbf{x}_N = (N-1)\int\Psi^*\Psi r_{12}^{-1} ds_1 d\mathbf{x}_2 \dots d\mathbf{x}_N + [N(N-1)/2 - (N-1)]\int\Psi^*\Psi r_{23}^{-1} ds_1 d\mathbf{x}_2 \dots d\mathbf{x}_N$ which reduces to

$$D_{rr}[\rho] = (N-1)\int b_2(\mathbf{r}_1,\mathbf{r}_2)r_{12}^{-1}d\mathbf{r}_2 + [N(N-1)/2 - (N-1)]\int b_3(\mathbf{r}_1,\mathbf{r}_2,\mathbf{r}_3)r_{23}^{-1}d\mathbf{r}_2 d\mathbf{r}_3 \qquad \text{(Eq.2.1.28)}$$

with $\int D_{rr}[\rho]d\mathbf{r}_1 = (N(N-1)/2)\int\Psi^*\Psi r_{12}^{-1} d\mathbf{x}_1 d\mathbf{x}_2 \dots d\mathbf{x}_N = (N(N-1)/2)\int b_2(\mathbf{r}_1,\mathbf{r}_2)r_{12}^{-1}d\mathbf{r}_1 d\mathbf{r}_2$ for ground and excited state states coming from the anti-symmetric property of $\Psi$. Notice the similarity between the integral vs. differential form in Eqs.2.1.19, 21 vs. Eqs.2.1.23, 28. An approximation for its measure [2.1.2] is (i=1 case)

$$V_{ee} \equiv (1/N)\int D_{rr}[\rho_0]\, d\mathbf{r}_1 \approx \int[2^{-1/3}(N-1)^{2/3}\rho_0^{4/3} + \text{correction}]d\mathbf{r}_1 \qquad \text{(Eq.2.1.29)}$$

for ground state which is not a very accurate approximation. The latter means that the "correction" has a larger value in comparison to the corresponding $F_{exchange}$ function above, but not a bad one. However, for i=2 case

$$V_{ee} \equiv (1/N)\int D_{rr}[b_2]d\mathbf{r}_1 d\mathbf{r}_2 = ((N-1)/2)\int b_2\, r_{12}^{-1}d\mathbf{r}_1 d\mathbf{r}_2 \qquad \text{(Eq.2.1.30)}$$

which is totally accurate and holds for ground and excited states as well. With the first terms in Eqs.2.1.26, 27 and 29 for $\rho_0$, the integrand of functional for the 1st HK for i=1 in Eq.2.1.22 reduces to an algebraic expression for $\rho_0$ [2.1.2, 81]. The N factor in Eqs.2.1.26, 27 and 29 drops via Eq.2.1.10 if these are substituted into Eq.2.1.23 – as Eq.2.1.24 shows. More importantly, the integrands of Eqs.2.1.26, 27 and 29 survives (as approximants) in Eq.2.1.23 for ground state. This fact will be investigated and analyzed in a separate work, although for the H-like atoms below some will be discussed for this hypothesis.

### 2.1.4. The exact density functional operator for N=1 electron systems (H-like atoms and general one-electron systems), and Slater determinant level density functional operator for N=2 electron systems

The one-electron equation (and the 2-electron equation for single-orbital Slater determinants) have been known for a very long time and have been in active use since the mid-1980's by Parr, Levy, Perdew and others [2.1.82-88]. These equations have even been (approximately) extended to larger systems. However, this problem has been treated in the context of the (integral form) of the HK theorems. Below, the aforementioned differential forms are considered with their connection to the integral forms, which – the author thinks – is worth to archive in detail. Consider the case N=1, M≥1, $Z_A$≥1 for ground and excited states. This is where one single electron is in the quantum force field of a nuclear frame. It is not in the general interest, except its M=1 subcase, the H-like atom. The latter is a basic problem in quantum mechanics. Solving it for $\Psi$ via Eq.2.1.1 is fully described in university textbooks; however, it is interesting with respect to solving it for $\rho$ from the nonlinear equation Eq.2.1.23. From Eq.2.1.1 with $\Psi = \alpha_1 f(\mathbf{r}_1)$, multiplying by $\Psi$ from the left, integrating for spin coordinate and considering f as real function ($\rho(\mathbf{r}_1) = f^2(\mathbf{r}_1)$), Eq.2.1.23 takes the particular form in this case as

$$D[N=1,\rho(\mathbf{r}_1)] \equiv -(1/4)\nabla_1^2\rho(\mathbf{r}_1) + (1/8)\rho(\mathbf{r}_1)^{-1}|\nabla_1 \rho(\mathbf{r}_1)|^2 + \rho(\mathbf{r}_1)v(\mathbf{r}_1) = E_{electr}\,\rho(\mathbf{r}_1) \quad \text{(Eq.2.1.31)}$$

where the first two terms containing the nabla operator are an evaluation of the kinetic part $D_\nabla[N=1, \rho(\mathbf{r}_1)] = -(1/2)f\nabla_1^2 f = -(1/2)\rho^{1/2}\nabla_1^2\rho^{1/2}$. At N=1, the $D_{rr}$ (or $V_{ee}$) term vanishes. This D in Eq.2.1.31 is an algebraically exact analytical sub-case (N=1) of the general D density operator (N≥1) in Eq.2.1.23. We emphasize here that Eq.2.1.31 is not an approximation, it is an exact form. For M=1, such as H-like atoms, the one-electron density of s, p, d, f, … orbitals, which are commonly found in many textbooks, are the analytical eigen-solutions of Eq.2.1.31. For example, the ground state 1s atomic orbital is described by the known $E_{electr,0} = -Z_1^2/2$ eigen-energy of Eqs.2.1.1 and 31 accompanied with the eigen-function $\Psi_0(\mathbf{x}_1) = C\,\alpha_1 \exp(-Z_1|\mathbf{r}_1|)$ for Eq.2.1.1 (the $\alpha_i$ is the spin function) and $\rho_0(\mathbf{r}_1) = C^2 \exp(-2Z_1|\mathbf{r}_1|)$ for Eq.2.1.31 (with $\langle\alpha_1|\alpha_1\rangle=1$), etc. Interestingly, the real value $C^2 > 0$ normalization constants of $\rho(\mathbf{r}_1)$ drops from both sides of Eq.2.1.31 like C in Eq.2.1.1 for $\Psi$, despite the fact that Eq.2.1.31 (similar to the more general Eq.2.1.23) is non-linear in $\rho$. In more detail, Eq.2.1.1 is linear (additive and homogen) in $\Psi$, while Eq.2.1.31 is non-linear in $\rho$ because non-additive, but still homogen (i.e. if $\rho$ is a solution, then const.$\rho$ is also a solution of Eq.2.1.31). However, for N>1, Eq.2.1.23 is even not homogen. With respect to the 2nd HK theorem, it is elementary to prove that if one takes an "educated pick" like $\rho_0 = C^2\exp(-b|\mathbf{r}_1|)$ instead of a full variational solution procedure (and keeping $N = \int\rho_0(\mathbf{r}_1)d\mathbf{r}_1 = 1$, which makes the constrain $C^2=b^3/8\pi$) the energy functional from Eq.2.1.31 by Eq.2.1.24 for M=1, $E_{electr,0} = \int D[\rho_0(\mathbf{r}_1)]d\mathbf{r}_1$, has the property that its energy minimum from $\partial E_{electr,0}/\partial b = 0$ is exactly at $b=2Z_1$, as expected. In this way the factor in the exponent of the one-electron density of the 1s orbital of the H-like atom has been recovered. The derivation is elementary: Place the nucleus in the origin. For M=1 Eq.2.1.24 converts Eq.2.1.31 to $0+C^2\pi/b - 4\pi C^2 Z_1/b^2 = E_{electr,0}$, which is a second order algebraic function with the constraint $C^2=b^3/8\pi$ mentioned above. This $E_{electr,0}$ has the minimum value indicated just above at $b= 2Z_1$ and $E_{electr,0}= -Z_1^2/2$. It also shows how the kinetic and potential terms together provide an energy variational minimum for the ground state. The zero integral of the first term of D in Eq.2.1.31 will be discussed in general below, see Eq.2.1.32.

The derivation of Eq.2.1.31 is similar to the Slater determinant level case below for N=2, so we show the scheme of derivation there. The point is, one can take advantage of that: 1., in $\int\Psi^*\nabla_1^2\Psi$ for kinetic term the $\Psi$ depends on $\mathbf{x}_1$ as $\alpha_1 f(\mathbf{r}_1)$ only and not on other electrons being N=1, so from Eq.2.1.7 $\rho(\mathbf{r}_1)$=const.$f^2(\mathbf{r}_1)$, 2., the electron-nucleus term is the simple Eq.2.1.20 with i=1, as well as a no electron-electron term. The N=1 strongly limits the usage of Eq.2.1.31, but theoretically, we can see an exact analytical form for D in the case when N=1. It is interesting which algebraic terms the Eqs.2.1.23-27 include for the general N≥1 case what Eq.2.1.31 for N=1 does not: The $\rho^{5/3}$ main term is necessary to the anti-symmetric relation between electrons (if N>1) but the other terms are only corrections. Note that for N>1, the 2nd, 3rd, ..., Nth electrons fall into one term by the anti-symmetric property of $\Psi$, see the terms with factor (N-1) in Eqs.2.1.25 and 27 – so the number of main algebraic groups is only two in t or $v_{en}$ irrespectively of the value of N. Comparing the kinetic (first two) terms in Eq.2.1.31 to Eq.2.1.26, we note that one can use $\int\nabla_1^2\rho(\mathbf{r}_1)d\mathbf{r}_1 = 0$ for any N and ground- and excited states in Eq.2.1.24. It means that only the Weizsacker term survives in the integration for N=1 among the two kinetic terms in Eq.2.1.31. The terms including $\nabla$ are necessary for the "cusp condition" [2.1.2], the nabla operator provides that the one-electron density does not "blow up" at the positions of nuclei. Seeing that the integral of $\nabla_1^2\rho$ is zero for ground- and excited state one-electron density when going from Eq.2.1.31 to the HK energy functional (Eq.2.1.24) via integrating the left hand side of Eq.2.1.31 and dividing it by N is as follows: Let us assume that $\Psi$ is real in Eq.2.1.1 and the notation $d\mathbf{o} \equiv ds_1 d\mathbf{x}_2 d\mathbf{x}_3..d\mathbf{x}_N$ (i.e. $d\mathbf{x} = d\mathbf{o}d\mathbf{r}_1$). Eq.2.1.7 yields $\rho(\mathbf{r}_1) \equiv \int\Psi^2 d\mathbf{o}$ for any N, M, $\{R_A,Z_A\}$, ground- and excited states. It follows that $(1/2)\partial^2\rho/\partial u_1^2 = \int [(\partial\Psi/\partial u_1)^2 + \Psi\, \partial\Psi^2/\partial u_1^2]d\mathbf{o}$ for u = x, y or z spatial coordinates. Using the rule of partial integration ($\int f'g = fg - \int fg'$) and the limit behavior $\Psi(u_1 = +/-\infty)=0$: $(1/2)\int\partial^2\rho/\partial u_1^2 d\mathbf{r}_1 = \int [(\partial\Psi/\partial u_1)^2 + \Psi\, \partial\Psi^2/\partial u_1^2]d\mathbf{o}d\mathbf{r}_1 = \int [(\partial\Psi/\partial u_1)^2 - (\partial\Psi/\partial u_1)^2]d\mathbf{o}d\mathbf{r}_1 = 0$, i.e.

$$\int\nabla_1^2\rho(\mathbf{r}_1)d\mathbf{r}_1 = 0 \qquad \text{(Eq.2.1.32)}$$

when integrating over the entire three-dimensional real space. (In the $\int\ldots d\mathbf{o}d\mathbf{r}_1$ above the one dimensional integral over $u_1$ was evaluated first or taken most inside.)

Technically Eqs.2.1.28-30 can be included in Eq.2.1.31, because its (N-1) factor eliminates the electron-electron $D_{rr}$ terms for N=1 (the correction is zero for N=1 in Eq.2.1.29), as it is expected for Eq.2.1.31. Again, Eqs.2.1.26 and 29 are approximate expressions for N≥1 while Eq.2.1.31 is accurate for the N=1. As a summary for Eq.2.1.31: it has a restriction as N=1, however, the operators in it are exact and not approximate, as well as holds for ground and excited states. Eq.2.1.10 provides $\int\rho(\mathbf{r}_1)d\mathbf{r}_1 = N=1$ now, and integrating both sides of Eq.2.1.31 yields

$$E_{electr,0} \le \int [(1/8)\rho_{0,trial}(\mathbf{r}_1)^{-1}|\nabla_1\rho_{0,trial}(\mathbf{r}_1)|^2 + \rho_{0,trial}(\mathbf{r}_1)v(\mathbf{r}_1)]d\mathbf{r}_1 \qquad \text{(Eq.2.1.33)}$$

for the HK energy functional, which is an exact form again, and its variational minimum by the 2nd HK theorem yields the ground state one-electron density and ground state energy. Also, in this special case when N=1, Eq.2.1.33 holds with equality if solutions ground- or excited states one-electron densities are substituted for $\rho_{0,trial}$ yielding the corresponding energy levels. Further restriction in Eqs.2.1.31-33 by M=1 yields the exact expressions for the H-like atoms.

For N=2 electron molecules with M≥1 and $Z_A \ge 1$, like He, $Li^+$, $H_2$, $LiH^{2+}$, $H_3^+$, etc., a Slater determinant (S) single reference level estimation (as used in HF-SCF) for the ground state in the vicinity of stationary points (if existing) can be expressed with an accurate analytical sub-case in Eqs.2.1.23-30. Unlike the one electron case above for M>1, here we have cases of interest like the molecules listed above, although many other two-electron molecular systems like highly ionized, unstable benzene ($C_6H_6^{40+}$) also belong here, even if they have no practical, only theoretical interest. This simple N=2 case shows a bit more about these DFT expressions. The scheme of derivation now is the same as it has yielded in Eq.2.1.31 above. In this formalism we approximate the wave function with a Slater determinant as $\Psi_0 \approx S \equiv sf_1f_2$, where $s= s(s_1,s_2)\equiv (\alpha_1\beta_2-\alpha_2\beta_1)$ is the spin function, and $f_1\equiv f(\mathbf{r}_1)$ and $f_2\equiv f(\mathbf{r}_2)$ is the single MO occupied by the two electrons (a real function, so f*=f). If the spin normalization is $\langle s|s\rangle=2$, and the f is an MO, then $\rho_{0,HF\text{-}SCF} = 2f^2(\mathbf{r}_1)$ if f is normalized as $\int f^2(\mathbf{r}_1)d\mathbf{r}_1=1$ to satisfy

Eq.2.1.10. The HF-SCF method [2.1.1] solves the problem using the VP. The index HF-SCF is used because in the minimization of HF-SCF method the starting stage is the approximation with a Slater determinant and this one-electron density has that typical algebraic form. Here (via HK theorems) one must minimize the HK functional in Eq.2.1.24 or Eq.2.1.22 (for i=1), but – at least for basis set limits – the two (HF-SCF procedure or minimization by 2nd HK) should be the same. For the sake of brevity the index (0,S) is used next rather than (0,HF-SCF), indicating that it is a ground state Slater level approximation with respect to Eq.2.1.23 or 24, i.e. the minimization is supposed to proceed by the 2nd HK, not the HF-SCF procedure. Now we find the actual form of Eq.2.1.23 for N=2 and $\Psi_0 \approx S$. It follows that $f_1=(\rho_{0,S}(\mathbf{r}_1)/2)^{1/2}$ and $\nabla_1^2\rho_{0,S} = 4f_1\nabla_1^2f_1 + 4|\nabla_1f_1|^2 = 4f_1\nabla_1^2f_1 + 2|\nabla_1\rho_{0,S}{}^{1/2}|^2$. Below, $D_{k,S}$ means the approximation of $D_k$ for N=2 indicated in the argument, where k= $\nabla$,Rr,rr and index S stands for "Slater determinant level". For N=2 the Eq.2.1.25 takes the form (notice there is no integration for $\mathbf{r}_1$ at the beginning, $\rho_{0,S} \equiv \rho_{0,S}(\mathbf{r}_1)$) $D_{\nabla,S}[N=2, \rho_{0,S}(\mathbf{r}_1)] = -(1/2)\langle s|s\rangle\int f_1^*f_2^*\nabla_1^2f_1f_2d\mathbf{r}_2 -(1/2)\langle s|s\rangle\int f_1^*f_2^*\nabla_2^2f_1f_2d\mathbf{r}_2 = -(f_1\nabla_1^2f_1)\int f_2^2d\mathbf{r}_2 - f_1^2\int f_2^*\nabla_2^2f_2d\mathbf{r}_2 = -f_1\nabla_1^2f_1 - f_1^2\int f_2^*\nabla_2^2f_2d\mathbf{r}_2 = -(1/4)\nabla_1^2\rho_{0,S} + (1/2)|\nabla_1\rho_{0,S}{}^{1/2}|^2 -(1/2)\rho_{0,S}\int((1/4)\nabla_1^2\rho_{0,S} - (1/2)|\nabla_1\rho_{0,S}{}^{1/2}|^2)d\mathbf{r}_1$, finally

$$D_{\nabla,S}[N=2, \rho_{0,S}(\mathbf{r}_1)] = -(1/4)\nabla_1^2\rho_{0,S}(\mathbf{r}_1) + (1/8)\rho_{0,S}{}^{-1}|\nabla_1\rho_{0,S}|^2 - B_{kin}\rho_{0,S}(\mathbf{r}_1) \qquad \text{(Eq.2.1.34)}$$

where $B_{kin} \equiv (1/2)\int((1/4)\nabla_1^2\rho_{0,S} -(1/2)|\nabla_1\rho_{0,S}{}^{1/2}|^2)d\mathbf{r}_1 = (-1/16)\int(\rho_{0,S}{}^{-1}|\nabla_1\rho_{0,S}|^2)d\mathbf{r}_1$ via Eq.2.1.32. It follows that $(1/N)\int D_{\nabla,S}[N=2, \rho_{0,S}(\mathbf{r}_1)]d\mathbf{r}_1 = \int(-(1/4)\nabla_1^2\rho_{0,S}(\mathbf{r}_1) + (1/8)\rho_{0,S}{}^{-1}|\nabla_1\rho_{0,S}|^2)d\mathbf{r}_1 = \int(1/8)\rho_{0,S}{}^{-1}|\nabla_1\rho_{0,S}|^2)d\mathbf{r}_1$, via Eq.2.1.32, as well as notice the similarity of this integral kinetic term (N=2) to the one (N=1) in Eq.2.1.31. For N=2 Eq.2.1.27 yields the simple

$$D_{Rr,S}[N=2, \rho_{0,S}(\mathbf{r}_1)] = v(\mathbf{r}_1)\rho_{0,S}(\mathbf{r}_1) + \langle s|s\rangle\int v(\mathbf{r}_2)f_1^2f_2^2d\mathbf{r}_2 =$$
$$= v(\mathbf{r}_1)\rho_{0,S}(\mathbf{r}_1) + B_{Rr}\rho_{0,S}(\mathbf{r}_1) \qquad \text{(Eq.2.1.35)}$$

where $B_{Rr} \equiv (1/2)\int v(\mathbf{r}_1)\rho_{0,S}(\mathbf{r}_1)d\mathbf{r}_1$. It is easy to see that $(1/N)\int D_{Rr,S}[N=2, \rho_{0,S}(\mathbf{r}_1)]d\mathbf{r}_1 = \int v(\mathbf{r}_1)\rho_{0,S}(\mathbf{r}_1)d\mathbf{r}_1$, as expected, not only for this approximate $\rho_{0,S}(\mathbf{r}_1)$, but the form for the true $\rho_0(\mathbf{r}_1)$ is recovered as well with the integration (see Eqs.2.1.19, 20 and 24 for N=2). For N=2 Eq.2.1.28 yields the nonlocal integral operator or functional $D_{rr,S}[N=2, \rho_{0,S}(\mathbf{r}_1)] = \langle s|s\rangle\int f_1^2f_2^2r_{12}^{-1}d\mathbf{r}_2 = 2\int f_1^2f_2^2r_{12}^{-1}d\mathbf{r}_2$, i.e.

$$D_{rr,S}[N=2, \rho_{0,S}(\mathbf{r}_1)] = (1/2)\int\rho_{0,S}(\mathbf{r}_1)\rho_{0,S}(\mathbf{r}_2)r_{12}^{-1}d\mathbf{r}_2 \qquad \text{(Eq.2.1.36)}$$

Finally, Eq.2.1.23 for N=2 and $\Psi_0 \approx S$ is

$$D[N=2, \rho_{0,S}(\mathbf{r}_1)] \approx D_{\nabla,S}[N=2, \rho_{0,S}(\mathbf{r}_1)] + D_{Rr,S}[N=2, \rho_{0,S}(\mathbf{r}_1)] +$$
$$+D_{rr,S}[N=2, \rho_{0,S}(\mathbf{r}_1)] = \rho_{0,S}(\mathbf{r}_1)\, E_{electr,0,S} \qquad \text{(Eq.2.1.37)}$$

with $\rho_{0,S}(\mathbf{r}_1)$ normalized to N=2, which was involved in the derivation. In comparison to Eq.2.1.31, the kinetic operator is the same with an additive term $-B_{kin}\rho_{0,S}(\mathbf{r}_1)$, so the case of electron-nuclear term with an additive term $B_{Rr}\rho_{0,S}(\mathbf{r}_1)$ as well – the $B_{kin}$ and $B_{Rr}$ are constants. In Eqs.2.1.36 - 37 the electron-electron term is an "accurate" form restricted on a Slater determinant level - the classical electrodynamic form was recovered. To approach the real ground state energy of the two-electron systems, the "exchange" correction is necessary, as the "correlation" correction is also needed for the kinetic part (first term in Eq.2.1.37) which is also an exact expression but, again, on a Slater determinant level only. Notice that this kind of derivation for Eqs.2.1.31 and 37 is not as simple if N > 2, because the analytical trick used in the kinetic part is more hectic even for a Slater determinant form. This is because more than one f-function (MO orbital) appear there in a linear combination (and it's not possible to take advantage of using $f(\mathbf{r}_1)=(\rho_{0,S}/2)^{1/2}$), although in the paragraph just before Eq.2.1.23 there is a note about the possibility of overcoming this problem.

Now Eq.2.1.24 also yields a similar (but now only Slater determinant level) HK energy functional from Eq.2.1.37 like the one in Eq.2.1.33, and the $\nabla_1^2\rho_{0,S}(\mathbf{r}_1)$ section drops again via Eq.2.1.32

$$E_{electr,0,S} \le \int[(1/8)\rho_{0,S,trial}(\mathbf{r}_1)^{-1}|\nabla_1\rho_{0,S,trial}(\mathbf{r}_1)|^2 + v(\mathbf{r}_1)\rho_{0,S,trial}(\mathbf{r}_1) + + (1/4)\int\rho_{0,S,trial}(\mathbf{r}_1)\rho_{0,S,trial}(\mathbf{r}_2)r_{12}^{-1}d\mathbf{r}_2]d\mathbf{r}_1 \quad \text{(Eq.2.1.38)}$$

Eqs.2.1.31 and 37 are the nonlinear DFT partial differential equations of the problem. Eqs.2.1.31 and 37 (like Eqs.2.1.33 and 38) have similar forms of course and could be expressed with one common equation for N=1 or 2 using a Kronecker delta. For N>2, the corresponding general explicit form of the DFT differential equation has more complexity but only through the additional correctional terms only; a few are indicated in Eqs.2.1.26 and 29. As a summary for Eq.2.1.37: it has a restriction like N=2 and $\Psi_0 \approx S$, and as a consequence of the latter, it approximates the ground state in the vicinity of stationary points (geometrical minimum or transition state) only. We can say that the operators in it are exact and not approximate, but again, with respect to the Slater determinant level only. Eq.2.1.10 was used in the derivation as $\int\rho_{0,S}(\mathbf{r}_1)d\mathbf{r}_1 = N = 2$. The HK energy functional via Eq.2.1.24, which is an exact form now in Eq.2.1.38 with respect to the Slater determinant level, provides the variational minimum by the 2nd HK theorem, and yields the ground state one-electron density and ground state energy. The latter state corresponds to the HF-SCF energy with a similar basis set for this N=2 electron problem.

As a note to the N=2 electron problem, if one uses the $b_2$ level instead of $\rho$, the expression in Eq.2.1.30 (for N≥1) is a totally accurate form in comparison to the approximate one (for N=2) in Eq.2.1.36 or the corresponding one in Eq.2.1.38. For generalization, one can start from Eq.2.1.1 for N≥2 and integrate up to $b_2(\mathbf{r}_1,\mathbf{r}_2)$ level: multiply by $\Psi^*$ from the left, integrate fully over $ds_1ds_2d\mathbf{x}_3...d\mathbf{x}_N$ (i.e. except for $\mathbf{r}_1$ and $\mathbf{r}_2$) and using the anti-symmetric property of $\Psi$ to obtain $D_\nabla[b_2(\mathbf{r}_1,\mathbf{r}_2)]$ analogously to Eqs.2.1.25 and 27. However, instead of two groups of terms, one obtain three. But, the equation corresponding to Eqs.2.1.22, 24 for the 2nd HK is

$$\int[(1/N)D_\nabla[b_2(\mathbf{r}_1,\mathbf{r}_2)] + v(\mathbf{r}_1)b_2(\mathbf{r}_1,\mathbf{r}_2) + ((N-1)/2)b_2(\mathbf{r}_1,\mathbf{r}_2)r_{12}^{-1}]d\mathbf{r}_1d\mathbf{r}_2 = E_{electr} \quad \text{(Eq.2.1.39)}$$

for ground- and excited state and N≥2 with $b_2$ normalised to N. In Eq.2.1.39 we left the kinetic term ($D_\nabla$) on a notation level only, however this 2-electron case is known to be exact from the work of Nagy [2.1.8] and it also follows from the work of Ayers [2.1.15]. The Slater determinant case has been treated in detail by Higuchi [2.1.93]. The last two terms on the left hand side of Eq.2.1.39 are, of course, totally accurate expressions for the nuclear-electron and electron-electron terms for ground- and excited states for Eq.2.1.24 in comparison to the corresponding approximate ground state level electron-electron term in Eq.2.1.38. Eq.2.1.39 provides a particular form for nuclear-electron and electron-electron terms as $V_{en} + V_{ee} = \int[v(\mathbf{r}_1) + ((N-1)/2)r_{12}^{-1}]b_2(\mathbf{r}_1,\mathbf{r}_2)d\mathbf{r}_1d\mathbf{r}_2$ which is a completely precise part for energy functional (for ground- and excited states and for N≥2). Also, it does not need an "exchange" correction for the $V_{ee}$ part. Eq.2.1.39 provides the functional for Eq.2.1.22 to apply the 2nd HK for ground state. In this way, the problem is reduced from 4N dimension to 6 spatial ones. However, as in CI or HF-SCF calculations, one x-anti-symmetric Slater determinant is not enough. Similarly, in the case of r-symmetric $b_2(\mathbf{r}_1,\mathbf{r}_2)$, one $g(\mathbf{r}_1)g(\mathbf{r}_2)$ product (generated by a single g function, a Hartree product) is not enough for an accurate (but still approximate) solution to Eq.2.1.39 or variational minimization of it for the ground state. Pistol [2.1.25] has proposed a dense basis for the N-representable, two-electron densities ($b_2$), in which all N-representable two-electron densities can be expanded, using positive coefficients along with the inverse problem of finding a representative wavefunction, giving the prescribed two-electron density. (This latter is extended to the case of $b_i$ densities [2.1.24] by the same author.) An additional problem comes from the $D_\nabla$ kinetic energy term in Eq.2.1.39: It is a nonlinear differential operator in $b_2$ as well as at present it is not completely known, but continuous investigation is focusing on this kinetic energy functional [2.1.8-10, 13, 89-90] and its variational procedures [2.1.14].

## 2.2. Conversion of the non-relativistic electronic Schrödinger equation to scaling correct moment functional of ground state one-electron density to estimate ground state electronic energy

### 2.2. Preliminary

The reduction of the electronic Schrödinger equation or its calculating algorithm from 4N-dimensions to a nonlinear, approximate density functional of a 3 spatial dimension one-electron density for an N electron system which is tractable in practice, is a long desired goal in electronic structure calculation. In a seminal work, Parr et al. suggested a well behaving density functional in power series with respect to density scaling within the orbital-free framework for kinetic and repulsion energy of electrons. The updated literature on this subject is listed, reviewed and summarized. Using this series with some modifications, a good density functional approximation is analyzed and solved via the Lagrange multiplier device. The introduction of a Lagrangian multiplier to ensure normalization is a new element in this part of the related, general theory. Its relation to Hartree-Fock and Kohn-Sham formalism is also analyzed for the goal to replace all the analytical Gaussian based two and four center integrals ($\int g_i(\mathbf{r}_1)g_k(\mathbf{r}_2)r_{12}^{-1}d\mathbf{r}_1 d\mathbf{r}_2$, etc.) to estimate electron – electron interactions with cheaper numerical integration. The Kohn-Sham method needs the numerical integration anyway for correlation estimation.

### 2.2.1. Introduction

The non-relativistic spinless fixed nuclear coordinate electronic Schrödinger equation (SE) in free space is capable of describing the electronic motion in molecular systems by providing the anti-symmetric wavefunction $\Psi(\{Z_A,\mathbf{R}_A\},\{\mathbf{x}_i\})$ and electronic energy $E_{total\ electr}= E_{electr}(\{\mathbf{R}_A,Z_A\}) + V_{nn}$ of the ground and excited states. $V_{nn}=\Sigma_{A=1,\dots,M}\Sigma_{B=A+1,\dots,M}Z_AZ_BR_{AB}^{-1}$, where $R_{Au}$ (u = x,y,z) are the M nuclear coordinates with nuclear charges $Z_A$, as well as $\mathbf{x}_i = (\mathbf{r}_i,s_i) = (x_i,y_i,z_i,s_i)$ are the N spin-space electronic coordinates (4N dimensions). For the commonly used ab initio calculations as configuration interactions (CI, for ground and excited states) and the faster Hartree-Fock Self Consistent Field (HF-SCF, for ground state) [2.2.1] longer time and larger disc space are still demanded, even for ground state $\Psi_0$ and $E_{electr,0}$, as well as convergence problems can rise at about N=10 and 500 respectively. The density functional theory (DFT) method, based on the Kohn-Sham (KS) formulation [2.2.2-3] effectively improves the "error" of the HF-SCF method (called correlation energy $E_{corr} \equiv E_{electr,0}-E_{HF\text{-}SCF/basis}$ [2.2.1, 4-5]), technically with some in-built [2.2.6-16] functionals during the SCF algorithm, called "exchange-correlation functionals" – not detailed here. (On the other hand, $E_{corr}$ can be estimated after the HF-SCF routine, for example with Møller-Pleset (MP) and many other methods [2.2.1] – also not detailed here.) Another thing, one should not forget about the basis set error and basis set superposition error [2.2.12]. However, the kinetic functional in the KS method is still the sum of the N nabla-square operators, so the computational costs remain similar to the HF-SCF method (3N dimensional in nature). It has long been desired in DFT, to reduce the dimensionality to 3. While the HF-SCF and KS methods are highly developed, there are still no tractable methods based solely on the 3 spatial dimension one-electron density.

The fascinating idea of moment-based density functionals is seductive: replace the thorny functional analysis that accompanies DFT with "function analysis" by writing the energy as a function (not a functional) of the moments of the electron density. This paper works along those lines. The energy functional for ground state based on scaling correct power series is reviewed and the standard Lagrange multiplier method is introduced in this relation, which ensures the normalization of the density, to solve and analyse these density functionals. We also discuss about the relation of these density functionals with the Kohn-Sham DFT and Hartree-Fock theory.

### 2.2.2. Review of the Energy Functional for Ground State Based on Scaling Correct Power Series

#### 2.2.2.a. The N-electron density functional and density integro-differential operator

The ground state N-normalized one-electron density, $\rho_0(\mathbf{r}_1)$, is the central variable in DFT. Since the density functional for $\rho_0(\mathbf{r}_1)$ is non-linear, its solution generally requires numerical integration as described and cited in refs.[2.2.14-15], not only in the correction terms as in KS formalism, but also in the main terms as well. In the one-electron density formulation of DFT, the energy functional (in the absence of external field other than the molecular frame) comes from

$$E_{electr}[\rho]= N^{-1}[\int D_{\nabla}(\rho(\mathbf{r}_1))d\mathbf{r}_1 +\int D_{rr}(\rho(\mathbf{r}_1))d\mathbf{r}_1] -\Sigma_{A=1,\dots,M}Z_A\int\rho(\mathbf{r}_1)r_{A1}^{-1}d\mathbf{r}_1 \equiv$$
$$\equiv N^{-1}\int D[\rho]d\mathbf{r}_1 \equiv F[\rho(\mathbf{r}_1)], \qquad \text{(Eq.2.2.1)}$$

where $r_{A1}\equiv |\mathbf{R}_A-\mathbf{r}_1|$ and the kinetic-, electron-electron-, and nuclear-electron energy terms [2.2.14-19] can be identified. (In the literature [2.2.2] the notation F is sometimes used in another way i.e. the energy functional of nuclear-electron attraction is not included in it, but added after as $F[\rho(\mathbf{r}_1)] -\Sigma_{A=1,\dots,M}Z_A\int\rho(\mathbf{r}_1)r_{A1}^{-1}d\mathbf{r}_1$.) For ground state electronic energy, the 2$^{nd}$ Hohenberg – Kohn (HK) theorem [2.2.2, 20] referring to the energy variation principle demands, the true electronic DFT functional satisfies the relation $E_{electr,0}[\rho_0] \leq E_{electr,0}[\rho_{0,trial}]$ for a trial, N-normalized, everywhere positive density $\rho_{0,trial}(\mathbf{r}_1)$, where $\rho_0$ is the true solution. The N-norm is

$$\int\rho_{0,trial}(\mathbf{r}_1)d\mathbf{r}_1 = N. \qquad \text{(Eq.2.2.2)}$$

The terms of N-electron DFT (differential or integro-differential [2.2.14]) operator (D) come from integrating both sides of the electronic SE containing the Hamiltonian H for all $\mathbf{x}_i$ except $\mathbf{r}_1$ after multiplying by the complex conjugate of the same j$^{th}$ excited state wave function from the left:

$$D[\rho]\equiv D_{\nabla}[\rho]+D_{Rr}[\rho]+D_{rr}[\rho]=\rho E_{electr} \qquad \text{(Eq.2.2.3)}$$

The disadvantage of D is its non-linearity. Notice that $N^{-1}$ in Eq.2.2.1 comes from integrating both sides of Eq.2.2.3 for the 3 dimensional space and the normalization $\int\rho d\mathbf{r}_1= N$. (In detail, one must be careful with the normalization when manipulating for Eq.2.2.1: while $\int\Psi_0^*\Psi_0 d\mathbf{x}_1\dots d\mathbf{x}_N=1$ stemming from "N over N is 1 in combinatorics for HF-SCF", the $\int\rho_0 d\mathbf{r}_1= \int(\int\Psi_0^*\Psi_0 ds_1 d\mathbf{x}_2\dots d\mathbf{x}_N)d\mathbf{r}_1= \int\Psi_0^*\Psi_0 d\mathbf{x}_1\dots d\mathbf{x}_N=N$ stemming from "N over 1 is N for DFT.) The peculiarity of D is that some of its terms can have zero integral [2.2.14] in the form of Eq.2.2.1, although it plays a part in shaping the $\rho$ via Eq.2.2.3. For H-like atoms (or an unstable system of a molecular frame with one electron) the sub-case of Eq.2.2.3 is the partial differential equation, $D[N=1,\rho(\mathbf{r}_1)] \equiv -(1/4)\nabla_1^2\rho(\mathbf{r}_1) + (1/8)\rho(\mathbf{r}_1)^{-1}|\nabla_1 \rho(\mathbf{r}_1)|^2 +\rho(\mathbf{r}_1)v(\mathbf{r}_1) = E_{electr}\, \rho(\mathbf{r}_1)$, for ground and excited states [2.2.14]. In practice, the main problem with D or F is that their exact analytical formula are unknown, there are only approximations for them, the latter are problematic in programming, and more importantly in chemical accuracy (i.e. to reach the 1 kcal/mol even in energy differences).

While the DFT formula for the nuclear–electron energy term (using notation $v(\mathbf{r}_1) \equiv -\Sigma_{A=1,\dots,M}Z_A r_{A1}^{-1}$ for "external potential"),

$$V_{ne}[\rho(\mathbf{r}_1)] \equiv N^{-1}\int D_{Rr}(\rho(\mathbf{r}_1))d\mathbf{r}_1 = -\Sigma_{A=1,\dots,M}Z_A\int\rho(\mathbf{r}_1)r_{A1}^{-1}d\mathbf{r}_1 = \int v(\mathbf{r}_1)\rho(\mathbf{r}_1)d\mathbf{r}_1 \qquad \text{(Eq.2.2.4)}$$

in Eq.2.2.1, is extremely simple and analytically 100% accurate, the other two in F are very difficult algebraically and only approximations are known. (In. ref.[2.2.14] the integral formula, $D_{Rr}[\rho] = \rho(\mathbf{r}_1)v(\mathbf{r}_1) + (N-1)\int d_2(\mathbf{r}_1,\mathbf{r}_2)v(\mathbf{r}_2)d\mathbf{r}_2$, is reported for the intergo-differential equation in Eq.2.2.3, where $d_2$ is the N-normalized two-electron density. There exists another equation which compares to Eq.2.2.4 with respect to its simplicity and also its importance "at the same time" in DFT, the famous electrostatic theorem of Feynman as a subcase of Hellmann–Feynman theorem [2.2.2, 21]: $\partial E_{electr}/\partial R_{Au} = \int\rho(\mathbf{r}_1)(\partial v(\mathbf{r}_1)/\partial R_{Au})d\mathbf{r}_1 = -Z_A\int\rho(\mathbf{r}_1)(u_1-R_{Au})r_{A1}^{-3}d\mathbf{r}_1$ with u=x, y or z to be used in $\partial E_{total\ electr}/\partial R_{Au} = \partial E_{electr}/\partial R_{Au} + \partial V_{nn}/\partial R_{Au}$ with straightforward partial derivative for $V_{nn}$.)

### 2.2.2.b. Scaling correct power series for kinetic and electron-electron repulsion density functionals

Parr et al. reported a power series [2.2.22] based on the rules of density scaling [2.2.2] for the other two terms than the nuclear-electron one in F: for kinetic energy in Eq.2.2.1 the series of coordinate homogeneous functional of degree two is

$$T[\rho(\mathbf{r}_1)] \equiv N^{-1}\int D_{\nabla}(\rho(\mathbf{r}_1))d\mathbf{r}_1 = \Sigma_{j=1,\ldots n} A_j[\int\rho^{[1+2/(3j)]}d\mathbf{r}_1]^j \quad \text{(Eq.2.2.5)}$$

, while the electron-electron repulsion energy term, the functional is of a degree one

$$V_{ee}[\rho(\mathbf{r}_1)] \equiv N^{-1}\int D_{rr}(\rho(\mathbf{r}_1))d\mathbf{r}_1 = \Sigma_{j=1,\ldots n} B_j[\int\rho^{[1+1/(3j)]}d\mathbf{r}_1]^j \quad \text{(Eq.2.2.6)}$$

(The density scaling, which is the base of Liu and Parr's work [2.2.22], is well discussed in the book of general theory in ref. [2.2.2], and will not be detailed here.) In ref.[2.2.14] the 100% accurate integral formula, $D_{rr}[\rho] = (N-1)\int d_2(\mathbf{r}_1,\mathbf{r}_2)r_{12}^{-1}d\mathbf{r}_2 + [N(N-1)/2 - (N-1)]\int d_3(\mathbf{r}_1,\mathbf{r}_2,\mathbf{r}_3)r_{23}^{-1}d\mathbf{r}_2d\mathbf{r}_3$, is reported for the intergo-differential equation in Eq.2.2.3, where $d_2$ and $d_3$ are the N-normalized two- and three-electron densities. These are symmetric (called r-symetric) in exchange of $\mathbf{r}_i$ and $\mathbf{r}_j$. Furthermore, the two-electron density functional, $V_{ee} \equiv (1/N)\int D_{rr}[d_2]d\mathbf{r}_1d\mathbf{r}_2 = ((N-1)/2)\int d_2r_{12}^{-1}d\mathbf{r}_1d\mathbf{r}_2$, is also 100% accurate analytically [2.2.14]: however, the N- representability is not so simple for $d_2$ and for $d_3$. The latter means that, when $d_2$ or $d_3$ is expanded into a series of e.g. Gaussian type orbital (GTO) in 6 or 9 dimensional $(\mathbf{r}_1,\mathbf{r}_2)$ or $(\mathbf{r}_1,\mathbf{r}_2,\mathbf{r}_3)$ space as r-symmetric function, one must ensure that it can be de-convoluted into an anti-symmetric 4N dimensional wavefunction (generally it is not necessarily possible). Furthermore, $\int d_3(\mathbf{r}_1,\mathbf{r}_2,\mathbf{r}_3)d\mathbf{r}_3 = d_2(\mathbf{r}_1,\mathbf{r}_2)$ and $\int d_2(\mathbf{r}_1,\mathbf{r}_2)d\mathbf{r}_2 = \rho(\mathbf{r}_1)$ hold.

Before we analyze Eqs.2.2.5-6 and their consequences further, we mention that while there has always been some work on "moment expansions" of the electron density, the work really started in earnest with the work of Agnes Nagy in the mid-1990's, and the subsequent work from the Parr group that this stimulated. The idea is incredibly attractive: one can rewrite every density functional as a function of the moments of the density. (In practice, it is a bit tricky, because one has to make sure the moments are complete; cf. ref. [2.2.23].) This allows one to replace the functional analysis in DFT with simple multivariate calculus, which is a huge formal advantage. Most of the work (the only exception we know of is a tiny bit of work from Parr [2.2.23]) assumes that quantities can be written as a linear function of the moments, though that is obviously an incorrect assumption, thought it is perhaps a useful approximation. The biggest drawback of these approaches is that most moment expansions (and especially most nonlinear moment expansions) are not size consistent. The biggest advantage of this approach is that it works well (if not excellently) and that there are beautiful mathematical results, including an explicit method for finding the exact universal density functional from the form of the density functional for one specific system [2.2.23-24]. Our contribution here fits into this context.

### 2.2.2.c. Truncation opportunities and the series constants in scaling correct power series for density functionals

Truncation j=1 in Eq.2.2.5 provides the classical Thomas-Fermi (TF) formula ($T \approx A_1\int\rho^{5/3}d\mathbf{r}_1$) as the main term for T with TF constant [2.2.2] $c_F = (3/10)(3\pi^2)^{2/3} = 2.871234 \approx A_1$. The rest, mostly in KS formalism, is approximated in the literature: with local, non-local, spin and spinless, gradient corrected DFT functionals for ground state. These contain the derivatives of $\rho_0$, and have completely different forms than Eq.2.2.5. For example, $T[\rho_0(\mathbf{r}_1)] \approx \int[c_F\rho_0^{5/3} + (\lambda/8)|\nabla_1\rho_0|^2/\rho_0 + \text{corr.terms}]d\mathbf{r}_1$ form is the so-called Weizsacker gradient correction [2.2.2, 16]. (In the TF+$\lambda$W theories, the estimation for $\lambda$ is between 1/9 and 1/5 [2.2.2, 16], however, a very popular choice, early on, was $\lambda$=1 [2.2.25-28].) We will not summarize the vast literature about it here, but as analyzed below, we mention that Eqs.2.2.5-6 have reality via the general property of functions capable to be expanded into series. The constants $A_j$ in Eq.2.2.5 can be subdivided as

$$A_1 \equiv c_{10}c_F \text{ and } A_j = c_{10}c_F a_j = A_1 a_j \quad \text{for } j=2,3,4,\ldots \qquad \text{(Eq.2.2.7)}$$

where the $c_{10}$ is supposed to correct the TF constant, and the others ($a_j$) are "behind" $A_1$ for higher terms without N-dependence. The $c_{10}$, $a_2$, $a_3$, $a_4$, … can come from parameter fitting, ($c_{10} > 0$ is not far from unity, and $|a_j| < c_{10}$ for j=2,3,4, …).

Truncation j=1 in Eq.2.2.6 gives the main term as $V_{ee} = B_1\int\rho^{4/3}d\mathbf{r}_1 + \text{corr}_1$ with $B_1 \approx 2^{-1/3}(N-1)^{2/3}$, mentioned and analyzed in ref.[2.2.2], however, it can also only be the main term of correction (Dirac exchange functional approximation [2.2.6, 18] with constant $B_{Dirac}$) if the main term is taken as the classical Coulomb repulsion energy as $V_{ee} = (1/2)\int\rho_0(\mathbf{r}_1)\rho_0(\mathbf{r}_2)r_{12}^{-1}d\mathbf{r}_1d\mathbf{r}_2 + \text{corr}_2$ with $\text{corr}_2 = B_{Dirac}\int\rho_0^{4/3}d\mathbf{r}_1$. The latter is a more accurate approximation, i.e. generally $|\text{corr}_2| < |\text{corr}_1|$, however, both $\text{corr}_i$ are necessary for accuracy. This coincidence is not accidental, since the Dirac formula is also a scaling correct power series truncated after the first term. $V_{ee}[\rho(\mathbf{r}_1)]$ scales one, but the classical Coulomb repulsion energy approximation, scales two, which is incorrect. It is the main source of correlation energy ($E_{corr}$), which is the major problem with respect to chemical accuracy in HF-SCF (due to the lack of formula) or KS (which does have a suitable but yet not perfect formula) methods, stemming from using only a single Slater determinant to approximate $\Psi_0$.

We also mention that, the classical Coulomb repulsion energy as the main algebraic term contains only the first (in fact second) powers of $\rho_0$, good for HF-SCF routine where a GTO basis set is used to make the integrations analytical in the approximation. The Dirac formula is one trial of the many which is designed to estimate its error ($\text{corr}_2$). It is considered in great detail in the literature. Recall again the local, non-local, spin and spinless, gradient corrected, hybrid, etc. exchange-correlation functionals in KS formalism. Historically, the promising approximations of $\text{corr}_2$ have made DFT successful in practice, but there is still no total control over its accuracy in different systems. The exact analytical form is unknown at the present time, there are only empirical formulas, parameterized and optimized mainly for ground states.

The constants $B_j$ in Eq.2.2.6 can be treated as

$$B_1 \equiv c_{20}(2^{-1/3}(N-1)^{(2/3)c200}) \text{ and } B_j = B_1 b_j \quad \text{for } j=2,3,4,\ldots \qquad \text{(Eq.2.2.8)}$$

where the $c_{20}$ is supposed to correct the expression $2^{-1/3}(N-1)^{(2/3)c200}$, and the others ($b_j$) are "behind" $B_1$ for higher terms without N-dependence. In ref.[2.2.2] $c_{200}$= 1, leaving the power simply as 2/3, but we are trying to correct this part too by tuning with the factor $c_{200}$ in later work. The $c_{200}$, $c_{20}$, $b_2$, $b_3$, $b_4$, … can come from a parameter fitting as well, ($c_{20}$, $c_{200} > 0$ are not far from unity, and $|b_j| < c_{20}$ for j=2,3,4,…).

We call the attention that it isn't strictly true that the exact form of the functional isn't known for these sorts of moment expansions. The exact form is known, but it is hopelessly complicated and, as pointed out by Ayers, contains terms that are not included in the simple series expansion in Eqs.2.2.5-6. Specifically, increasingly complicated ratios of moments appear [2.2.23-24]. While the approach is very elegant, the results in those papers [2.2.23-24] are much less favorable than those of Liu, Nagy, and Parr [2.2.29-30], probably because much the dataset being fit was much larger. The modern literature on density moments in DFT is quite small, with only a few important researchers (Nagy, Parr, and some others) have published yet results [2.2.23-24, 29-43].

### 2.2.2.d. The magnitude of the series constants of scaling correct power series for density functionals

In ref.[2.2.22], the series on the right hand side of Eq.2.2.6 is used for exchange energy [2.2.2] (as a part of $E_{corr}$ in HF-SCF formalism) or the similar magnitude exchange correlation energy [2.2.3] (to describe Coulomb and Fermi holes in KS formalism), however, here we use it to estimate the entire $V_{ee}$. In this way, if we do not truncate too soon, Eqs.2.2.5-6 help to avoid the problem of $E_{corr}$, as well as the algorithm becoming simpler since one does not have to deal with messy derivatives and non-local integrals. In ref.[2.2.22] the formulas in Eqs.2.2.5-6 were tested with HF-SCF one-electron densities, $\rho_{0,HF\text{-}SCF}$, and among many conclusions, the most important thing for us now is that 3-4 terms may be enough for chemical accuracy, and in accord, the absolute value of the coefficients decrease rapidly. Here we use these formulas as direct substitution into Eq.2.2.1 and solve them for ground state, but we emphasize that Eqs.2.2.4-6 hold for excited sates as well. The rapid decrease of $A_j$ and $B_j$ are not surprising if one recognizes that a crude and more precise estimation (see 2.2.Appendix) for the magnitude of power terms in Eqs.2.2.5-6 is $[\int\rho^{[1+a/(3j)]}d\mathbf{r}_1]^j \sim [\int\rho d\mathbf{r}_1]^j = N^j$ and

$$[\int\rho^{[1+a/(3j)]}d\mathbf{r}_1]^j \approx G^j \text{ with } G(x) \equiv N^x(N^3/\pi)^{x-1}/x^3 \text{ and } x \equiv 1+a/(3j), \qquad \text{(Eq.2.2.9)}$$

respectively, where a = 1 or 2 and j>1, as well as notice that G(1)=N, – i.e. these increase rapidly with N, which is large in calculations for molecules. Another hypothesis is that a replacement of $a_j$ and $b_j$ in Eqs.2.2.7-8 with $a_jN^{1-j}$ and $b_jN^{1-j}$ for j=2,3,4, … may be better (i.e. in this way $a_j$ and $b_j$ are more independent from N), because Eqs.2.2.5-6 contain larger powers of N in view of Eq.2.2.9. (View in the perspective of dimensional analysis that functional $\int\rho^{5/3}d\mathbf{r}_1$ approximates T and $\int\rho^{4/3}d\mathbf{r}_1$ approximates $V_{ee}$, while functional $\int\rho d\mathbf{r}_1$ gives N.) An answer for this will be given via tests on real systems. Also, we must mention that N-dependent functionals are not size consistent. Hard to find a good reference for that (though it is obvious), for example, it was mentioned by Parr in his work on the Fermi-Amaldi model [2.2.44].

### 2.2.2.e. Density functionals in scaling correct power series form versus partial differential equation to describe molecular systems

Replacing a partial differential eigenvalue equation with a functional containing algebraic equation can be perilous, but recall the truth that $E_{electr}$ in SE depends only on the $\{\mathbf{R}_A, Z_A\}_{A=1,2,\ldots M}$ molecular frame (the basic, original inspiration of the HK theorems). In this way, for a power series expansion, e.g. with ρ (in which DFT states that it contains all the properties), it is just a question of the quality of the power series that has been chosen. We point out that the HF-SCF and CI methods (see Fock matrix, secular equation, or advanced devices based on series expansion, etc.) obtain roots (energy values) from a $k^{th}$ order determinant transformed from SE. This also corresponds to a $k^{th}$ order algebraic equation, so from this view the form examined here should not be considered unusual. We mention that Parr et al. [2.2.2, 45] recognized that in F for ground state, the problem of finding the electronic structure of molecules reduces to treat some algebraic expressions for the $2^{nd}$ HK theorem back in 1979. However, due to the early stages of computers, problems of accuracy and finding a convenient method to locate the extremum, it has not moved into a focus of interest. Mostly, HF-SCF level $\rho_{0,HF\text{-}SCF}(\mathbf{r}_1)$ functions were used to test these kinds of DFT functionals.

Expanding with the Weizsacker term, Handy et al. [2.2.18] have tested the non-KS formalism DFT functionals by expanding the $\rho_0(\mathbf{r}_1)$ with a gaussian basis set. Before and more generally, similar approaches have been examined by Liu and Parr [2.2.22], however, they only focused on atoms and correlation, here we also examine molecules, as well as we consider Eqs.2.2.5-6 as the main and correction terms together. Most importantly [2.2.22], Parr introduced a genius form of expansion in $\rho$, which is correct in density scaling. The related ideas of the contracted Schrödinger equation by Nakatsuji [2.2.46] and March's density differential equation [2.2.47] should also be taken into account. These latter two papers, which are more than thirty years old, have established an idea to reduce the dimensionality of the electronic Schrödinger equation, but up until today, the main task is to work out a tractable algorithm that overcomes the difficulty stemming from its non-linear nature.

The N-representability (meaning that anti-symmetric wave function exists which generates this $\rho$ via $\rho(\mathbf{r}_1)= \int\Psi^*\Psi ds_1 d\mathbf{x}_2 \ldots d\mathbf{x}_N$, most importantly for ground state) is simple [2.2.48-49] in one-electron DFT, where in fact there is no N-representability problem, however, one must approximate the exact energy functional (F or D). In this N-representability problem, we cite Garrod and Percus for the pair density (first attempt, [2.2.50]), Davidson (explicit demonstraton, [2.2.51]), Pistol (lattice model solution, [2.2.52]), and Ayers (real-space solution [2.2.53]), as well as there is a review by Davidson [2.2.54]. According to our particular problem here, the best references for the N-representability of the one-electron distribution function (the normal electron density) are refs.[2.2.55-56]. The advantage of the refs.[2.2.17, 48-49] is that they demonstrate that even without the (quite simple) constraints on the electron density, one can minimize the energy, provided that the functionals are defined appropriately. Below, a model is introduced wherein the HF-SCF or KS orbitals will be completely eliminated from the DFT formulation and the density can be solved directly from these DFT functionals. It has been a commonly desired task [2.2.17] and this work targets that task. More precisely, the only real disadvantage of KS orbitals in DFT is their 3N dimensional nature in spatial space, otherwise, by using KS orbitals one regains a one-electron picture from a many electron DFT problem where electron correlation is included. The form and energies of KS orbitals are the basis of many qualitative rationalizations of DFT results.

Here we perform the first ever variational calculation with a moment functional (to our knowledge) and have several interesting, provocative, and even controversial ideas on how the method might be applied. There has been a lot of work on orbital-free DFT, those methods are effective, but not very accurate, see details on this in refs.[2.2.57-60]. Finally, Eqs.2.2.4-6 are not restricted to the vicinity of stationary points on the potential energy surface, and do not suffer with the open or closed shell programming complexities that are present in HF-SCF or KS methods.

### 2.2.3. Lagrangian for Scaling Correct Power Series Energy Functional to Estimate Ground State Electronic Energy, its solution, analysis and discussion

Now, we are at the main part of our work. Using the "Lagrange's method of undetermined multiplier" for the 2nd HK theorem, we must minimize the functional $L^* = E_{electr}[\rho_0] - \lambda(\int\rho_0(\mathbf{r}_1)d\mathbf{r}_1 - N)$ with respect to ground state one-electron density, $\rho_0$, where we emphasize the ground state with subscript zero. The $\lambda$ is the Lagrange multiplier, providing that the density is normalized to N electrons as constrain. Using Eqs.2.2.1-6 it takes the form

$$L^* = \Sigma_{j=1,\dots n} A_j[\int\rho_0^{[1+2/(3j)]}d\mathbf{r}_1]^j + \Sigma_{j=1,\dots n} B_j[\int\rho_0^{[1+1/(3j)]}d\mathbf{r}_1]^j + \int v(\mathbf{r}_1)\rho_0(\mathbf{r}_1)d\mathbf{r}_1 - \lambda(\int\rho_0(\mathbf{r}_1)d\mathbf{r}_1 - N) \qquad \text{(Eq.2.2.10)}$$

In HF-SCF there are also constrains for all pairs of molecular orbitals (MO) to get them ortonormal, here we have only one constraint: the N-normalization. (To be more precise, we also need to force the density to be nonnegative, e.g., by writing it as the square of some other function, see a particular choice in Eq.2.2.18 below.) Therefore, we set the first variation in $L^*$ equal to zero

$$0 = \delta L^* = \int\{\Sigma_{j=1,\dots n} (1+2/(3j))jA_j [\int\rho_0^{[1+2/(3j)]}d\mathbf{r}_1]^{j-1}\rho_0^{2/(3j)} + \Sigma_{j=1,\dots n} (1+1/(3j))jB_j[\int\rho_0^{[1+1/(3j)]}d\mathbf{r}_1]^{j-1}\rho_0^{1/(3j)} + v(\mathbf{r}_1) - \lambda\}\delta\rho_0(\mathbf{r}_1)d\mathbf{r}_1 \qquad \text{(Eq.2.2.11)}$$

where we have integrals to evaluate inside the integrand. Since $\delta\rho_0$ is arbitrary, it follows that the quantity in the curly brackets must be zero. It yields

$$\Sigma_{j=1,\dots n}\{(1+2/(3j))jA_j[\int\rho_0^{[1+2/(3j)]}d\mathbf{r}_1]^{j-1}\rho_0^{2/(3j)} + (1+1/(3j))jB_j[\int\rho_0^{[1+1/(3j)]}d\mathbf{r}_1]^{j-1}\rho_0^{1/(3j)}\} + v(\mathbf{r}_1) = \lambda \qquad \text{(Eq.2.2.12)}$$

which is a 3 spatial dimension integral equation. Eq.2.2.12 is a substitute for the 4N spin-orbit dimension partial differential electronic Schrödinger equation, and the ground state electronic energy is just $E_{electr,0}(\{\mathbf{R}_A,Z_A\})/N \equiv \lambda$. The $\lambda$ is called the chemical potential. (More precisely, the electronic chemical potential is the partial derivative $\partial E_{electr,0}/\partial N$, which is more sensible if N is large.) The larger the n, the more accurate Eq.2.2.12 is, and hopefully it converges fast. (We mention that there are other ways to choose terms to the exact answer in the moment expansion [2.2.23].) Recall that in the HF-SCF formalism the single Slater determinant is a very good but not a very precise form of approximation, the drawback of HF-SCF, that is, it needs correction (correlation calculation) to reach chemical accuracy even for energy differences. In Eqs.2.2.11-12 the series expansion of $\Psi_0$ via $\rho_0$ ala Parr can be taken as arbitrarily accurate with increasing n.

#### 2.2.3.1. Semi-analytical solutions for truncated scaling correct power series functionals or Lagrangian

##### 2.2.3.1.1. First order truncation

It is useful to consider the truncations for Eq.2.2.12. If n=1, then

$$(5/3)A_1\rho_0^{2/3} + (4/3)B_1\rho_0^{1/3} + v(\mathbf{r}_1) \approx \lambda \equiv E_{electr,0} / N \qquad \text{(Eq.2.2.13)}$$

This equation, which is a crude approximation for the solution of Eq.2.2.3 or the SE for ground state, has been considered in detail in ref.[2.2.15]. Although it does have some flaws (see below), it maintains some positive properties, e.g. it approximates absolute ground state electronic energy values quite well for atoms with $\{Z_A < 11$ and $2 < N < Z_A +2\}$ and molecules built of these atoms. For atoms, it predicts [2.2.15] ionization potential better in some cases than e.g. the HF-SCF/6-31G*. For atoms, and irrespective of the nuclear frame of equilibrium geometry molecules, it provides [2.2.15] a very close value to the virial theorem value: 2. It should also be noticed that such comparisons (HF and definitely a too small basis set) are not relevant, because actual calculations in practice try to use larger and larger basis sets. However, many researchers agree that functionals should be equally suitable for smaller basis sets too. Generally, one should use a relatively small basis to start with and put more emphasis on the empirical parameterization. It is an appealing idea to assume that the parameterization performed within a small basis expansion set can absorb some deficiencies of the basis limitations itself (see p.108 in ref.[2.2.3]). The latter has also been confirmed as a side result in a new correlation calculation method published in ref. [2.2.61]. Actually, at this point in this section we show a basis set free algorithm, but in later truncations for more accurate results below, basis set will be necessary.

The algorithm to solve Eq.2.2.3 for this truncation is as follows. With the substitution $z \equiv \rho_0^{1/3}$, Eq.2.2.13 is a second order algebraic equation, and can be solved for $z(\mathbf{r}_1, \lambda_{approx})$, providing the $\rho_{0,approx} = z^3(\mathbf{r}_1, \lambda_{approx})$. (We draw attention to the fact that $v(\mathbf{r}_1)$ does not appear in the kinetic and electron-electron Hamiltonian or DFT operator explicitly: however, $\rho$ includes it implicitly as $\rho = \rho(v(\mathbf{r}_1))$, a known functional relationship, see ref.[2.2.2] – it is satisfied via the approximate Eq.2.2.13.) For Eq.2.2.13, it is important and convincing to mention some early work of March's [2.2.47] who derived the $\rho_0(x_1) = \text{const.}(\lambda - v(x_1))^{1/2}$ for independent fermions in one dimension (which is exact in those very simple conditions as well as $\lambda$ being the chemical potential). The energy functional in Eq.2.2.10 in this case (n=1) is

$$E_{electr}[\rho_0] \approx E_{electr,0,approx} \equiv \int (A_1\rho_0^{5/3} + B_1\rho_0^{4/3} + v(\mathbf{r}_1)\rho_0) d\mathbf{r}_1 \qquad \text{(Eq.2.2.14)}$$

and $\rho_{0,approx}(\mathbf{r}_1) = Cz^3(\mathbf{r}_1, \lambda_{approx})$ is supposed to be substituted for $\rho_0$ in the integrand, where C fixes $N = C\int z^3(\mathbf{r}_1, \lambda_{approx}) d\mathbf{r}_1$ to be satisfied in every step. Integral in Eq.2.2.14 depends on $\lambda$ such as exhibiting one well defined minimum, and the numerical solution for $\partial E_{electr}[\rho_0]/\partial\lambda = 0$ yields the approximation for ground state electronic energy (recall the 2nd HK theorem). This completes the procedure indicated in the title of this section. All the integral evaluations must be numerical. Its two parameters, $c_{10}$ and $c_{20}$ via Eqs.2.2.7-8, were fitted [2.2.15] to ground state electronic energies of CI atomic ions. The limit and integral behavior of model $\rho_0$ from Eq.2.2.13 is as follows. For a peak at $\mathbf{R}_A$, the integral $\int Z_A^{3/2} R_{A1}^{-3/2} d\mathbf{r}_1 = Z_A^{3/2} \int |\mathbf{r}_1|^{-3/2} d\mathbf{r}_1 = 4\pi Z_A^{3/2} \int u^2 u^{-3/2} du = (8\pi/3)(Z_A r_{max})^{3/2}$ over a sphere with radius $r_{max}$ around $\mathbf{R}_A$, i.e. finite, although the integrand value is infinite at $\mathbf{R}_A$. Similarly holds for other algebraic powers of model $\rho_0$ appearing for integration in Eq.2.2.14. However, because the "ring off" at around a radial $r_{max}$ value via the discriminant in Eq.2.2.13 (that is a 2nd order algebraic equation for $\rho_0^{1/3}$), the integral in Eq.2.2.14 is finite in the algorithm. Computer investigations have shown that this internal $r_{max}$ value in the calculation is about 3-4 times the van der Waals' radius of atoms in a molecule. Although the energy integral is finite in Eq.2.2.14, one drawback of model $\rho_0$ in Eq.2.2.13 is that $\lim_{\mathbf{r}1 \to \mathbf{R}A}[\rho_{0,approx}] = \infty$, instead of an expected finite value as has just been mentioned. Recall e.g. the analytic atomic 1s solution for H-like atoms.

The flaws of truncation at n=1 can be summarized as follows: 1. The normalization constant, C, is not 1 (it was introduced after the solution of a second order equation), but about 0.46, however, it has at least a very small dependency on $(Z_A, N)$ of atoms and nuclear frame $(\{\mathbf{R}_A, Z_A\}, N = \Sigma Z_A)$ of (at least neutral or close to neutral) molecules. 2. The $\rho_{0,approx}$ depends on certain power of $v(\mathbf{r}_1)$ yielding infinite values at any nuclei $\mathbf{R}_A$, and as it is characteristic in certain DFT approximations, it can not show the shell structure for atoms, it is only a decaying function. 3. The value of $\lambda_{approx}$ at minimum $(\lambda_{approx,min})$ multiplied by N, and the integral $(E_{electr,0,approx})$ at this $\lambda_{approx,min}$ has to be the same, i.e. they have to be self-consistent, however, instead [2.2.15], $E_{electr,0,approx}/(N\lambda_{approx,min}) \approx 3$, showing a marginally stronger dependency on the nuclear frame than C above. 4. The check for virial theorem for atoms and equilibrium molecules gives values between 1.95-2.05, which is a bit off the expected theoretical value 2.00. 5. If atoms with atomic charge Z > 10 are involved in the molecular system, the calculated electronic energy value is absolutely invalid, it means that powers belonging to n=1 are not enough. 6. It can not account for chemical bond, for example calculating energy of atomization yields that known stable molecules are not stable via Eq.2.2.14; it is in accord with the known weakness of TF functional if it stands alone for kinetic energy – again, the truncation at n=1 is too early.

We also note, that the energy functional in Eq.2.2.14 is a known, well-established expression [2.2.2] as first approximation. The $\rho_{0,\text{approx}}$ from Eq.2.2.13 provides an educated guess for trial one-electron density that was new in ref.[2.2.15], and new here is that how it relates to the Lagrangian. Another way to originate Eq.2.2.13 is by integrating SE yields $\int(\Psi^* H\Psi - E\Psi^*\Psi)d\mathbf{x}_1...d\mathbf{x}_N = 0$, and if – trivially – one substitutes a true solution $\Psi$ (more specifically the ground state $\Psi_0$) into the left hand side, the integral is zero because the integrand itself is a zero function. Actually, it has a more rigorous internal relationship because an integral can be zero too if the integrand is not a zero function. This integral form of SE also leads to the true DFT functional with the device of reducing the variables of integration mentioned in Eq.2.2.1. Now, start with the approximate Eq.2.2.14 as established in the literature and rearrange it as $\int(NA_1\rho_0^{5/3} + Nv(\mathbf{r}_1)\rho_0 + NB_1\rho_0^{4/3} - \rho_0 E_{\text{electr},0,\text{approx}})d\mathbf{r}_1 \approx 0$ with back-substitution (or extension) of $N=\int\rho_0 d\mathbf{r}_1$. All terms in the integrand are supposed to follow the individual energy terms (kinetic, etc.). In this way one can suppose, that the integrand in this case is also an approximate zero function, and we have recovered a similar equation to Eq.2.2.13 for expressing $\rho_{0,\text{approx}}$, if one divides with $\rho_0$ and N. The $\lambda \equiv E_{\text{electr},0}/N$ correspondence can be recognized. (Notice that fitting parameters, $c_{10}$ in $A_1$ and $c_{20}$ in $B_1$, can absorb 5/3 and 4/3 respectively in Eq.2.2.13 as was done in ref.[2.2.15].) This derivation is a bit more complex than it looks at first take, some more details can be found in ref.[2.2.14]: For example, there can be additive terms in the integrand which individually yield zero integral value, although they are not zero functions, see equation 32 in ref.[2.2.14]. Consequently, these terms do not show up in F but shapes the $\rho_0$ in an equation like Eq.2.2.13 for Eq.2.2.14. Generally speaking, it is just another relationship between the exact DFT functional F and the exact DFT integro-differential operator D mentioned above. Eq.2.2.13 suffers from the crude truncation (n=1) after a rigorous and exact derivation yielding Eq.2.2.12, but Eqs.2.2.13-14 at least show explicitly how the DFT functional and its approximate solution behave as functions.

### 2.2.3.1.2. Second order truncation

Truncation of Eq.2.2.12 at n=2 yields

$$(5/3)A_1\rho_0^{2/3} + (8/3)A_2[\int\rho_0^{4/3}d\mathbf{r}_1]\rho_0^{1/3} + (4/3)B_1\rho_0^{1/3} + (7/3)B_2[\int\rho_0^{7/6}d\mathbf{r}_1]\rho_0^{1/6} + v(\mathbf{r}_1) \approx \lambda \quad \text{(Eq.2.2.15)}$$

With substitution $u \equiv \rho_0^{1/6}$, one yields the integral-equation for $u(\mathbf{r}_1)$ as

$$(5/3)A_1u^4 + (8/3)A_2[\int u^8 d\mathbf{r}_1]u^2 + (4/3)B_1u^2 + (7/3)B_2[\int u^7 d\mathbf{r}_1]u + v(\mathbf{r}_1) \approx \lambda \quad \text{(Eq.2.2.16)}$$

The procedure should be similar to truncation at n=1, however, it is much more difficult to solve this equation for $u \equiv \rho_0^{1/6}$ than Eq.2.2.13 for $z \equiv \rho_0^{1/3}$. But obviously, Eq.2.2.16 is more flexible than Eq.2.2.13, i.e. it provides a more realistic $u^6 \equiv \rho_{0,\text{approx}}(\mathbf{r}_1, \lambda)$ in accord with the fact that the series in Eqs.2.2.5-6 converge rapidly [2.2.22]. Furthermore, because it is not an algebraic equation like Eq.2.2.13, but a relation between functions and their integrals (or functions and their derivatives), the cusp condition for $\rho_{0,\text{approx}}$ is better satisfied, e.g. it yields finite value at any nuclei, $\mathbf{R}_A$. It can be simplified crudely as $\int u^8 d\mathbf{r}_1 \equiv \int\rho_0^{8/6}d\mathbf{r}_1 \approx \int u^7 d\mathbf{r}_1 \equiv \int\rho_0^{7/6}d\mathbf{r}_1 \approx (\int\rho_0 d\mathbf{r}_1) = N$ or $N^{8/6 \text{ or } 7/6}$, or more realistically as G(x=4/3) and G(x=7/6), respectively, according to Eq.2.2.9. With the later, Eq.2.2.16 degrades to

$(5/3)A_1u^4 + [(8/3)G(x=4/3)A_2+ (4/3)B_1]u^2 +(7/3)G(x=7/6)B_2u +v(\mathbf{r}_1) \approx \lambda$, (Eq.2.2.17)
which is a 4th order algebraic equation in u, a more powerful equation than Eq.2.2.13, which was 2nd order in z. It can be solved analytically because the general analytic solution exists up to a 4th order algebraic equation: however, like z from Eq.2.2.13, u via Eq.2.2.17 contains certain positive powers of $v(\mathbf{r}_1)$, crudely represented as $\rho_{0,approx} \sim v(\mathbf{r}_1)^c$, which suffers again from the unrealistic cusp $\lim_{\mathbf{r}1\to\mathbf{R}A} [\rho_{0,approx}] = \infty$. On the other hand, the analytic solution of a 4th order algebraic equation is via the 3rd order algebraic equation, and Eq.2.2.17 does not have the problem of negative discriminant (artificial error) for some far away positions from the nuclear frame as Eq.2.2.13 has. The better properties of Eq.2.2.17 to predict electronic energy will be reported in a later work. Notice the fine detail that the truncation n=1 of Eq.2.2.12 it only yields an algebraic equation (second order, Eq.2.2.13) suffering from e.g. the wrong cusp description beside the not adequate accuracy, while truncation n=2 (or higher) of Eq.2.2.12 yields integral (or differential) equation (Eq.2.2.15) which is more flexible to describe properties, e.g. cusps, it is also more accurate.

### 2.2.3.1.3. Larger than second order truncation

Equations 13 and 17 reveal that Eqs.2.2.10-12 need numerical integration and the power series in it should go up to at least n=4 in the truncation to accurately describe shell structure, ground state electronic energy ($E_{electr,0}$) and ground state one-electron density ($\rho_0(\mathbf{r}_1)$), as a function of nuclear frame ($\{\mathbf{R}_A, Z_A\}$) and number of electrons (N). Numerically solving Eq.2.2.15 has similar, at least, not fewer programming complexities than the more accurate Eq.2.2.12, so, one should evaluate the latter for more accuracy wherein the n is a tuning variable for accuracy. Eqs.2.2.13 and 17 can show approximately how the true algebraic form of $\rho_0(\mathbf{r}_1)$ may analytically behave, what is less visible by the numerical solution of Eq.2.2.12. $E_{electr,0}(\{\mathbf{R}_A,Z_A\},N)$ via Eq.2.2.12 is supposed to be accurate not only in the vicinity of stationary points but in the van der Waals regions as well, and for open and closed shell molecular systems since spin pairing effect does not come up in this method in contrast to HF-SCF and post HF-SCF methods. Analytical integration may possibly be used for Gaussian type atomic orbital (GTO) basis set, see chapter 5 below. If numerical integration is chosen, the Slater type atomic orbital (STO) basis set can also be used, a more realistic choice, since it provides faster convergence. The parameters $c_{10}$, $a_2$, $a_3$, $a_4$, … and $c_{20}$, $c_{200}$, $b_2$, $b_3$, $b_4$, … entered in Eqs.2.2.7-8 must be fitted to e.g. CI atomic and atomic ion ground state energies, which are supposed to be transferable [2.2.15] for molecular systems at any place on the potential energy surface for ground state. Of course, accurately known molecular $E_{electr,0}$ values can also be used for fitting procedure, e.g. stationary point G2 values, however these are not totally accurate in contrast to atomic $E_{electr,0}$ values from CI calculations or measurements, of which accuracy is far below the chemical accuracy. Eq.2.2.13 (n=1) needed fit [2.2.15] for $A_1$ and $B_1$ or equivalently for $c_{10}$ and $c_{20}$, Eq.2.2.15 (n=2) needs fit for $A_1$, $A_2$, $B_1$ and $B_2$ or equivalently for $c_{10}$, $a_2$, $c_{20}$ and $b_2$ as well as $c_{200}$ is unity or additional fitting parameter for Eqs.2.2.13 and 15. If truncation is at n> 2, see chapter 3.2 below, the $c_{10}$, $a_2$, …, $a_n$, $c_{20}$, $c_{200}$, $b_2$, …, $b_n$ parameters need to be fitted. It will be detailed in a later paper, the theoretical foundation is described in chapters below.

We also mention that, it is pretty well known that e.g. the ionization potential can be well approximated using the moment expansion. However, if one considers a long series of atoms, with very different electron numbers, the density-moment expansion stops working as well [2.2.31]. In the literature there are opinions that, first, it is difficult to expand the Coulomb energy in terms of moments. For example, in the study of Tran, there are impressive results but the results are far from the sub-milli-Hartree accuracy that is needed in practical computations of the Coulomb energy [2.2.32], and that work only treats the absolute simplest case – atoms. Second, and more importantly, the moment expansion (at least the linear moment expansion [2.2.23]) does not necessarily converge. Not every functional can be exactly expressed as a simple power series of the moments, even trying to reproduce a simple functional (like the Weizsäcker kinetic energy, or the Coulomb energy). One must keep these in mind when we suggest alternative functional in Eq.2.2.25 below. However, the promising results in ref.[2.2.15] on atoms and molecules indicate the opportunities in this direction.

### 2.2.3.2. Numerical solution for scaling correct power series functional at larger truncations

#### 2.2.3.2.1. LCAO approximation of one-electron density to start the minimization

As was just analyzed, Eq.2.2.10 must be solved numerically for the minimum (extremum) because Eq.2.2.16 and the higher n-truncated cases of Eq.2.2.12 cannot be solved analytically. For this purpose, we have to proceed further with Eq.2.2.11. The density can be expanded as a linear combination of atomic orbitals (LCAO) where the basis, $\{b_k(\mathbf{r}_1)\}_{k=1\ldots L}$, is consisted of e.g. L Cartesian, $x^a y^b z^c \exp(-\alpha r_{A1}{}^i)$ STO (i=1) or GTO (i=2) basis functions (or contracted basis functions), a wisely chosen bunch, grouped and centered on each nuclei (as in HF-SCF or KS methods for MO's). A good choice for this form is

$$\rho_0(\mathbf{r}_1) \approx (\Sigma_{k=1\ldots L}\, d_k\, b_k(\mathbf{r}_1))^2 \qquad \text{(Eq.2.2.18)}$$

,a function which is positive everywhere, as required by the 2$^{nd}$ HK. If L is large enough and the basis set is wisely chosen, the true $\rho_0$ will be approximated correctly. One must at least consider the concept of "minimal basis" [2.2.1]. Recall the form of HF-SCF or KS one-electron density [2.2.1] with N/2 (>1, e.g. closed shell) ortonormal molecular orbitals, $\rho_0(\mathbf{r}_1) \approx 2\Sigma_{i=1\ldots N/2}\, [\Sigma_{k=1\ldots L1}\, c_{ik}\, b_k(\mathbf{r}_1)]^2 \geq 0$, wherein the $\{c_{ik}\}$ set, also called LCAO coefficients, contains $L_1(N/2)$ elements, and the square brackets contain the i$^{th}$ MO, called $f_i(\mathbf{r}_1)$, see also Eq.2.2.23 below. Though there are only $L_1$ square terms in it, $(2\Sigma_i c_{ik}{}^2) b_k{}^2$, running via index k and the $2\Sigma_i c_{ik}{}^2$ corresponds to $d_k{}^2$, but more cross terms, $b_k b_j$, if $L=L_1$, in comparison to Eq.2.2.18. (For example, if N=4 and $L=L_1=2$, it yields $\rho_0 \approx d_1{}^2 b_1{}^2 + d_2{}^2 b_2{}^2 + 2 d_1 d_2 b_1 b_2$ by Eq.2.2.18, i.e. the weight of cross term (inter-nuclear electron density, $b_1 b_2$) is fixed by square term coefficients $d_1$ and $d_2$. On the other hand, the HF-SCF density (just mentioned or Eq.2.2.23 below) provides $2(c_{11}{}^2+c_{21}{}^2) b_1{}^2 + 2(c_{12}{}^2+c_{22}{}^2) b_2{}^2 + 4(c_{11}c_{12}+c_{21}c_{22}) b_1 b_2$, i.e. there are four coefficients to weight the three terms, i.e. the inter-nuclear electron density can be tuned more independently from the weight of cusps ($b_1{}^2$ and $b_2{}^2$). Notice, that in this simple example Eq.2.2.18 requires 2 parameters ($d_1$, $d_2$) to fit vs. 4 parameters ($c_{11}$, $c_{12}$, $c_{21}$, $c_{22}$) via Eq.2.2.23; to improve the flexibility of the former we must allow for the fact $L>L_1$. Notice also, that Eq.2.2.23 builds the parts of electron density (cusps $b_1{}^2$ and $b_2{}^2$ and bond $b_1 b_2$) via 4 parameters, although 3 would be enough as in $C_1 b_1{}^2 + C_2 b_2{}^2 + C_3 b_1 b_2$ for the DFT central variable on this basis set $\{b_1,b_2\}$ level.) In this way one should accept $L>L_1$, but first, one should use STO and numerical integration instead of GTO with analytical integration employed by HF-SCF or KS. This allows the basis set to have fewer elements, i.e. with a lower value L. Secondly, using the HF-SCF or KS methods N/2 or (N+1)/2 pieces of MO's must be approximated, while here there is only one quantity, the $\rho_0$. Finally, if L is large enough, Eq.2.2.18 is a good approximation. There is another way to choose the form than Eq.2.2.18: $\rho_0(\mathbf{r}_1) \approx \Sigma_{k=1\ldots L}\Sigma_{j=k\ldots L}(c_{kj}\, b_k(\mathbf{r}_1) b_j(\mathbf{r}_1))$ with symmetric $c_{kj}= c_{jk}$ property, containing L(L+1)/2 terms, i.e. more cross terms. However, one must ensure that it provides everywhere positive one-electron density which is more difficult than in the case of Eq.2.2.18. In Eq.2.2.18, the right hand side is obviously $\geq 0$, only the L needs to be increased for more accuracy. (For example, $[\Sigma_{k=1\ldots L}\Sigma_{j=k\ldots L}(c_{kj}\ b_k(\mathbf{r}_1)b_j(\mathbf{r}_1))]^2$ is a way to ensure positive function values or the form in Eq.2.2.23 itself, but in respect to programming it has more difficult indexing than Eq.2.2.18. Like the approximate Slater form of the wavefunction in HF-SCF, this model serves to approximate one-electron density in DFT, and along with the choice of basis set, both are crucial points for effective calculation.)

#### 2.2.3.2.2. Numerical recipe for direct minimization

Inserting Eq.2.2.18 into Eq.2.2.10, and taking the derivative with respect to $d_k$ and $\lambda$, Eq.2.2.11 reformulates as

$$0 = \partial L^*/\partial d_i = \Sigma_{j=1,\ldots n}\ (1+2/(3j))jA_j\ [\int\rho_0^{[1+2/(3j)]}d\mathbf{r}_1]^{j-1}\int\rho_0^{2/(3j)}\rho_{0i})d\mathbf{r}_1$$
$$+ \Sigma_{j=1,\ldots n}\ (1+1/(3j))jB_j[\int\rho_0^{[1+1/(3j)]}d\mathbf{r}_1]^{j-1}\int\rho_0^{1/(3j)}\rho_{0i}d\mathbf{r}_1\ + \int(v(\mathbf{r}_1) - \lambda)\rho_{0i}d\mathbf{r}_1 \quad \text{(Eq.2.2.19)}$$
$$0 = \partial L^*/\partial\lambda = N - \int\rho_0(\mathbf{r}_1)d\mathbf{r}_1 \quad \text{(Eq.2.2.20)}$$

for i=1…L. Using Eq.2.2.18, the partial derivatives are simply

$$\rho_{0i} \equiv \partial\rho_0(\mathbf{r}_1)/\partial d_i = 2\ b_i(\mathbf{r}_1)\ (\Sigma_{k=1\ldots L}\ d_k\ b_k(\mathbf{r}_1)),\ \text{and}\ \rho_{0im} \equiv \partial^2\rho_0(\mathbf{r}_1)/\partial d_i\partial d_m = 2\ b_i(\mathbf{r}_1)b_m(\mathbf{r}_1) \quad \text{(Eq.2.2.21)}$$

i.e. the second and third indices refer to the partial derivatives. As mentioned [2.2.22], a truncation at n= 4 in Eq.2.2.10 is adequate. The system in Eqs.2.2.19-21 is non-linear, so e.g. the "steepest descent (gradient)" method can be employed. This method needs the second derivatives or Jacobian $\partial^2L^*/(\partial d_i\partial d_m)$ for all i,m= 1,…,L+1, where $d_{L+1} \equiv \lambda$. The Jacobian matrix is ((L+1)x(L+1) dimensional, with element at row i and column m as $W_{im} \equiv \partial^2L^*/(\partial d_i\partial d_m)$ a straightforward 2nd derivative.

Eqs.2.2.10 and 18 yield the powers for the recently defined LCAO parameters in Eq.2.2.18, $d_k$. In $L^*$, the $d_k$ parameters obtain the integer and non-integer power values between 1 and maximum 2(1+2/(3n))n= 2(n+2/3)= 9.333 for n= 4; roughly and generally 2n+1. It means, that the $L^*$, that we have to optimize via Eqs.2.2.19-20, is an L+1 dimensional polynomial with parameter vector $\{d_k\}_{k=1\ldots L+1}$ with roughly the degree of about 2n+1 if truncation at j=n is taken – and hopefully, the truncation n=4 will provide a flexible enough function to calculate ground state electronic energy and one-electron density for molecular systems. The coefficients to $d_k$ come from integrating certain powers of linear combinations of the basis functions $b_k(\mathbf{r}_1)$, see note on the non-integer powers as well as GTO and STO basis sets in this respect below. Also, see Eq.2.2.19 for the algebraic position of $v(\mathbf{r}_1)$, where the $\int b_k(\mathbf{r}_1)b_i(\mathbf{r}_1)r_{A1}^{-1}d\mathbf{r}_1$ kind of integral comes up, but the $\int b_k(\mathbf{r}_1)b_i(\mathbf{r}_2)r_{12}^{-1}d\mathbf{r}_1d\mathbf{r}_2$ kind (also characteristic in HF-SCF or KS method) does not.

### 2.2.3.3. On some expected behaviors of the Lagrangian

Note must be made on the asymptotic (far from the nuclei) behavior of the density: From the general theory [2.2.3], as well as it was discussed above, the $\rho_0(\mathbf{r}_1)$ must 1. be a non-negative function of only the three spatial variables, 2. vanish at infinity ($\rho_0(\mathbf{r}_1 \to \infty) = 0$), and 3. integrate to the total number of electrons (Eq.2.2.2). The first property is ensured with the right hand side of Eq.2.2.18, the second is ensured with e.g. a nuclear centered GTO or STO basis set, and the third is ensured with Eq.2.2.10 via $\lambda$. However, a finer relationship [2.2.3] is its asymptotic exponential decay for large distances from all nuclei, that is $\rho_0(\mathbf{r}_1) \sim \exp[-2\ \mathrm{sqrt}(2I)\ |\mathbf{r}_1|]$, where I is the exact first ionization energy of the system. This latter can be easily ensured with e.g. an STO basis set, and the LCAO coefficients are supposed to yield the constant value, 2 sqrt(2I) as well as the large enough value of n in Eqs.2.2.5-6 is important in this respect.

For the question, how do these series mathematically converge, the answer can come from refs.[2.2.15, 22]. Evidences have been shown [2.2.15] that the main parts of different energies come from j=1 in Eqs.2.2.5-6 or 10, and the convergence is very fast [2.2.22] thereafter: a truncation at n= 4 or 5 may enough for chemical accuracy.

With respect to the spin states or spin polarization (measured through the spin-polarization parameter as $\xi \equiv (\rho_{0,\alpha}-\rho_{0,\beta})/\rho_0$ with $\rho_0=\rho_{0,\alpha}+\rho_{0,\beta}$ [2.2.3, 62]), Eq.2.2.10, solved e.g. via Eqs.2.2.19-20, describes the one having the lowest, i.e. the ground state energy, inherent in the Lagrangian method and 2$^{nd}$ HK theorem. This means that, the choice, what HF, post HF, and KS methods have in this respect, e.g. to enforce singlet vs. triplet spin state calculation, for example $1s^2 2s^2 2p_x^2$ (excited state) vs. $1s^2 2s^2 2p_x^1 2p_y^1$ (ground state in agreement with Hund's rule) carbon atom, is not available here; Eq.2.2.10 always provides the ground state only. But on the other hand, basic problem present in HF, post HF, and KS methods with increasing bonds lengths or atom-atom distance inside a molecule toward transforming non-stable molecule or system with van der Waals distances, that is, for example stable $H_2$ molecule (Spin= ½ - ½= 0, approximate wave function= $(\alpha_1\beta_2-\alpha_2\beta_1)f(\mathbf{r}_1)f(\mathbf{r}_2)$) vs. well but not infinitively separated two H atoms (e.g. Spin= ½ + ½ =1, recall RHF, UHF modes etc.), is not a problem in Eq.2.2.10, it is supposed to handle any change in inter-nuclear distances in the system under consideration continuously. At this point we call the attention that although Eqs.2.2.3-6 are valid for ground and excited states too, but Eq.2.2.10, or its solution via e.g. Eqs.2.2.19-20, is on the calculation track for ground state only, (recall that the HK theorems apply to ground states).

Dobson [2.2.63-64] have shown, among others, that van der Waals complexes can be accurately accounted by $\rho_0(\mathbf{r}_1)\rho_0(\mathbf{r}_2)h(\mathbf{r}_1,\mathbf{r}_2)$ kernels, which use only the density and not its derivatives – notice that this kernel description is formally the definition of the Coulomb hole. Approximations leading to Eq.2.2.10 also use only $\rho_0$, but with using local functionals, capable to account for correlation effects [2.2.22].

### 2.2.4. Two serious tests have already been made for the scaling correct power series energy functionals

The calculation and proof test on atoms and molecules in ref.[2.2.15] for n=1 in Eqs.2.2.5-6 leading to Eq.2.2.13, which is $(5/3)NA_1\rho_0^{5/3} + (4/3)NB_1\rho_0^{4/3} + Nv(\mathbf{r}_1)\rho_0 \approx \rho_0 E_{electr,0}$ via a small reformulation to get comparable expression to the one reported in ref.[2.2.15], has yielded that 1: $(5/3)NA_1 = 1.4433781907\ N\ c_F$; notice that $A_1 \equiv c_{10}c_F$ (Eq.2.2.7) so $(5/3)c_{10} = 1.4433781907$, (to avoid confusion, the entire product $(5/3)c_{10}$ here was called $c_{10}$ in ref. [2.2.15], i.e. $(5/3)c_{10}^{here} = c_{10}^{ref.[2.2.15]}$), 2: $(4/3)NB_1 = 0.8374131087\ N\ 2^{-1/3}(N-1)^{(2/3)}$, notice that $B_1 \equiv c_{20}(2^{-1/3}(N-1)^{(2/3)})$ (Eq.2.2.8 with $c_{200}$ =1) so $(4/3)c_{20} = 0.8374131087$, (to avoid confusion, the entire product $(4/3)c_{20}$ here was called $c_{20}$ in ref. [2.2.15], i.e. $(4/3)c_{20}^{here} = c_{20}^{ref.[2.2.15]}$). Calculation on ionisation potentials of atoms is demonstrated on 2.2.Figure 1. Important, in ref.[2.2.15] a direct calculation for $\rho_0$ and $E_{electr,0}$ was done with a non-HF-SCF one-electron density, where the latter is the second order algebraic solution for $\rho_0$ via Eq.2.2.13. The weakness of this fit from ref.[2.2.15] is that n was truncated early, namely at n=1, as detailed in section 2.2.3.1 above.

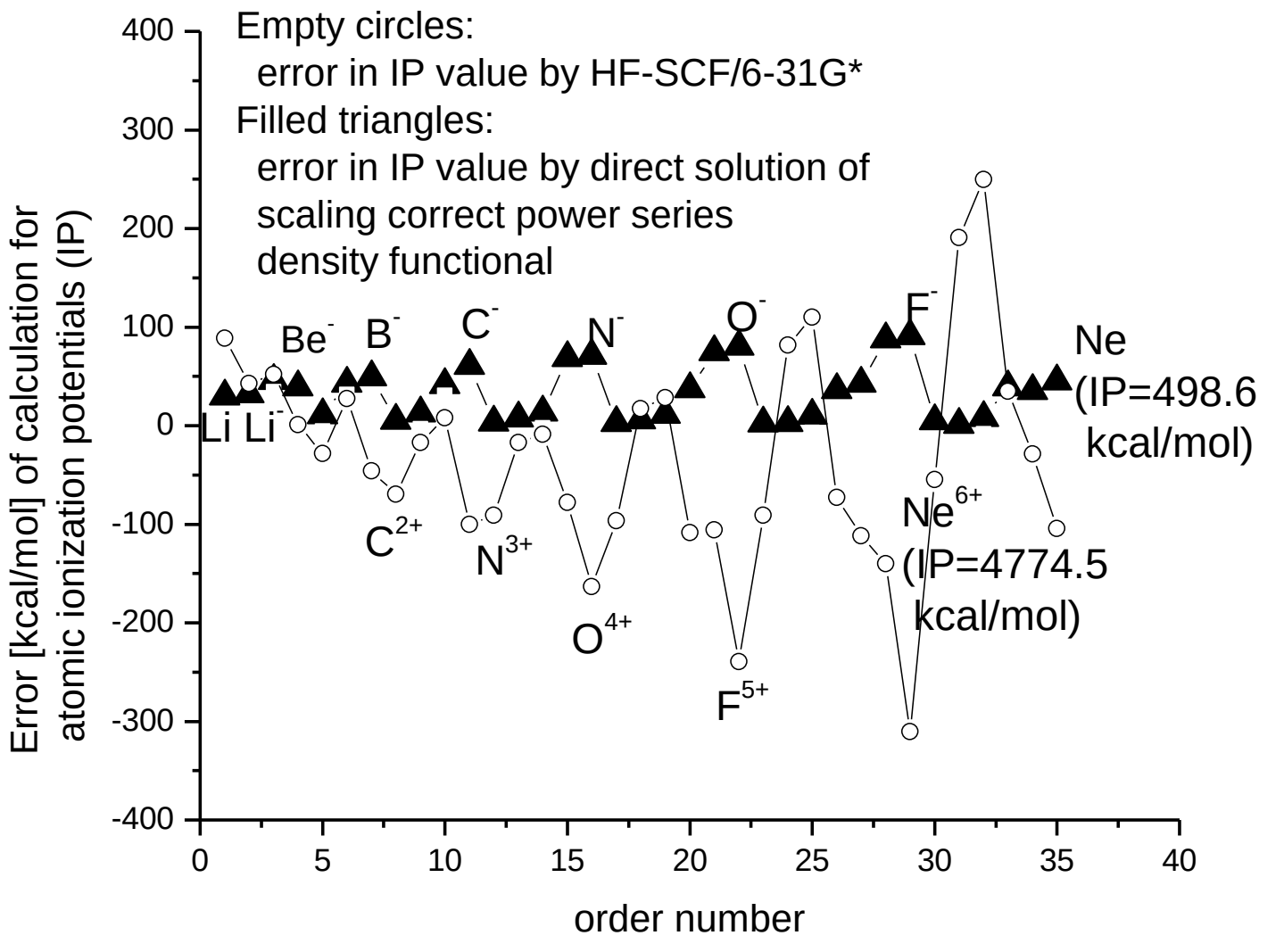

2.2.Figure 1.: Error of calculation for atomic ionization potentials (IP for $A \rightarrow A^+$) by HF-SCF/6-31G*, and Eq.2.2.10 (n=1, basis set free calculation) with optimized parameters from ref.[2.2.15], ordered with increasing atomic number, Z, and number of electrons, N; (the huge IP values for $Ne^{6+}$ and Ne are marked for comparison).

Another calculation and proof test on atoms was made in ref.[2.2.22] for n=3 in Eqs.2.2.5-6: however, they used HF-SCF one-electron density, and accurate $E_{electr,0}$ to fit the parameters of Eqs.2.2.5-6, i.e. not a direct calculation for $\rho_0$ and $E_{electr,0}$, but a fit after an ab intio calculation. Another difference is that instead of Eq.2.2.6, they used the form $J[\rho_0]+\Sigma_{j=1,\dots n}C_{xj}[\int\rho_0^{[1+1/(3j)]}d\mathbf{r}_1]^j$, where $J[\rho_0]=C_J[\int\rho_0^{6/5}d\mathbf{r}_1]^{5/3}$. In this way the classical Coulomb repulsion energy $J[\rho_0]$ was modeled with a DFT form (i.e. with a functional of $\rho_0$) and the scaling correct series (Eq.2.2.6) was used to estimate the related part of correlation energy. (Recall the Dirac form mentioned above for comparison.) In this way T and J were high values, while $C_{xj}$'s served for only a correction. The fitted constants they have obtained are $A_1$= 3.26422, $A_2$= -0.02631, $A_3$= 0.00498, $C_J$ = 1.0829, $C_{x1}$= -0.85238, $C_{x2}$= 0.004911, $C_{x3}$= -0.000074. Although this fit in ref.[2.2.22] was suggested for correlation calculation after a HF-SCF routine, and its credence was demonstrated, its artifact in the view of this work is that it was not a direct calculation for $\rho_0$ for the fit. Furthermore, we draw attention to the fact that the form, $V_{ee}= C_J[\int\rho_0^{6/5}d\mathbf{r}_1]^{5/3} + \Sigma_{j=1,\dots n}C_{xj}[\int\rho_0^{[1+1/(3j)]}d\mathbf{r}_1]^j$, used instead of Eq.2.2.6 is another proper power series, and Eq.2.2.10 can be changed accordingly (that is: the sum for $B_j$ has to be replaced by this sum for $C_J$ and $C_{xj}$).

We should also mention the classical example known for decades [2.2.2-3] and indicated above, with respect to this parameter value and fitting: Slater's approximation of HF exchange energy $\int\varepsilon_x(\rho(\mathbf{r}_1))\rho(\mathbf{r}_1)d\mathbf{r}_1 \approx C_x\int\rho(\mathbf{r}_1)^{4/3}d\mathbf{r}_1$, where $C_x = -(9/8)(3/\pi)^{1/3}\alpha$ and is called the $X\alpha$ method. It depends solely on the local values of the electron density, and $\alpha$ is an adjustable, semi-empirical parameter. It has enjoyed a significant amount of popularity among physicists, but has never made much impact on chemistry. This 4/3-power law of electron density was obtained from two completely different approaches [2.2.3]: Slater (based on the potential of a uniformly charged sphere from standard electrostatics with radius corresponding to the Fermi hole), Bloch in 1929, and Dirac (as named above and as it is cited among chemists) in 1930 - using the concept of uniform electron gas, a fictitious model system of constant electron density. Typical values obey $0.666 < \alpha < 1$ and depend on molecular frame (N, {$\mathbf{R}_A$, $Z_A$}) slightly i.e. $\alpha$ stays in this interval: however, taking only an average value from this interval can destroy the chemical accuracy (1 kcal/mol) even for differences of ground state electronic energy in the outcome of the calculation for different systems. Exact mathematical form for this small functional dependence or fluctuation is unknown, but a well-established fact is that a major part of it is described by this 4/3-power formula. The rest can be described by higher power terms via Eq.2.2.6. Similarly, as mentioned above, in their model Thomas and Fermi [2.2.2-3] have arrived at the $T_{TF} = c_F\int\rho^{5/3}d\mathbf{r}_1$, very simple expression for the kinetic energy based on the uniform electron gas also.

As it has been demonstrated in refs.[2.2.15 and 22] fit to existing ground state (e.g. CI) atomic, atomic ions and (e.g. G2 or G3) molecular energies are feasible for parameters in Eqs.2.2.5-6. The best next step is a parameter fit for Eqs.2.2.5-6 with direct calculation for $\rho_0$ and $E_{electr,0}$ and n>3 as described above in section 2.2.3 and based on Eq.2.2.10 – it is our plan and will be reported in a later work.

We must mention some other parametrizations and tests: Burke and coworkers have recently developed semi-classical approaches, for example for the kinetic energy of one-dimensional model finite systems the leading corrections to local approximations as a functional of the potential have been derived [2.2.65]. Furthermore, condition on the Kohn–Sham kinetic energy and modern parametrization of the Thomas–Fermi density was elaborated by them [2.2.66], being the recovery of the correct expansion yields a condition on the Kohn–Sham kinetic energy that is important for the accuracy of approximate kinetic energy functionals for atoms, molecules, and solids – see also the discussion in section 2.2.5 in this relation.

## 2.2.5. Relation to HF-SCF and Kohn-Sham formalism, and replacing all the time consuming gaussian based two and four center integrals

### 2.2.5.1. Comparing the energy functionals

Classically, to solve the SE for ground state electronic energy, $E_{electr,0}$, and normalized, anti-symmetric $\Psi_0$ with the help of the variation principle, one must minimize the energy functional $E[\Psi_{0,trial}] = \langle\Psi_{0,trial}|H|\Psi_{0,trial}\rangle$, where H is the electronic Hamiltonian (used also at the beginning for Eq.2.2.1) with the known bra-ket notation [2.2.1-2]. The HF approximation [2.2.1-2] uses a single Slater determinant for $\Psi_{0,trial}$, denoted by $S_{trial}$, obtaining

$$E_{HF}[S_{trial}] = \Sigma_{i=1,\dots,N} \int\psi_i^*(\mathbf{x}_1)[(-½)\nabla_1^2 + v(\mathbf{r}_1)]\psi_i(\mathbf{x}_1)d\mathbf{x}_1 + (½)\Sigma_{i,j=1,\dots,N} (J_{ij}-K_{ij}) \qquad \text{(Eq.2.2.22)}$$

where the Coulomb integral is $J_{ij} = \iint\psi_i(\mathbf{x}_1)\psi_i^*(\mathbf{x}_1)[r_{12}^{-1}]\psi_j^*(\mathbf{x}_2)\psi_j(\mathbf{x}_2)d\mathbf{x}_1d\mathbf{x}_2$, and the exchange integral is $K_{ij} = \iint\psi_i^*(\mathbf{x}_1)\psi_j(\mathbf{x}_1)[r_{12}^{-1}]\psi_i(\mathbf{x}_2)\psi_j^*(\mathbf{x}_2)d\mathbf{x}_1d\mathbf{x}_2$. In Eq.2.2.22, the $\psi_i$ are the orto-normalized MO's approximated with LCAO using a GTO basis to be able to solve the integrals analytically, they also have pair-wise the same spatial part to build up $S_{trial}$. The latter means that there is a common spatial function, f, such as $\psi_1(\mathbf{x}_1) = \alpha_1 f(\mathbf{x}_1)$ and $\psi_2(\mathbf{x}_2) = \beta_2 f(\mathbf{x}_2)$ for i=1,2, g for i=3,4, and so on, where f,g,… are orto-normalized also. A systematic notation for them is {$f_1, f_2, \dots, f_{(N/2)}$ or $f_{(N+1)/2}$} for even and odd N respectively. In this way the ground state one-electron density (via $\rho_0 \equiv \int\Psi_0^*\Psi_0 ds_1 d\mathbf{x}_2\dots d\mathbf{x}_N \approx \int S_{opt}^* S_{opt} ds_1 d\mathbf{x}_2\dots d\mathbf{x}_N$) is

$$\rho_{0,Slater,trial} = 2\Sigma_{i=1,\dots,N/2} f_i^2 \text{ or } 2\Sigma_{i=1,\dots,(N-1)/2} f_i^2 + f_{(N+1)/2}^2. \qquad \text{(Eq.2.2.23)}$$

Above we have used the notation $\rho_{0,HF-SCF}$ for this, meaning the optimized one - electron density, but now we want to emphasize the Slater determinant formalism included during the optimization. The main cases [2.2.1] abbreviated as RHF, ROHF and UHF, etc. will not be detailed further now. Eq.2.2.22 is decomposed to the so called HF or Fock differential equations and with standard computer routines the minimization problem can be treated to find the LCAO parameters for all HF molecular orbitals $f_i$. Because a single Slater determinant is only an approximation for the $\Psi_0$, $E_{electr,0} < E_{HF}[S_{opt}]$ (see, variation principle), and the difference comes from the basis set error and correlation energy mentioned above. The latter, called $E_{corr}$, is calculated after [2.2.1-2] the HF-SCF routine.

The KS theory [2.2.2-3], based on DFT, corrects this error during (i.e. not after) the algorithm using the single determinant form via the functional

$$E_{KS}[\rho_{0,Slater,trial}] = -\Sigma_{i=1,\ldots,N/2}\int f_i^*(\mathbf{r}_1)\nabla_1^2 f_i(\mathbf{r}_1)d\mathbf{r}_1 + \int v(\mathbf{r}_1)\rho_{0,Slater,trial}(\mathbf{r}_1)d\mathbf{r}_1 + (1/2)\int\rho_{0,Slater,trial}(\mathbf{r}_1)\rho_{0,Slater,trial}(\mathbf{r}_2)r_{12}^{-1}d\mathbf{r}_1 d\mathbf{r}_2 + E_{xc}(\rho_{0,Slater,trial}) \quad \text{(Eq.2.2.24)}$$

for the even N in the sum and the corresponding one for the odd N. Comparing Eq.2.2.22 and 24, the terms with nablas are basically the same (before and after integration over spins), actually it is a main idea in KS formalism. The latter means that the functional in Eq.2.2.24 does not only contain one-electron density, as it should in DFT (e.g. in Eq.2.2.10), but it also contains one-electron orbitals - overcoming the difficulties of not knowing the peculiar form of kinetic energy functional. The terms with the external (mostly nuclear frame) potential, v, in Eq.2.2.24 is also basically the same as the single determinant based approximation in Eq.2.2.22. However, the terms with $r_{12}^{-1}$ have basically different forms in Eq.2.2.22 vs. 24 even though they yield similar values in comparison to the magnitude of $E_{electr,0}$. The term $E_{xc}$ (exchange-correlation) [2.2.1-3] in Eq.2.2.24 is an extra device in comparison to Eq.2.2.22, and according to DFT it can correct the error that Eq.2.2.22 makes. Actually, the main idea in KS formalism comes into effect during the SCF routine, and for this reason the HF orbitals from Eq.2.2.22 and KS orbitals from Eq.2.2.24 are not the same: however, they are close to each other. Similarly, the final correlation energy and basis set error, the $E_{corr}$ and $E_{xc}$ values are also close to each other at least on the same basis set level, and about 1-2 % of $E_{electr,0}$. Eq.2.2.24 is decomposed to the so called KS differential equations, and with standard computer routines the minimization problem can be solved to find the LCAO parameters for all KS molecular orbitals $f_i$. Here we do not address the problem of the single Slater determinant RHF vs. UHF behavior in the vicinity of stationary points vs. dissociating or van der Waals region etc. that Eq.2.2.22 has, but Eq.2.2.24 can treat better.

A great technical advantage of KS formalism was that all the previously existing HF-SCF routines in the history of computation chemistry could be modified easily to handle any or both of Eq.2.2.22 or Eq.2.2.24. Knowing a very good form approximating the exact $E_{xc}$ in Eq.2.2.24, the $E_{electr,0} \approx E_{KS}[\text{optimized } \rho_{0,Slater}]$ would hold very accurately. Without details, the acceptable approximate forms of $E_{xc}$ in Eq.2.2.24 embody the following properties focusing on the subject of this work: 1, it provides algebraic variation properties, but not necessarily variation with respect to $E_{electr,0}$, 2, it is designed (e.g. in its parameter fit for approximate $E_{xc}$) to Slater determinant or Eq.2.2.23, although $\rho_{0,trial}$ can possess other algebraic forms, see e.g. the solution of Eq.2.2.13 [2.2.15] and Eq.2.2.17 or Eq.2.2.18, 3, there are some simple but important mathematical properties [2.2.3] that $E_{xc}$ or parts of it should provide, for example, the two-electron density is factorized as $d_2(\mathbf{x}_1,\mathbf{x}_2)= \rho(\mathbf{x}_1)\rho(\mathbf{x}_2)(1+f(\mathbf{x}_1,\mathbf{x}_2))$, where f is called the correlation factor, and of course it strongly relates to the correlation energy, and theoretically $\int\rho(\mathbf{x}_2)f(\mathbf{x}_1,\mathbf{x}_2)d\mathbf{x}_2 = -1$, a property that an approximation must have - at least approximately, etc.. For property 1, recall DFT concerning the variation of the true functional $E_{electr,0} \leq E[\rho_{0,trial}]=T[\rho_{0,trial}]+ V_{ee}[\rho_{0,trial}]+ V_{ne}[\rho_{0,trial}]$ with true T, $V_{ne}$ and $V_{ee}$ functionals and the minimum at the true N-normalized $\rho_0$, as opposed to the fact that $E_{xc}$ in Eq.2.2.24 is only an approximate functional in practice. For property 2, recall that a theoretically correct $E_{xc}$ for true $\rho_0$ re-corrects the error made by the previous terms in Eq.2.2.24, but $\rho_0$ is approximated with a Slater form, so its correction has to be provided also. Furthermore, the basis set error is always present in practice. As we have emphasized, $E_{xc}$ in Eq.2.2.24 is not exactly known, only approximate forms are available and tested. We do not summarize the vast literature about it, but we do mention that no overall approximate form is yet known which provides the chemically accurate calculations: geometry optimums, energy differences, vibronic frequencies, dipol moments, van der Waals forces, etc. for any system. Instead, each existing and accepted functional is good for certain groups of chemical systems and problems only, but fails for some others. For this reason, different functionals are used in different systems or problems, that is not adequate scientifically. We have also mentioned above that the suggested and accepted approximate forms for $E_{xc}$ in the literature include derivatives (gradients) of the one-electron densities (or spin densities), in contrast, here we deal with scaling correct power series including the main (j=1) and correction terms (j>1) in Eqs.2.2.5-6.

Based on the previous parts of this work, an alternative functional to the ones in Eqs.2.2.22 and 24 used in HF-SCF and KS routines, respectively, is

$$E_{SCMF-1}[\rho_{0,Slater,trial}] = -\Sigma_{i=1,\dots,N/2}\int f_i^*(\mathbf{r}_1)\nabla_1^2 f_i(\mathbf{r}_1)d\mathbf{r}_1 + \int v(\mathbf{r}_1)\rho_{0,Slater,trial}(\mathbf{r}_1)d\mathbf{r}_1 + \Sigma_{j=1,\dots,n}\{C_j[\int\rho_{0,Slater,trial}{}^{[1+2/(3j)]}d\mathbf{r}_1]^j + B_j[\int\rho_{0,Slater,trial}{}^{[1+1/(3j)]}d\mathbf{r}_1]^j\} \quad \text{(Eq.2.2.25)}$$

for the even N in the sum and the corresponding for odd N, as well as SCMF stands for "scaling correct moment functional". In Eq.2.2.25 the terms with $C_j$ originate from Eq.2.2.5 knowing that (unlike Eq.2.2.5) it does not approximate the entire kinetic energy, T, but only the correction to the first sum with nabla (now it is used as in the original idea from Parr et al. [2.2.22]). For this reason, the values of $C_j$ are different from $A_j$ in Eq.2.2.5, but presumably $C_1$, $C_2$, $C_3$, … are similar to $A_2$, $A_3$, $A_4$, … respectively, in magnitude. (Compare it algebraically to the aforementioned Dirac exchange term - an algebraic form that can be a correction term as well as a main term, depending how one uses it.) The terms with $B_j$ in Eq.2.2.25 account for the entire electron-electron repulsion energy from Eq.2.2.6, and being a scaling correct power series, it is supposed to account accurately if n is large enough and the accurate $\rho_0$ is used, and no correction like $E_{xc}$ (exchange correlation, Fermi and Coulomb hole, etc.) is needed. The only adjustment needed in the values of C and B coefficients with respect to Eqs.2.2.5-6 is that in Eq.2.2.25 the Slater type one-electron density in Eq.2.2.23 is used, not the real one as in Eqs.2.2.5-6 – although with a larger basis set, the difference may be negligible. According to practice [2.2.22], n should go up to 4 or 5 to reach chemical accuracy.

In the Fock equations associated to HF-SCF or KS method the two-electron operators ($r_{ij}^{-1}$) are reduced to one-electron operators via some standard non-local integration technique. In this way, algebraically a Slater determinant is a 100% accurate form on the way to finding the antisymmetric solution for the system of Fock equations [2.2.3]. With $E_{xc}$ in Eq.2.2.24 the Fock equations own the "perturbation" toward a solution to hit the value of the ground state electronic energy of its stem equation - the electronic Schrödinger equation, more accurately. In Eq.2.2.25 there are only one-electron terms and operators, so for the associated Fock equations, a Slaterian form for the solution is also adequate in the beginning too. The role of $E_{xc}$ in Eq.2.2.24 corresponds to the role of terms with coefficients B for j>1 or 2 in Eq.2.2.25, and terms with coefficients C provide even more improvement (namely in the form of kinetic operators used). An accepted drawback of $E_{xc}$ in Eq.2.2.24 in the literature, is that it cannot be improved systematically, while the scaling correct power series in Eq.2.2.25 provides systematic improvement by the increasing n. Moreover, in the next chapter we analyze that the integration needed in Eq.2.2.25 puts us on the road to improve upon Eq.2.2.22-24; the key point is that there is no term with $r_{ij}^{-1}$.

In the literature there are opinions that it is difficult to expand the Coulomb energy in terms of moments, although the above mentioned $C_J[\int\rho_0^{6/5}d\mathbf{r}_1]^{5/3}$ main term does the job good [2.2.22] with three correctional terms, and even the $B_1\int\rho^{4/3}d\mathbf{r}_1$ main term in Eq.2.2.6 performs remarkable [2.2.15] without correctional terms. We mention that some researchers strictly say that, expanding the Coulomb term in moments is ridiculous. It may work for atoms, or for molecules near equilibrium. But it can never work for a system like the stretched HF dimer because the 1/r electrostatic repulsion between electrons on the fragments is missed. Of course, this is a practical point (not a mathematical point): it just suggests that (mathematically) the moment expansion converges very (perhaps infinitely) slowly. By this reason, instead of Eqs.2.2.6 or 10, the alternative form (compare what KS uses, $V_{ee}\approx(1/2)\int\rho_0(\mathbf{r}_1)\rho_0(\mathbf{r}_2)r_{12}^{-1}d\mathbf{r}_1d\mathbf{r}_2+\int\varepsilon_{xc}\rho_0 d\mathbf{r}_1$, the origin of the huge literature on exchange-correlation energy) is

$$V_{ee}[\rho(\mathbf{r}_1)]= (1/2)\int\rho_0(\mathbf{r}_1)\rho_0(\mathbf{r}_2)r_{12}^{-1}d\mathbf{r}_1d\mathbf{r}_2+\Sigma_{j=1,\dots n}B_j[\int\rho_0^{[1+1/(3j)]}d\mathbf{r}_1]^j \quad \text{(Eq.2.2.26)}$$

In this way, the classical Coulomb term is the major one, and the entire set of coefficients B falls into correction terms - compare to Eq.2.2.6 where $B_1$ was connected to the major term and $B_2$, $B_3$, … were the correctional ones. It can be used to develop Eq.2.2.25, more, another alternate energy functional to Eqs.2.2.22, 24 and 25, and to the parts in Eq.2.2.10 without $\lambda$ is

$$E_{SCMF-2}[\rho_0] = \Sigma_{j=1,\dots n} A_j[\int\rho_0^{[1+2/(3j)]}d\mathbf{r}_1]^j + \int v(\mathbf{r}_1)\rho_0(\mathbf{r}_1)d\mathbf{r}_1 + (1/2)\int\rho_0(\mathbf{r}_1)\rho_0(\mathbf{r}_2)r_{12}^{-1}d\mathbf{r}_1d\mathbf{r}_2+\Sigma_{j=1,\dots n}B_j[\int\rho_0^{[1+1/(3j)]}d\mathbf{r}_1]^j \quad \text{(Eq.2.2.27)}$$

with e.g. the use of Eq.2.2.18. The expressions in Eqs.2.2.10 and 19 can be changed accordingly. The additional term entering to Eq.2.2.19, by the replacement of Eq.2.2.6 with Eq.2.2.26, is $\int\rho_0(\mathbf{r}_2)(\partial\rho_0(\mathbf{r}_1)/\partial d_i)r_{12}^{-1}d\mathbf{r}_2d\mathbf{r}_1$. Notice that Eq.2.2.27 contains only $\rho_0$ at this point and Slaterian form is not a restriction. The only major restriction coming up by this classical Coulombic term is that GTO basis must be used for all terms (no way to use STO) in the corresponding expression to Eq.2.2.19, as well as numerical integration cannot be used for this term (containing $r_{12}^{-1}$) but analytical one; but of course the other terms can be evaluated only numerically. However, the known, relatively good long-range behavior of this major non-local functional in Eq.2.2.26 (firs term) is well established in a light contrast to the local functional in Eq.2.2.6. There is no nabla terms in Eq.2.2.27, so this form algebraically is rather belong to the ones in section 2.2.3.2 with respect to solution algorithm and DFT, like Eq.2.2.10, and unlike Eqs.2.2.22, 24-25.

### 2.2.5.2. In relation to numerical integration and programming

It is quite obvious that existing HF-SCF routines solving Eq.2.2.22 can easily be modified to solve Eq.2.2.25, as it was possible for the KS formalism in Eq.2.2.24. The advantage of Eq.2.2.25 is that expensive analytical integration for terms containing $r_{12}^{-1}$ are not necessary as opposed to Eqs.2.2.22 and 24, the most time consuming procedure, despite the fact that subroutines for these analytical integrals, i.e. $r_{12}^{-1}$ in the integrand multiplied with GTO's, are highly developed today in practice. On the other hand, the necessary tools of numerical integration for the nonlinear $E_{xc}$ (Eq.2.2.24) are already built in existing codes using KS formalism. The terms with C and B coefficients in Eq.2.2.25 can also be calculated numerically and accurately without larger additional programming input. Numerical integration is the first choice for the terms with $B_j$ and $C_j$ in Eq.2.2.25, because of its not-integer powers. Furthermore, since the numerical integration used in these tasks is very accurate, the first two terms in Eq.2.2.25 - kinetic and nuclear-electron attraction, can also be shifted to the numerical integration subroutine, making the program structure simpler. Recall, that in the case of GTO basis set these two terms are traditionally evaluated analytically. In this way, even the faster, more powerful STO basis set can be used. (Notice that for integrals with $r_{ij}^{-1}$ in the integrand, the 6 dimensional $\int\ldots d\mathbf{r}_1 d\mathbf{r}_2$ must be evaluated after all algebraic reductions, not so, for the simpler 3 dimensional $\int\ldots d\mathbf{r}_1$, e.g. for the kinetic and nuclear-electron terms. For analytic integration one had to switch from STO to GTO basis set in the HF-SCF method, but with the KS method, the additional non-linear $E_{xc}$ term has entered the arena, and it cannot be integrated analytically, even though it only needs the $\int\ldots d\mathbf{r}_1$ and the use of the GTO basis set cannot counterbalance the non-linearity. As a consequence, numerical integration is necessary for this part: however, with computational chemistry problems, fast numerical integration is not available for $\int\ldots d\mathbf{r}_1 d\mathbf{r}_2$, not so, in the case of $\int\ldots d\mathbf{r}_1$, but Eq.2.2.25 is free of $r_{ij}^{-1}$.) It must be emphasized that "numerical integration for all integrals" has an important effect on computation time, i.e. the computation time in this case is proportional to the number of nuclei (M) in contrast to $N^c$ characteristic in HF-SCF or KS routines (c = 2 to 4), where N is the number of electrons. Recall that M<<N in practically important systems, also recall the study in ref.[2.2.15]. The use of STO or GTO basis with numerical integration for all terms in Eq.2.2.25 and the fit for C and B parameters will be reported in later work.

We must mention that many powerful multicenter integration schemes, based on density fitting (close to proposal above), have been developed since, see e.g. review chapters 7.3–7.6 in ref.[2.2.3] and references therein, as well as Ahlrichs et al. [2.2.67-68] and Parrinello et al. [2.2.69].

### 2.2.5.3. In relation to analytical integration

If one wants to stay with analytical integration, avoiding the numerical, our note on it is as follows for terms with C and B coefficients in Eq.2.2.25, or A and B coefficients in Eqs.2.2.10 or 27: The fractional (i.e. not integer) power, $\rho_{?0,Slater,trial}^c$ in Eq.2.2.25 with Eq.2.2.23 or $\rho_0^c$ in Eqs.2.2.10 or 27 with Eq.2.2.18 takes the values c= 5/3, 8/6, … and 4/3, 7/6, … up to a truncation, where the c's are between 1 (j or n $\to \infty$) and 5/3= 1.6667 (j=1). The $\rho_0^c$ can be expanded with the help of e.g. a truncated Taylor series containing integer powers instead of c's: in this way, the use of GTO basis, allows analytical integration. It is important to note that the Taylor series must be expanded with a region of values of $\rho_0$, and not with a particular value of it. Here we briefly mention the way to analytical integration. Using the least square device $\int[\rho_0^c - (\Sigma_{i=1\ldots k}\, a_i \rho_0^i)]^2 d\rho_0$ = minimum, where the integration is from 0 to R, the $a_i$ coefficients can be obtained, as well as k=4-6 is enough according to our preliminary tests. The one-electron density is always positive, so the interval for integration starts from zero (far away from the molecule or at the nods, if any, inside the molecular frame), while the maximum value (R) is what a one-electron density can pick up. The latter is at the nuclei with maximum $Z_A$ in the nuclear frame. (In the view of "atoms in molecule" and "core electrons" concepts, recall that in H-like atoms the radial part of 1s wavefunction is $R_{10}=2Z_A^{3/2}\exp(-Z_A r_1)$, so the maximal value (R) is $\max(\rho_0)\sim R_{10}^2\sim Z_A^3$ – i.e. magnitudes larger than the values at bonds or inter-atomic regions.) The arising difficulty is, that generally $\rho_0$ has high sharp spherical-like peaks at the nuclei and much smoother curves and lower values on a graph, for example in the case of an equilibrium molecular system. As a consequence, weighted least square is more suitable. The above form is an adequate choice, since the value of c is not far from the integer powers (i) present, as well as that, it has a similar monotonity to the integer powers in the expression. In this way $\rho_{0,Slater,trial}^{c}$ is replaced with $\Sigma_{i=1...k}a_i\rho_{0,Slater,trial}^{i}$ in Eq.2.2.25 and similarly in Eqs.2.2.10 and 27, and with integer powers analytical integration is possible, since the product of GTO type functions (via the sum in $\rho_{0,Slater,trial}$) is also GTO type: however, an arising problem may cancel this opportunity, e.g. in the case of, let us say, 100 or 1000 or more GTO basis functions in a basis set, the 4-6$^{th}$ power of their sum (see Eqs.2.2.18 or 23) generate an enormous number of terms to sum up. In contrast, the numerical integration needs to sum up these 100 or 1000 terms only and taking the c$^{th}$ power of that value thereafter. Finally, we must state that numerical integration is the only choice when Eqs.2.2.5-6 are involved in the functional, however, the faster STO basis set can be used.

We must mention one other way for integration: Numerical integration on a finite grid (see e.g. refs.[2.2.14-15] or chapter 7.4 in ref.[2.2.3]) may have a disadvantages, mostly due to the 'numerical noise' inherent in this approach. To get rid of these problems it is possible to have grid-free implementations to compute terms like the ones in Eqs.2.2.5-6 or $E_{xc}$ in Eq.2.2.24. A well-known fact from linear algebra is that a function of a matrix which is expressed in an orthonormal basis can be evaluated by first diagonalizing the matrix, then applying the function on the diagonal elements and finally transforming the matrix back to its original basis. An illustration of this simple procedure for the functional $\rho_0^{4/3}$ can be found in chapter 7.5 in ref.[2.2.3], see also the related references therein.

## 2.2. Summary

The contribution of addressing moment functionals in a true variational method is very interesting, important and useful, and it should have been done long ago - at least the author thinks that beside the many related research referenced, this has not been done yet in a complete discussion, and this work has targeted to do that. After summarising the scaling correct power series or moment functionals for the different energy terms in the electronic Schrödinger equation, the Lagrangian method was applied first in the literature (to the author's knowledge) for variational solution of the ground state with restricting the N-normalization of the one-electron density. Possible semi-analytical solutions were discussed for some early truncations, as well as feasible numerical recipe was described for any high level later truncation. Reporting some promising preliminary calculations and results, the method was compared with the Hartree-Fock-SCF and Kohn-Sham methods on theoretical ground along with the discussion of opportunities for analytical vs. numerical integration – the inclusion or substitution of the crucial classical Coulombic term was also discussed.

## 2.2. Appendix

In Eq.2.2.9 the first, weaker approximation is $[\int\rho^{[1+a/(3j)]}d\mathbf{r}_1]^j \approx [\int\rho d\mathbf{r}_1]^j = N^j$, which is fine for a large j, but for the smallest index j=1, the largest power $\int\rho^{5/3}d\mathbf{r}_1 \approx N$ is not accurate enough. The idea is reasonable for large j, because $1 \leq 1+a/(3j) \leq 5/3= 1.6667$ and $\lim_{j\to\infty}(1+a/(3j))= 1$ for both a=1 and 2 with rigorous monotonity. However, the number of electrons in a system is generally high, recall e.g. that N=10 for $CH_4$, so an additional fact to this very weak approximation is that although the power in the integrand decrease with j, but the integral is on power j – counterbalancing the decreasing a/(3j). Being $\rho\geq 0$, a better approximation is $\int\rho^{[1+a/(3j)]}d\mathbf{r}_1 \approx [\int\rho d\mathbf{r}_1]^{[1+a/(3j)]} = N^x$. An even more accurate approximation is as follow. The exp(-2Zr$_1$) is an atomic 1s orbital, and let us approximate the decay of ρ with it (at 2-3 van der Waals distances from the molecule), as well as we can use $Z = \Sigma Z_A = N$ for a molecule owing peaks at nuclei. Take the known integral equality $\int (Z^3/\pi)^x \exp(-2Zxr_1) d\mathbf{r}_1 = (Z^3/\pi)^{x-1}/x^3$ with the extension $\int N^x (Z^3/\pi)^x \exp(-2Zxr_1) d\mathbf{r}_1 = N^x (Z^3/\pi)^{x-1}/x^3 = N^x (N^3/\pi)^{x-1}/x^3$ for $x=1+a/(3j)$. For example, for N=20 and considering a 1s density for 20 electrons $\rho(\mathbf{r}_1) = N(Z^3/\pi)\exp(-2Zr_1) = N(N^3/\pi)\exp(-2Nr_1) = (1.6/\pi)10^5\exp(-40r)$ the $\int \rho d\mathbf{r}_1 = N = 20$ and $\int \rho^{4/3} d\mathbf{r}_1 \approx 313$ in contrast to N=20 or $N^{4/3} = 20^{4/3} = 54.29$. An $N^x (N^3/\pi)^{x-1}/x^3$ with $x=1+a/(3j)$ is more accurate than N or $N^x$ for $\int \rho^{[1+a/(3j)]} d\mathbf{r}_1$.

## 2.3. Participation of electron-electron repulsion energy operator in the non-relativistic electronic Schrödinger equation via the coupling strength parameter along with generalizing the Hund's rule, the emblematic theorems virial-, Møller-Plesset-, Hohenberg-Kohn-, Koopmans-, Brillouin theorem and configuration interactions formalism

### 2.3. Preliminary

No detailed analysis has yet been published on the ratio or participation of electron-electron repulsion energy ($V_{ee}$) in total electronic energy – apart from virial theorem and the highly detailed and well-known algorithm $V_{ee}$, which is calculated during the standard HF-SCF and post-HF-SCF routines. Using a particular modification of the SCF part in the Gaussian package we have analyzed the ground and the excited state solutions of the electronic Hamiltonian $H_\nabla+H_{ne}+ aH_{ee}$ via the coupling strength parameter "a". Technically, this modification was essentially a modification of a single line in an SCF algorithm, wherein the operator $r_{ij}^{-1}$ was overwritten as $r_{ij}^{-1} \to a*r_{ij}^{-1}$, and used "a" as input. The most important findings are: that the repulsion energy $V_{ee}(a)$ is a quasi-linear function of "a", as well as the statement and analysis of the extended 1st Hohenberg-Kohn theorem ($\Psi_0(a=1) \Leftrightarrow H_{ne} \Leftrightarrow Y_0(a=0)$) and its consequences in relation to "a". The latter allows an algebraic transfer from the simpler solution of case a=0 (where the single Slater determinant is the accurate form) to the realistic wanted case a=1. Moreover, the case a=0 generates an orto-normalized set of Slater determinants which can be used as a basis set for CI calculations.

### 2.3.1. Introduction

#### 2.3.1.a A different hierarchy than the adiabatic connection (AC)

The non-relativistic, spinless, fixed nuclear coordinate electronic Schrödinger equation (SE) for a molecular system containing M atoms and N electrons with nuclear configuration $\{\mathbf{R}_A, Z_A\}_{A=1,\dots,M}$ in free space is

$$H(a=1)\Psi_k= (H_\nabla+ H_{ne}+ H_{ee})\Psi_k= E_{electr,k}\Psi_k \qquad \text{(Eq.2.3.1)}$$

where $\Psi_k$ and $E_{electr,k}$ are the $k^{th}$ excited state (k=0,1,2,...) anti-symmetric wave function (with respect to all spin-orbit electronic coordinates $\mathbf{x}_i \equiv (\mathbf{r}_i,s_i)$) and electronic energy, respectively, as well as the electronic Hamiltonian operator contains the sum of: kinetic energy, nuclear–electron attraction and electron–electron repulsion operators. These operators are spinless operators; the spin coordinates are introduced in $\Psi_k$ only. (The use of spin in this way is a handicap being Eq.2.3.1 non-relativistic. If atoms with the atomic number $Z_A > 16$ appear in the molecule, additional relativistic correction is certainly required for adequate accuracy.) If external forces apply, its operator may contain spin coordinates, however, we have not considered those cases in this work. The most popular calculations [2.3.1-2] for deducing important physical properties from Eq.2.3.1 as stationary points (geometry minimums and transition states), vibronic frequencies, rotations, van der Waal's interactions, excited states, reaction barriers, reaction heats, etc. are the expensive but accurate configuration interactions (CI) method for ground and excited states, and the less accurate but faster and less memory taxing Hartree–Fock self consistent field (HF-SCF) method for a ground state with or without correlation corrections. The CI works for any nuclear geometry, while the HF-SCF is only for the vicinity of stationary points where the assumption of the spin-pairing effect is plausible via a single Slater determinant.

As indicated in Eq.2.3.1, the Hamiltonian can be extended with coupling strength parameter "a" as $H(a)= H_\nabla+ H_{ne}+ aH_{ee}$ of which only a=1 makes physical sense. The a=0 case mathematically provides a good starting point to solve the very important problem when

a=1, as well as discussing other "a" values (as side properties) is interesting from a theoretical point of view. In this extended Hamiltonian H(a) the dimensionless coupling strength parameter "a" scales the electron-electron interaction energy, $V_{ee}(a)$, with this simple and precise definition. It has already been shown [2.3.13] that, for example, it is capable to correct the HF-SCF energy remarkably well with scaling it a bit below unity. On the other side, the role of coupling strength parameter is also known in the literature which defines the "adiabatic connection (AC) Hamiltonian", wherein the Hamiltonian is extended similarly, not purely with operators as above, but in the context of Kohn-Sham (KS) formalism (with the help of one-electron density, etc.). Langreth and Perdew [2.3.3-4] as well as Gunnarsson and Lundqvist [2.3.5] established the AC formalism in the mid 1970s, which attempts to compute the ground state correlation energy ($E_{corr}$) using the KS determinant as a reference, of which the algebraic origin is that only a=0 case has a single Slater determinant form solution, all the other a≠0 does not; also, the ground state is targeted primarily with it. To estimate consequence of AC in energy, the random phase approximation (RPA) [2.3.6-8] is one of the oldest non-perturbative methods for computing the ground state correlation energy of many-electron systems, and, e.g. the first bloom of RPA was in solid-state physics; see also refs. [2.3.9-10]. Important to emphasize that, in AC the coupling strength parameter connects the KS system to physical system of interacting electrons (H(a=1)), while in this work it connects an unphysical system (no electron repulsion, H(a=0)) to the system treated at the mean-field Hartree-Fock level and above; (so, the two hierarchies should not be confused). Below we analyze the behavior of case a=0 with our hierarchy defined here (different than the AC), and its effect on ground and excited states, not only on ground state correlation effects.

The $\mathbf{R}_A$ and $\mathbf{r}_i$ notate the Cartesian coordinates of the $A^{th}$ nucleus with nuclear charges $Z_A$ and $i^{th}$ electron, respectively, in the molecular system with A= 1,…,M, i=1,…,N, as well as the spin-orbit coordinates denoted as $\mathbf{x}_i \equiv (s_i, \mathbf{r}_i)$. $E_{total\ electr,k}$ includes the nuclear repulsion terms: ($V_{nn}$) and $E_{electr,k}$ (electronic energy) and, notates it without: $E_{total\ electr,k} = E_{electr,k} + \Sigma_{A=1,\dots,M} \Sigma_{B=A+1,\dots,M} Z_A Z_B / R_{AB}$. $R_{Ai}$, $R_{AB}$ and $r_{ij}$, the distances between constituting particles. The electronic potential energy surface (PES), the total electronic energy, $E_{total\ electr,k}$, parametrically depends on the nuclear coordinates. For electronic ground state energy, the HF-SCF procedure minimizes the energy functional $<S_0|H_\nabla|S_0> + <S_0|H_{ne}|S_0> + <S_0|H_{ee}|S_0> > <\Psi_0|H|\Psi_0> \equiv E_{electr,0}$ for a normalized single Slater determinant approximate wave function (denoted as $S_0$) with constrain so that its molecular orbitals (MO) are ortho-normalized, approximating the three energy terms: kinetic ($T \equiv <\Psi_0|H_\nabla|\Psi_0>$), electron-nuclear attraction ($V_{ne} \equiv <\Psi_0|H_{ne}|\Psi_0>$) and electron-electron repulsion energy ($V_{ee} \equiv <\Psi_0|H_{ee}|\Psi_0>$). The $<S_0|H|S_0>$ can never reach the value: $E_{electr,0}$ (variation principle), causing about a 1% non-negligible energy error, known as correlation energy ($E_{corr}$). Correlation effects can be calculated by density functional theory (DFT) [2.3.2, 11] during the HF-SCF algorithm or by wave function based methods performed afterwards, e.g. by MP2, MP3, CCSD, etc. [2.3.12]. Another variational effect, causing energy increase in the calculation, is the basis set error, i.e., not reaching the basis set limit, in practice. By this reasoning the HF-SCF/basis indicates the particular basis set used.

Before our talk, we emphasize that the coupling strength parameter [2.3.11], we manipulate with, is used in DFT for investigating the "exchange (Fermi) and correlation (Coulomb) hole". In short, these holes are those, which are connected to the error created by the HF-SCF and the DFT based Kohn-Sham (KS) methods [2.3.2, 11] try to re-correct during the routine using the one-electron density, or the wave function methods [2.3.1] which try to re-correct after the routine using the wave functions. The error stems from the use of one single Slater determinant to approximate the ground state wave function when it is effected by the operator $1/r_{ij}$ in the Hamiltonian, and is responsible for the exchange (Fermi hole) error, and correlation (Coulomb hole) error (estimated as $E_{corr}:= E_{xc}<0$) in the calculation of $<\Psi_0|H_{ee}|\Psi_0>$ with approximation $<S_0|H_{ee}|S_0>$. However, in this work we use the coupling strength parameter to investigate the entire term $<\Psi_0|H_{ee}|\Psi_0>$. We note that there is another error stemming from the use of $S_0$ in calculating the kinetic energy, $<S_0|H_\nabla|S_0>$, to approximate $<\Psi_0|H_\nabla|\Psi_0>$, that is about a magnitude less than $E_{xc}$ and has an opposite sign. Furthermore, physicists [2.3.2] divide this problem as $E_{corr}:= E_x+E_c$, where the $E_x$ accounts for the error stemming from $<S_0|H_\nabla|S_0>$ and $E_c$ from $<S_0|H_{ee}|S_0>$.

The standard HF-SCF routine was modified with a few simple program lines, which can be done in any of the existing SCF subroutines: Those lines of the SCF subroutine (particularly in Monstergauss 1981, a very early version of Gaussian package [2.3.12] and used for all calculations in this work) were modified, which calls for the subroutine to calculate the two and four center integrals (known as K and J integrals [2.3.1-2]) for $<S_0|H_{ee}|S_0>$ with particular molecular orbitals (MOs). Simply, the seed term $r_{ij}^{-1}$ was overwritten with $a*r_{ij}^{-1}$, and the parameter "a" was programmed as input. Essentially it was a simple modification in one line only as variable $\rightarrow$ parameter*variable. The a=1 leaves the operator $H_{ee}$ in full effect in our work in a regular way, while a=0 totally switches it off for our purpose indicated in the title. In this way, this input parameter "a" assumes the role of the coupling strength parameter. (It was just the same as in ref.[2.3.13] for a totally different purpose, particularly for correlation calculation, based on a very different theoretical point of view.)

The HF-SCF subroutine in Gaussian98 and 09 [2.3.12] has yielded the same HF-SCF/basis (a=1) energy values up to micro-hartree as the modified Monstergauss - an important technical test - but the HF-SCF/6-31G* with a=0 or 1, did not converge for some molecules in Monstergauss, persistent for N> 34 or order number > 88 on related figures below because the early version Monstergauss did not contain the (convergence improving) DIIS device at that time, while smaller basis set, i.e., HF-SCF/STO-3G with a=0 or 1, did not have the convergence problem and, the Gaussian 98 or 09 does not have this convergence problem at all. We also note that the operator $1/r_{ij}$ with a=0 was switched off - technically the standard basic calculations with it were performed during the runs but not used, i.e. they were discarded; for this reason no total, only a partial quantitative result is reported here on the CPU time and disc space saving with Eq.2.3.2 in comparison to Eq.2.3.1, which is obviously substantial for larger systems. To switch off the calculations with $1/r_{ij}$ completely, one would need a few more "if statements" in the program, however, here we focus on the elucidation of the rich, valuable effect caused by coupling strength parameter "a".

Below we extend the regular notation "HF-SCF/basis" to "HF-SCF/basis/a", wherein the latter indicates the value of the coupling strength parameter "a" used beside the basis set in the standard HF-SCF algorithm, extended with a coupling strength parameter as input described above. The a=1 is the standard HF-SCF/basis for Eq.2.3.1 and a=0 is for Eq.2.3.2, and there can be other values for "a" to manipulate [2.3.footnotes 1].

With this device, the mode a=0 solved the equation as a counter part of Eq.2.3.1:

$$(H_\nabla + H_{ne})Y_k= e_{electr,k}Y_k \qquad \text{(Eq.2.3.2)}$$

for ground state k=0 (or lowest lying enforced spin multiplicity state) with a Slater determinant for $Y_0$ with HF-SCF/basis/a=0 algorithm. Moreover, even excited states (k>0) can be obtained by HF-SCF/basis/a=0 algorithm for Eq.2.3.2, see the trick later in this work, which is definitely not feasible for Eq.2.3.1 by only using HF-SCF/basis/a=1 algorithm. The eigenvalue pairs, ($e_{electr,k}$, $Y_k$), - of course - differ from Eq.2.3.1, and are notated differently. In Eq.2.3.2 $t\equiv<Y_0|H_\nabla|Y_0>$ and $v_{ne}\equiv<Y_0|H_{ne}|Y_0>$ are the ground state kinetic and nuclear-electron attraction energies, respectively. A very important result we mention "in medias res" is: The $S_0$ obtained by HF-SCF/basis/a=1 performed on Eq.2.3.1 is very close to $Y_0$ by HF-SCF/basis/a=0 performed on Eq.2.3.2, moreover, calculating $Y_0$ in this way is not restricted to the vicinity of the stationary point, detailed below. As will be analyzed in section 2.3.2, no effect corresponding to correlation effect rises up in Eq.2.3.2, the Slater determinant is an adequate form for ground state anti-symmetric $Y_0$, more, for $Y_k$ too, only the problem of not reaching the basis set limit elevates the energy governed by the variation principle. (Recall again that when the single Slater determinant $S_0$ approximates the not single determinant $\Psi_0$, it creates the correlation error beside the basis set error.) Eq.2.3.1 has physical sense while Eq.2.3.2 does not, but Eq.2.3.2 provides a very rich pro-information for Eq.2.3.1, as will be demonstrated in this work. Furthermore, the HF-SCF/basis/a=0 calculation for Eq.2.3.2 is faster, more stable and less memory taxing in comparison to HF-SCF/basis/a=1 for Eq.2.3.1. We name Eq.2.3.2 the "associated partial differential equation" to Eq.2.3.1, and, as the main point in this work, we describe the mathematical and computing connection between them. We also note that adding the function $aH_{ee}$ to the solved Eq.2.3.2 to generate somehow the solution of Eq.2.3.1 is in fact a well known device in the theory of ordinary differential equations starting from the elementary homogeneous (e.g. $y''+y=0$) vs. non-homogeneous case (e.g. $y''+y=f(x)$), which is definitely not analyzed and considered in computational chemistry for Eq.2.3.1 in the way we discuss here.

**2.3.1.b. The aim of section 2.3**

In this study the effect (to be more particular, the quasi-linear effect) of the electron-electron repulsion term on the total electronic energy is studied from its total neglect (a=0) to its full strength (a=1), calculus of function series $\{Y_k\}$ is detailed in view of coupling strength parameter (a), the most natural estimation stemmed by $Y_k(a\equiv 0)$ for true (a=1) ground and excited state electronic energy (see Eqs.2.3.29 and 48 later) is derived along with the accurate variation equation for ground state (see Eq.2.3.38 later), wherein not anti-symmetric, but the easier symmetric function (w) has to be sought, as well as this orthogonal basis set $\{Y_k\}$ provides simpler Hamiltonian matrix for different level CI calculations in its off-diagonal elements (see Eq.2.3.46 later) along with an opportunity to avoid the restriction from Brillouin's theorem, (even the SCF convergence originated from $1/r_{12}$ is eliminated from the algorithm).

Summarizing the frequent notations for easier reading: The ($y_k(a)$,$enrg_{electr,k}(a)$)) is the k-th eigenvalue pair of electronic Hamiltonian H(a), most importantly, we use distinguishing notations for a=0 (Eq.2.3.2) as ($Y_k$,$e_{electr,k}$) and for a=1 (Eq.2.3.1) as ($\Psi_k$,$E_{electr,k}$), as well as $S_0$ (generally $s_0(a)$) is a single determinant approximation for $\Psi_0$ (generally for $y_0(a)$) via HF-SCF/basis/a=1 (generally with a) energy minimizing algorithm, (k=0,1,2,…, $enrg_{electr,k}(a)\leq enrg_{electr,k+1}(a)$). As analyzed and used below, $y_k(a=0)= Y_k$ has a single determinant form solution, while $y_k(a\neq 0)$ do not, and $E_{electr,0}$(method) approximates $E_{electr,0}$ by a certain method (HF-SCF, KS, CI, etc.).

**2.3.2. Theory for calculating ground state via Eq.2.3.2**

**2.3.2.a. The electronic Schrödinger equation (Eq.2.3.1, a=1) versus the "totally non-interacting reference system" (TNRS, Eq.2.3.2, a=0)**

Similarly to Eq.2.3.1, by definition, we ask $Y_k$ for Eq.2.3.2 to be anti-symmetric and well behaving (vanishing at infinity and square-integrable), normalized as $<Y_k|Y_k>=1$, and the ground state one-electron density is defined as $\rho_0(\mathbf{r}_1,a=0)= N\int Y_0^* Y_0 ds_1 d\mathbf{x}_2 \ldots d\mathbf{x}_N$, where "TNRS" stands for "totally non-interacting reference system" (name explained below) in analogy to the corresponding anti-symmetric and well behaving $\Psi_k$: $<\Psi_k|\Psi_k>=1$ and $\rho_0(\mathbf{r}_1,a=1)= N\int\Psi_0^*\Psi_0 ds_1 d\mathbf{x}_2 \ldots d\mathbf{x}_N$, respectively [2.3.footnotes 2]. A trivial property is that for N=1, i.e. for H-like atoms (M=1) and molecular frame with one electron (M>1), Eqs.2.3.1 and 2 overlap or identical. Furthermore and more importantly, certain theorems for Eq.2.3.1 hold for Eq.2.3.2 as well. Most importantly, Eq.2.3.2 is a linear partial differential equation, the variation principle holds, and the 1st ("$\rho_0(\mathbf{r}_1,a=0)$ of TNRS defines $Y_0$ and the nuclear frame") and 2nd ("variation principle for $\rho_0(\mathbf{r}_1,a=0)$ of TNRS in the DFT functional stemmed by Eq.2.3.2") Hohenberg – Kohn (HK) theorems hold in an analogue sense which will be set out in more detail later.

The density functional for the nuclear–electron term is the simple 100 % accurate $-\Sigma_{A=1,\ldots,M}Z_A\int R_{A1}^{-1}\rho_{trial}(\mathbf{r}_1)d\mathbf{r}_1$, i.e., the same form for both Eqs.2.3.1 and 2. Also, the same form for the kinetic term holds for both Eqs.2.3.1 and 2 – however, only approximate forms and not exact forms are known as yet. It is also obvious that with the a=0 mode (switching the effect of operator $H_{ee}$ off), Eq.2.3.2 can also be treated with the HF-SCF/basis/a algorithm (this mode indicated in section 2.3.1 is the correct treatment), and the energetically lowest lying eigenvalue pair ($e_{electr,0}$,$Y_0$) corresponds to ($E_{electr,0}$, $\Psi_0$). Furthermore, $E_{electr,0} >> e_{electr,0}$ for any molecular system (in stationer or non-stationer geometry), and the large difference mainly stems from the lack of $V_{ee}(a=1)$ when a=0. Moreover, the ground state versus the energetically lowest lying state with an enforced spin multiplicity feature is also the same as in the HF-SCF/basis/a treatment of Eqs.2.3.1 and 2 – recall the example of neutral ($1s^2$,$2s^2$,$2p^2$) carbon atom open shell (triplet, ground state) versus closed shell (lowest lying singlet) states, and so on. However, if spin-spin interaction is not considered via Coulomb repulsion, Hund's rule does not apply for Eq.2.3.2 (a=0) itself, for example, but extension and approximation in e.g. Eq.2.3.29 below sets it back on the right track, that is, e.g. the triplet has lower energy than the singlet, as in the HF-SCF/basis/a=1 for Eq.2.3.1 (p.103 of ref. [2.3.1] - also set out in more detail later).

On the other hand, there are major mathematical differences between Eqs.2.3.1 and 2 aside from the visible inclusion of or omitted operator $H_{ee}$. For both, operator $H_\nabla$ makes them differential equations, which is generally necessary - philosophically speaking - to describe a physical phenomenon, and for both, $H_{ne}$ defines the nuclear frame, and the molecular system for them. However, operator $H_{ee}$ is very special in Eq.2.3.1 in the sense that algebraically it is the "simplest" term, but in contrast, as it has turned out in the history of computational chemistry, it introduces the most difficult effect [2.3.14] in HF-SCF computation, known as the non-classical Coulomb effect. It is difficult to treat for HF-SCF/basis/a routine via "correlation calculation" or "exchange correlation DFT" devices after or during, respectively [2.3.1-2, 11]. $H_{ee}$ operator is responsible for the fact that a single Slater determinant $S_0$ for $\Psi_0$ in Eq.2.3.1 is not enough for total accuracy, although in the vicinity of stationary points on the PES, it provides a good approximation, and it can provide many characteristic properties of the ground state eigenvalue. In contrast, a single Slater determinant form is adequate for Eq.2.3.2 not only for the ground, but also for excited states, and the HF-SCF/basis/a=0 with basis set limit accurately calculates the eigenvalue pairs ($e_{electr,k}$, $Y_k$) for ground and excited states.

The manipulation with Slater determinants in HF-SCF theory is well established [2.3.1], but some textbook properties must be overviewed, since a new aspect is described, that is, we make an allowance for Eq.2.3.1 to be replaced by Eq.2.3.2. In detail, Eq.2.3.2 is $(H_\nabla + H_{ne})Y_k = (-(1/2)\Sigma_{i=1,\dots,N}\nabla_i^2 - \Sigma_{i=1,\dots,N}\Sigma_{A=1,\dots,M}Z_A R_{Ai}^{-1})Y_k = \Sigma_{i=1,\dots,N}h_i = e_{electr,k}Y_k$, where $h_i \equiv -(1/2)\nabla_i^2 - \Sigma_{A=1,\dots,M}Z_A R_{Ai}^{-1}$ is the one-electron operator widely used [2.3.1] in HF-SCF theory for Eq.2.3.1 as well. In the Fock or Kohn Sham equations [2.3.1-2, 11] Eq.2.3.1 is decomposed to the one-electron equations

$$(h_i + aV_{ee,eff}(\mathbf{r}_i))\ \phi_i(\mathbf{r}_i) = \varepsilon_i\ \phi_i(\mathbf{r}_i) \qquad \text{(Eq.2.3.3)}$$

where $\phi_i(\mathbf{r}_i)$ is the $i^{th}$ MO, and technically $\phi_i$ counts the MOs with the idex i, so the notation is reducible from $(h_i, \phi_i(\mathbf{r}_i), \varepsilon_i)$ to $(h_1, \phi_i(\mathbf{r}_1), \varepsilon_i)$ mathematically. $V_{ee,eff}$ is the effective potential from electron-electron repulsion; (other habit [2.3.11] is that $H_{ne}$ is shifted algebraically into $V_{ee,eff}$, and called $V_{eff}$, but we do not use that here). Importantly, because $1/r_{ij} \rightarrow a/r_{ij}$ change for operator $H_{ee}$ was made in the algorithm, the parameter "a" entered linearly to $V_{ee,eff}$ in Eq.2.3.3. $V_{ee,eff}$ in Eq.2.3.3 is expressed with the known J and K integrals in HF-SCF theory, or $V_{ee,eff}(\mathbf{r}_i) = \int\rho_0(\mathbf{r}_2,KS)r_{i2}^{-1}d\mathbf{r}_2 + V_{xc}(\mathbf{r}_i)$ in Kohn-Sham formalism (the first term is the classical Coulomb term, the second is the non-classical Coulomb term for "exchange-correlation"). $V_{ee,eff}(\mathbf{r}_i)$ is the term where the N equation in Eq.2.3.3 is coupled (a=1 or generally a≠0). (The $\rho_0$ depends on $\Psi_0$ which is approximated with a single Slater determinant containing all other $\phi_j$, j=1,…,N and j≠i.) Another property is that MOs are ortho-normal, that is $<\phi_i|\phi_j> = \delta_{ij}$. In Eq.2.3.3 the operator seed $1/r_{ij}$ is reduced to the variable, $\mathbf{r}_i$ via performing the integrations, and virtually all equations in Eq.2.3.3 depend on one-electron. It is in fact coupled, though virtually not coupled, so the 100% adequate anti-symmetric solution for the equation system in Eq.2.3.3 (but not for Eq.2.3.1) is a Slater determinant, and this system is known as: "non-interacting reference system" [2.3.11], as is well known. System in Eq.2.3.3 is commercially programmed [2.3.12] by the standard HF-SCF or Kohn – Sham formalism, and if a=0 is set in the input, a special modification in the algorithm for this work, Eq.2.3.3 reduces to:

$$h_1\phi_i(\mathbf{r}_1) \equiv (-(1/2)\nabla_1^2 - \Sigma_{A=1,\dots,M}Z_A R_{A1}^{-1})\ \phi_i(\mathbf{r}_1) = \varepsilon_i\ \phi_i(\mathbf{r}_1) \qquad \text{(Eq.2.3.4)}$$

System Eq.2.3.4 is the Fock equation system for Eq.2.3.2. However, we are at a reduction where a single Slater determinant as anti-symmetric solution is not only 100 % adequate for Eq.2.3.4, but also for Eq.2.3.2. The reason for this is: that all operators are one-electron operators, the two electron operators with seed elements $1/r_{ij}$ are cancelled by a=0, that is, Eq.2.3.4 describes a non-coupled system. For this reason, we call Eqs.2.3.2 or 4: TNRS, distinguishing them from the "non-interacting reference system" above. More simply, Eq.2.3.4 should not be considered as an equation system containing N equations enumerated by i ($h_i\phi_i = \varepsilon_i\phi_i$), but in fact it is a single eigenvalue equation ($h_1\phi_i = \varepsilon_i\phi_i$). Eigenvalues of Eq.2.3.4 are $(\varepsilon_i, \phi_i(\mathbf{r}_1))$ for i=1,2,…∞, the i=1 is the lowest lying state of Eq.2.3.4 and it is the lowest lying MO for Eq.2.3.2 in its k=0 ground state. The single Slater determinant for Eq.2.3.2 is accomplished for N electrons from the eigenvaues of Eq.2.3.4, just as in the basic HF-SCF theory. Notice that the HF-SCF/basis/a algorithm (at any "a") is accomplished in such an algebraic way that it keeps the MOs ortho-normal in $s_0(a)$ during the optimizing algorithm, particularly in $S_0 = s_0(a=1)$, so for $Y_0 = s_0(a=0)$, in great accord that the eigenfunctions of linear Eq.2.3.4 are mathematically ortho-normal.

Lemma: Eq.2.3.4 with the value of N and Eq.2.3.2 are equivalent. (More, it holds for the case too, when mathematically, one needs a symmetric $Y_k$, instead of an anti-symmetric one.) For a moment, let us use a simpler and more comprehensive notation so as not to get lost in the jungle of indexes: From Eq.2.3.4 let we have $\phi_i$= f, g, h for i=1,2,3 with state/MO energies $\varepsilon_1 \le \varepsilon_2 \le \varepsilon_3$, respectively, and N=3. (We name these MOs, after the corresponding HF-SCF correlated or un-correlated ones via Eq.2.3.1 (a=1).) With these, some {anti-symmetric eigenfunction (wave function), energy eigenvalue (electronic energy)} solution pairs of Eq.2.3.2 are

$$Y_0= |\alpha_1 f(\mathbf{r}_1), \beta_2 f(\mathbf{r}_2), \alpha_3 g(\mathbf{r}_3)> \text{ and } e_{electr,0}= 2\varepsilon_1 + \varepsilon_2 \quad \text{(Eq.2.3.5)}$$

$$Y_{k'}= |\alpha_1 f(\mathbf{r}_1), \alpha_2 g(\mathbf{r}_2), \beta_3 g(\mathbf{r}_3)> \text{ and } e_{electr,k'}= \varepsilon_1 + 2\varepsilon_2 \quad \text{(Eq.2.3.6)}$$

$$Y_{k''}= |\alpha_1 f(\mathbf{r}_1), \alpha_2 g(\mathbf{r}_2), \alpha_3 h(\mathbf{r}_3)> \text{ and } e_{electr,k''}= \varepsilon_1 + \varepsilon_2 + \varepsilon_3 \quad \text{(Eq.2.3.7)}$$

where|.,.,> is the standard Bra-ket notation for Slater determinants, we will use these later for simple demonstrations during our discussion. The electronic energy of the system in Eq.2.3.2 is the sum of energy levels (states of Eq.2.3.4 or MOs of Eq.2.3.2) is generally speaking weighted as populated.

$$e_{electr,k}= \Sigma_{i=1,\dots} n_i \, \varepsilon_i \quad \text{(Eq.2.3.8)}$$

where $n_i$ is the population of the $i^{th}$ energy level: 0, 1 or 2, the lattermost is with opposite spins. Of course, $\Sigma_{i=1,\dots} n_i= N$ must hold. Eq.2.3.5 is the ground state k=0, because there is no other way to get lower $e_{electr,k}$ now, and a degenerate state to Eq.2.3.5 is $Y_0= |\alpha_1 f(\mathbf{r}_1), \beta_2 f(\mathbf{r}_2), \beta_3 g(\mathbf{r}_3)>$. The spin multiplicities are 2(1/2 –1/2 +1/2)+1=2, 2, 4 in Eqs.2.3.5-7, respectively. For excited states, the k is numbered as k'<k'' if $\varepsilon_1 + 2\varepsilon_2 < \varepsilon_1 + \varepsilon_2 + \varepsilon_3$, and no spin-spin interaction is taken into account. As it is clear, the excited states of Eq.2.3.2 can also be described or accomplished as shown in Eqs.2.3.6-7, and a single Slater determinant is a 100 % accurate form for these too, as for ground state. In general contrast, a single Slater determinant e.g., LUMO for the excited state is an even a worse approximation than the approximation $S_0$ for ground state, both by HF-SCF/basis/a=1 for Eq.2.3.1. The analogue of Eq.2.3.8 between MO energies ($\varepsilon_i$) and ground state electronic energy ($E_{electr,0}$(a≠0)) is not held in the context of HF-SCF/basis/a=1 approximation or Kohn-Sham formalism, that is $E_{electr,0}$(HF-SCF or KS/basis/a=1) ≠ $\Sigma_{i=1,\dots} n_i \, \varepsilon_i$ for deepest possible filling, where $\varepsilon_i$'s are from Eq.2.3.3 with a=1 (more generally if a≠0): some cross terms must be subtracted [2.3.1]. However, what must be subtracted, that goes to zero if a→0. The definition of restricted (RHF) and unrestricted (UHF) form of Slater determinants also lose their necessity here in Eq.2.3.2, while these are a certain handicap in HF-SCF approximation for Eq.2.3.1 (to get lower energy from the variation principle allowing more LCAO parameters), recall again that the single determinant is an accurate form of solution for Eq.2.3.2, but only an approximate solution for Eq.2.3.1. (The $\phi_i$'s are eigenfunctions of Eq.2.3.3 if a=0 for Eqs.2.3.2 and 4 with k≥0, while they are not when a≠0, particularly when a=1 in Eq.2.3.3 for Eq.2.3.1 with k=0, only an energy minimization.) Numerical example for RHF vs. UHF in relation to a=1 vs. a=0 will be exhibited below in a separate section.

In textbook style, we note that Eqs.2.3.2 or 4 can be solved analytically for N=M=1, those are the famous atomic orbitals (AO, and called 1s, 2s, 2p, etc.) and energies ($-Z^2/2$ for 1s, etc.) for H-like or one-electron atoms in ground and excited states; actually, this case overlaps with Eq.2.3.1. However, for molecular frame with N≥1, M>1 the HF-SCF/basis/a algorithm with a=0 input is a perfect way to numerically describe the $\phi_i$ eigenfunctions of Eqs.2.3.3, 4 or MOs of Eq.2.3.2 with linear combination of atomic orbitals (LCAO), since there is no analytical solution for $\phi_i$'s, but the single determinant is an accurate form for any k≥0 to mix $\phi_i$'s or their LCAO approximations. (For a≠0 in Eq.2.3.3, nor analytical solutions for $\phi_i$'s exist, neither is the single determinant an accurate form for Eq.2.3.1 especially with k=0, but the HF-SCF/basis/a is a famous and useful approximation especially if a=1.) For a value of N and multiplicity 2S+1= $2\Sigma s_i$+1 in the regular way, the HF-SCF/basis/a=0 algorithm calculates the lowest lying N/2 or (N+1)/2 energy values ($\varepsilon_i$) and MOs ($\phi_i$), the latter with LCAO expansion of the basis set for Eqs.2.3.2 or 4. (The $N=Z_A=Z_B=M-1=1$, for example, specifies the TNRS for the famous hydrogen-molecule-ion ($H_2^+$) problem via Eq.2.3.4, and it is known that there is no analytic solution even for this the simplest case, only recursive formulas.) As is known, the regular HF-SCF/basis/a=1 algorithm optimizes an $S_0$ single determinant energetically for Eq.2.3.1 keeping MOs of $S_0$ ortho-normal during the optimization. This enforced ortho-normalization is also used in HF-SCF/basis/a=0 mode for Eqs.2.3.2 or 4 to obtain $Y_0$, in agreement that eigenvectors of Eq.2.3.4 (which are MOs of $Y_0$) is ortho-normal set coming purely from the mathematical nature of Eq.2.3.4.

Finally, important for Eqs.2.3.2 or 4 is that, 1., the HF-SCF/basis/a=0 (the "a-value modified" commercial HF-SCF/basis/a=1 algorithm) is technically perfectly adequate for solving these equations, and 2., for any molecular geometry on the PES, not only at the vicinity of stationary points (which is a serious restriction of HF-SCF/basis/a≠0 case) the HF-SCF/basis/a=0 gives a mathematically adequate result, stemming from the fact that for a=0, the single determinant is an accurate form of the solution, only basis set error present. The $Y_k$'s of Eq.2.3.2 have much better mathematical properties than $S_0$ for $\Psi_0$ of Eq.2.3.1, however, $S_0$ has been about ready in practice for a long time, while $Y_k$'s are not, and have to be converted in order to be able to be used for $\Psi_k$'s of Eq.2.3.1, this will be outlined below step by step. A Slater determinant is a 100% accurate form of solution ($Y_0= y_0(a=0)$), that is, there is no Coulomb hole, because there is no electron-electron interaction at all, and no Fermi hole, because the anti-symmetric property is provided 100% by a Slater determinant.

### 2.3.2.b. Spin states in TNRS (Eq.2.3.2, a=0)

The spin states must be commented upon for Eqs.2.3.2 and 4: The Hamiltonian does not contain any spin coordinates in Eqs.2.3.1-2 and hence both, $S_{op}^2$ and $S_{op,z}$ total spin operators commute with it:

$$[H_\nabla + H_{ne} + aH_{ee}, S_{op}^2 \text{ or } S_{op,z}] = 0, \qquad \text{(Eq.2.3.9)}$$

what we have extended with the coupling strength parameter "a"; index "op" stands for: "operator". Consequently, the exact eigenfunctions, not single determinant $\Psi_k$ of Eq.2.3.1 (a=1) or single determinant $Y_k$ of Eq.2.3.2 (a=0) are also eigenfunctions of the two spin operators [2.3.1], which in the case of a=0

$$S_{op}^2\, Y_k = S(S+1)Y_k\ , \qquad \text{(Eq.2.3.10)}$$

$$S_{op,z}\, Y_k = M_S Y_k\ , \qquad \text{(Eq.2.3.11)}$$

where S and $M_S$ are the spin quantum numbers describing the total spin $S= \Sigma_{i=1\ldots N}\, s_i$ and its z component of an N electron (TNRS, a=0) state $Y_k$.

For Eq.2.3.10, the single determinants are simple cases, particularly $Y_0$, because for a closed shell $Y_k$, the $S_{op}^2\, Y_k=0$ is a pure singlet [2.3.1], but an open shell $Y_k$ is generally not eigenfunction of $S_{op}^2$ [2.3.1], except when all the open shell electrons have parallel spins, e.g. in Eqs.2.3.5-7. However, as in the case of general single determinants (for example, excitation from $S_0$ using e.g. LUMO for CI), now in the case of single determinants $Y_k$, if $Y_k$ is not a pure spin state single determinant via Eq.2.3.2, e.g. (unlike Eq.2.3.7 the) $|\alpha_1 f(\mathbf{r}_1), \alpha_2 g(\mathbf{r}_2), \beta_3 h(\mathbf{r}_3)\rangle$, spin adapted configuration can be formed by taking appropriate linear combinations (thanks to that operator $H_\nabla + H_{ne} + aH_{ee}$ in Eqs.2.3.1-2 is linear), which is a bit of a hectic task for N>2, see p. 103 of ref.[2.3.1], but systematically, it can be performed.

On the other hand, for Eq.2.3.11 any single (open or closed shell) determinant, particularly, $Y_k$, is always eigenfunction of $S_{op,z}$ [2.3.1], and

$$S_{op,z}\, Y_k = ((N_\alpha - N_\beta)/2)Y_k = M_S Y_k\ . \qquad \text{(Eq.2.3.12)}$$

Note, that in this section about spins, not only $Y_0$, but all $Y_k$ ground and excited states (k=0,1,2,…) are commented upon in Eqs.2.3.9-12, and from this point of view, the approximation in Eq.2.3.48 later in this work will have greater importance.

Since excited state $Y_k$ determinants have come up in the discussion of spins, we draw attention again to the fact that Eqs.2.3.2 and 4 yield correct mathematical ground and excited states single determinant forms (via algorithm HF-SCF/basis/a=0, suffering from basis set error for all k=0,1,2,…), see more details below in section 2.3.3, while HF-SCF/basis/a≠0 for approximating the ground state, e.g. of Eq.2.3.1 (a=1, k=0), suffering not only from basis set, but a correlation error too. The latter relatively well approximates the ground state, and the first excited state LUMO may also be considered as relatively good approximation, but higher excited states (coming technically from the algorithm) can only be considered a mathematical orto-normalized basis set formed of MOs, (used in CI calculations as a pre-calculation to obtain a mathematical basis), but physically they are useless. In other words, HF-SCF/basis/a=1 approximates relatively well the ground state of Eq.2.3.1 (k=0) with $S_0$, mainly, if a correlation calculation follows or is included, but excited states (k>0) which stem technically from the algorithm should not be trusted physically, this is a well known fact.

### 2.3.2.c. General functional links between the electronic Schrödinger equation (Eq.2.3.1, a=1) and the TNRS (Eq.2.3.2, a=0) along with the values of coupling strength parameter 'a', estimation for ground state $E_{electr,0}$ with TNRS

An important link between Eq.2.3.1 and Eq.2.3.2 comes from $<\Psi_0|H_{ee}|Y_0>= <\Psi_0|H - (H_\nabla + H_{ne})|Y_0>= <\Psi_0|H|Y_0> -<\Psi_0|(H_\nabla + H_{ne})|Y_0>= <Y_0|H|\Psi_0> -e_{electr,0}<\Psi_0|Y_0>= E_{electr,0}<Y_0|\Psi_0> -e_{electr,0}<\Psi_0|Y_0>$, where the hermitian property of H (and its three parts) was used. Finally,

$$E_{electr,0}= e_{electr,0} + <\Psi_0|H_{ee}|Y_0>/<\Psi_0|Y_0>= e_{electr,0} + (N(N-1)/2)\xi, \quad \text{(Eq.2.3.13)}$$

bearing in mind that $\Psi_0$ and $Y_0$ are both anti-symmetric, and $\xi\equiv <\Psi_0|r_{12}^{-1}|Y_0>/<\Psi_0|Y_0>$. The analysis and test on 149 molecular G3 ground state electronic energy [2.3.15-16] (with G3 equilibrium geometry, neutral charge ($\Sigma Z_A= N$) molecules and selection $\max(Z_A)\leq 10$) will follow in section 2.3.3 wherein the figures exhibit the behavior of $\xi$ and ratio $(E_{electr,0} - e_{electr,0})/e_{electr,0}$ as a function of molecular frame seeded in operator $H_{ne}$. These two quantities are quasi-constants, which is surprising at first glance, but we call it the virial theorem, that is, $(V_{nn}+ V_{ne}+ V_{ee})/T= -2= (V_{nn}+ v_{ne})/t \equiv (V_{nn}+ <Y_0|H_{ne}|Y_0>)/<Y_0|H_\nabla|Y_0>$ holds exactly on atoms, atomic ions and equilibrium/transition state geometry molecules, i.e., the value 2 is invariant on the nuclear frame seeded in $H_{ne}$ and N in molecular systems. (Non-equilibrium molecules obey a slightly more complex virial equation [2.3.2], not detailed here.) It is important to emphasize the significant difference between $V_{ee}\equiv <\Psi_0|H_{ee}|\Psi_0>= (N(N-1)/2)<\Psi_0|r_{12}^{-1}|\Psi_0>$, as the electron-electron repulsion energy term in the sum $E_{electr,0}= T+V_{ne}+V_{ee}$, and the corresponding value from Eq.2.3.13, $<\Psi_0|H_{ee}|Y_0>/<\Psi_0|Y_0>= (N(N-1)/2)<\Psi_0|r_{12}^{-1}|Y_0>/<\Psi_0|Y_0>$, as the energy increase by electron-electron repulsion between the two Hamiltonians in Eq.2.3.1 (electron-electron interaction is on) and Eq.2.3.2 (electron-electron interaction is switched off). Because the nuclear-nuclear repulsion energy, $V_{nn}$, is added after the calculation, that is cancelled in the difference in Eq.2.3.13, and as a consequence: $E_{total\ electr,0} - e_{total\ electr,0}= E_{electr,0} - e_{electr,0}$. (Notice that the divisor $<\Psi_0|\Psi_0>$ comes up in $V_{ee}$ if it is not normalized to unity, making the algebraic analogy even closer between $V_{ee}$ and $(N(N-1)/2)\xi$.) A more general expression than Eq.2.3.13 is hold between k and k' excited states, coming from the same one line derivation:

$$E_{electr,k}= e_{electr,k'} + (N(N-1)/2)\langle\Psi_k|r_{12}^{-1}|Y_{k'}\rangle/\langle\Psi_k|Y_{k'}\rangle\ , \qquad \text{(Eq.2.3.14)}$$

and Eqs.2.3.13 and 14 forecast the generalization of 1st HK theorem, detailed later.

Further relations are the obvious $E_{electr,k} > e_{electr,k}$, because $1/r_{ij} \geq 0$ always, but for the sake of chemical accuracy (1 kcal/mol), the $E_{electr,0} >> e_{electr,0}$ is more plausible algebraically. For further relations, one can start from the variation principle: Let the normalized solution of Eq.2.3.2, the $Y_0$, be a trial for Eq.2.3.1, and one gets $E_{electr,0} \leq \langle Y_0|H|Y_0\rangle= \langle Y_0|H_{ee}|Y_0\rangle + \langle Y_0|H_\nabla+H_{ne}|Y_0\rangle= \langle Y_0|H_{ee}|Y_0\rangle + e_{electr,0}$, that is

$$E_{electr,0} \leq e_{electr,0} + \langle Y_0|H_{ee}|Y_0\rangle\ . \qquad \text{(Eq.2.3.15)}$$

The reverse situation, when $\Psi_0$, the solution of Eq.2.3.1 is a trial function for Eq.2.3.2, one gets the simpler

$$e_{electr,0} \leq \langle\Psi_0|H_\nabla+H_{ne}|\Psi_0\rangle\ . \qquad \text{(Eq.2.3.16)}$$

Equality holds for both in Eqs.2.3.15-16 in the trivial case N=1, because there $\Psi_0(N=1)= Y_0(N=1)$. From Eq.2.3.1, it separates as $\langle\Psi_0|H|\Psi_0\rangle= \langle\Psi_0|H_\nabla+H_{ne}|\Psi_0\rangle + \langle\Psi_0|H_{ee}|\Psi_0\rangle= E_{electr,0}$, and the right hand side is majored by Eq.2.3.15 as $\langle\Psi_0|H_\nabla+H_{ne}|\Psi_0\rangle + \langle\Psi_0|H_{ee}|\Psi_0\rangle \leq e_{electr,0} + \langle Y_0|H_{ee}|Y_0\rangle$, and with Eq.2.3.16 one obtains

$$\langle\Psi_0|H_{ee}|\Psi_0\rangle \leq \langle Y_0|H_{ee}|Y_0\rangle\ . \qquad \text{(Eq.2.3.17)}$$

The counterpart of Eq.2.3.15 comes from Eq.2.3.16 with an extension as $e_{electr,0} + \langle\Psi_0|H_{ee}|\Psi_0\rangle \leq \langle\Psi_0|H_\nabla+H_{ne}|\Psi_0\rangle + \langle\Psi_0|H_{ee}|\Psi_0\rangle= \langle\Psi_0|H|\Psi_0\rangle= E_{electr,0}$ which is

$$E_{electr,0} \geq e_{electr,0} + \langle\Psi_0|H_{ee}|\Psi_0\rangle\ . \qquad \text{(Eq.2.3.18)}$$

In summary the full relation is

$$\begin{aligned} &e_{electr,0} << (e_{electr,0}+\langle\Psi_0|H_{ee}|\Psi_0\rangle) \leq \\ &\leq E_{electr,0}= (e_{electr,0}+\langle\Psi_0|H_{ee}|Y_0\rangle/\langle\Psi_0|Y_0\rangle) \leq \\ &\quad \leq (e_{electr,0}+\langle Y_0|H_{ee}|Y_0\rangle) \end{aligned} \qquad \text{(Eq.2.3.19)}$$

which extends Eq.2.3.17 as

$$\langle\Psi_0|H_{ee}|\Psi_0\rangle) \leq \langle\Psi_0|H_{ee}|Y_0\rangle/\langle\Psi_0|Y_0\rangle \leq \langle Y_0|H_{ee}|Y_0\rangle\ . \qquad \text{(Eq.2.3.20)}$$

The relationships in Eq.2.3.13 to 19 can be developed further with the Hellmann–Feynman theorem [2.3.2] which says $\partial E_{electr,k}/\partial\lambda= \langle\Psi_k|\partial H(\lambda)/\partial\lambda|\Psi_k\rangle$ with normalization $\langle\Psi_k|\Psi_k\rangle=1$, and $H(\lambda)$ is the Hamiltonian in Eq.2.3.1 developed or extended with parameter $\lambda$ among its terms in addition to the already existing ones e.g., nuclear coordinates and atomic charges. Right now a continuous (and linear) variable (parameter) between Eqs.2.3.1 and 2 is the coupling strength parameter $\lambda$=a as $aH_{ee}= a\Sigma_{i=1,\ldots,N}\Sigma_{j=i+1,\ldots,N}\ r_{ij}^{-1}$, where a=1 yields Eq.2.3.1 and a=0 yields Eq.2.3.2, but other values for "a" are also possible; constantly bearing in mind that only a=1 has physical sense or reality. Considering the ground state (k=0), emphasizing parameter "a" in the argument of eigenvalue and eigenfunction as ($enrg_{electr,0}(a)$, $y_0(a)$), and with $\partial H(a)/\partial a= \partial(aH_{ee})/\partial a= H_{ee}$, it follows that

$$\partial enrg_{electr,0}(a)/\partial a= \langle y_0(a)|H_{ee}|\ y_0(a)\rangle= (N(N-1)/2)\ \langle y_0(a)|r_{12}^{-1}|y_0(a)\rangle\ , \qquad \text{(Eq.2.3.21)}$$

where the anti-symmetric property of $y_0$ was used, and obviously, $enrg_{electr,0}(a=0)= e_{electr,0}$, $enrg_{electr,0}(a=1)= E_{electr,0}$, $y_0(a=0)= Y_0$ and $y_0(a=1)= \Psi_0$. Integrating Eq.2.3.21 with respect to the coupling strength from a=0 to a=1 yields

$$E_{electr,0} - e_{electr,0}= (N(N-1)/2) \int_{[0,1]}\langle y_0(a)|r_{12}^{-1}|y_0(a)\rangle da\ . \qquad \text{(Eq.2.3.22)}$$

Comparing Eq.2.3.13 and Eq.2.3.22 one obtains:

$$\int_{[0,1]}\langle y_0(a)|r_{12}^{-1}|y_0(a)\rangle da= \langle\Psi_0|r_{12}^{-1}|Y_0\rangle/\langle\Psi_0|Y_0\rangle \qquad \text{(Eq.2.3.23)}$$

where the interval [0,1] used in the integration, can also be straightforwardly extended to the general interval $[a_1,a_2]$. (The $y_0(a)$ is normalized for all "a", but $\langle\Psi_0|Y_0\rangle$ in the denominator of Eq.2.3.13 or 23 is not necessarily unity.) Eq.2.3.23 is a certain "integral average relationship": the value of integral can be expressed with the endpoint wave functions in the Bra-ket notation. In fact, Eqs.2.3.13, 22-23 are also direct consequence of the integral Hellmann-Feynman theorem.

Define the ground state one and two-electron densities associated with parameter "a" in the usual way as $b_0(\mathbf{r}_1,\mathbf{r}_2,a)= (N(N-1)/2)\int y_0(a)^* y_0(a)ds_1ds_2d\mathbf{x}_3\ldots d\mathbf{x}_N$ and $\rho_0(\mathbf{r}_1,a)= N\int y_0(a)^* y_0(a)ds_1d\mathbf{x}_2d\mathbf{x}_3\ldots d\mathbf{x}_N$ providing the normalization $\iint b_0(\mathbf{r}_1,\mathbf{r}_2,a)d\mathbf{r}_1d\mathbf{r}_2= ((N-1)/2)\int\rho_0(\mathbf{r}_1,a)d\mathbf{r}_1= N(N-1)/2$. ($\int$ is over the 3 dimensional spatial space. The $\iint b_0(\mathbf{r}_1,\mathbf{r}_2,a)d\mathbf{r}_1d\mathbf{r}_2$ gives the number of electron pairs in the system, while $\int\rho_0(\mathbf{r}_1,a)d\mathbf{r}_1$ gives the number of electrons for any value of a, just as well known for a=1. For excited states $b_k(\mathbf{r}_1,\mathbf{r}_2,a)$ and $\rho_k(\mathbf{r}_1,a)$ are defined analogously with $y_k(a)$.) With these, the right hand side of Eq.2.3.21 can be written as

$$\partial enrg_{electr,0}(a)/\partial a= \iint b_0(\mathbf{r}_1,\mathbf{r}_2,a)\, r_{12}^{-1}d\mathbf{r}_1d\mathbf{r}_2= v_{ee}(a)/a \ , \qquad \text{(Eq.2.3.24)}$$

where $v_{ee}(a)\equiv \Sigma_{i=1,\ldots,N}\Sigma_{j=i+1,\ldots,N}\iint y_0^*(a)y_0(a)(a/r_{ij})d\mathbf{x}_1d\mathbf{x}_2\ldots d\mathbf{x}_N= a(N(N-1)/2)\ \langle y_0(a)|r_{12}^{-1}|y_0(a)\rangle= a\langle y_0(a)|H_{ee}|y_0(a)\rangle$ with normalized $y_0$, and with the physical $v_{ee}(a=1)= V_{ee}$, for example. Notice that the linear multiplier "a", present in $v_{ee}$, has disappeared from Eq.2.3.24, it only has an effect inside, inherited by the normalized $y_0(a)$. Furthermore, Eq.2.3.22 can be written alternatively as

$$E_{electr,0} - e_{electr,0}= \int_{[0,1]} \iint b_0(\mathbf{r}_1,\mathbf{r}_2,a)\, r_{12}^{-1}d\mathbf{r}_1d\mathbf{r}_2da \qquad \text{(Eq.2.3.25)}$$

via Eq.2.3.24. We mention the theoretical derivations by Wilson [2.3.17], Politzer et al. [2.3.18] and ref. [2.3.2] on p.17 obtaining analogue expression to Eq.2.3.25 for a completely different purpose and completely different goal, when they scaled the nuclear charges in operator $H_{ne}$. The algebraic analogy comes from formally applying the DFT and Hellmann–Feynman theorem. Also, Eq.2.3.13 is known for its ground state in the discussion on RPA [2.3.19-20], but is described in more detail here. (Applying the classical approximation used in the Kohn-Sham method [2.3.2] for Eqs.2.3.24 and 25 - causing the huge problem and literature of exchange-correlation and self-interaction – it yields the $\partial enrg_{electr,0}(a)/\partial a \approx (1/2)\iint\rho_0(\mathbf{r}_1,a)\rho_0(\mathbf{r}_2,a)r_{12}^{-1}d\mathbf{r}_1d\mathbf{r}_2$ and $E_{electr,0} - e_{electr,0} \approx (1/2)\int_{[0,1]}\iint\rho_0(\mathbf{r}_1,a)\rho_0(\mathbf{r}_2,a)r_{12}^{-1}d\mathbf{r}_1d\mathbf{r}_2da$, but we do not address these approximations further. We call attention to the fact that Eq.2.3.25 does not have an exchange-correlation effect by the use of $b_0$ two-electron density [2.3.21-22].) Furthermore, the second derivative is interesting from Eq.2.3.21 for real valued $y_0(a)$

$$\partial^2 enrg_{electr,0}(a)/\partial a^2= N(N-1)\ \langle y_0(a)|r_{12}^{-1}|\ \partial y_0(a)/\partial a\rangle=$$
$$= \iint(\partial b_0(\mathbf{r}_1,\mathbf{r}_2,a)/\partial a)r_{12}^{-1}d\mathbf{r}_1d\mathbf{r}_2= (\partial/\partial a)\iint b_0(\mathbf{r}_1,\mathbf{r}_2,a)r_{12}^{-1}d\mathbf{r}_1d\mathbf{r}_2 \qquad \text{(Eq.2.3.26)}$$

Besides the fact that Eqs.2.3.21-26 are interesting theoretically, their actual values and numerical behavior have importance for molecular systems. Below we will demonstrate that $\partial enrg_{electr,0}(a)/\partial a$ is nearly constant, i.e., $enrg_{electr,0}(a)$ exhibits quasi-linear, more accurately quasi-linear with simple curvature behavior. It is less linear with increasing basis set, as it is displayed for an extended interval for $-0.1 \le a \le 1.1$ on 2.3.Figure 1 detailed below. (Negative coupling means attractive electrons, also considered in the literature with AC [2.3.23].) At a=0, the electron-electron repulsion does not have to be calculated, but there exists the non-vanishing value $v_{ee}(a)/a$, see middle term of Eq.2.3.21 and 21. We emphasize that case a=0 has no correlation effect, because $H_{ee}$ is cancelled by $aH_{ee}$, even the kinetic term has no problem since the wave function has a single determinant form. If $\partial enrg_{electr,0}(a)/\partial a$ is a quasi-constant, then from Eq.2.3.22 a "linear approximation" is $E_{electr,0} \approx e_{electr,0} + \langle Y_0|H_{ee}|Y_0\rangle$, and only Eq.2.3.2 has to be solved to approximate the ground state of Eq.2.3.1. See further reasoning below leading to Eq.2.3.29, as well as section 2.3.3.b later for numerical example showing that LCAO coefficients are close to each other between $Y_0$ and $\Psi_0$, i.e. $\Psi_0$ can be approximated first degree by $Y_0$ in Eq.2.3.13. We emphasize that it is only the first step of approximation toward more accurate ones. The term $\langle Y_0|H_{ee}|Y_0\rangle$ has to be calculated after the HF-SCF/basis/a=0 iteration, but no such calculation during. Technically, it can be achieved by a HF-SCF/basis/a=0 (converged) calculation (yielding $Y_0$ and $e_{electr,0}$), followed by a HF-SCF/basis/a=1 calculation. However, (!) in the latter, the starting LCAO parameter is not a standard initial guess matrix, commented on in more detail in section 2.3.3.e below, but the LCAO coefficient matrix from the previously converged a=0 calculation, and only the first (or $0^{th}$) iteration has to be taken, that is, the $e_{electr,0}+ <Y_0|H_{ee}|Y_0>$ energy value, and one can stop the HF-SCF/basis/a=1 algorithm at the beginning, - a simple technical manipulation in computing.

2.3.Figure 1.:

Plot of electronic energy denoted as $enrg_{electr,0}(a)$ in Eq.2.3.28 to show that it depends quasi-linearly on coupling strength parameter “a”, the $enrg_{electr,0}(a)$ is approximated with $s_0(a)$ from HF-SCF/basis/a algorithm suffering from basis set and correlation error. The slope at a=0 is the exact $<Y_0|H_{ee}|Y_0>/<Y_0|H_\nabla+H_{ne}|Y_0>$ suffering from basis set error only. The figure a shows the variance of slope on nuclear frame and N, (e.g. not monotonic with N), figure b shows that larger basis yields larger curvature, as well as the linear regression on figure b represents how the curve in case of STO-3G basis deviates from straight line. Notice that, Hamiltonian $H_\nabla+H_{ne}+aH_{ee}$ depends on coupling strength parameter linearly, but the quasi-linear behavior of $enrg_{electr,0}(a)$ instead of exact-linear does not come from basis set error only, this fact is forced by Eq.2.3.30. The quasi-linear behavior manifests that Eq.2.3.2 has rich pre-information for Eq.2.3.1 in computation chemistry.

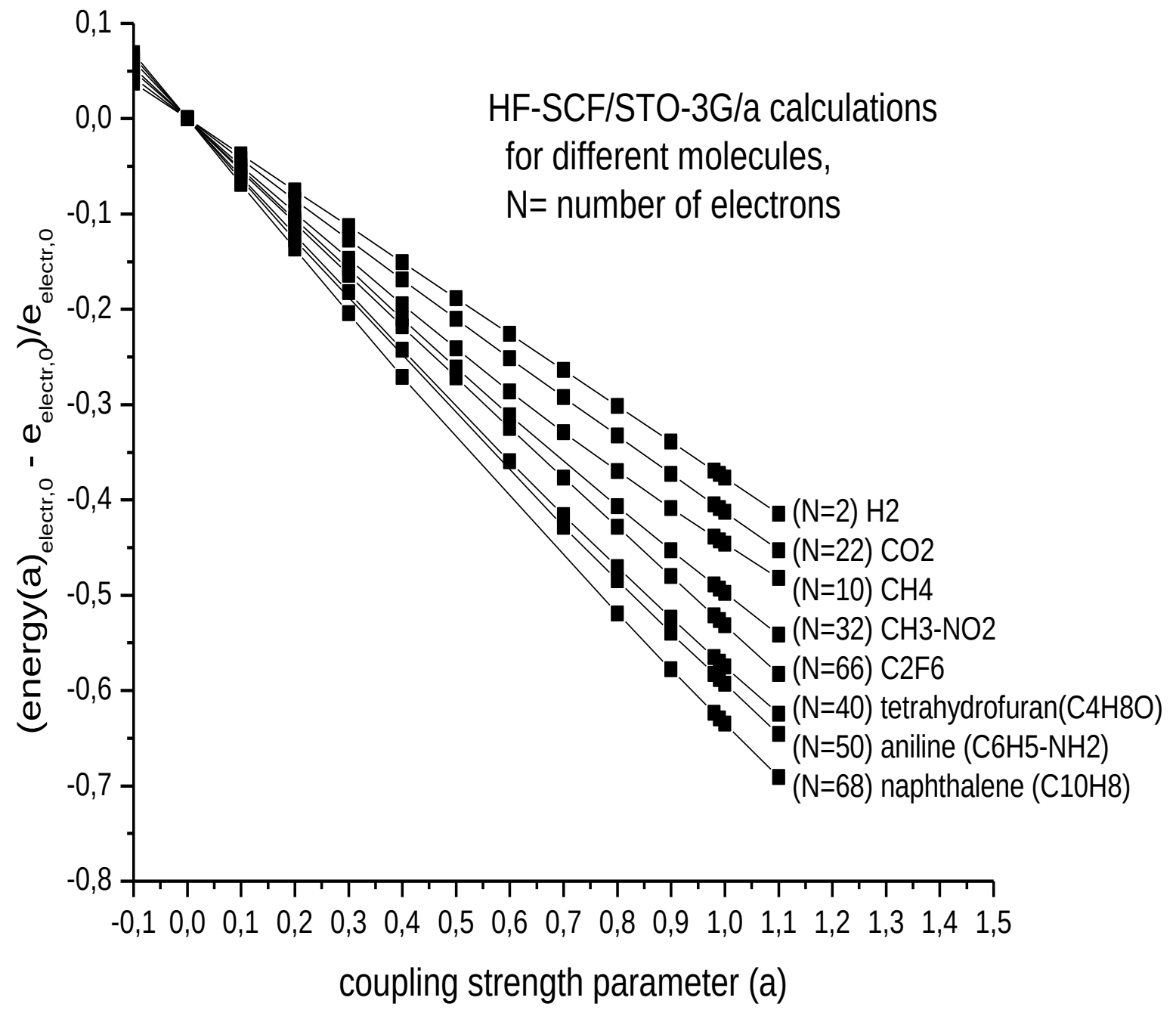

2.3.Figure 1.a

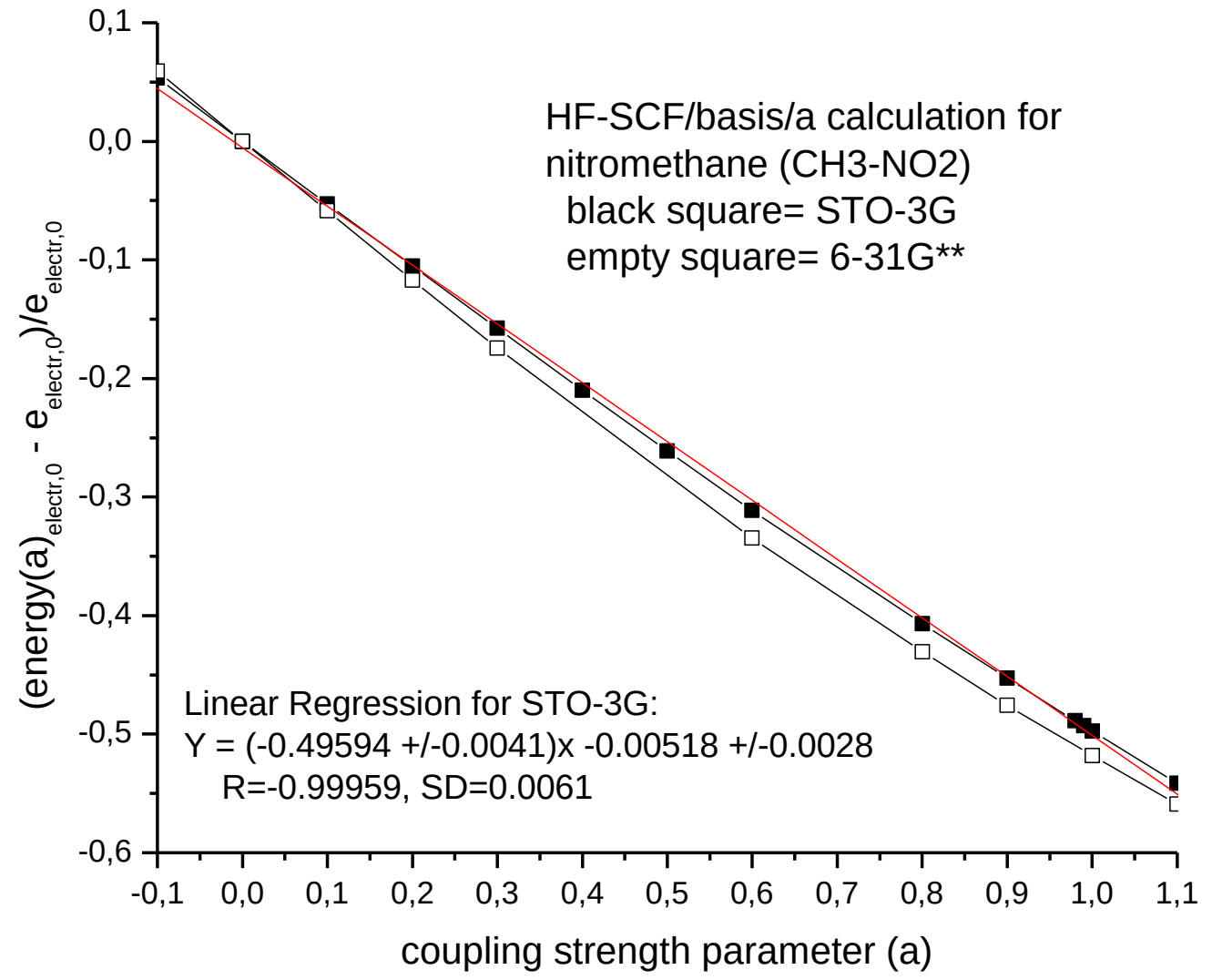

2.3.Figure 1.b

For generalization, we must note that, changing from interval [0,1] to $[a_1,a_2]$ is a bit more complex since a=0 "annihilates" and a=1 "un-changes" algebraic terms, yielding simpler expressions. However, we started with these values, because a=1 has physical reality (Eq.2.3.1) and a=0 is the TNRS. The more general form of Eq.2.3.13 comes from $\langle y_0(a_2)|(a_2-a_1)H_{ee}|y_0(a_1)\rangle = \langle y_0(a_2)|H(a_2)-H(a_1)|y_0(a_1)\rangle = \langle y_0(a_2)|H(a_2)|y_0(a_1)\rangle - \langle y_0(a_2)|H(a_1)|y_0(a_1)\rangle = (enrg_{electr,0}(a_2) - enrg_{electr,0}(a_1))\langle y_0(a_2)|y_0(a_1)\rangle$, from which

$$enrg_{electr,0}(a_2) = enrg_{electr,0}(a_1) + \\ + (N(N-1)/2)\langle y_0(a_2)|(a_2-a_1)r_{12}^{-1}|y_0(a_1)\rangle/\langle y_0(a_2)|y_0(a_1)\rangle \quad \text{(Eq.2.3.27)}$$

where $enrg_{electr,0}(a_i)$ is the ground state electronic energy, (similarly true for excited states), and $\xi(a_1,a_2) \equiv \langle y_0(a_2)|(a_2-a_1)r_{12}^{-1}|y_0(a_1)\rangle/\langle y_0(a_2)|y_0(a_1)\rangle$. Eq.2.3.27 is a generalization of case $[a_1,a_2]$= [0,1] in Eq.2.3.13. If $a_1=a_2$ or N=1 (no electron-electron pair) in Eq.2.3.27, then the second term on the right is zero and the equation becomes a triviality, as expected. If $a_1=0$ and $a_2=a$, the coupling term is in effect as

$$enrg_{electr,0}(a) = e_{electr,0} + (N(N-1)/2)\langle y_0(a)|ar_{12}^{-1}|Y_0\rangle/\langle y_0(a)|Y_0\rangle\ . \quad \text{(Eq.2.3.28)}$$

In the one dimensional domain of the variable coupling strength parameter "a", the a=0 is a singular point in the sense that it is the point of TNRS. However, the latter does not hold at any values when $a\neq 0$, i.e., those $y_0(a)$ wave functions are not a single determinant, as is well known for $\Psi_0 = y_0(a=1)$, and the deviation is more pronounced as the variable "a" deviates more from zero.

From Eq.2.3.27, the $\lim_{a1\to a2=a}$ for expression $(enrg_{electr,0}(a_2) - enrg_{electr,0}(a_1))/(a_1-a_2) = (N(N-1)/2)\langle y_0(a_2)|r_{12}^{-1}|y_0(a_1)\rangle/\langle y_0(a_2)|y_0(a_1)\rangle$ yields the same just as before in Eqs.2.3.21 or 24. Note, that although $v_{ee}(a=0)=0$ is a triviality, but the $(v_{ee}(a)/a)|_{a=0} \neq 0$ and, as a consequence, the $\partial enrg_{electr,0}(a)/\partial a|_{a=0} = (N(N-1)/2)\langle Y_0|r_{12}^{-1}|Y_0\rangle \neq 0$ also hold, see Eq.2.3.24 and 2.3.Figure 1. On 2.3.Figure 1 the relative $(enrg_{electr,0}(a)- e_{electr,0})/e_{electr,0} = \langle y_0(a)|aH_{ee}|Y_0\rangle/(\langle y_0(a)|Y_0\rangle\langle Y_0|H_\nabla+H_{ne}|Y_0\rangle)$ value from Eq.2.3.28 is plotted, wherein the $enrg_{electr,0}(a)$ is approximated with $s_0(a)$ from HF-SCF/basis/a algorithm, i.e., the $\langle s_0(a)|aH_{ee}|Y_0\rangle/(\langle s_0(a)|Y_0\rangle\langle Y_0|H_\nabla+H_{ne}|Y_0\rangle)$ value. The slope at a=0 is $\partial((enrg_{electr,0}(a)- e_{electr,0})/e_{electr,0})/\partial a|_{a=0} = (\partial enrg_{electr,0}(a)/\partial a)|_{a=0}/e_{electr,0} = \langle Y_0|H_{ee}|Y_0\rangle/\langle Y_0|H_\nabla+H_{ne}|Y_0\rangle$ which does not vary strongly as "a" evolves, as the figure shows. Continuing the derivation with "a" in Eq.2.3.21 or 24 yields the forecasting first approximation relationship. The variable "a" is not among the integral variables in <|>, i.e., not among $d\mathbf{x}_1\ldots d\mathbf{x}_N$, so it can be shifted into the integrand, that is $\partial/\partial a\int = \int\partial/\partial a$, and we arrive at the expression as in Eq.2.3.26 above. However, the second derivative in Eq.2.3.26 is a "small value" only, because from normalization $1= \langle y_0(a)|y_0(a)\rangle \Rightarrow 0= \partial 1/\partial a= \partial\langle y_0(a)|y_0(a)\rangle/\partial a= 2\langle y_0(a)|\partial y_0(a)/\partial a\rangle$, and this weighting implies that $\langle y_0(a)|H_{ee}|y_0(a)\rangle = V_{ee}(a)$ is a "large value", while $\langle y_0(a)|r_{12}^{-1}|\partial y_0(a)/\partial a\rangle$ in Eq.2.3.26 is a "small" one. As a consequence, a small (about zero) value $\partial^2 enrg_{electr,0}(a)/\partial a^2$ in Eq.2.3.26 yields that Eqs.2.3.27-28 are about straight lines with respect to the coupling strength variable "a", as demonstrated via computation in 2.3.Figure 1. This means that Eq.2.3.28 can be substituted with an equation of line with the help of Eq.2.3.21 or 24 as $enrg_{electr,0}(a) \approx e_{electr,0} + [(\partial enrg_{electr,0}(a)/\partial a)|_{a=0}]a = e_{electr,0} + a[(N(N-1)/2)\langle Y_0|r_{12}^{-1}|Y_0\rangle]$, and as indicated above, at a=1

$$E_{electr,0} \approx E_{electr,0}(TNRS) \equiv e_{electr,0} + (N(N-1)/2)\langle Y_0|r_{12}^{-1}|Y_0\rangle\ . \quad \text{(Eq.2.3.29)}$$

The right side is more generally, $e_{electr,0} + (v_{ee}(a)/a)|_{a=0}$ for which the corresponding part of Eq.2.3.19 gives the lower boundary. Eq.2.3.29 comes from when the $Y_0$ trial function is substituted for $\langle Y_0|H|Y_0\rangle$ instead of an energy optimized trial $S_0$, but we have provided a more detailed analysis here. More accurate expressions will be analyzed as functions in the sections below theoretically and computationally along with particular examples. A validation of Eq.2.3.29 is manifest in 2.3.Figure 1.

On the right hand side of Eq.2.3.29 the $(N(N-1)/2)\langle Y_0|r_{12}^{-1}|Y_0\rangle = 2J - K$, after expanding the determinants where J and K are the known Coulomb- and exchange integrals well known

in HF-SCF formalism, only here the MOs belong to $Y_0(a=0)$ (and not $S_0(a=1)$). Eq.2.3.29 can be rewritten taking the idea from KS formalism as $(N(N-1)/2)\langle Y_0|r_{12}^{-1}|Y_0\rangle \approx ½\int\rho_0(\mathbf{r}_1,a=0)\rho_0(\mathbf{r}_2,a=0)r_{12}^{-1}d\mathbf{r}_1d\mathbf{r}_2$, very generally, $(N(N-1)/2)\langle y_k(a)|r_{12}^{-1}|y_k(a)\rangle = \int b_k(\mathbf{r}_1,\mathbf{r}_2,a)r_{12}^{-1}d\mathbf{r}_1d\mathbf{r}_2 \approx ½\int\rho_k(\mathbf{r}_1,a)\rho_k(\mathbf{r}_2,a)r_{12}^{-1}d\mathbf{r}_1d\mathbf{r}_2$. (The latter expression is formally the classical Coulomb repulsion, but the inclusion of a non-point charge cannot make it accurate, it still suffers from exchange-correlation deficiency as Eq.2.3.29 itself, which immediately tells us that if one omits the term $r_{12}^{-1}$ from the integrals changing the repulsion operator to electron counting: $N(N-1)/2 \approx N^2/2$, indeed it cannot be accurate.)

One other expression stemming from the variation principle has to be emphasized: the $S_0$ from HF-SCF/basis/a=1 for Eq.2.3.1 is energetically better than $Y_0$ from HF-SCF/basis/a=0 for Eq.2.3.2, when one uses this $Y_0$ for Eq.2.3.1, that is,

$$E_{electr,0} < \langle S_0|H|S_0\rangle \leq e_{electr,0} + (N(N-1)/2)\langle Y_0|r_{12}^{-1}|Y_0\rangle , \quad \text{(Eq.2.3.30)}$$

where the equality may come up when small e.g., STO-3G basis set is used. Eq.2.3.30 is an extension of Eq.2.3.15. The error (correlation) of the middle part with $S_0$ in Eq.2.3.30 stems from the fact that $\Psi_0$ is approximated with incorrect wave function form, namely with $S_0$. The left part is a known relation (variation principle) in Eq.2.3.30, while the expression on the right hand side for $Y_0$ comes from first perturbation and not from energy minimization, so the right side relationship between expressions containing $S_0$ vs. $Y_0$ comes from a variation principle, but at least the LCAO coefficients vary slowly between $Y_0(a=0)$ and $S_0(a=1)$. (Again, LCAO parameters in correct functional form $Y_0$ come from solving Eq.2.3.2 numerically, while in the incorrect functional form $S_0$ the LCAO parameters come from the energy minimization of $\langle S_0|H|S_0\rangle$ for Eq.2.3.1 (restricted by the known ortho-normalization for MOs).)

**2.3.2.d. Generalization of Møller-Plesset perturbation theory (1934) in relation to coupling strength parameter, another justification of perspective in Eq.2.3.29**

Here we mention the known and widely used Møller-Plesset (MP) perturbation theory [2.3.1] in the context of an HF-SCF/basis/a=1 case based on Eq.2.3.3. It is quite obvious, that for any value of "a", this theory applies just as in the case of a=1, a generalization in this work in relation to the coupling strength parameter. However, a=0 is a special case; the essential observation in MP perturbation theory (a≠0) is that, all Slater determinants formed by exciting electrons forming the occupied to the virtual orbitals are also eigenfunctions of Eq.2.3.3 with an eigenvalue equal to the sum of the one electron energies of the occupied spin-orbitals, so a determinant formed by exciting from the $p^{th}$ spin-orbital in the Hartree-Fock ground state into the $r^{th}$ virtual spin-orbital only canges the MO in the Slater determinant, and the eigenvalue changes to $E_{electr,0} \to E_{electr,0} + \varepsilon_p - \varepsilon_r$, which is somewhat surprising but true for a≠0, and more so for a=1 but, trivial in Eq.2.3.2 (a=0), as discussed above, see Eq.2.3.8. The typical corrections in MP theory take the form: $|\langle s_0(a)|ar_{ij}^{-1}|s_{0,pq}^{rs}(a)\rangle|^2/(\varepsilon_p+\varepsilon_q-\varepsilon_r-\varepsilon_s)$ in the second order correction, wherein the $s_{0,pq}^{rs}(a)$ is the determinant from $s_0(a)$ e.g., the p and q spin-orbitals are changed to the excited (virtual) r and s ones, as well as this term is summed up for all i<j electrons (N) and all r<s available (i.e. calculated) virtual spin-orbitals. This expression is extended with the coupling strength parameter "a" and particularly for a=0 all the MP corrections cancel (because of the term $ar_{ij}$), which means that Eq.2.3.2 does not need any correction, because the Slater or single determinant is an accurate wave function form. However, for practical use, it needs manipulation to relate Eq.2.3.2 to Eq.2.3.1 somehow. In this manipulation, e.g., Eq.2.3.29 includes the electron-electron repulsion as a simplification for a main term, but it must be refined further to acheive the correlation effect. The latter is not the subject of this work, we only illustrate the rich relationship of Eq.2.3.2 to Eq.2.3.1. MP theory, for example, may provide some clue for this refinement for Eq.2.3.29.

In relation to the MP theory, the $s_0$, which includes the effect of $ar_{12}^{-1}$ somehow (a≠0), but not precisely, the MP tries to correct it to approach $y_0(a)$ most importantly $\Psi_0(a=1)$ as possible in terms of energy; however, $Y_0(a=0)$ is not affected by $r_{12}^{-1}$ at all. MP theory corrects what HF-SCF/basis/a≠0 makes in approximating electron-electron repulsion energy, i.e.,the error coming from $<S_0|H_{ee}|S_0> \approx <\Psi_0|H_{ee}|\Psi_0>$ when e.g., a=1, based on the Rayleigh-Schrödinger perturbation theory. The a=1 case in the literature has been generalized here above for general a≠0 value on theoretical grounds. One point is fundamental in applying MP for a=0 case: The first order MP energy correction [2.3.1] is $E_0^{(1)}= <s_0(a\neq 0)|aH_{ee}-aV_{aa,eff}|s_0(a\neq 0)>$, i.e., only the difference is in the core (see Eq.2.3.3 for $aV_{aa,eff}$). If a=0, the $E_0^{(1)}=0$, i.e., Eq.2.3.2 needs no correction but, if we want to relate Eq.2.3.2 to Eq.2.3.1, then "$aH_{ee}$ is not approximated by $aV_{aa,eff}$", but "$H_{ee}$ is approximated crudely with zero", because a=0, so $E_0^{(1)}= <Y_0(a=0)|H_{ee}-0|Y_0(a=0)|>$, i.e., the full term is in the core, exactly what Eq.2.3.29 has from another point of view via Eq.2.3.13 above. Further corrections to Eq.2.3.29 for higher accuracy can be done with the exact MP2 analogue $|<Y_0|r_{12}^{-1}|Y_{0,pq}^{rs}>|^2/(\varepsilon_p+\varepsilon_q-\varepsilon_r-\varepsilon_s)$ terms (as well as MP3, MP4 etc.) if one uses the MP method to correct.

We have just generalized the MP perturbation theory for $H_\nabla+ H_{ne}+ aH_{ee}$ extended with the general value "a", particularly for Eq.2.3.29 (a=0) which switches $H_\nabla+H_{ne}$ to $H_\nabla+H_{ne}+H_{ee}$. An interesting similarity known in HF-SCF/basis/a=1, that is, 1st order MP is only the HF-SCF energy itself, an overlap between HF-SCF and MP theory, and corresponding here to that, in TNRS the 1st order perturbation to Eq.2.3.2 to approximate Eq.2.3.1 for k=0 ground state is only Eq.2.3.29, although the latter can be deduced if a $Y_0$ trial function is substituted into $E_{elects,0}=<\Psi_0|H|\Psi_0>$ for $\Psi_0$. Numerical tests of further MP perturbation corrections for Eq.2.3.2 to switch it to Eq.2.3.1 will be discussed in another work. Finally, we mention that it is said that, although MP is the standard way how a perturbation theory or correlation calculation must be accomplished, but MP is not the best among correlation calculations (e.g., DFT provides us with something better nowadays), problematic for higher systems [2.3.24], or particularly, e.g., "the transition metal chemistry is a graveyard for UHF-based MP methods" [2.3.11], and so on. Our hypothesis for this difficulty is that a LUMO in HF-SCF/basis/a algorithm for the (N+1)th electron in an N-electron system is calculated from the N electron repulsion only if a=1, generally if a≠0, instead of N+1, but an excitation to LUMO comes from the lowest lying N electrons, which has an impact on calculating LUMO physically and plausibly, it is like the possible (but not inevitable weakness of a general extrapolation in comparison to interpolation on the same system. However, in the special case, when a=0, Eq.2.3.2 does not have this impact, at least mathematically (see the concept of "virtual N" in section 2.3.4.a.)

The one-electron density from Eq.2.3.2 to Eq.2.3.1 can also be perturbed as the electronic energy for the ground state above. (A clue for this: Adding weighted $C_{pqrs}\int Y_0^* Y_{0,pq}^{rs} ds_1 d\mathbf{x}_2 \dots d\mathbf{x}_N$ terms to $\rho_0(\mathbf{r}_1,a=0)$ to correct on "MP2 level" looks plausible, and the integration of this corrected $\rho_0(\mathbf{r}_1,a=0)$ gives $\int\rho_0(\mathbf{r}_1,a=0)d\mathbf{r}_1+ \Sigma C_{pqrs}<Y_0|Y_{0,pq}^{rs}>= N+\Sigma 0= N$, since $\{Y_k\}$ is ortho-normal set. In this way, the corrected $\rho_0(\mathbf{r}_1,a=0)$ does not have to be re-normalized, its shape is corrected by "add-subtract" design which keeps the norm, but its derivative changes: $\nabla_1[\rho_0(\mathbf{r}_1,a=0)+ C_{pqrs}\int Y_0^* Y_{0,pq}^{rs} ds_1 d\mathbf{x}_2 \dots d\mathbf{x}_N] \neq \nabla_1\rho_0(\mathbf{r}_1,a=0)$, which are necessary for better kinetic energy estimation, see Eq.2.3.32 (t>T).) 2.3.Figure 2 demonstrate that in cases (dotted line on figure b) do manifest. The KS formalism uses the approximation $<\Psi_0|H_{ee}|\Psi_0>\approx \frac{1}{2}\int\rho_0(KS/basis/a=1,\mathbf{r}_1)\rho_0(KS/basis/a=1,\mathbf{r}_2)r_{12}^{-1}d\mathbf{r}_1 d\mathbf{r}_2$ for the a=1 case, known well in literature, and corrects it with DFT denoted as $E_{xc}[\rho]$ (both during SCF), as well as with better efficiency [2.3.11] than MP (after SCF). In this KS formalism the exchange correlation energy estimation, $E_{xc}[\rho]$, is worked out for KS or HF-SCF/basis/a=1 and ground state density and energy, but for $a\neq 1$ cases the generalization is beyond the scope of this work. However, as a first try the behavior of $E_{xc}[\rho_k(\mathbf{r}_1,a=0)]$, i.e., some good working functionals should be tested when ground and excited state one-electron densities (or their corrected forms) are substituted into them. (Notice that, which is obvious at this point, like the $<S_0|H_{ee}|S_0>$, the $<Y_0|H_{ee}|Y_0>$ first approximates $<\Psi_0|H_{ee}|\Psi_0>$ in TNRS in Eq.2.3.29. The LCAO coefficients are close in $S_0$ and $Y_0$, but not the same, so for the one-electron densities formed from $S_0$ vs. $Y_0$, see the demonstration below with 2.3.Figure 2.)

Finishing this section from a mathematical point of view, if an additional operator was in effect beside $H_{ee}$ referring to e.g., external forces, the algorithm or procedure is exactly the same as the one leading to Eq.2.3.29 and its discussion; the operator $H_{ee}$ must be changed or extended accordingly.

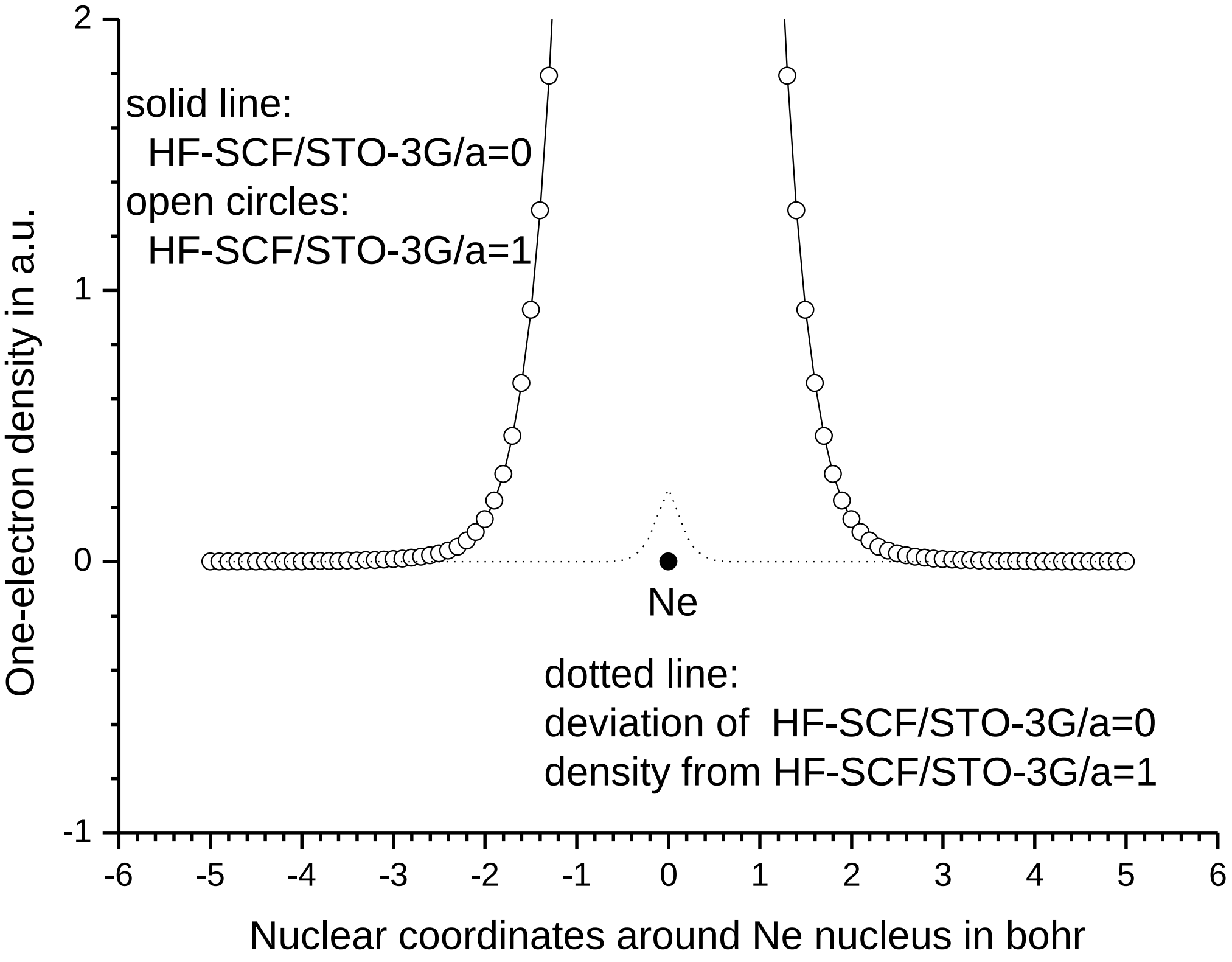

2.3.Figure 2.a.: One-electron density of Ne (neon, N=10), positioned at the origin. The about 240.04 peak value at the origin differs within 0.1% error by the two calculations. The small difference between the two curves indicates that the LCAO coefficients are close to each other between TNRS (a=0) and the interacting system where a=1.

2.3.Figure 2.: Different levels of calculations compared: $\rho(\mathbf{r}_1$,HF-SCF/STO-3G/a=0)= $10\int Y_0^{*}Y_0 ds_1 d\mathbf{x}_2 \ldots d\mathbf{x}_{10}$, $\rho(\mathbf{r}_1$,HF-SCF/STO-3G/a=1)= $10\int S_0^{*}S_0 ds_1 d\mathbf{x}_2 \ldots d\mathbf{x}_{10}$ and $\rho(\mathbf{r}_1$,B3LYP/STO-3G), certainly a=1 in B3LYP [2.3.12]. Notice that in $\rho(\mathbf{r}_1$,HF-SCF/STO-3G/a=0) the $V_{ee}$ is not included, that is the TNRS, while in the other two, an approximate $V_{ee}$ is included somehow.

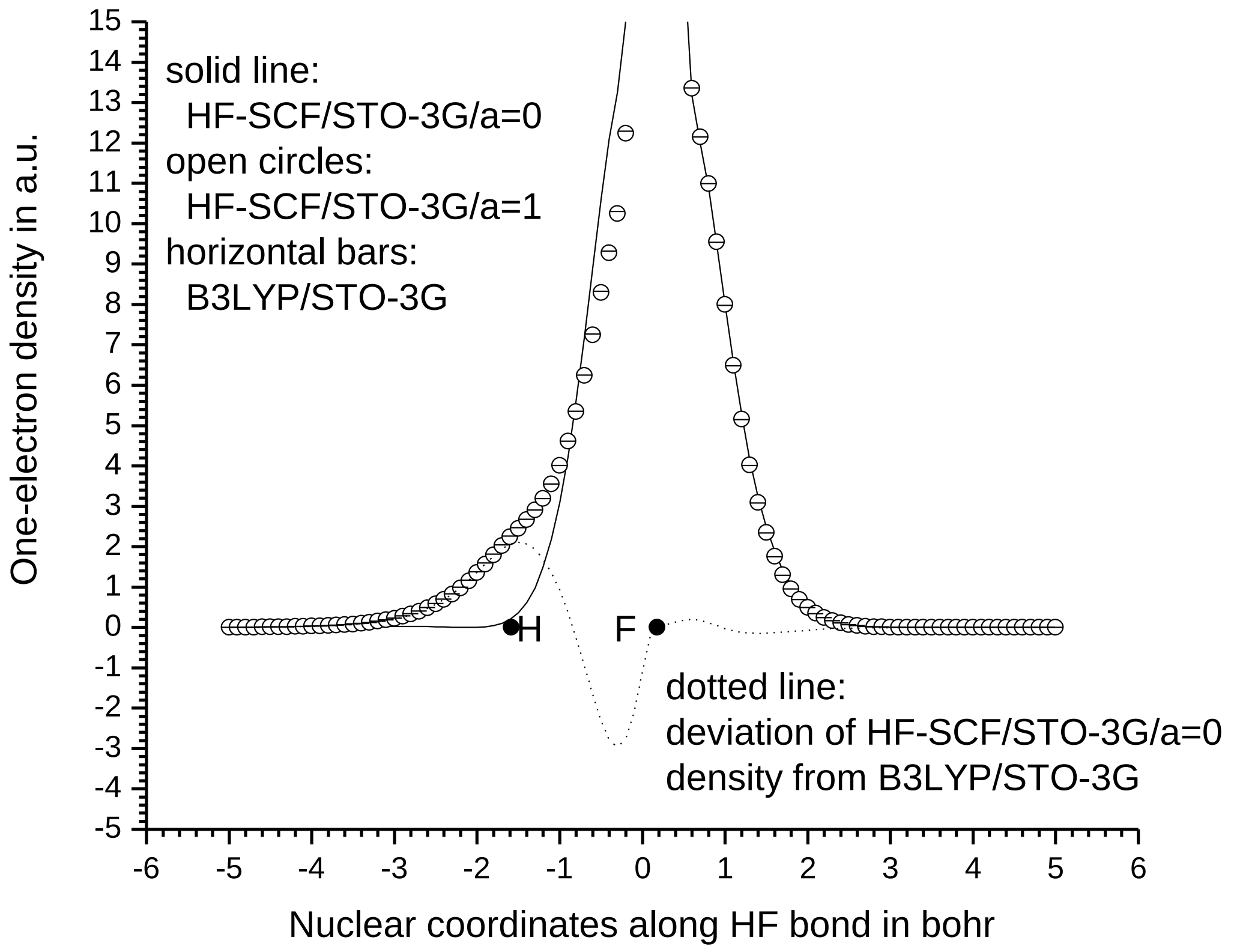

2.3.Figure 2.b.: One-electron density of HF (hydrogen-fluorid, N=10) molecule, the molecule is positioned along the z-axis and the origin is at the center of mass. Certain part of the ρ($\mathbf{r}_1$,HF-SCF/STO-3G/a=0) curve runs together with the other two (ρ($\mathbf{r}_1$,HF-SCF/STO-3G/a=1) and ρ($\mathbf{r}_1$,B3LYP/STO-3G)), e.g. the part just behind the F atom, where dotted line is almost zero, there the inclusion or neglecting the electron-electron interaction has no strong effect on density, while on the H atom ($Z_H=1<<Z_F=9$) the deviation is pronouncing. The about 184.50 peak value differs within 0.02% error by the three calculations. B3LYP method is chosen for extra comparison beside HF-SCF/STO-3G/a=1, because it includes the correlation effect fairly. Dotted line is the difference ρ($\mathbf{r}_1$,B3LYP/STO-3G)- ρ($\mathbf{r}_1$,HF-SCF/STO-3G/a=0), this kind of difference is similar what is used e.g. in RPA. The ratio ρ($\mathbf{r}_1$,B3LYP/STO-3G)/ρ($\mathbf{r}_1$,HF-SCF/STO-3G/a=0) with basis set error and very good correlation estimation approximates the $[w_{DFT}(\mathbf{r}_1)]^2$ in Eq.2.3.37. In the vicinity of H atom, the ρ($\mathbf{r}_1$,HF-SCF/STO-3G/a=0) is visibly steeper (∂ρ($\mathbf{r}_1$,HF-SCF/STO-3G/a=0)/∂z is larger) than of the other two, yielding higher (integral value) kinetic energy (t(a=0) vs. T(a=1)) described in Eq.2.3.32.

**2.3.2.e. Comparing how the virial theorem works in the electronic Schrödinger equation (Eq.2.3.1, a=1) vs. the TNRS (Eq.2.3.2, a=0), and its generalization via coupling strength parameter 'a'**

The virial theorem with our extension (coupling strength) parameter "a", also holds and reads as

$$(V_{nn} + \langle y_0(a)|H_{ne}|y_0(a)\rangle + a\langle y_0(a)|H_{ee}|y_0(a)\rangle)/\langle y_0(a)|H_{\nabla}|y_0(a)\rangle = -2 \quad \text{(Eq.2.3.31)}$$

for atoms ($V_{nn}$=0) and stationary (equilibrium or transition state) molecules for any values of "a", not only for a=0 and a=1. (It holds outside of the interval [0,1] too, however, extreme values can blow up the calculation.) For a=1, $(V_{nn} + V_{ne} + V_{ee})/T= -2$, while for a=0, $(V_{nn}+v_{ne})/t= -2$. Because $E_{total\ electr,0}= T+V_{ne}+V_{ee}+V_{nn}$ and $e_{total\ lectr,0}= t+v_{ne}+V_{nn}$, the virial theorem provides for atoms ($V_{nn}$=0) and stationary molecules that $E_{total\ electr,0}= -T$ and $e_{total\ electr,0}= -t$. While Eq.2.3.13 holds anywhere on the PES; the simpler form of virial theorem in Eq.2.3.31 is restricted to atoms and stationary points. As a consequence, Eq.2.3.13 can be expanded with the virial theorem with a restriction of Eq.2.3.31 as

$$t-T= E_{total\ electr,0}-e_{total\ electr,0}= E_{electr,0}-e_{electr,0}= \langle\Psi_0|H_{ee}|Y_0\rangle/\langle\Psi_0|Y_0\rangle= (N(N-1)/2)\xi\ . \quad \text{(Eq.2.3.32)}$$

From Eq.2.3.19 we have $E_{electr,0} - e_{electr,0} >> 0$, and by Eq.2.3.32 this yields t >> T, i.e., the kinetic energy is higher in the TNRS as opposed to the interacting one. Without or with decreased electron-electron repulsion (tuned by "a"), the electron cloud shrinks to the nuclei, and in accord with the known fact, closer to the nucleus - the kinetic energy is higher. Important "back restrictions" of Eq.2.3.32 is that we have to elevate the condition of the points of stationary molecules on PES, only atoms strictly obey Eq.2.3.32. The reason is: that if parameter "a" alters, the stationary geometries for t, T, or equivalently for $e_{electr,0}$ and $E_{electr,0}$, are not the same; but still the t > T is true for any geometry. This argument via Eq.2.3.32 is valid not only for the endpoint of interval [0,1], but also holds for the end points of any interval $[a_1,a_2]$. This will be looked at in more detail with respect to LCAO parameters in section 2.3.3 along with how to relate t to T. (For relativistic Hamiltonian, see ref. [2.3.25].)

As a simple example of Eq.2.3.32, consider the He atom ($V_{nn}$=0, N=2, Z=2) and using the textbook $E_{electr,0}(1s,N=1)=-Z^2/2$ in hartree, the TNRS (Eq.2.3.2, N=2, a=0) or HF-SCF/AO(1s)/a=0 yields $Y_0= (\alpha_1\beta_2-\alpha_2\beta_1)1s(\mathbf{r}_1)1s(\mathbf{r}_2) \Rightarrow$ t(TNRS, exact)= $(Z^2/2)N$= 4 and $v_{ne}$(TNRS, exact)= $-Z^2N$= -8 $\Rightarrow -(V_{nn}+v_{ne})/t$= 2 indeed, and $e_{electr,0}$= 4-8= -t= -4, while t≈ t(TNRS, HF-SCF/STO-3G/a=0)= 3.863497 suffers basis set error. On the other hand, HF-SCF/STO-3G/a=1 with $\Psi_0 \approx S_0$ (bearing basis set and correlation error) yields an approximation for $E_{electr,0}= T+ V_{ne}+ V_{ee}$ the -2.807784= 2.823526 -6.687023+ 1.055713, for $E_{electr,0}$= -T the 2.807784≈ 2.823526, and for the theoretical $-(V_{ne}+V_{ee}+V_{nn})/T$=2 the 1.9944. Compare t= 4≈ 3.863497 to T≈ 2.823526, for which the t-T≈ 1.04 and $(E_{electr,0}-e_{electr,0})/e_{electr,0}$= (t-T)/t≈ -0.27 are in agreement with 2.3.Figure 3.c (below).

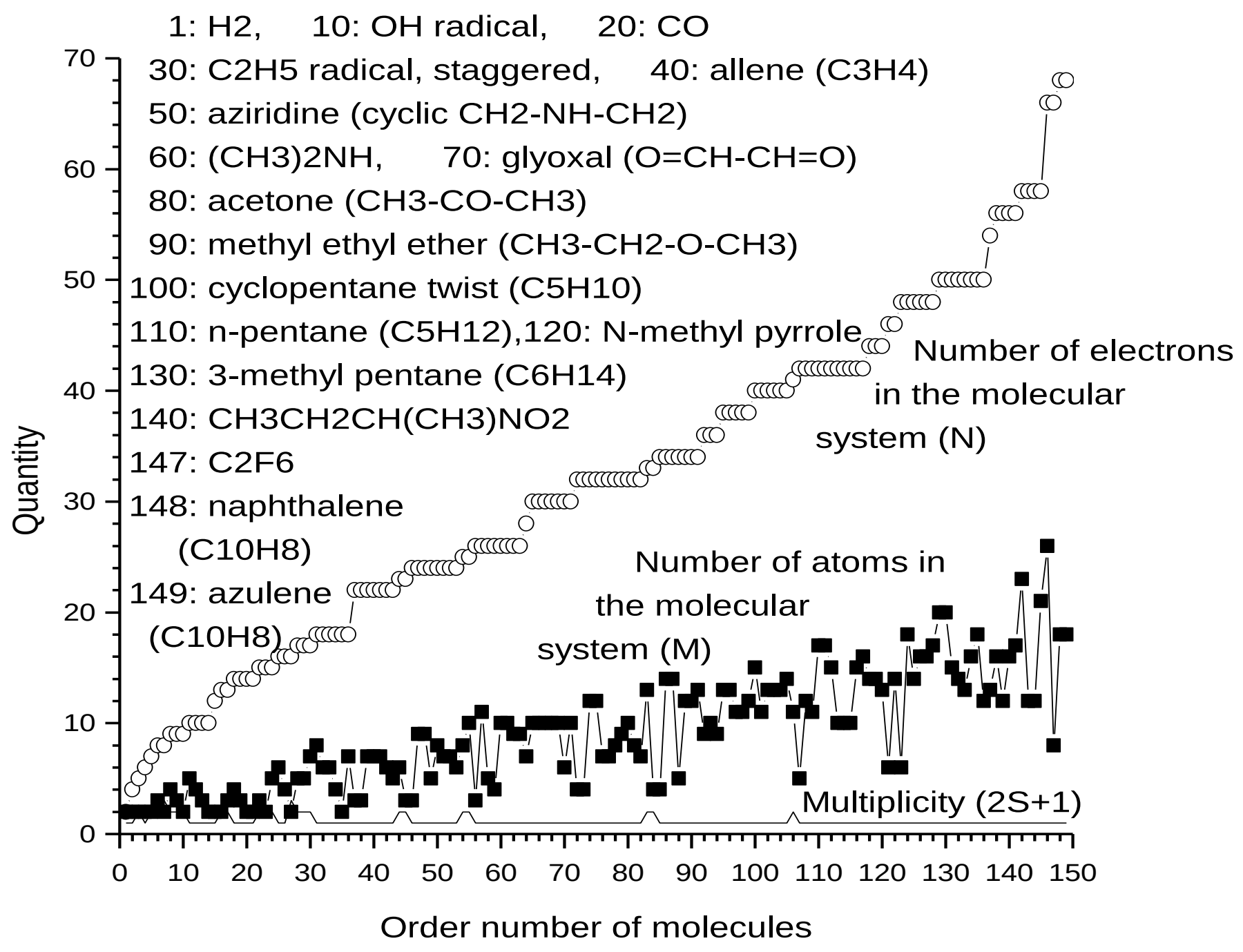

2.3.Figure 3.a.: Some data of the nuclear frame of 149 neutral molecules from the G3 set [2.3.15-16] (selected as maximal $Z_A < 11$ in the system) for 2.3.Figures 3.b-d to identify them in our analysis.

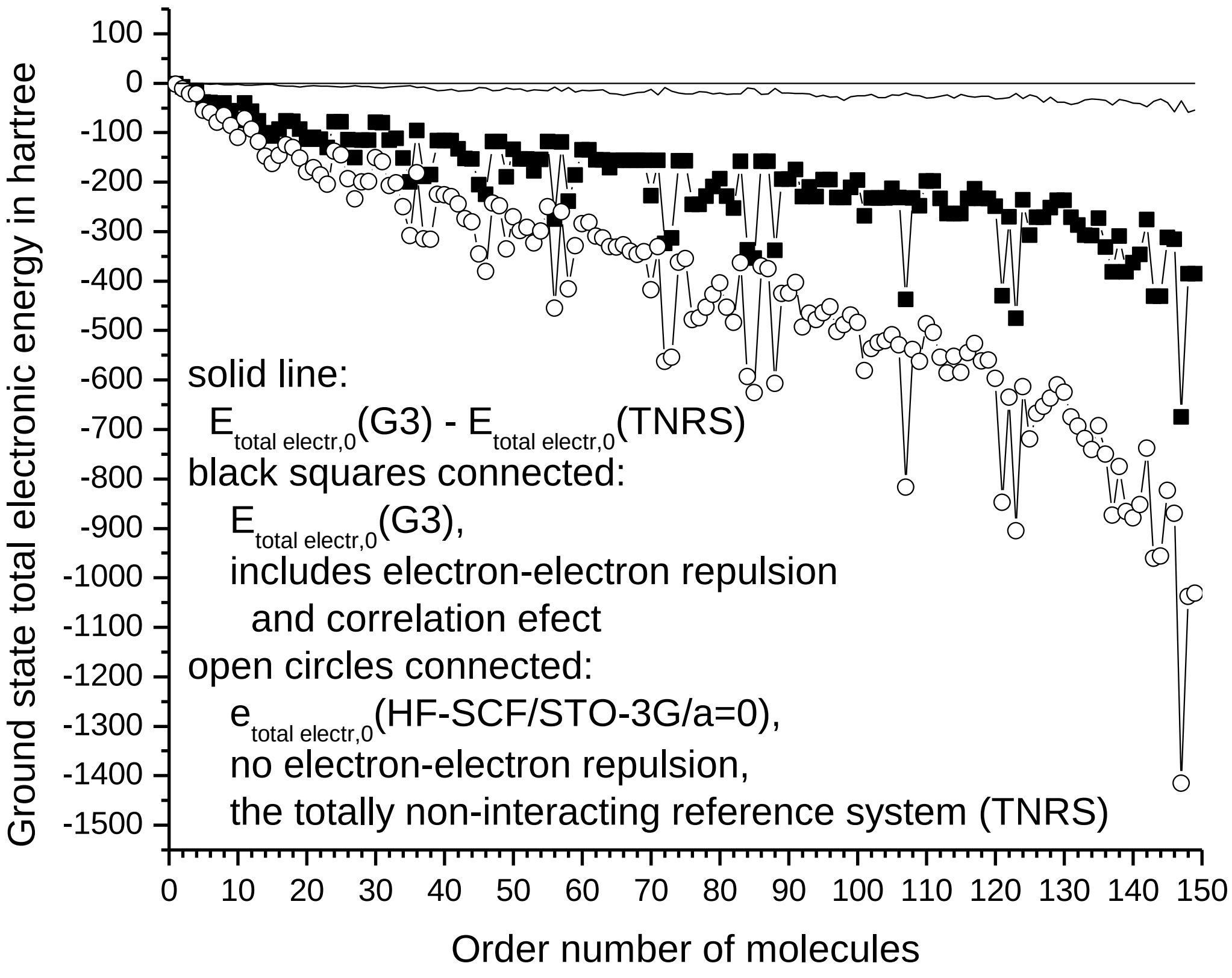

---

2.3.Figure 3.b.: Ground state total electronic energy of molecules is plotted as function of the order number of molecules shown on 2.3.Figure 3.a. The order number is chosen for N to be monotonic, so local peaks come from very different molecules having same (or close) N values but different ground state total electronic energies, i.e. the shape of the curve itself has no particular meaning, the important message is that the two curves (black squares and open circles) run together like the same fingerprint. The related curve for open circles with larger 6-31G** basis set would yield lower energy values by about 2% (basis set error improvement), and would be almost at the same position for eyes, not plotted. Solid line is the deviation $E_{total\ electr,0}(G3) - E_{total\ electr,0}(TNRS)$ via first approximation in Eq.2.3.29 which brings the open circle values (Eq.2.3.2 with small basis set error) remarkably back to black square ones (Eq.2.3.1 with G3 estimation).

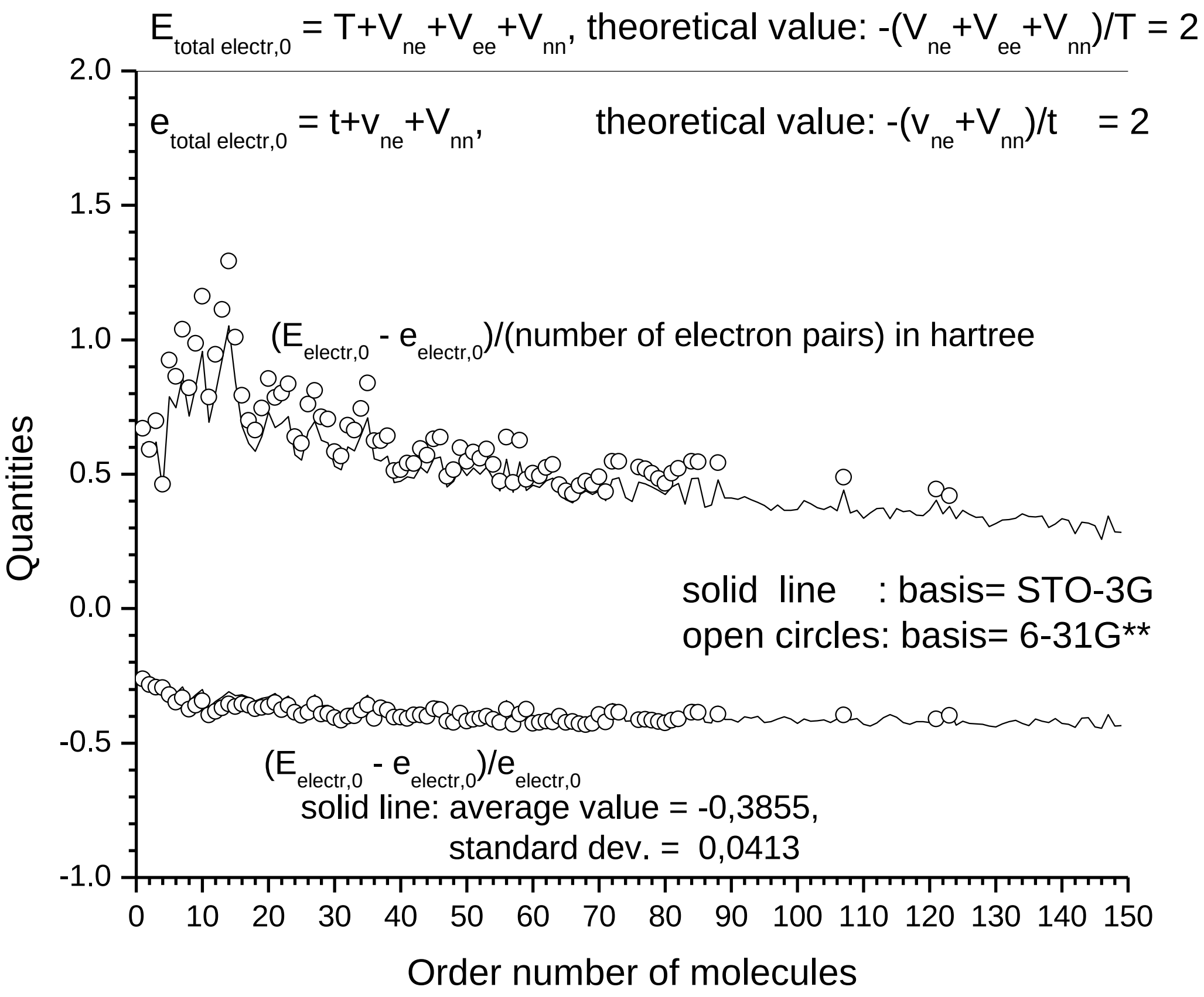

2.3.Figure 3.c.: $E_{electr,0}$ is via G3 calculation [2.3.15-16] (including correlation effect and correction of basis set error), while calculation of $e_{electr,0}$ is from HF-SCF/basis/a=0 for Eq.2.3.2 (suffering from basis set error). This plot exhibits the quasi-constant behaviour of $\xi$ [hartree] in Eq.2.3.13 and ($E_{electr,0}$(G3) - $e_{electr,0}$)/$e_{electr,0}$ approximation for Eq.2.3.39. Notice that the robust change in 2.3.Figure 3.b has disappeared in this re-plot in relation to energy.

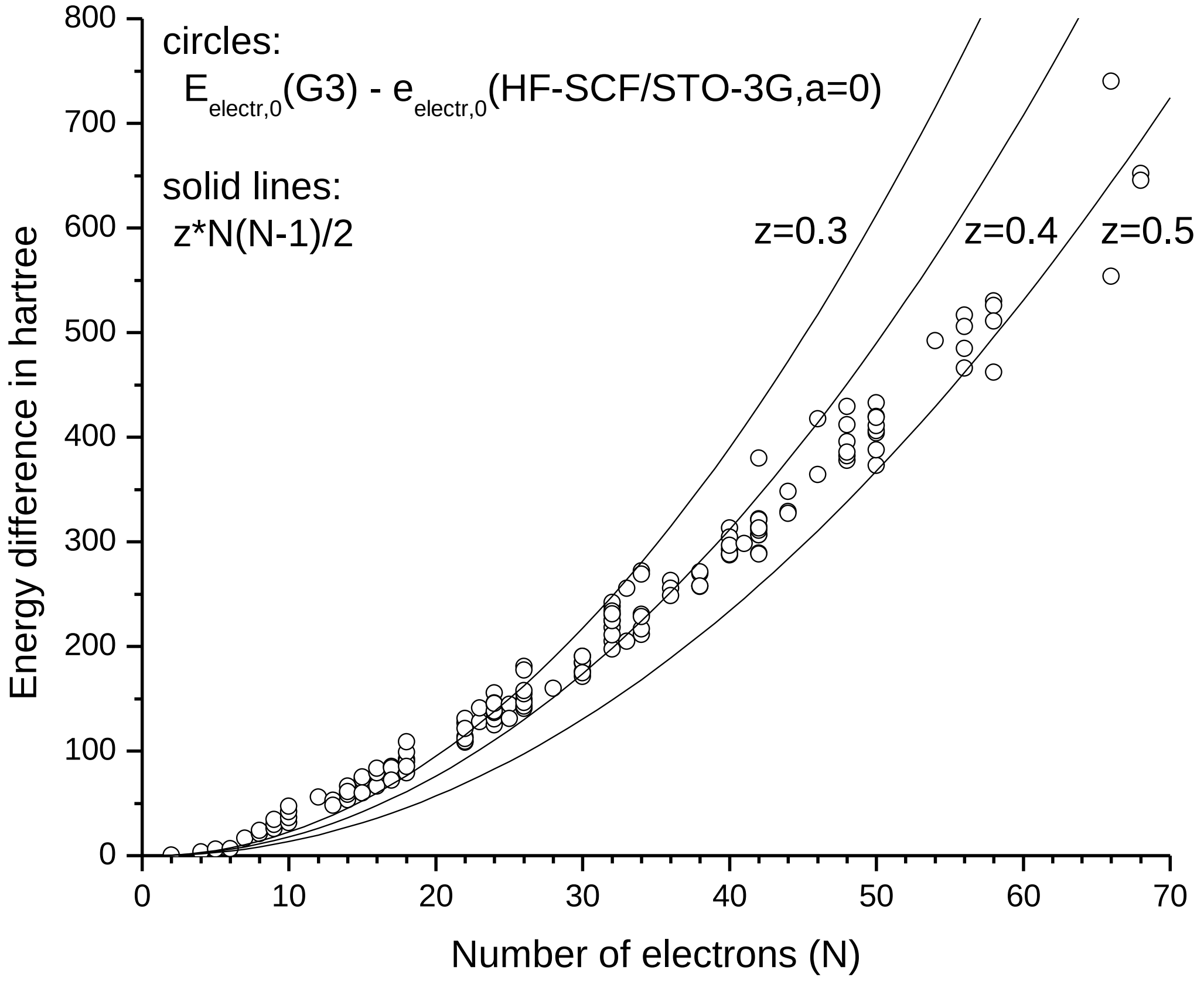

2.3.Figure 3.d.: Re-plot of data from a-c: $E_{electr,0}$ - $e_{electr,0}$ as a function of number of electrons (N) in neutral equilibrium geometry molecules from the G3 set to show the quasi-constant character of ξ on nuclear frame (curve parameter z scans some values to guide the eyes) introduced in Eq.2.3.13.

### 2.3.2.f. Generalization of the 1st Hohenberg-Kohn theorem (1964) from the electronic Schrödinger equation (Eq.2.3.1, a=0) to a general coupling constant, paying particular attention to TNRS (Eq.2.3.2, a=0)

In section 2.3.3 we will demonstrate that the HF-SCF optimized Slater determinant $S_0$ as an approximation for $\Psi_0$ (in Eq.2.3.1) and the $Y_0$ (from Eq.2.3.2) do not differ significantly in respect to LCAO coefficients (providing the same basis set is used), the large energy difference ($E_{electr,0}$ - $e_{electr,0}$) stems from whether the electron-electron repulsion term is added (Eq.2.3.1) or not (Eq.2.3.2), as approximated e.g., in Eq.2.3.29.

The approximation in Eq.2.3.29 will also be tested and commented upon, but theoretically we analyze another accurate functional link between $\Psi_0$ and $Y_0$ after Eq.2.3.13. Eq.2.3.13 makes a perfect accurate link theoretically, but is not very useful in practice, because it contains the uncalculated $\Psi_0$ on the right hand side, while Eq.2.3.29 belongs to typical and practical forms, but unfortunately, not accurate enough. As indicated above, the 1st HK theorem [2.3.2, 11], (that is, $\rho_0 \Rightarrow \{N, Z_A, R_A\} \Rightarrow H \Rightarrow \Psi_0 \Rightarrow E_{electr,0}$ and all other properties, while the opposite way such as $\Psi_0 \Rightarrow \rho_0$, $E_{electr,0}$ and all other properties is obvious), provides that $\Psi_0 \Leftrightarrow H \Leftrightarrow H_\nabla + H_{ne} \Leftrightarrow Y_0$, i.e., $\Psi_0 \Leftrightarrow Y_0$, which is more generally $\rho_0(\mathbf{r}_1,a) \Leftrightarrow \rho_0(\mathbf{r}_1,a=0)$ or $\rho_0(\mathbf{r}_1,a=1)$. The inclusion of the coupling strength parameter "a" makes it more general, even for two different values of "a" as

$$\rho_0(\mathbf{r}_1,a_1) \quad \text{or} \quad y_0(a_1) \Leftrightarrow \rho_0(\mathbf{r}_1,a_2) \quad \text{or} \quad y_0(a_2)\,. \qquad \text{(Eq.2.3.33)}$$

in practice, the most important for a DFT establishment in this work:

$$\rho_0(\mathbf{r}_1,a=0) \text{ from } H_\nabla+H_{ne} \Leftrightarrow E_{electr,0} \text{ from } H_\nabla+H_{ne}+H_{ee}\ , \qquad \text{(Eq.2.3.34)}$$

for example, recall again the weak approximation in Eq.2.3.29.

The one-electron density from the wave function is a basic definition for any value of "a", but we mention the old and known decomposition in the opposite way [2.3.2]: from density $\rho_0(\mathbf{r}_1,a)$ to an approximate single Slater determinant $s_0(a)$. Recall that $y_0(a)$ and $s_0(a)$ $\rightarrow Y_0(a=0)$ if $a\rightarrow 0$, wherein the functional error such as the single determinant $s_0(a)$ approximates a non-single determinant function that $y_0(a)$ eliminates, while a basis set error can remain in both. Furthermore, the generalization of a 2nd HK theorem is in fact trivial, since $H_\nabla+H_{ne}+aH_{ee}$ is linear not only for the a=1 (source of 2nd HK) but also for a≠1. The HK theorems or their corresponding statements for excited states are still problematic [2.3.26].

### 2.3.2.g. Particular functional link between the electronic Schrödinger equation (Eq.2.3.1, a=1) and the TNRS (Eq.2.3.2, a=0) focusing on the ground state (k=0)

In section 2.3.4 a CI expansion will be outlined, but for the ground state an immediate algebraic link is obvious, at least in the vicinity of stationary points in the ground state. For example, the Quantum Monte Carlo method typically employs a trial wave function which is a single Slater determinant times a Jastrow pair-correlation factor [2.3.10, 27-28]. Because the LCAO coefficients do not differ significantly (at least not in the vicinity of stationary points, see below in section 2.3.3), we can assume that there exists $w(\mathbf{r}_1,\mathbf{r}_2,\ldots\mathbf{r}_N)$ r-symmetric (i.e., for the exchange of any $\mathbf{r}_i$ and $\mathbf{r}_j$) such that improving $\Psi_0 \approx S_0$ by $\Psi_0 = w(\mathbf{r}_1,\mathbf{r}_2,\ldots\mathbf{r}_N)Y_0$ is possible, where $S_0(a=1)$ and $Y_0(a=0)$ are the HF-SCF/basis/a single Slater determinant approximation for Eq.2.3.1 and a solution for Eq.2.3.2, respectively. Hypothetically, for Eq.2.3.1 $wY_0$ can be a better approximation than $S_0$. The form of w must be a wise approximation while its analytical form is unknown, because if e.g., w is chosen only as $w=\Pi_{i=1\ldots N}p(\mathbf{r}_i)$ with a high enough quality LCAO for p, the $Y_0$ becomes energetically better, but remains a Slater determinant belonging to a better basis set. But this $(\Pi_{i=1\ldots N}p(\mathbf{r}_i))Y_0$ still cannot totally reach $\Psi_0$, because of its single determinant nature, and in this way $Y_0$ can approach $S_0$ more closely or improve only, but that can be done simply with a=1 with a better basis set, see 2.3.Appendix 1 for additional remarks.

If w is good enough, $wY_0$ may approach $\Psi_0$ more efficiently than $S_0$. More generally, and extending with coupling strength parameter 'a' and supposing basis set limit, the equality can (hypothetically) hold with r-symmetric w in such a way that

$$y_0(a)= w(\mathbf{r}_1,\mathbf{r}_2,\ldots\mathbf{r}_N,a)Y_0 , \qquad \text{(Eq.2.3.35)}$$

that is, how $y_0(a\neq0)$, and particularly $y_0(a=1)=\Psi_0$ and $y_0(a=0)=Y_0$ connect via w. To suppose the inclusion of the spin-orbit coordinate $\mathbf{x}_i$ and x-symmetricity, that is, w/r to exchange of any $\mathbf{x}_i$ and $\mathbf{x}_j$, in w is not necessary, since $Y_0$ already contains the spin coordinates, and in this way, the execution of spin algebra in $wY_0$ as well as in $|wY_0|^2$ would yield a contradiction. Furthermore,

$$\rho_0(\mathbf{r}_1,a=1)= N\int|\Psi_0|^2 := N\int w^2|Y_0|^2= N\int w^2(\int|Y_0|\, ds_1\ldots ds_N)d\mathbf{r}_2\ldots d\mathbf{r}_N$$

$$\text{vs. } \rho_0(\mathbf{r}_1,a=0)= N\int|Y_0|^2 ds_1 d\mathbf{x}_2\ldots d\mathbf{x}_N \qquad \text{(Eq.2.3.36)}$$

also indicates that $w(\mathbf{r}_1,\ldots,\mathbf{r}_N)$ is such that it corrects $Y_0$ to attain the physically accurate set and plausible $\rho_0(\mathbf{r}_1,a=1)$, as an additional functional to Eq.2.3.34; (:= stands for "let it be equal"). We mention again that $S_0$ and $y_0(a=0)=Y_0$ are single Slater determinants, while $\Psi_0$ or the general $y_0(a\neq0)$ are not single Slater determinant wave functions. In this hypothesis leading to Eq.2.3.35, Eq.2.3.36 is even more plausible since $\Psi_0$ and $Y_0$ are well-behaving, so by definition w is the real value - real variable function which makes Eq.2.3.36 hold. Finally, we can suppose that with the basis set limit and w containing r-symmetric spatial coordinates, Eq.2.3.35 holds between the x-anti-symmetric $y_0(a)$ and $Y_0$. Trivial case: if a=0 the w=1. At least the r-dependence (i.e., spin-independence) of w is in accord with the fact that operator $H_{ee}$ (which transfers the $Y_0$ to $y_0(a\neq0)$) contains only spatial coordinates. The normalization reads as $\langle y_0(a)|y_0(a)\rangle= \langle wY_0|wY_0\rangle= 1= \langle Y_0|Y_0\rangle$. Note, that changing from $Y_0$ to $wY_0$ may need renormalization. A DFT correspondent or alternative of w in Eq.2.3.35 notated as $w_{DFT}(\mathbf{r}_1)$ exists provided by Eq.2.3.33, and acting as a functional link between the real and TNRS one-electron densities as

$$\rho_0(\mathbf{r}_1,a=1)= [w_{DFT}(\mathbf{r}_1)]^2\rho_0(\mathbf{r}_1,a=0) \qquad \text{(Eq.2.3.37)}$$

with normalization $N= \int\rho_0(\mathbf{r}_1,a=1)d\mathbf{r}_1= \int[w_{DFT}(\mathbf{r}_1)]^2\rho_0(\mathbf{r}_1,a=0)d\mathbf{r}_1$.

A more detailed note is made on possible forms to approximate w in 2.3.Appendix 1 and 2 until we have information on its exact analytical form, if that exists. The square in Eq.2.3.37 ensures its required everywhere positive value, as well as its being in accord algebraically with Eq.2.3.36. The variation equation of w is as follows: In Eq.2.3.35 w should be a well behaved (aside from $\langle w|w\rangle=\infty$) r-symmetric function, but $wY_0$ must definitely be a well behaved x-anti-symmetric function, and using normalization constraint $\langle wY_0|wY_0\rangle=1$, the variation equation can be obtained from Eq.2.3.A1 of 2.3.Appendix 1 after multiplying with $(wY_0)^*$ from the left and integrating as

$$enrg_{electr,0}(a)= e_{electr,0} -(N/2)\langle wY_0|Y_0\nabla_1{}^2 w\rangle -N\langle wY_0|\nabla_1 Y_0\nabla_1 w\rangle +$$

$$+ a\langle wY_0|H_{ee}|wY_0\rangle \qquad \text{(Eq.2.3.38)}$$

in comparison to the approximate Eq.2.3.29 or the accurate but not practical Eq.2.3.28. In Eq.2.3.38 the equality holds by the minimizing r-symmetric w via a variation principle. If a=0, then w(a=0)=1 and $enrg_{electr,0}(a=0)= e_{electr,0}$, reducing to Eq.2.3.2, if a=1 then $enrg_{electr,0}(a=1)= E_{electr,0}$ in Eq.2.3.38 for Eq.2.3.1, as well as if $\Psi_0\approx Y_0$ i.e., $w(a=1)\approx1$ crude approximation is taken, then Eq.2.3.38 reduces to Eq.2.3.29 (simply because $\nabla_1 w=0$).

In Eq.2.3.38, the $e_{electr,0}$ via Eq.2.3.2 (a=0) suffers from a basis set error only, while $enrg_{electr,0}(a)$ of Eq.2.3.1 extended with "a" suffers from a basis set error and a correlation error if only an HF-SCF/basis/a algorithm is used and not the CI or DFT related methods. But calculating $enrg_{electr,0}(a)$ via Eq.2.3.38, the additive terms to $e_{electr,0}$ on the right hand side provide the full electron-electron repulsion energy including correlation effects (seeded by $H_{ee}$), if solved for accurate r-symmetric w, because $y_0(a)=w(a)Y_0$, and the terms seeded by nablas correct the kinetic and electron-nuclear attraction parts. Comparing Eq.2.3.38 to Eq.2.3.29 for the most important case a=1, the $N\{(\langle wY_0|Y_0\nabla_1^2 w\rangle/2 + \langle wY_0|\nabla_1 Y_0\nabla_1 w\rangle\}$ converts (corrects) the $t+v_{ne}$ to $T+V_{ne}$; below in Eq.2.3.42, a PS expansion takes care of it. Again, the $Y_0$ in Eq.2.3.38 stems from a pre-calculation.

As a summary of section 2.3.2, many interesting and useful properties and equations in relation to coupling strength parameter as well as the a-value extended HF-SCF/basis/a algorithm have been introduced. The HF-SCF/basis/a=0 case, yields the exact single Slater determinant wave function form ($Y_k(a=0)$) for both ground and excited states in Eq.2.3.2 suffering only from basis set error, while the HF-SCF/basis/a≠0 case yields the approximate single determinant form ($s_0(a)$) for the non-single Slater determinant wave functions ($y_0(a)$; most importantly $S_0(a=1)\approx y_0(a=1)=\Psi_0$) for the ground state in Eq.2.3.1, suffering, not only from basis set error, but requiring a correlation energy calculation too. Emphasis must be put on the fact that, HF-SCF/basis/a=1 still yields very useful $S_0$ and energy results in practice, mainly if a correlation calculation is included or follows, avoiding the lengthy CI method. We emphasize also that case a=1 makes only physical sense ($\Psi_0$), but DFT provides the existence of a link between $Y_k$ and $\Psi_k$. Very interestingly, the LCAO coefficients change "slowly" with "a", as will be demonstrated for a=0 vs. 1 next.

### 2.3.3. Computation properties of TNRS (Eq.2.3.2, a=0) for modeling real molecular systems

#### 2.3.3.a. The quasi-constant property of $\langle y_0(a)|ar_{12}^{-1}|Y_0\rangle/\langle y_0(a)|Y_0\rangle$ in Eq.2.3.28, particularly of $\xi\equiv\langle\Psi_0|r_{12}^{-1}|Y_0\rangle/\langle\Psi_0|Y_0\rangle$ in Eq.2.3.13, illustrated with neutral ground state G3 molecules

The first fundamental property is that the TNRS described by Eq.2.3.2 has similar LCAO coefficients as Eq.2.3.1 via HF-SCF for the same molecular system (on the same basis level, of course), Furthermore, the LCAO coefficients vary slowly with the coupling strength parameter "a". The second fundamental property is that the value of $\xi$ is roughly a quasi-constant in Eq.2.3.13, irrespective of a molecular frame and the number of electrons in the atoms and is at least in equilibrium molecular systems. These two properties indicate that Eq.2.3.2 is a useful tool to solve Eq.2.3.1 computationally as a starting part of algorithm. We have proven this quasi-constant behavior of $\xi$ theoretically and hypothetically above in part, and in this section we demonstrate it with computation on many molecular systems. We emphasize that the equations connecting Eq.2.3.1 and Eq.2.3.2 derived at in section 2.3.2 are exact expressions.

We have selected 149 molecules from the G3 set for testing and exhibiting the behavior of $\xi$. Their size is exhibited by the number of electrons (N) and number of atoms (M) along with their multiplicity (2S+1) in 2.3.Figure 3.a. All molecular systems are neutral, and in equilibrium geometry, some molecules are marked in the inset. The platos on the curve N mean the same number of electrons in different molecules, as well as that the order number is chosen for N to be monotonic. The ground state total electronic energies of these molecules are exhibited in 2.3.Figure 3.b. The message of this pair of curves is that $E_{electr,0}$ (from the G3 calculation [2.3.15-16] with the inclusion of electron – electron repulsion, more, correlation effect, i.e., a=1 or Eq.2.3.1) and $e_{electr,0}$ (from HF-SCF/STO-3G/a=0 calculation, which is without electron – electron repulsion, i.e., TNRS by a=0 or Eq.2.3.2) run together with same shape and monotony, and the difference (the electron-electron repulsion) is quasi-linear with respect to the number of electron pairs N(N-1)/2, detailed in 2.3.Figures 3.c-d. The two curves in 2.3.Figure 3.b run together like the same fingerprints supporting that Eq.2.3.2 has rich pre-information for Eq.2.3.1.

The basic message of the plot on 2.3.Figure 3.b-d is to show that in spite of the very different molecular frames involved and energy values provided by Eq.2.3.1 vs. Eq.2.3.2, i.e., when the electron-electron repulsion is included (a=1) vs. not (a=0) for neutral and equilibrium geometry molecular systems, Eq.2.3.2 rigorously follows Eq.2.3.1 and $\xi$ in Eq.2.3.13 is a quasi-constant. The latter indicates somehow that the LCAO parameters do not change robustly between Eqs.2.3.1 and 2, that is, one of the most important quantities in computation chemistry, the chemical bond is already indicated by Eq.2.3.2. From the robust change in energy plotted on 2.3.Figure 3.b, two relatively constant quantities are extracted and plotted in 2.3.Figure 3.c supporting the theory described in sections 2.3.1-2.3.2: The upper curve in 2.3.Figure 3.c shows the difference ($E_{electr,0}$ - $e_{electr,0}$) related to the number of electron pairs (N(N-1)/2), i.e., the $\xi$ in Eq.2.3.13 as a function of the order number of molecules chosen on 2.3.Figure 3.a. In fact, this plot shows the change of $\xi$ as a function of a nuclear frame. The lower curve in 2.3.Figure 3.c shows the ratio ($E_{electr,0}$ - $e_{electr,0}$)/$e_{electr,0}$, which if multiplied 100 times gives the percentage of the ground state electronic energy of a molecular system; it increases from the TNRS state when it “switches” to the realistic interacting system, its average value for 149 G3 molecules and its standard deviation is also shown. It is obvious from the notation, but we call attention to the fact that by this ratio, neither $E_{electr,0}$ nor $e_{electr,0}$ contain the term $V_{nn}$ because we focus on a comparative analysis of the solutions of Eqs.2.3.1-2; in practice one is generally interested in the value of $E_{total\ electr,0}$ and $e_{total\ electr,0}$. 2.3.Figure 3.c plots the difference between these two electronic energies and $V_{nn}$ drops anyway. More importantly, this average value is useful for showing its quasi-constant behaviour, although not useful enough for using as a constant in particular molecular calculations, because it far exceeds the chemical accuracy. If it was a rigorous constant, $E_{electr,0}$ could be extrapolated simply and directly from the $e_{electr,0}$ of Eq.2.3.2. At the very top of the plot in 2.3.Figure 3.c, the famous theoretical constant value, 2 (invariant to nuclear frame), from the virial theorem for equilibrium geometry molecules are also shown to compare their rigorous value 2 to the quasi constant upper ($\xi$) and lower

$$(E_{electr,0}-e_{electr,0})/e_{electr,0}= (E_{electr,0}/e_{electr,0}) -1= = <\Psi_0|H_{ee}|Y_0>/[<\Psi_0|Y_0><Y_0|H_{\nabla}+H_{ne}|Y_0>] \quad \text{(Eq.2.3.39)}$$

curves or values (via Eq.2.3.13) for these neutral equilibrium molecules. Both curves, but mainly the lower, are visibly quasi-invariant on nuclear frame if one compares the robust change in energy in 2.3.Figure 3.b and the definitely non-robust changes around the values $0.4 < \xi/\text{hartree} < 1$ of the upper curve and $-0.4 < (E_{electr,0}-e_{electr,0})/e_{electr,0} < -0.3$ (standard deviation of about 0.04 which decreases with increasing N) of the lower curve in 2.3.Figure 3.c. 2.3.Figure 3.d also shows the quasi-constant behavior of $\xi$ in large scale and from another point of view than in 2.3.Figure 3.c, i.e., represented as a curve parameter. The theoretical Eq.2.3.39 contains the exact values and functions of Eqs.2.3.1-2, but note, that ($E_{electr,0}$(G3)-$e_{electr,0}$)/$e_{electr,0}$ is plotted in 2.3.Figure 3.b, because only the accurate G3 level calculations are available. However, by Eq.2.3.29 or by $\Psi_0 \rightarrow Y_0$ as $a \rightarrow 0$, Eq.2.3.39 reduces to

$$(E_{electr,0}(\text{TNRS})-e_{electr,0})/e_{electr,0}= <Y_0|H_{ee}|Y_0>/<Y_0|H_{\nabla}+H_{ne}|Y_0> , \quad \text{(Eq.2.3.40)}$$

a value on the right hand side of which the nominator is determined by a denominator, that is, solving Eq.2.3.2 provides the $Y_0$ and the nominator can be evaluated. In this way the right hand side in Eq.2.3.40 is an integral value generated by $H_{ne}$ via Eq.2.3.2, which has already arisen above in 2.3.Figure 1. Furthermore, the ratio of Eqs.2.3.39 and 40 gives $(E_{electr,0}-e_{electr,0})/(E_{electr,0}(TNRS)-e_{electr,0})= \langle\Psi_0|H_{ee}|Y_0\rangle/[\langle\Psi_0|Y_0\rangle\langle Y_0|H_{ee}|Y_0\rangle]$ (Eq.2.3.41) wherein the right hand side targets the interesting ratio between the real (a=1) and TNRS (a=0) systems in relation to electron-electron repulsion energy. A plot of Eq.2.3.40 using 149 G3 molecules (with G3 level calculation for approximating $E_{electr,0}$) and the TNRS integral property in Eq.2.3.41 are exhibited in 2.3.Figure 4 showing quasi-independent behavior on the nuclear frame determined by the operator $H_{ne}$. Note that, in both equations the nuclear frame is the only parametric variable in these two ratios of wave functions.

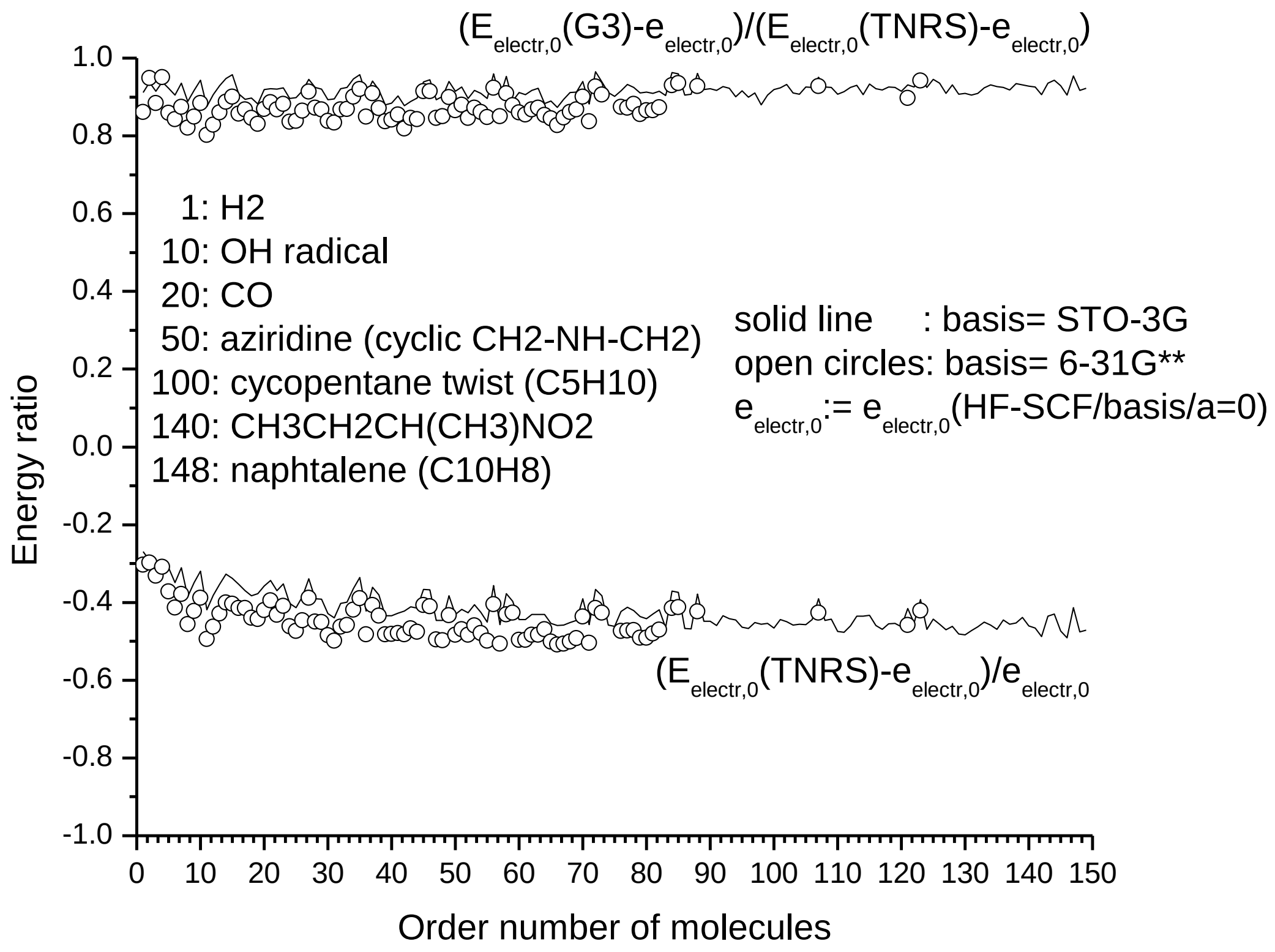

2.3.Figure 4.: Energy ratios plotted as order number of G3 molecules listed in 2.3.Figure 3.a. These ratios are equal to certain ratios of integral values of wave functions of real $\Psi_0$(a=1) and TNRS $Y_0$(a=0) described in Eqs.2.3.40-41. Notice the quasi-constant behavior, mainly in the upper curve, wherein the small $H_2$ (N=2) and large naphthalene (N=68) have about the same ratio in this respect. Upper curve is basically the ratio when one partially replaces $\Psi_0$ with $Y_0$ in calculating electron-electron interaction energies, see Eq.2.3.41. The ($E_{electr,0}$(G3 versus TNRS) - $e_{electr,0}$)/$e_{electr,0}$ ratios are plotted on 2.3.Figure 3.c vs. here.

In this section we have demonstrated the behavior of $(N(N-1)/2)\langle\Psi_0|r_{12}^{-1}|Y_0\rangle/\langle\Psi_0|Y_0\rangle$ as the counterpart of the similar but not the same quantity $V_{ee} \equiv (N(N-1)/2)\langle\Psi_0|r_{12}^{-1}|\Psi_0\rangle$, in Eqs.2.3.1 and 13 introduced via $\xi$. The $\xi N(N-1)/2$ is the exact difference between the ground state electronic energy ($E_{electr,0}$) of the real energy operator or Hamiltonian, $H_\nabla + H_{ne} + H_{ee}$, with a ground state wave function $\Psi_0$ and the ground state energy ($e_{electr,0}$) of energy operator $H_\nabla + H_{ne}$ with ground state eigenfunction $Y_0$, while $V_{ee}$ is the energy part of electron-electron repulsion in the $E_{electr,0}$ value related to the operator $H_{ee}$. The $\xi$ has a quasi-constant behavior as a function of the nuclear frame $\{R_A,Z_A\}_{A=1toM}$ containing M nuclei and N electrons, tested on ground state, stationary, and neutral ($\Sigma Z_A=N$) molecules. Compare this quasi constant $\xi$ to the rigorous constant value 2 of virial theorem plotted in 2.3.Figure 3.c, as well as the robust change in the value of $E_{electr,0}$, both on an increasing molecular frame and N (2.3.Figure 3.a-b). This supports the quasi-linear behaviour of $\xi$ described in sections 2.3.1-2.3.2, recall the particular equations in Eqs.2.3.24-25 and 29 used for an immediate estimation (extrapolation) from Eq.2.3.2 to Eq.2.3.1. The bottom line for the next section then is that: the quasi-constant nature of $\xi$ means that the converged or stationary LCAO coefficients by HF-SCF/basis/a routine do not vary significantly with coupling strength parameter "a".

### 2.3.3.b. The LCAO coefficients in ground state (k=0) of electronic Schrödinger equation (Eq.2.3.1, a=1) in comparison to the TNRS (Eq.2.3.2, a=0)

First of all, it is useful to show a detailed typical outcome (LCAO coefficients, MO energies, electronic ground state energy) of a HF-SCF/basis/a calculation for some molecular systems on the road to exploring the relation of Eq.2.3.1 to Eq.2.3.2. Because the matrix of LCAO coefficients is relatively large in respect to the space allowed for publishing, we have chosen small systems with the small STO-3G basis set: Neon (Ne) atom and hydrogen-fluoride (HF) molecule. However, these small sizes do not restrict us from exhibiting characteristic behaviors; recall such properties as for example, C-C distances; functional groups and many others in molecules can show molecular size quasi-independent characteristics. The STO-3G basis reflects the Slater type atomic orbitals (STO) intellectually very well at the cost of lower energy accuracy, and the inherently called 1S, 2S, 2PX, 2PY, 2PZ contracted bases functions are linear combinations of Gaussian type atomic orbitals (GTO) based on some rigorous definitions [2.3.12]. The HF-SCF/STO-3G/a calculation for Ne atom by Monstergauss for Eq.2.3.1 (a=1) has yielded the matrix of LCAO coefficients:

```
1CLOSED SHELL SCF, NUCLEAR REPULSION ENERGY IS 0.000000000 HARTREES
0CONVERGENCE ON DENSITY MATRIX REQUIRED TO EXIT IS  1.0000D-05
0  CYCLE      ELECTRONIC  ENERGY           TOTAL  ENERGY         CONVERGENCE
EXTRAPOLATION
    1       -126.604525025      -126.604525025
    2       -126.604525025      -126.604525025   1.81460D-16
0AT TERMINATION TOTAL ENERGY IS      -126.604525  HARTREES
1MOLECULAR ORBITALS                          5 OCCUPIED MO
                             1          2          3          4          5
    EIGENVALUES---   -32.21252   -1.70610   -0.54305   -0.54305   -0.54305

  1  1  NE    1S       0.99501   -0.26941    0.00000    0.00000    0.00000
  2  1  NE    2S       0.01978    1.03065    0.00000    0.00000    0.00000
  3  1  NE    2PX      0.00000    0.00000    0.00000    1.00000    0.00000
  4  1  NE    2PY      0.00000    0.00000    0.00000    0.00000   -1.00000
  5  1  NE    2PZ      0.00000    0.00000   -1.00000    0.00000    0.00000
```

while for Eq.2.3.2 (a=0, no electron-electron interaction, i.e., TNRS)

```
1CLOSED SHELL SCF, NUCLEAR REPULSION ENERGY IS 0.000000000 HARTREES
0CONVERGENCE ON DENSITY MATRIX REQUIRED TO EXIT IS  1.0000D-05
0  CYCLE      ELECTRONIC  ENERGY          TOTAL  ENERGY        CONVERGENCE
EXTRAPOLATION
    1       -182.113502106      -182.113502106
    2       -182.113502106      -182.113502106   0.00000D+00
0AT TERMINATION TOTAL ENERGY IS      -182.113502  HARTREES
1MOLECULAR ORBITALS                          5 OCCUPIED MO
                       1           2           3           4           5
    EIGENVALUES---   -49.42500  -10.95959  -10.22405  -10.22405  -10.22405

  1  1  NE    1S       1.00094    0.24650    0.00000    0.00000    0.00000
  2  1  NE    2S      -0.00389   -1.03083    0.00000    0.00000    0.00000
  3  1  NE    2PX      0.00000    0.00000    1.00000    0.00000    0.00000
  4  1  NE    2PY      0.00000    0.00000    0.00000   -1.00000    0.00000
  5  1  NE    2PZ      0.00000    0.00000    0.00000    0.00000    1.00000
```

The "NUCLEAR REPULSION ENERGY" is what we notate in this work with $V_{nn}$, and the "EIGENVALUES" are the MO energies in hartree in both lists ($\varepsilon_i$ in Eq.2.3.3 with a=1 or 0). The "TOTAL ENERGY" and "ELECTRONIC ENERGY" are the $E_{total\ electr,0}$ and $E_{electr,0}$ in the first list and $e_{total\ electr,0}$ and $e_{electr,0}$ in the second list in hartree, respectively.

In the case of a=0, where atom (M=0) is considered, the MOs are the known accurate text book one-electron atomic orbitals, e.g., for 1s of Ne (Z=10) that is at the accuratel known energy level $-Z^2/2=-50$ h in comparison to the approximated -49.425 h in the list (basis set error), as well as the linear combination 1.00094(1S) -0.00389(2S) of GTO functions (from column 1) is supposed to approximate the accurately known normalized 1s STO atomic orbital, exp(-10$\mathbf{r}_1$), etc.. The 2-5$^{th}$ MOs are the atomic $-Z^2/(2n^2)= -12.5$ h for Z=10 and n=2 (2s, 2p), the deviation from this value stems from the weakness of STO-3G (basis set error again), but at least the $p_x$, $p_y$ and $p_z$, like MOs (3-5), have self consistent degenerates i.e. the same values of -10.22405 h on top of this the ±1 factored LCAO values indicate pure p orbitals as has to be; sign alteration comes from irrelevant phase factor, furthermore the exact 1 absolute value LCAO coefficients stem from the rigidity of a small STO-3G basis set. (Using a larger basis set, the HF-SCF/6-31G**/a=0 MO energies are closer to the –50 h (MO 1) and –12.5 h (MO 2-5) values: -49.94645, -12.14391, -11.83664, -11.83664, -11.83664, yielding "TOTAL ENERGY" -195.200544 h.) Notice that these one-electron (N=1) energies are reproduced (up to basis set error) by our manipulated commercial HF-SCF/basis/a=0 routine for an N=10 electron system showing the expected relationship between Eqs.2.3.4 and 2 via Eqs.2.3.5-8. As indicated in Eq.2.3.19, $e_{electr,0}$= -182.113502 h << $E_{electr,0}$= -126.604525 h ("much larger" refers to chemical accuracy) even in a small system like this. The sets of values of MO energy levels are very different between a=1 vs. a=0: {-32.21252, -1.70610, …} vs. {-49.42500, -10.95959, …}, because these were made before adjustment e.g., by Eq.2.3.29. Eq.2.3.8 holds as 2(-49.42500 -10.95959 -10.22405 -10.22405 -10.22405)= -182.113502 h in the case of a=0, but not if a=1, because 2(-32.21252 -1.70610 -0.54305 -0.54305 -0.54305)= -71.09554 $\neq$ -126.604525 h, - as I have mentioned, the latter is generally well known in HF-SCF theory. The most important message we get by comparing cases a=1 to a=0 in the lists above, i.e., comparing Eq.2.3.1 and Eq.2.3.2, is that the LCAO coefficients are very similar - a manifesting property. It means that the LCAO coefficients build up very closely to Eq.2.3.1 via Eq.2.3.2 with the HF-SCF/basis/a routine, so for "a" values between. Comparing the above LCAO columns 2-5 to each other, i.e., the 2-5$^{th}$ MOs in the above lists

a=1 vs. a=0, the irrelevant difference is the order (compare the 3rd, 4th and 5th MOs between the two sets) as well as phase factors or signs, e.g., {-0.26941(1S)+1.03065(2S)} vs. {0.24650(1S)-1.03083(2S)} for the 2nd MO in the two sets.

Given that LCAO coefficients are close to each other in the cases a=0 vs. 1, it would be useful to consider starting with Eq.2.3.2 in the HF-SCF routines, that is, to set up a=0 i.e., switching off the effect of operator $H_{ee}$, and when SCF convergence is complete for Eq.2.3.2 which is always one step, (see below), the switch to a=1 would continue with these LCAO coefficients to find the stationer LCAO coefficients for Eq.2.3.1 (physically realistic system). We emphasize again that a=0 switches off the calculation for electron-electron interactions; (described by operator $H_{ee}$), the most expensive part in HF-SCF calculations. That is, the CPU time and disc space as well as the convergence problem can be reduced. The converged (one step) HF-SCF/basis/a=0 yields $e_{electr,0}$ and $Y_0$ of Eq.2.3.2, followed by the switch to a=1 starting with the LCAO coefficients from a converged HF-SCF/basis/a=0, and the beginning of iteration (one step) HF-SCF/basis/a=1 yields the value of the right hand side of Eq.2.3.29, an interesting approximation which can be tuned to increase the accuracy by the coupling strength parameter "a" itself with slightly less value than unity, (see details below) or by applying another correction ensured by the generalized 1st HK in Eq.2.3.34. Another way is to let the HF-SCF/basis/a=1 continue, which requires many steps in practice and the system will converge to $S_0$, (see Eq.2.3.30) in this respect. It is an alternative way of convergence from the Harris initial values used in practice, see section 2.3.3.e below.

Continuing the note on the LCAO coefficients for a=1 vs. a=0 yields, for example, that the lowest lying 1s-like MO in the interacting system (a=1 or Eq.2.3.1) is approximated as: $\phi_1(a=1,\mathbf{r}_1,Ne)$= 0.99501(1S)+0.01978(2S), while in the TNRS (a=0 or Eq.2.3.2) it is: $\phi_1(a=0,\mathbf{r}_1,Ne)$= 1.00094(1S)-0.00389(2S). It means that both are essentially an 1S function for a=1 and 0 (tuned with a linear combination of GTO's in STO-3G basis), and it contributes to the one-electron density with a function such as $(\phi_1)^2$ in the algebra of HF-SCF theory. However, there is a finer detail: From Eq.2.3.3, e.g., the closed shell $\rho(\mathbf{r}_1$,HF-SCF/basis/a)= $2\Sigma_{i=1 \text{ to } N/2}\phi_i^2(\mathbf{r}_1,a)$ density does not seem to vary greatly with "a", most importantly between a=0 and a=1, because of their similar LCAO coefficients. The $\phi_i$'s are expanded in LCAO, and the enforced normalization in the algorithm, $\int\rho(\mathbf{r}_1$,HF-SCF/basis/a)$d\mathbf{r}_1$=N for any "a", makes the change via "a" even less visible. That is, the shape of $\rho(\mathbf{r}_1$,HF-SCF/basis/a) does not change drastically with "a"; its integral properties change even less, manifesting in its normalization which is fixed to N. For example, the classical electron-electron energy approximation $\int\rho(\mathbf{r}_1$,HF-SCF/basis/a)$\rho(\mathbf{r}_2$,HF-SCF/basis/a)$r_{12}^{-1}d\mathbf{r}_1d\mathbf{r}_2$ in DFT or its alternative in HF-SCF formalism with J and K integrals does not change drastically either, LCAO phase factors drop by squares, a property important in Eq.2.3.29. However, the t(HF-SCF/basis/a)= -$\Sigma_{i=1\ldots N/2}$ $\langle\phi_i(\mathbf{r}_1,a)|\nabla_1^2|\phi_i(\mathbf{r}_1,a)\rangle$= $\Sigma_{i=1\ldots N/2}$ $\langle\nabla_1\phi_i(\mathbf{r}_1,a)|\nabla_1\phi_i(\mathbf{r}_1,a)\rangle$ kinetic energy (recall the notations t(a=0) and T(a=1)) can yield a more pronounced difference between a=0 and a=1, because the slopes differ. As has just been described for 1S like MO: $\phi_1(\mathbf{r}_1$,a,Ne) is steeper if a=0 than if a=1, in agreement with the general relationship established in Eq.2.3.32. Recall that the (accurate) one-electron density, $\rho(\mathbf{r}_1,a)$, defines the Hamiltonian H(a) by the 1st Hohenberg-Kohn theorem, and it particularly defines H(a=1) of Eq.2.3.1 and H(a=0)= $H_\nabla+H_{ne}$ of Eq.2.3.2. Importantly, the HF-SCF/basis/a=0 algorithm for Eq.2.3.2 brings the TNRS density $\rho_0(\mathbf{r}_1$,a=0) close to the HF-SCF/basis/a=1 density $\rho(\mathbf{r}_1$,HF-SCF/basis/a=1) of Eq.2.3.1, which is obviously useful for Kohn –Sham formalism, as well as for post-HF-SCF methods or correlation calculations. In the case of Ne atom, the a= 0 vs. 1 value densities are plotted on 2.3.Figure 2.a for comparison.

Finally, we emphasize that the above findings and conclusions do not depend on the size of a molecular system, not detailed here for the sake of brevity, but carefully tested on a large number of systems. The most important task in theoretical chemistry is to describe the chemical bond, so a molecular system is also exhibited here as an example after the atom Ne. For the hydrogen-fluorid molecule (MP2(full)/6-31G* geometry, $E_{total\ electr,0}$(MP2 level)= -100.1841 Hartree [2.3.12]), the HF-SCF/STO-3G/a=1 for Eq.2.3.1 yields:

```
1CLOSED SHELL SCF, NUCLEAR REPULSION ENERGY IS 5.099731703 HARTREES
0CONVERGENCE ON DENSITY MATRIX REQUIRED TO EXIT IS  1.0000D-05
0 CYCLE   ELECTRONIC ENERGY      TOTAL ENERGY    CONVERGENCE   EXTRAPOLATION
    1       -103.453458282       -98.353726579
    2       -103.658442376       -98.558710673   4.81239D-02
    3       -103.671344215       -98.571612512   1.00099D-02
    4       -103.671920720       -98.572189017   2.13556D-03      4-POINT
    5       -103.671934950       -98.572203247
    6       -103.671950402       -98.572218699   5.80744D-06
0AT TERMINATION TOTAL ENERGY IS       -98.572219  HARTREES
1MOLECULAR ORBITALS                      5 OCCUPIED MO
                          1          2          3          4          5          6
    EIGENVALUES---   -25.90153   -1.46601   -0.58015   -0.46365   -0.46365    0.61156

  1  1   F    1S       0.99472   -0.24986    0.08063    0.00000    0.00000    0.08298
  2  1   F    2S       0.02247    0.94095   -0.42420    0.00000    0.00000   -0.53979
  3  1   F    2PX      0.00000    0.00000    0.00000    0.28444   -0.95869    0.00000
  4  1   F    2PY      0.00000    0.00000    0.00000    0.95869    0.28444    0.00000
  5  1   F    2PZ     -0.00283   -0.08462   -0.70026    0.00000    0.00000    0.82101
  6  2   H    1S      -0.00558    0.15494    0.52694    0.00000    0.00000    1.07402
```

while the HF-SCF/STO-3G/a=0 for Eq.2.3.2 yields:

```
1CLOSED SHELL SCF, NUCLEAR REPULSION ENERGY IS 5.099731703 HARTREES
0CONVERGENCE ON DENSITY MATRIX REQUIRED TO EXIT IS  1.0000D-05
0 CYCLE   ELECTRONIC ENERGY      TOTAL ENERGY    CONVERGENCE   EXTRAPOLATION
    1       -151.075395174      -145.975663471
    2       -152.831334744      -147.731603041   0.00000D+00
0AT TERMINATION TOTAL ENERGY IS      -147.731603  HARTREES
1MOLECULAR ORBITALS                      5 OCCUPIED MO
                          1          2          3          4          5          6
    EIGENVALUES---   -40.59236   -9.55517   -8.81672   -8.72571   -8.72571   -4.49671

  1  1   F    1S       1.00121    0.23152    0.08800    0.00000    0.00000    0.03901
  2  1   F    2S      -0.00549   -1.03159   -0.35933    0.00000    0.00000   -0.40485
  3  1   F    2PX      0.00000    0.00000    0.00000   -0.03804   -0.99928    0.00000
  4  1   F    2PY      0.00000    0.00000    0.00000   -0.99928    0.03804    0.00000
  5  1   F    2PZ      0.00024    0.44410   -0.94971    0.00000    0.00000    0.26910
  6  2   H    1S       0.00188    0.20530   -0.09439    0.00000    0.00000    1.18497
```

MOs 1-5 are occupied pair-wised with opposite spins by the N=10 electrons in a ground state, the 5$^{th}$ is the highest occupied MO (HOMO) and the 6$^{th}$ MO is the virtual lowest unoccupied MO (LUMO) in both lists. (The LUMO and higher MOs are not listed in the case of Ne above.) The LUMO in HF-SCF approximation can be handled relatively easily for qualitative discussions, but one must be careful in a quantitative argument. We will comment upon the excited state a=1 vs. a=0 later in this work, for the time being, we mention that, the 6$^{th}$, unoccupied MO (LUMO) also has a similar LCAO coefficient in the case of a=1 and a=0, just as the other 1-5$^{th}$ MOs aside from phase factors. Ne is a central symmetric system, while the rod shaped hydrogen-fluorid defines direction, the latter was positioned along the z axis, and as a consequence it is reflected in the approximate atomic 2p like MOs (MO 3, 4 and 5 in both cases a=1 and 0), and in relation to our talk, now the order number of the MOs are the same in comparison to a=1 and a=0, but sign change happened, e.g. in the 2$^{nd}$ MO: sgn(0.94095) vs. sgn(-1.03159). The HOMO and LUMO play important roles in chemical reactions, and the case of a=1 vs. a=0 will be discussed in section 2.3.3. An important feature is, that in case of the hydrogen-fluorid molecule the TNRS (HF-SCF/basis/a=0) already indicates the bond (now by shifting the LCAO from -1 value) in the same way as the regular HF-SCF/basis/a=1 does- the latter is well known. Recall that, in the case of Ne, both (a=0 and 1) yield three equivalent p orbitals, hence, no polarization because there is no bond, while with hydrogen fluorid both (a=0 and 1) yield 2 equivalent $p_x$ and $p_y$ as well as a different $p_z$ along the bond, which is more apparent in the LCAO list above than in 2.3.Figure 2.b.

The similar LCAO coefficients for approximating Eq.2.3.1 with HF-SCF/basis/a=1 (suffering correlation effects and basis set error) vs. approximating Eq.2.3.2 with HF-SCF/basis/a=0 (suffering from a basis set error only but before adjustment with the inclusion of $V_{ee}$) supports the quasi-linear behavior which has led to Eq.2.3.29, that is, the right hand side of Eq.2.3.21 is quasi-independent of the value of the coupling strength parameter "a".

### 2.3.3.c. A quick power series estimation for the ground state (k=0) of the electronic Schrödinger equation (Eq.2.3.1, a=1) from TNRS (Eq.2.3.2, a=0) starting from $E_{electr,0}$(TNRS) in Eq.2.3.29

The quick estimation to convert Eq.2.3.2 to Eq.2.3.1 commences from Eq.2.3.29, because it accounts for the big part of the large difference in energy value between Eqs.2.3.1 and 2, as 2.3.Figure 3.b demonstrates, wherein the three equations (Eqs.2.3.1, 2 and 29) are plotted for comparison. The quick method we have chosen is the "moment expansion" of the electron density [2.3.22, 29], this work started in earnest with the work of Agnes Nagy in the mid-1990s [2.3.30], and the subsequent work from the Parr group that this stimulated. The idea is incredibly attractive: one can rewrite every density functional as a function of the moments of the density, making sure the moments are complete [2.3.29]. This allows one to replace the functional analysis in DFT with a simple multivariate calculus, which is a huge formal advantage. Most of the work assumes that quantities can be written as a linear function of the moments.

Among moments the most famous are the Thomas-Fermi ($T\approx c_F\int\rho_0^{5/3}d\mathbf{r}_1$) or Weizsacker approximation ($T\approx (1/8)\int|\nabla_1\rho_0|^2/\rho_0 d\mathbf{r}_1$), Dirac formula ($V_{ee}\approx B_D\int\rho_0^{4/3}d\mathbf{r}_1$), as well as Parr terms $c_{AB}(\int\rho_0^A d\mathbf{r}_1)^B$ in the power series wherein easy formulas for A and B keeps it scaling correct [2.3.2] up to infinity, separately for T and $V_{ee}$, as well as see the work of Carter et al. on the orbital-free DFT [2.3.31-34]. Similarly, Kristyan approximated the correlation effect by a partial [2.3.35-38] and a full [2.3.13] integration of wave functions, the latter is the simple $E_{corr}\approx k_c\langle S_0|H_\nabla|S_0\rangle+k_{ee}\langle S_0|H_{ee}|S_0\rangle$ yielding remarkable results, wherein one can recognize that $a=1+k_{ee}$ is the coupling strength parameter with the task to correct the HF-SCF/basis/a=1 calculation to better approach the energy value $E_{electr,0}$ in Eq.2.3.1. Common in these formulas: 1, the first terms, come from plausible assumptions and derivations, but secondary and higher terms definitely necessary for chemical accuracy, (a manifest example is that only the Thomas-Fermi approximation for T fails to describe chemical bonds), 2, first terms with proper (but generally empirical) parameters can be used to account for the entire term T or $V_{ee}$, or just for their correction, depending (quite surprisingly) on how the user wants to define them. For example, in $(1+k_{ee})\langle S_0|H_{ee}|S_0\rangle$ [2.3.13] for the set of 149 G3 molecular energies the unity accounts for the entire term $V_{ee}$ and $k_{ee}$ for correction, also, if all energy correction is only attributed to electron–electron interaction, the $a= 1+k_{ee}= 0.99353272$ with a 6-31G** basis, here we notate as HF-SCF/6-31G**/a=0.99353272, improves the average deviation of HF-SCF/6-31G**/a=1 from 0.7851 h to 0.1255 h on average), 3, a plausible series in principle converges to the accurate energy value, but as a drawback, probably coming from the imperfect parametrization, only small power terms (2 to 4) can account for a large pool of molecular systems, which, while increasing the power decrease the range of molecular systems in terms of accuracy probably the latter is responsible, that moment expansion has not come before DFT correlation calculations. The parametrization of moment expansion is not as rigorous mathematically as e.g., the wave function based MP method (see accurate coefficients $1/(\varepsilon_p+\varepsilon_q-\varepsilon_r-\varepsilon_s)$ in section 2.3.2.d), as well as not being as fortunate as in DFT formulas wherein the expressions are more compact and not sums (see Generalized Gradient Approximations (GGA) formulas).

In addition to the above, for example, the $(1+k_{ee})<S_0|H_{ee}|S_0>$ correction [2.3.13] can be done during SCF routine or after, negative $k_{ee}$ decreases the energy, as is to be expected from the $E_{corr}$ via variation principle. Based on these sound working devices, here we extend the form in Eq.2.3.29 as

$E_{electr,0} \approx E_{electr,0}$(TNRS with $L^{th}$ order PS expansion) $\equiv e_{electr,0} + \Sigma_{j=1\ldots L}(a_j t^j + b_j v_{ne}{}^j + c_j z^j) =$

$= E_{electr,0}(\text{TNRS}) + (a_1 t + b_1 v_{ne} + (c_1-1)z) + \Sigma_{j=2\ldots L}(a_j t^j + b_j v_{ne}{}^j + c_j z^j)$ , (Eq.2.3.42)

where the pre-calculated $t \equiv <Y_0|H_\nabla|Y_0>$, $v_{ne} \equiv <Y_0|H_{ne}|Y_0>$ and $z \equiv a<Y_0|H_{ee}|Y_0>$ integrals are used, as well as $c_1 z = z+(c_1-1)z$ being used to show its more visible extension from Eq.2.3.29 - PS stands for power series. In Eq.2.3.29 and Eq.2.3.42 the most important a=1 is involved. In Eq.2.3.42 if $a \neq 1$, then $enrg_{electr,0}(a)$ replaces $E_{electr,0}(a=1)$, and if a=0 then $a_j=b_j=c_j=0$ for all j, as well as coupling strength parameter "a" is not to be confused with PS coefficients $a_j$. In section 2.3.2.d the MP perturbation uses the excited states ($Y_k$) in the expansion ("vertical" algebraic way), while Eq.2.3.42 uses only $Y_0$ ("horizontal" algebraic way), notice that the latter is practically instant in respect to computation, while the former can be time consuming.

We have obtained the coefficients in Eq.2.3.42 by least square fitting to 149 ground state G3 molecular energies to minimize the average absolute deviation. The solid line in 2.3.Figure 3.b (Eq.2.3.29 with HF-SCF/STO-3G/a=0) is greatly improved and is energetically plotted as squares and triangles in 2.3.Figure 5 (Eq.2.3.42 with HF-SCF/STO-3G/a=0 and L= 2 and 3), it is better than the HF-SCF/STO-3G/a=1 regular calculation, so the correlation effect is somehow accounted for by Eq.2.3.42. We emphasize that in this work no more serious correlation calculation is considered than Eq.2.3.42, we just want to demonstrate the way to use Eq.2.3.2 for solving Eq.2.3.1. It is also out of scope, that if PS coefficients in Eq.2.3.42 are transferable i.e., the same, between ground (k=0) and excited (k>0) states, based on the algebraic fact that one single determinant (the $Y_k$) is included only in Eq.2.3.42, a fortunate algebraic determinant property may provide this. We have obtained the following values for PS coefficients in Eq.2.3.42 for the second order (L=2) case as

$a_1$= -0.761233, $b_1$= -0.448435, $c_1$= 0.430207
$a_2$= 2.270220E-004, $b_2$= -5.068453E-005, $c_2$= 1.678742E-004

, while the third order (L=3) PS coefficients are

```
a1= -0.853118,       b1= -0.519268,       c1=  0.289831
a2=  5.224182E-004, b2= -1.321651E-004, c2=  6.744563E-004
a3= -2.026111E-007, b3= -2.221198E-008, c3= -4.823247E-007.
```

The average absolute deviation in h and % and the maximum absolute deviation in h from G3 data are:

```
L=2 in Eq.2.3.42     : 1.615905 h or 1.02 %,  7.015398 h
L=3 in Eq.2.3.42     : 1.563234 h or 1.06 %,  7.270620 h
HF-SCF/STO-3G/a=1: 3.497650 h or 1.88 %, 11.976560 h.
```

As can be seen, the HF-SCF/basis/a=1 calculation is improved, so correlation effects are accounted for by Eq.2.3.42, as well as the L=3 level not improving much over L=2 ("small power" property, see above). Larger L can yield not-realistic values for coefficients, a known problem that can happen in least square fit with PS expansion. The latter means that, e.g. the L= 2 and 3 level coefficients are realistic in that they correct the different energies with main terms, which are the j=1 terms in Eq.2.3.42: Negative $a_1$ and $|a_1|<1$ necessary to subtract a part of kinetic energy away requited by Eq.2.3.32, so for $b_1$ to keep the virial theorem hold, as should $0<c_1<1$ be to satisfy Eq.2.3.20, (also, the unidentical Eq.2.3.41 with its approximate value of 0.8-0.9 in 2.3.Figure 4 shows plausible correspondence with $c_1$), and indeed the least square fit has provided that these relationships hold as well as optimizing the energy deviation of Eq.2.3.42 from G3 data to a minimum.

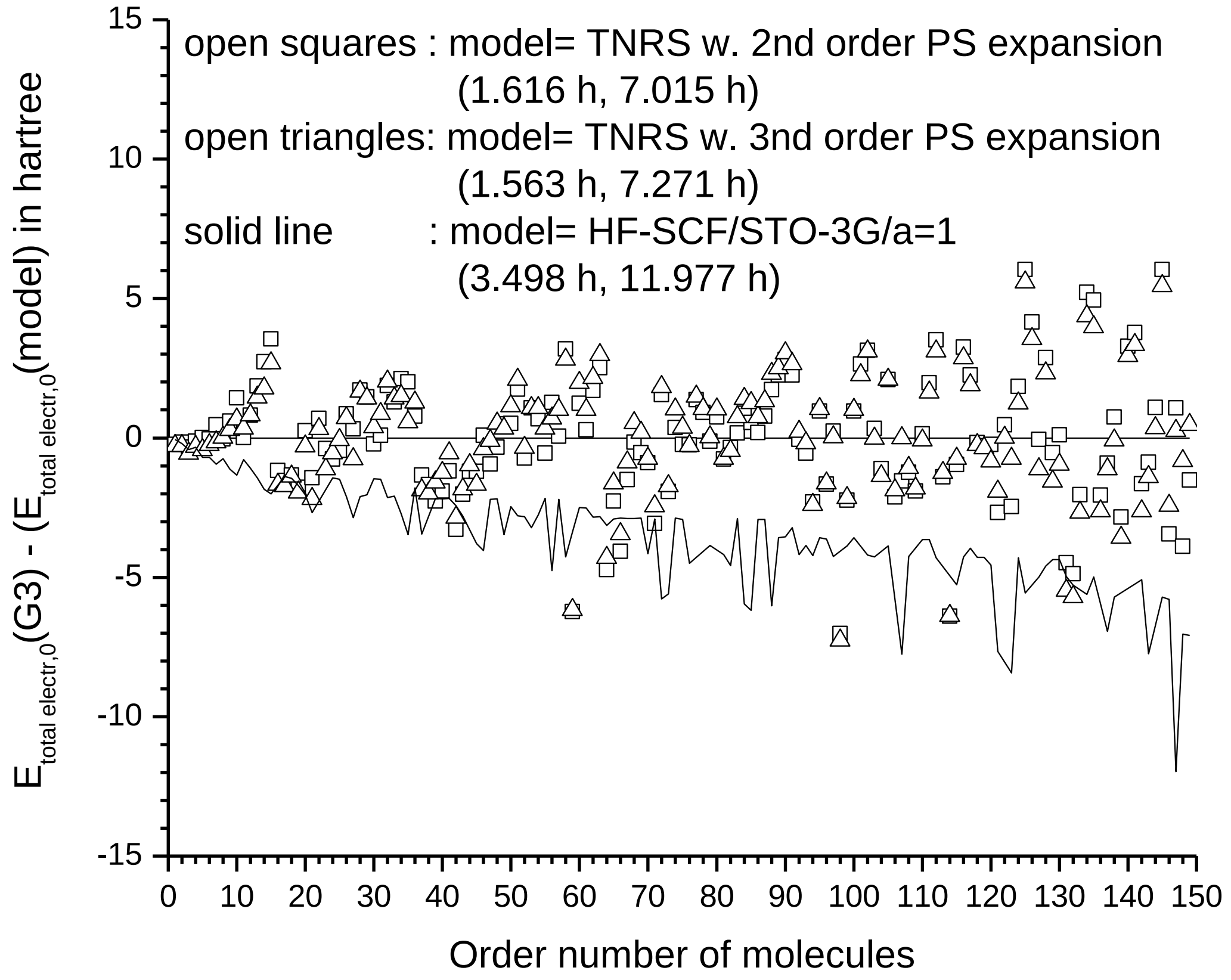

2.3.Figure 5.: $E_{total\ electr,0}$(G3)- $E_{total\ electr,0}$(model) energy differences plotted as order number of G3 molecules listed in 2.3.Figure 3.a. In case of model= HF-SCF/STO-3G/a=1, it is the $E_{corr}$(HF-SCF/STO-3G/a=1) in relation to accurate G3 calculation. The models TNRS (Eq.2.3.29 with HF-SCF/STO-3G/a=0) corrected with 2$^{nd}$ and 3$^{rd}$ order power series (PS) expansion in Eq.2.3.42 are the result of a "quick" correlation calculation showing how close these bring the values $e_{electr,0}$ back to values $E_{electr,0}$ shown on 2.3.Figure 3.b. As can be seen, the 3$^{rd}$ order does not yield much better improvement over 2$^{nd}$ order power series expansion for $E_{corr}$, belonging to typical problems of this method. While HF-SCF/STO-3G/a=1 is variational ($E_{corr}$<0), the other two are not (points are above and below the zero line), as known in DFT. Least square fit to total electronic energy deviations from the set of 149 G3 molecular data has yielded the values for coefficients in Eq.2.3.42 listed in section 2.3.3.c, numbers in parenthesis are the "average absolute deviation" and "maximum absolute deviation" in the fit.

**2.3.3.d. Note on the RHF/UHF mode in HF-SCF/basis/a=0 (TNRS, Eq.2.3.2)**

Note should be taken of the UHF mode, which is the most common molecular orbital method for open shell molecules where the numbers of electrons in two spins are not equal. For example, let us consider the triplet carbon atom ($1s^2 2s^2 2p_x^{\,1} 2p_y^{\,1}$, 2S+1=3); the UHF mode HF-SCF/STO-3G/a=1 yields MO energies in hartree:

4 occupied MO with $\alpha$ spin: -10.93172 -0.72837 -0.32803 -0.32803 0.23600

2 occupied MO with $\beta$ spin: -10.88869 -0.46668 0.30854 0.38107 0.38107

$E_{electr,0}$(HF-SCF): -37.198393, for theoretical S(S+1) = 2, a value of 2.0000 is obtained at STO-3G (thanks to minimal basis), but spin-contamination is 2.0048 at 6-31G**.

The UHF mode HF-SCF/STO-3G/a=0 TNRS yields MO energies in hartree:

4 occupied MO with $\alpha$ spin: -17.76299 -3.92916 -3.67049 -3.67049 -3.67049

2 occupied MO with $\beta$ spin: -17.76299 -3.92916 -3.67049 -3.67049 -3.67049

$e_{electr,0}$(HF-SCF): -50.725273, for theoretical S(S+1) = 2, a value of 2.0000 is also obtained at STO-3G and 6-31G**, i.e., no spin-contamination even at larger basis level. The UHF and RHF mode is the same in the case of HF-SCF/STO-3G/a=0, i.e., in the case of TNRS, that is, the coupled Roothaan equations, known as the Pople–Nesbet–Berthier equations fall back to simple Roothaan equations. This is, because the electron-electron repulsion is responsible for the spatial part of MO split in UHF (e.g. -10.93172 and -10.88869, etc.) to get deeper energy via a variation principle in the case of a=1, more generally in the case of a≠0. The UHF method in a=1 (more generally a≠0) mode in principle has this drawback.

In view of philosophy, the UHF virtually contradicts that in $\mathbf{x}_i$ and $\mathbf{x}_j$ spin-orbitals, for example, the spin coordinates $s_i$ and $s_j$ are enough to differ to satisfy the Pauli's exclusion principle. In the HF-SCF/basis/a=1 numerical example above, the spatial parts of MOs were allowed to split a bit to reach deeper energy, However, many scientists opposed to this split, i.e. to UHF calculations based on that the $S_0$ single determinant, have an even worse form theoretically in the UHF mode than in the RHF mode, in this respect. The profit on deeper energy via UHF vs. RHF mode in the HF-SCF/basis/a=1 calculation can be counterbalanced by the DFT which applies the $E_{xc}[\rho$(HF-SCF/basis/a=1)] functional during or after the algorithm, having a non-variational nature.

**2.3.3.e. Number of steps in convergence when performing HF-SCF/basis/a=0 (TNRS, Eq.2.3.2), and the LCAO values from it as starting values**

The energy values during the convergence steps are also listed for Ne and hydrogen-fluorid above. The hydrogen-fluorid is a slightly more complex system than the Ne, and the number of steps in the HF-SCF/STO-3G/a convergence already manifests: six were necessary for Eq.2.3.1 (a=1) and two for Eq.2.3.2 (a=0), which makes an example for the faster convergence mentioned above in the HF-SCF approximation for Eq.2.3.1 vs. Eq.2.3.2 mainly if Eq.2.3.29 type or more sophisticated, however, one step approximations are used. Actually, the a=0 case has a deeper property in this respect, convergence in HF-SCF/basis/a=0 needs only two steps only more exactly one, after setting up an initial guess for LCAO parameters, the eigensolver yields the $Y_0$ in the next step, irrespectively of molecular size. Of course, starting with the commonly used Harris approximation for initial LCAO parameters [2.3.39-40] for performing HF-SCF/basis/a=1 and finishing the convergence, or starting with LCAO from a converged one step HF-SCF/basis/a=0 and finishing the convergence, the final $E_{electr,0}$(HF-SCF/basis/a=1) and LCAO parameters will be strictly the same via the variation principle kept by the subroutine of Gaussian or Monstergauss, see Eq.2.3.30. However, we place emphasis on the fact that the HF-SCF/basis/a solution for Eq.2.3.2 (a=0) can be achieved in basically one step for molecules of any size via the HF-SCF algorithm, while for a≠0, the number of convergence is always more than one that increases with molecular size, and larger molecular systems may have problems such as break down in convergence in later steps, experienced since long in practice (a=1). This one step is a benefit if e.g. the quick Eq.2.3.29 or Eq.2.3.42 follows to finish the calculation.

The Harris approximation [2.3.39-40] is based on the following empirical property: the density of a system comprising closed-shell atoms or molecules is approximated by overlapping the HF-SCF densities of the free atoms (or molecules), and the energy is then calculated using the Thomas-Fermi approximations [2.3.11, 39] for the electrostatic and kinetic energy terms and for the exchange-correlation energy, as well as this, one can derive a simple expression for the binding energy for given geometry. Diagonalizing the Harris functional [2.3.40] for the initial guess is the default for all HF-SCF/basis/a=1 and DFT calculations in Gaussian 2009 and earlier versions [2.3.12] after Monstergauss 1981.

However, in this work we do not focus on describing a non-empirical initial guess for the LCAO coefficients via Eq.2.3.2, but many other, more important properties via the coupling strength parameter going far beyond. It should be emphasized that the difference that the Harris approximation makes - a crude initial guess for one-electron density, $\rho_0(\mathbf{r}_1,a=1)$, using spherical atoms in a molecular frame, while Eq.2.3.2 yields the $\rho_0(\mathbf{r}_1,a=0)$ of TNRS which includes the stem of the chemical bond and density around the atoms in the molecule deformed from the atomic spherical shape in a molecular environment as they have to be for further processing. Again, we use the HF-SCF terminology when referring to Eq.2.3.2, i.e. the eigenvalue problem of the core Hamiltonian. There is no Fockian, neither self consistency in this context.

#### 2.3.3.f. Generalization of Koopmans' theorem (1934) for the electronic Schrödinger equation (Eq.2.3.1, a=1) with a general coupling constant, paying particular attention to TNRS (Eq.2.3.2, a=0)

The Koopmans' theorem [2.3.1] states that according to the closed-shell HF-SCF theory, the first ionization energy of a molecular system is equal to the negative of the orbital energy of the highest occupied molecular orbital (HOMO). Seemingly it is trivial, but in practice, if a system is given by $H_{ne}$ and N, one does not have to make two HF-SCF/basis/a=1 calculations for a ($S_0$, $E_{electr,0}$(HF-SCF)) pair with N and N-1, and taking the energy difference for estimating ionization energy, but one calculation for N is enough, because one of its intrinsic energy values, of the HOMO, is about the same (but not exactly the same) as the difference in $E_{total\ electr,0}$(HF-SCF) for N vs. N-1. The equilibrium geometry encapsulated in $H_{ne}$ and $H_{nn}$, differs slightly between N and N-1, because there is a shrink in the lowest lying doubly occupied MOs in $S_0$ if N is reduced to N-1 by the stronger effect of the nuclear frame (which slightly expands) if the number of electrons decreases, - Koopmans' theorem comes from a purely mathematical derivation in HF-SCF formalism. Here we introduce a similar mathematical situation in which: if a system is given by operator $H_{nn}$ or $H_{ne}$ and N, Eq.2.3.2 determines Eq.2.3.1 "somehow" via the coupling strength parameter. After the relatively easy algorithm which solves Eq.2.3.2 (with HF-SCF/basis/a=0), there exists an algorithm transferring the energy to that of Eq.2.3.1 (more accurate than Eq.2.3.29 above or Eq.2.3.48 below) which may be substitute for the almost century old misery of correlation or CI calculations.

Going back to Koopmans' theorem, its generalization is that it holds for any coupling strength parameter "a", the proof [2.3.1] is exactly the same; moreover, it is trivial for Eq.2.3.2 (a=0), because in $Y_0$ the MOs from Eq.2.3.4 do not change if N decreases to N-1, which is not the case if $a\neq 0$, as well as this, see more details below; what is more, it holds for open-shell systems as well if a=0. As an example, in the above calculation for Ne, the ionization energy is $-f_5(a=1)=$ 0.54305 hartree (as can be seen in the output) from HF-SCF/STO-3G/a=1 calculation for Eq.2.3.1, the accurate CI calculation for Eq.2.3.1 or the experimental values are 0.7946 or 0.79248 hartree, respectively. The error from STO-3G basis set is large (0.79248 -0.54305= 0.24943), because this basis set is modest concerning energy differences (on PES with same N) but suffering from larger error for absolute energy values e.g., HOMO. On the other hand, $-\phi_5$(a=0)= 10.22405 hartree from a HF-SCF/STO-3G/a=0 for Eq.2.3.2, from the list above, the much larger, nonphysical value comes from not involving $V_{ee}$ in TNRS, so electron-electron repulsion energy does not pull the HOMO to higher energy level but, importantly, the right hand side of Eq.2.3.29 returns the ionization potential value to 0.54305 as with $-f_5$(a=1), the accurate back transfer up to 5 digits is accidental and originates from the now fortunate, rigid STO-3G basis set, but does show the power of the Eq.2.3.29. HF-SCF/basis/a=1 case, more precisely spin-unrestricted KS, for open shell can be found in ref. [2.3.41].

**2.3.3.g. The Hund's rule (1927) in relation to TNRS**

Some general spin and Hund's rule related properties for Eqs.2.3.1-2 were discussed in section 2.3.2.a. In atomic physics, Hund's rules refer to a set of three rules, which are used to determine the term symbol that corresponds to the ground state of a multi-electron atom. It was first empirically established and then later proven in HF-SCF theory [2.3.1], but generally it has not yet been proven for Eq.2.3.1 only for HF-SCF/basis/a=1. The first rule is especially important in chemistry, where it is often referred to simply as: Hund's rule. For a given electron configuration, the term with maximum multiplicity has the lowest energy. Therefore, the term with lowest energy is also the term with maximum S. It tells us something about how the electronic structure builds up as N increases. However, contradictions in the quantum mechanical explanation of the periodic table may arise [2.3.42-43].

An important test for estimation in Eq.2.3.29 and its finer refinements above, which is a forecast for excited states as well for Eq.2.3.48 below, is how Eq.2.3.29 obeys Hund's rule. In relation to the coupling strength parameter "a", the case of a=0 manifests for Hund's rule in comparison to other properties or emblematic theorems, since this Hund behavior annihilates in the ground state eigenvalue ($e_{electr,0}$,$Y_0$), moreover, in excited states too. (It means that in degenerate states, e.g., atomic p orbitals, the high spin fill up is energetically the same as the lower spin fill up.) A representative calculation for first row neutral elements can be found in 2.3.Table 1, where column with Eq.2.3.1 in the head is a conventional HF- SCF/basis/a=1 calculation, it obeys Hund's rule, as it is well known, i.e., all high spins are more stable (see the square bracket values). Column with Eq.2.3.2 in the head does not obey Hund's rule, (see the zeros in the square brackets), as is mentioned above, while the column with Eq.2.3.29 in the head obeys Hund's rule again, and is close to the values in the column with Eq.2.3.1 in the head, indicating that approximation in Eq.2.3.29 is a promising and plausible first approximation. Even the order of the calculated values for the energy gap between high and low spin states agree between columns with Eqs.2.3.1 and 29 in the head in 2.3.Table 1: carbon has the smallest and nitrogen has the largest energy gap. Additionally, using the smaller STO-3G basis set, this energy gap shown in square brackets in 2.3.Table 1 for C, N and O atoms is 0.10881, 0.16447, 0.14233 hartree, respectively for both columns with Eqs.2.3.1 and 29 in the head, i.e., there is no difference between the two columns in these values. The reason for the latter is that the STO-3G basis set contains one branch of Gaussians and is not flexible enough to change the LCAO parameters in this respect, yet yields reasonable values; an overlap like this is characteristic of TNRS via HF-SCF/basis/a=1 for Eqs.2.3.1 vs. HF-SCF/basis/a=0 for Eq.2.3.29 with the minimal basis set STO-3G, and not an accidental coincidence. From an analytical point of view, Hund's rule applies if $a \neq 0$, emblematic property in the case of real (a=1) systems, but as $a \to 0$; the energy gap between high and low spin states also goes to zero and Hund's rule annihilates in this respect.

2.3.Table 1.: HF-SCF/6-31G**/a=0 and 1 energies for high and low spin states in hartree to test Eq.2.3.29 in relation to Hund's rule. In the square brackets the important energy differences between the high and low spin states of the same atom (negative sign means more stablility); X= number of convergence steps in HF-SCF in that column, Y= sources of error in that column. For the last three energy columns for partial comparison, the high spin CI calculations [2.3.44] have also been listed in hartree for C, N and O atoms under the multiplicity 2S+1, respectively.

| Atom | Configuration over $[1s^2 2s^2]$ | 2S+1, $E_{electr,0}$(CI) | a=1, Eq.2.3.1, $E_{electr,0} \approx \langle S_0\|H\|S_0\rangle$ | a=0, TNRS, $e_{electr,0}$ from Eq.2.3.2 | a=0, $E_{electr,0}$(TNRS) from Eq.2.3.29 |
|---|---|---|---|---|---|
| C | $2p_x 2p_y$ | 3, triplet, -37.8450 | -37.680860 [-0.09230] | -53.106285 [0] | -35.971284 [-0.14335] |
| C | $2p_x^2$ | 1, singlet | -37.588558 | -53.106285 | -35.827936 |
| N | $2p_x 2p_y 2p_z$ | 4, quadruplet, -54.5893 | -54.385442 [-0.13966] | -77.929276 [0] | -52.145336 [-0.22481] |
| N | $2p_x^2 2p_y$ | 2, dublet | -54.245778 | -77.929276 | -51.920527 |
| O | $2p_x^2 2p_y 2p_z$ | 3, triplet, -75.0674 | -74.783934 [-0.12733] | -109.338617 [0] | -71.698628 [-0.19431] |
| O | $2p_x^2 2p_y^2$ | 1, singlet | -74.656604 | -109.338617 | -71.504319 |
| X | - | - | 7-11 | 1 | 1 |
| Y | - | - | basis set, correlation | basis set | basis set, correlation |

**2.3.4. Theory for calculating excited states via TNRS (Eq.2.3.2, a=0)**
**2.3.4.a. TNRS ground and excited states $\{Y_k, e_k\}$ from Eq.2.3.2 (a=0, k=0,1,2,… in one step) using atomic basis set, as an orto-normalized Slater determinant basis set for CI calculations on Eq.2.3.1 (a=1, k=0,1,2,…)**

In section 2.3.3 we have demonstrated that the solution for TNRS (Eq.2.3.2, a=0) can be technically obtained via the standard HF-SCF algorithm with the device that the coupling strength parameter "a" is programmed as input in a fast and stable way notated as HF-SCF/basis/a=0. Beside the very interesting and important fact that the HF-SCF/basis/a LCAO parameters for a=0 (TNRS Eq.2.3.2) and a=1 (Eq.2.3.1) are close to each other (aside from some possible phase factors or degeneracy), the LCAO coefficients of TNRS can be obtained in only one step with HF-SCF/basis/a=0 algorithm for Eq.2.3.2, irrespective of system size (a small demonstration is given in row X in 2.3.Table 1). In contradiction, the HF-SCF/basis/a=1 LCAO coefficients of a real system (or the mathematical a≠0 cases) can only be obtained through many steps; operator $H_{ee}$ is responsible for this, and the number of steps dramatically increases with the number of electrons (N), more precisely, with system size. This increase in computation demand is demonstrated in a little more detail in section 2.3.3.b with a small system; compare the 2 (basically 1) vs. 6 steps in the case of hydrogen-fluorid. A finer detail is that if a convergence problem arises from system size when a=1 and large N, the breakdown never happens in the first 2 steps, but usually much later, at least this is our experience in practice. We mention again that for Eq.2.3.2 (a=0) the solutions $Y_k$ (k=0,1,2,. . .) have an exact Slater determinant form, recall section 2.3.2.a in particular Eqs.2.3.5-8, while for other coupling strength parameter values (a≠0) the form of solutions, $y_0(a)$, are not single determinant forms creating the correlation effect if single determinant ($s_0(a)$) is used to approximate $y_0(a)$. It is also known that for real (a=1) systems the HF-SCF/basis/a=1 approximation with basis set and correlation error, yields physically trustable MOs up to LUMO, though higher MOs must be considered with caution. In more detail, estimation HF-SCF/basis/a≠0 for ground state provides the virtual orbital LUMO (≡LUMO+1) as a byproduct, which can be used as a weak estimation of the first excited state, but LUMO+2, LUMO+3, etc. cannot be used for this purpose, alone they would be mathematically very weak to estimate the eigenvalues of operator $H_\nabla+H_{ne}+aH_{ee}$, and only good for constructing ortho-normal basis sets for CI calculations in a next step (the $s_0(a) \rightarrow s_{0,p}{}^q(a)$, etc. manipulation). However, the HF-SCF/basis/a=0 MOs are mathematically correct up to N/2 or (N+1)/2 and over- suffering from basis set error only. (These LUMO+1,2,… orbitals are created as a byproduct over the lowest lying N/2 or (N+1)/2 MOs, as well as require high enough AOs to be used in the basis set, and more importantly, they can have physical meaning after a further plausible process which can transfer them from Eq.2.3.2 to Eq.2.3.1.) Generalized AC in ensemble DFT for excited states using minimal basis can be found in ref. [2.3.45].

The manipulation using the Slater determinant CI theory is well established [2.3.1], but as in section 2.3.2.a, some textbook properties must be overviewed. To obtain ground ($Y_0$) and excited ($Y_k$) states of Eq.2.3.2 via the HF-SCF/basis/a=0 algorithm is as follows: The ground state $Y_0$ can be calculated as described in section 2.3.3; it solves Eq.2.3.4 first for some $\phi_i$ states, and sets up the ground state ($Y_0$) for Eq.2.3.2 as Eq.2.3.5. With e.g. by product LUMO which satisfies Eq.2.3.4 as the other MOs, the excited states ($Y_k$) can also be set up as Eqs.2.3.6-7. For excited states one has to provide basis set adequate for higher $\phi_i$ (i > N/2 or (N+1)/2) states: for example, for the close to neutral (charge -1, 0 or 1) small hydrogen-fluorid molecule (section 2.3.3.b), e.g. the STO-3G or 6-31G*, etc. basis sets include the 1s, 2sp and in the latter the 3spd AOs [2.3.12] for approximating $\Psi_0$ in Eq.2.3.1 using HF-SCF/basis/a=1 algorithm (suffering with basis set and correlation error), these basis sets were worked out for these kinds of systems. However, these basis sets can also be used for HF-SCF/basis/a=0 algorithm (causing basis set error only) for approximating $Y_0$ (Eq.2.3.2) or $\phi_i$ (Eq.2.3.4) being the 2p the highest occupied AO in fluoride participating in the hydrogen-fluoride molecule. Calculating higher TNRS excited states for hydrogen-fluoride via Eqs.2.3.2 or 4, the 4spdf, 5spdfg, etc. AOs are also necessary, depending how high the values k or i are targeted in $Y_k$ (or $\phi_i$). (This must also be provided if (a different kind of) CI algorithm is performed on Eq.2.3.1.) The HF-SCF/basis/a generated $\{Y_k(a=0)\}$ determinant basis set can be used for CI calculations, as the $\{S_k(a=1)\}$ in practice, the linear algebra is exactly the same, but the algebraic forms do differ a slightly, of course, and this will be elaborated upon next.

In calculating $Y_k$ of Eq.2.3.2, one has to apply the standard way of linear algebra for energy Hamiltonian [2.3.1] which requires to compute the matrix elements $\langle b_i|h_1|b_j\rangle$ for i,j=1,2,…,K, where $\{b_1(\mathbf{r}_1), b_2(\mathbf{r}_1),\dots b_K(\mathbf{r}_1)\}$ is an adequate, atom-centered AO basis set [2.3.12]. The K eigenvalues (MO energies) and eigenvectors (wave functions) of this KxK Hamiltonian matrix approximates the lowest lying K eigenstates: the orbital energy values, $\{\varepsilon_i\}_{i=1..K}$, and ortho-normal wave functions, $\{\phi_i(\mathbf{r}_1)\}_{i=1..K}$, of Eq.2.3.4. This is what we call a one step algorithm (section 2.3.3.e), because the eigensolver is used only once. Now a=0, but recall that in HF-SCF/basis/a≠0 algorithm every step after the initial estimation needs eigensolver, what HF-SCF or KS do during a typical (a=1) SCF device [2.3.1-2]. As a result, the $\phi_i(\mathbf{r}_1)$ wave functions (of Eq.2.3.4 and the so-called MOs of Eq.2.3.2) are expressed in LCAO in the basis set chosen, and are ortho-normal required by Eqs.2.3.2 and 4. Like in Eqs.2.3.5-7, the eigenstate energy values $\{e_{elekt,k}\}_{k=0,1,..L-1}$ along with the set of single Slater determinant wave functions $\{Y_k\}_{k=0,1,..L-1}$ can be accomplished systematically by mixing $\phi_i$ (i= 1.2,…,N/2 or (N+1)/2, …K) and using Eq.2.3.8 as in the standard algebra with a Slater determinant for case a=1, i.e. for $\{S_k\}$ [2.3.1], but there the corresponding equation to Eq.2.3.8 is not as simple. The L=(2K)!/(N!(2K-N)!), since all $\phi_i$ can be non-, singly- or doubly (and oppositely) occupied by $\alpha$ or $\beta$ spins.

What is important is the orthogonal property

$$<\phi_i(\mathbf{r}_1)|\phi_j(\mathbf{r}_1)>= <Y_i(\mathbf{x}_1,\ldots, \mathbf{x}_N)|Y_j(\mathbf{x}_1,\ldots, \mathbf{x}_N)>= \delta_{ij} \qquad \text{(Eq.2.3.43)}$$

where obviously, the Bra-ket, <|>, integration means 3 and 4N dimensions, respectively. Normalization $N<Y_i|Y_i>= \int\rho_i(\mathbf{r}_1,a=0)d\mathbf{r}_1= N$ also holds, as a conventional definition for i[th] excited state. Property in Eq.2.3.43 for orbital set $\{\phi_i(\mathbf{r}_1)\}$ and determinant set $\{Y_k\}$ comes from the hermitian and linear nature of the operators in Eq.2.3.4 and Eq.2.3.2, respectively, that is,

$$\varepsilon_j\delta_{ij}= <\phi_i|h_1|\phi_j>= <\phi_j|h_1|\phi_i>= \varepsilon_i\delta_{ji} \qquad \text{(Eq.2.3.44)}$$

and,

$$e_{electr,j}\delta_{ij}= <Y_i|H_\nabla + H_{ne}|Y_j>= <Y_j|H_\nabla + H_{ne}|Y_i>= e_{electr,i}\delta_{ji} \qquad \text{(Eq.2.3.45)}$$

as well as for basis set elements $<b_i|h_1|b_j>= <b_j|h_1|b_i>$. In this chapter, the subject of section 2.3.2.a is generalized from ground state (as in Eq.2.3.5) to excited states (as in Eqs.2.3.6-7). The normalization in Eq.2.3.43 is just a matter of using a proper constant multiplier with $\phi_i$ or $Y_k$. The anti-symmetric orto-normalized Slater determinant basis set $\{Y_k\}$ from Eq.2.3.2 (a=0) via HF-SCF/basis/a=0 algorithm which is also complete i.e., any anti-symmetric function for interchanging any ($\mathbf{x}_i$, $\mathbf{x}_j$) pair in 4N dimensional ($\mathbf{x}_1,\ldots, \mathbf{x}_N$) space that can be expanded with them, can be used for solving Eq.2.3.1 for ground and excited states with determinant expansion of $\Psi_k$, like the ortho-normal $\{S_k\}$ set from HF-SCF/basis/a=1 in practice.

For an LCAO estimation by HF-SCF/basis/a=0, the main step is the above diagonalisation of matrix $<b_i|h_1|b_j>$. Of course, the computation time increases for this one main step with the size of Hamiltonian matrix, that is, with system size. The program to solve the eigenvalue problem $<b_i|h_1|b_j>$ is straightforward, but an HF-SCF/basis/a=0 algorithm (from commercial programs modified with the coupling strength parameter "a" as input) can conveniently be used.

### 2.3.4.b. Tricking the HF-SCF/basis/a=0 algorithm to obtain excited states, $Y_k$, beside the ground state $Y_0$ of TNRS (Eq.2.3.2, a=0)

If higher states than $Y_0$ are required, one can still use the existing HF-SCF/basis/a=0 codes, but one has to do a trick with changing the charge of system because only LUMO+1 or LUMO+2 come out purely as a byproduct, and no higher excited states. That is, simply increasing N on the same nuclear frame $\{R_A,Z_A\}_{A=1,\ldots,M}$ encapsulated in $H_{ne}$ by using the correct multiplicity, although the latter allows greater freedom than in the case a=1; a demonstration of this follows by continuing the discussion on the LCAO coefficient matrix of the equilibrium hydrogen-fluoride molecule listed in section 2.3.3.b. For this molecule, the HF-SCF/STO-3G/a=0 calculation for Eq.2.3.2 or 4 was processed in neutral ($\Sigma Z_A$=N=10) and singlet ($1+2\Sigma s_i=1$) mode, as usual. It calculates the lowest lying N/2= 5 doubly occupied molecule orbitals (MOs). That is, this routine calculates $Y_0$ for Eq.2.3.2 as output by listing the five MOs and their energy eigenvalues belonging to $Y_0$, which are, at the same time, the wave functions $\{\phi_1,..,\phi_5\}$ and eigenenergies $\{\varepsilon_1,..,\varepsilon_5\}$ of Eq.2.3.4. Calculating higher MO or $\phi_i$, one simply has to increase the number of electrons, e.g. adding -1 charge to the molecule (N=11), and using correct multiplicity (here $1+2\Sigma s_i= 2$). This HF-SCF/STO-3G/a=0 calculation for Eq.2.3.2 or 4 yields exactly the same LCAO coefficients and energy eigenvalues as for the neutral (N=10) hydrogen-fluoride listed in section 2.3.3.b, because it is the TNRS (a=0) calculation; only instead of 5 doubly occupied MO, there are 5 doubly occupied MOs plus 1 singly occupied 6[th] MO, which is the $\phi_6$ of Eq.2.3.4. As expected, the LUMO virtual orbital (N=10, 6[th] MO) and the HOMO (N=11, 6[th]) uppermost occupied MOs are exactly the same, which is definitely not true if it is not a TNRS i.e., $a\neq0$ wherein all $\{\varepsilon_i, \phi_i\}$ changes if N changes. (The $E_{total\ electr,0}$ is different, of course, -147.731603 hartree (N=10, see above in section 2.3.3.b) and -152.228311 hartree (N=11) now, because Eq.2.3.8 has one more term.) To calculate the first $\{\phi_1,..,\phi_8\}$ states with orbital energies $\{\varepsilon_1,..,\varepsilon_8\}$ of Eq.2.3.4, one has to set up N=15 (charge -5) and multiplicity 2 or N=16 (charge -6) and multiplicity 1 in the HF-SCF/basis/a=0 routine for Eq.2.3.4, both yield the same results, only the occupation of HOMO $\phi_8$ is different: $\phi_8^1$ in the former and $\phi_8^2$ in the latter. And of course, the first 5 states are the same again as for N=10 and 11 above, however, one must be aware of the basis set chosen at this point, see next paragraph. As the HF-SCF/basis/a=1 names the LUMO+1,2,… as "virtual" MOs, this tricking above with N could have the name "virtual" N. The fact that a $[HF]^{-6}$ for the calculation of TNRS states by HF-SCF/basis/a=0 occupied up to i=8 is not a stable molecule in nature in any geometry is irrelevant now, because this is a trick only to obtain higher TNRS states of Eq.2.3.2. If the first two virtual MOs (LUMO+1,+2) are enough, the original N does not have to be increased.

Processing this outcome further, the neutral (N=10) HF-SCF/basis/a=0 ground state hydrogen-fluoride (wherein the $\{\phi_1,..,\phi_5\}$ are doubly occupied and all levels above are empty) the above calculated $\{\phi_6, \phi_7, \phi_8\}$, i.e., the LUMO+1,+2 and +3 can be used e.g. for single-double-excitation-CI, wherein the electron-electron interaction is taken care of by the eigenvalues adequately large CI matrix to estimate the desirable and the physically plausible lowest lying solutions of Eq.2.3.1 better than Eq.2.3.29, see below for more details. However, attention has to be paid to the basis set. In this example for $\phi_7$ and higher states, an STO-3G basis set (which is defined originally for commercial Gaussian package [2.3.12] for close to neutral molecules and is widely used in practice), containing 6 elements (see the 6 rows of LCAO coefficients for hydrogen-fluoride in section 2.3.3.b) is not adequate for calculating a progressively negatively charged molecule, because only six independent vectors of LCAO coefficients exist in the space spanned by the STO-3G basis. Even the quality of the first virtual state wave function $\phi_6$ with orbital energy $\varepsilon_6$ is questionable with STO-3G basis, because altogether, the AO of $1s^1$ of H atom and $1s^2 2s^2 p^5$ of F atom form five saturated (doubly occupied) MOs, that is, N=10 electrons on $\{\phi_1,..,\phi_5\}$ states. It means that a minimal basis set must contain higher AOs, i.e., basis functions 2sp on H and 3spd on F for an adequate description of $\phi_6$ and linear independency for $\phi_7$ in this calculation - the only known criteria for this manner of choosing a basis set. Recall the old textbook example of $H_2O$ which is mistakenly predicted to be linear, if only the 1s and 2s AOs are in a basis set lacking the 2p orbitals. Furthermore, large and high quality basis sets are available [2.3.12]; the simplest STO-3G has been chosen here for easy discussion. Larger size, higher quality bases sets yield more accurate orbital wave functions and energies, of course, but at the cost of longer computation.

We emphasize that generating the set $\{Y_k\}$ with a HF-SCF/basis/a=0 algorithm is simpler, faster (one step), more effective (larger k) and more convenient than generating set $\{S_k\}$ with an HF-SCF/basis/a=1, although, the author knows perfectly well that the latter is effectively used and widely tested in practice. Additionally, as outlined above, in the case of a=1 the LUMO and up bear the properties of $S_0$, passing it to the ortho-normal basis set for the CI calculation it generates, but only $S_0$ has a really close relationship to Eq.2.3.1, while on the other hand, in the case of a=0 all $Y_k$ are the solution of Eq.2.3.2, which is a pre-equation to Eq.2.3.1.

**2.3.4.c. The ground (k=0) and excited (k>0) states via the electronic Schrödinger equation (Eq.2.3.1, a=1) using the nuclear frame generated orto-normalized Slater determinant basis set $\{Y_k\}$ from TNRS (Eq.2.3.2, a=0) for different levels of CI calculation, estimation for excited states $E_{electr,k}$ with TNRS**

In section 2.3.3.b we have demonstrated that the LCAO coefficients of TNRS obtained by HF-SCF/basis/a=0 algorithm for Eq.2.3.2, suffering only from basis set error, but the correct form of wave function, the single determinant is used, and of an HF-SCF/basis/a=1 approximation for Eq.2.3.1 suffering not only from basis set error but the lack of correlation estimation too, stemming from the use of the inadequate single determinant form are close to each other - on the same basis level of course. To obtain more accurate ground and lowest lying excited states, or to simply obtain the ground state more accurately than the HF-SCF/basis/a=1 algorithm can provide, the different levels of the CI methods can be used, using excited Slater determinant N-electron basis functions beside the one in the ground state in the many determinant expansion for $\Psi_k$ of Eq.2.3.1. In the literature the basis set $\{S_k\}$ is obtained from $S_0$ and LUMO+1, 2 of HF-SCF/basis/a=1 algorithm [2.3.1]. Here we will not go into the extensive literature of correlation calculations on the HF-SCF method (post HF-SCF methods) [2.3.1] and during HF-SCF method (KS formalism [2.3.2]) or general DFT methods [2.3.2] nor the different versions of CI methods [2.3.1]. We shall only mention that generally, CI calculations are the most accurate and most time consuming methods, as well as it is *ab initio*, i.e., they do not use empirical parameters, only physical constants like Planck's constant (h), etc.. Its error can come only from the lower quality of a generally AO basis set $\{b_k(\mathbf{r}_1)\}$ for MOs to expand with LCAO coefficients, and the truncation of determinant basis set (in the range of simpler to more complicated, i.e., full CI). In practice $\{S_k\}$ is used, but the soon to be introduced $\{Y_k\}$ can also be used, and in principle, it is also a mathematically correct way to obtain wave functions $\Psi_{k'}$ via expansion with them. (We mention the plane wave expansion (PWE) vs. the LCAO expansion, useful for describing crystals vs. molecules, though not detailed here.)

Using Eqs.2.3.43-45, the standard way of expanding [2.3.1] anti-symmetric wave functions $\Psi_{k'}$ of Eq.2.3.1 using the basis set $\{Y_k\}$ from Eq.2.3.2 is the linear combination $\Psi_{k'}= \Sigma_k c_k(k')Y_k$, where beside the TNRS ground state $Y_0$, the TNRS $Y_k$ with k> 0 are the single, double, triple, etc. excited N-electron Slater determinants. This determinant expansion and treatment of the called multi-determinant representation of the exact k'$^{th}$ excited state wave function $\Psi_{k'}$ (k'=0,1,…) of Eq.2.3.1 is a very well known procedure, described in ref.[2.3.1], using the $\{S_k\}$ determinant basis from an HF-SCF/baisis/a=1 algorithm. However, we now introduce the use of the TNRS basis set $\{Y_k\}$ from Eq.2.3.2 via HF-SCF/basis/a=0 algorithm and analyze in this work. We draw your attention to the fact that the $\{Y_k\}$ from solving Eq.2.3.2 is a nuclear frame ($H_{ne}$) generated basis set, containing strong and pure information about the nuclear frame via Eq.2.3.2. The determinant set, $\{S_k\}$, is generated from $S_0$ by HF-SCF/basis/a=1 which approximates (energy minimization) the ground state $\Psi_0$ of Eq.2.3.1. $S_k$'s are worse approximations for excited states $\Psi_k$ than the $S_0$ for $\Psi_0$, but the HF-SCF/basis/a=0 generated Slater determinants are the correct ground and excited state solutions of Eq.2.3.2 as an alternative. $S_0$ suffers from basis set error and correlation effect, while $Y_k$ suffers only from basis set error, but $S_0$ is just about physically plausible, while $Y_k$ must be converted to have any physical meaning, – notice that $Y_k$ invokes ground and excited states in contrast to the mere ground state $S_0$. By the principles of linear algebra, changing basis set should not be a problem, mainly from the point of view discussed above, inasmuch as LCAO coefficients do not vary greatly with the value of the coupling strength parameter "a" (section 2.3.3.b) namely, between a= 0 and 1. What is important is that both, $\{S_k\}$ and $\{Y_k\}$, are ortho-normal N-electron determinant basis set for CI, and both are adequate to expand $\Psi_{k'}$ wave functions with them, they behave as basis set change in relation to each other.

The particular generation of basis sets is as follows. In section 2.3.3.b both, $Y_0$ and $S_0$, have been demonstrated for the Ne atom and the hydrogen-fluoride molecule. For Ne, only the lowest 5 occupied MOs are listed, but for hydrogen-fluoride, 5 occupied and 1 unoccupied (LUMO, the $6^{th}$) are also exhibited. The $6^{th}$ MO in these examples is ready to be used to generate single excited HF-SCF/STO-3G Slater determinants, $\{S_k\}$ as $S_0 \rightarrow S_k \equiv S_{0,i}^{6}$, etc. with i=1,2,…5 in CI calculations for Eq.2.3.1 (a=1), but the $\{Y_k\}$ as $Y_0 \rightarrow Y_k \equiv Y_{0,i}^{6}$, etc. (i.e. of Eq.2.3.2, a=0) determinant basis set is also ready to be used in CI calculations for Eq.2.3.1 (a=1), too. For example, the ground state Eq.2.3.5, and single excited states Eq.2.3.6 and Eq.2.3.7 (single excited states, because the configuration of one electron is changed in respect of Eq.2.3.5) are the textbook examples for generating excited elements in the determinant basis set $\{Y_k\}$ obtained from HF-SCF/basis/a=0 or $\{S_k\}$ from $S_0$ from the HF-SCF/basis/a=1 algorithm. We discuss here $\{S_k\}$ and $\{Y_k\}$ parallel to show the position of $\{Y_k\}$ in the business.

Standard linear algebra provides the set of eigenstates $\{\Psi_{k''}, E_{electr,k''}\}$ of Eq.2.3.1 by expanding $\Psi_{k''}$ in basis $\{Y_k\}$: one must diagonalize the Hamiltonian matrix $<Y_{k'}|H_\nabla + H_{ne} + H_{ee}|Y_k>$ as the second main step, just like the first main step, the diagonalization of $<b_i|h_1|b_j>$ for the set of eigenstates $\{Y_k, e_{electr,k}\}$ of Eq.2.3.2 (HF-SCF/basis/a=0) with tricking (virtual) N as described above if necessary. Using the properties of Eq.2.3.2 as

$$<Y_{k'}|H_\nabla+H_{ne}+aH_{ee}|Y_k>= e_{electr,k} <Y_{k'}|Y_k> + a<Y_{k'}|H_{ee}|Y_k> \quad \text{(Eq.2.3.46)}$$

,where we have extended it with the coupling strength parameter "a" to be more general. With Eq.2.3.43, the diagonal elements (k'=k) reduce to

$$<Y_k|H_\nabla+H_{ne}+aH_{ee}|Y_k>= e_{electr,k} + a(N(N-1)/2)<Y_k|r_{12}^{-1}|Y_k> . \quad \text{(Eq.2.3.47)}$$

Importantly, the generalization of Eq.2.3.29 has been obtained (using Eq.2.3.14) making the link between case a=0 and 1 for ground (k=0) and excited (k>0) states as:

$$E_{electr,k} \approx E_{electr,k}(TNRS) \equiv e_{electr,k} + (N(N-1)/2)<Y_k|r_{12}^{-1}|Y_k> . \quad \text{(Eq.2.3.48)}$$

Eq.2.3.48 is based again on the knowledge that LCAO parameters do not vary greatly between $\Psi_k$ and $Y_k$, although it was only exhibited above for $\Psi_0$ and $Y_0$. However, Eq.2.3.48 gives exact diagonal elements, as well as again being the first approximation from the Rayleigh-Schrödinger perturbation theory as was used above for k=0. We emphasize that approximation Eq.2.3.48 has to be regarded with caution until it is fully tested, its k=0 bottom case in Eq.2.3.29 is at least tested here. HF-SCF/basis/a=1 mode provides ground state $S_0$ and $E_{electr,0}$(HF-SCF), like the weaker HF-SCF/basis/a=0 extended with Eq.2.3.29, but HF-SCF/basis/a=0 mode can provide more, the excited states in Eq.2.3.48 as a simple estimation – of course all these can be continued with CI calculations or DFT corrections for higher accuracy. Furthermore, if one orders $E_{electr,k}$ as usual as $E_{electr,k} \leq E_{electr,k+1}$ for k=0,1,2,…, it must be proved that $Y_k$ belongs to $\Psi_k$, i.e., that no energy value switch or cross correspondence is in a (k,k+1) pair, for example, which is a plausible hypothesis, ($S_k$ is problematic in this respect, because one has to stop around LUMO). If "a" differ from zero by small (infinitesimal) value $\delta a$, the $Y_k$ and $y_k(a=0+\delta a)$ obviously belong to each other in relation to a running k. (The "≤" necessary in relation to energy, the "<" is not enough, because TNRS can remove degeneracy gaps, see column with Eq.2.3.2 in the head in 2.3.Table 1, moreover, $\Psi_k$ characteristically can have degeneracy.) Recall that in practice, based on the HF-SCF/basis/a=1 algorithm, i.e., the standard HF-SCF/basis method, calculating an excited state is more problematic than the ground state, that is, the LUMO+1 and up has to be regarded with caution. In contrast, the simple right hand side of Eq.2.3.48 can be calculated with the HF-SCF/basis/a=0 and the trick with the molecular charge as was demonstrated in section 2.3.4.a is plausible for any k as a first approximation, but needs refinement.

The quality of estimation in Eq.2.3.48 will be tested in a later work. However, applying the idea in section 2.3.2.d, the MP first correction for state k of Eq.2.3.2 to Eq.2.3.1 is just Eq.2.3.48. Furthermore, as k=0 is corrected with states k=1, 2, etc., similarly, the k>0 can be corrected with 0,1,2,…,k+1, k+2, etc. states, that is, with the MP2 analogue $|<Y_k|r_{12}^{-}$

$^1|Y_{k,pq}{}^{rs}>|^2/(\varepsilon_p+ \varepsilon_q- \varepsilon_r- \varepsilon_s)$ terms, wherein r and s can mean de-excitation too. The latter is another justification for Eq.2.3.48 besides the idea of the Hamiltonian matrix, so Eq.2.3.48 is a plausible approximation as a first guess. Recalling that for a symmetric matrix the sum of its eigenvalues is equal to its trace, i.e., the sum of the diagonal elements, the diagonal elements in Eq.2.3.47 can be summed up for k yielding another relationship. (Notice that when e.g., MP2 terms are correct, the factors $(\varepsilon_p+ \varepsilon_q- \varepsilon_r- \varepsilon_s)^{-1}$ have positive signs in de-excitation and negative in excitation determinants.)

Applying Eq.2.3.43 again, the off-diagonal elements (k'≠k) reduce to

$$<Y_{k'}|H_\nabla+H_{ne}+aH_{ee}|Y_k> = e_{electr,\, k\, or\, k'}<Y_{k'}|Y_k> + a<Y_{k'}|H_{ee}|Y_k> =$$
$$= a(N(N-1)/2)<Y_{k'}|r_{12}{}^{-1}|Y_k> \quad \text{if } k'\neq k, \quad \text{(Eq.2.3.49)}$$

that is, the off diagonal elements contain the electron-electron interaction only, which means qualitatively, that the purely Coulomb operator off-diagonal elements correct the deviations via diagonalization that the diagonal elements (Eq.2.3.47 or Eqs.2.3.29 and 48) make as error to finally obtain the desired $E_{electr,k}$. Notice again that the $(Y_k, e_{electr,k})$ pairs (k=0,1,2,…) in Eq.2.3.48 can be obtained by one main step (eigensolving $<b_i|h_1|b_j>$ by a HF-SCF/basis/a=0 algorithm) after calculating the integrals in Hamiltonian matrix elements and following with the diagonal coulomb integrals $<Y_k|r_{12}{}^{-1}|Y_k>$. The latter comes into being after mixing up $Y_k$ as demonstrated in Eqs.2.3.5-7, irrespectively of the size of the molecular system. (Irrespectively means that this is one main step in the algorithm, but of course, the size of Hamiltonian matrix increases by minimal or higher quality basis sets reflecting the molecular frame and N.) Anyway, Eq.2.3.48 with e.g., an effective and quick DFT correction would greatly cancel out the robust CI method for excited states.

In practice, where the set of Slater determinants in use is not the $\{Y_k\}$ by HF-SCF/basis/a=0, but $\{S_k\}$ by HF-SCF/basis/a=1, the off-diagonal elements corresponding to Eq.2.3.49 contain orbital energies of MOs too (see Table 4.1 on p.236 of ref.[2.3.1]), while the corresponding orbital energies (the $\varepsilon_k$'s in Eq.2.3.4) are missing in Eq.2.3.49. In this mathematical formalism, those $e_{electr,k}$ are enough to be included in Eq.2.3.47 only. The latter may indicate that there is strong correspondence and order between indexing in $\{Y_k\}$ and $\{\Psi_k\}$ mentioned above. It is obvious between $Y_0$ and $\Psi_0$, but true for $Y_k$ and $\Psi_k$ for any k in which the degeneracy can slightly modify, e.g., in 2.3.Table 1 a branch of $\{Y_k\}$ corresponds to a branch of $\{\Psi_k\}$ and is known as the "possible 2p configurations," in that case.

Not surprisingly, the matrix in Eq.2.3.46 is diagonal for a=0, because the set of wave functions $\{Y_k\}$ is expressed trivially with itself, and anyway, in this case Eq.2.3.1 reduces to Eq.2.3.2. Analogously, Eq.2.3.46 is diagonal for a=1 (more generally for any "a") if the orto-normalized eigenfunction set $\{\Psi_k\}$ from Eq.2.3.1 (more generally $\{y_k(a)\}$) is used as basis set, however this basis set is the one we are looking for (now for a≠0 and in practice particularly for a=1), the most important set in quantum mechanics when a=1, known analytically for only one-electron atoms, M=N=1 – for which, actually, the value of "a" is irrelevant and Eqs.2.3.1 and 2 overlap via N=1. If the off-diagonal elements (Eq.2.3.49) are neglected, the matrix in Eq.2.3.46 diagonalizes to Eqs.2.3.47 or 48, so approximations Eqs.2.3.29 and 48 are theoretically supported now via linear algebra, the word "approximation" cannot be over emphasized, see $E_{total\ electr,0}$(G3)-$E_{total\ electr,0}$(TNRS) plotted with a solid line in 2.3.Figure 3.b, which is remarkable but, far beyond chemical accuracy. If Eq.2.3.49 is set to be zero on the right, the coupling strength parameter can be used to empirically re-correct this simplification via an empirical value (shifting "a" back from a=1), that is, with slightly less unity than has been demonstrated in section 2.3.3.c to estimate cases for Eq.2.3.1, yet more sophisticated corrections (correlation) are necessary (MP theory was mentioned as a possibility above , for example,), if one wants to avoid the eigensolving (CI method) Eq.2.3.46.

The Coulombic terms $<Y_{k'}|r_{12}{}^{-1}|Y_k>$ in Eqs.2.3.47 and 49 generate many products, although the orthogonality of MO set $\{\phi_i\}$ from Eq.2.3.4 generating $\{Y_k\}$ makes many

cancellation as it is well known (see again Table 2.4 on p.70 or Table 4.1 on p.236 of ref.[2.3.1]) so, calculating these matrix elements is in fact not difficult though, time consuming. One important thing must be mentioned. The spin related properties and manipulations are exactly the same in the case of $\{Y_k\}$ as in the case of $\{S_k\}$, since only the LCAO coefficients differ somehow, but the size of the determinants and the topology (energy ladder) is the same, and must be taken into account in the same way. Eq.2.3.48 is a lucky form, because only one single determinant is involved, so its spin state is obvious, see Eq.2.3.10-12 in section 2.3.2.b in this respect. The only problem is that Eq.2.3.48 is not accurate enough, so a correction from DFT for excited states would be useful using e.g., $\rho_k(\mathbf{r}_1)$ from $Y_k$. However, if one wants to use a kind of CI correction, i.e., use off diagonal elements in Eq.2.3.49 for correction, then the spin situation must be taken into account. For example, let us suppose that one is interested in the singlet states of a molecule. In this case those $Y_k$ determinants must be eliminated from the determinant expansion of which $M_S \neq 0$ in Eq.2.3.12, i.e., which are not singlets, - just like in the routine HF-SCF/basis/a=1 calculations, for obtaining the lowest lying triplet/singlet energy of an atom via manipulating the input multiplicity for $S_0$. The spin algebra for Eqs.2.3.47 and 49 is exactly the same as for basis set $\{S_k\}$, described in ref.[2.3.1], and not repeated here for the sake of brevity. Only the two simplest spin-adapted cases are mentioned when N=even in $Y_0$ obtained from HF-SCF/basis/a=0, the doubly excited singlet $Y_{p(\alpha)p(\beta)}^{r(\alpha)r(\beta)}$, wherein $(\alpha,\beta)$ electron pair from p orbital below LUMO are promoted to r orbital over HOMO with the same $(\alpha,\beta)$ spin configuration as indicated in brackets, as well as the singly excited singlet configuration:

$$2^{-1/2}(Y_{p(\alpha)}^{r(\alpha)} + Y_{p(\beta)}^{r(\beta)}) \ . \qquad \text{(Eq.2.3.50)}$$

Notice that in Eq.2.3.50 both terms alone are also diagonal elements in Eq.2.3.48, but not pure spin states.

### 2.3.4.d. Generalization of Brillouin's theorem (1934) in relation to coupling strength parameter "a"

It is important to mention Brillouin's theorem [2.3.1, 46] too. To avoid becoming lost in the jungle of notations, we summarize, or list the notations from above as initial conditions for generalized Brillouin's theorem: $y_k(a)$ is the exact $k^{th}$ excited state solution of Eqs.2.3.1 and 2, extended with coupling strength parameter "a" knowing that $y_k(a=0)$ have a single determinant form while $y_k(a\neq 0)$ have non-single determinant form, $Y_k \equiv y_k(a=0)$ can be obtained via the HF-SCF/basis/a=0 algorithm for any k; $\Psi_k \equiv y_k(a=1)$ is the physical wave function for ground (k=0) and excited (k>0) states of molecular systems ( one of the ultimate goal in computation chemistry), as well as HF-SCF/basis/a=1 (the regular HF-SCF/basis) algorithm provides the famous approximation $S_0 \approx \Psi_0$, along some lowest lying excited states $S_k$, called N-electron Slater determinants. In this chapter, we discuss the $s_0(a)$ single Slater determinant HF-SCF/basis/a calculation, of which $s_0(a) \approx y_0(a)$, and particularly, $y_0(a=0) = s_0(a=0) \equiv Y_0$ and $\Psi_0 \approx s_0(a=1) \equiv S_0$ holds. Starting from lowest lying state $Y_0$ or $S_0$, one can make singly excited Slater determinant [2.3.1] basis elements to describe $\Psi_k$ for lower k values, most importantly for k=0 by replacing a spin-orbital HOMO level or below (call it b) to a spin-orbital LUMO level or higher (call it r), denoted as $Y_{0,b}^{r}$ and $S_{0,b}^{r}$, respectively. (For example, the three, f, g and h, MOs in Eqs.2.3.5-7 split into six spin-orbitals ($\alpha$f, $\beta$f, $\alpha$g, $\beta$g, $\alpha$h, $\beta$h) in counting columns in the basis set elements $Y_k$ and analogously $S_k$ – a known method.) Brillouin's theorem states that $\langle S_0|H|S_{0,b}^{r}\rangle = 0$ as a consequence of the HF-SCF/basis/a=1 algorithm [2.3.1]. For this reason, extending $S_0$ with only singly excited determinants to improve $\Psi_0$ or improve $\Psi_0$ and estimate $\Psi_1$ is impossible, the doubly excited determinants $S_{0,bc}^{rs}$ are necessary and are the most important corrections to $\Psi_0$, more exactly the $\{S_0, \{ S_{0,b}^{r}\}, \{ S_{0,bc}^{rs}\}\}$ basis set. (Although these Brillouin matrix elements are zero, the singly excited $S_{0,b}^{r}$ do have an effect on $\Psi_0$ via Hamilton matrix elements as $< S_{0,b}^{r}|H|S_{0,bc}^{rs}>$.) With the language of linear algebra, the $s_0(a\neq 0)$ approximate solution in this integral product does what eigenfunctions can do typically: annihilates the operator in the core and the product has an exact eigenvalue, particularly zero.

A trivial extension of Brillouin's theorem for cases HF-SCF/basis/a (which approximates $y_0(a)$ by single determinant $s_0(a)$) is formally the same, that is

$$<s_0(a)|H_\nabla+H_{ne}+aH_{ee}|s_{0,b}^{r}(a)>= 0 \,, \qquad \text{(Eq.2.3.51)}$$

and the proof is the same. Eq.2.3.51 for a=0, i.e. for Eq.2.3.2 and its generated $\{Y_k\}$ eigenfunction set (as a newly introduced candidate basis set for CI treatment for Eq.2.3.1) tells us only that the triviality such as $<Y_0|H_\nabla+ H_{ne}|Y_{0,b}^{r}>=0$, although the more general $<Y_{k'}|H_\nabla+H_{ne}|Y_k>= 0$ is also true of Eq.2.3.49 with a=0 for $k'\neq k$, where indices k and k' count the ground ($Y_0$), singly ($Y_{0,b}^{r}$), doubly ($Y_{0,bc}^{rs}$), … n-touply excited Slater determinants as well, because $Y_k$'s are eigenfunctions. Like Hund's rule annihilates at a=0, see above, Brillouin's theorem becomes a triviality, because $s_0(a=0)$ becomes equal to $Y_0$, that is, an approximate form becomes an exact form. Eq.2.3.51 for eigenvalues trivially yields $<y_{k'}(a)|H_\nabla+H_{ne}+aH_{ee}|y_k(a)>= 0$ also for the wider range $k'\neq k$, because $y_k(a)$'s are eigenfunctions. The Brillouin theorem (a=1 in Eq.2.3.51) and its extension wherein "a" can be any in Eq.2.3.51 tells us more, because $s_0(a)$ and $s_{0,b}^{r}(a)$ are not eigenfuntions of $H_\nabla+H_{ne}+aH_{ee}$, yet these matrix elements are still zero, a characteristic property from the HF-SCF/basis/a algorithm. As in the discussion on Eq.2.3.50 above, the right hand side of Eq.2.3.49 is zero if a= 0, or zero if $Y_{k'}$ or $Y_k$ differ in three or more spin-orbitals. For example, with $a\neq 0$ in Eq.2.3.49, the $Y_{0,bcde}^{rspq}$ and $Y_{0,bcde}^{rsvw}$ differ in only two spin-orbitals, and do not yield zero for the right hand side of Eq.2.3.49 or the left hand side of Eq.2.3.50. In this way, Eq.2.3.51 reduces to subcases of Eq.2.3.49 if a=0, but Eq.2.3.49 with $a\neq 0$ tells us even more than Eq.2.3.51, the reason being that the operator in Eq.2.3.51 and in wave functions have the same "a" values, while in Eq.2.3.49 the operator contains a value of "a", but the wave function is $Y_k\equiv y_k(a=0)$ for k and k' i.e., two different "a" values are involved. An important consequence of this is that, for Eq.2.3.1, the $\{Y_0, \{Y_{0,b}^{r}\}\}$ truncated basis set generated by Eq.2.3.2 (using the minimal, singly-excitated ones) can already be used as a basis to estimate $\Psi_0$ better than e.g., Eq.2.3.29, even to estimate $\Psi_1$ also by the eigenvectors of the Hamiltonian matrix. This means that, it can provide the large part of correlation energy, and the doubly excited determinants do not have to be calculated to save computer time and disc space unless one needs more accurate results or higher excited states. Again, in the literature the CI calculation is based on HF-SCF/basis/a=1 generated $\{S_0, \{S_{0,b}^{r}\}, \{S_{0,bc}^{rs}\}\}$ or a higher basis set to solve Eq.2.3.1, while here, we are talking about the HF-SCF/basis/a=0 generated $\{Y_0, \{Y_{0,b}^{r}\}\}$ or higher basis set to solve Eq.2.3.1.

It is important to mention that, to stop at single excited determinants in set $\{Y_k\}$ can be restricted by e.g., symmetry reason. For example, even the simple $H_2$ molecule (p.63 of ref.[2.3.1]) owns the property that (gerade HOMO and ungerade LUMO) restricts the wave function approximation by excluding single excited determinants (that is, producing zero CI Hamiltonian matrix elements). It yields the $\Psi_0 \approx c_0Y_0 + c_{12}^{34}Y_{12}^{34}$ simplest improvement over $c_0Y_0$ e.g. in basis $\{Y_k\}$ as is well known for $S_0$ and $S_{12}^{34}$, i.e., the simplest CI approximation needs double excited determinant.

## 2.3.Conclusions

The coupling strength parameter extended Hamiltonian $H(a)= H_\nabla+ H_{ne}+ aH_{ee}$ has been analyzed in relation with and compared to the mathematical TNRS (a=0) and the physical (a=1) cases, many exact equations have been derived in relation to „a", as well as this algorithms have been outlined on how to convert the electronic energy from case a=0 to a=1 in computation. The HF-SCF/basis/a algorithm with input „a" value was used for demonstrating computations. Emblematic theorems have been extended from a=1 to general a≠1 cases: Most important, is the extension of 1st HK as $Y_0(a=0) \Leftrightarrow H_{ne} \Leftrightarrow \Psi_0(a=1)$ manifesting as $E_{electr,0}= e_{electr,0} + <\Psi_0|H_{ee}|Y_0>/<\Psi_0|Y_0>$, it is interesting that Hund's rule or the necessity of an RHF/UHF mode annihilate as $a\to 0$. The correct single determinant form solutions $\{Y_k(a=0)\}$ suffering from basis set error only, can always be obtained in one step by HF-SCF/basis/a=0 in contrast to the not correct single determinant form $S_0(a=1)$ from HF-SCF/basis/a=1 in relation to the non-single determinant $\Psi_0$, which always needs more steps to converge. $\{Y_k(a=0)\}$ provides us with a mathematically correct, ortho-normal, well-behaving basis set for CI calculations, and the single excited determinants from $Y_0$ have more freedom than that which is restricted by Brillouin theorem if $S_0$ is used to generate basis set for CI. The first approximation $E_{electr,k}\approx E_{electr,k}(TNRS)\equiv e_{electr,k} + (N(N-1)/2)<Y_k|r_{12}^{-1}|Y_k>$ (coming from trial function via the variation principle for k=0) for ground (k=0) and excited (k>0) states has been well founded via extending the MP perturbation theory and the linear algebra of the CI theory, but of course finer correlation calculation (DFT, MP, etc.) or eigensolving certain levels of the CI matrix is necessary for chemical accuracy. This expression constitutes the diagonal elements of the TNRS-CI matrix, while the off-diagonal elements are simply the $<Y_{k'}|H_{ee}|Y_k>$. For ground state, the buildup of LCAO parameters was analyzed as a function of „a", and it was demonstrated for the prediction that the electronic energy builds up quasi-linearly between $(Y_0,e_{electr,0})$ and $(\Psi_0,E_{electr,0})$ as $a=0\to 1$ opposing to the strictly linearity in operator $aH_{ee}$.

## 2.3.Appendices:

### 2.3.Appendix 1.: General equation for w in Eq.2.3.35

Substitute $wY_0$ into Eq.2.3.1 after extension $H_{ee}\to aH_{ee}$ along with using $\nabla_1^2(wY_0)= w(\nabla_1^2Y_0)+ 2\nabla_1Y_0\nabla_1w+ Y_0\nabla_1^2w$ and Eq.2.3.2, it yields, for a pre-fixed "a" value: $-(1/2)Y_0 \Sigma_{i=1,\dots,N}\nabla_i^2w -\Sigma_{i=1,\dots,N}\nabla_iY_0\nabla_iw + aH_{ee}wY_0= (enrg_{electr,0}(a)- e_{electr,0})wY_0$. It provides the variation equation, Eq.2.3.38, as well as an equation for r-symmetric w, if $Y_0$ is separated term by term. The $Y_0$ has an exact Slater determinant form, all N! terms are different but algebraically equivalent, so it is enough to consider the first, or the one from diagonal. For example, for N=2 the $Y_0= |\alpha_1f_1,\beta_2f_2>= \alpha_1\beta_2f_1f_2 - \alpha_2\beta_1f_1f_2$ splits to N!=2 parts, and equality holds for the terms with the same spin parts, now both terms yield the same: $(1/2)(\nabla_1^2w+ \nabla_2^2w) -(\nabla_1\ln f_1\nabla_1w+ \nabla_2\ln f_2\nabla_2w) + aw/r_{12}= (enrg_{electr,0}(a) - e_{electr,0})w$ after dividing by the spatial part $f_1f_2$. As a particular example, consider a non-relativistic atom (1≤Z≤18, M=a=1) with N=2 electrons: $Y_0(a=0)$ contains $\phi_i(1s)= f_i=2Z^{3/2}\exp(-Z|\mathbf{r}_i|)$ with $\varepsilon_i=-Z^2/2$ in a.u. for i=1,2 (no basis set error), yielding the exact

$$-(1/2)(\nabla_1^2+\nabla_2^2)w +Z(\nabla_1|\mathbf{r}_1|\nabla_1w+\nabla_2|\mathbf{r}_2|\nabla_2w) +w/r_{12}= (E_{electr,0}+Z^2)w.$$

### 2.3.Appendix 2.: Analytical solution for w in Eq.2.3.35 for the simplest case of (M=0, N=2, a=1)

To have a feeling about w, we mention that for a=1 and N=2 the core spatial equation for w is the $[-(1/2)(\nabla_1^2+\nabla_2^2) + r_{12}^{-1}]z= \varepsilon z$ eigenvalue equation, and its analytical solution with the smallest $\varepsilon$ value is $z(\mathbf{r}_1,\mathbf{r}_2)= \exp(r_{12}/2)$ with $\varepsilon=-0.25$, mentioned earlier in ref.[2.3.47].

This core equation is accidentally the same as the a=1, N=2, M=0 case of Eq.2.3.1 too. Its spatial part shows why simple i.e., single, Hartree product $z= \exp(r_{12}/2)\approx p(\mathbf{r}_1)p(\mathbf{r}_2)$ cannot account accurately, moreover, Slater determinants cannot be an analytic solution.

To have a taste of how the necessity of a correlation calculation manifests itself in the anti-symmetrised approximation $\Psi_0 \equiv (\alpha_1\beta_2 - \alpha_2\beta_1)\exp(r_{12}/2) \approx S_0 \equiv (\alpha_1\beta_2 - \alpha_2\beta_1)p(\mathbf{r}_1)p(\mathbf{r}_2)$, let us approximate the exponential spatial part with one gaussian function (basis set error, here an STO-1G, suffering function shape error), yielding $\exp(r_{12}/2) \approx c.\exp(r_{12}^2/2) = c.\exp(½((x_1-x_2)^2+(y_1-y_2)^2+(z_1-z_2)^2)) = c.\exp(r_1^2/2)\ \exp(r_2^2/2)\ \exp(-(x_1x_2+y_1y_2+z_1z_2))$. In the latter triple product, the correspondence $p(\mathbf{r}_i) := \exp(r_i^2/2)$ can be made for i=1,2 (recall the 1s type AOs), and the role of r-symmetric w in Eq.2.3.35 can be recognized as $w(\mathbf{r}_1,\mathbf{r}_2,a=1, N=2, M=0) := \exp(-(x_1x_2+y_1y_2+z_1z_2))$ acting as a correlation function. Its energy equivalent is the correlation energy, approximating the accurate $E_{corr}$ from $[-(1/2)(\nabla_1^2+\nabla_2^2) + r_{12}^{-1}](\exp(r_{12}/2) - p(\mathbf{r}_1)p(\mathbf{r}_2))$. Approximating, because the gaussian basis set approximation was involved, as well as for $E_{corr}$ normalized <|> integral average must be calculated, since $p(\mathbf{r}_1)p(\mathbf{r}_2)$ is not an eigenfuntion. The $\exp(-(x_1x_2+y_1y_2+z_1z_2))$ accounts for the Fermi and Coulomb hole, the STO-GTO type difference $\exp(r_{12}/2)-c.\exp(r_{12}^2/2)$ is responsible for the basis set error, which goes to basis set limit if a lot of GTO is used (notice that here $\Psi_0$ is approximated, but in practice it is the $S_0$), and $\exp(r_{12}/2)-p(\mathbf{r}_1)p(\mathbf{r}_2)$ is responsible for the basis set and correlation error. In this simple example a quasi-accurate $E_{corr}$ can be evaluated, because the accurate wave function is known, but for physically important cases (N>1, M>0) in Eq.2.3.1, the $\Psi_0$ is unknown. (Notice that $\exp(r_{12}/2)$ is not well behaving since its integral over $d\mathbf{r}_1 d\mathbf{r}_2$ is infinite.) A much more sophisticated model than $-(1/2)(\nabla_1^2+\nabla_2^2) + r_{12}^{-1}$, called uniform electron gas (defined as a large $N\to\propto$ in a cube of volume $V\to\propto$, but finite $\rho=N/V$, throughout which there is a uniform spread of positive charge sufficient to make the system neutral), has led to very serious correlation calculations, see refs.[2.3.48-51].

**2.3.Footnotes:**

1.a: S.Kristyan: 16th International Conference on Density Functional Theory and its Applications, CELEBRATING THE 50TH ANNIVERSARY OF THE KOHN-SHAM THEORY, August 31 - September 4, 2015, Debrecen, Hungary, Conference Abstract p.108, http://dft2015.unideb.hu/home

1.b.: S.Kristyan: The 8th Molecular Quantum Mechanics Conference, June 26 - July 1, 2016, Uppsala, Sweden, Abstract 0018, http://conference.slu.se/mqm2016/

1.c.: S.Kristyan: Theory and Applications of Computational Chemistry Conference, Aug. 28 – Sept. 2, 2016, Univ. of Washington in Seattle, USA, Abstract W44, http://www.tacc2016.org/

2.: $Y_k$ is well behaving via Eq.2.3.2 for any molecular frame $H_{ne}$, only the zero trivial solution has not-finite integral. Well behaving property is important to ask for $S_0$, which is not a correct determinant form for ?$_0$.

**2.3.References:**

## 2.4. Analytic evaluation of Coulomb integrals for one, two and three-electron operators, $R_{C1}^{-n}R_{D1}^{-m}$, $R_{C1}^{-n}r_{12}^{-m}$ and $r_{12}^{-n}r_{13}^{-m}$ with n, m=0,1,2

### 2.4. Preliminary

In the title, where R stands for nucleus-electron and r for electron-electron distances, the (n,m)=(0,0) case is trivial, the (n,m)=(1,0) and (0,1) cases are well known, fundamental milestone in integration and widely used in computation chemistry, as well as based on Laplace transformation with integrand $\exp(-a^2t^2)$. The rest of the cases are new and need the other Laplace transformation with integrand $\exp(-a^2t)$ also, as well as the necessity of a two dimensional version of Boys function comes up in case. These analytic expressions (up to Gaussian function integrand) are useful for manipulation with higher moments of inter-electronic distances, for example in correlation calculations.

### 2.4.1.Introduction

The title is a bit mathematically compact in the sense that for one-electron density, ρ, the $\int\rho(1)R_{C1}^{-n}R_{D1}^{-m}d\mathbf{r}_1$, $\int\rho(1)\rho(2)R_{C1}^{-n}r_{12}^{-m}d\mathbf{r}_1d\mathbf{r}_2$ and $\int\rho(1)\rho(2)\rho(3)r_{12}^{-n}r_{13}^{-m}d\mathbf{r}_1d\mathbf{r}_2d\mathbf{r}_3$ include, 1., The first two are not, but the last one is invariant to an interchange of n and m, 2., Trivial case: if (n,m)=(0,0) then these reduce to $(\int\rho(1)d\mathbf{r}_1)^i= N^i$ for i=1,2 and 3, resp., 3., The well known case: if (n,m)=(1,0) or (0,1) then these are the $\int\rho(1)R_{C1}^{-1}d\mathbf{r}_1$ and $\int\rho(1)\rho(2)r_{12}^{-1}d\mathbf{r}_1d\mathbf{r}_2$ integrals, 4., The rest values for (n,m) are new, and constitute the topic of this work. Furthermore, integrals such as $\int\rho(1)\rho(2)\rho(3)\rho(4)r_{12}^{-n}r_{34}^{-m}d\mathbf{r}_1d\mathbf{r}_2d\mathbf{r}_3d\mathbf{r}_4= (\int\rho(1)\rho(2)r_{12}^{-n}d\mathbf{r}_1d\mathbf{r}_2)(\int\rho(1)\rho(2)r_{12}^{-m}d\mathbf{r}_1d\mathbf{r}_2)$, etc. break up to simpler elements and fall into the cases discussed. As n and m scan the values 0,1,2, these integrals are one, two or three-electron Coulomb integrals, and next in the Introduction we list the new cases in order of their physical importance, as well as expressions will be derived not for ρ, but for primitive Gaussians, $G_{Ai}$, (because the real or any physically realistic model $\rho(1)\geq 0$ can be well approximated as linear combination of a well chosen set $\{G_{A1}\}$). The standard definitions, - especially for distances - are listed at the end, which helps to guide the reader.

### 2.4.1.a. Two and three-electron Coulomb integrals for electron-electron interactions with $r_{12}^{-n}r_{13}^{-m}$

The Coulomb interaction between two charges in classical physics is $Q_1Q_2r_{12}^{-n}$, and is one of the most important fundamental interactions in nature. The power "n" has the rigorous value 2 describing the force, while as a consequence, the n=1 yields the energy. In quantum physics and computation chemistry [2.4.1-3], the HF-SCF and post-HF-SCF, DFT as well as CI main theories, all based on Slater determinants, provide approximate, but formally similar expressions [2.4.4-5] for electron-electron interactions, wherein the exact theory says that the Coulomb interaction energy is represented by the two-electron energy operator $r_{12}^{-1}$.

Using GTO functions, which is

$$G_{Ai}(a,nx,ny,nz)\equiv (x_i-R_{Ax})^{nx}(y_i-R_{Ay})^{ny}(z_i-R_{Az})^{nz}\exp(-a|\mathbf{r}_i-\mathbf{R}_A|^2) \qquad \text{(Eq.2.4.1)}$$

with a>0 and nx, ny, nz ≥0 benefiting its important property such as $G_{Ai}(a,nx,ny,nz)G_{Bi}(b,mx,my,mz)$ is also (a sum of) GTO, the Coulomb interaction energy for molecular systems is expressed finally with the linear combination of the famous integral

$$\int G_{A1}G_{B2}\,r_{12}^{-1}d\mathbf{r}_1d\mathbf{r}_2. \qquad \text{(Eq.2.4.2)}$$

(In Eq.2.4.1 we use double letters for polarization powers i.e., nx, ny and nz to avoid "indice in indice", nx=0,1,2,… are the s, p, d-like orbitals, etc.. Especially in DFT, one ends up with more complex correction terms than Eq.2.4.2, but for main terms the seed is Eq.2.4.2.) The analytic evaluation [2.4.1, 6] of the integral in Eq.2.4.2 has been fundamental and a mile stone in the history of computation chemistry.

In view of the general and extreme power of series expansion (trigonometric Fourier, polynomial Taylor, Pade, etc.) in numerical calculations, however, practically the

$$\int G_{A1}G_{B2}\, r_{12}^{-2}d\mathbf{r}_1 d\mathbf{r}_2 \quad \text{as well as} \quad \int G_{A1}G_{B2}G_{C3}r_{12}^{-n}\, r_{13}^{-m}d\mathbf{r}_1 d\mathbf{r}_2 d\mathbf{r}_3 \qquad \text{(Eq.2.4.3)}$$

with n,m=1,2 important terms have come up in computation chemistry, what we can call higher moments with respect to inter-electronic distances $r_{ij}$, though their analytical evaluations have not been provided yet. To understand why these terms can have importance in computation, we recall the KS-DFT [or its origin, HF-SCF] formalism for electron-electron repulsion energy approximation with one-electron density as $V_{ee}\approx (1/2)\int\rho(\mathbf{r}_1)\rho(\mathbf{r}_2)r_{12}^{-1}d\mathbf{r}_1 d\mathbf{r}_2$ [or with normalized singe Slater determinant $S_0$ as $V_{ee}\approx (N(N-1)/2)\int S_0{}^*S_0 r_{12}^{-1}ds_1 d\mathbf{r}_1 \ldots ds_N d\mathbf{r}_N$], and the corresponding operator formalism, wherein $\rho$ is the sum of square of KS-MO's [or analogously with HF-SCF MO's], and the KS or HF-SCF MO's are the LCAO with GTO: These two (about the same, but not exactly the same) integrals for $V_{ee}$, which is one of the basic ideas in KS [or known as 2J(Coulomb integral)-K(exchange integral) in HF-SCF] formalism, account for almost the 99% of the Coulomb interaction energy (aside from basis set error and model error in $\rho$ [or S] itself), although the rest (called exchange-correlation energy, $E_{xc}$, for which e.g. the very weak $B_{Dirac}\int\rho^{4/3}d\mathbf{r}_1$ holds [or correlation energy, $E_{corr}$, for which e.g. the MP theory, the one of the earliest ones, has provided very remarkable approximation]) must also be approximated to reach chemical accuracy (1 kcal/mol). Among many-many ideas in the, as yet not-completely worked out, though very advanced correlation calculation, higher moments of inter-electronic distances, indicated in the title or Eq.2.4.3 have come up as candidate terms in the estimation.

We also mention, that instead of manipulating with the power of $r_{ij}$ in Eq.2.4.2, like in Eq.2.4.3, another algebraic way to use terms in correlation calculation is the moment expansion of $\rho$, as e.g. for the rough local moment expansion $V_{ee}\approx 2^{-1/3}(N-1)^{2/3}\int\rho(\mathbf{r}_1)^{4/3}d\mathbf{r}_1$ for main term itself, see review by Kristyan in ref. [2.4.7]. However, local moment expansion methods face to the problem of very slow convergence [2.4.8]. The weakness in it is the local operator vs. non-local operator, and it seems that a key to improve the existing Coulomb energy approximations in this directions is the use of e.g.

$$\{ \int[\rho(\mathbf{r}_1)]^p[\rho(\mathbf{r}_2)]^q r_{12}^{-1}d\mathbf{r}_1 d\mathbf{r}_2 \}^t \qquad \text{(Eq.2.4.4)}$$

non-local moment expansion for correlation effects (even for all kinds of approximation just mentioned above for $V_{ee}$ with a bit different parametrization for each), which is not considered yet in this literature [2.4.7]. A convenient immediate property of Eq.2.4.4 is that with GTO functions analytical evaluation is possible without any extra expressions if p and q are integer, since again, product of Gaussians is also Gaussians and analytical evaluation of Eq.2.4.2 is well known. (Also, Eq.2.4.4 is symmetric with the interchange of p and q, as well as proper choice of $t\neq 1$, Eq.2.4.4 can be made "scaling correct" depending on the p and the q, a useful property [2.4.2, 7, 9].) Taking things even a step further, one can combine Eqs.2.4.3 and 4 in relation to p, q, n and m and test the possible benefit in a correlation calculation.

Integrals in Eq.2.4.3 belong mathematically to the so called "explicitly correlated R12 theories of electron correlation", which bypass the slow convergence of conventional methods [2.4.1-2] by augmenting the traditional orbital expansions with a small number of terms that depend explicitly on the inter-electronic distance $r_{12}$. However, only approximate expressions are available for evaluation, for example, the equation, numbered 52 in ref. [2.4.10], suggests for the second one in Eq.2.4.3 that

$<ijm|r_{12}^{-1}r_{13}^{-1}|kml> \approx \Sigma_p <ij|r_{12}^{-1}|pm><pm|r_{12}^{-1}|kl>$, (Eq.2.4.5)

where the bracket notation [2.4.1-2] is used along without reducing product Gaussians to single Gaussians, as well as the GTO basis set {p} for expansion has to be a "good quality" for adequate approximation.

### 2.4.1.b. One-electron Coulomb integrals for nuclear-electron interactions with $R_{C1}^{-n}$

After mentioning a possible way above to correct for $V_{ee}$, wherein, for example, the HF-SCF or KS level ρ is expanded with a linear combination of Gaussians one can use a similar kind of power expansion for $V_{ne}$, that is, the term

$$\int G_{A1}R_{C1}^{-2}d\mathbf{r}_1 \quad \text{(Eq.2.4.6)}$$

similarly expands the opportunities to correct for $V_{ne}$. Recall [2.4.7] that (unlike the aforementioned HF-SCF and KS approximate expressions for $V_{ee}$, the) $V_{ne}= -\Sigma_{C=1,\dots,M}Z_C\int\rho(\mathbf{r}_1)R_{C1}^{-1}d\mathbf{r}_1$ is an exact equation for $V_{ne}$, the only error entering is that not the exact but HF-SCF, KS, etc. approximations are used for ρ in practice. By this reason the terms in Eq.2.4.6 has less importance to correct for $V_{ne}$ than Eq.2.4.3 to correct for $V_{ee}$. However, Eq.2.4.6 is still mathematically important, because it is a prerequisite (see below) to evaluate Eq.2.4.3 analytically, the aim and topic of this work. Analog expression of Eq.2.4.4 in relation to Eq.2.4.6 is the

$$\{ \int\rho^p R_{C1}^{-n}d\mathbf{r}_1 \}^t \quad \text{(Eq.2.4.7)}$$

which can be discussed analogously.

Furthermore, if derivatives appear, such as $\int(\partial\rho(\mathbf{r}_1)/\partial x_1)^p R_{C1}^{-n}d\mathbf{r}_1$, $\int(\partial\rho(\mathbf{r}_1)/\partial x_1)^p\rho(\mathbf{r}_2)^q\, r_{12}^{-n}d\mathbf{r}_1 d\mathbf{r}_2$ or many other algebraic possibilities (recall that derivatives of ρ are used frequently even by empirical reasons in DFT [2.4.3, 11], e.g. in the generalized gradient approximations), and ρ is given as linear combination of Gaussians, analytical evaluation of Eq.2.4.3 and Eq.2.4.6 are fundamental building blocks for analytical integral evaluation, since not only the products, but the derivatives of Gaussians in Eq.2.4.1 are Gaussians.

### 2.4.1.c More general one-electron and the mixed case two-electron Coulomb integrals with $R_{C1}^{-n}R_{D1}^{-m}$ and $R_{C1}^{-n}r_{12}^{-m}$, respectively

These cases come up not only mathematically after the above cases, but in computation for electronic structures as well. Not going into too much details, we outline one way only as example: Applying the Hamiltonian twice for the ground state wave function simply yields $H^2\Psi_0= E_{electr,0}H\Psi_0= E_{electr,0}^2\Psi_0$, or $<\Psi_0|H^2|\Psi_0>= E_{electr,0}^2$. The $H^2$ preserves the linearity and hermetic property from operator H, and if e.g. HF-SCF single determinant $S_0$ approximates $\Psi_0$ via variation principle from $<S_0|H|S_0>$, the approximation $(<S_0|H^2|S_0>)^{1/2}\approx E_{electr,0}$ is better than $<S_0|H|S_0>\approx E_{electr,0}$, coming from basic linear algebraic properties of linear operators for the ground state. However, $H^2$ yields very hectic terms, the $H_{ne}^2$, $H_{ne}H_{ee}$ and $H_{ee}^2$ products show up, for example, yielding Coulomb operators belonging to the types in the title. Using $<S_0|H^2S_0>= <HS_0|HS_0>$, the right side keeps the algorithm away from operators like $\nabla_1^2\, r_{12}^{-1}$ at least.

### 2.4.2.a One-electron spherical Coulomb integral for $R_{C1}^{-2}$

Now $R_{C1}\equiv|\mathbf{R}_C-\mathbf{r}_1|$ and $R_{P1}\equiv|\mathbf{R}_P-\mathbf{r}_1|$, and we evaluate the one-electron spherical Coulomb integral for $G_{P1}(p,0,0,0)= \exp(-p\, R_{P1}^2)$ in Eq.2.4.1 analytically, i.e. the

$$V_{P,C}^{(n)}\equiv \int_{(R3)} \exp(-p\, R_{P1}^2)\, R_{C1}^{-n}\, d\mathbf{r}_1\,, \quad \text{(Eq.2.4.8)}$$

for which n=1 is well known and 2 is a new expression below. The idea comes from the Laplace transformation for n= 1 and 2 respectively as

$$R_{C1}^{-1} = \pi^{-1/2} \int_{(-\infty,\infty)} \exp(-R_{C1}^2 t^2)dt , \quad \text{(Eq.2.4.9a)}$$

$$R_{C1}^{-2} = \int_{(-\infty,0)} \exp(R_{C1}^2 t)dt = \int_{(0,\infty)} \exp(-R_{C1}^2 t)dt , \quad \text{(Eq.2.4.9b)}$$

wherein notice the two different domains for integration. In this way (using 2.4.Appendixes 1-2 after the e.g. middle part in Eq.2.4.9b) the $V_{P,C}^{(2)} = \int_{(-\infty,0)}\int_{(R3)}\exp(-p\, R_{P1}^2)\exp(R_{C1}^2 t)\, d\mathbf{r}_1 dt = \int_{(-\infty,0)} \int_{(R3)}\exp(pt(p-t)^{-1}R_{CP}^2)\exp((t-p)R_{S1}^2)\, d\mathbf{r}_1 dt = \int_{(-\infty,0)} (\pi/(p-t))^{3/2} \exp(pt(p-t)^{-1}R_{CP}^2)dt$. Using $u:=t/(p-t)$ changes the domain t in $(-\infty,0) \rightarrow$ u in $(-1,0)$, $V_{P,C}^{(2)} = \pi^{3/2}p^{-1/2}\int_{(-1,0)} (u+1)^{-1/2}\exp(p\, R_{CP}^2 u)du$, and using $w:= (u+1)^{1/2}$ changes the domain u in $(-1,0) \rightarrow$ w in $(0,1)$ and yields

$$V_{P,C}^{(2)} = (2\pi^{3/2}/p^{1/2}) \int_{(0,1)} \exp(p\, R_{CP}^2 (w^2-1))dw = (2\pi^{3/2}/p^{1/2})e^{-v}F_0(-v) , \quad \text{(Eq.2.4.10)}$$

where $F_0(v)$ is Boys function with $v\equiv p\, R_{CP}^2$. For Eq.2.4.10 the immediate minor/major values come from $1\leq \exp(pR_{CP}^2 w^2) \leq \exp(v\equiv pR_{CP}^2)$ if $0\leq w\leq 1$ as

$$0 < \exp(-v) < [p^{1/2}/(2\pi^{3/2})]\, V_{P,C}^{(2)} < 1, \quad \text{(Eq.2.4.11)}$$

and for a comparison, we recall the well known expression for n=1

$$V_{P,C}^{(1)} = (2\pi/p) \int_{(0,1)} \exp(-p\, R_{CP}^2\, w^2)dw = (2\pi/p)F_0(v) \quad \text{(Eq.2.4.12)}$$

with immediate minor/major values

$$0 < \exp(-v) < [p/(2\pi)]\, V_{P,C}^{(1)} < 1. \quad \text{(Eq.2.4.13)}$$

Note that point $\mathbf{R}_S$ can be calculated by the m=2 case in 2.4.Appendix 2, but its particular value drops, because integral value in 2.4.Appendix 1 is invariant by shifting a Gaussian in R3 space. Eqs.2.4.11 and 13 tell that up to normalization factor with p, the $V_{P,C}^{(1)}$ and $V_{P,C}^{(2)}$ are in same range, roughly in (0,1). The ratio of the two is easily obtained when $R_{CP}=0$, then the integrands become unity, and

$$V_{P,C}^{(2)}(R_{CP}=0)/ V_{P,C}^{(1)}(R_{CP}=0)= (2\pi^{3/2}/p^{1/2})/(2\pi/p)= (\pi p)^{1/2} \quad \text{(Eq.2.4.14)}$$

as well as for n=1 and 2 the lim $V_{P,C}^{(n)}=0$ if $R_{CP}\rightarrow\infty$.

Note that the integral is the type $\int\exp(-w^2)dw$ in Eq.2.4.12, a frequent expression coming up in physics, but contrary, the $\int\exp(w^2)dw$ has come up in Eq.2.4.10. The latter is infinite on domain $(0,\infty)$, otherwise similar algebraic blocks have come up in Eqs.2.4.8-13 for n=1 vs. 2, which is not surprising; but, the evaluation of $F_0(v)$ differs significantly from $F_0(-v)$. Integration in Eq.2.4.12 can be related to the “erf” function (i.e. for $F_0(v>0)$) in a calculation which is standard in programming, but lacks analytical expression, as well as the “erf” is inbuilt function in program languages like FORTRAN. However, integration in Eq.2.4.10 cannot be related to any inbuilt function like “erf”, but its evaluation numerically belongs to standard devices, mainly because the integrand is a simple monotonic elementary function.

Note that, 1., The algebraic keys are in Eq.2.4.9 and 2.4.Appendix 2 to evaluate Eq.2.4.8 analytically - up to Gaussian function $\exp(\pm w^2)$ in the integrand. If not GTO but STO is used in Eq.2.4.1, i.e. not $R_{P1}^2$ but $R_{P1}$ shows up in the power of Eq.2.4.8, the evaluation for the corresponding integral in Eq.2.4.8 is far more difficult, stemming from the fact that the convenient device in 2.4.Appendix 2 cannot be used. A simple escape route is to use the approximation $\exp(-pR_{P1})\approx \Sigma_{(i)}c_i G_{P1}(a_i,0,0,0)$, which is well known in molecular structure calculations, see the idea of STO-3G basis sets and higher levels in which one does not even need many terms in the summation but, in fact in this way, one loses the desired complete analytical evaluation for the original integral $\int_{(R3)}\exp(-pR_{P1})R_{C1}^{-n}\, d\mathbf{r}_1$. 2., In Eq.2.4.9 the power correspondence in the integrand and integral value for n=1 vs. 2 is $R_{C1}^{-1} \leftrightarrow R_{C1}^2$ vs. $R_{C1}^{-2} \leftrightarrow R_{C1}^2$, what is the seed of trick for analytical evaluation, and may indicates the way for further generalizations. 3., Fast, accurate and fully numerical integration for one-electron Coulomb integrals in Eq.2.4.8 is available for any n≥1 integer and non-integer values of n, the general numerical integral scheme is widely used in DFT correlation calculations based on Voronoi polygons, Lebedev spherical integration and Becke’ scheme [2.4.12-21] in R3. However, this numerical process is definitely not applicable for two and three-electron Coulomb integrals in R6 or R9, respectively because it is extremely slow in computation; the reason being that the at least K=1000 points for numerical integration becomes $K^2$ or $K^3$, respectively, that is, the computation time is K or $K^2$ times longer, respectively.

### 2.4.2.b One-electron non-spherical Coulomb integral for $R_{C1}^{-2}$

If the more general $G_{P1}(p,nx,ny,nz)$ is used, Eq.2.4.8 generates the analytical evaluation as a seed, and no further trick needed than Eq.2.4.9, the only formula necessary is how to shift the center of polynomials (2.4.Appendix 3). We use the notations ${}^{full}V_{P,C}^{(n)}$ and $V_{P,C}^{(n)}$, the former stands for any (spherical and non-spherical, nx+ny+nz≥0) quantum number, while the letter denotes the simplest spherical (1s-like) case, nx=ny=nz=0. Before we outline the evaluation for

$${}^{full}V_{P,C}^{(2)} \equiv \int_{(R3)} G_{P1}(p,nx1,ny1,nz1)\, R_{C1}^{-2}\, d\mathbf{r}_1 , \qquad \text{(Eq.2.4.15)}$$

we should mention that Gaussians in Eq.2.4.1 are called, more precisely, "Cartesian Gaussian", and alternatively, the "Hermite Gaussians" have also been defined, see 2.4.Appendix 4. All integrands in Eqs.2.4.1-3 generate Gaussians, but Hermite Gaussians are special linear combination of Cartesian Gaussians owing good recurrence and overlap relations which are not reviewed here. 2.4.Appendix 2 shows that the product of two Cartesian Gaussians yields an overlap distribution, and the Gaussian product rule reduces two-center integrals to one-center integrals, (see how the $\mathbf{R}_S$ enters between Eq.2.4.9 and Eq.2.4.10). However, in the case of non-spherical (nx+ny+nz>0) Gaussians in Eq.2.4.1, large summation of Cartesian monomials is needed to evaluate the integration of non-spherical Cartesian Gaussians in Eq.2.4.15, after applying Appendices 2-3 to locate that $\mathbf{R}_S$ between $\mathbf{R}_P$ and $\mathbf{R}_C$. (Recall that the spherical nx+ny+nz=0 case in Eq.2.4.10 has only one compact term.) In practice, one expands Cartesian overlap distributions in Hermite Gaussians to evaluate ${}^{full}V_{P,C}^{(1)}$ analytically, and utilizes the simpler integration properties of Hermite Gaussians. Instead of recalling and using these relations, we yet use the former, since the summation is still compact; but in practice the latter is faster, tested and known for ${}^{full}V_{P,C}^{(1)}$. But of course, both yield the same final values.

In Eq.2.4.15 we use triple letters (nx1, etc.) which benefits in use below to distinguish between electrons 1 and 2. Using Eq.2.4.A.3.1 for POLY($x_1$,P,S,nx1), POLY($y_1$,P,S,ny1), POLY($z_1$,P,S,nz1) and Eq.2.4.9, the ${}^{full}V_{P,C}^{(2)} = \Sigma_1 \int_{(R3)}\int_{(-\infty,0)} (x_S-x_P)^{nx1-i1}(x_1-x_S)^{i1}\, (y_S-y_P)^{ny1-j1}(y_1-y_S)^{j1}\, (z_S-z_P)^{nz1-k1}(z_1-z_S)^{k1} \exp(-p\, R_{P1}^2)\exp(t\, R_{C1}^2)\, dt d\mathbf{r}_1$. Using Eq.2.4.A.2.3, $\exp(-p\, R_{P1}^2)\exp(t\, R_{C1}^2) = \exp(pt\, R_{CP}^2/(p-t))\exp((t-p)R_{S1}^2)$, where $x_S=(px_P-tx_C)/(p-t) \Rightarrow x_S-x_P=t(x_P-x_C)/(p-t)$ and so for x an y, so the integrand becomes $(t/(p-t))^{n1-m1}\, (x_P-x_C)^{nx1-i1}(x_1-x_S)^{i1}\, (y_P-y_C)^{ny1-j1}(y_1-y_S)^{j1}\, (z_P-z_C)^{nz1-k1}(z_1-z_S)^{k1}\, \exp(pt\, R_{CP}^2/(p-t))\, \exp((t-p)R_{S1}^2)$ with short hand abbreviations (for sum and multiplication operators)

$$\Sigma_1 \equiv \Sigma_{i1=0}^{nx1}\Sigma_{j1=0}^{ny1}\Sigma_{k1=0}^{nz1} \binom{nx1}{i1}\binom{ny1}{j1}\binom{nz1}{k1} \quad \text{for even i1, j1, k1 only} \qquad \text{(Eq.2.4.16a)}$$

$$n1 \equiv nx1+ny1+nz1 \qquad \text{(Eq.2.4.16b)}$$

$$m1 \equiv i1+j1+k1 \qquad \text{(Eq.2.4.16c)}$$

$$\Gamma_1 \equiv \Gamma((i1+1)/2)\, \Gamma((j1+1)/2)\, \Gamma((k1+1)/2) \qquad \text{(Eq.2.4.16d)}$$

$$D \equiv (x_P-x_C)^{nx1-i1}(y_P-y_C)^{ny1-j1}\, (z_P-z_C)^{nz1-k1} \qquad \text{(Eq.2.4.16e)}$$

which allows for integrating out for $\mathbf{r}_1$ via 2.4.Appendix 1, yielding ${}^{full}V_{P,C}^{(2)} = \Sigma_1\Gamma_1 D \int_{(-\infty,0)} (t/(p-t))^{n1-m1}\, (p-t)^{-(m1+3)/2} \exp(pt\, R_{CP}^2/(p-t))\, dt$. Important warning is that summation in Eq.2.4.16a is for even i1, j1, k1= 0,2,4,6,… only, coming from the property of odd powers in 2.4.Appendix 1. Using u:=t/(p-t) and thereafter $w^2$:=u+1 changes the domain t in $(-\infty,0) \rightarrow$ u in $(-1,0) \rightarrow$ w in (0,1), and one ends up with

$$^{full}V_{P,C}^{(2)}= 2\Sigma_1\Gamma_1 D\, p^{-(m1+1)/2} \int_{(0,1)} (w^2-1)^{n1-m1} w^{m1} \exp(p\, R_{CP}^2(w^2-1))\, dw \,. \quad \text{(Eq.2.4.17a)}$$

If n1=0, then Eq.2.4.17a reduces to Eq.2.4.10 as expected ($\Gamma(1/2)$ product provides the $\pi^{3/2}$). For example, in a full d-orbital case (nx1=ny1=nz1=2) Eq.2.4.17a has $2^3$= terms via Eq.2.4.16a for one primitive Gaussian in Eq.2.4.15, which is not so bad, and the integrand falls to more than one terms but, by using Hermite Gaussians, the calculation is more effective. However, Eq.2.4.17a is very compact mathematically and just embeded do-loops in programming. Furthermore, since m1 is always even via Eq.2.4.16a, it yields that integrand in Eq.2.4.17a is always linear combination of $w^{2L} \exp(p\, R_{CP}^2(w^2-1))$ for L=0,1,2,…, i.e. Boys function can be recalled again as in Eq.2.4.10, that is, $e^{-v}F_L(-v)$ with $v\equiv p\, R_{CP}^2$.

The expression for n=1 (in $R_{C1}^{-n}$) comes out in analogous way with the help of the known substitution $w^2:= t^2/(p+t^2)$, and the final integral is

$$^{full}V_{P,C}^{(1)}= 2p^{-1}\pi^{-1/2}\Sigma_1\Gamma_1 D\, p^{-m1/2}\int_{(0,1)}(-w^2)^{n1-m1} (1-w^2)^{m1/2} \exp(-p\, R_{CP}^2\, w^2)\, dw. \quad \text{(Eq.2.4.17b)}$$

Eq.2.4.17b reduces to Eq.2.4.12 if n1=0 in Eq.2.4.16b as expected, and since powers of $w^2$ appear, it makes the linear combination of Boys functions $F_L(v)$ with $v\equiv p\, R_{CP}^2$. De-convolution of Boys functions from $F_L(\pm v)$ to $F_0(\pm v)$ can be found in 2.4.Appendix 5. Note that D in Eq.2.4.16e dynamically provides signs.

**2.4.2.c One-electron spherical Coulomb integral for $R_{C1}^{-n}R_{D1}^{-m}$ with n, m=1,2**

As in Eq.2.4.10 vs. Eq.2.4.17, the non-spherical case is an extension of simplest spherical case with a hectic but obvious systematic summation with respect to powers (or quantum numbers) nx, ny and nz, as well as polynomials of spatial coordinates, however, the algebraic seed is Eq.2.4.10 itself. The summation is the same kind in this section and in the rest of the article too, so to save space, we derive the simplest spherical cases only.

We evaluate analytically the one-electron spherical Coulomb integral

$$V_{P,CD}^{(n,m)}\equiv \int_{(R3)}\exp(-pR_{P1}^2)R_{C1}^{-n}R_{D1}^{-m}d\mathbf{r}_1 \,. \quad \text{(Eq.2.4.18)}$$

Depending on the value of (n,m), the proper one of Eq.2.4.9 must be picked. Let us take the example of (n,m)= (1,2). Using Eq.2.4.9a and e.g. the far right side in Eq.2.4.9b for D as $R_{D1}^{-2}= \int_{(0,\infty)} \exp(-R_{D1}^2 u)du$, as well as 2.4.Appendixes 1-2, $V_{P,CD}^{(1,2)}= (\pi^{-1/2})\int_{t=(-\infty,\infty)}\int_{u=(0,\infty)}[\int_{(R3)}\exp(-g\, R_{W1}^2)d\mathbf{r}_1] \exp(-f/g)dudt= (\pi^{-1/2})\int_{t=(-\infty,\infty)}\int_{u=(0,\infty)} [(\pi/g)^{3/2}] \exp(-f/g)dudt$, the location of $\mathbf{R}_W$ is irrelevant again, and finally

$$V_{P,CD}^{(1,2)}= \pi\int_{t=(-\infty,\infty)}\int_{u=(0,\infty)}g^{-3/2}\exp(-f/g)dudt \quad \text{(Eq.2.4.19a)}$$

$$g\equiv p + t^2 +u \quad \text{(Eq.2.4.19b)}$$

$$f\equiv p\, t^2 R_{PC}^2 +p\, u\, R_{PD}^2 +u\, t^2 R_{CD}^2 \,. \quad \text{(Eq.2.4.19c)}$$

Like for Eq.2.4.10 or Eq.2.4.12, by simple substitution one can end up with $\int_{(0,1)}\int_{(0,1)}(\dots)dtdu$ integration on unit square. This integration can be done numerically, see section 2.4.3.d. The algorithm is straightforward for other cases of (n,m).

**2.4.3. Two and three-electron spherical Coulomb integrals**

**2.4.3.a Two-electron spherical Coulomb integral for $r_{12}^{-2}$, the (n,m)=(2,0) or (0,2) case**

For $G_{P1}(p,0,0,0)$ and $G_{Q2}(q,0,0,0)$ in Eq.2.4.1 we evaluate the two-electron spherical Coulomb integral analytically, i.e.:

$$V_{PQ}^{(n)}\equiv \int_{(R6)} \exp(-p\, R_{P1}^2) \exp(-q\, R_{Q2}^2)\, r_{12}^{-n}\, d\mathbf{r}_1 d\mathbf{r}_2, \quad \text{(Eq.2.4.20)}$$

for which n=1 is well known and 2 is a new expression below. Re-indexing Eq.2.4.10 for C→2 and R→r (i.e. electron 2 takes the role of nucleus C algebraically) yields

$$V_{P,C}^{(2)} = \int_{(R3)} \exp(-p\, R_{P1}^2)r_{12}^{-2}\, d\mathbf{r}_1 = (2\pi^{3/2}/p^{1/2}) \int_{(0,1)} \exp(p\, R_{P2}^2 (w^2-1))dw \,, \quad \text{(Eq.2.4.21)}$$

and similarly, Eq.2.4.12 yields

$$V_{P,C}^{(1)} = \int_{(R3)} \exp(-p\, R_{P1}^2)r_{12}^{-1}\, d\mathbf{r}_1 = (2\pi/p) \int_{(0,1)} \exp(-p\, R_{P2}^2\, w^2)dw \,, \quad \text{(Eq.2.4.22)}$$

which – depending on n - integrates $\mathbf{r}_1$ out from Eq.2.4.20. Eq.2.4.21 must be used for the new one only, yielding $V_{PQ}^{(2)} = (2\pi^{3/2}/p^{1/2})\int_{(0,1)}\int_{(R3)} \exp(pR_{P2}^2(w^2-1) -qR_{Q2}^2)d\mathbf{r}_2dw$. Recalling 2.4.Appendixes 1-2 yields $V_{PQ}^{(2)} = (2\pi^{3/2}/p^{1/2})\int_{(0,1)} \exp(pq(w^2-1)R_{PQ}^2/(p-pw^2+q)) \int_{(R3)}\exp(-(p-pw^2+q)R_{S2}^2)d\mathbf{r}_2dw = (2\pi^3/p^{1/2})\int_{(0,1)} (p-pw^2+q)^{-3/2} \exp(pq(w^2-1)R_{PQ}^2/(p-pw^2+q)) dw$, which yet does not show that p and q are equivalent, but they are, and the location of $\mathbf{R}_S$ is irrelevant again. At this point elementary numerical integration can be performed again, or more elegantly using $u:= (1-w^2)/(p-pw^2+q)$ thereafter $w^2:= 1-u(p+q)$ changes the domain w in (0,1) $\rightarrow$ u in $(0,(p+q)^{-1})$ $\rightarrow$ w in (0,1). Finally, with $v\equiv pqR_{PQ}^2/(p+q)$

$$V_{PQ}^{(2)} = 2\pi^3(pq)^{-1/2}(p+q)^{-1}\int_{(0,1)} \exp(v(w^2-1))dw = (2\pi^3(pq)^{-1/2}(p+q)^{-1})e^{-v}F_0(-v) \quad \text{(Eq.2.4.23)}$$

where $F_0(v)$ is the Boys function, and the immediate minor/major values come from $1\leq \exp(vw^2) \leq \exp(v)$ if $0\leq w\leq 1$ as

$$0 < \exp(-v) < [(pq)^{1/2}(p+q)/(2\pi^3)]V_{PQ}^{(2)} < 1. \quad \text{(Eq.2.4.24)}$$

For comparison, we recall the well known expression for n=1 as

$$V_{PQ}^{(1)} = (2\pi^{5/2}/(pq)) \int_{(0,c)} \exp(-pqR_{PQ}^2 w^2)dw \quad \text{(Eq.2.4.25)}$$

with $c\equiv(p+q)^{-1/2}$ in the integration domain, and it can be expressed with Boys or with "erf" functions, and the immediate minor/major values (from w:=c/0 in the integrand)

$$0 < \exp(-v) < [pq(p+q)^{1/2}/(2\pi^{5/2})]V_{PQ}^{(1)} < 1. \quad \text{(Eq.2.4.26)}$$

In Eqs.2.4.23-26 the expressions are symmetric to interchange of p and q, as expected. The ratio of the two is easily obtained when $R_{PQ}=0$, then the integrands become unity, and

$$V_{PQ}^{(2)}(R_{PQ}=0)/ V_{PQ}^{(1)}(R_{PQ}=0)= (2\pi^3(pq)^{-1/2}(p+q)^{-1}/(2c\pi^{5/2}/(pq))= (\pi pq/(p+q))^{1/2} \quad \text{(Eq.2.4.27)}$$

as well as for n=1 and 2 the lim $V_{PQ}^{(n)}=0$ if $R_{PQ}\rightarrow\infty$.

**2.4.3.b Two-electron spherical Coulomb integral for the mixed term $R_{C1}^{-n}r_{12}^{-m}$ with n, m=1,2**

Taking n=m=1 as an example: $\int_{(R6)}\exp(-pR_{P1}^2) \exp(-qR_{Q2}^2) R_{C1}^{-1}r_{12}^{-1}d\mathbf{r}_1d\mathbf{r}_2= \int_{(R3)}\exp(-pR_{P1}^2) [\int_{(R3)}\exp(-qR_{Q2}^2) r_{12}^{-1} d\mathbf{r}_2] R_{C1}^{-1}d\mathbf{r}_1 = \int_{(R3)}\exp(-pR_{P1}^2) [(2\pi/q) \int_{(0,1)} \exp(-qR_{Q1}^2 u^2)du] R_{C1}^{-1}d\mathbf{r}_1 = (2\pi/q)\int_{(0,1)}\int_{(R3)}\exp(-pR_{P1}^2-qR_{Q1}^2u^2) R_{C1}^{-1}d\mathbf{r}_1du = (2\pi^{1/2}/q) \int_{u=(0,1)}\int_{t=(-\infty,\infty)} \int_{(R3)}\exp(-pR_{P1}^2-qR_{Q1}^2u^2-R_{C1}^2t^2) d\mathbf{r}_1dtdu = (2\pi^{1/2}/q) \int_{(0,1)}\int_{(-\infty,\infty)} [\int_{(R3)}\exp(-g R_{W1}^2) d\mathbf{r}_1] \exp(- f/g) dtdu= (2\pi^{1/2}/q) \int_{(0,1)}\int_{(-\infty,\infty)} [\pi/g]^{3/2} \exp(-f/g) dtdu$, using Eq.2.4.31 below (for parameter m, instead of n), Eq.2.4.9a and 2.4.Appendixes 1-2, as well as the location of $\mathbf{R}_W$ drops again. Finally

$$\int_{(R6)}\exp(-pR_{P1}^2)\exp(-qR_{Q2}^2)R_{C1}^{-1}r_{12}^{-1}d\mathbf{r}_1d\mathbf{r}_2=(2\pi^2/q)\int_{u=(0,1)}\int_{t=(-\infty,\infty)} g^{-3/2}\exp(-f/g)\, dtdu \quad \text{(Eq.2.4.28a)}$$

$$f\equiv pqR_{PQ}^2u^2+pR_{PC}^2t^2+qR_{QC}^2u^2t^2 \quad \text{(Eq.2.4.28b)}$$

$$g\equiv p+qu^2+ t^2 \quad \text{(Eq.2.4.28c)}$$

wherein u and t are not equivalent. As with Eq.2.4.10 or Eq.2.4.12, the range $\int_{(-\infty,\infty)}$ for t can be converted to $\int_{(0,1)}$ by simple substitution to end up with $\int_{(0,1)}\int_{(0,1)}(\ldots)dtdu$ integration on unit square. Alternatively, $(2\pi/q)\int_{(0,1)} [\int_{(R3)}\exp(-pR_{P1}^2 -qR_{Q1}^2u^2) R_{C1}^{-1}d\mathbf{r}_1] du= (2\pi/q) \int_{(0,1)} [\int_{(R3)}\exp(-(p+qu^2) R_{W1}^2) R_{C1}^{-1}d\mathbf{r}_1] \exp(-pq R_{PQ}^2u^2 /(p + qu^2)) du= (2\pi/q) \int_{(0,1)} [(2\pi/(p+qu^2))\int_{(0,1)}\exp((p+qu^2)R_{WC}^2t^2)dt] \exp(-pqR_{PQ}^2u^2/(p+qu^2)) du$ , using Eq.2.4.31 below, 2.4.Appendix 2 and Eq.2.4.12 for the square bracket yielding immediately $\int_{(0,1)}\int_{(0,1)}(\ldots)dtdu$, wherein $\mathbf{R}_W= (p\mathbf{R}_P+qu^2\mathbf{R}_Q)/(p+qu^2)$ has a different role than above and does not drop. Finally, by using the Boys function

$$\int_{(R6)}\exp(-pR_{P1}^2)\exp(-qR_{Q2}^2)R_{C1}^{-1}r_{12}^{-1}d\mathbf{r}_1d\mathbf{r}_2=(4\pi^2/q)\int_{(0,1)}F_0(gR_{WC}^2)g^{-1}\exp(-f/g)du \quad \text{(Eq.2.4.29a)}$$

$$f\equiv pqR_{PQ}^2u^2 \quad \text{(Eq.2.4.29b)}$$

$$g\equiv p+qu^2 . \quad \text{(Eq.2.4.29c)}$$

Note that $R_{WC}$ in $F_0$ in Eq.2.4.29a depends on u as $gR_{WC}^2= (p+qu^2)|\mathbf{R}_W-\mathbf{R}_C|^2= |p\mathbf{R}_P+qu^2\mathbf{R}_Q - g\mathbf{R}_C|^2$. The $w^2:= u^2/(p+qu^2)$, changing the domain u in (0,1) $\rightarrow$ w in $(0,(p+q)^{-1/2})$, reduces the

exponential part of the integrand in Eq.2.4.29a to $\exp(-pqR_{PQ}^2\, w^2)$, but the algebraic complexity becomes even worse in the other terms in the integrand. Eq.2.4.29 vs. Eq.2.4.28 shows us something about the two dimensional version of the Boys function, see section 2.4.3.d, i.e. how the two dimensional integral in Eq.2.4.28 can be reduced to one dimensional, although the Boys functions still as a notation for non-analytic integration, so Eq.2.4.29 is just an "embedding" with respect to Eq.2.4.28. Integrations in Eqs.2.4.28 and 29 can be done numerically, see section 2.4.3.d. The algorithm is straightforward for other cases of (n,m).

### 2.4.3.c Three-electron spherical Coulomb integral for $r_{12}^{-n}r_{13}^{-m}$ with n,m=1,2

For three totally different $G_{Ai}(a,0,0,0)$ in Eq.2.4.1 we evaluate the two-electron spherical Coulomb integral analytically as follows: the

$$V_{PQS}^{(n,m)} \equiv \int_{(R9)} \exp(-p\, R_{P1}^2) \exp(-q\, R_{Q2}^2) \exp(-s\, R_{S3}^2)\, r_{12}^{-n} r_{13}^{-m}\, d\mathbf{r}_1 d\mathbf{r}_2 d\mathbf{r}_3\,. \quad \text{(Eq.2.4.30)}$$

Eqs.2.4.21 and 22 provide the key substitutions for integrating out with $\mathbf{r}_2$ and $\mathbf{r}_3$, because the integrand can be separated for electrons 2 and 3 as $\exp(-pR_{P1}^2)\, [\exp(-qR_{Q2}^2) r_{12}^{-n}]\, [\exp(-sR_{S3}^2) r_{13}^{-m}]$. For example, for n=m=1, Eq.2.4.22 must be applied twice for $\mathbf{r}_2$ and $\mathbf{r}_3$ with index change (1,2)→(2,1) and (1,2)→(3,1), respectively, since $\mathbf{r}_1$ is the "common variable" in the denominator:

$$V_Q^{(n=1)} = \int_{(R3)} \exp(-qR_{Q2}^2) r_{12}^{-1} d\mathbf{r}_2 = (2\pi/q) \int_{(0,1)} \exp(-qR_{Q1}^2\, u^2) du = (2\pi/q) F_0(qR_{Q1}^2), \quad \text{(Eq.2.4.31)}$$

$$V_S^{(m=1)} = \int_{(R3)} \exp(-sR_{S3}^2) r_{13}^{-1} d\mathbf{r}_3 = (2\pi/s) \int_{(0,1)} \exp(-sR_{S1}^2\, t^2) dt = (2\pi/s) F_0(sR_{S1}^2). \quad \text{(Eq.2.4.32)}$$

Eqs.2.4.30-32 and 2.4.Appendixes 1-2 yield $(qs/(4\pi^2)) V_{PQS}^{(1,1)} = \int_{(0,1)}\int_{(0,1)}\int_{(R3)} \exp(-pR_{P1}^2 - qR_{Q1}^2 u^2 - sR_{S1}^2 t^2)\, d\mathbf{r}_1 du dt = \int_{(0,1)}\int_{(0,1)} [\int_{(R3)} \exp(-g\, R_{W1}^2)\, d\mathbf{r}_1] \exp(-f/g) du dt = \int_{(0,1)}\int_{(0,1)} [\pi/g]^{3/2} \exp(-f/g) du dt$, and finally

$$V_{PQS}^{(1,1)} = (4\pi^{7/2}/(qs)) \int_{(0,1)}\int_{(0,1)} g^{-3/2} \exp(-f/g) du dt \quad \text{(Eq.2.4.33a)}$$

$$f \equiv pqR_{PQ}^2 u^2 + psR_{PS}^2 t^2 + qsR_{QS}^2 u^2 t^2\,, \quad \text{(Eq.2.4.33b)}$$

$$g \equiv p + qu^2 + st^2\,. \quad \text{(Eq.2.4.33c)}$$

The location of point $\mathbf{R}_W$ is irrelevant again in the case of 1s-like functions. This integration can be done numerically, see section 2.4.3.d, which is still more stable and more reliable than Eq.2.4.5 because the latter is basis set choice dependent and much more complex. In Eq.2.4.33 the q and s are equivalent, but they are not equivalent with the role of p, as expected, coming from the role of electron 1 vs. {2 and 3} in Eq.2.4.30. For n and/or m=2 cases not Eq.2.4.22 but Eq.2.4.21 must be applied analogously to evaluate Eq.2.4.30, the algorithm is straightforward again.

Inclusion of the Boys function can come if the expressions with t and u are not used from Eqs.2.4.31-32, but instead the far right sides with Boys functions yielding $V_{PQS}^{(1,1)} = (4\pi^2/(qs)) \int_{(R3)} F_0(qR_{Q1}^2)\, F_0(sR_{S1}^2) \exp(-pR_{P1}^2) d\mathbf{r}_1$. For this we have not used 2.4.Appendix 1 yet. As well as this, the Boys function shows up in its integrand as in Eq.2.4.29, and its argument depends on electron coordinate. At this point the aforementioned accurate DFT numerical integration [2.4.12-21] can be used again, since the space R9 in Eq.2.4.30 has been reduced to R3. However, to develop this further, analytically, one should use the definition of the Boys function leading the equation back to Eq.2.4.33 to be tractable.

The way to Eq.2.4.33 was to apply Eqs.2.4.31-32, then 2.4.Appendixes 1-2, yielding two dimensional integral on the unit square. Another way, analogous to Eq.2.4.29 yielding one dimensional integral on the unit segment is to apply only Eq.2.4.31 and not Eq.2.4.32 or vice versa, then 2.4.Appendixes 1-2, and then Eq.2.4.25: It yields $(q/(2\pi)) V_{PQS}^{(1,1)} = \int_{(0,1)} [\int_{(R6)} \exp(-pR_{P1}^2 - qR_{Q1}^2 u^2) \exp(-sR_{S3}^2)\; r_{13}^{-1} d\mathbf{r}_1 d\mathbf{r}_3]\, du = \int_{(0,1)} [(2\pi^{5/2}/((p+qu^2)s)) \int_{(0,c)} \exp(-(p+qu^2)sR_{VS}^2$ $w^2)dw]\ \exp(-pqu^2R_{PQ}^2/(p+qu^2))du$, where $\mathbf{R}_V \equiv (p\mathbf{R}_P + qu^2\mathbf{R}_Q)/(p+qu^2)$ and does not drop, since it depends on u. Finally,

$$V_{PQS}^{(1,1)} = (4\pi^{7/2}/(qs)) \int_{(0,1)} h(u)\, g^{-1}\exp(-f/g)\, du \qquad \text{(Eq.2.4.34a)}$$

$$h(u) \equiv \int_{(0,c)}\exp(-g\, s\, R_{VS}^2\, w^2)dw \qquad \text{(Eq.2.4.34b)}$$

$$c \equiv (g+s)^{-1/2} \qquad \text{(Eq.2.4.34c)}$$

$$f \equiv pqR_{PQ}^2u^2 \qquad \text{(Eq.2.4.34d)}$$

$$g \equiv p+qu^2\ . \qquad \text{(Eq.2.4.34e)}$$

Eq.2.4.34 is a one dimensional integral in contrast to the two dimensional integrals in Eq.2.4.33, both yield the same value for $V_{PQS}^{(1,1)}$, of course, as well as this h(u) in Eq.2.4.34b is the pre-stage of Boys function $F_0$ as in Eq.2.4.25. Here again as in section 2.4.3.b, Eq.2.4.34 can be considered as the two dimensional version of Boys function wherein a one dimensional Boys function is in the integrand. Again, Eq.2.4.34 is only the "embedding" form of Eq.2.4.33. Section 2.4.1.c outlines a way how Coulomb operator $r_{12}^{-n}r_{13}^{-m}$ can come up; see its cardinality in 2.4.Appendix 6.

### 2.4.3.d The two dimensional Boys function, its pre-equation and integration

This case comes up if a spatial coordinate appears "twice", like electron 1 in the main title of this work for n, m>0, see Eqs.2.4.19, 28 and 33. As in Eqs.2.4.10, 12, 23 or 25 the $\int_{(0,1)}(\ldots)dt$, now again by simple substitution one can end up with $\int_{(0,1)}\int_{(0,1)}(\ldots)dtdu$ integration on unit square if necessary, which can be done numerically with a simple standard device on the unit square. These two dimensional integrals can be developed further, like the one dimensional integral in Eqs.2.4.10, 12, 23, 25, see Eq.2.4.28 vs. Eq.2.4.29 and Eq.2.4.33 vs. Eq.2.4.34, as examples. It yields the extension of the one dimensional Boys function to its two dimensional version (Eqs.2.4.29 or 34), which is not worked out and analyzed yet in the literature and will be looked at in a separate work. Note the close algebraic similarity or in fact the same type in Eqs.2.4.19, 28 and 33.

If we consider the right hand side of Eq.2.4.29a or Eq.2.4.34a as a kind of two dimensional Boys function, one can see that a one dimensional Boys function appears in its integrand. We draw attention to the fact, that at the beginning, i.e. in "seed equations" Eqs.2.4.10 and 12 we obtained the one dimensional Boys function $F_0$ via the term $g^{-3/2}\exp(-f/g)$ in the integrand as a pre-equation, (recall the derivation in middle stage e.g. as $V_{P,C}^{(2)} = \pi^{3/2}\int_{(-\infty,0)}g^{-3/2}\exp(f/g)dt$ with $f \equiv pR_{CP}^2t$ and $g \equiv p-t$), and when the two dimensional cases came up, the same term showed up in the integrand again, but instead of function set {f(t), g(t)}, the {f(u,t), g(u,t)}, see Eqs.2.4.19, 28 and Eq.2.4.33. The $g^{-3/2}\exp(f/g)$ is the core part of integrands for all cases in the main title of this work. Finer property is that, $f=f((-u)^K,(-t)^L)$ and $g=g((-u)^K,(-t)^L)$ are 2nd and 1st order polynomials, respectively, with respect to $(-u)^K$ and $(-t)^L$, where K, L = 1 or 2; wherein the middle part of Eq.2.4.9b has been used, alternatively, with the far right side of Eq.2.4.9b the $-u \rightarrow u$ and $-t \rightarrow t$ transformations should be done in this sentence. The K, L= 1 generates $\exp(w^2)$, while the 2 generates $\exp(-w^2)$ type Gaussians in the integrand.

### 2.4.Conclusions

Analytical evaluation of Coulomb one-electron integral, $\int_{(R3)} \exp(-pR_{P1}^2)R_{C1}^{-n}\, d\mathbf{r}_1$, has yielded $(2\pi^{3/2}/p^{1/2})e^{-v}F_0(-v)$ for n=2 in comparison to the known $(2\pi/p)F_0(v)$ for n=1, where $F_0$ is the Boys function with $v \equiv pR_{CP}^2$, and these expressions generate the formulas not only for higher quantum numbers (non-spherical or nx+ny+nz > 0 cases), but for two and three-electron Coulomb integrals as well, as indicated in the title. The equations derived help to evaluate the important Coulomb integrals $\int\rho(1)R_{C1}^{-n}R_{D1}^{-m}d\mathbf{r}_1$, $\int\rho(1)\rho(2)R_{C1}^{-n}r_{12}^{-m}d\mathbf{r}_1d\mathbf{r}_2$ and $\int\rho(1)\rho(2)\rho(3)r_{12}^{-n}r_{13}^{-m}d\mathbf{r}_1d\mathbf{r}_2d\mathbf{r}_3$ for n, m=0, 1, 2 in relation to powers of distances among the elements in the set of electrons and nuclei.

**2.4.Appendix 1:** For m= 1 and 2, the $\int_{(0,\infty)} x^n \exp(-ax_1^m)dx_1= \Gamma[(n+1)/m]/(m\, a^{(n+1)/m})$ holds for a>0. If m=2 and n=0 $\Rightarrow \int_{(R3)}\exp(-ar_1^2)d\mathbf{r}_1= (\int_{(-\infty,\infty)}\exp(-ax_1^2)dx_1)^3=(\pi/a)^{3/2}$. If m=2 $\Rightarrow \int_{(-\infty,\infty)} x^n \exp(-ax_1^2)dx_1= \Gamma[(n+1)/2]/a^{(n+1)/2}$ for even n, but zero if n is odd. The gamma function is $\Gamma[n+1]= n!$ for n=0,1,2,…, with $\Gamma[1/2]= \pi^{1/2}$ and $\Gamma[n+1/2]= 1x3x5x…(2n-1)\,\pi^{1/2}/2^n$ for n=1,2,… . The $\mathrm{erf}(x)\equiv 2\pi^{-1/2}\int_{(0,x)} \exp(-w^2)dw$, for which $\mathrm{erf}(\infty)=1$.

**2.4.Appendix 2**: The product of two Gaussians, $G_{J1}(p_J,0,0,0)$ with J=1,…,m=2 is another Gaussian centered somewhere on the line connecting the original Gaussians, but a more general expression for m>2 comes from the elementary

$$\Sigma_J\, p_J\, R_{J1}^2 = (\Sigma_J\, p_J)\, R_{W1}^2 + (\Sigma_J\Sigma_K\, p_J\, p_K\, R_{JK}^2)/(2\Sigma_J\, p_J) \qquad \text{(Eq.2.4.A.2.1)}$$

$$\mathbf{R}_W \equiv (\Sigma_J\, p_J\, \mathbf{R}_J)/(\Sigma_J\, p_J) \qquad \text{(Eq.2.4.A.2.2)}$$

where $\Sigma_{J \text{ or } K}\equiv \Sigma_{(J \text{ or } K=1 \text{ to } m)}$ and $R_{J1} \equiv |\mathbf{R}_J-\mathbf{r}_1|$ for $\exp(\Sigma_J\, c_J)= \Pi_{(J=1 \text{ to } m)}\exp(c_J)$, keeping in mind that $R_{JJ}=0$, and the m centers do not have to be collinear. For m=2, this reduces to

$$p\, R_{P1}^2 + q\, R_{Q1}^2 = (p+q)\, R_{W1}^2 + pqR_{PQ}^2/(p+q) \qquad \text{(Eq.2.4.A.2.3)}$$

yielding the well known and widely used

$$G_{P1}(p,0,0,0)\, G_{Q1}(q,0,0,0)= G_{W1}(p+q,0,0,0)\exp(-pqR_{pq}^2/(p+q))\, . \qquad \text{(Eq.2.4.A.2.4)}$$

We also need the case m=3, which explicitly reads as

$$p\, R_{P1}^2 + q\, R_{Q1}^2 + s\, R_{S1}^2 = (p+q+s)\, R_{W1}^2 + (pqR_{PQ}^2+psR_{PS}^2+qsR_{QS}^2)/(p+q+s)\, . \qquad \text{(Eq.2.4.A.2.5)}$$

Only the $G_{W1}(p+q+s,0,0,0)$ depends on electron coordinate $\mathbf{r}_1$ in Eq.2.4.A.2.4-5, not the other multiplier, indicating that the product of Gaussians decomposes to (sum of) individual Gaussians, (s=0 reduces Eq.2.4.A.2.5 to Eq.2.4.A.2.4).

**2.4.Appendix 3:** Given a single power term polynomial at $\mathbf{R}_P$, we need to rearrange or shift it to a given point $\mathbf{R}_S$. For variable x, this rearrangement is $(x-x_P)^n= \Sigma_{i=0 \text{ to } n}\, c_i\, (x-x_S)^i$, which can be solved systematically and immediately for $c_i$ by the consecutive equation system obtained from the 0,1,…$n^{th}$ derivative of both sides at x:= $x_S$, yielding

$$\mathrm{POLY}(x,P,S,n) \equiv (x-x_P)^n= \Sigma_{i=0 \text{ to } n}\, ({}^n{}_i)(x_S-x_P)^{n-i}\, (x-x_S)^i\ , \qquad \text{(Eq.2.4.A.3.1)}$$

where $({}^n{}_i)=n!/(i!(n-i)!)$. If $x_S=0$, it reduces to the simpler well known binomial formula as $(x-x_P)^n= \Sigma_{i=0 \text{ to } n}\, ({}^n{}_i)(-x_P)^{n-i}x^i$.

**2.4.Appendix 4:** The Hermite Gaussians are defined as

$$H_{Ai}(a,t,u,v)\equiv (\partial/\partial R_{Ax})^t(\partial/\partial R_{Ay})^u(\partial/\partial R_{Az})^v\exp(-a|\mathbf{r}_i-\mathbf{R}_A|^2)\, , \qquad \text{(Eq.2.4.A.4.1)}$$

and $H_{Ai}(a,2,0,0)= (\partial/\partial R_{Ax})^2\exp(-a\, R_{Ai}^2)= (\partial/\partial R_{Ax})[-2a\, (R_{Ax} -x_i)\, \exp(-a\, R_{Ai}^2)]= -2a\, \exp(-a\, R_{Ai}^2)+ 4a^2(R_{Ax} -x_i)^2\exp(-a\, R_{Ai}^2)= -2aG_{Ai}(a,0,0,0)+ 4a^2G_{Ai}(a,2,0,0)$ is an example that Hermite Gaussians are linear combination of Cartesian Gaussians.

**2.4.Appendix 5:** De-convolution of Boys functions from $F_L(v)\equiv \int_{(0,1)} \exp(-vt^2)t^{2L}dt$ to $F_0(v)= \int_{(0,1)} \exp(-vt^2)dt$ for v>0 and v≤0 comes from the help of partial integration ($\int f'g=[fg]-\int fg'$) on interval [0,1] with $f'=t^M$, M≠-1 and $g=\exp(-vt^2)$, and K:=M+2 thereafter. After elementary calculus:

$$2v\int_{(0,1)}t^K \exp(-vt^2)dt = (K-1)\int_{(0,1)}t^{K-2} \exp(-vt^2)dt - \exp(-v) \qquad \text{(Eq.2.4.A.5.1)}$$

for K=0,-1, ±2, ±3, ±4,…, i.e. any integer except 1, and v is any real number, i.e. v>0 and v≤0. (For K=1 the $2v\int_{(0,1)} t\exp(-vt^2)dt = 1-\exp(-v)$ by $\int g'\exp(g(t))dt=\exp(g(t))$.) In Boys functions the K=2L ≥0 is even, so K=1 is jumped, and with K:=2L+2 Eq.2.4.A.5.1 yields

$$2vF_{L+1}(v)= (2L+1)F_L(v) - \exp(-v) . \qquad \text{(Eq.2.4.A.5.2)}$$

The value of L recursively goes down to zero, and the value of $F_0(v)$ is needed only at the end. The v=0 case is trivial and the v>0 is well known in the literature but, the v<0 cases are also needed for cases described in the main title of this work.

**2.4.Appendix 6:** The cardinality in the set generated by electron-electron repulsion operator $H_{ee}^2= (\Sigma_{i=1..N} \Sigma_{j=i+1...N} r_{ij}^{-1})^2$ comes from elementary combinatorics. $H_{ee}$ contains $\binom{N}{2}=N(N-1)/2$ and $H_{ee}^2$ contains $N^2(N-1)^2/4$ terms. In relation to integration with single Slater determinant, it contains three kinds of terms: $r_{12}^{-2}$, $r_{12}^{-1}r_{13}^{-1}$ and $r_{12}^{-1}r_{34}^{-1}$ as

$$\langle S^*|H_{ee}^2|S\rangle= \binom{N}{2}\{\langle S^*|r_{12}^{-2}|S\rangle +2(N-2)\langle S^*|r_{12}^{-1}r_{13}^{-1}|S\rangle +\binom{N-2}{2}\langle S^*|r_{12}^{-1}r_{34}^{-1}|S\rangle\} . \qquad \text{(Eq.2.4.A.6.1)}$$

The control sum $\binom{N}{2} +2(N-2)\binom{N}{2} + \binom{N}{2}\binom{N-2}{2} = N^2(N-1)^2/4$ holds, as well as the magnitude of cardinality of individual terms on the right in Eq.2.4.A.6.1 are $N^2$, $N^3$ and $N^4$, respectively.

**2.4.Common notations, abbreviations and definitions** (watch for upper/lower cases in definitions for distances distinguishing between electrons and nuclei)

CI = configuration interactions

DFT = density functional theory

f, g= functions with different variables in integrands

$F_L(v) \equiv \int_{(0,1)} \exp(-vt^2)\, t^{2L}\, dt$, the Boys function, L=0,1,2,…

GTO = primitive Gaussian-type atomic orbital, the $G_{Ai}(a,nx,ny,nz)$ in Eq.2.4.1

$H\equiv H_\nabla+H_{ne}+H_{ee}$= non-relativistic electronic Hamiltonian for the sum of kinetic motion, and nuclear-electron and electron-electron interactions, respectively

HF-SCF = Hartree-Fock self-consistent field

KS = Kohn-Sham

LCAO = linear combination of atomic orbitals

MO = molecular orbital

MP = Møller-Plesset

N = number of electrons in the molecular system

$Q_i$ = charge of classical particle i

R3 = 3 dimension spatial space as domain for $\int_{(R3)}\ldots d\mathbf{r}_1\equiv\int_{(-\infty,\infty)}\int_{(-\infty,\infty)}\int_{(-\infty,\infty)}\ldots dx_1dy_1dz_1$

R6= R3xR3= 6 dimension domain for $\int_{(R3xR3)}\ldots d\mathbf{r}_1 d\mathbf{r}_2$, as well as R9=R3xR3xR3

$\mathbf{R}_A \equiv (R_{Ax}, R_{Ay}, R_{Az})$ or $(x_A, y_A, z_A)$ = 3 dimension position (spatial) vector of nucleus A

$R_{AB} \equiv |\mathbf{R}_A-\mathbf{R}_B|= ((R_{Ax}-R_{Bx})^2+(R_{Ay}-R_{By})^2+(R_{Az}-R_{Bz})^2)^{1/2}$ = nucleus-nucleus distance

$R_{Ai} \equiv |\mathbf{R}_A-\mathbf{r}_i|= ((R_{Ax}-x_i)^2+(R_{Ay}-y_i)^2+(R_{Az}-z_i)^2)^{1/2}$ = nucleus-electron distance

$\mathbf{r}_i\equiv (x_i,y_i,z_i)$ = 3 dimemsion position (spatial) vector of electron i

$r_{ij} \equiv |\mathbf{r}_i-\mathbf{r}_j|= ((x_i-x_j)^2+(y_i-y_j)^2+(z_i-z_j)^2)^{1/2}$ = electron-electron distance

$\rho(\mathbf{r}_1)$= one-electron density, here only ground state mentioned, more precisely the $\rho_0(\mathbf{r}_1)$, the index short hand holds as $\rho(i)\equiv \rho(\mathbf{r}_i)$ for electron i=1,2,…,N

S = single Slater determinant to approximate the ground state wave function, more precisely the $S_0(s_1\mathbf{r}_1,\ldots,s_N\mathbf{r}_N)$; not to be confused with point S at $\mathbf{R}_S$

SE = non-relativistic electronic Schrödindger equation

$s_i = \alpha$ or $\beta$ spin of electron i

STO = primitive Slater-type atomic orbital, Eq.2.4.1 with change $|\mathbf{r}_i-\mathbf{R}_A|^2 \rightarrow |\mathbf{r}_i-\mathbf{R}_A|$

STO-3G = STO is approximated with linear combination of three GTO
v = function variable, its values are $pR_{CP}^2$, $pqR_{PQ}^2/(p+q)$, etc.
$V_{ee}$ = electron-electron repulsion energy in SE
$V_{ne}$ = nuclear-electron attraction energy in SE
$Z_A$ = nuclear charge of nucleus A=1,2,..M atoms in a molecular system
()*= complex conjugate

## 2.4.References:

# 3. Computation devices and case studies in quasi-hierarchically ordered thesis (English and Hungarian)

In this chapter I list my most important results and equations as thesis followed by a short description or outline of the particular subject. These are grupped as subjects belonging and based on each other, as well as they are listed in logical order, however, they can read independently as well. If one needs more details or references, those can be found in the corresponding articles containing all informations in the year of publication (see my list of publication) for which the year is indicated in the title of the particular thesis. This is the reason, that figures and tables are not numbered in this chapter, only the thesis, as well as smaller overlaps can occure. Most important equations and statements are highlighted with yellow colour, less but still important statements are underlined. Thesis with greater importance and not detailed above is described in more detail.

## 3.1 Developing theories in computation chemistry (correlation energy in focus)

**Thesis-1-Theory: Charge-exchange in gas-surface reactions, case of Na/W**

Theory of near-resonant charge-exchange has been worked out for Na/W gas-surface collision, based on the eikonal formalism, short wavelength-, as well as Padé approximation, with using accurate model diabatic potential surface representation to predict the charge-exchange probability during scattering and desorption, useful in experimental surface analysis to estimate the integral value of diabatic coupling between the atomic and ionic states. Diabatic potentials have been modeled in analytic forms for the Na/W system.

= = = = = =

Rezonancia közeli töltés átadás elméletét dolgoztam ki a Na/W gáz-felület ütközés esetére, eikonal formalizmusra, rövid hullámhossz-, valamint Padé approximációra alapozva, felhasználva pontos modell diabatikus potenciál felület reprezentációt a töltés átadás valószínűségének előre jelzésére a szóródás és deszorpció alatt, mely hasznos a kísérleti felület analízisben a diabatikus kapcsolási állandó integrális értékének becslésére az atomi és ionos állapotok között. Modelleztem a Na/W rendszer diabatikus potenciáljainak analitikus formáit.

**Sandor Kristyan, J.A.Olson:**
**International Journal of Quantum Chemistry, 56 (1995) 51-69**

= = = = = =

Representative equations/tables/figures:

Since more than one electronic state (neutral, ionic, excited) is necessary to describe this system, this is a non-adiabatic process. In this near-resonant charge-exchange the Auger transitions can be neglected. Near-resonant charge-exchange can take place when the ionization energy of the atom is close to the work function of the metal surface a typical example is when sodium (Na) atom collides with tungsten (W) surface. (The Auger effect is a physical phenomenon in which the filling of an inner-shell vacancy of an atom is accompanied by the emission of an electron from the same atom. When a core electron is removed, leaving a vacancy, an electron from a higher energy level may fall into the vacancy, resulting in a release of energy. Although most often this energy is released in the form of an emitted photon, the energy can also be transferred to another electron, which is ejected from the atom; this second ejected electron is called an Auger electron.) Simple algebraic derivation shows that the diagonal elements are the effective potential energy surfaces governing the nuclear motion and the off-diagonal elements cause transitions between the potential energy surfaces. For a system initially being in state i [$P_i(t=0)= 1$], the crude probability (literature) of transition to state j is $P_j^f \equiv P_j(t=t_1) \approx P_j^{fm} \equiv \exp(-c'/v_{init}) = \exp(-c/(E_{kin}^{init})^{0.5})$, where $v= dR/dt$ is the velocity of the particle perpendicular to the surface (e.g., in case of gas-surface reactions), $t_1$ is a time after the particle has passed the coupling region $V_{12}^d$, (see figures below, f= ”final”, m= “a model”). It holds for adsorbed particle, when it gains enough kinetic energy to desorb and to pass the coupling region once when leaving the surface. The literature constant is $\sim(V_{ij}^d)^2/(dV_{jj}^d/dR)$, a positive value. The corresponding constant ($\alpha$) in this work shows a similar dependence, but the approximate equations are much more flexible and describe the computational solution of a Na atom colliding with a W surface much better.

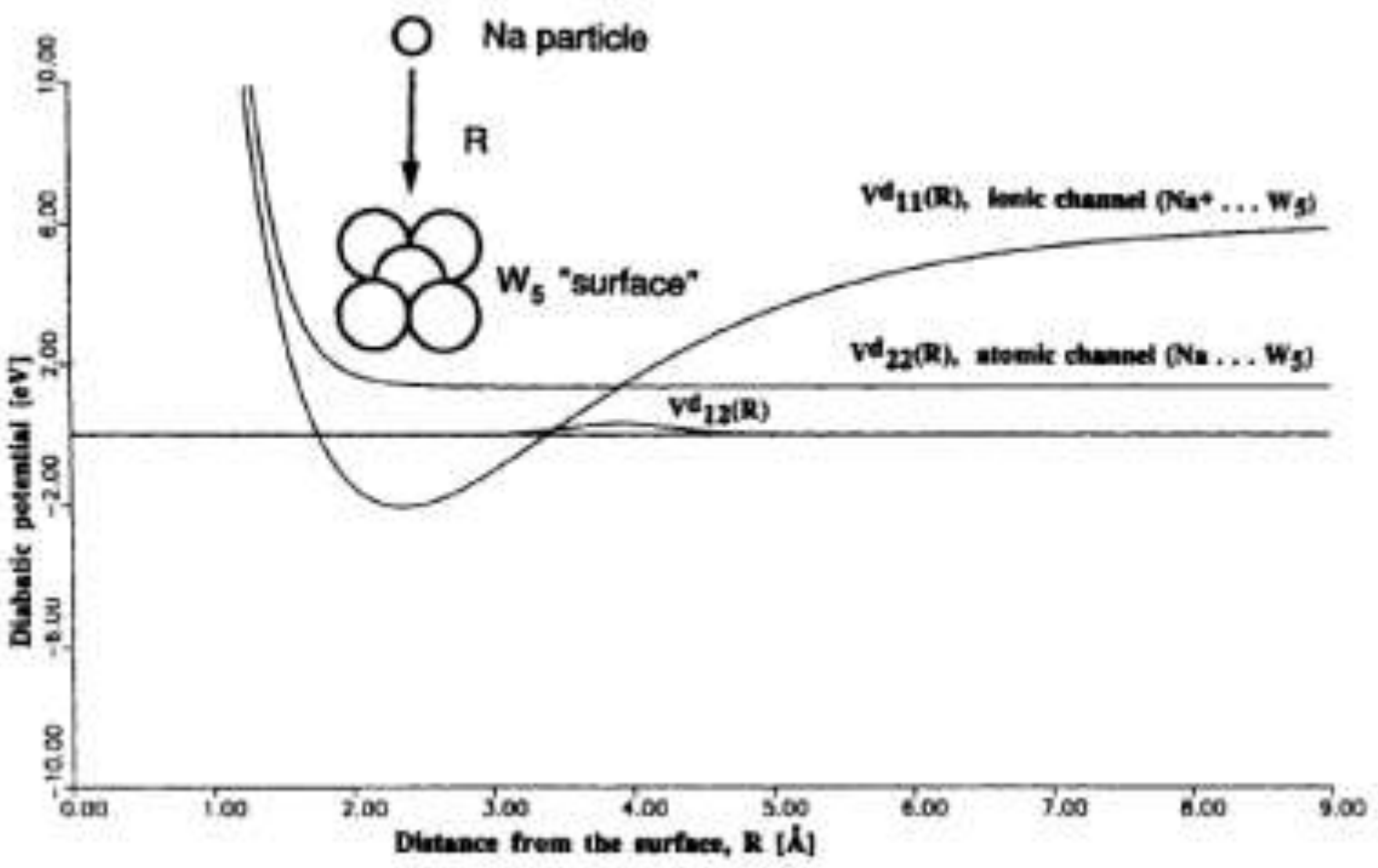

Diabatic model potentials of the Na/W system, R is the trajectory of Na particle. During the collision/ desorption between the Na particle and the W(100) surface (represented by 5 W atoms), the Na nucleus is subjected to $V_{ij}^{d}$ potentials.

$V_{12}^{d}=V_{21}^{d}$, as well as $V_{ij}^{d}$ are fitted to "diatomics in molecules metod".

Adiabatic potentials avoid crossings, but diabatic potentials do not. The diabatic couplings in the vicinity of the avoid crossing have dominant values of generally smooth functions and make the description of collisional charge-exchange easier. The other advantage of using diabatic potentials is that the potential curves belong to a definite character. For example, in figure above $V_{11}^{d}(R)$ represents the ionic character or ionic channel of the system and $V_{22}^{d}(R)$ represents the atomic character or atomic channel. In the case of adiabatic potentials, the ionic and atomic characters are mixed and a function of distance R.

Let $|\Phi(R)\rangle = |\Psi^{a}(R)\rangle\phi(R)$ for the wavefunction in the total Schrödinger equation, where $\phi(R)$ is an nx1 column matrix, the amplitude is written in the form $\phi(R)= \chi(R)\exp(2\pi iS(R)/h)$, where $\chi$ is an nx1 column matrix. This form is called common eikonal since the eikonal S(R) is the same for all electronic channels, dS/dR= momentum of the particle, the short wavelength approximation allows $d^2\chi/dR^2\approx 0$. The $c_j$ are the complex amplitudes in $\chi(R)= c(t)\exp(if(t)/h)$, where the function f(t) is necessary in the derivation, $c= (c_1,\ldots,c_n)$ and t is time. The c is an nx1 complex vector $c_j=c_j^{R}+ic_j^{I}$, giving the probability that the system is in state (or channel) j by $P_j = |c_j|^2$, (complex amplitudes are necessary to describe moments of nuclei, while real numbers are enough to describe electronic potentials by electrons). The energy of the particle is $E= P^2/2m+ c_lc_1^{*}V_{11}^{d}+ (c_2c_1^{*}+c_lc_2^{*})V_{12}^{d}+ c_2c_2^{*}V_{22}^{d}$ = const. of the motion, where * denotes the complex conjugate, the kinetic energy of the particle $E_{kin} = P^2/2m$ is a function of the distance from the surface R, because $V_{ij}^{d}(R)$ is so. Finally, the equation system to solve with computation is $\{dc_1/dt= -i(V_{11}^{d}c_1+V_{12}^{d}c_2)$ & $dc_2/dt= -i(V_{21}^{d}c_1 + V_{22}^{d}c_2)\}$, which separates to four equations via its real and imaginary parts.

Figure below shows that the two probability jumps of $P_j(t)$ are large (almost unity) at smaller energies, e.g., at $E_{kin}^{init}$ = 25 eV, and opposite (depending on the initial phase angle of $c_j(t)$: up-down and down-up), while at larger, e.g., at $E_{kin}^{init}$ = 300 eV, there may be two subsequent jumps (down-down and up-up). During a collision there are two probability jumps, as is exhibited in figure below, but during desorption (starting from P=0), there is only one jump since the desorbing particle passes the coupling region only once.

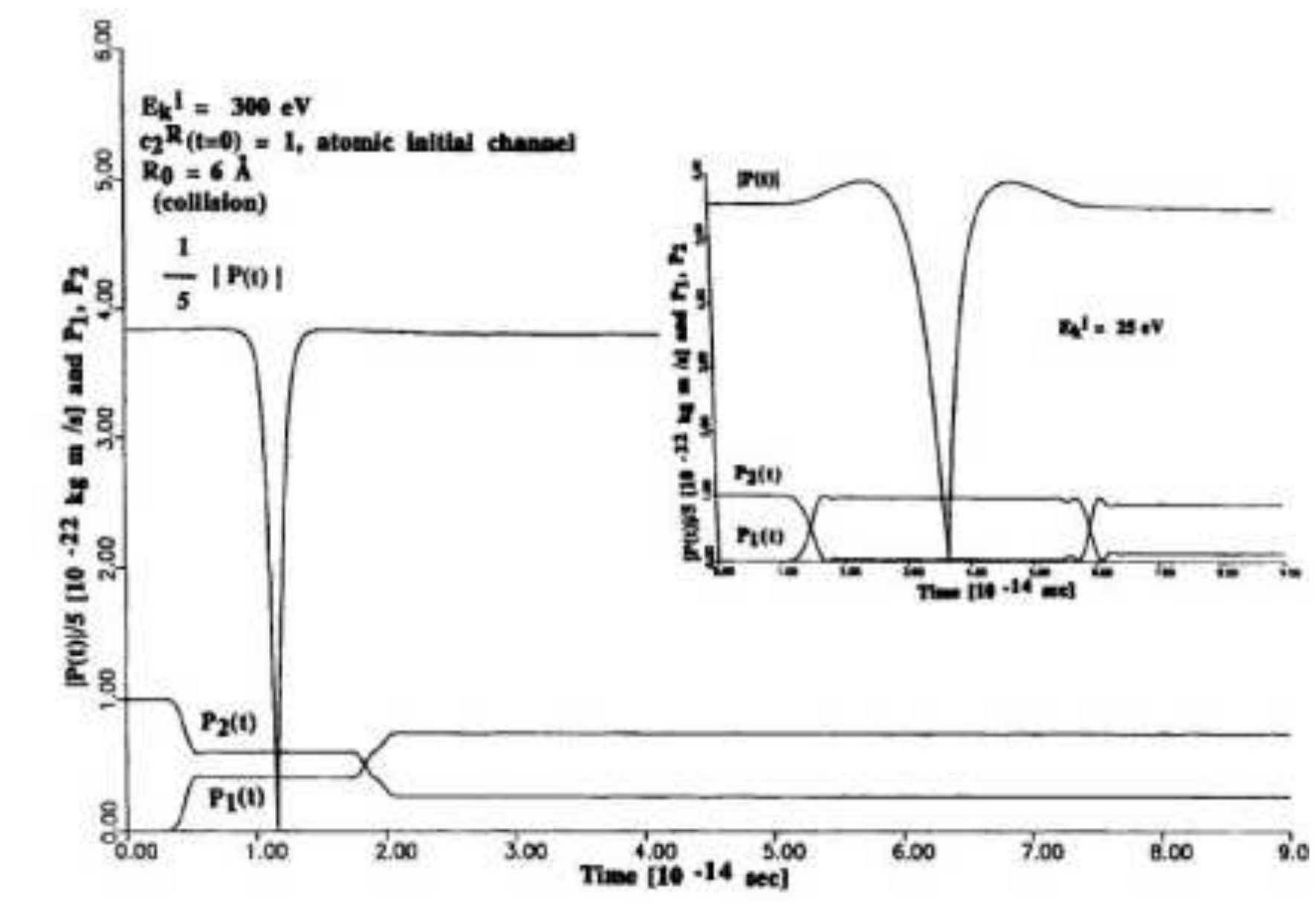

Computational solution for the Na momentum, |P(t)|, and the probability of ionic channel, $P_1(t)$, and atomic channel, $P_2(t)$, as a function of time during scattering. (Division by 5 is to provide a more compact view only.)

A probability jump occurs in the value of $P_j(t)$ when the particle passes the coupling region (where $V_{12}{}^d(R)$ is sigruficantly not zero), forming the link between the atomic and ionic electronic channels in the charge-exchange process (atom $\rightarrow$ ion or ion $\rightarrow$ atom).

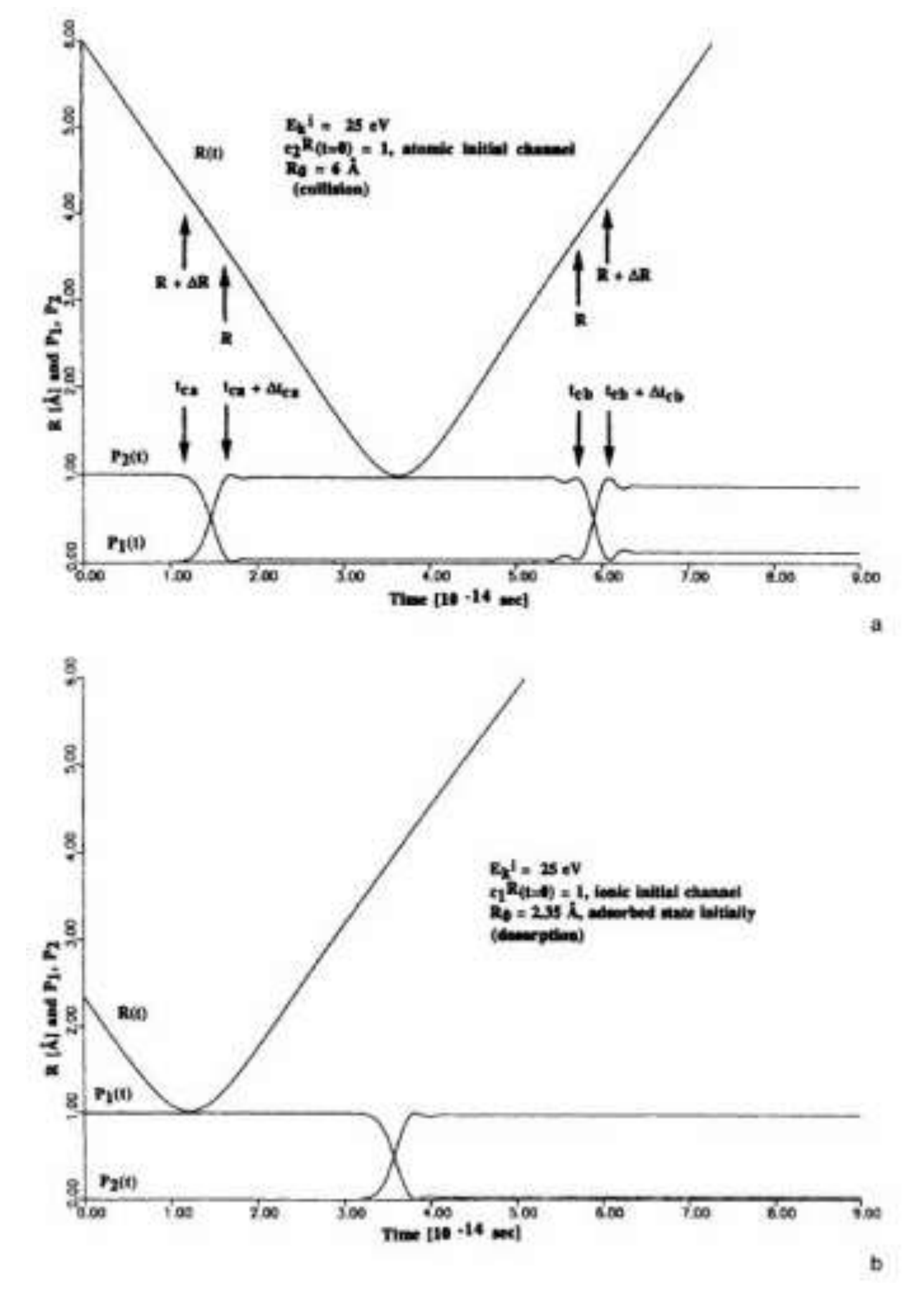

Computational solution for the distance of Na from the surface, R(t), and the probability of ionic (j=1) channel and atomic (j=2) channel, $P_j(t)$, as a function of time.

The ion yield (# of back-scattered ionized atoms / # of incident atoms) is $P_1(t=\infty) = 1 - P_2(t=\infty)$; the conservation of probability, $P_1(t) + P_2(t) = 1$, is satisfied at any time during the computation.

Computational solutions for the final probability of ionic channel, $P_1^f \equiv P_1(t=\infty)$, are exhibited below as a function of initial kinetic energy $E_{kin}^{init}$ (perpendicular to the surface) and the approximate expressions, $P_1^{fm}$, for comparison (m= "a model").

Figure a below shows the case of an incoming colliding Na particle, $P_2(t=0)=1$:

- if $E_{kin}^{init} < \approx 8$ eV: $Na \rightarrow Na^+$ almost totally ($P_1^f \approx 1$) and adsorbs on the valley of $V_{11}^d(R)$,
- if $\approx 8$ eV $< E_{kin}^{init} < \approx 18$ eV: it can desorb, $Na \rightarrow Na^+ \rightarrow Na$ happens almost totally and $P_1^f \approx 0$,
- if $E_{kin}^{init} > \approx 18$ eV: the upper envelope of $P_1^f$ increases up to $E_{kin}^{init} \approx 150$ eV.

Figure b below shows the case of an adsorbed $Na^+$ particle, $P_1(t=0)=1$:

- if $E_{kin}^{init} < \approx 4$ eV: it vibrates in the valley of $V_{11}^d(R)$, it cannot desorb and cannot pass the coupling region, $P_1^f \approx 1$,
- if $E_{kin}^{init} > \approx 4$ eV just a little: it desorbs, pass $V_{12}^d$ region slowly, $Na^+ \rightarrow Na$ occurs and $P_1^f \approx 0$,
- if $E_{kin}^{init} >> \approx 4$ eV: it passes $V_{12}^d$ region faster, leaving less time for charge-exchange and the desorbing $Na^+$ remains an ion, i.e. $P_1^f$ is gets closer and closer to unity as $E_{kin}^{init}$ increases.

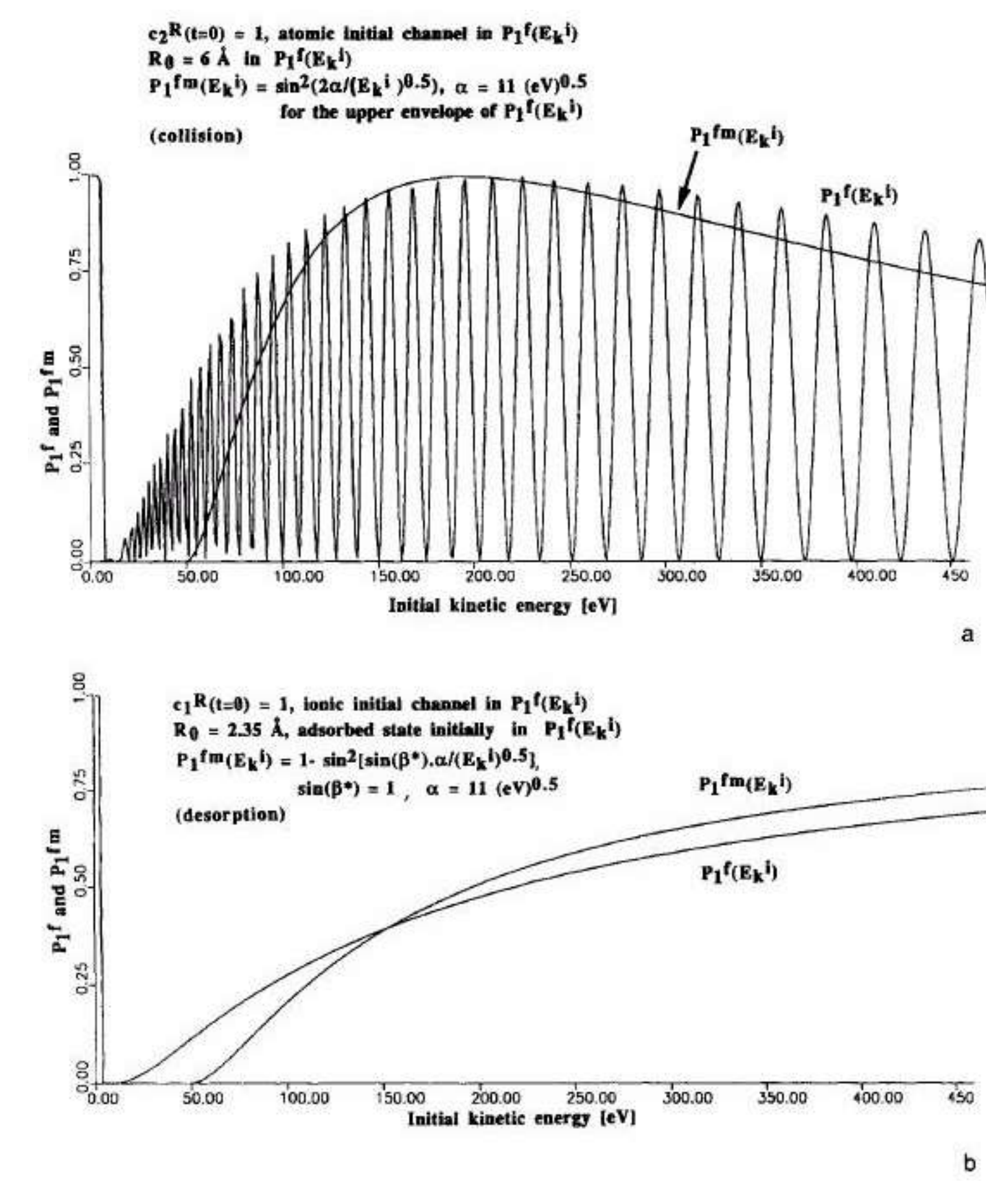

The oscillation in figure a is a consequence of the initial value of the phase angle $\theta_0$ in $c_j$. The physically realistic solution is that if one takes average values with $\theta_0$, i.e., $P_1^f$ must be calculated at all equidistant e.g. 20-40 grid points $\theta_0$, in the interval $[0, 2\pi]$ and averaged, it gives a curve very close to the average value of the upper and lower envelopes.

The CPU time to pick up figures a or b is 2 min/Mflops at any given $E_{kin}^{init}$ value. (Speed of micro (super) computers were 1 (300) MFlops in 1995, the fastest six-core PC processor reached 109 gigaFlops in 2010.)

The equation describing the upper envelope on figure a above is

$$P_1^{fm} = \sin^2[2\alpha/(E_{kin}^{init})^{0.5}] \quad \text{for} \quad 2\alpha/(E_{kin}^{init})^{0.5} \leq \pi ,$$

$$\alpha \equiv (m/2)^{0.5} \int_{[0,\infty]} V_{12}^d(R)dR$$

for the probability that an atomic particle $[P_2(t<t_{ca})=1]$ collides with the surface and it leaves the surface in ionic state, fitting the $\alpha$ gives an experimental value for $\int_{[0,\infty]} V_{12}^d(R)dR$. With our Na/W model potentials <u>$\alpha = 13.86$ (eV)$^{0.5} \approx 14$ (eV)$^{0.5}$</u>.

If the initial state of the adsorbed particle is an ion [$P_1(t=0)=1$, see figure b above], the final probability for leaving the surface as an ion (i.e., without charge-exchange) after gaining some kinetic energy $E_{kin}^{init}$ out from the surface is

$$P_1^{fm}= 1-\sin^2[\sin(\beta^*)\alpha/( E_{kin}^{init})^{0.5}] ,$$

where $\sin(\beta^*)$ is near to unity. (Unphysical model-oscillation of $P_1^{fm}$ comes from the property of the function $\sin(l/x)$ at small $E_{kin}^{init}$, not plotted in figure a-b above, – although at least it indicates the initial steep jump from value 1.0… .)

Another way is based on the Padé approximation (w/o details, while Taylor series are based on $T(a,n,x)\equiv\Sigma_{i=0\ldots n}a_ix^i$, the (n,m) level Padé approximation is based on $T(a,n,x)/(1+T(b,m,x))$ providing more flexibility): The probability that an adsorbed ion desorbs as an ion (i.e., without charge-exchange) from the surface, after gaining some kinetic energy $E_{kin}^{init}$ out from the surface is (in this case the lowest Padé function is (0,2) level)

$$P_1^{fm}= 1/(1 + \alpha^2/E_{kin}^{init}) .$$

A larger $\alpha$ means stronger coupling between the atomic and ionic states, yielding smaller $P_1^f$, i.e., the desorbing ion from the surface [$P_1(t=0)=1$] is subjected to a stronger charge transfer ($Na^+ \rightarrow Na$) when passing the coupling region, causing a smaller ion yield. On the other hand, larger $E_{kin}^{init}$ means that the ion passes the coupling region fast, leaving less time for charge transfer, causing a larger ion yield $P_1^f$ (see figure above and below).

When an atom collides with the surface, the probability that it leaves the surface in ionic state after the collision can be obtained from the desorbing case just above via $P_1^{fm}:= 2P_1^{fm}(1-P_1^{fm})$, factor 2 comes from the two independent a→a→b and a→b→b events, yielding

$$P_1^{fm}= 2\alpha^2E_{kin}^{init}/(\alpha^2 + E_{kin}^{init})^2 .$$

A nice test for self-consistency is the estimation for the location of maximum from $dP_1^{fm}/dE_{kin}^{init}=0$ yielding

$$E_{kin}^{init}(max)= \alpha^2 = (½)m(\int_{[0,\infty]}V_{12}^d(R)dR)^2 \underline{\approx 196\ eV}$$

in agreement with the figure-a above with function value $P_1^{fm}(E_{kin}^{init}(max)=\alpha^2)=0.5$, independent of $\alpha$ and as an average of upper and lower envelope; on the other hand, it provides a model estimate for $\alpha$ from experimental maximum.

A higher degree Padé approximation is (in this case this Padé function is (0,4) level)

$$P_1^{fm}= 1/[1 + \alpha^2(E_{kin}^{init})^{-1} + (\alpha^*)^4(E_{kin}^{init})^{-2}] ,$$

$$\alpha^*\equiv (m/2)^{0.5} \int_{[0,\infty]}\{\ (V_{12}^d)^2[\ 8(V_{12}^d)^2-( V_{11}^d-V_{22}^d)^2\ ]/12\ \}^{0.25}dR$$

for the probability that an adsorbed ion desorbs from the surface as an ion (i.e., without charge-exchange) after gaining some initial kinetic energy $E_{kin}^{init}$. It provides a very excellent approximation for $P_1^f$, see figure below. (Higher level Padé functions than (0,4) encounter higher derivatives of $V_{ij}^d$ and, unfortunately, faces strange neglections, not detailed.)

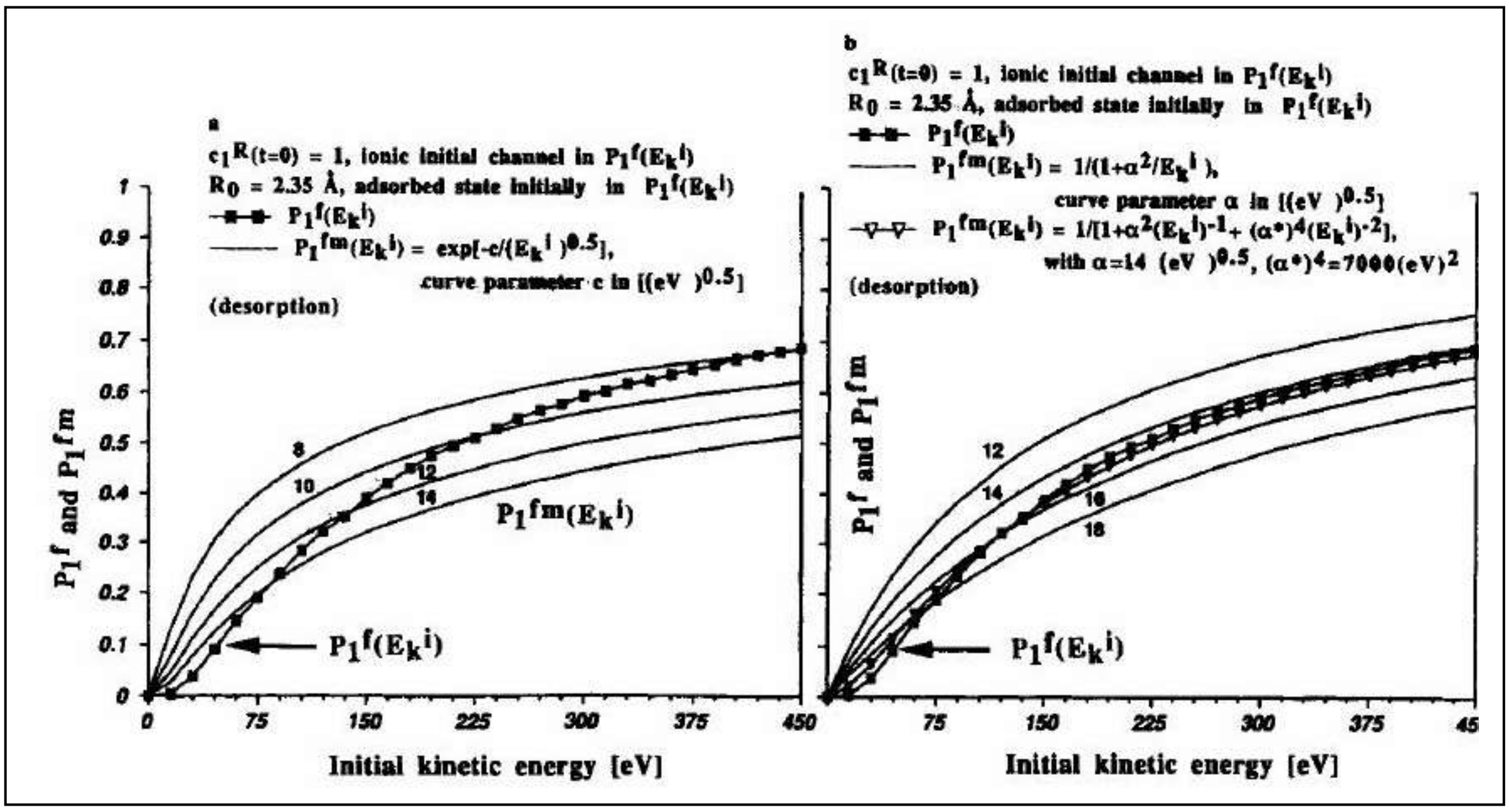

Above, the computational solution for the final probability of ionic channel, $P_1^f$, and its different approximate solutions, $P_1^{fm}$, with different curve parameters are plotted for comparison and parametric estimation. The same $P_1^f(E_{kin}^{init})$ is plotted on a and b (solid squares, as well as the segment $P_1^f(E_{kin}^{init} <\approx 4\ eV) = 1$, when the particle cannot desorb is not shown). On figure a, the crude literature estimation cannot follow the computation solution with one parameter value for c, (candidate values c= 8, 10, 12, and 14 $(eV)^{0.5}$ are plotted). On figure b, Padé approximants are plotted: The (0,2) level with $\alpha$= 12, 14, 16, and 18 $(eV)^{0.5}$ yielding estimate $14<\alpha<18$, and the fitted (0,4) level to the computational $P_1^f$ as

$$\underline{P_1^{fm} = 1/[1+(14)^2 eV(E_{kin}^{init})^{-1}+(7000)(eV)^2(E_{kin}^{init})^{-2}],}$$

for this Na/W system, the superiority of the latter is visible.

**Thesis-2-Theory: Semilocal density functional theory and London dispersion forces, test on noble gases $X_2$, X=He, Ne, and Ar**

The reproduction of the interatomic potential in noble gases $X_2$ (X=He, Ne, and Ar) by Kohn-Sham theory was investigated using a density functional program which can perform counterpoise corrections for both, basis sets and numerical integration. It has been found that, none of the functional considered accounts successfully for the dispersion interaction: The Becke exchange and the Becke-Lee-Yang-Parr (B-LYP) exchange-correlation functionals yield a purely repulsive potential after counterpoise correction. The Dirac-Slater (D-S) functional gives minima which are too deep, at internuclear distances which are too short, particularly for $He_2$ and $Ne_2$. The experimental repulsive potential is reproduced best by D-S calculations while the B-LYP results are close to the SCF ones.

= = = = = =

$X_2$ nemesgáz rendszerek (X= He, Ne, Ar) inter-atomos potenciáljának Kohn-Sham elmélet szerinti reprodukálhatóságát vizsgáltam egy sűrűség funkcionál program (Pulay Péter “TX90” nevű programja) segítségével, mely számította a “counterpoise corrections (egyensúlyi korrekciót)” a bázis készletre és a numerikus integrálásra egyaránt. Azt találtam, hogy egyik vizsgált funkcionál sem tudott megfelelően számot adni a diszperzios kölcsönhatásokért: A Becke “exchange (kicserélődési)” és Becke-Lee-Yang-Parr (B-LYP) “exchange-correlation (kicserélődési-korrelációs)” funkcionálok gyenge taszítási potenciált eredményeztek a “counterpoise correction” alkalmazása után. A $He_2$ és $Ne_2$ rendszerek esetében a Dirac-Slater (D-S) funkcionál túl mély minimumokat és túl rövid magtávolságokat eredményezett. A kísérleti taszítási potenciált legjobban a D-S számítások reprodukálták, míg a B-LYP eredmények az SCF számítások közelében maradtak.

= = = =

Representative equations/tables/figures:

London dispersion forces (instantaneous dipole–induced dipole forces) are a type of force acting between atoms and molecules, and are part of the van der Waals forces. They can act between molecules without permanent multipole moments. They are: 1., inter-molecular forces, 2., much weaker than intra-molecular forces, 3., always attractive forces between atoms or molecules, 4., their origin comes from electrons/nuclei of one atom/molecule interacting with electrons/nuclei of another atom/molecule, 5., although they are weak, but they do not cancel, 6., play an important role in conformational preferences of larger molecules. Because they are a purely correlation effect, they cannot be reproduced at the Hartree-Fock level in supermolecule calculations. Density functional theory (DFT) describes electron correlation effects in ground-state molecules remarkably well at reasonable computational cost. Exact density functional theory would, of course, include all correlation effects, including the dispersion force.

Density functional theories back in 1994 approximated the exact exchange-correlation functional $\varepsilon_{xc}$ by a local functional of the electron density, or the density and density gradient (semilocal). Nevertheless, the prevailing (and characteristically optimistic) opinion in the DFT community appeared to be that the dispersion interaction is somehow accounted for. However, several deficiencies exist in the semilocal exchange-correlation functional. Due to the weak nature of the dispersion interaction, and the fact that all DFT programs employ some sort of numerical integration scheme which limits the numerical accuracy of the results, as well as many methods use further approximations when treating large molecules or solids, a test is useful for answering the question: How semilocal density functional theory accounts for the London dispersion forces?

Three functional presented are thought to be characteristic of the most used DFT methods in 1994: 1., the simple Dirac-Slater exchange (D-S, based roughly on $C\int\rho^{4/3}(\mathbf{r}_1)d\mathbf{r}_1$), 2., the Becke gradient correction added to 1. (B), and 3., the correlation functional of Lee, Yang and Parr, as formulated by Miehlich et al. added to 2. (B-LYP).

A major problem for calculating intermolecular potential functions is the basis set superposition error (BSSE) usually corrected by counterpoise (CP) technique (A+B $\leftrightarrow$ AB referenced as each calculated in the full dimer basis at the same geometry as the dimer). In DFT, the CP raises special problems, as the numerical integration scheme depends on the positions and charges of the nuclei: In our method we used not only the dimer basis set, but also the dimer integration scheme to evaluate monomer energies for the CP technique. Two different basis sets were used: a triple-zeta, augmented with diffuse functions (denoted as TZ), and a set of three polarization functions and two second polarization function added to TZ (basis TZP).

Figure below shows the potential energy curves in the vicinity of van der Waals minimum for (a) $He_2$ (b) $Ne_2$ and (c) $Ar_2$, as a function of internuclear distance, (O) with and (square) without counterpoise correction. The best present estimates of the true potential curves, based on a theoretical model carefully fitted to a number of experimental data, are given for $He_2$ (HFD-ID), $Ne_2$ (HFD-B) and $Ar_2$ (HFDID1) by R. A. Aziz at al. For $He_2$, the results of an accurate *ab initio* (CI) calculation is also included. The SCF, B, and D-S curves are shifted by (a) 100, -100, -200 μhartree, (b) and (c) 200, -200, -400 μhartree, resp., for a better view. Only the results with the TZP basis are shown, the TZ results are quite similar. Our calculated results uncorrected for BSSE show minima, although at the wrong position. However, these minima vanish after the CP correction for all methods but the Dirac-Slater exchange. As seen, the DFT with CP is much larger than in the SCF case, and shows a different geometry dependence because it corrects also for the grid superposition error, arising from the change in the numerical integration procedure in the presence of another atom. Numerical experiments in which the basis set for a single atom was retained but the numerical integration was modified as in the presence of a second atom show indeed that the grid superposition error is dominant in our procedure (integration scheme by Lebedev's method).

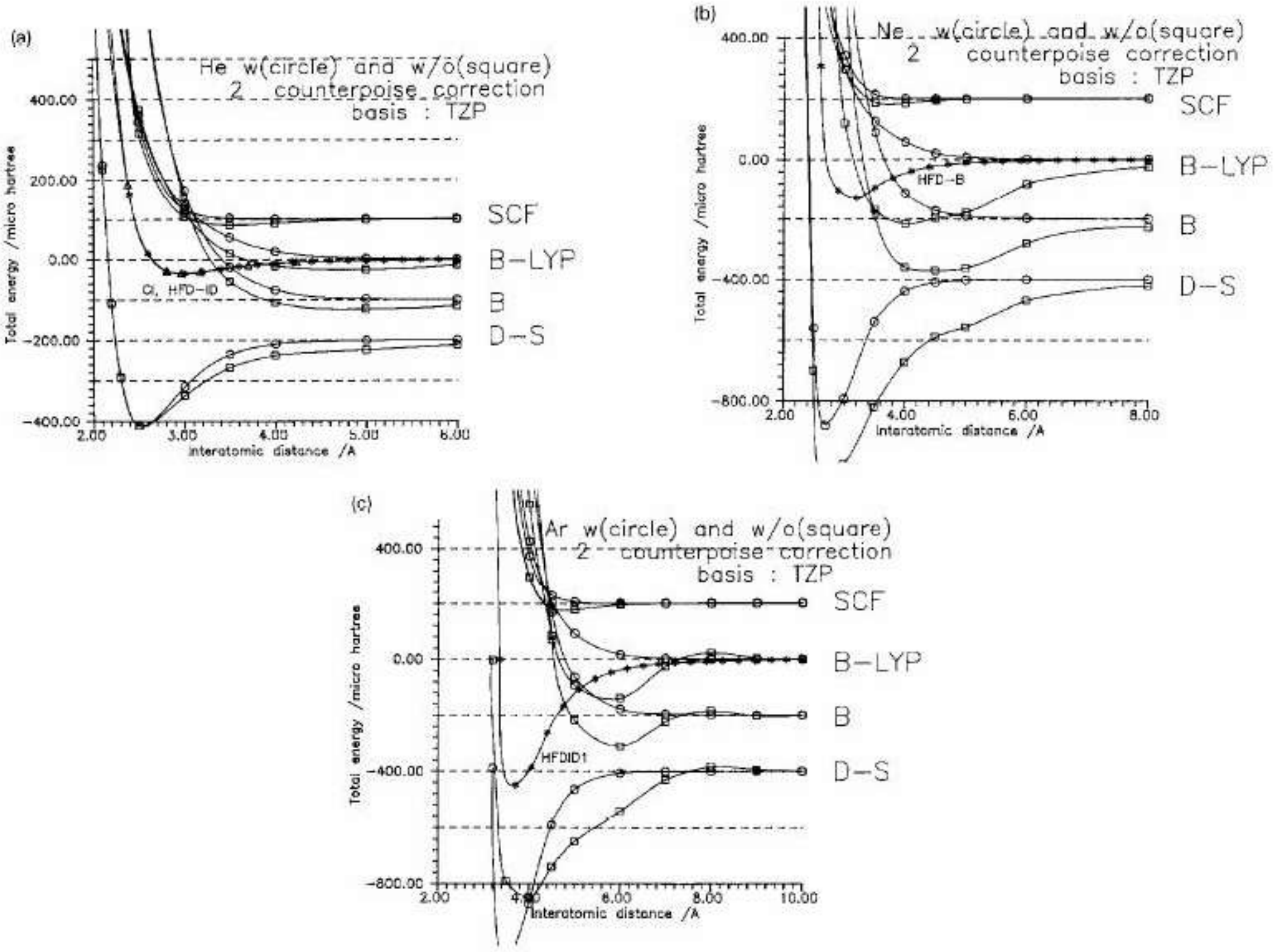

The potential curves corrected for BSSE are purely repulsive for the SCF, B and B-LYP methods. The simplest exchange-correlation potential, D-S, shows a van der Waals minimum which, compared to the experimental values is many times too deep for $He_2$, about three times too deep in $Ne_2$, and only about 1.5 times too deep for $Ar_2$. The D-S minimum rises too fast on the long R side. This rise is probably exponential and originates from an overlap effect, unlike the true van der Waals attraction which is dominated by an $R^{-6}$ term at long distances. Positions and energies of the minima (Angstom) of the diatomic potential energy (μhartree) curves after CP:

| Molecule | Method | Internucl.dist. | Total energy |
|---|---|---|---|
| He...He | expt. | 2.970 | -34.581 |
| | D-S/TZ | 2.616 | -167.619 |
| | D-S/TZP | 2.560 | -214.643 |
| Ne...Ne | expt. | 3.091 | -133.797 |
| | D-S/TZ | 2.780 | -487.717 |
| | D-S/TZP | 2.780 | -502.150 |
| Ar...Ar | expt. | 3.758 | -453.605 |
| | D-S/TZ | 3.676 | -649.502 |
| | D-S/TZP | 3.628 | -767.997 |

Comparing the experimental and calculated repulsive parts of the potentials, <u>all methods perform fairly for the repulsive branches</u>: The B-LYP and SCF methods are very similar, the experimental curves lie between the D-S and the B-LYP (and SCF) curves but closer to the former. This latter is very prominent for Ar, where the D-S results are closest to experiment.

The best performance is provided by the simplest method, the D-S exchange, therefore, DFT theories back in 1994 are probably not useful for the investigation of weakly interacting systems. In view of the good performance of modern density functional methods for the bulk of correlation effects, it is of considerable interest to develop hybrid methods which include the dispersion energy in DFT calculations, a prediction in 1994 which was experienced true in the next two decades. Conventional correlation results calculations can be formulated in terms of localized orbitals, and in this form the correlation energy is largely transferable and can be parametrized.

Meanwhile, this result has gotten into the review book:

Wolfram Koch, Max C. Holthausen

**A Chemist's Guide to Density Functional Theory**

Second Edition

depend only on a local electron density. In order to describe London interactions, a fully nonlocal functional must be applied and a local density functional is in principle not capable of describing this long-range, nonlocal correlation effect. Accordingly, some standard functionals, while correctly describing the short-range repulsion, were found to completely fail in the description of the attractive branches in the potentials of van der Waals complexes like $He_2$, $Ne_2$ or $Ar_2$. Numerical applications show in fact minima on the potential energy surfaces of such systems, although too deep and at the wrong positions, but these usually vanish after correcting for basis set superposition errors. The presence of actual minima has been attributed to overlapping densities, which decay exponentially in r, and not to a physically correct description of true dispersion interactions dominated by the long range fluctuating dipole ($1/r^6$) term (Kristyán and Pulay, 1994). Pérez-Jordá and Becke, 1995, investigated the performance of the SVWN and BP86 functionals as well as Becke's two hybrid approaches for the description of the $He_2$, $Ne_2$, $Ar_2$, HeNe, HeAr, and NeAr rare gas dimers. Also this group found a strong overbinding for the LDA with minima located at too short distances, and only repulsive interactions for the GGA and the related hybrid functional. Interestingly, the half-and-half approach not including any gradient corrections provided a quite reasonable description of the potential shapes, but this approach gave minima which were too shallow. From the latter finding it appears that some DFT models

237

**Thesis-3-Theory: Semilocal density functional theory and the ground-state total energy of highly ionised atoms**

Three generally used exchange-correlation functions (good for most other properties) were investigated in order to discern their ability to reproduce ground-state total energies of highly ionized atoms as well as the sum of their first two ionization energies. Total ground-state energies of closed shell atoms with N electrons and Z atomic numbers were considered for $2 \le N \le Z \le 18$, and N=2,4,6,8,10. The sum of the first two ionization energies, $I1+I2$, is calculated for closed shell atoms with Z=2,4,6,8,10. The density functional theory (DFT) methods investigated are remarkably successful in accounting for the ground-state total energy of the ionized states of atoms, although their accuracy significantly varies with the positive charge of the ionized atom. Interestingly, the conventional Hartree–Fock self-consistent field (HF-SCF) method is more ‘‘rigid’’ with respect to this type of variance in accuracy. The Becke gradient corrected exchange function gives good results, but the Becke exchange with the Lee–Yang–Parr correlation function is better. However, there are some ionized states of atoms for which even the best density functional methods do not exceed the accuracy of the conventional Hartree–Fock SCF method. The simple Dirac–Slater functional gives poor results. The comparison of these methods to accurate *ab initio* calculations and experimental data are reported in detail. Interestingly, the accuracy of these methods (as a function of the degree of ionization) may reflect the shell structure of the atom.

= = = = = =

Három általánosan használt “exchange-correlation (kicserélődési-korrelációs)” funkcionált vizsgáltam, hogy kimutassam hogyan képesek reprodukálni az alap állapotú, erősen ionizált atomok totális energiáit, valamint az első két ionizációs energia összegét. Az N elektronos és Z atomszámú, zárt héjú atomok totális alap állapot energiáit vizsgáltam $2 \le N \le Z \le 18$ és N=2,4,6,8,10 esetekben. Az első két ionizációs energia összegeket ($I1+I2$), így tehát zárt héjú atomok esetében, Z=2,4,6,8,10 értékeknél számoltam. A vizsgált sűrűség funkcionál elmélet (density functional theory (DFT)) módszerek rendkívül sikeresen adtak számot az atomok ionizált állapotának alap állapotú totális energiáiról, bár a pontosságuk jelentősen változik az ionizált atom pozitív töltésével. Érdekes módon, a pontosságra nézve, a konvencionális Hartree–Fock önkonzistens (HF-SCF) módszer “merevebb” e tekintetben. A Becke gradiens korrigált “exchange (kicserélődési)” funkcionál jó eredményt ad, de a Becke “exchange (kicserélődés)” a Lee–Yang–Parr korrelációs funkcionállal jobb. Néhány ionizált atomi állapot esetében azonban, a legjobb DFT módszerek sem szárnyalják túl a konvencionális Hartree–Fock SCF módszer pontosságát. Az egyszerű Dirac–Slater funckcionál szegényes eredményt ad. E munkámban szintén részletezem ezen módszerek összehasonlítását pontos *ab initio* számításokkal és kísérleti adatokkal. Érdekes módon, e módszerek pontosságában (mint az ionizációs fok függvénye) reflektálni látszódik az atomok héj szerkezete.

**Sandor Kristyan: Journal of Chemical Physics, 102 (1995) 278-284**

= = = = = =

Representative equations/tables/figures:

Modern DFT describes electron correlation effects in ground-state molecules remarkably well at reasonable computational cost, however, they are approximate, more, they are worked out (and parametrized) primarily for neutral or close to neutral molecules. However, highly ionized states (e.g. of atoms) have not yet been fully tested, while reliably exact *ab initio* calculations for highly ionized atoms have been reported in the literature up to N =18 and Z =28, as well as experimental first (I1) and second (I2) ionization energies have long been known accurately.

Estimations of the total ground-state energies, $E_{total\ electr,0}(N,Z)$, in hartrees by different methods (HF-SCF, D-S, B, B-LYP) are shown for ionized atoms with Z nuclear charge and N electrons compared to the exact calculations, $E_{total\ electr,0}(N=2,Z,CI)$, via their deviations (using Pulay' TX90 program extended with DFT functionals by Pulay and Kristyan, as well as the 6-31G** basis sets, except for Mg (Z=12), where the double zeta basis set by Huzinaga was used).

For N=2 electrons:

| Z | HF-SCF | D-S | B | B-LYP |
|---|---|---|---|---|
| 2 | 0.048564 | 0.189083 | 0.049727 | 0.005879 |
| 3 | 0.044377 | 0.274011 | 0.050189 | 0.002560 |
| 4 | 0.045772 | 0.363661 | 0.057292 | 0.008096 |
| 5 | 0.047111 | 0.453343 | 0.064142 | 0.014292 |
| 6 | 0.048570 | 0.543448 | 0.071105 | 0.020996 |
| 7 | 0.050251 | 0.633658 | 0.078148 | 0.027960 |
| 8 | 0.052156 | 0.724164 | 0.085490 | 0.035318 |
| 9 | 0.054293 | 0.814859 | 0.093070 | 0.042978 |
| 10 | 0.056746 | 0.905897 | 0.100842 | 0.050827 |
| 11 | 0.063490 | 0.999296 | 0.111854 | 0.061829 |
| 12 | 0.058323 | 1.086963 | 0.113252 | 0.063187 |
| 13 | 0.069550 | 1.182441 | 0.128855 | 0.079177 |
| 14 | 0.072961 | 1.274463 | 0.137731 | 0.088204 |
| 15 | 0.076295 | 1.366428 | 0.146638 | 0.097292 |
| 16 | 0.079993 | 1.458738 | 0.155863 | 0.106651 |
| 17 | 0.083839 | 1.551222 | 0.165239 | 0.116140 |
| 18 | 0.087875 | 1.643891 | 0.174789 | 0.125795 |

HF-SCF= $E_{total\ electr,0}(2,Z,HF\text{-}SCF) - E_{total\ electr,0}(2,Z,CI)$, simple Hartree-Fock SCF,

D-S = $E_{total\ electr,0}(2,Z,D\text{-}S) - E_{total\ electr,0}(2,Z,CI)$, simple Dirac – Slater exchange,

B = $E_{total\ electr,0}(2,Z,B) - E_{total\ electr,0}(2,Z,CI)$, Becke gradient corr. added to D-S,

B-LYP = $E_{total\ electr,0}(2,Z,B\text{-}LYP) - E_{total\ electr,0}(2,Z,CI)$, corr. Lee-Yang-Parr added to B,

where $E_{total\ electr,0}(2,Z,CI)$ is an *ab initio* CI calculation by E. R. Davidson et al. (1991),

e.g. $E_{total\ electr,0}(N=2,Z=2,CI)= 2.903724$ and $E_{total\ electr,0}(N=2,Z=18,CI)= 312.907186$.

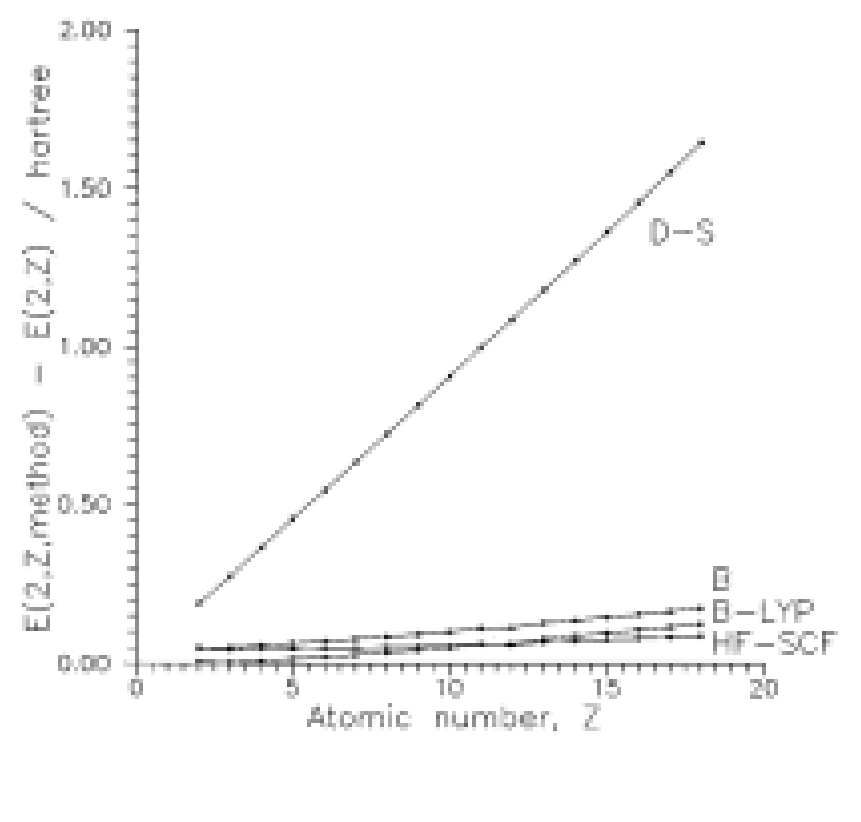

The graphical presentation of the deviations in HF-SCF and three DFT methods shown in table above for N= 2 electron higly inized atoms.

The graphical presentation of the deviations in HF-SCF and three DFT methods for N>2 electron higly inized atoms are:

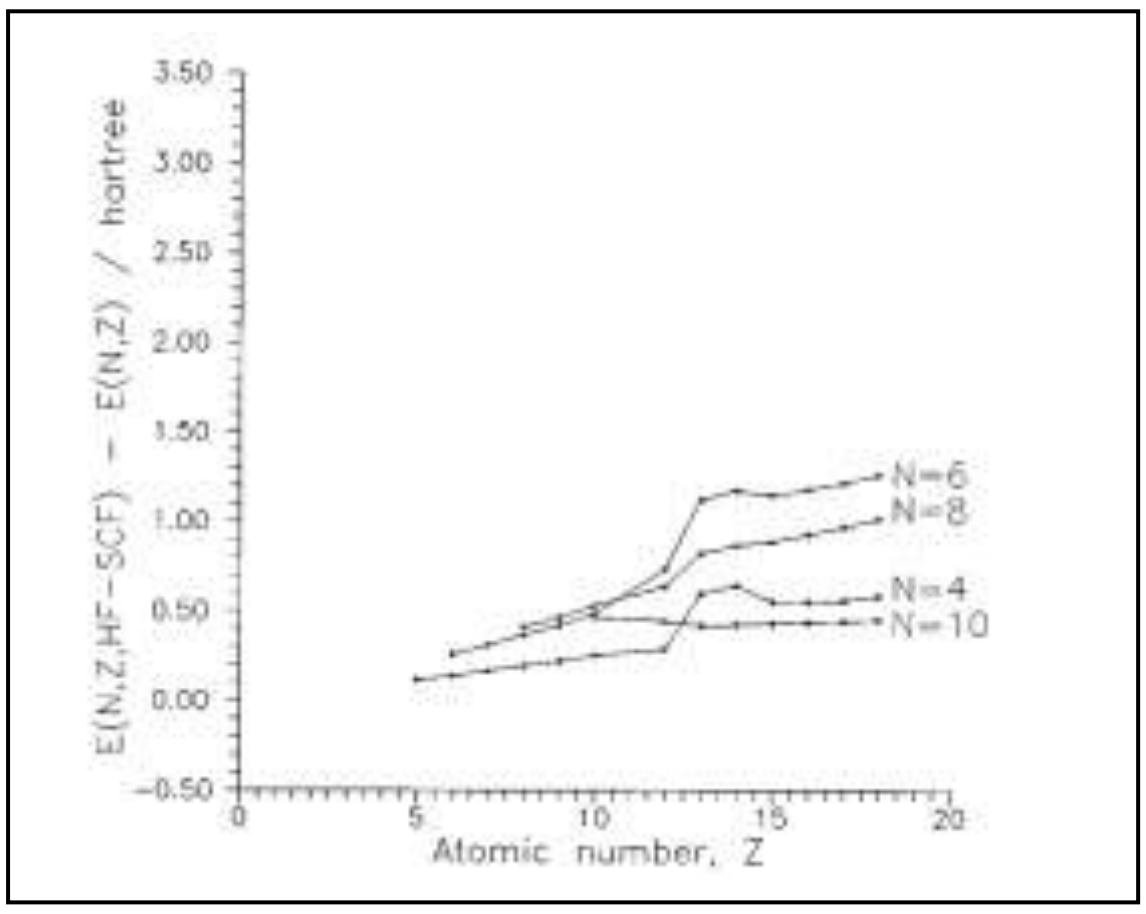

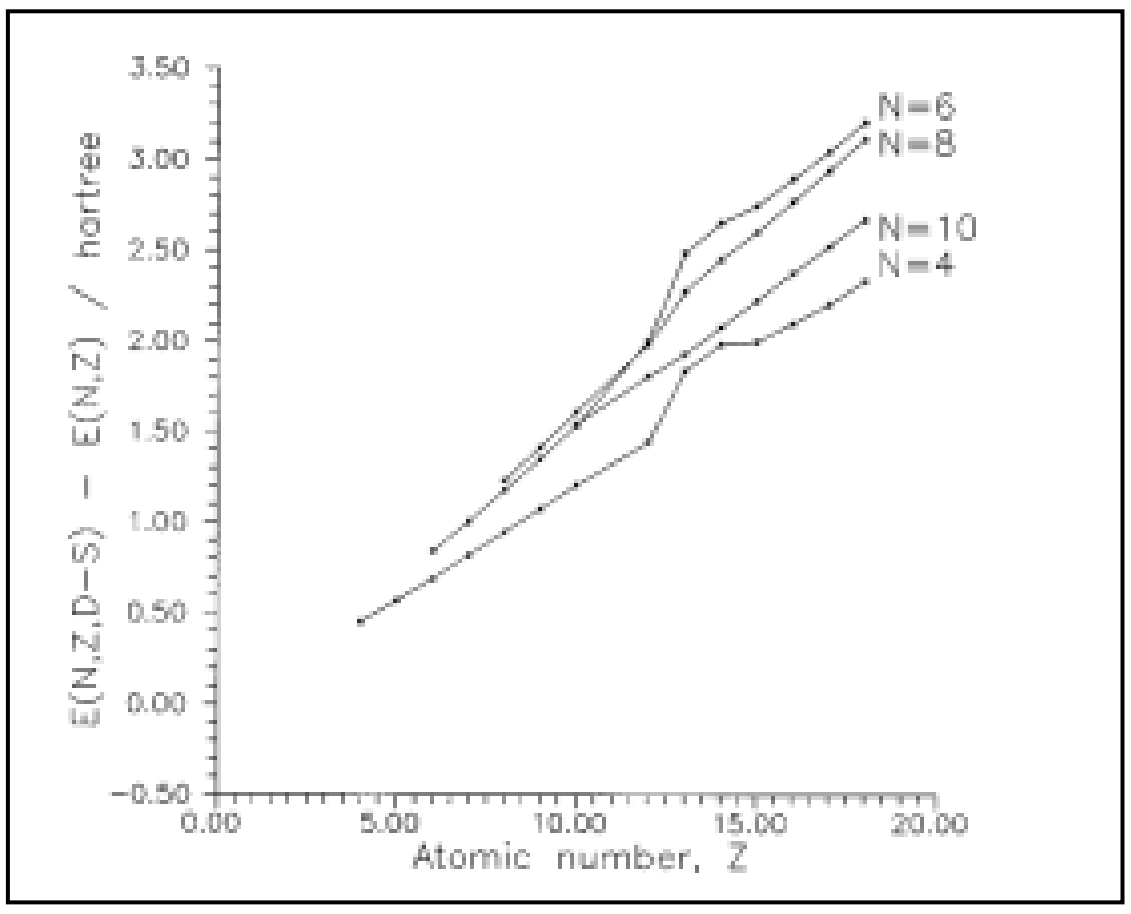

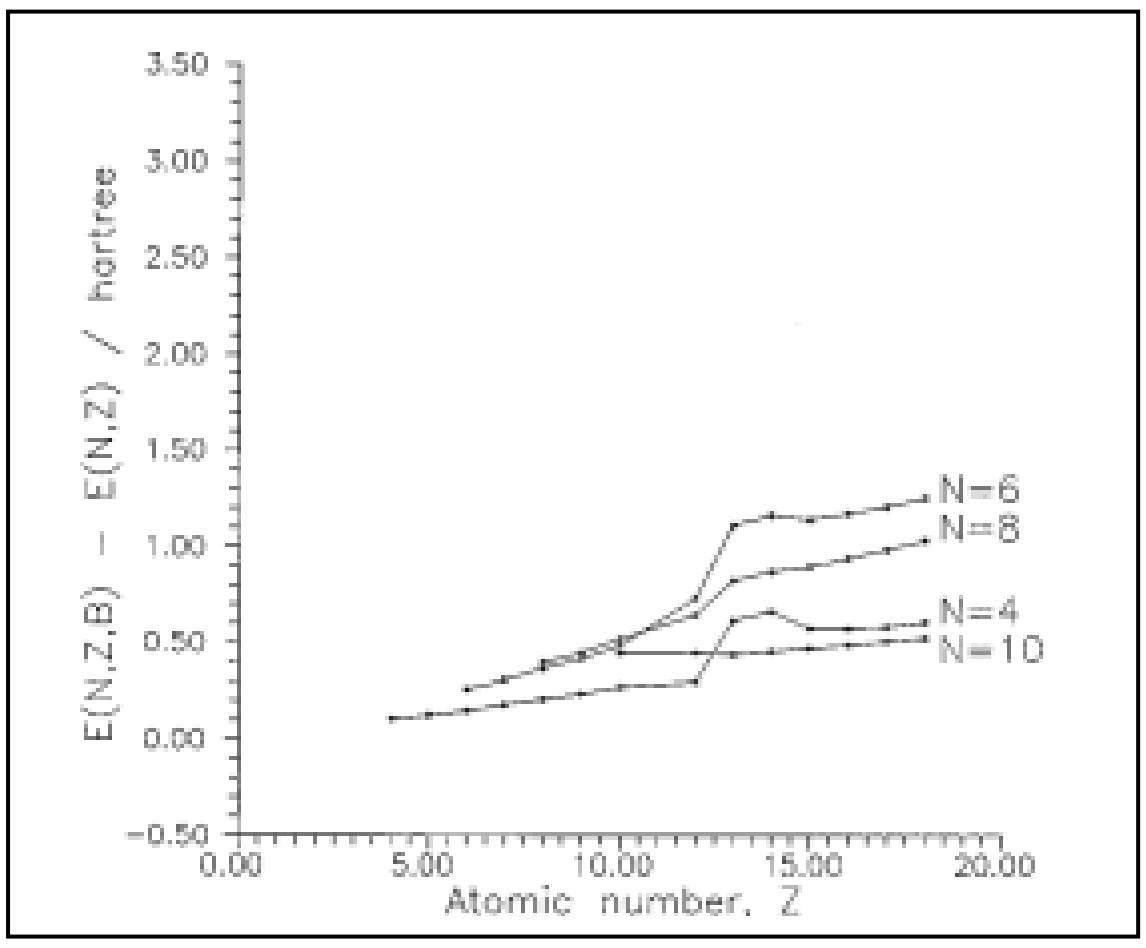

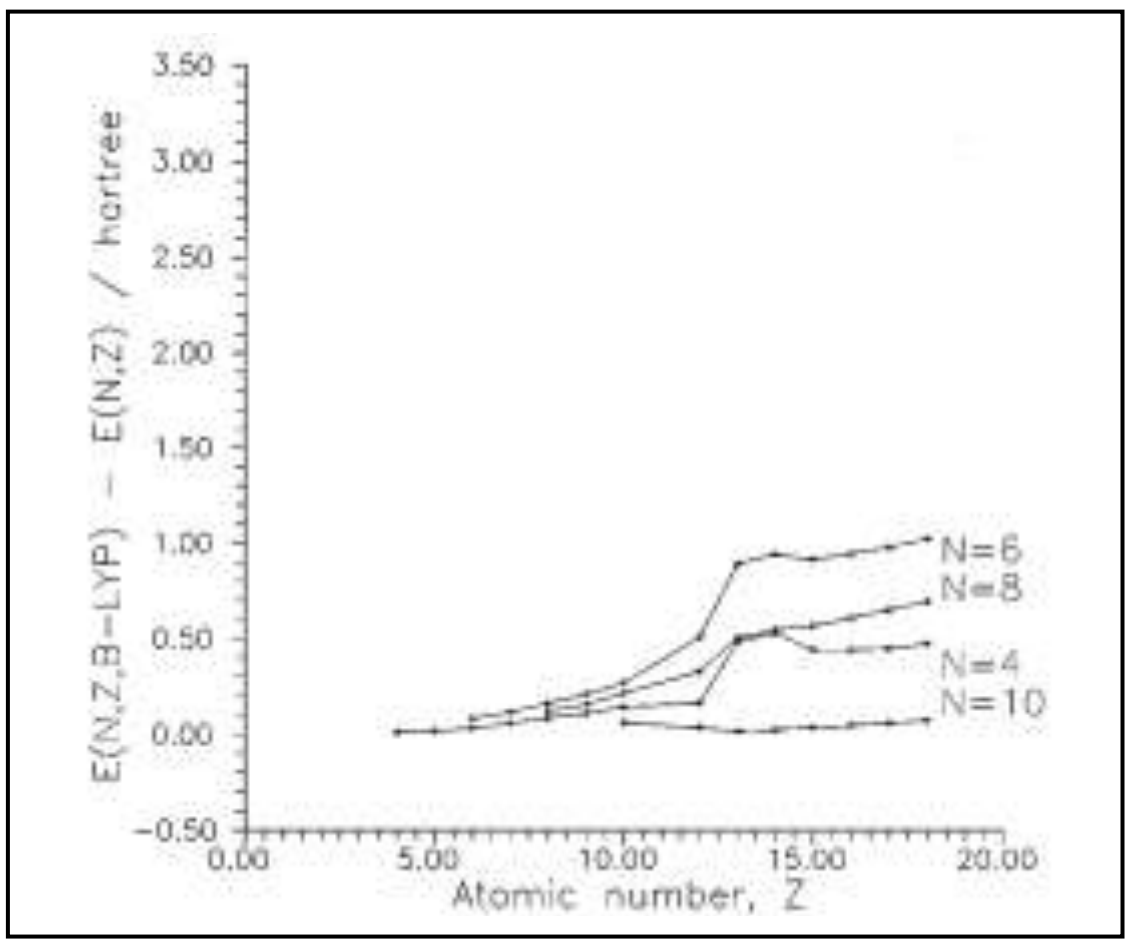

On the figures above a peak can be observed at Z =11 (Na) with N =4, 6, 8 electrons, but not if N=2 or 10. Adequate explanation for it was not found. It may be a convergence and not a numerical integration error, since it shows already in case of HF-SCF. (As it is known HF-SCF calculations use Gaussian basis functions (e.g., 6-31G** here) making all integrals analytical, but DFT methods (e.g., D-S, B, B-LYP here) use numerical integration only, and therefore must encounter numerical errors.)

Notice that figures above exhibit errors of absolute (full) energy values in estimating $E_{total\ electr,0}(N,Z)$, while considering ionization energies, particularly the sum of first ionization energy (I1: $A \rightarrow A^+$) and second ionization energy (I2: $A^+ \rightarrow A^{++}$) as

$$I1+I2= E_{total\ electr,0}(N=Z-2,Z,)-E_{total\ electr,0}(N=Z,Z),$$

allows us to investigate errors in estimating relative (energy difference) values, which is always more important in chemistry. Indeed, the different DFT methods do not deviate so much from each other, that is, errors(estimating I12) < errors(estimating $E_{total\ electr,0}(N,Z)$) exhibited below. (Calculating I1 and I2 individually, needs open and closed shell calculations, which is more problematic than closed shell calculations only in general, however, we test here certain DFT methods on closed shells (N=even, singlet) only. Also, no zero point energy (ZPE) in case of atoms which would make the test of DFT methods more complex.)

The sum of first two ionization energies of atoms, I1+I2 (hartree) as function of Z by different methods (a), and their deviation (hartree) from experimental (Expt) values (b), I12≡ I1+I2 is a short hand notation, and I12(method) is an estimation for it:

(a)

| Z | I12(HF-SCF) | I12(D-S) | I12(B) | I12(B-LYP) | I12(Exact) | I12(Expt) |
|---|---|---|---|---|---|---|
| 2 | 2.855160 | 2.714641 | 2.853997 | 2.897845 | 2.903724 | 2.902904 |
| 4 | 0.641114 | 0.926495 | 0.964036 | 1.010153 | 1.011794 | 1.011658 |
| 6 | 1.189342 | 1.155903 | 1.199318 | 1.259671 | 1.310100 | 1.309348 |
| 8 | 1.746764 | 1.736479 | 1.759440 | 1.833027 | 1.792100 | 1.790671 |
| 10 | 2.365541 | 2.367555 | 2.374005 | 2.456065 | 2.301000 | 2.297433 |

(b)

| Z | HF-SCF | D–S | B | B-LYP | Exact |
|---|---|---|---|---|---|
| 2 | -0.047744 | -0.188263 | -0.048907 | -0.005059 | 0.000820 |
| 4 | -0.370544 | -0.085163 | -0.047622 | -0.001505 | 0.000136 |
| 6 | -0.120005 | -0.153445 | -0.110029 | -0.049677 | 0.000752 |
| 8 | -0.043907 | -0.054192 | -0.031231 | 0.042356 | 0.001429 |
| 10 | 0.068 108 | 0.070122 | 0.076573 | 0.158632 | 0.003567 |

I12(HF-SCF)= $E_{total\ electr,0}$(N=Z-2,Z,HF-SCF)-$E_{total\ electr,0}$(N=Z,Z,HF-SCF)

I12(D–S) = $E_{total\ electr,0}$(N=Z-2,Z,D-S) -$E_{total\ electr,0}$(N=Z,Z,D-S)

I12(B) = $E_{total\ electr,0}$(N=Z-2,Z,B) -$E_{total\ electr,0}$(N=Z,Z,B)

I12(B-LYP) = $E_{total\ electr,0}$(N=Z-2,Z,B-LYP) -$E_{total\ electr,0}$(N=Z,Z,B-LYP)

I12(Exact) = $E_{total\ electr,0}$(N=Z-2,Z,CI) -$E_{total\ electr,0}$(N=Z,Z,CI)

I12(Expt) = Sum of the experimental value of the first two ionization energies, I1+I2

HF-SCF = I12(HF-SCF)-I12(Expt)

D–S = I12(D-S) -I12(Expt)

B = I12(B) -I12(Expt)

B-LYP = I12(B-LYP) -I12(Expt)

Exact = I12(Exact) -I12(Expt)

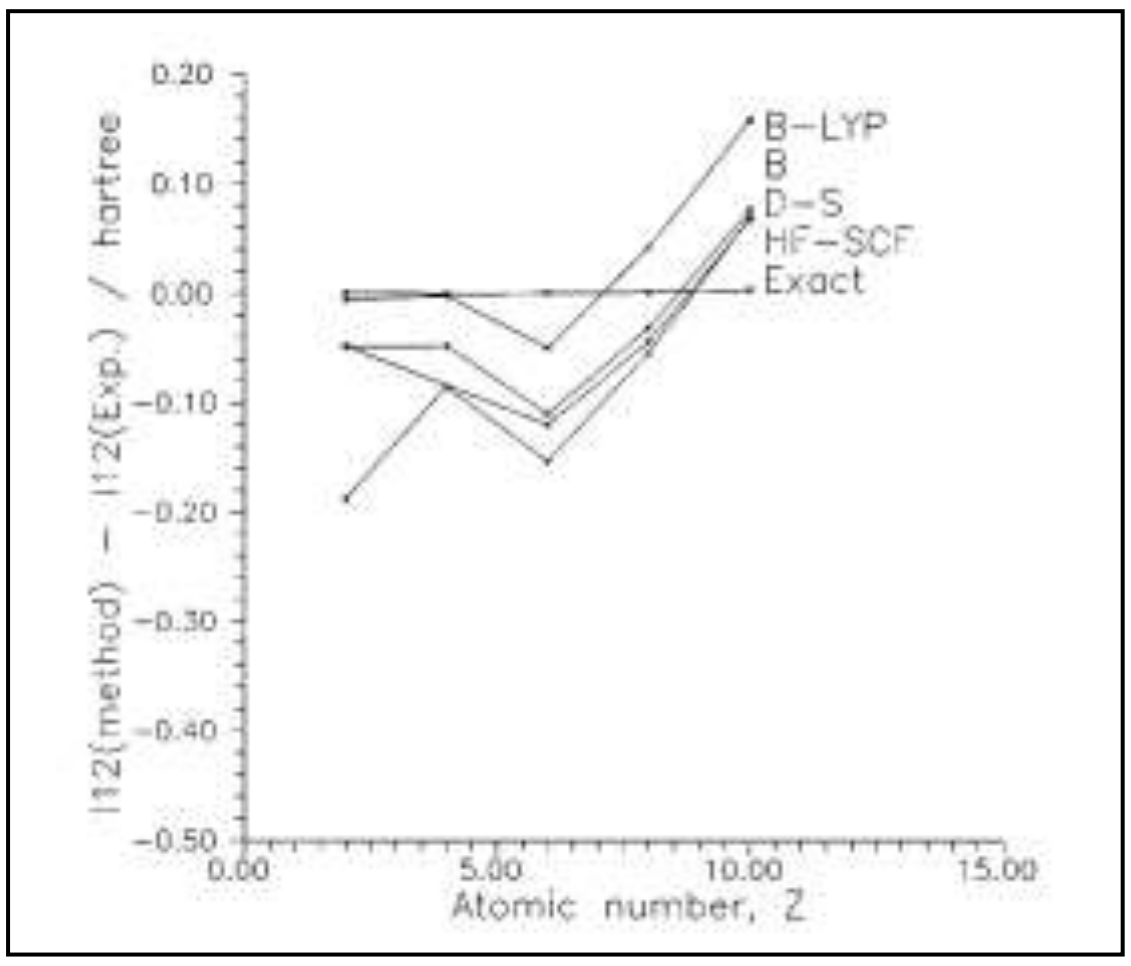

As a summary, the local and gradient-corrected local (semilocal) DFT investigated (D-S, B, B-LYP) for ionized atoms perform well for neutral atoms, but systematically fail as the positive charge of the iso-electronic (N is kept constant and Z increases) atom increases, which is, interestingly, less characteristic in the case of HF-SCF; recall that DFT is supposed to correct HF-SCF. The quality of DFT methods in estimating $E_{total\ electr,0}(N,Z)$ or I12:

B-LYP > B > D-S.

In contrast, for the description of dispersion forces (in weakly interacting systems) between noble atoms ($He_2$, $Ne_2$, and $Ar_2$) the simplest D-S method gives the best, though not satisfactory, results. As the degree of ionization decreases (Z=constant, N increases), the shell structure of the atom affect the accuracy of the different DFT methods for these closed shell (N is even, singlet) calculations: The accuracy is generally worst at N =6, bad at N =8, but always better at N =4 (completely filled s orbitals, $1s^2 2s^2$), and N =10 (completely filled sp orbitals, $1s^2 2s^2 2p^6$). Only the B-LYP method approaches chemical accuracy (1 kcal/mol≈ 0.0016 hartree) for estimating $E_{total\ electr,0}(N,Z)$ in the cases of (N=2, Z=2,3,4), (N=4, Z=4) and (N=10, Z=11).

Finally, DFT theories back in 1995 are probably not useful for the investigation of highly ionized atoms, which was a limitation of those exchange-correlation functionals in DFT. The well-performing B-LYP formulas for neutral atoms and molecules have come out only after a prolonged development. If the weakness in describing highly ionized atoms is corrected, the other weaknesses (calculating dispersion forces or NMR chemical shift) of DFT may also be rectified, and the DFT methods would fully display their advantage in comparison to exact calculations (which contain configuration interactions): same accuracy with less computational time.

**Thesis-4-Theory: Basis set choice in density functional theory for electronic structures of molecules**

The effect of basis set in Kohn-Sham theory on estimating correlation energy has been investigated using Pulay' TX90 program extended with DFT functionals by Pulay and Kristyan. Lee-Yang-Parr (LYP) density functional estimation, MP2 perturbation calculation and Davidson's accurate configuration interactions calculations for correlation energies of atoms (with $2 \leq N \leq Z \leq 18$) as well as water, ammonia and pyrrole were used for this test. Interestingly, in certain cases the DFT estimation with the simple 3-21G basis set has produced the same results as the huge triple-zeta basis set with polarization functions within the chemical accuracy (1 kcal/mol).

= = = = = =

Vizsgáltam a bázis készlet hatását a Kohn-Sham elméletben a korrelációs energia becsléseknél a Pulay' TX90 kód felhasználásával, melynek DFT funkcionálokkal való kiegészítését Pulayval programoztam be. Tesztnek a Lee-Yang-Parr (LYP) sűrűség funkcionál közelítést, az MP2 perturbációs számítást, és Davidson pontos konfigurációs kölcsönhatások módszerét használtam a korrelációs energiák számítására atomok ($2 \leq N \leq Z \leq 18$) valamint víz, ammónia és pirrol esetében. Érdekes módon, bizonyos esetekben a DFT közelítés az egyszerű 3-21G bázis készlettel hasonló eredményeket produkál a kémiai pontosságra nézve (1 kcal/mol), mint a nagy tripla-zeta bázis készlet polarizációs függvényekkel.

**Sandor Kristyan: Chemical Physics Letters, 247 (1995) 101-111 and 256 (1996) 229-240**

= = = = = =

Representative equations/tables/figures:

The estimation of correlation energy ($E_{corr}$) with rapid subroutines based on (generally non-variational) density functional theory (DFT) is a link between the calculations generated by the Hartree-Fock self-consistent field (HF-SCF) method (quick, variational, requiring relatively low disc space, but only for ground state and cannot provide chemical accuracy) and the configuration interactions (CI) method (variational, providing accurate results for ground and excited states, but very lengthy and require much more disc space). The Møller-Plesset (MP2, MP3, MP4, etc.) is also a (perturbation method) link, but beside it is not fast enough, it is non-variational, and worst, fails to provide chemical accyracy in many cases, (the full reason is still a dilemma). The source of error in HF-SCF calculations is that it cannot account well for the correlation between electrons, and this problem starts when even two electrons are involved only in the molecular system, but on the other side, the not-complete basis set also produces error. The effect of basis set is investigated inherited in these three famous methods (HF-SCF, MP2, DFT (particularly LYP)), the CI (practically totally accurate) is used as control for atoms.

It is known in HF-SCF/basis calculations since long, that, 1.: For energy differences, the large part of error from using a smaller basis set cancels, more, even part of the error from correlation effects can cancel, recall for example when not the accurate value of an energy barrier, but a relation only (e.g. if an ortho- or meta-conformation is more reactive) has to be established, 2.: Reaching the HF limit is difficult in case of larger molecules, problems start at molecules containing more than as low as M≈5 atoms. Fortunately, on the other hand it is also known that, high quality DFT estimations for correlation energy can be achieved even with poor basis sets, i.e. one does not have to reach the HF limit with the HF-SCF one-electron density, however, there is no larger quantitative comparative study in the literature about this theoretical/computational phenomenon.

An example for the magnitudes: the HF-SCF limit of neutral Ar atom in hartree is $E_{electr,0}$(N=Z=18,HF)= -526.817513 with correlation energy $E_{corr}$(N=Z=18,CI)= -0.72216. Although it is only about 0.14% error, it is much higher (≈450 times) than the chemical accuracy. The basis sets used in this test with increasing size and power/quality are:

ST0-3G < 3-21G < 6-31G** < TZ (triple-zeta with diffuse functions) < TZP (a set of three polarization functions and two second polarization functions (3D+2F) added to TZ).

ATOMS:

We consider the ratio $E_{corr}$(method/basis)/$E_{corr}$(CI) with increasing basis set for cosed shell atoms (N=2(ls$^2$), N=4(ls$^2$2s$^2$), N=10(1s$^2$2s$^2$2p$^6$), N=12 (1s$^2$2s$^2$2p$^6$3s$^2$), N=18 (1s$^2$2s$^2$ 2p$^6$3s$^2$3p$^6$) , if the method is better and better along with the basis set, it must approaches unity; the $E_{corr}$(CI) by Davidson is related to the HF-SCF/basis value.

The ratio of the MP2 with Biosym TZ+2P basis set (≈TZ or TZP in quality) and CI level correlation energies, $E_{corr}$(N,Z,MP2)/$E_{corr}$(N,Z,Cl), for closed shell atoms are listed below. Inspecting the values, the MP2 can account for about the 0.30-0.45= 30-45% of correlation energy with this relatively large basis set. In detail:

| Atom | Z | N=2 | N=4 | N= 10 | N= 12 | N= 18 |
|---|---|---|---|---|---|---|
| He | 2 | 0.7128333 | | | | |
| Li | 3 | 0.3277565 | | | | |
| Be | 4 | 0.3106459 | 0.3912166 | | | |
| B | 5 | 0.3184627 | 0.4522357 | | | |
| C | 6 | 0.3228704 | 0.4583343 | | | |
| N | 7 | 0.3254529 | 0.4643798 | | | |
| 0 | 8 | 0.3248598 | 0.4655492 | | | |
| F | 9 | 0.3251952 | 0.4676578 | | | |
| Ne | 10 | 0.3258178 | 0.4704872 | 0.6528470 | | |
| Na | II | 0.3093416 | 0.4056081 | 0.2333967 | | |
| Mg | 12 | 0.3121526 | 0.4246433 | 0.2268409 | 0.2504355 | |
| Al | 13 | 0.3510121 | 0.4594083 | 0.3301369 | 0.3628658 | |
| Si | 14 | 0.3469853 | 0.4650570 | 0.3931555 | 0.4413351 | |
| P | IS | 0.3428903 | 0.4655264 | 0.3469882 | 0.4042806 | |
| S | 16 | 0.3435972 | 0.4675429 | 0.3165706 | 0.3776204 | |
| CI | 17 | 0.4375402 | 0.4699948 | 0.2181131 | 0.2941331 | |
| Ar | 18 | 0.3455275 | 0.4738414 | 0.2903960 | 0.3519147 | 0.4196467 |

The ratio of the LYP with different, decreasing size and quality basis sets from the top and CI level correlation energies, $E_{corr}$(N,Z,LYP)/$E_{corr}$(N,Z,CI), for closed shell atoms to consider the effect of basis set in DFT calculations are listed below. Inspecting the values,

```
TZP     basis: atoms without parenthese,average= 0.9078 (1.0183),
TZ      basis: atoms in parentheses (), average= 0.9042 (1.0183),
6-31G** basis: atoms in parentheses )(, average= 0.9051 (1.0157),
        no data for  Mg (Z = 12),
3-21G   basis: atoms in parentheses [], average= 0.8951 (0.9949),
STO3G   basis: atoms in parentheses {}, average= 0.9052 (1.0097):
Atom  z     N=2         N=4         N= 10       N= 12       N= 18
```

| | | | | | | |
|---|---|---|---|---|---|---|
| He | 2 | 0.9719221 | | | | |
| Li | 3 | 1.0038143 | | | | |
| Be | 4 | 1.0087662 | 0.9611185 | | | |
| B | 5 | 1.0059875 | 0.9173574 | | | |
| C | 6 | 1.0012474 | 0.8656930 | | | |
| N | 7 | 0.9962243 | 0.8143724 | | | |
| O | 8 | 0.9916007 | 0.7665221 | | | |
| F | 9 | 0.9873775 | 0.7224743 | | | |
| Ne | 10 | 0.9839121 | 0.6826046 | 0.9638554 | | |
| Na | 11 | 0.9807118 | 0.5938884 | 1.0066231 | | |
| Mg | 12 | 0.9768802 | 0.5041015 | 1.0347101 | 1.0007614 | |
| Al | 13 | 0.9755962 | 0.5317408 | 1.0541145 | 1.0488329 | |
| Si | 14 | 0.9734919 | 0.5015649 | 1.0685101 | 1.0583831 | |
| P | 15 | 0.9715427 | 0.5311597 | 1.0791562 | 1.0635492 | |
| S | 16 | 0.9696386 | 0.5094742 | 1.0879471 | 1.0657929 | |
| CI | 17 | 0.9683718 | 0.4889659 | 1.0953457 | 1.0661660 | |
| Ar | 18 | 0.9677965 | 0.4696907 | 1.1017821 | 1.0650961 | 1.0274075 |
| - - - - - - | | | | | | |
| (He) | 2 | 0.97192213 | | | | |
| (Li) | 3 | 1.00381428 | | | | |
| (Be) | 4 | 1.00876615 | 0.96111847 | | | |
| (B) | 5 | 1.00598754 | 0.91735741 | | | |
| (c) | 6 | 1.00124741 | 0.86569303 | | | |
| (N) | 7 | 0.99622429 | 0.81437238 | | | |
| (O) | 8 | 0.99160069 | 0.76652213 | | | |
| (F) | 9 | 0.98737754 | 0.72247427 | | | |
| (Ne) | 10 | 0.98391213 | 0.68260456 | 0.96385543 | | |
| (Na) | 11 | 0.98071183 | 0.55364404 | 1.00662307 | | |
| (Mg) | 12 | 0.97688024 | 0.49675509 | 1.03471014 | 1.00058284 | |
| (Al) | 13 | 0.97559623 | 0.46740820 | 1.05411447 | 1.04883290 | |
| (Si) | 14 | 0.97349193 | 0.43465546 | 1.06851012 | 1.05838307 | |
| (P) | 15 | 0.97154265 | 0.53115970 | 1.07915618 | 1.06354916 | |
| (S) | 16 | 0.96963862 | 0.50947421 | 1.08794706 | 1.06579290 | |
| (Cl) | 17 | 0.96837185 | 0.48896588 | 1.09534569 | 1.06616599 | |
| (Ar) | 18 | 0.96779649 | 0.46969066 | 1.10178212 | 1.06509608 | 1.02740748 |
| - - - - - - | | | | | | |
| )He( | 2 | 0.97280165 | | | | |
| )Li( | 3 | 1.00412712 | | | | |
| )Be( | 4 | 1.00874299 | 0.85069826 | | | |
| )B( | 5 | 1.00486331 | 0.91818267 | | | |
| )C( | 6 | 0.99866433 | 0.86606412 | | | |
| )N( | 7 | 0.99237039 | 0.81015801 | | | |
| )0( | 8 | 0.98635009 | 0.75779998 | | | |
| )F( | 9 | 0.98077736 | 0.71231532 | | | |
| )Ne( | 10 | 0.97638680 | 0.67053968 | 0.96243658 | | |
| )Na( | 11 | 0.97686543 | 0.64222421 | 1.00651769 | | |
| )Al( | 13 | 0.96539610 | 0.56353424 | 1.05395272 | 1.04874598 | |
| )Si( | 14 | 0.96117868 | 0.54145375 | 1.06814553 | 1.05764111 | |
| )P( | 15 | 0.95631827 | 0.51008990 | 1.07918739 | 1.06328976 | |
| )S( | 16 | 0.95294849 | 0.48708769 | 1.08761922 | 1.06596585 | |
| )CI( | 17 | 0.95031647 | 0.46549210 | 1.09463397 | 1.06727782 | |
| )Ar( | 18 | 0.94797568 | 0.44628255 | 1.10033067 | 1.06718775 | 1.02766582 |
| - - - - - - | | | | | | |
| [He] | 2 | 0.96694923 | | | | |
| [Li] | 3 | 0.97915823 | | | | |
| [Be] | 4 | 0.97623566 | 0.96316659 | | | |
| [B] | 5 | 0.97135935 | 0.91404505 | | | |
| [C] | 6 | 0.97216898 | 0.86703786 | | | |
| [N] | 7 | 0.97449163 | 0.82126526 | | | |
| [O] | 8 | 0.97664225 | 0.77881260 | | | |
| [F] | 9 | 0.98033206 | 0.73923974 | | | |
| [Ne] | 10 | 0.98433077 | 0.70268803 | 0.95624542 | | |
| [Na] | 11 | 0.88703368 | 0.63836707 | 0.99507466 | | |
| [Mg] | 12 | 0.89046382 | 0.60049664 | 1.02542168 | 0.99447148 | |
| [Al] | 13 | 0.89447858 | 0.56687424 | 1.04505768 | 1.04696364 | |
| [Si] | 14 | 0.90030059 | 0.53636491 | 1.06109856 | 1.05691979 | |
| [P] | 15 | 0.90702446 | 0.50962209 | 1.07313828 | 1.06224017 | |
| [S] | 16 | 0.91391077 | 0.48585066 | 1.08202403 | 1.06388139 | |
| [Cl] | 17 | 0.92115056 | 0.46434878 | 1.08954650 | 1.06383121 | |
| [Ar] | 18 | 0.92851552 | 0.44490935 | 1.09599881 | 1.06247713 | 1.02736455 |
| - - - - - - | | | | | | |
| {He} | 2 | 0.97148920 | | | | |
| {Li} | 3 | 1.00116695 | | | | |
| {Be} | 4 | 1.00713520 | 1.00932739 | | | |
| {B} | 5 | 1.00527692 | 0.92849777 | | | |
| {C} | 6 | 0.99820931 | 0.84846310 | | | |
| {N} | 7 | 0.99094624 | 0.78668576 | | | |
| {O} | 8 | 0.98400950 | 0.74064325 | | | |
| {F} | 9 | 0.97733506 | 0.69943078 | | | |
| {Ne} | 10 | 0.97140149 | 0.66364506 | 0.96214265 | | |
| {Na} | 11 | 0.96830715 | 0.62990975 | 1.00123776 | | |
| {Mg} | 12 | 0.96201783 | 0.60140124 | 1.03137458 | 1.06383679 | |
| {Al} | 13 | 0.95552002 | 0.57596427 | 1.05180945 | 1.05608534 | |
| {Si} | 14 | 0.94936737 | 0.55177358 | 1.06604913 | 1.05093825 | |
| {P} | 15 | 0.94349532 | 0.52845407 | 1.07740200 | 1.05073811 | |
| {S} | 16 | 0.93778883 | 0.50670407 | 1.08636482 | 1.04984941 | |
| {CI} | 17 | 0.93130197 | 0.48697544 | 1.09271034 | 1.04376274 | |
| {Ar} | 18 | 0.92601719 | 0.46760702 | 1.09920247 | 1.04534555 | 1.02163028 |

The average of values listed for the five sections (marked with same style parentheses) of the table, followed by a value in parentheses means: average with excluding cases N= 4 from this set; multiplying by 100 gives the percentage immediately the LYP methods can account for $E_{corr}$. Since for cases N= 4 the LYP estimations are weaker (visibly differs from other columns in related parts of the table) than the other cases, two types of averages are listed: including and excluding cases N= 4 in the branch. (Another dependency of DFT methods on shell structure of atoms was also reported in Kristyan's previous Thesis/work on highly ionized atoms.) The comparison of these average values shows that practically the LYP equations, as a DFT estimation for correlation, are about basis independent for atoms from the first three rows of the periodic table with respect to those basis sets considered.

The above two tables reveal that MP2 with large basis set accounts for 30-45% (strong under shot) of $E_{corr}$ for closed shell ground state atoms with $2 \leq N \leq Z \leq 18$, while this is 95-110% (slight under and over shot) in cae of LYP with large and small basis sets, except the case $4 = N \leq Z \leq 18$, where the LYP accuracy decreases from about 96 to 47% as Z increases, as well as MP2 does not show that anomaly at N=4 electrons what LYP does. (Recall that not-producing the full value of $E_{corr}$ does not necessarily mean wrong prediction for energy differences, manifesting e.g. in “molecular mechanics” (MM) method, which is amazingly good for energy differences, while meaningless for absolut values.) Moreover, the smaller basis set STO-3G is best among the considered ones (excluding the case N= 4), with respect to its average being the closest to the value 1.0. An explanation for why a smaller basis set can provide slightly better DFT approximation ($E_{corr}$(N,Z,LYP)) than the larger one for atoms can be that finer (here larger in size and quality) basis set can follow the finer structure of the one-electron (e.g. HF-SCF) density, but in using it as input in the generally algebraically difficult DFT functionals (here LYP) for estimating correlation energy, the numerical integration scheme cannot follow these, and computational inaccuracy can occur. (Recall that HF-SCF typically reproduces the shell structure reasonably, while DFT generally fails (smoothes or nivellates) in this respect.)

Among the MP2 approximations (a wavefunction method), column N=4 is the slightly best, slightly improving with Z, however, among the LYP ones (a DFT method and in case of all basis sets) column N=4 is manifestingly worst and manifestingly worstening with Z.

MOLECULES:

The effect of basis set on ground state total energy estimation, $E_0$(HF-SCF), and DFT correlation energy estimation, $E_{corr}$(LYP), was tested using Program TX90 again, wherein the HF-SCF one-electron density was used as input to the LYP equations. The CPU times for calculating $E_{corr}$(LYP) values listed below make sense with the computer speed only: IBM-RISC/6000 (about 25 Mflops). The ΔHF and ΔLYP are the decrease of $E_0$(HF-SCF) and $E_{corr}$(LYP) estimations from the previous (weaker) basis set, resp., recall chemical accuracy 1kcal/mol≈ 0.0016 hartree/particle. Molecules are neutral and geometry optimized with (Biosym) MM; CI data were not available.

| Basis | $-E_0$(HF-SCF) (hartree) | ΔHF (hartree) | $-E_{corr}$(LYP) (hartree) | ΔLYP (hartree) | CPU time (s) |
|---|---|---|---|---|---|
| water($H_2O$): | | | | | |
| ST03G | 73.2411861 | | 0.3485372 | | 1.10 |
| 3-21G | 74.1244891 | 0.8833030 | 0.3508774 | 0.0023402 | 1.44 |
| 6-31G** | 74.5663458 | 0.4418566 | 0.3518635 | 0.0009860 | 9.27 |
| TZ | 74.6370701 | 0.0707244 | 0.3520074 | 0.0001439 | 19.57 |
| TZP | 74.6575124 | 0.0204423 | 0.3521750 | 0.0001676 | 1730.37 |
| Methane($CH_4$): | | | | | |
| ST03G | 37.0087351 | | 0.3183894 | | 2.22 |
| 3-21G | 37.6186075 | 0.6098725 | 0.3280186 | 0.0096292 | 3.19 |
| 6-31G** | 37.8719103 | 0.2533028 | 0.3281765 | 0.0001579 | 26.90 |
| TZ | 37.9325540 | 0.0606437 | 0.3292887 | 0.0011122 | 60.39 |
| pyrrole($C_4NH_5$): | | | | | |
| ST03G | 193.0934140 | | 1.3652068 | | 62.23 |
| 3-21G | 196.0189820 | 2.9255680 | 1.3865679 | 0.0213610 | 105.31 |
| 6-31G** | 197.4021568 | 1.3831747 | 1.3865110 | -0.0000568 | 1018.15 |
| TZ | 197.7509811 | 0.3488243 | 1.3894012 | 0.0028901 | 2795.57 |

The $-E_{corr}$(MP2,TZ+2P,Biosym), $E_{corr}$(LYP,6-31G**,TX90)/$E_{corr}$(MP2,TZ+2P,Biosym), CPU time for MP2 calculation and the required disc space for MP2 calculation, resp., are:

```
water  : 0.259060 hartree, 1.358230,     11.93 s,   2 Mbytes,
methane: 0.200440 hartree, 1.637277,     42.11 s,   6 Mbytes,
pyrrole: 0.879357 hartree, 1.576734, 53 min 44 s, 354 Mbytes.
```

Conclusions from cases of ATOMS and MOLECULES:

$E_{corr}$(LYP,basis,TX90)/$E_{corr}$(MP2,TZ+2P,Biosym) ratio for an atomic particles with (N,Z) is ≈1.5-4, and the corresponding ratio is ≈1.5 for molecules listed in tables above: The MP2 calculation is much less accurate than the LYP for estimating $E_{corr}$, at least in absolute value. The LYP shows very little basis dependence among the five types of basis sets considered with respect to the chemical accuracy (except case LYP for atoms with N=4). It should be kept in mind, because the MP2 calculations need considerable more CPU time, disc space and more complex programming than the LYP, as well as before year about 2000, the commercial program packages treated the MP2 as a classic reliable one over any other estimations, while after year about 2000 the winer is B3LYP.

The HF-SCF/basis approximation for $E_{total\ electr,0}$ is variational and the chemical accuracy cannot be reached with smaller than TZ or TZP basis. On the other hand, although the LYP approximation for correlation energy, $E_{corr}$(LYP), is not variational, a 3-21G basis set can produce almost the same value as the TZ or TZP basis set for small molecules within the chemical accuracy, the CPU times listed manifest for the importance of this property. Negative ΔLYP value, e.g. for pyrrole, is either an artificial computation round error because of the too small basis dependence of DFT equations or a non-variational property.

**Thesis-5-Theory: Introducing RECEP for correlation energy calculations**

A simple quasi-linear relationship has been introduced (first in the literature) between the number of electrons, N, participating in any molecular system and the correlation energy as

$$-0.035(N-1) > E_{corr}[\text{hartree}] > -0.045(N-1)$$

independently form the size of molecular system (!) and developed to estimate correlation energy more accurately immediately after *ab initio* calculations by using the partial charges of atoms in molecular systems, easily obtained from Hartree-Fock self-consistent field (HF-SCF) calculations. The method is compared to the well-known B3LYP, MP2, CCSD and G2M methods, as well as correlation energy estimations for negatively (-1) charged atomic ions are also commented.

= = = = =

Egyszerű qvázi-lineáris összefüggést állapítottam meg (elsőként az irodalomban) a molekuláris rendszerek elektron száma, N, és korrelációs energiája között, miszerint

$$-0.035(N-1) > E_{corr}[\text{hartree}] > -0.045(N-1)$$

függetlenül a molekula rendszer méretétől (!), amit tovább fejlesztve lehetővé vált a korrelációs energia pontosabb becslése közvetlenül *ab initio* számítások után az által, hogy felhasználjuk hozzá az atomok parciális töltéseit az adott molekula rendszerben, mely könnyen hozzáférhető “Hartree-Fock self-consistent field” (HF-SCF) számítások esetén. A módszert összehasonlítottam az ismert B3LYP, MP2, CCSD és G2M módszerekkel, valamint a negatívan (-1) töltött atomi ionok korrelációs energia becslésére szintén kitértem.

**Sandor Kristyan: Chemical Physics, 224 (1997) 33-51**

= = = = = =

Representative equations/tables/figures:

In this study the primary goal was to show the “global behavior” of the magnitude of correlation energy on electron content (N) and atomic partial charges (the latter is a consequence of electron negativity) of atoms in molecular systems. From the definition of correlation energy, $E_0(CI) = E_0(\text{HF-SCF/basis}) + E_{corr}(N,\{Z_A,R_A\}_M, \text{basis})$ the $E_{corr}$ must be a function of N, M nuclear coordinates with charges ($\{Z_A,R_A\}_M$), and basis set, wherein (full) configuration interactions (CI) is the accurate many-determinantal expanded solution for ground state, while the HF-SCF/basis one is the optimized single-determinantal approximation on a basis level used or chosen.

Statement: In molecular systems, $E_{corr}$ mainly and quasi-linearly depends on electron content, N, as

$$-0.035(N-1) > E_{corr}[\text{hartree}] > -0.045(N-1) ,$$

and depends much less on nuclear coordinates, $\{Z_A,R_A\}_M$, however, dependence on $\{Z_A,R_A\}_M$ cannot be neglected with respect to chemical accuracy (1 kcal/mol ≈ 0.0016 hartree/particle). This latter manifests in case of any potential barrier or transition state, where N is (always) conserved, and the energy difference ($\Delta E_0$(HF-SCF/basis)) between the two sides soleily comes from the change in $\{Z_A,R_A\}_M$ at a given N, so for $\Delta E_{corr}$. (Well known and interesting is that basis set errors in $E_{corr}$ have fortunate cancellation in energy differences in many cases.)

The CI (figure a, all $2 \leq N \leq Z$(atomic number)$\leq 18$) and B3LYP (figure b, Z-N=-1) estimations of atomic $E_{corr}$ are plotted as function of N to support equation above:

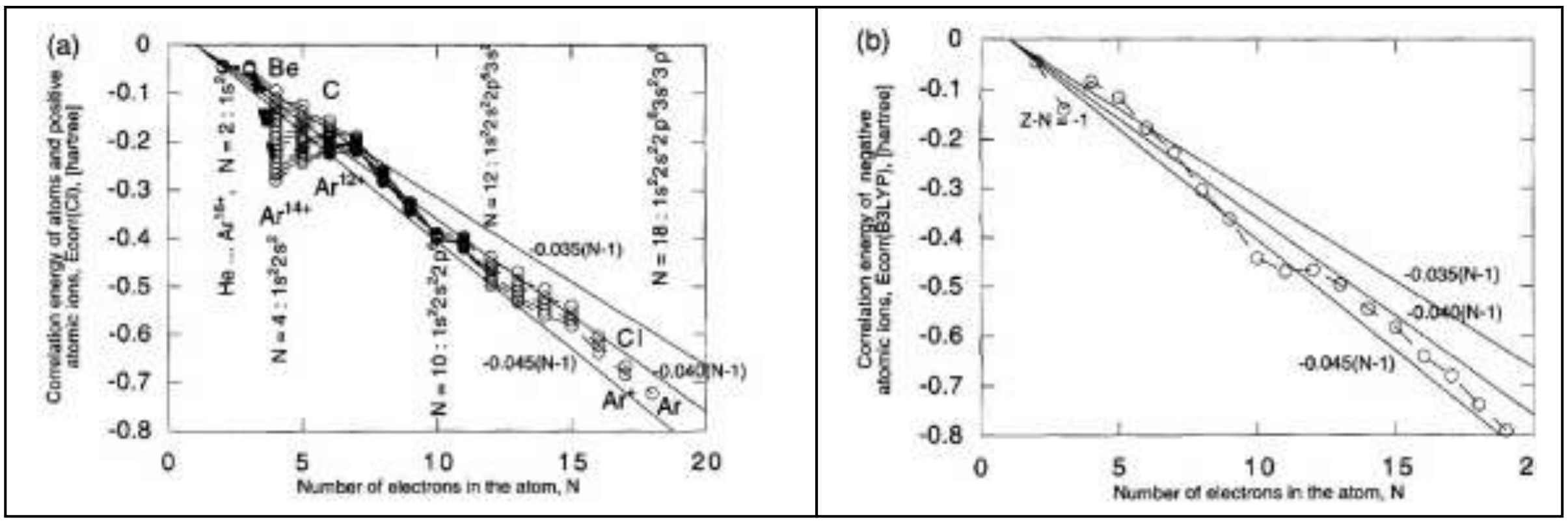

$E_{corr}$(method) vs. N plots: in CI calc. the positive atomic ions are exhibited and iso-Z values connected (a, dashed lines), in B3LYP calc. the -1 charged atomic anions with $1 \leq Z \leq 18$ are exhibited (b, dashed line), 6-311+G(3df,2p) basis set is used (a,b). Both plots empirically justify 1$^{st}$ eq. by the help of the straight solid lines with scanning slopes (except in the neighborhood N=3-6 (a), the case of extremally charged atoms). (CI calculations for negative ions were missing in literature up to 1997 as well as $He^{-1}$, $Ne^{-1}$, $Mg^{-1}$, $Ar^{-1}$ do not exist in free space, although technically HF-SCF ground state energy and B3LYP correlation calculation can be performed. Dianions, $A^{-2}$, do not exists in free space, but can be stable in media, e.g. $SO_4^{+2}$). B3LYP is primarily suggested for neutral and positively charged systems, but used now for $A^-$. For Z>18 relativistic effect needs to be involved, as vell as there is no correlation if N=1. The CI plot shows the shell structure effect on $E_{corr}$ as visible breaks at N=2, 4, 10, 12, however, the closed (N is even) vs. open shell (N is odd) alternation is not visible in contrast to that correlation calculations are ‘easier’ for cosed shells. For large molecules, macroscopic media, crystals, metals, etc. 1$^{st}$ eq. becomes: $-(\partial E_{corr}/\partial N)_{\{ZA,RA\}M} \approx 0.035 \div 0.045$.

Further empirical rule is that the 1$^{st}$ eq. or these plots are transferable to molecular systems (nuclear frame independence property), as the following calculation represents this: Considering the $E_{corr}$(LYP)/(N-1) values of methane, water, ammonia, alanine, glycine, pyrrole neutral molecules and their highly ionized states all at their neutral equilibrium geometries with TX90 program [P.Pulay - G.Fogarasi et al. 1982-1990, as well as DFT subroutines were written by Kristyan in 1993-1994 for it] on an IBM machine a weighted least square fit has yielded

$$E_{corr} \approx a(N-1) \text{ with } -a=-0.03829 \text{ hartree}$$

for 1$^{st}$ eq. with about 2.3% deviation. (The basis set level in these HF-SCF calculations corresponded to a 6-31G* one in Gaussian program.) Test was also done for the $H+HNO_3$ system potential energy surface, see here later, where three transition states can be found, and 1$^{st}$ eq. was found to be hold with the help of B3LYP, MP2, CCSD and G2M sophisticated methods as well as the improved ESP-0 method of 1$^{st}$ eq. (using Gausian92/DFT with a bit more accurate basises than 6-31G*). Important is that 1$^{st}$ eq. is hold for transition states as well, that is, 1$^{st}$ eq. holds for stationary points (equilibrium and transition states). Constant in eqs.1-2 is basis set dependent value and particularly belongs to 6-31G* or similar level, its basis set dependence has been shown in other work of mine.

1$^{st}$ eq. has size consistency problem (like HF-SCF and unlike CI). For example, if an N-electron molecule dissociates into two entities containing $N_1$ and $N_2$ electrons ($N=N_1+N_2$), the discrepancy between the two estimations for $E_{corr}$ is $a(N-1) - [a(N_1-1)+a(N_2-1)] = a(N-1) -$

a(N-2) = a > chemical accuracy, although for large N the a(N-1) ≈ a(N-2) still holds for the global property.

The first improvement of 1st eq., notated as $E_{corr}$(ESP-i), was done with the help of electrostatic charges or potential (ESP), where i is the molecular charge as i≡ $\Sigma_{A=1...M}Z_A$-N= $\Sigma_{A=1...M}$ESP-i(A), (the latter side must hold for the ESP partial charges on atoms), and with the electron content on atom A defined as $N_A$= $Z_A$ - ESP-i(A):

$$E_{corr}(ESP\text{-}i) = \Sigma_{A=1...M}\, E_{corr}(N_A, Z_A)\ .$$

For atoms (M=1), i= $Z_A$-N and $N_A$=N, and e.g. for neutral (i=0) $NO_2$ with HF-SCF/6-311G(d,p) level wave function by Gaussian92/DFT, the ESP-0(A) = 0.523942 and -0.261971 on N and (two equivalent) O atoms, resp. (M=3). $Z_A$ is always integer, but $N_A$ is generally not if M>1. In this approximation $E_{corr}(N_A,Z_A)$ is the HF-SCF/basis atomic correlation energy if $N_A$ is integer ($N_A$=$N_A$*), and if not (practically this is the case for M>1), the linar interpolation is

$$E_{corr}(N_A, Z_A)\equiv\ [(N_A^*+1)-N_A]E_{corr}(N_A^*,Z_A) + (N_A-N_A^*)E_{corr}(N_A^*+1,Z_A),$$

where $N_A$* is the integer part of $N_A$, (e.g. 3.14*=3). The $E_{corr}(N_A^*,Z_A)$ is the HF-SCF/basis atomic correlation energy with the help of the known atomic (i=0,1,2,… i.e $N_A^*\le Z_A$) CI ground state energies from the literature and the calculated B3LYP/basis (i<0 i.e. $N_A^*>Z_A$), see the figures above. $E_{corr}(N_A,Z_A)$=0 is taken for $0\le N_A\le 1$ (in practice $0<N_A\le 1$ occures only, otherwise if $N_A$=0, that $Z_A$ does not belong to that chemical system), the most frequent case for H atoms (e.g. in $CH_4$, etc.), but in the 1 anstrom nuclear distance unstable $LiCl^{18+}$ the $0<N_A<<1$ on Li, since all the 2 electrons are practically on 1s of Cl, that is, for the correlation energy of this sytem $E_{corr}$(ESP-18 for $LiCl^{18+}$)≈ Ecorr(CI for $Cl^{15+}$). The latter shows the advantage of 3rd eq., but the zero contribution of partially positive H atom in molecular systems is a source of error, that is improved in later works.

Notes on eqs.3-4.:

1.: Only small finite number of atoms come up, e.g. in organic- or biochemistry, mostly the H, C, N, O and S for which $N_A$* (or $N_A$*+1) is integer, while enormous number of molecules can be built of these atoms, and always owning $N_A$ values nearby, i.e. $|N_A-Z_A| < 1$ if i=0 (neutral covalent molecules); as well as neutral and anion molecules have ± partial charges, while more and more positive molecules (kations) have only positive partial charges.

2.: There are about half dozen popular partial charge definitions in the literature based on different physical views, they give similar, but different values, so 3rd eq. is not unique, but can be accurate if a particular partial charge is chosen. (One of the most important tests for a partial charge definition is to reproduce the experimental dipole moment.)

3.: No spin dependence in 3rd eq., e.g. LYP estimation does have. However, there is one more hidden property in 4th eq. what should be considered: For example, for carbon atom, the ($1s^2,2s^2,2p_x^1,2p_y^1$) triplet state is the ground state, and not the singlet ($1s^2,2s^2,2p_x^2$) to use for an $E_{corr}(N_A^*,Z_A)$ value, although in molecular environment, the carbon atom is very likely in singlet bound state. The true values is somewhere between, e.g. a fitted parameter, it has been targeted and developed in later work.

4.: 3rd eq. has a better size consistency than 1st eq., e.g. if $NH_3$ dissociates to 3H+N, the latter system has the exact correlation energy.

5.: 3rd eq. is in principle not-variational (like the popular and sophisticated B3LYP as well), but variational-like, at least, that is, it always stayes around the real $E_{corr}$, and never diverge, (the ESP is a plausible partial charge subroutine).

6.: $E_{corr}(N_A^*,Z_A)$ values belong to spherical atoms, while when atoms in molecular environment or bond, their electron cloud is not spherical, a source of error in 4th eq.. However, the $\{N_A\}_{A=1...M}$ set of values is definitely a measure or description of the molecular system, that is, they implement the effect of nuclear frame what eqs.1-2 lack or miss, so eqs.3-4 improves 1st eq., e.g. in estimating potential barriers, the most important quantity targeted by computation chemistry. This fortunate property comes from that $E_{corr}$ is an integrated property, so the partial charge.

Test of $E_{corr}$(ESP-i) estimation (3rd eq.) of correlation energy for the $H+HNO_3$ reaction system: The scheme is

$H+HNO_3(T1) \rightarrow H_2+NO_3$,

$H+HNO_3(T2) \rightarrow (ON)OH_2(T_3) \rightarrow OH+cisHONO(T_5) \rightarrow H_2O+NO_2$ ,

where $T_1,T_2,T_3,T_5$ are transition states ($T_4$ transition state exist on the potential energy surface but not involved here, named and described by us in the literature). $E_{corr}$(method)/(N-1) vs. N is plotted below (along with 2nd eq.) – notice that G2M is the most accepted method and ESP-0 (neutral molecules so i=0) is pretty close to it. Notice that MP2 and CCSD acccounts only for valence electrons ($N_v$), i.e. only parts of $E_{corr}$, as well as LYP overestimates the $E_{corr}$.

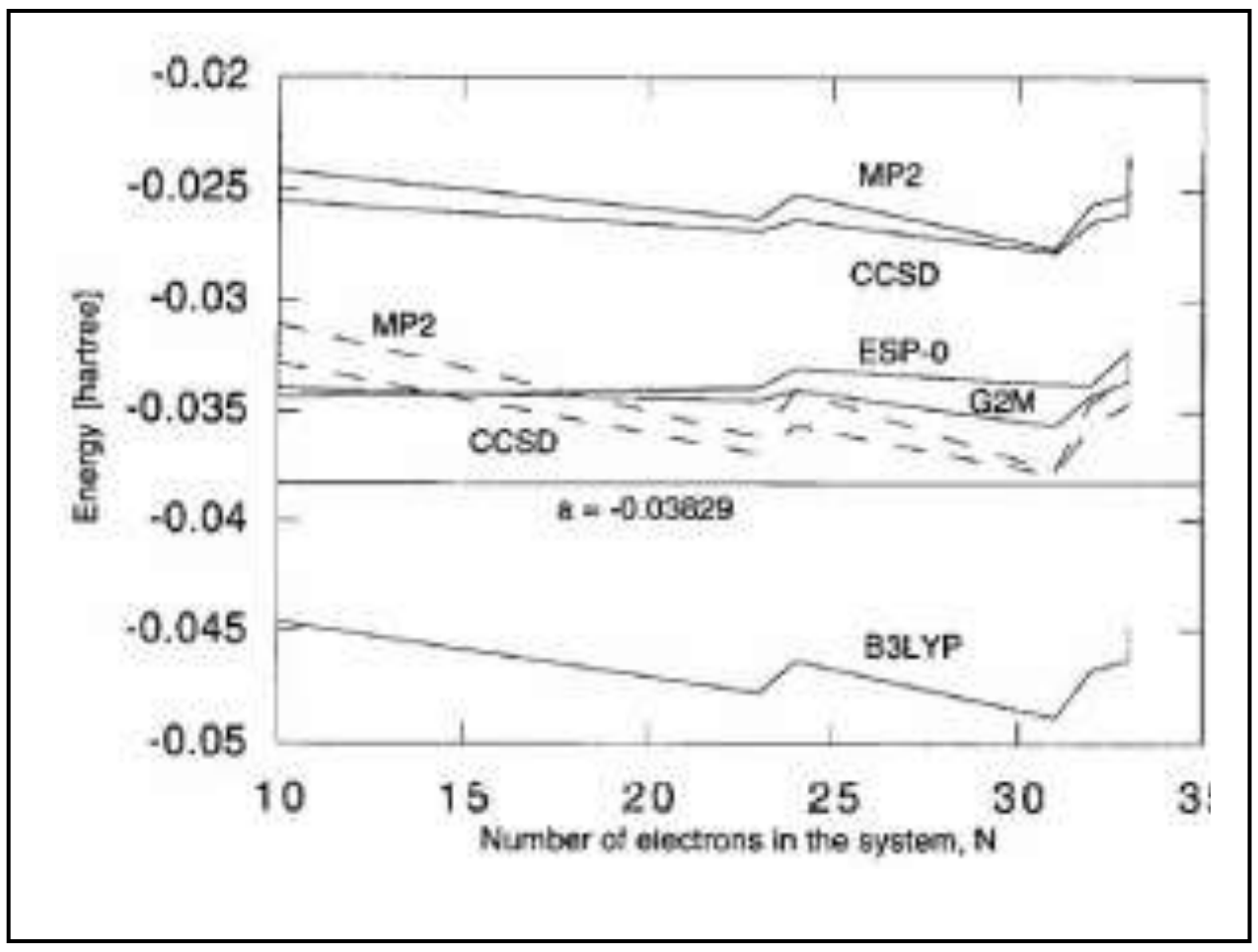

Solid lines are $E_{corr}$(method)/(N-1), dashed lines are $E_{corr}$(method)/($N_v$-1), where $N_v$ is the number of valence electrons, so for MP2 and CCSD the dashed lines make more sense w/r to eqs.1-2.

The difference between the two MP2 and two CCSD curves w/r to themself may indicate that core electrons (frozen core) should not be considered as intact w/r to $E_{corr}$, because dashed MP2 and CCSD are more parallel to B3LYP (total correlation) than the solid MP2 and CCSD (valence correlation).

Notice that $E_{corr}(B3LYP)/E_{corr}(MP2) \approx N/N_v \approx 1.5\text{-}2$ supports 1st eq..

Notice that ESP-i method is instant after HF-SCF/basis, while all other methods are time and disc space consuming, except B3LYP, but that is not instant either.

Comment on the term N-1 in 1st eq.: It is plausible in that at N=1 (no correlation) it reproduces $E_{corr}=0$. Correlation energy ($E_{corr}$, coming from the inadequate treatment of Slater determinant for electron-electron interactions) comes from the N(N-1)/2 electron pairs, but one electron ‘feels’ only N-1 neighbors around. Eqs.1-2 are derived empirically, and not theoretically, but in the view of the established central role of ground state one-electron density, $\rho(\mathbf{r}_1)$, in DFT, the one-electron operator, $h_1(\mathbf{r}_1) \equiv -(1/2)\nabla_1^2 + (1/2)\Sigma_{j=2,\dots,N} r_{1j}^{-1} - \Sigma_{A=1,\dots,M} Z_A R_{A1}^{-1}$, may support eqs.1-2 in the sense that the N-1 multiplier in 1st eq. originates from $\Sigma_{j=2,\dots,N}$, even the $E_{corr}(B3LYP)/E_{corr}(MP2) \approx N/N_v \approx 1.5\text{-}2$ on figure above support this.

**Thesis-6-Theory: Development in RECEP**

The feasibility of extremely rapid estimation of correlation energy, $E_{corr}$(ESP-i), a simple linear relationship using the HF-SCF basesd partial charge distribution in closed-shell molecules has been analyzed. This method has been further refined by introducing fitted quasi-universal “atomic correlation energies (parameters) in molecular environment”, renamed as RECEP (rapid estimation of correlation energy from partial charges) and analyzed for 18 molecules and ions, as well as the new method has been compared to the B3LYP, CCSD and G2 methods.

= = = = = =

Analizáltam a különösen gyors korrelációs energia közelítés, nevezett $E_{corr}$(ESP-i) alkalmazhatóságát, mely egy egyszerű lineáris összefüggés ami felhasználja a HF-SCF alapon számított parciális tölteseloszlást zárt héjú molekuláris rendszerekben. E módszert tovább finomítottam úgynevezett illesztett qvázi-univerzális “molekula környezetbeli atomi korrelációs energiák (paraméterek)” definiálásával és beillesztésével, újra neveztem mint RECEP (rapid estimation of correlation energy from partial charges = korrelációs energia gyors közelítése parciális töltések felhasználásával) és analizáltam 18 molekula és ion esetében, valamint ezen új módszert összehasonlítottam a B3LYP, CCSD és G2 módszerekkel.

**<u>Sandor Kristyan - Gabor I. Csonka: Chemical Physics Letters, 307 (1999) 469-478</u>**

= = = = = =

Representative equations/tables/figures:

While HF theory is well-defined and unique for closed-shell molecules, several versions of HF theory are used for open-shell molecules. Correlation energy for an open-shell molecule is usually defined with respect to unrestricted Hartree–Fock (UHF) theory where the spatial orbitals are different for $\alpha$ and $\beta$ spins. However, some authors prefer to define it with respect to restricted Hartree–Fock (RHF) theory where the spatial orbitals for $\alpha$ and $\beta$ spins are identical. Sometimes, it may be convenient to replace the HF approximation in the above definition with another well-defined approximation such as a multiconfiguration reference function. Here, only closed-shell molecules are studied and RHF used for such molecules. The electrons with antiparallel and parallel spins must be clearly differentiated because the HF method provides a good description for the electrons with parallel spins (Pauli exclusion) and poorer (average) description for the antiparallel spins.

Here, the GAUSSIAN 94 program was used and the 6-311+G(2d,p) basis set was selected for all methods. HF–SCF, full and frozen core CCSD(T), G2 and B3LYP calculations were performed. The natural population analysis (NPA) charges were calculated via the HF–SCF method using the G2 geometries, and the geometries were reoptimized in the B3LYP calculations. The reason to change the ESP (electrostatic potential partial charges) to NPA is that NPA is based on *ab initio* calculation and bond theory, while ESP is a semi-classical definition (classical point charge is moved around the molecular electron cloud). For various estimations of the molecular correlation energy, the following notations are used: $E_{corr}$(method) with method= CCSD, G2, B3LYP, RECEP, while $E_{corr}$ itself denotes the accurate complete-CI value. For atoms, the number of electrons (N), atomic charge (Z) and method is noted in the argument, e.g. E(CI,N,Z), and sometimes the spin states.

Like in case of ESP, the renamed and developed RECEP also starts with

$$E_{corr}(RECEP) = \Sigma_{A=1 \ldots M} E_{corr}(N_A, Z_A)$$

,i.e. the two basic assumptions, which are in fact theorems in it, can be summarized as follows: the correlation energy is the sum ($\Sigma_{A=1\ldots M}$) of the atomic correlation energies ($E_{corr}(N_A, Z_A)$) and the atomic correlation energies can be estimated from the partial charges. The proof of the first theorem is quite simple, and it has been strongly supported empirically by the author using ESP charges as well: The "gradient vector field analysis (see Bader's theory) of the electron density ($\rho(\mathbf{r}_1)$)" provides that the molecules can be cut apart to virial atoms. The zero flux surfaces* of the electron density give the borders for these virial atoms and in these atomic volumes the virial theorem is fulfilled. The total energy can be calculated from the sum of the virial atomic energies. (*= with an example of biological shape instead of $\rho(\mathbf{r}_1)$: the neck has this property separating the head from torso…) The proof of the second theorem is very difficult and we shall use it as a work hypothesis.

Partial charges are mathematical constructions that may help chemists to establish empirical rules. We recall here only four atomic charge definitions. First we mention the widely used Mulliken charges. The deficiencies of this method are well-known: for example the Mulliken charges are oscillating with respect to the basis set increase and do not show convergent behavior, so a rule of thumb is to use the small STO-3G basis set. However its great advantage is that it comes immediately from the educated distribution of LCAO parameters among the atoms in molecular systems. The charges, derived from the electrostatic potential (ESP) and from natural population analysis (NPA) show considerably better stability with respect to the increase of the basis set. Finally, we mention the charges derived from Bader's population analysis, coming directly from the integration of $\rho(\mathbf{r}_1)$ in the domains of virial atoms. Its disadvantage is its hectic algorithm, which are rarely available in commertional programs, as well as beside it suffers numerical instabilities, there exist examples where charges are assigned to spatial domains without nuclei.

The value of e.g. $E_{corr}$(LYP) change only a little if we introduce HF–SCF limit electron density. Similar observations have been made for the LSDA electron density, which is successfully used for correlation energy calculations in gradient corrected functionals. This observation originates in the fact that the $E_{corr}$ is an integrated quantity with respect to the $\rho(\mathbf{r}_1)$, thus more accurate electron density causes little change in the $E_{corr}$, recall the constarin $\int\rho(\mathbf{r}_1)d\mathbf{r}_1= N$. With respect to 1st eq., the latter luckily means that ratio of electron content around any two atoms, the $N_A:N_B$, changes much slower with basis set, than the $E_0$ or $\rho(\mathbf{r}_1)$. (If non-integrated quantities, e.g. $\nabla\rho(\mathbf{r}_1)$ in gradient corrections, appear in $E_{corr}$(method), the change in basis set, and as a consequence in $\rho(\mathbf{r}_1)$, can have strong or stronger effect.)

The CI values were used for $E_{corr}$(ESP), but it is not the best choice for 1st eq.. The reason is that the atoms change their spin state in molecules, so using the open-shell high spin multiplett correlation atomic energies would provide some bias. For example, in case of $CH_4$, the partial charge on C is between 0 and -1, yielding electron content $6<N_A<7$. The accurate CI calculations provide the correlation energies for high spin states, i.e. $E_{corr}$(CI, $N_A$=6, $Z_A$=6, triplet, i.e. $1s^2\ 2s^2\ 2p_x\ 2p_y$) and $E_{corr}$(CI, $N_A$=7, $Z_A$=6, quartet, i.e. $1s^2\ 2s^2\ 2p_x\ 2p_y\ 2p_z$) values for a linear interpolation. However, in a closed-shell (at equilibrium or close to equilibrium) methane there is no unpaired electron around the carbon atom; the molecule is singlet. Unfortunatelly, the correlation energy is very sensitive to spin pairing effects because the opposite spin electrons have different correlation energy than the parallel spin electrons.

Thus instead of using the correlation energy of high spin atomic states the energy of the excited or low spin states is proposed, e.g. $E_{corr}$(CI, $N_A$=6, $Z_A$=6, singlet, i.e. $1s^2\ 2s^2\ 2p_x^2$) and $E_{corr}$(CI, $N_A$=6, $Z_A$=7, doublet, i.e. $1s^2\ 2s^2\ 2p_x^2\ 2p_y$). It is seemingly contradictory to use excited state parameters in approximating ground state molecules with 1st eq., however, we mention that the CI method is superior to the HF–SCF method in approximating ground state ($E_0$) value, because excited Slater determinants are also used beside the ground one.

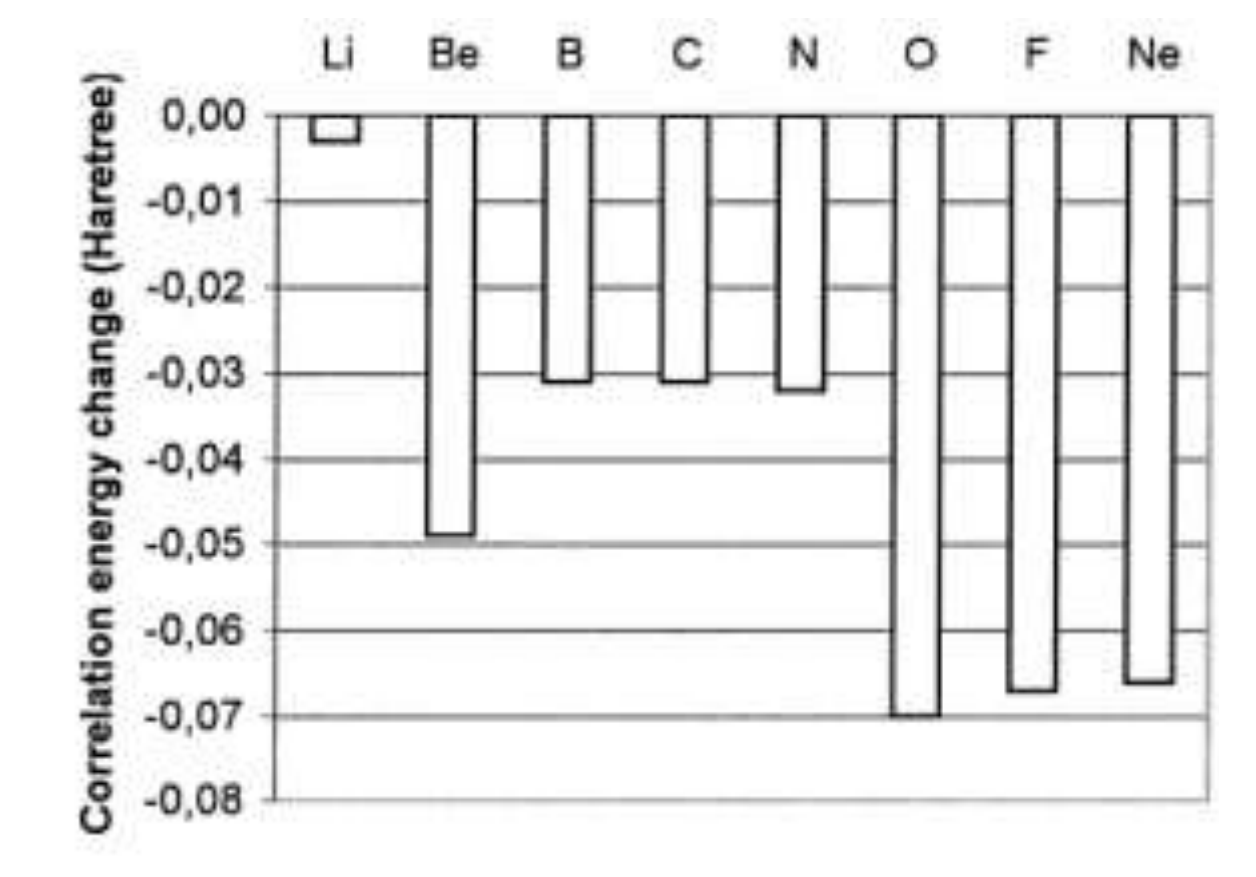

The exact correlation energy (CI from literature and HF-SCF/6-311+G(2d,p)) differences between the neighboring elements going from left to right in the first row of the periodic table are shown, e.g. Li shows the correlation energy difference between He and Li, Be shows the difference between Li and Be, etc.. This chart shows the important spin pairing effect in correlation energy (e.g. the pairing in 2s shell in Be, and in 2p shell from the oxygen atom (cf. Hund's rule) is manifesting).

Atomic correlation energies in hartree to use in 1st eq.:

| Atom | N | Z | $E_{corr}$(N,Z,CI, low spin) | $E_{corr}$(N,Z,B3LYP, low spin) |
|---|---|---|---|---|
| H | 2 | 1 | -0.0395 | -0.0432 |
| He | 2 | 2 | -0.0420 | -0.0531 |
| Li | 2 | 3 | -0.0435 | -0.0491 |
| | 3 | 3 | -0.0453 | -0.0593 |
| Be | 2 | 4 | -0.0443 | -0.0425 |
| | 3 | 4 | -0.0474 | -0.0601 |
| | 4 | 4 | -0.0943 | -0.0994 |
| B | 4 | 5 | -0.1113 | -0.1060 |
| | 5 | 5 | -0.1249 | -0.1316 |
| | 6 | 5 | -0.1640 | -0.1765 |
| C | 5 | 6 | -0.1388 | -0.1400 |
| | 6 | 6 | -0.1754 | -0.1911 |
| | 7 | 6 | -0.2087 | -0.2258 |
| | 8 | 6 | -0.2839 | -0.2883 |
| N | 6 | 7 | -0.1856 | -0.2005 |
| | 7 | 7 | -0.2143 | -0.2373 |
| | 8 | 7 | -0.2877 | -0.3035 |
| | 9 | 7 | -0.3314 | -0.3622 |
| O | 7 | 8 | -0.2202 | -0.2445 |
| | 8 | 8 | -0.2839 | -0.3079 |
| | 9 | 8 | -0.3314 | -0.3619 |
| | 10 | 8 | -0.4080 | -0.4513 |
| F | 9 | 9 | -0.3245 | -0.3599 |
| | 10 | 9 | -0.3995 | -0.4430 |
| Ne | 10 | 10 | -0.3905 | -0.4338 |

In the table bove, the high spin and low spin $E_{corr}$(N,Z,B3LYP) for N= 6, 7, and 8 electronic systems were compared, the changes are approximately -0.019, -0.026, and -0.026 hartree, respectively. The exact CI values were corrected with these (high-to-low-spin-corection) values for N=6, 7, and 8, named for the first row atoms and ions, forcing them to be in singlet or doublet states, because accurate CI calculations of low spin states were not available in 1999:

$E_{corr}$(N,Z,CI, low spin)≡ $E_0$(N,Z,CI, high spin) - $E_0$(N,Z,HF-SCF, high spin) + "high-to-low-spin-correction", furthermore,

$E_{corr}$(N,Z,B3LYP, low spin)≡ $E_0$(N,Z,B3LYP, low spin) - $E_0$(N,Z,HF-SCF, low spin) values were also calculated; also, for HF-SCF and B3LYP the RHF and 6-311+G(3df,2pd) basis set was used.

<u>$E_{corr}$(RECEP-c) denotes</u> here the RECEP calculated with using $E_{corr}$(N,Z,CI, low spin) and NPA charges for atomic $E_{corr}(N_A, Z_A)$ interpolated as in $E_{corr}$(ESP-i), (c stands for "corrected exact" or CI), and <u>$E_{corr}$(RECEP-d) denotes</u> here the RECEP using $E_{corr}$(N,Z,B3LYP, low spin), (d stands for DFT).

Our experience is that instead of full-CI or B3LYP, e.g. CCSD(T) or other sophisticated methods, as well as instead of NPA other sophisticated partial charge methods can also be used for $E_{corr}(N_A, Z_A)$, detailed in other works.

Notice that, 1., e.g. singlet for a neutral nitrogen atom is fictitious, since only doublet or quartet exists in atomic state, 2., these atomic correlation parameters are calculated in free space to use in molecular environment, 3., all these may eliminate the dilemma of the use of UHF vs. RHF correlation energy.

The hydrogen atoms require special attention because their partial charges fall frequently in $0<N_A<1$, and using zero atomic correlation values in this range is not plausible. In ESP-i method, the algorithm for $E_{corr}(N_A, Z_A)$ was a.: for $0<N_A<1 \Rightarrow E_{corr}(0\le N_A\le 1, Z_A=1) = 0$, because 1 electron has no correlation,

b.: for $1\le N_A<2 \Rightarrow$ linear interpolation of $E_{corr}(N_A=1, Z_A=1) = 0$
and $E_{corr}(N_A=2, Z_A=1$,B3LYP, low spin (singlet))= -0.0432 hartree,

c.: In RECEP method the algorithm for $E_{corr}(N_A, Z_A)$ is the case b, but for the wider $0<N_A<2$.

Cases a-b yield $E_{corr}$(ESP-i) = 0 for $H_2$ molecule (wherein $N_A$=1 on both atoms via ESP or NPA charges as well), which is essentially incorrect, but case c yields reasonable $E_{corr}$(RECEP) correlation energy for the $H_2$ molecule as well (-0.0432/2-0.0432/2=-0.0432h).

<u>Test of RECEP estimation for correlation energy</u>: Test 1.: Important 10-electron systems were selected first, the protonated, neutral and deprotonated water and ammonia, and the methane and deprotonated methane. In these systems (G2 opt. geom.) the correlation energy changes systematically depending on the number of protons and lone pairs:

$$CH_4, CH_3^-, NH_4^+, NH_3, NH_2^-, OH_3^+, OH_2, OH^-.$$

All correlation calculations (CCSD-full and frozen core, G2, B3LYP, RECEP-c and d) show the <u>same tendencies</u>: e.g. the deprotonation of the neutral molecules changes the G2 correlation energies by -0.015±0.002 hartree, the protonation of one of the free lone pairs change the correlation energies by about +0.0055 hartree.

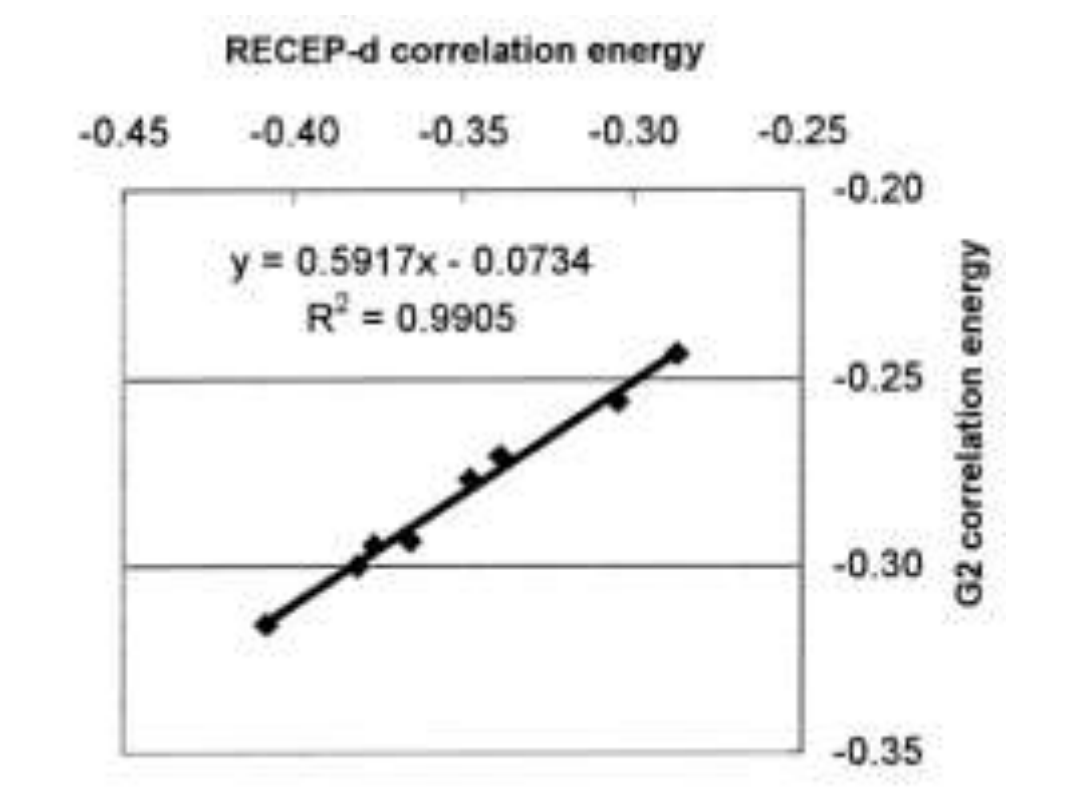

There is an excellent agreement ($R^2$ stat. corr. ≈ unity, 0.99) between the RECEP-d and G2 correlation energies. Notice that the G2 method does not include the core correlation, thus the RECEP-d correlation energies are considerably more negative by about 40 – 80 millihartrees, see the 0.0734 h stat. anal. const. Also notice that G2 is a computationally expensive method while the RECEP is instant, (after HF-SCF and NPA calc., it can be calculated even on a pocket calculator…).

Test 2.: For larger scope of molecules: NPA charges on the non-hydrogen atom in the molecule in a.u.; RECEP-c, RECEP-d, and G2 correlation energies in hartree; the molecules are listed in monotonic order with N, and if N is the same, with G2 correlation energy values:

| Molecule | N | NPA | RECEP-c | RECEP-d | G2 |
|---|---|---|---|---|---|
| LiH | 4 | 0.812 | -0.0799 | -0.0901 | -0.0343 |
| BeH2 | 6 | 1.218 | -0.1107 | -0.1258 | -0.0696 |
| BH3 | 8 | 0.434 | -0.1645 | -0.1946 | -0.1232 |
| CH4 | 10 | -0.711 | -0.2646 | -0.2868 | -0.2433 |
| CH3- | 10 | -1.335 | -0.2870 | -0.3043 | -0.2558 |
| NH4+ | 10 | -0.824 | -0.3181 | -0.3388 | -0.2706 |
| NH3 | 10 | -1.031 | -0.3282 | -0.3478 | -0.2766 |
| NH2- | 10 | -1.512 | -0.3397 | -0.3657 | -0.2933 |
| OH3+ | 10 | -0.777 | -0.3451 | -0.3763 | -0.2945 |
| HF | 10 | -0.564 | -0.3755 | -0.4162 | -0.2951 |
| OH2 | 10 | -0.927 | -0.3492 | -0.3812 | -0.2998 |
| F- | 10 | -1.000 | -0.3995 | -0.4430 | -0.3150 |
| OH- | 10 | -1.362 | -0.3718 | -0.4080 | -0.3155 |
| LiF | 12 | -0.976 | -0.4413 | -0.4904 | -0.3064 |
| C2H2 | 14 | -0.225 | -0.3966 | -0.4312 | -0.3408 |
| B2H6 | 16 | 0.017 | -0.3693 | -0.3926 | -0.2724 |
| C2H4 | 16 | -0.339 | -0.4395 | -0.4774 | -0.3544 |
| C2H6 | 18 | -0.510 | -0.4839 | -0.5251 | -0.3745 |

On H atoms, the NPA can be obtained from symmetry and molecular charge using the NPA values on non-hydrogen atoms listed; in case of LiF the NPA refers about Fluor atom. Figure below shows the linear fits for the RECEP and G2 relative correlation energies from table above. (In order to decrease the effect of the missing core correlation from the G2, the correlation energies were recalculated relative to LiH molecule, (relative means: values deviating with respect to LiH).

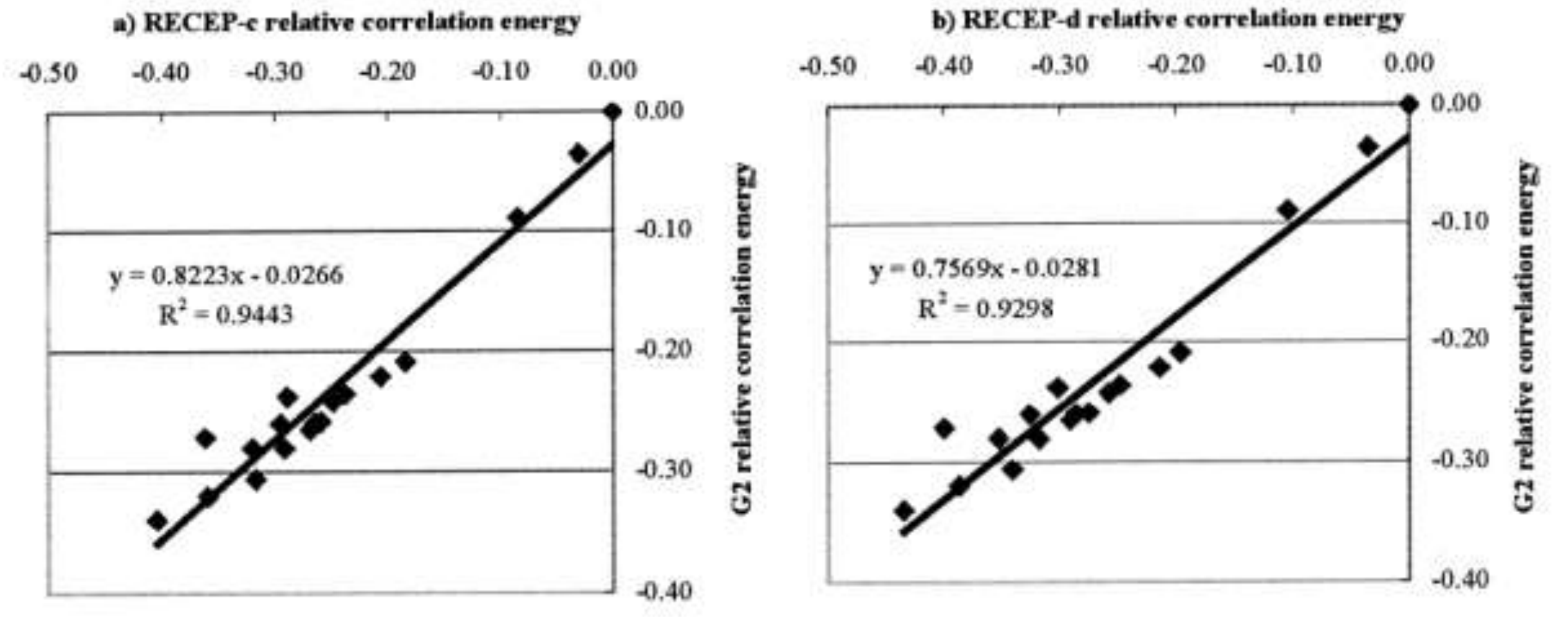

The quasi-linear dependence of the correlation energy on the partial charges (particularly on NPA) in molecules has been developed and analyzed in 1st eq.. It is literally a DFT method based on a totally different technique than the known other methods, to give a taste, NPA comes from integrating (operator effect on) $\rho$, or e.g. like in the famous LDA (exchange) operator and approximation $E_{corr}(LDA)= \int O_{LDA}(\rho)d\mathbf{r}_1$, where $O_{LDA}(\rho) \equiv$ cont.$\rho^{4/3}$, for which the integral goes (switches) to N if $O_{LDA}(\rho) \to O_{identity}(\rho) \equiv \rho$, the 1st eq. behaves in same way.

**Thesis-7-Theory: Fitting atomic correlation parameters in RECEP, test on Gaussian-2 total energies**

The accuracy of the RECEP method has been increased considerably by the use of fitted atomic correlation parameters. RECEP is an extremely rapid, practically prompt calculation of the total correlation energy of closed shell ground-state neutral molecules at stationary (minimums and transition states) points after a HF-SCF calculation. The G2 level correlation energy and HF-SCF charge distribution of 41 closed-shell neutral molecules (composed of H, C, N, O, and F atoms) of the G2 thermochemistry database were used to obtain the fitted RECEP atomic correlation parameters (with 1.8 kcal/mol standard deviation). Four different mathematical definitions of partial charges, as a multiple choice, were used to calculate the molecular correlation energies. The best results were obtained using the natural population analysis (NPA), although the other three are also recommended for use. Test was done on a different, non-overlapping set of other 24 molecules from the G2 thermochemistry database showing 2.3 kcal/mol standard deviation, which means that the fitted RECEP parameters are transferable. Extension for charged molecules, radicals, and molecules containing other atoms is straightforward. Numerical example as a recipe is also provided.

= = = = = =

A RECEP módszer pontosságát jelentősen javítottam illesztett atomi korrelációs paraméterek bevezetésével. A RECEP egy rendkívül gyors, azonnali számítás a zárt héjú, alap állapotú, semleges molekulák teljes korrelációs energia becslésére stacionárius pontok (minimumok és átmeneti állapotok) közelében egy a HF-SCF számítást követően. Összesen 41 zárt héjú, semleges, H, C, N, O, és F atomokból álló molekula G2 szintű korrelációs energiáját valamint HF-SCF töltés eloszlását használtam a G2 termokémiai adatbázisból a RECEP atomi korrelációs paraméterek illesztéséhez (szórás= 1.8 kcal/mol). Négy különböző matematikai definíción alapuló parciális töltést használtam, mint választási lehetőséget, a molekuláris korrelációs energia számítására. A legjobb eredményt a „natural population analysis (NPA)" szolgáltatta, bár a másik hármat szintén javasolhatom használatra. Tesztet végeztem különböző, nem átfedő, 24 db G2 termokémiai adatbázisból származó molekulán, amely 2.3 kcal/mol szórást mutatott, tehát megállapítható, hogy az illesztett RECEP paraméterek transzferálhatók. A módszer egyértelműen kiterjeszthető töltött molekulákra, gyökökre, és más atomokat is tartalmazó molekulákra. A numerikus számítási receptet leközöltem.

**Sandor Kristyan - Gabor I. Csonka:**
**Journal of Computational Chemistry, 22 (2001) 241-254**

= = = = = =

Representative equations/tables/figures:

HF-SCF/6-311CG(2d,p) and G2 calculations were performed, using the G2 geometries. The ChelpG (widely used in molecular mechanics programs), Merz–Kollman (MK), Mulliken (the ancient one based on LCAO parameters), and natural population analysis (NPA, very good basis set convergence property) partial charges, easily available in commertial programs, were calculated from the HF-SCF wave function using the Gaussian 94 program package.

$E_{corr}(\text{RECEP}) = \Sigma_{A=1\ldots M} E_{corr}(N_A, Z_A)$ is supposed to be used for molecules in the vicinity of equilibrium nuclear configurations and transition states, i.e., in chemical bond, but not for van der Waals or London dispersion forces; anyway, HF-SCF calculation is not supposed to be done in those regions. Important to note that the "RECEP atomic correlation energies or parameters", $E_{corr}(N_A,Z_A)$, in a molecule are necessarily different from the atomic correlation energies $E_{corr}$(CI,N,Z, spin state) of the corresponding atoms in free space, although they have similar values, or at least, trends.

Important is how to calculate the $E_{corr}(N_A, Z_A)$: The noninteger $N_A$ can be treated by linear interpolation between nearby integer values ($N \leq N_A \leq N+1$) as proposed previously. Initial guess was that these terms for interpolation can be derived from the CI correlation energy of ground state, high spin atomic ions in free space ($E_{corr}$(ESP-i)), then it was improved by using lowest lying low spin (ground or excited) states ($E_{corr}$(RECEP), non-fitted parameters), but the accuracy was not enough. However, these parameters can be calculated for the whole periodic system (although for atoms with $Z > 18$ some ralativistic correction is necessary), e.g. for boron (B) compounds or carbon (C) compounds it can be done in the same way as long as HF-SCF/basis calculation is available. With "RECEP fitted atomic correlation parameters", see below, the accuracy can be improved strongly, but one needs molecule set to use for fit for which accurate, e.g. G2, CCSD or else can be done as pre-calculation. However, in practice one can find this set easily for the above mentioned carbon compounds, but not really for boron compounds, and this is a limitation of fitted RECEP parameters. For using fitted atomic correlation parameters, the notation is not changed, $E_{corr}$(RECEP) has been used in literature, but in the discussion the word "fit" must be mentioned.

Fitting procedure for $E_{corr}(N_A, Z_A)$ provides the optimal values for the "RECEP atomic correlation parameters" using 41 known G2 molecular correlation energies for a larger set of small molecules. These fitted parameters in hartree (noted as RECEP-fit) approaches chemical accuracy for molecular correlation energies for closed-shell, singlet, ground-state neutral molecules in the vicinity of stationary points:

| Atom | N | Z | Exactb | Exactc | B3LYPd | ChelpG | MK | Mulliken | NPA |
|---|---|---|---|---|---|---|---|---|---|
| H | 2 | 1 | -0.0395 | -0.0395 | -0.0432 | -0.0406 | -0.0408 | -0.0397 | -0.0376 |
| C | 4 | 6 | -0.1264 | -0.1264 | n.a. | n.a. | n.a. | -0.1171 | -0.1105 |
| | 5 | 6 | -0.1388 | -0.1388 | -0.1400 | -0.1385 | -0.1381 | -0.1423 | -0.1387 |
| | 6 | 6 | -0.1564 | -0.1754 | -0.1911 | -0.1653 | -0.1650 | -0.1660 | -0.1659 |
| | 7 | 6 | -0.1827 | -0.2087 | -0.2258 | -0.1868 | -0.1864 | -0.1866 | -0.1909 |
| N | 6 | 7 | -0.1666 | -0.1856 | -0.2005 | -0.2259 | -0.2265 | -0.2515 | -0.2227 |
| | 7 | 7 | -0.1883 | -0.2143 | -0.2373 | -0.2281 | -0.2280 | -0.2240 | -0.2259 |
| | 8 | 7 | -0.2617 | -0.2877 | -0.3035 | -0.2333 | -0.2333 | -0.2302 | -0.2351 |
| | 9 | 7 | n.a. | n.a. | -0.3622 | n.a. | n.a. | n.a. | -0.3804 |
| O | 8 | 8 | -0.2579 | -0.2839 | -0.3079 | -0.2692 | -0.2690 | -0.2712 | -0.2703 |
| | 9 | 8 | -0.3314 | -0.3314 | -0.3619 | -0.2743 | -0.2738 | -0.2646 | -0.2790 |
| F | 9 | 9 | -0.3245 | -0.3245 | -0.3599 | -0.2901 | -0.2903 | -0.2879 | -0.2892 |
| | 10 | 9 | -0.3995 | -0.3995 | -0.4430 | -0.2956 | -0.2940 | -0.3048 | -0.3061 |

n.a. = not available

Exactb = $E_{corr}$(N,Z,CI, high spin) for ground-state atoms in free space (RHF);
n.a.: double anions are not stable in free space.

Exactc= $E_{corr}$(N,Z,CI, low spin) corrected atomic CI correlation energies:
-0.019 hartree for N=6, -0.026 hartree for N=7 and 8, as a quick additive adjustment to estimate low spin singlet or doublet (not necessarily ground state) in free space (CI is n.a). Using these is called RECEP-c.

B3LYPd= $E_{corr}$(N,Z,B3LYP, low spin) for singlet or doublet (not necessarily ground state) low spin state in free space, a difference of B3LYP and HF-SCF/6-311CG(3df,2pd). Using these is called RECEP-d.

Exactb-c-d are true atomic correlation energies in free space, but the values under partial charges are RECEP atomic correlation parameters only to estimate molecular correlation energies.

Exactb-c-d values can be used with any partial charge but give only moderate results.

Values under the four types of partial charge are the optimized parameters, calculated with a linear fit to the correlation energies of the 41 G2 molecules, as well as these are restricted to use the partial charge indicated, but provide accurate results, more, all these four have about the same quality, and can provide the chemical accuracy, (n.a.: because partial charge values requiring these ($N_A$, $Z_A$) value pairs did not occur in the used 41 molecules during the linear fit. (However, a proper larger set can target these and other, unlisted values.) Using these for Ecorr($N_A$, $Z_A$) in the case of different partial charges (for a multiple choice) is called RECEP-fit atomic correlation parameters. It is also important that the HF-SCF/6-311+G(2d,p) level partial charges (and HF-SCF energies) are recommended to calculate, i.e., in this way the deviation from the basis set limit is incorporated in $E_{corr}$(RECEP) as well.

Statement: Comparing "atomic correlation energies in free space" under columns Exactb-c-d to "atomic correlation energies in molecular environment" under columns named after partial charged used reveal that these values are very close to each other at any (N,Z) coordinate.

The abacus or recipe of RECEP method (c, d, fit) is represented via the case of methyl-nitrite ($CH_3$-O-N=O) using fitted RECEP atomic correlation parameters from table above:

| $Z_A$ | NPA | $N_1$ | $N_A$ | $N_2$ | $E_{corr}(N_1,Z_A)$ | $E_{corr}(N_2,Z_A)$ | $E_{corr}(N_A,Z_A)$ |
|---|---|---|---|---|---|---|---|
| 6 | -0.133 | 6 | 6.133 | 7 | -0.1659 | -0.1909 | -0.1692 |
| 8 | -0.490 | 8 | 8.490 | 9 | -0.2703 | -0.2790 | -0.2746 |
| 1 | 0.171 | 0 | 0.829 | 2 | 0.0 | -0.0376 | -0.0156 |
| 1 | 0.165 | 0 | 0.835 | 2 | 0.0 | -0.0376 | -0.0157 |
| 1 | 0.165 | 0 | 0.835 | 2 | 0.0 | -0.0376 | -0.0157 |
| 7 | 0.504 | 6 | 6.496 | 7 | -0.2227 | -0.2259 | -0.2243 |
| 8 | -0.382 | 8 | 8.382 | 9 | -0.2703 | -0.2790 | -0.2736 |

NPA charge in a.u. was calculated at the HF-SCF/6-311+G(2d,p) level,
$E_{corr}(N_A, Z_A)$ is linearly interpolated using $E_{corr}(N_1,Z_A)$ and $E_{corr}(N_2,Z_A)$

$N_A = Z_A - NPA_A$, $N_1 < N_A < N_2 \equiv N_1+1$, while for hydrogen ($Z_A$=1) the $N_1$=0 and $N_2$ = 2.

$E_{corr}$($N_1$ or $N_2$, $Z_A$) are from column NPA of previous table.

Finally: $E_{corr}$(RECEP-fit)= $\Sigma_{(A)}E_{corr}(N_A,Z_A)$= -0.9886 hartreee, this value has to be added to the HF-SCF/6-311+G(2d,p) value, cf. G2 value in table for 41 molecules below. $E_{corr}$(G2)= -0.9875 hartree, and Ecorr(G2)- Ecorr(RECEP-fit)= 0.001131 hartree≈ 0.7 kcal/mol, an excellent agreement. Notice that a HF-SCF calculation with partial charge calculation (now NPA) on this molecule is the only *ab initio* calculation, and it demands much less disc space and CPU time than, for example, a G2 calculation. This calculation of $E_{corr}$(RECEP-fit) after can be done even on a pocket calculator. (The accuracy of RECEP method has been introduced here; thus, the problems arising from the geometry optimization is not addressed.)

Table below shows the result for the set of 41 neutral closed-shell molecules composed of H, C, N, O, and F atoms, selected from the G2 thermochemistry database. Those molecules were selected for which the most reliable experimental heat of formations were available, their $E_{corr}(G2)$ values were used to obtain the fitted $E_{corr}(N_A,Z_A)$ RECEP atomic correlation parameters, see the four columns headed by partial charges above. The improvement of RECEP-fit over RECEP-d,c is manifesting:

| | Molecule | $E_0$(HF-SCF) hartree | $E_{corr}$(G2) hartree | RECEP-d kcal/mol | RECEP-c kcal/mol | RECEP-fit kcal/mol |
|---|---|---|---|---|---|---|
| 1 | CH4 | -40.2102 | -0.2433 | 27.2 | 13.0 | 1.3 |
| 2 | NH3 | -56.2150 | -0.2767 | 44.5 | 32.0 | 0.0 |
| 3 | H2O | -76.0527 | -0.2999 | 50.8 | 30.8 | -0.9 |
| 4 | HF | -100.0526 | -0.3063 | 68.7 | 43.2 | 0.4 |
| 5 | C2H2 | -76.8422 | -0.3698 | 38.4 | 16.6 | 1.5 |
| 6 | H2C=CH2 | -78.0584 | -0.4064 | 44.4 | 20.4 | 3.0 |
| 7 | H3C—CH3 | -79.2541 | -0.4480 | 48.0 | 21.7 | 1.7 |
| 8 | HCN | -92.8979 | -0.4031 | 36.8 | 14.4 | 1.9 |
| 9 | H2C=O | -113.9033 | -0.4617 | 54.8 | 29.2 | 1.7 |
| 10 | CH3—OH | -115.0815 | -0.5028 | 68.0 | 36.4 | 0.8 |
| 11 | H2N—NH2 | -111.2174 | -0.5149 | 65.3 | 39.4 | -0.3 |
| 12 | HO—OH | -150.8235 | -0.5684 | 75.7 | 40.6 | 0.2 |
| 13 | CO2 | -187.6892 | -0.6835 | 82.3 | 30.1 | -0.2 |
| 14 | CF4 | -435.7780 | -1.3054 | 240.1 | 107.9 | 0.6 |
| 15 | COF2 | -311.7100 | -0.9954 | 160.3 | 68.7 | -0.3 |
| 16 | N2O | -183.7207 | -0.7274 | 40.6 | -1.9 | -1.5 |
| 17 | NF3 | -352.6474 | -1.1020 | 156.1 | 75.4 | 1.1 |
| 18 | C2F4 (F2C =CF2) | -473.5672 | -1.4763 | 245.1 | 142.8 | -1.8 |
| 19 | CF3CN | -428.6011 | -1.4111 | 206.7 | 101.8 | -1.3 |
| 20 | Propyne (C3H4) | -115.8984 | -0.5744 | 60.1 | 26.4 | 2.3 |
| 21 | Allene (C3H4) | -115.8970 | -0.5739 | 59.2 | 26.2 | 2.3 |
| 22 | Cyclopropene (C3H4) | -115.8553 | -0.5800 | 55.5 | 21.7 | -1.8 |
| 23 | Propylene (C3H6) | -117.1082 | -0.6132 | 64.6 | 28.6 | 2.3 |
| 24 | Cyclopropane (C3H6) | -117.0916 | -0.6175 | 62.3 | 26.3 | -0.2 |
| 25 | Propane (C3H8) | -118.2994 | -0.6551 | 67.6 | 29.4 | 0.7 |
| 26 | Trans-butadiene | -154.9667 | -0.7793 | 80.2 | 34.4 | 2.2 |
| 27 | 2-butyne | -154.9525 | -0.7795 | 81.3 | 35.5 | 2.6 |
| 28 | C4H6 | -154.9303 | -0.7843 | 77.8 | 32.0 | -0.6 |
| 29 | Bicyclobutane | -154.9120 | -0.7916 | 73.0 | 27.2 | -5.3 |
| 30 | Cyclobutane (C4H8) | -156.1390 | -0.8257 | 80.5 | 32.5 | -2.2 |
| 31 | Isobutene (C4H8) | -156.1577 | -0.8219 | 83.5 | 35.6 | 0.5 |
| 32 | Trans-butane (C4H10) | -157.3446 | -0.8626 | 87.0 | 36.7 | -0.6 |
| 33 | Isobutane (C4H10) | -157.3451 | -0.8649 | 85.8 | 35.6 | -1.9 |
| 34 | Spiropentane (C5H8) | -193.9673 | -0.9957 | 94.9 | 37.1 | -4.3 |
| 35 | Benzene (C6H6) | -230.7633 | -1.1134 | 111.1 | 45.7 | 1.1 |
| 36 | H2CF2 | -237.9779 | -0.7722 | 134.6 | 79.8 | 1.7 |
| 37 | HCF3 | -336.8798 | -1.0393 | 185.5 | 104.3 | 0.8 |
| 38 | H3C—NH2 | -95.2473 | -0.4811 | 60.3 | 35.3 | -1.2 |
| 39 | CH3—CN | -131.9605 | -0.6062 | 58.5 | 26.0 | 3.2 |
| 40 | CH3—NO2 | -243.7359 | -0.9920 | 88.3 | 30.5 | -0.6 |
| 41 | CH3—O—N=O | -243.7366 | -0.9875 | 89.9 | 31.7 | 0.7 |

RECEP-d, RECEP-c, RECEP-fit= $E_{corr}$(G2)- $E_{corr}$(RECEP-d,-c,-fit)
$E_0$(HF-SCF)= HF-SCF/6-311+G(2d,p) level total energies,
#27: Dimethylacetylene (2-butyne), #28: Methylenecyclopropane (C4H6)

$E_0(G2) = E_0(HF\text{-}SCF) + E_{corr}(G2)$ total energy for ground electronic state $\Rightarrow$

The energy differences directly provide the deviation compared to the G2 energy:

$E_{corr}(G2) - E_{corr}(RECEP\text{-}d,\text{-}c,\text{-}fit) = E_0(G2) - E_0(RECEP\text{-}d,\text{-}c,\text{-}fit)$ and

$E_0(RECEP\text{-}d,\text{-}c,\text{-}fit) = E_0(HF\text{-}SCF) + E_{corr}(RECEP\text{-}d,\text{-}c,\text{-}fit)$ .

The basis set error is incorporated into the Ecorr(RECEP-fit) correlation energy as well.

The geometry optimization was on MP2/6-31G(d) level.

The HF/6-311+G(2d,p) NPA charges were used for all three columns in these correlation calculations, so column NPA in the two tables above was used for the RECEP-fit.

For RECEP-d and RECEP-c, any partial charge can be used, but they do not reach chemical accuracy, while RECEP-fit does.

Obviously, all the zero point energies and thermal corrections are left out, because these are not necessary for the present purpose.

The HF-SCF basis set limit calculations are very demanding computationally, so moderate basis set was chosen.

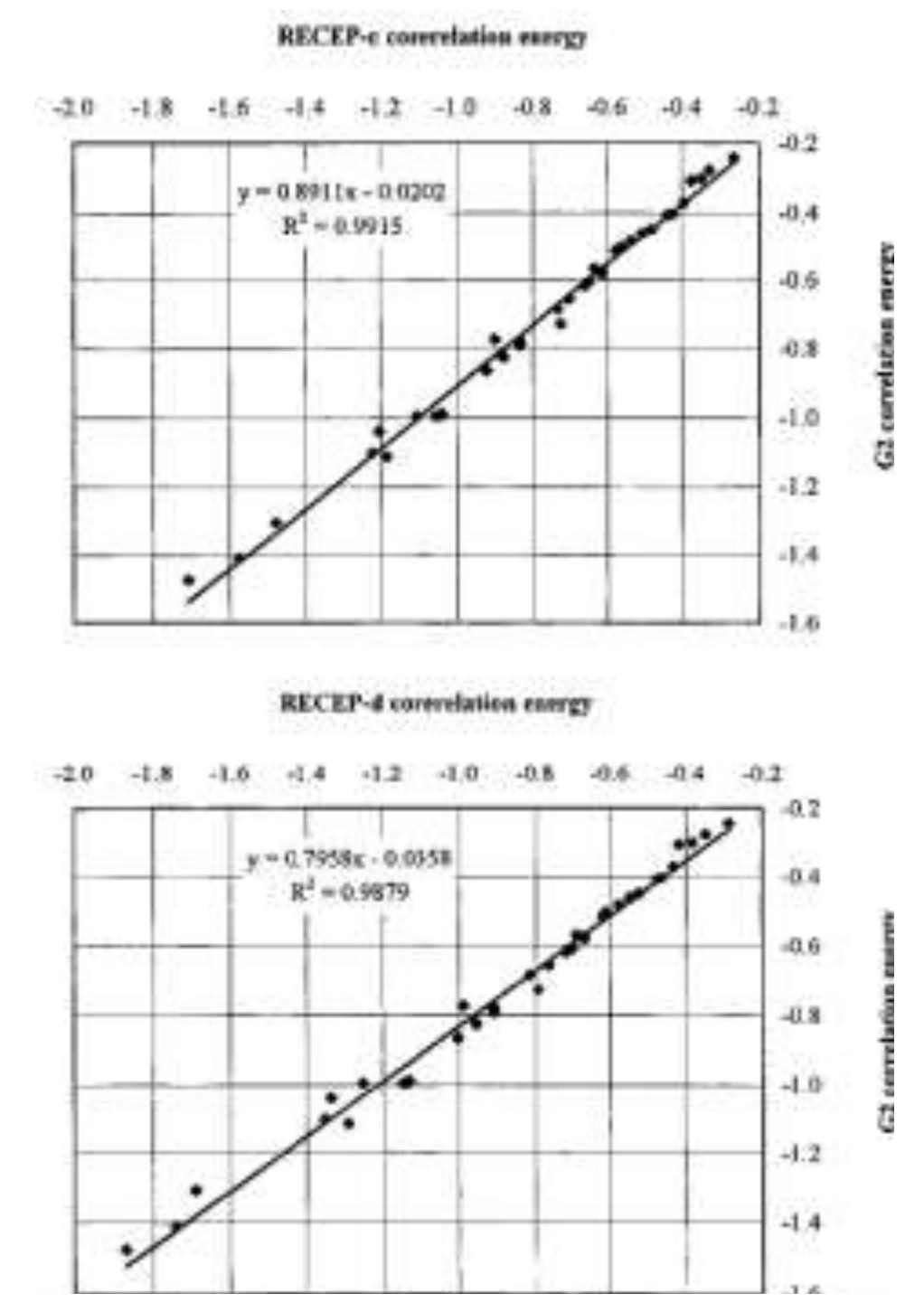

Statistical correlation between $E_{corr}$(RECEP-c,-d) and $E_{corr}$(G2) in hartree listed in table above. In spite of large differences, a very good linear relationship exists ($R^2$ =0.99), the essential difference is that in average:

$|E_{corr}(G2)/E_{corr}(RECEP\text{-}c)| = 0.8911$

$|E_{corr}(G2)/E_{corr}(RECEP\text{-}d)| = 0.7958.$

Using the linear equation parameters in this figure, the largest difference is reduced to 40 kcal/mol and the standard deviation is about 18 kcal/mol, a considerable improvement compared to the errors in table above showing columns RECEP-c and –d, however an *a posteriori* method such as RECEP-c and –d still canot reach the chemical accuracy.

The table above shows the superior property of RECEP-fit over RECEP-c and –d, but its transferability to other system not included in the set for fit must be demonstrated:

The set of 24 molecules (not included in the linear fit) to test the (self-consistency or transferability of) RECEP atomic correlation parameters in the NPA column in the first table in this thesis above:

| Molecule | $E_0$(HF-SCF) hartree | $E_{corr}$(G2) hartree | RECEP-fit kcal/mol |
|---|---|---|---|
| 1 HCOOH (formic acid) | -188.8266 | -0.7230 | -0.4 |
| 2 HCOOCH3 (methyl formate) | -227.8588 | -0.9296 | -1.2 |
| 3 CH3CONH2 (acetamide) | -208.0454 | -0.9055 | -1.7 |
| 4 C2H4NH (aziridine) | -133.0800 | -0.6519 | -3.0 |
| 5 NCCN (cyanogen) | -184.6243 | -0.7785 | -1.4 |
| 6 (CH3)2NH (dimethylamine) | -134.2834 | -0.6882 | -1.6 |
| 7 CH3CH2NH2 (*trans*) | -134.2953 | -0.6882 | -2.0 |
| 8 CH2CO (ketene) | -151.7713 | -0.6284 | 1.9 |
| 9 C2H4O (oxirane) | -152.9137 | -0.6748 | -2.4 |
| 10 CH3CHO (acetaldehyde) | -152.9638 | -0.6665 | 2.2 |
| 11 HCOCOH (glyoxal) | -226.6589 | -0.8878 | 2.2 |
| 12 CH3CH2OH (ethanol) | -154.1321 | -0.7091 | 0.3 |
| 13 CH3OCH3 (dimethylether) | -154.1148 | -0.7088 | 0.3 |
| 14 CH2=CHF (vinyl fluoride) | -176.9412 | -0.6734 | 2.0 |
| 15 CH2=CHCN (acrylonitrile) | -169.8097 | -0.7733 | 2.5 |
| 16 CH3COCH3 (acetone) | -192.0208 | -0.8731 | 1.7 |
| 17 CH3COOH (acetic acid) | -227.8859 | -0.9276 | 0.4 |
| 18 CH3COF (acetyl fluoride) | -251.8768 | -0.9340 | 0.9 |
| 19 (CH3)2CHOH (isopropanol) | -193.1819 | -0.9181 | -1.8 |
| 20 C2H5OCH3 | -193.1655 | -0.9154 | -0.5 |
| 21 (CH3)3N | -173.3212 | -0.8984 | -4.3 |
| 22 C4H4O (furan) | -228.6888 | -1.0114 | -4.6 |
| 23 C4H5N (pyrrole) | -208.8676 | -0.9900 | -5.5 |
| 24 C5H5N (pyridine) | -246.7597 | -1.1468 | -0.3 |

RECEP-fit= $E_{corr}$(G2)- $E_{corr}$(RECEP-fit); $E_0$(HF-SCF)= HF-SCF/6-311+G(2d,p) total energy.

The criteria used to obtain the fitted atomic $E_{corr}(N_A, Z_A)$ values was a simple least square fit for $\mathrm{Min}[\Sigma_{i=1\ldots149}(E_{corr}(G2)_i - E_{corr}(RECEP)_i)^2]$ using the linear behavior on $E_{corr}(N_A, Z_A)$. It compensates for the errors arising from the spin pairing effects, the partial charges, and limited basis set used for HF-SCF energy calculation, as well as transferable. It approximates sufficiently the G2 total energy after a simple HF-SCF partial charge and energy calculation, and is four to five orders of magnitude faster than the expensive G2 calc.. (Different parametrization is necessary for calculating correlation energy of radicals.) The root-mean-square deviations from the G2 total energy for the 41 molecules listed above in tables are 2.0, 2.0, 2.1, and 1.8 kcal/mol for ChelpG, MK, Mulliken and NPA charges, resp..

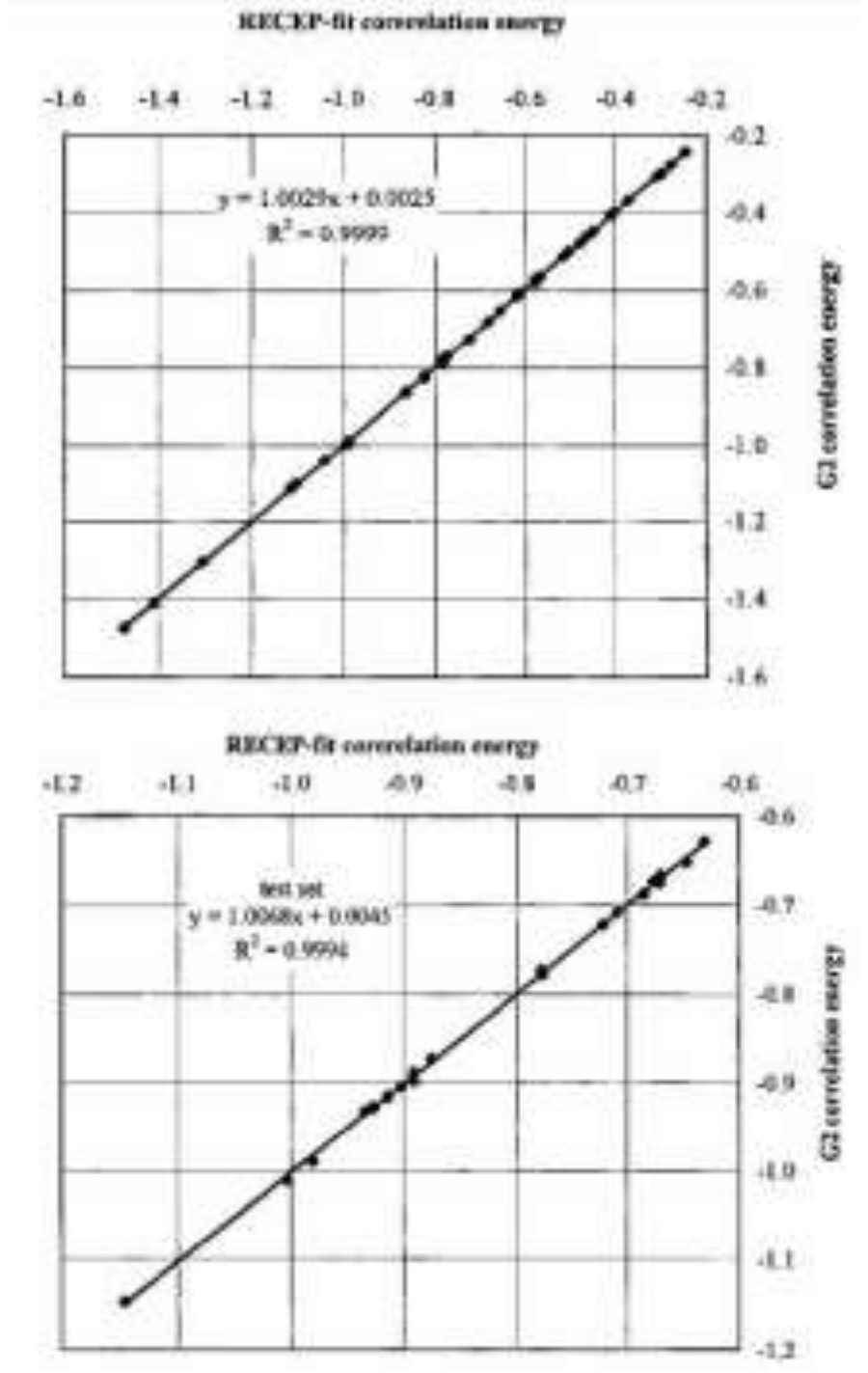

Statistical data for correlation energies $E_{corr}$(RECEP-fit) vs. $E_{corr}$(G2) in hartree listed in tables above, upper: training set of 41 molecules, lower: test set of 24 molecules. The agreement between the G2 and the RECEP-fit correlation energies is excellent:

$$|E_{corr}(G2)/E_{corr}(RECEP\text{-}fit)| = 1.00$$

up to two decimal.

In summary, a comparison of these statistics and tables shows that RECEP-c and RECEP-d do not achieve the chemical accuracy, while RECEP-fit does, more precisely, the RECEP-fit results approximate with mean absolute difference (MAD) < 2 kcal/mol the prestigious G2 results.

Note: The G2 total energies are of high quality and easily available, although further improvements are necessary. Benchmark quality *ab initio* total atomization energies (TAE) became recently available for small neutral molecules up to trans-butadiene and benzene; providing an opportunity to judge the quality of the HF-SCF/6-311+G(2d,p) and the G2 total energies and the correlation energy from these two: The G2 correlation energy accounts up to about 79% with statistical correlation of the excellent value $R^2$ = 0.9995 to the best estimate. The G2 method provides a reliable thermochemistry; however, the correlation energies derived from G2 calculations should be scaled up. This is in line with RECEP-c and -d results (cf. figures above) and with the recent G3 total energies available on the Internet.

**Thesis-8-Theory: Reproducing Gaussian-3 total energy with RECEP**

Gaussian-3 ground state total electronic energies have been approximated using single point HF-SCF total electronic energies plus the correlation energy corrections calculated from HF-SCF partial atomic charges according to the "rapid estimation of correlation energy from partial charges" (RECEP-3) method as an improvement over RECEP-2 (renamed from RECEP) which was based on G2/97 thermocemistry database. The best results to calculate the molecular correlation energies were obtained using natural population analysis (NPA), the other three partial charge definitions used ChelpG, Merz-Kollman (MK) and Mulliken provide slightly worse results. The overall root-mean-square deviation of the fitted RECEP-3 energies from Gaussian-3 energies for 65 molecules is 1.76 kcal/mol. The root-mean-square deviation of calculated RECEP-3 enthalpies of formation from experimental enthalpies of formation for the 65 molecules is 2.17 kcal/mol. The current fitted RECEP-3 parameters are recommended to estimate total correlation energies of closed-shell ground-state neutral molecules at the vicinity of stationary points on the potential-energy surface composed of H, C, N, O, and F atoms.

= = = = = =

Gaussian-3 szintű alap állapot totális elektronikus energiákat közelítettem egypontos (adott geometriára vonatkozó) Hartree–Fock ön-konzisztens (HF-SCF) totális elektronikus energiák, plusz korrelációs energia korrekciók felhasználásával, utóbbit a HF-SCF parciális atomi töltések segítségével a "rapid estimation of correlation energy from partial charges" (RECEP-3) módszerem alapján, mely egy tovább fejlesztése a RECEP-2 (RECEP-ről átnevezve) módszernek. Utóbbi a G2/97 termokémiai adatbázisra volt alapozva. A molekuláris korrelációs energia értékekre a legjobb eredményt a „natural population analysis" (NPA) segítségével kaptam, a másik három parciális töltés definíció, mint alternatíva, nevezetesen a ChelpG, Merz-Kollman (MK) és Mulliken kicsit gyengébb eredményt szolgáltatott. Az illesztett RECEP-3 energiák átlagos négyzetes eltérése a Gaussian-3 energiáktól 65 molekula esetében 1.76 kcal/mol, míg a számolt RECEP-3 képződési entalpiák átlagos négyzetes eltérése a kísérleti értékektől a 65 molekula esetében 2.17 kcal/mol. A jelen illesztett RECEP-3 paramétereket javasolni tudom a totális korrelációs energiák közelítésére zárt héjú, alap állapotú, valamint H, C, N, O, és F atomokból álló molekulák esetében a potenciál felület stacionárius pontjai környezetében.

**Sandor Kristyan - Adrienn Ruzsinszky - Gabor I. Csonka:**
**Journal of Physical Chemistry A, 105 (2001) 1926-1933**
= = = = = =

Representative equations/tables/figures:

The so-called composed (or extrapolation) method, Gaussian-2 (G2) or Gaussian-3 (G3), has improved the very poor convergence and scaling with the basis-set increase in conventional expensive correlation methods. G3 theory achieves significantly improved accuracy compared to that of G2 theory, as well as G3 theory requires fewer computational resources than G2 theory. Both are well documented in literature, but still both need larger computation effort. (E.g.: G3 theory uses $2^{nd}$ order perturbation eq. geom. [MP2(FU)/6-31G(d)] and ZPE [HF/6-31G(d)] followed by a series of Møller-Plesset single point calculations at [MP2(FU)/G3large, MP4(FC)/6-31G(d), MP4(FC)/6-31+G(d), MP4(FC)/6-31G(2df,p)], and quadratic configuration interaction [QCISD(T,FC)/6-31G(d)], where FU= full $2^{nd}$ order and FC= frozen core approximation.

The recent density functional theory (DFT) methods use considerably faster algorithms (BPW91, B3PW91, and B3LYP) and provide good estimation for the correlation energy, as well as show a basis-set convergence similar to that of the HF-SCF method. However, there is no simple way to improve the results (cf. the use of semi empirical functionals), and numerical instabilities might occur as well (cf. numerical integrals).

RECEP theory is a radically different approach to calculate correlation energy very rapidly (almost instant) and effectively, requiring a simple HF-SCF energy calculation at the equilibrium geometry and a partial charge calculation on constituting atoms. RECEP-2 is a multi-linear fitting procedure for the RECEP atomic correlation parameters to reproduce the G2 total electronic energies from HF-SCF/6-311+G(2d,p) single-point total electronic energies and from the partial charges. The impressing superiority of the results of G3 theory over the results of G2 theory inspired the author to provide the called RECEP-3 parameters obtained from reproducing the G3 total electronic energies with using MP2(FU)/6-31G(d) geometries and single-point HF-SCF/6-311+G(2d,p) calculation. These calculations require several orders of magnitude less CPU time than the G3 calculations. The abacus or recipe is the same for RECEP-2 and -3, and instant after the HF-SCF routine.

In RECEP theory, the partial charges are used to estimate correlation energy ($E_{corr}$) which is quasi-linear with the number of electrons. Partial charges are essentially mathematical constructions that serve to reflect the electron content around the selected atom of the molecule. Not physically measurable quantities, because they depend on the definition of the partition scheme of the electron density, but e.g. they can be defined to reproduce the measurable dipole moment and electrostatic potential, as well as they are successfully applied to identify the electron-rich (nucleophile) and electron-poor (electrophile) functional groups of molecules.

$E_{par}$(N, Z, method) and $E_{fitpar}$(N, Z, method, charge definition,L) RECEP atomic correlation parameters (hartree) for $E_{corr}$(RECEP, method, charge definition, L) estimation for molecular correlation energies:

| | N | Z | CI | ls-CI | ls-B3LYP | G2 NPA 41 fit | G3 MK 41 fit | G3 ChelpG 41 fit | G3 Mulliken 41 fit | G3 NPA 41 fit | G3 NPA 65 fit |
|---|---|---|---|---|---|---|---|---|---|---|---|
| H | 2 | 1 | -0.0395 | -0.0395 | -0.0432 | -0.0376 | -0.0419 | -0.0417 | -0.0398 | -0.0374 | -0.0381 |
| C | 4 | 6 | -0.1264 | -0.1264 | -0.1079 | -0.1105 | n.a. | n.a. | -0.1515 | -0.1466 | -0.1487 |
| | 5 | 6 | -0.1388 | -0.1388 | -0.1400 | -0.1387 | -0.1802 | -0.1808 | -0.1821 | -0.1796 | -0.1783 |
| | 6 | 6 | -0.1564 | -0.1754 | -0.1911 | -0.1659 | -0.2094 | -0.2098 | -0.2106 | -0.2103 | -0.2111 |
| | 7 | 6 | -0.1827 | -0.2087 | -0.2258 | -0.1909 | -0.2322 | -0.2323 | -0.2357 | -0.2392 | -0.2361 |
| N | 6 | 7 | -0.1666 | -0.1856 | -0.2005 | -0.2227 | -0.2700 | -0.2696 | -0.2659 | -0.2640 | -0.2641 |
| | 7 | 7 | -0.1883 | -0.2143 | -0.2373 | -0.2259 | -0.2740 | -0.2741 | -0.2690 | -0.2721 | -0.2721 |
| | 8 | 7 | -0.2617 | -0.2877 | -0.3035 | -0.2351 | -0.2805 | -0.2805 | -0.2801 | -0.2833 | -0.2850 |
| O | 8 | 8 | -0.2579 | -0.2839 | -0.3079 | -0.2703 | -0.3161 | -0.3163 | -0.3184 | -0.3181 | -0.3171 |
| | 9 | 8 | -0.3314 | -0.3314 | -0.3619 | -0.2790 | -0.3237 | -0.3240 | -0.3265 | -0.3295 | -0.3298 |
| F | 9 | 9 | -0.3245 | -0.3245 | -0.3599 | -0.2892 | -0.3399 | -0.3397 | -0.3373 | -0.3396 | -0.3399 |
| | 10 | 9 | -0.3995 | -0.3995 | -0.4430 | -0.3061 | -0.3446 | -0.3460 | -0.3588 | -0.3575 | -0.3572 |

41, 65= 41 molecules from the G2 dataset have been increased to 41+24= 65 molecules from G3 dataset, L= # of molecules used in the fit: 41 or 65,

fit= multilinear least square fitting procedure as in RECEP-2 with 6-311+G(2d,p) basis set to G2 or G3 data set to obtain the RECEP-2 or 3 parameters, resp., last two columns differ by increasing the molecule set from 41 to 65 molecules,

N = number of electrons, Z = nuclear charge,

CI = correlation energies for ground-state atoms in free space,
ls-CI = estimated low-spin sate CI atomic correlation energies: -0.019 hartree correction for 6 electronic systems and -0.026 hartree correction for 7 and 8 electronic systems, low-spins are necessary because "atoms are in quasi low-spin states in molecules",
ls-B3LYP= low-spin state B3LYP atomic correlation energies (e.g., for carbon, the $1s^2 2s^2 2p_x^2$ singlet low-spin state) calculated as a difference of B3LYP/6-311+G(3df,2pd) and HF/6-311+G(3df,2pd) energies,
n.a. = not available= it was not necessary to calculate for the molecules, that is, out of range for partial charges to occure in practice,
G2 = fit to G2 energies called RECEP-2 parameters (RECEP renamed) to compare,
G3 = fit to G3 energies called RECEP-3 parameters as an improvement over RECEP-2,
$E_{par}$(N, Z, method) = atomic correlation energies= $E_{corr}$(method=CI or ls-CI or ls-B3LYP), to weight out molecular correlation energies, $E_{corr}$(RECEP), by the RECEP abacus, for H atom the background is green in the table for better groupping in view.
$E_{fitpar}$(N, Z, method, charge definition,L)= fitted, fictive atomic correlation energies, to to weight out molecular correlation energies, $E_{corr}$(RECEP), by the RECEP abacus, for H atom the background is yellow in the table for better grouping in view.

Statements: Table above contains the atomic RECEP parameters, all colums are capable to reproduce molecular correlation energies, $E_{corr}$, via RECEP abacus (which is a weighting with HF-SCF/basis partial charges on atoms). Columns with green colours are "simple atomic correlation energies", they can be exended simply to any atoms up to Z=18, (where relativistic effects starts to be pronounced), as $E_{corr}$(method,N,Z)= $E_{total\ electr,0}$(method,N,Z) - $E_{total\ electr,0}$(HF-SCF/basis,N,Z). Columns with yellow colours are (least square) "fitted atomic correlation energies" valid in RECEP theory, it can be extended with increasing the L= 41 or 65 size set including other atoms (Z) needed, creating a bit larger inconvenience over columns with green colours to aquire more atomic RECEP parameters. Accuracy strongly increases from left: The most accurate molecular $E_{corr}$ is provided by the last column.

Table above shows the following:

1.: low-spin B3LYP parameters < low-spin CI values,
2.: fitted RECEP-2 (G2, NPA,41) are closer to low-spin CI than to high-spin CI values, as expected by the low-spin states of atoms in molecular environment,
3.: fitted RECEP-3 (G3) parameters < corresponding RECEP-2 (G2) parameters as a consequence of the lower G3 energy,
4.: $E_{fitpar}(N,Z,NPA,41) \approx E_{fitpar}(N,Z,NPA,65)$ as a convergence to increased molecule set,
5.: G3 correlation parameters are similar for the MK and ChelpG partial charges,
6.: G3 correlation parameters are similar for the Mulliken and NPA partial charges,
7.: Somewhat larger difference can be observed between the ChelpG and NPA charges,
8.: fitted atomic RECEP correlation energy parameters (and the quality of the results) depend very slightly on the charge definition, thus the method is not sensitive to the partial charges used in the fitting procedure, but for a smaller increase in accuracy it should be distinguished.

The use of the HF-SCF partial charges, derived from the one-electron density, $\rho_{HF\text{-}SCF}(\mathbf{r})$, for molecular electron correlation effects ($E_{corr}$) can be justified readily in agreement that DFT correlation energy functionals provide adequate correlation energy using a relatively low-quality basis set, as well as that the $E_{corr}$ in DFT is an integrated quantity with respect to the

ρ(**r**), and the one-electron density integrates to the number of electrons in any case, that is, more accurate electron density causes a relatively small change in $E_{corr}$.

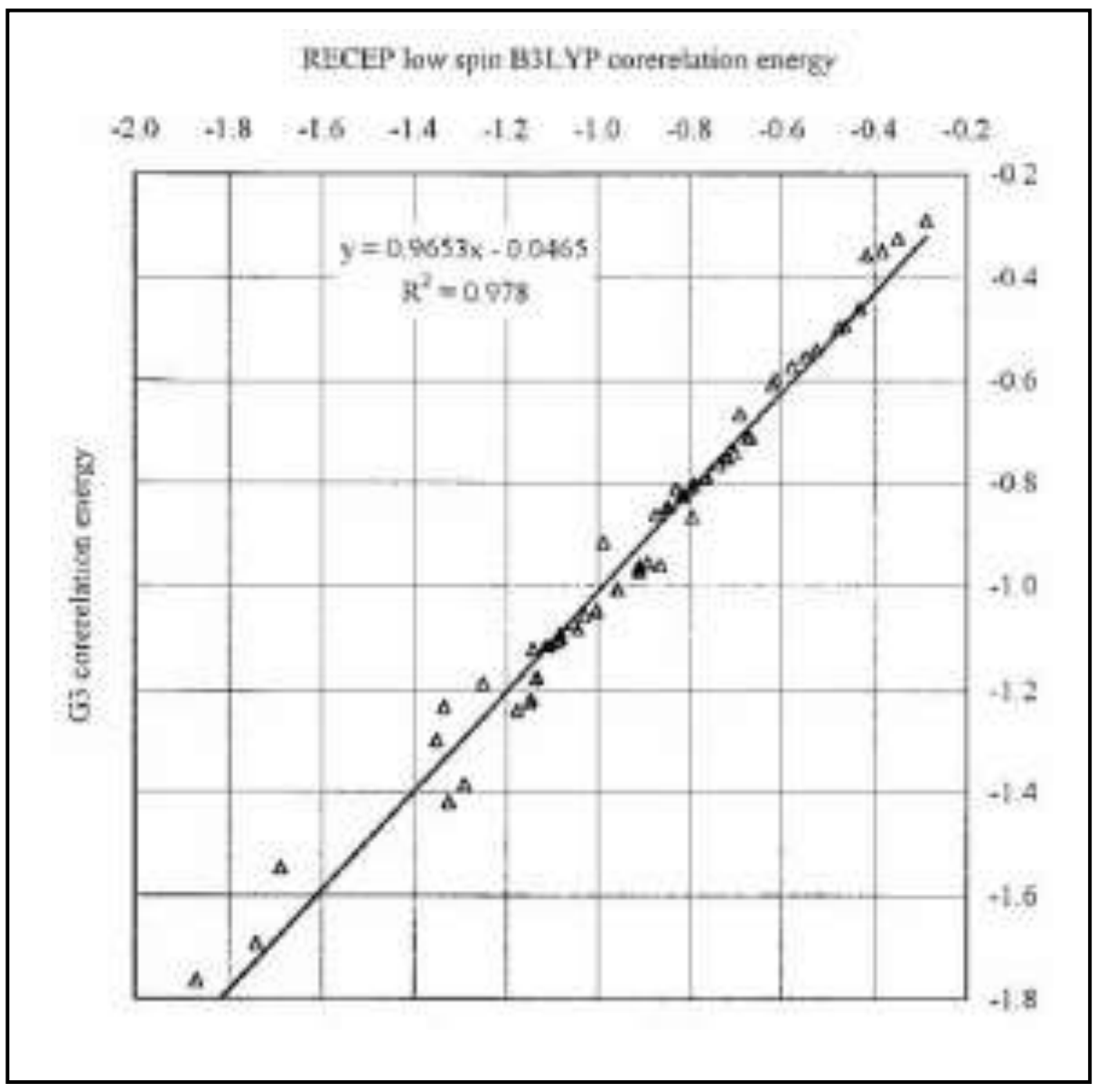

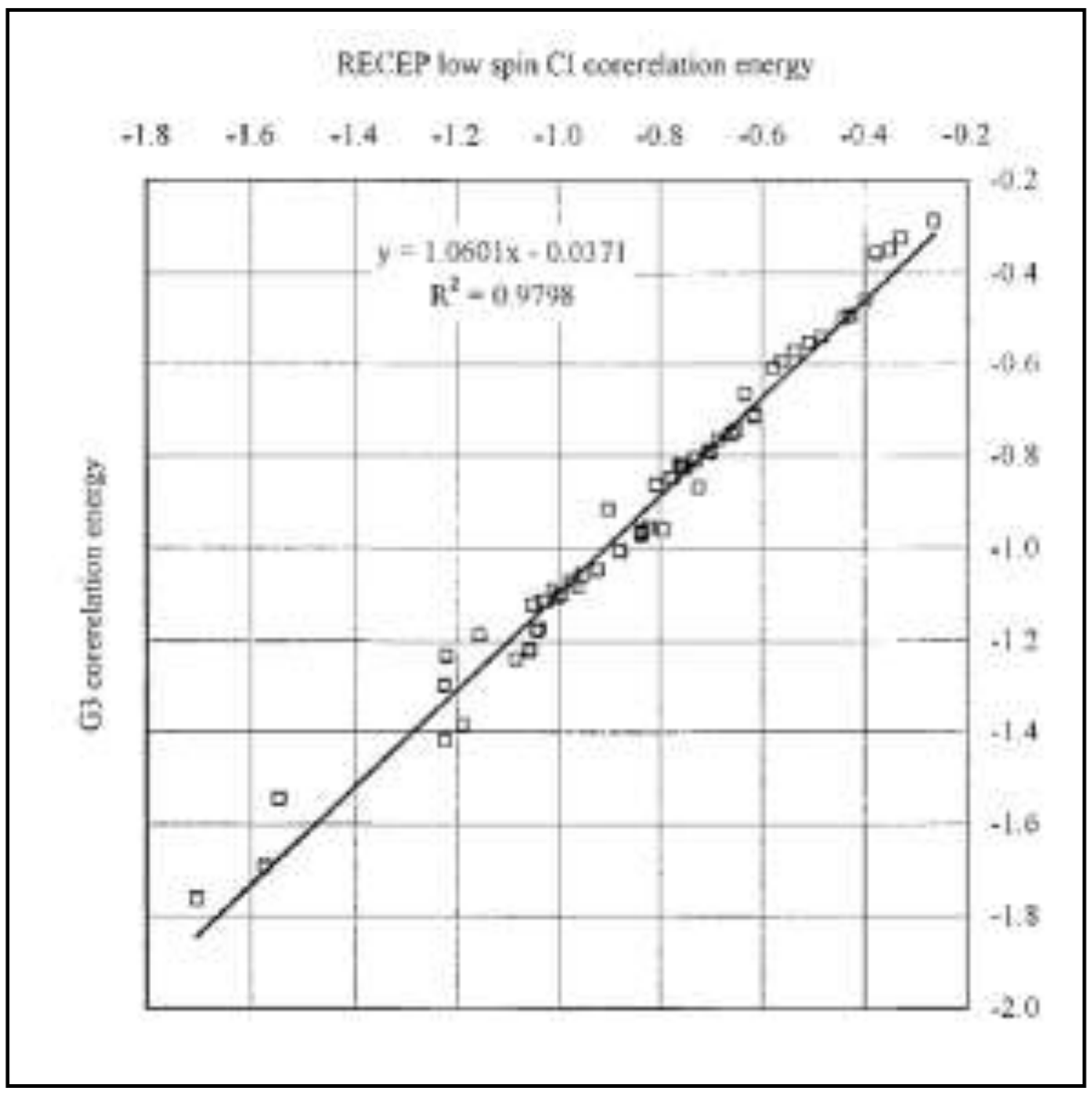

$E_{corr}$(RECEP-2, ls-B3LYP, NPA, renamed from RECEP-d) and $E_{corr}$(RECEP-2, ls-CI, NPA, renamed from RECEP-c) correlation energy vs. $E_{corr}$(G3).

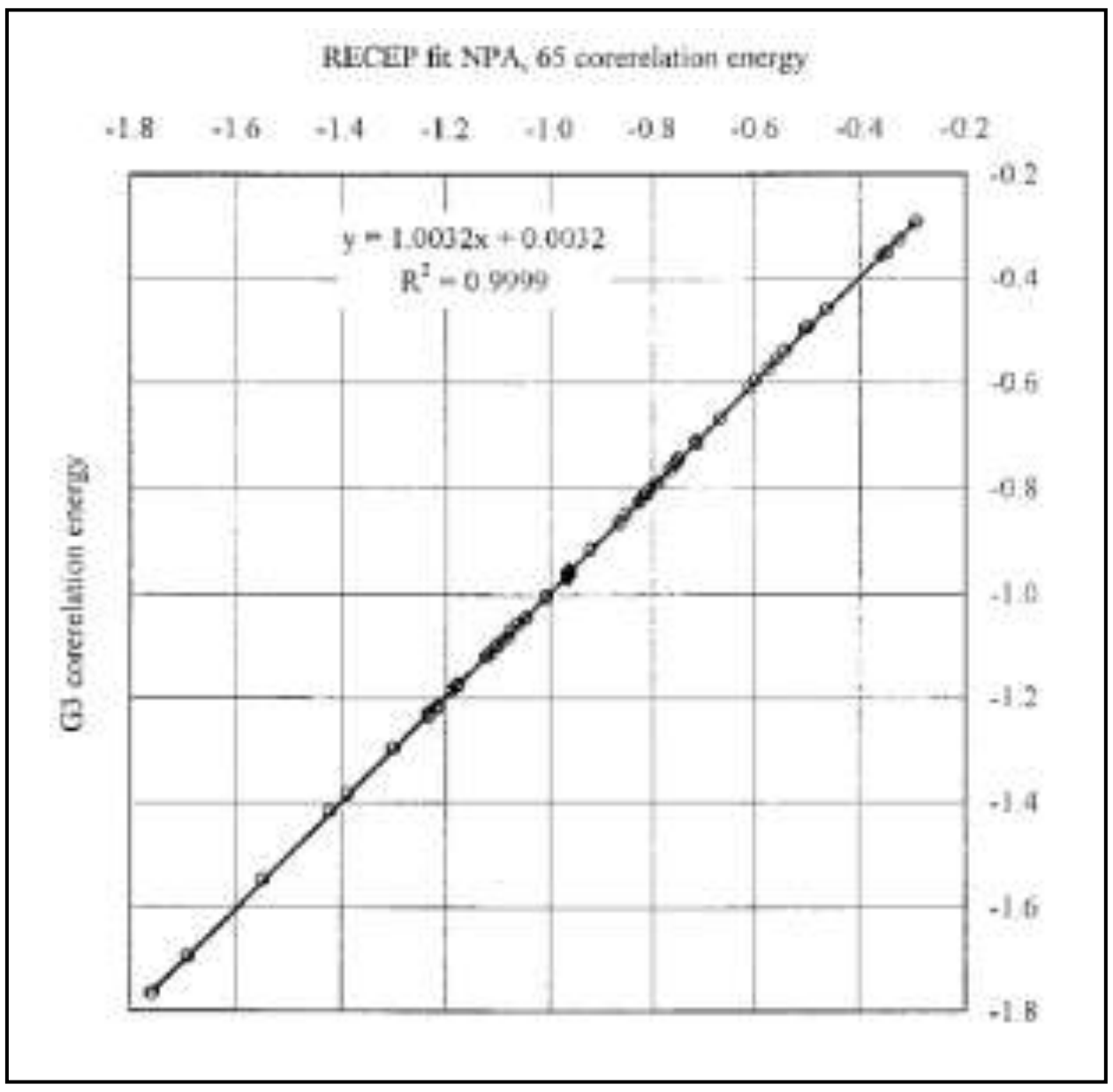

$E_{corr}$(RECEP-3, fit to G3, NPA, 65) correlation energy vs. $E_{corr}$(G3).

It shows (by inspecting these three plots) that fitted RECEP atomic correlation energies are superior to simple (raw) atomic correlation energies (e.g. by B3LYP or CI) to estimate molecular correlation energies in RECEP theory. The best results were obtained with NPA partial charges in comparison to MK, ChelpG and Mulliken ones, not shown. See numerical values in table below.

$E_{total\ electr,0}$(HF-SCF/6-311+G(2d,p)) and the G3 or G3 level molecular correlation energies, $E_{corr}$(Gi)= $E_{total\ electr,0}$(Gi) - $E_{total\ electr,0}$(HF/6-311+G(2d,p)) with i=2,3, in hartree and the deviations between RECEP estimation and G2 or G3 total electronic energies (dev, kcal/mol) for 65 molecules from the G2/97 database (optimized MP2(FU)/6-31G(d) geometries) using the RECEP parameters in table above:

| | Molecule | $E_{0,tot.electr}$(HF-SCF) hartree | $E_{corr}$(G3) hartree | ls-B3LYP NPA dev kcal/mol | ls-CI NPA dev | fit G2 41 NPA dev | fit G3 41 NPA dev | fit G3 65 NPA dev | $\Delta H_f^0$ kcal/mol | error |
|---|---|---|---|---|---|---|---|---|---|---|
| 1 | methane (CH4) | -40.2102 | -0.2901 | -2.1 | -16.3 | 1.3 | 1.4 | 0.9 | -17.9 | 1.2 |
| 2 | ammonia (NH3) | -56.2150 | -0.3251 | 14.2 | 1.8 | 0.0 | 0.0 | 0.0 | -11.0 | -0.8 |
| 3 | water (H2O) | -76.0527 | -0.3499 | 19.6 | -0.5 | -0.9 | -0.7 | -0.4 | -57.8 | -0.7 |
| 4 | hydrogen-fluorid | -100.0526 | -0.3574 | 36.7 | 11.3 | 0.4 | 0.3 | 0.3 | -65.1 | 0.5 |
| 5 | acetylene (C2H2) | -76.8422 | -0.4601 | -18.0 | -39.8 | 1.5 | 1.5 | 1.8 | 54.2 | 1.1 |
| 6 | ethylene (H2CdCH2) | -78.0584 | -0.4979 | -12.8 | -36.8 | 3.0 | 2.7 | 2.8 | 12.5 | 3.0 |
| 7 | ethane (H3C-CH3) | -79.2541 | -0.5405 | -9.8 | -36.1 | 1.7 | 1.5 | 1.2 | -20.1 | 1.5 |
| 8 | HCN | -92.8979 | -0.4936 | -19.8 | -42.1 | 1.9 | 2.1 | 3.0 | 31.5 | 3.2 |
| 9 | H2C=O | -113.9033 | -0.5538 | -2.8 | -28.4 | 1.7 | 1.7 | 2.0 | -26.0 | 2.6 |
| 10 | methanol (CH3-OH) | -115.0815 | -0.5971 | 9.1 | -22.6 | 0.8 | 0.9 | 1.9 | -48.0 | 2.0 |
| 11 | hydrazine(H2N-NH2) | -111.2174 | -0.6097 | 6.0 | -19.9 | -0.3 | -0.4 | 1.7 | 22.8 | -0.4 |
| 12 | HO-OH | -150.8235 | -0.6668 | 14.2 | -20.9 | 0.2 | 0.0 | -0.3 | -32.5 | -1.5 |
| 13 | CO2 | -187.6892 | -0.8225 | -4.5 | -37.3 | -0.2 | -0.3 | -0.9 | -94.1 | 0.3 |
| 14 | CF4 | -435.7780 | -1.5469 | 89.2 | -1.0 | 0.6 | 0.7 | 1.1 | -223.0 | 1.9 |
| 15 | COF2 | -311.7100 | -1.1860 | 41.2 | -20.6 | -0.3 | -0.6 | -0.7 | -149.1 | -4.1 |
| 16 | N2O | -183.7207 | -0.8673 | -46.8 | -89.3 | -1.5 | -2.0 | -2.3 | 19.6 | -4.0 |
| 17 | NF3 | -352.6474 | -1.2967 | 34.5 | -46.3 | 1.1 | 1.0 | 1.3 | -31.6 | 1.3 |
| 18 | F2C=CF2 | -473.5672 | -1.7634 | 65.6 | -36.7 | -1.8 | -1.9 | -2.7 | -157.4 | 2.2 |
| 19 | CF3CN | -428.6011 | -1.6935 | 30.2 | -61.8 | -1.3 | -0.9 | -0.8 | -118.4 | 0.9 |
| 20 | propyne (C3H4) | -115.8984 | -0.7106 | -25.0 | -58.7 | 2.3 | 2.3 | 2.4 | 44.2 | 2.2 |
| 21 | allene (C3H4) | -115.8970 | -0.7103 | -26.1 | -59.1 | 2.3 | 2.1 | 2.0 | 45.5 | 2.4 |
| 22 | cyclopropene(C3H4) | -115.8553 | -0.7152 | -29.0 | -62.8 | -1.8 | -1.4 | -0.9 | 66.2 | -3.2 |
| 23 | propylene (C3H6) | -117.1082 | -0.7503 | -21.1 | -57.1 | 2.3 | 2.2 | 2.2 | 4.8 | 2.3 |
| 24 | cyclopropane(C3H6) | -117.0916 | -0.7539 | -23.0 | -59.0 | -0.2 | 0.2 | 0.1 | 12.7 | -0.6 |
| 25 | propane (C3H8) | -118.2994 | -0.7932 | -18.7 | -56.9 | 0.7 | 0.5 | 0.2 | -25.0 | 0.5 |
| 26 | trans-butadiene | -154.9667 | -0.9611 | -33.4 | -79.2 | 2.2 | 1.9 | 2.4 | 26.3 | 2.1 |
| 27 | 2-butyne | -154.9525 | -0.9618 | -32.7 | -78.4 | 2.6 | 2.3 | 2.5 | 34.8 | 2.1 |
| 28 | meth-c.prop.(C4H6) | -154.9303 | -0.9657 | -35.6 | -81.4 | -0.6 | -0.5 | -0.2 | 47.9 | 1.3 |
| 29 | bicyclobutane | -154.9120 | -0.9718 | -39.6 | -85.4 | -5.3 | -4.5 | -4.1 | 51.9 | -6.7 |
| 30 | cyclobutane (C4H8) | -156.1390 | -1.0076 | -33.2 | -81.2 | -2.2 | -2.0 | -1.8 | 6.8 | -1.8 |
| 31 | isobutene (C4H8) | -156.1577 | -1.0045 | -30.6 | -78.5 | 0.5 | 0.5 | 0.4 | -4.0 | 0.4 |
| 32 | tr-butane(C4H10) | -157.3446 | -1.0463 | -27.8 | -78.1 | -0.6 | -0.9 | -1.0 | -30.0 | -0.6 |
| 33 | isobutane (C4H10) | -157.3451 | -1.0484 | -28.8 | -79.1 | -1.9 | -2.0 | -2.2 | -32.1 | -1.9 |
| 34 | spiropentane(C5H8) | -193.9673 | -1.2223 | -46.7 | -104.5 | -4.3 | -3.9 | -3.5 | 44.3 | -3.9 |
| 35 | benzene (C6H6) | -230.7633 | -1.3851 | -58.8 | -124.1 | 1.1 | 0.4 | 1.6 | 19.7 | 1.0 |
| 36 | H2CF2 | -237.9779 | -0.9165 | 44.4 | -10.3 | 1.7 | 1.4 | 1.5 | -107.7 | 2.2 |
| 37 | HCF3 | -336.8798 | -1.2319 | 65.2 | -8.1 | 0.8 | 0.9 | 0.6 | -166.6 | 1.1 |
| 38 | H3C-NH2 | -95.2473 | -0.5744 | 2.0 | -23.0 | -1.2 | -1.4 | 0.3 | -5.5 | -0.7 |
| 39 | CH3-CN | -131.9605 | -0.7428 | -26.9 | -59.3 | 3.2 | 3.3 | 3.3 | 18.0 | 3.5 |
| 40 | CH3-NO2 | -243.7359 | -1.1784 | -28.2 | -86.0 | -0.6 | -0.4 | -0.6 | -17.8 | -0.6 |
| 41 | CH3-ONO | -243.7366 | -1.1735 | -26.3 | -84.5 | 0.7 | 1.0 | 1.2 | -15.9 | 1.0 |
| 42 | HCOOH(formic acid) | -188.8266 | -0.8632 | 5.6 | -34.1 | -0.4 | 0.0 | -0.3 | -90.5 | -0.2 |
| 43 | HCOOCH3 | -227.8588 | -1.1148 | -6.6 | -58.0 | -1.2 | -1.2 | -1.0 | -85.0 | 0.6 |
| 44 | CH3CONH2 | -208.0454 | -1.0906 | -5.6 | -50.8 | -1.7 | -1.3 | -1.1 | -57.0 | -2.2 |
| 45 | C2H4NH(aziridine) | -133.0800 | -0.7887 | -16.7 | -52.5 | -3.0 | -2.6 | -0.8 | 30.2 | -2.0 |
| 46 | NCCN (cyanogen) | -184.6243 | -0.9598 | -60.6 | -103.4 | -1.4 | -1.8 | -1.0 | 73.3 | -1.3 |
| 47 | (CH3)2NH | -134.2834 | -0.8265 | -10.9 | -48.8 | -1.6 | -1.8 | 0.0 | -4.4 | -0.9 |
| 48 | CH3CH2NH2(trans) | -134.2953 | -0.8271 | -6.6 | -43.7 | -2.0 | -2.2 | -0.6 | -11.3 | -0.6 |
| 49 | CH2CO (ketene) | -151.7713 | -0.7661 | -18.4 | -50.0 | 1.9 | 2.1 | 0.3 | -11.4 | 1.1 |
| 50 | C2H4O (oxirane) | -152.9137 | -0.8125 | -9.8 | -50.9 | -2.4 | -2.2 | -0.6 | -12.6 | -0.6 |
| 51 | CH3CHO | -152.9638 | -0.8045 | -10.0 | -46.5 | 2.2 | 2.5 | 1.9 | -39.7 | 2.0 |
| 52 | HCOCOH (glyoxal) | -226.6589 | -1.0713 | -12.1 | -60.4 | 2.2 | 2.7 | 2.5 | -50.7 | 3.3 |
| 53 | CH3CH2OH | -154.1321 | -0.8494 | 0.3 | -42.8 | 0.3 | 0.2 | 0.9 | -56.2 | 1.0 |
| 54 | CH3OCH3 | -154.1148 | -0.8480 | -2.9 | -46.3 | 0.3 | 0.0 | 1.6 | -44.0 | 2.0 |
| 55 | CH2=CHF | -176.9412 | -0.8139 | 9.2 | -34.8 | 2.0 | 1.7 | 1.7 | -33.2 | 2.9 |
| 56 | CH2=CHCN | -169.8097 | -0.9547 | -40.7 | -83.3 | 2.5 | 2.2 | 2.9 | 43.2 | 1.3 |
| 57 | CH3COCH3 (acetone) | -192.0208 | -1.0567 | -18.1 | -65.6 | 1.7 | 2.4 | 1.0 | -51.9 | 1.1 |
| 58 | CH3COOH | -227.8859 | -1.1137 | -1.3 | -51.9 | 0.4 | 1.0 | -0.2 | -103.4 | -0.3 |
| 59 | CH3COF | -251.8768 | -1.1212 | 11.9 | -43.7 | 0.9 | 1.1 | -0.2 | -105.7 | -0.1 |
| 60 | (CH3)2CHOH | -193.1819 | -1.1041 | -10.3 | -64.5 | -1.8 | -1.9 | -1.7 | -65.2 | -1.2 |
| 61 | C2H5OCH3 | -193.1655 | -1.1006 | -12.0 | -66.8 | -0.5 | -1.0 | 0.4 | -51.7 | 1.5 |
| 62 | (CH3)3N | -173.3212 | -1.0821 | -24.9 | -75.4 | -4.3 | -4.7 | -2.7 | -5.7 | -2.5 |
| 63 | C4H4O (furan) | -228.6888 | -1.2387 | -40.7 | -98.0 | -4.6 | -4.3 | -3.8 | -8.3 | -4.3 |
| 64 | C4H5N (pyrrole) | -208.8676 | -1.2167 | -45.4 | -101.1 | -5.5 | -5.2 | -3.2 | 25.9 | -4.4 |
| 65 | C5H5N (pyridine) | -246.7597 | -1.4184 | -58.0 | -122.2 | -0.3 | -0.8 | 1.2 | 33.6 | 1.0 |

#28: methylene-cyclopropane, #56: acrylonitrile, #55: vinyl-fluoride,
#41: methyl-nitrite

Quality of RECEP-3 calculation for ground state total electronic energies (see table above):

dev = $E_{corr}$(method) - $E_{corr}$(RECEP, method, charge definition, L) in kcal/mol, because $E_{total\ electr,0}$(HF-SCF/6-311+G(2d,p)) values cancel:
dev= $E_{total\ electr,0}$(method)- $E_{total\ electr,0}$(RECEP) also.

Root-mean-square deviation (RMSD= $[\Sigma(x_i\text{-}x_{avrg})^2/n]^{1/2}$) of fitted RECEP, G3, NPA, and n=41 total electronic energies from G3 ones for the selected n=41 molecules is 1.72 kcal/mol, the average absolute deviation (MAD= $\Sigma|x_i\text{-}x_{avrg}|/n$) is 1.38 kcal/mol. RMSD of the same type of energies for the n=24 test molecules (from #42 to #65 in the table) is 2.32 kcal/mol, the MAD is 1.97 kcal/mol.

RMSD of fitted RECEP-3, G3, NPA, and n=65 molecules from G3 total electronic energies is 1.76 kcal/mol and MAD is 1.43 kcal/mol.

Quality of RECEP-3 calculation for enthalpies of formation (see table above):

$\Delta H_f^0$ = experimental enthalpies of formation at 298 K,

error = made by RECEP to estimate $\Delta H_f^0$ using $E_{total\ electr,0}$(RECEP-3,fit to G3, NPA, 65).

The atomization energies of the 65 molecules were obtained from the $E_{total\ electr,0}$(RECEP, G3, NPA, 65), from the G3 atomic energies, and from the corrected HF/6-31G(d) zero-point vibration energies (ZPVE). The enthalpy of formation at $\Delta H_f^0$(0 K), was calculated as a difference of the sum of the atomic enthalpies of formation and the atomization energy of the molecule. The necessary thermal corrections for $\Delta H_f^0$(298 K) were calculated from HF/6-31G(d) vibration analysis.

RMSD of the G3 level error in $\Delta H_f^0$(298 K) (not shown in the table above) from experimental is 1.15 kcal/mol, MAD is 0.74 kcal/mol for the 65 selected molecules, while RMSD of the best (G3,NPA,65) RECEP-3 error in $\Delta H_f^0$(298 K) listed in the last column in table above from the experimental values (column before the last) is 2.17 kcal/mol, MAD is 1.75 kcal/mol.

Judging the G3 calculation itself:

There is a simple linear relationship between HF/6-311+G(2d,p) and HF-SCF limit total electronic energies:

$$E_{total\ electr,0}(\text{HF-SCF limit}) = 1.000\ 178\ E_{total\ electr,0}\ (\text{HF/6-311+G(2d,p)})$$

within ±5 kcal/mol error bar, RMSD= 2 kcal/mol, and MAD= 1.4 kcal/mol, based on literature data for 17 molecules as $H_2$, CH, $CH_3$, $CH_4$, $NH_3$, $H_2O$, HF, $C_2H_2$, $C_2H_4$, CO, $N_2$, $H_2CO$, $O_2$ (large dev. 4.8 kcal/mol), $F_2$, $CO_2$, trans-butadiene, benzene (large dev. -4.2 kcal/mol), a formula which can be used to improve RECEP-3, for example.

The analysis of these 17 molecules show that for ground state total electronic energies the relationship CCSD(T) basis-set limit < G3 << G2 holds. Comparison of the HF-SCF/6-311+G(2d,p) basis set level $E_{corr}$(G3) with the infinite basis set CCSD(T) correlation energy yields that Ecorr(G3)/Ecorr(inf. basis set CCSD(T))≈ 0.97 with excellent statistical correlation between the two correlation energies:

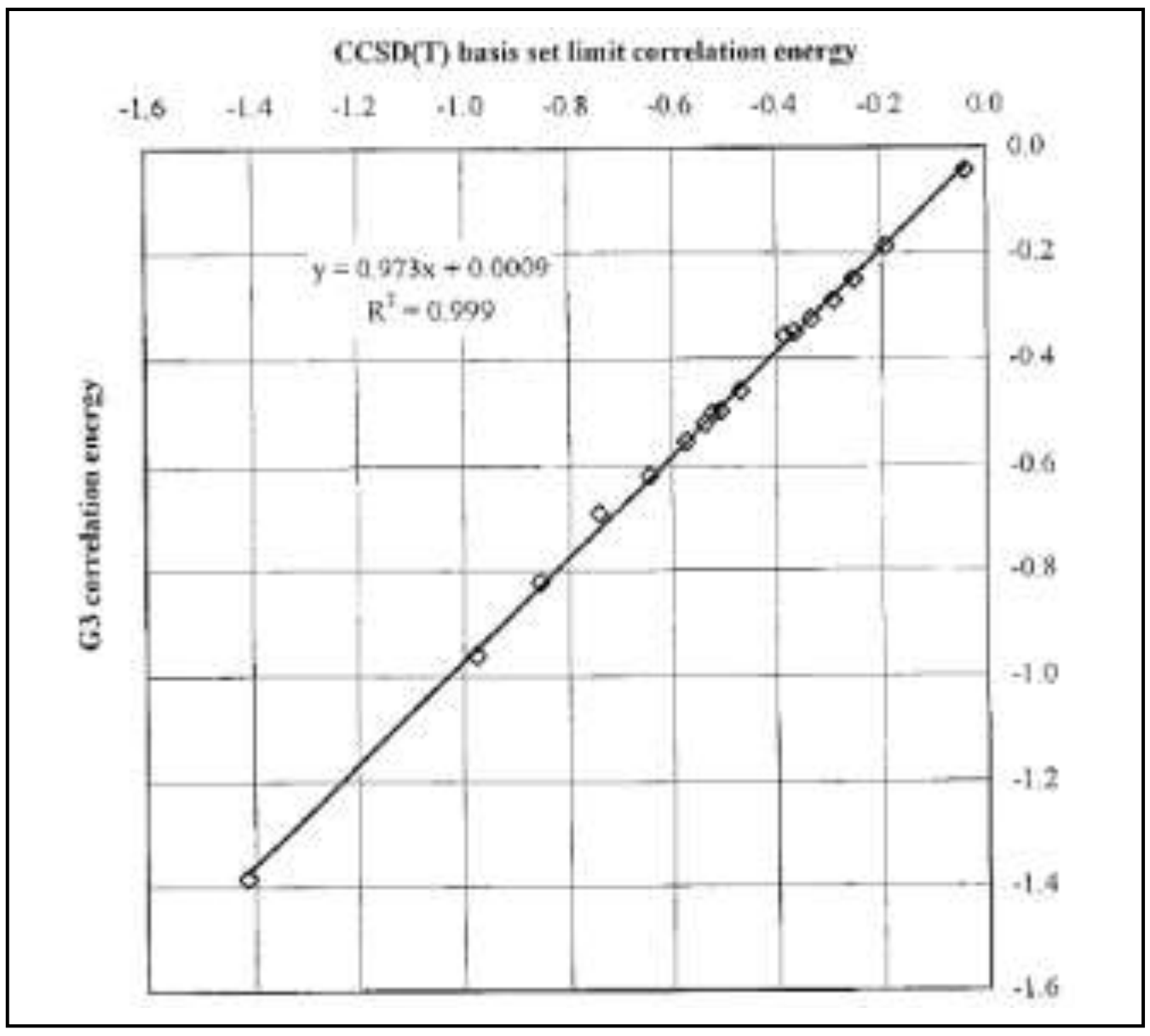

**Thesis-9-Theory: Nearly experimental quality total electronic energy with small basis set RECEP/6-31G(d)**

Gaussian-3 ground state total electronic energies have been approximated using single point HF-SCF/6-31G(d) energies and partial charges by “rapid estimation of correlation energy from partial charges” (RECEP) method. 65 closed shell, neutral and ground state molecules (composed of H, C, N, O, and F atoms) of the G2/97 thermochemistry database were selected. The main feature is that the larger basis set error can be incorporated into the atomic parameters, so the method has been renamed as “rapid estimation of basis set error and correlation energy from partial charges” (REBECEP). With these parameters a rather accurate energy values can be obtained from a small basis set HF-SCF/6-31G(d) calculation in the vicinity of stationary points. The average absolute deviation of the best REBECEP enthalpies of formation from the experimental ones is 1.39 kcal/mol for the test set.

= = = = = =

Gaussian-3 szintű alap állapot totális elektronikus energiákat közelítettem egypontos HF-SCF/6-31G(d) szintű energiákkal és parciális töltésekkel a “rapid estimation of correlation energy from partial charges” (RECEP) módszerem alapján. Erre a célra, hatvanöt, zárt héjú, semleges, alap állapotú, valamint H, C, N, O, és F atomokat tartalmazó molekulát választottam a G2/97 termokémiai adatbázisból. A legfontosabb tulajdonság, hogy a nagy bázis készlet hiba belefoglalható az atomi paraméterekbe, így a módszert átneveztem “rapid estimation of basis set error and correlation energy from partial charges” (REBECEP) módszerré. Ezekkel a paraméterekkel elég pontos energia értékek kaphatók egy kis bázisú HF-SCF/6-31G(d) számításból a stacionárius pontok közelében. A legjobb REBECEP képződéshők átlagos abszolút eltérése a kísérleti értékektől 1.39 kcal/mol.

**Sandor Kristyan - Arienn Ruzsinszky - Gabor I. Csonka:**
**Theoretical Chemistry Accounts, 106 (2001) 319-328**
= = = = = =

Representative equations/tables/figures:

Geometry optimization is not addressed, so the MP2(FU)/6-31G(d) equilibrium geometry is used again for the sake of consistency, because the G3 total energies are approximated and these energies were calculated at this geometry. The use of a considerably smaller, namely the 6-31G(d) basis set is analyzed, which increases the speed of HF-SCF calculations considerably, for example, the HF-SCF/6-31G(d) calculation of phenol and similar-sized molecules is more than 1 order of magnitude faster than the corresponding HF-SCF/6-311+G(2d,p) calculation. For larger molecules, it extends considerably the use of RECEP-type methods in relation to speed.

The use of the small 6-31G(d) basis set introduces considerable basis set error into the total energy, but the RECEP abacus is the same.

Obtaining the total electronic energy from experimental results for RE(BE)CEP:

For a given molecule of M atoms, the so-called experimental ground-state total electronic energy can be obtained as $E_{total\ electr}$(molecule , expt.)=

$\Delta H^0_f$(molecule, expt.) - $E_{ZPE}$(molecule, G3) - $E_{therm}$(molecule, G3)

-$\Sigma_{Atom=1...M}$ [$\Delta H^0_f$(Atom, expt.) - $E_{total\ electr}$(Atom, G3) - $E_{therm}$(Atom, G3)],

where $\Delta H^0_f$ is the experimental standard enthalpy of formation, $E_{ZPE}$ is the zero-point energy, $E_{therm}$ is the differences between the enthalpy at 298.15 K and the energy at 0 K as used in the G3 calculations; it obviously depends on the method of G3 and ZPE calculations. Obviously, $E_{corr}$(expt., basis)= $E_{total\ electr}$(expt.) - $E_{total\ electr}$(HF-SCF/basis).

The $E_{total\ electr}$(molecule , expt.) and $E_{total\ electr}$(G3) are counterparts, so the $E_{corr}$(expt., basis) and $E_{corr}$(G3, basis). The RECEP enthalpy of formation is calculated as

$\Delta H^0_f$(molecule, RECEP, expt., charge def.)=
$E_{total\ electr}$(molecule, RECEP, expt., charge def.) + $E_{ZPE}$(molecule, G3) + $E_{therm}$(molecule, G3) + $\Sigma_{Atom=1...M}$ [$\Delta H^0_f$(Atom, expt.) - $E_{total\ electr}$(Atom, G3) - $E_{therm}$(Atom, G3)],
i.e. it uses $E_{total\ electr}$(M, RECEP, expt., charge def.) instead of $E_{total\ electr}$(M, RECEP, G3, charge def.). The former approximates the experimental enthalpy of formation better than the latter; expt. vs. G3 in the argument refers what is used for RECEP parameter fit.

Basis set error incorporation into RE(BE)CEP:

The uniformly defined 6-31G(d) basis set speed up the HF-SCF calculation, but introduces considerable basis set error. Although the basis set error energy is usually treated at molecular level, here it is implicitly partitioned among the atoms and the atomic partitions are merged with the atomic correlation energy parameters, yielding the fitted parameters.

Fitted $E_{fitpar}$(N, Z, method, charge definition) REBECEP atomic parameters (hartree) for calculating $E_{corr}$(REBECEP, method, charge definition) for closed shell, neutral molecules in the vicinity of stationary points from HF-SCF/6-31G(d) results. N is the number of electrons, Z is the nuclear charge, $E_{corr}$(G3)= $E_{total\ electr}$(G3)- $E_{total\ electr}$(HF-SCF/6-31G(d)) of 65 molecules were used to obtain these atomic parameters via least square fit, n.a. is not available (no such partial charge value occurred). The (N=9, Z=7) nitrogen is not stable in gas phase (double anion), but one must recognize that this table lists REBECEP atomic energy parameters for atoms in a molecular environment (i.e. in a bond):

| Atom | N | Z | G3 Mulliken | G3 NPA | Expt. Mulliken | Expt. NPA |
|---|---|---|---|---|---|---|
| H | 2 | 1 | -0.0358 | -0.0330 | -0.0344 | -0.0315 |
| C | 4 | 6 | -0.1013 | -0.1092 | -0.1003 | -0.1129 |
| C | 5 | 6 | -0.1692 | -0.1697 | -0.1651 | -0.1666 |
| C | 6 | 6 | -0.2190 | -0.2200 | -0.2190 | -0.2203 |
| C | 7 | 6 | -0.2603 | -0.2588 | -0.2638 | -0.2613 |
| N | 6 | 7 | -0.2651 | -0.2650 | -0.2625 | -0.2662 |
| N | 7 | 7 | -0.2901 | -0.2866 | -0.2910 | -0.2866 |
| N | 8 | 7 | -0.3160 | -0.3173 | -0.3191 | -0.3206 |
| N | 9 | 7 | n.a. | -0.3847 | n.a. | -0.3823 |
| O | 8 | 8 | -0.3354 | -0.3402 | -0.3362 | -0.3396 |
| O | 9 | 8 | -0.3740 | -0.3742 | -0.3758 | -0.3763 |
| F | 9 | 9 | -0.3658 | -0.3672 | -0.3677 | -0.3655 |
| F | 10 | 9 | -0.4234 | -0.4203 | -0.4210 | -0.4226 |

The influence of the basis set (6-31G(d) vs. 6-311+G(2d,p)) on the partial charges:

The NPA partial charges are quite similar for the two basis sets, but the Mulliken partial charges show considerable differences as expected: For example, for 104 C atoms in different molecules, the average deviation of the 6-311+G(2d,p) Mulliken partial charges from the 6-31G(d) Mulliken partial charges is -0.098 au (with 0.192 rms deviation). The corresponding value for NPA charges is 0.043 au (with 0.087 rms deviation). For oxygen and nitrogen atoms the similarity of the NPA charges is even better, 0.010 or 0.015, respectively. Finally, the NPA partial charges usually providing the best RECEP results are not very sensitive to the basis set used, while Mulliken charges do not behave so well. Since there is a proportionality with the charges one can expect that the RECEP energy is less sensitive to the basis set extension effects than the correlation energy calculated in a classical way (e.g. CCSD(T)).

The only *ab initio* calculation required is a HF-SCF/6-31G(d) total electronic energy calculation followed by an almost instant NPA or Mulliken partial charge analysis. The average absolute deviation of $\Delta H^0_f$(molecule, REBECEP, expt., NPA) enthalpies of formation from the experimental enthalpies of formation is 1.39 kcal/mol for the test set of 65 neutral enthalpies, the average absolute deviation of $E_{total\ electr}$(molecule, REBECEP, G3, NPA) energies from the G3 total energies is 1.32 kcal/mol, the average absolute deviation of the G3 enthalpies of formation from the experimental enthalpies of formation is 0.74 kcal/mol for the selected 65 molecules, as well as the corresponding REBECEP results obtained using Mulliken charges provide 1.7 kcal/mol average absolute deviation.

Experimental enthalpies of formation, $\Delta H^0_f$, HF-SCF/6-31G(d) level ground-state total electronic energies, $E_{total\ electr}$(method), as well as G3 and experimental energy corrections, $E_{corr}$(method) for the 65 molecules selected from the G2/97 database. MP2(FU)/6-31G(d) equilibrium geometries are used as in G3 method, as well as the $E_{corr}$(method) includes the correlation energies and the 6-31G(d) basis set errors:

| No. | Molecule | $\Delta H^0_f$(expt.) (kcal/mol) | $-E_{tot.electr}$ HF-SCF (h) | $-E_{corr}$ G3 (h) | $-E_{corr}$ expt. (h) |
|---|---|---|---|---|---|
| 1 | Methane (CH4) | -17.9 | 40.1951 | 0.3052 | 0.3048 |
| 2 | Ammonia (NH3) | -11.0 | 56.1838 | 0.3563 | 0.3575 |
| 3 | Water (H2O) | -57.8 | 76.0098 | 0.3928 | 0.3933 |
| 4 | Hydrogen fluoride (HF) | -65.1 | 100.0023 | 0.4077 | 0.4073 |
| 5 | Acetylene (C2H2) | 54.2 | 76.8156 | 0.4867 | 0.4878 |
| 6 | Ethylene (H2C=CH2) | 12.5 | 78.0311 | 0.5252 | 0.5249 |
| 7 | Ethane (H3C-CH3) | -20.1 | 79.2286 | 0.5661 | 0.5656 |
| 8 | Hydrogen cyanide (HCN) | 31.5 | 92.8702 | 0.5213 | 0.5210 |
| 9 | Formaldehyde (H2C=O) | -26.0 | 113.8638 | 0.5934 | 0.5924 |
| 10 | Methanol (CH3OH) | -48.0 | 115.0342 | 0.6444 | 0.6442 |
| 11 | Hydrazine (H2N-NH2) | 22.8 | 111.1680 | 0.6591 | 0.6624 |
| 12 | Hydrogen peroxide (HO-OH) | -32.5 | 150.7601 | 0.7302 | 0.7321 |
| 13 | Carbon dioxide (CO2) | -94.1 | 187.6284 | 0.8833 | 0.8813 |
| 14 | CF4 | -223.0 | 435.6416 | 1.6834 | 1.6821 |
| 15 | COF2 | -149.1 | 311.6104 | 1.2857 | 1.2911 |
| 16 | N2O | 19.6 | 183.6631 | 0.9249 | 0.9276 |
| 17 | NF3 | -31.6 | 352.5320 | 1.4122 | 1.4121 |
| 18 | C2F4 (F2C=CF2) | -157.4 | 473.4117 | 1.9189 | 1.9111 |
| 19 | CF3CN | -118.4 | 428.4722 | 1.8224 | 1.8197 |
| 20 | Propyne (C3H4) | 44.2 | 115.8619 | 0.7471 | 0.7475 |
| 21 | Allene (C3H4) | 45.5 | 115.8602 | 0.7472 | 0.7465 |
| 22 | Cyclopropene (C3H4) | 66.2 | 115.8218 | 0.7487 | 0.7523 |
| 23 | Propylene (C3H6) | 4.8 | 117.0707 | 0.7878 | 0.7877 |
| 24 | Cyclopropane (C3H6) | 12.7 | 117.0585 | 0.7870 | 0.7882 |
| 25 | Propane (C3H8) | -25.0 | 118.2633 | 0.8293 | 0.8288 |
| 26 | trans-1,3-Butadiene | 26.3 | 154.9181 | 1.0096 | 1.0103 |
| 27 | Dimethylacetylene (2-butyne) | 34.8 | 154.9066 | 1.0077 | 1.0084 |
| 28 | Methylenecyclopropane (C4H6) | 47.9 | 154.8866 | 1.0095 | 1.0072 |
| 29 | Bicyclobutane | 51.9 | 154.8708 | 1.0130 | 1.0172 |
| 30 | Cyclobutane (C4H8) | 6.8 | 156.0964 | 1.0501 | 1.0501 |
| 31 | Isobutene (C4H8) | -4.0 | 156.1098 | 1.0524 | 1.0524 |
| 32 | trans-Butane (C4H10) | -30.0 | 157.2979 | 1.0930 | 1.0923 |
| 33 | Isobutane (C4H10) | -32.1 | 157.2984 | 1.0951 | 1.0947 |
| 34 | Spiropentane (C5H8) | 44.3 | 193.9172 | 1.2724 | 1.2730 |
| 35 | Benzene (C6H6) | 19.7 | 230.7020 | 1.4464 | 1.4474 |
| 36 | Difluoromethane (H2CF2) | -107.7 | 237.8948 | 0.9996 | 0.9985 |
| 37 | Trifluoromethane (HCF3) | -166.6 | 336.7692 | 1.3425 | 1.3416 |
| 38 | Methylamine (H3C-NH2) | -5.5 | 95.2091 | 0.6126 | 0.6141 |
| 39 | Acetonitrile (CH3-CN) | 18.0 | 131.9225 | 0.7808 | 0.7805 |
| 40 | Nitromethane (CH3-NO2) | -17.8 | 243.6538 | 1.2606 | 1.2606 |
| 41 | Methylnitrite (CH3-O-N=O) | -15.9 | 243.6596 | 1.2505 | 1.2508 |
| 42 | Formic acid (HCOOH) | -90.5 | 188.7586 | 0.9312 | 0.9311 |
| 43 | Methyl formate (HCOOCH3) | -85.0 | 227.7857 | 1.1879 | 1.1854 |
| 44 | Acetamide (CH3CONH2) | -57.0 | 207.9735 | 1.1625 | 1.1642 |
| 45 | Aziridine (C2H4NH) | 30.2 | 133.0373 | 0.8314 | 0.8333 |
| 46 | Cyanogen (NCCN) | 73.3 | 184.5778 | 1.0062 | 1.0067 |
| 47 | Dimethylamine [(CH3)2NH) | -4.4 | 134.2380 | 0.8719 | 0.8733 |
| 48 | trans-Ethylamine (CH3CH2NH2) | -11.3 | 134.2468 | 0.8756 | 0.8756 |
| 49 | Ketene (CH2CO) | -11.4 | 151.7218 | 0.8157 | 0.8145 |
| 50 | Oxirane (C2H4O) | -12.6 | 152.8654 | 0.8608 | 0.8608 |
| 51 | Acetaldehyde (CH3CHO) | -39.7 | 152.9135 | 0.8548 | 0.8546 |
| 52 | Glyoxal (HCOCOH) | -50.7 | 226.5864 | 1.1439 | 1.1424 |
| 53 | Ethanol (CH3CH2OH) | -56.2 | 154.0743 | 0.9072 | 0.9070 |
| 54 | Dimethyl ether (CH3OCH3) | -44.0 | 154.0634 | 0.8994 | 0.8987 |
| 55 | Vinyl fluoride (CH2=CHF) | -33.2 | 176.8807 | 0.8744 | 0.8725 |
| 56 | Acrylonitrile (CH2=CHCN) | 43.2 | 169.7620 | 1.0024 | 1.0050 |
| 57 | Acetone (CH3COCH3) | -51.9 | 191.9599 | 1.1176 | 1.1175 |
| 58 | Acetic acid (CH3COOH) | -103.4 | 227.8071 | 1.1926 | 1.1928 |
| 59 | Acetyl fluoride (CH3COF) | -105.7 | 251.7949 | 1.2031 | 1.2028 |
| 60 | 2-Propanol [(CH3)2CHOH) | -65.2 | 193.1139 | 1.1721 | 1.1714 |
| 61 | Methyl ethyl ether(C2H5OCH3) | -51.7 | 193.1033 | 1.1627 | 1.1609 |
| 62 | Trimethylamine [(CH3)3N) | -5.7 | 173.2682 | 1.1350 | 1.1346 |
| 63 | Furan (C4H4O) | -8.3 | 228.6224 | 1.3052 | 1.3060 |
| 64 | Pyrrole (C4H5N) | 25.9 | 208.8059 | 1.2784 | 1.2803 |
| 65 | Pyridine (C5H5N) | 33.6 | 246.6938 | 1.4844 | 1.4847 |

Energy deviations (kcal/mol):

| No. | Molecule | $\Delta H^0_f$, expt.-G3 | $E_{total\ el.}$(method)- $E_{total\ el.}$(REBECEP,method,charge def.) | | | |
|---|---|---|---|---|---|---|
| | | | G3 NPA | G3 Mulliken | expt. NPA | expt. Mulliken |
| 1 | Methane (CH4) | 0.3 | 0.1 | 0.5 | 0.2 | 0.8 |
| 2 | Ammonia (NH3) | -0.8 | 0.0 | -2.9 | 0.0 | -2.5 |
| 3 | Water (H2O) | -0.3 | -1.7 | -2.3 | -1.3 | -2.1 |
| 4 | Hydrogen fluoride (HF) | 0.2 | -2.2 | -2.2 | -1.8 | -2.3 |
| 5 | Acetylene (C2H2) | -0.7 | -1.7 | 0.2 | -2.1 | 0.1 |
| 6 | Ethylene (H2C=CH2) | 0.2 | -0.4 | 0.6 | -0.3 | 0.9 |
| 7 | Ethane (H3C-CH3) | 0.3 | 0.9 | 1.0 | 1.0 | 1.2 |
| 8 | Hydrogen cyanide (HCN) | 0.2 | 1.9 | 4.2 | 2.4 | 5.1 |
| 9 | Formaldehyde (H2C=O) | 0.6 | -0.4 | 1.0 | -0.5 | 1.3 |
| 10 | Methanol (CH3-OH) | 0.1 | 0.5 | 0.4 | 0.7 | 0.5 |
| 11 | Hydrazine (H2N-NH2) | -2.1 | 0.5 | 2.2 | 0.2 | 2.0 |
| 12 | Hydrogen peroxide (HO-OH) | -1.2 | 0.3 | -2.7 | -0.4 | -2.7 |
| 13 | Carbon dioxide (CO2) | 1.2 | -3.8 | -2.4 | -2.0 | -1.9 |
| 14 | CF4 | 0.9 | 2.3 | 1.8 | 3.6 | 1.8 |
| 15 | COF2 | -3.4 | -2.3 | -1.0 | -5.3 | -5.2 |
| 16 | N2O | -1.7 | -3.5 | -1.2 | -4.3 | -2.1 |
| 17 | NF3 | 0.1 | 0.8 | 1.9 | 0.2 | 1.9 |
| 18 | C2F4 (F2C=CF2) | 4.9 | -3.7 | -4.7 | -2.2 | -2.0 |
| 19 | CF3CN | 1.8 | 1.1 | -2.1 | 1.9 | -1.0 |
| 20 | Propyne (C3H4) | -0.2 | 0.6 | 1.0 | 0.7 | 1.3 |
| 21 | Allene (C3H4) | 0.5 | -0.7 | -1.0 | 0.0 | -0.2 |
| 22 | Cyclopropene (C3H4) | -2.2 | -2.3 | -2.0 | -4.4 | -4.1 |
| 23 | Propylene (C3H6) | 0.0 | 0.1 | 0.5 | 0.2 | 0.6 |
| 24 | Cyclopropane (C3H6) | -0.7 | 1.3 | 1.6 | 0.7 | 1.1 |
| 25 | Propane (C3H8) | 0.3 | 0.5 | 0.1 | 0.6 | 0.2 |
| 26 | trans-1,3-Butadiene | -0.4 | -1.3 | -0.3 | -1.5 | -0.4 |
| 27 | Dimethylacetylene(2-butyne) | -0.4 | 2.4 | 2.0 | 2.5 | 2.0 |
| 28 | Methylenecyclopropane (C4H6) | 1.5 | 0.0 | -0.5 | 1.8 | 1.2 |
| 29 | Bicyclobutane | -2.6 | -2.4 | -2.5 | -4.7 | -4.9 |
| 30 | Cyclobutane (C4H8) | 0.0 | 0.1 | -0.4 | 0.2 | -0.5 |
| 31 | Isobutene (C4H8) | 0.0 | -0.8 | -1.6 | -0.7 | -1.6 |
| 32 | trans-Butane (C4H10) | 0.4 | -0.3 | -1.1 | -0.1 | -0.9 |
| 33 | Isobutane (C4H10) | 0.2 | -1.4 | -2.2 | -1.3 | -2.2 |
| 34 | Spiropentane (C5H8) | -0.4 | -0.1 | -0.7 | -0.1 | -0.8 |
| 35 | Benzene (C6H6) | -0.6 | 2.1 | 2.0 | 2.3 | 1.9 |
| 36 | Difluoromethane (H2CF2) | 0.7 | 0.5 | 1.8 | -0.6 | 0.9 |
| 37 | Trifluoromethane (HCF3) | 0.5 | 2.0 | 3.5 | 0.6 | 2.0 |
| 38 | Methylamine (H3C-NH2) | -1.0 | -0.8 | -0.2 | -0.8 | -0.5 |
| 39 | Acetonitrile (CH3-CN) | 0.2 | 4.0 | 4.6 | 4.5 | 5.3 |
| 40 | Nitromethane (CH3-NO2) | 0.0 | -0.7 | -2.7 | 0.4 | -1.9 |
| 41 | Methylnitrite (CH3-O-N=O) | -0.2 | 3.0 | 1.1 | 3.4 | 1.5 |
| 42 | Formic acid (HCOOH) | 0.1 | 0.3 | 1.2 | -0.1 | 1.1 |
| 43 | Methyl formate (HCOOCH3) | 1.6 | 0.0 | 1.6 | 0.5 | 2.4 |
| 44 | Acetamide (CH3CONH2) | -1.1 | 0.6 | 0.6 | 0.2 | 0.1 |
| 45 | Aziridine (C2H4NH) | -1.2 | -0.4 | 1.0 | -0.9 | 0.5 |
| 46 | Cyanogen (NCCN) | -0.3 | -0.8 | -4.2 | -0.8 | -4.3 |
| 47 | Dimethylamine [(CH3)2NH) | -0.9 | -0.6 | 0.4 | -1.1 | -0.2 |
| 48 | trans-Ethylamine (CH3CH2NH2) | 0.0 | -0.7 | -0.7 | 0.3 | -0.1 |
| 49 | Ketene (CH2CO) | 0.8 | -1.7 | -0.9 | -1.5 | -0.2 |
| 50 | Oxirane (C2H4O) | 0.0 | 0.2 | -0.1 | 0.0 | -0.3 |
| 51 | Acetaldehyde (CH3CHO) | 0.1 | 1.1 | 2.0 | 0.7 | 2.0 |
| 52 | Glyoxal (HCOCOH) | 0.9 | 1.7 | 2.5 | 1.6 | 3.0 |
| 53 | Ethanol (CH3CH2OH) | 0.1 | 0.9 | 0.0 | 1.2 | 0.1 |
| 54 | Dimethyl ether (CH3OCH3) | 0.4 | 0.9 | 0.9 | 0.7 | 0.7 |
| 55 | Vinyl fluoride (CH2=CHF) | 1.2 | -0.6 | 0.6 | -0.1 | 1.3 |
| 56 | Acrylonitrile (CH2=CHCN) | -1.6 | 2.1 | 3.0 | 0.9 | 2.0 |
| 57 | Acetone (CH3COCH3) | 0.0 | 1.7 | 2.1 | 1.4 | 2.0 |
| 58 | Acetic acid (CH3COOH) | -0.1 | 2.2 | 2.3 | 1.8 | 2.0 |
| 59 | Acetyl fluoride (CH3COF) | 0.1 | 0.6 | 1.9 | 0.0 | 1.3 |
| 60 | 2-Propanol [(CH3)2CHOH) | 0.5 | -0.9 | -2.2 | -0.3 | -1.9 |
| 61 | Methyl ethyl ether(C2H5OCH3) | 1.1 | 0.9 | 0.1 | 1.5 | 0.5 |
| 62 | Trimethylamine [(CH3)3N] | 0.2 | -2.8 | -1.4 | -2.6 | -1.2 |
| 63 | Furan (C4H4O) | -0.5 | -2.9 | -3.2 | -3.3 | -3.6 |
| 64 | Pyrrole (C4H5N) | -1.2 | -2.7 | -3.8 | -2.7 | -4.0 |
| 65 | Pyridine (C5H5N) | -0.1 | 3.0 | 3.5 | 3.7 | 4.0 |

Last four columns: Deviation of total electronic energy $E_{total\ electr}$(G3 or experimental) from corresponding REBECEP results in the case of two partial charges (NPA or Mulliken). Notice that these deviations are not only the total electronic energy deviations but $\Delta H^0_f$ deviations as well, because $E_{ZPE}$ and $E_{therm}$ of the molecules are taken from G3 calculations. Summing $\Delta H^0_f$(Expt.) - $\Delta H^0_f$(G3); i.e. column Expt-G3, and
$E_{total\ el.}$(G3) - $E_{total\ el.}$ (REBECEP, G3, charge def.) yields directly
$\Delta H^0_f$(Expt.) - $\Delta H^0_f$ (REBECEP, G3, charge def.).

Table above reveals that NPA charges provide systematically better results than the Mulliken charges with any REBECEP parameterization, and this is in agreement with our earlier G2 and G3 results. This can be explained by the fact that Mulliken charges might show rather unphysical behavior with respect to the choice of basis set, while the NPA charges are free of such problems.

**Thesis-10-Theory: Performance of REBECEP method on the G3/99 test set**

Experimental enthalpies of formation have been approximated using single point HF-SCF total electronic energies plus the REBECEP energy corrections (calculated from the HF-SCF partial atomic charges and optimized atomic energy parameters). The performance of the method was tested on 50 closed shell neutral molecules from the G3/99 thermochemistry database plus urea, composed of H, C, N, O, and F atoms. The predictive force of the method has been demonstrated, because these larger molecules were not used for the optimization of the atomic parameters. We used the earlier RECEP-3 [HF/6-311+G(2d,p)] and REBECEP [HF/6-31G(d)] atomic parameter sets obtained from the G2/97 thermochemistry database containing small molecules together with natural population analysis (NPA) and Mulliken partial charges. The best results were obtained using the NPA charges, although the Mulliken charges also provide useful results. The rms deviations from the experimental enthalpies of formation for the selected 51 molecules are 1.15, 3.96, and 2.92 kcal/mol for Gaussian-3, B3LYP/6-11+G(3df,2p), and REBECEP [NPA] enthalpies of formation, resp., the corresponding average absolute deviations are 0.94, 7.09, and 2.27 kcal/mol, resp.. The REBECEP method performs considerably better for the 51 test molecules with a moderate 6-31G(d) basis set than the B3LYP method with a large 6-311+G(3df,2p) basis set.

= = = = = =

Kísérleti képződéshőket közelítettem egypontos HF-SCF totál elektronikus energiák felhasználásával és REBECEP energia korrekciókkal, (utóbbit HF-SCF parciális atomi töltések és optimalizált atomi energia paraméterek segítségével). A módszer teljesítőképességét 50 zárt héjú semleges molekulán (G3/99 termokémiai adatbázis) és urea molekulán teszteltem, az atomi összetétel H, C, N, O, és F volt. A módszer előrejelzési erejét demonstrálja, hogy ezek a nagyobb molekulák nem szerepeltek az atomi paraméterek optimalizálása során. Felhasználtam a korábbi RECEP-3 [HF/6-311+G(2d,p)] és REBECEP [HF/6-31G(d)] atomi paramétereket, melyeket még a G2/97 termokémiai adatbázis kis molekulái felhasználásával nyertem a „natural population analysis" (NPA) és Mulliken parciális töltések segítségével. A legjobb eredmények az NPA töltések felhasználásával kaphatók, bár a Mulliken töltések szintén hasznos eredményeket szolgáltatnak. A képződéshők rms eltérései a kísérletitől az 51 teszt molekula esetében 1.15, 3.96, és 2.92 kcal/mol-nak adódtak rendre a Gaussian-3, B3LYP/6-11+G(3df,2p), és REBECEP [NPA] szintű képződéshők esetében, míg az átlagos abszolút eltérések rendre 0.94, 7.09, és 2.27 kcal/mol voltak. A REBECEP módszer láthatóan jobban teljesít az 51 teszt molekula esetében a moderált 6-31G(d) bázis készlet felhasználásával, mint a B3LYP módszer a nagy 6-311+G(3df,2p) bázis készlettel.

**Sandor Kristyan - Adrienn Ruzsinszky - Gabor I. Csonka:**
**Theoretical Chemistry Accounts, 106 (2001) 404-411**

= = = = = =

Representative equations/tables/figures:

The advantage of the REBECEP method is that the very expensive calculation for the accurate total electronic ground state energy is not necessary, only the relatively cheap HF-SCF/basis, because the $E_{corr}$(method, basis set) can be estimated from atomic partial charges for closed shell ground state covalent molecules in the vicinity of stationary points by an inexpensive atom-by-atom method. Even the basis set error can be corrected, which allows to use smaller basis in HF-SCF/basis calculation, makig the algothm even faster. (The geometry optimization is not a part of this method.) In this way, one can predict, for example, G2 or G3 quality total energy and experimental thermochemistry results, and the speed of the method increases considerably by more than 1 order of magnitude e.g. in the case of aniline without loss of precision.

The G2/97 thermochemistry database was extended and relatively large molecules were included, making $E_{total\ electr.}$(G3) available for public, and was tested against 376 test energies (222 neutral enthalpies of formation, 88 ionization potentials, 58 electron affinities, and eight proton affinities) in the full G3/99 test set by its creators. This provides a good opportunity to test the performance of RECEP-3 and REBECEP parameter sets, especially because these new molecules were not used during the establishment of RECEP. The criterion for choosing the 50 molecules for the subset of the G3/99 pluss urea (#25) test set was that their experimental enthalpies of formation at 298 K have a quoted uncertainty of ±1 kcal or less, although this is not necessarily a guarantee for the accuracy of the experimental data. (All the selected molecules contain at least N= 30 electrons and the largest contains 68 electrons or 10 non-hydrogen atoms, as well as the Gaussian 98 program was used for all the calculations.) The corresponding energies, experimental enthalpies of formation with the quoted experimental errors are shown in the two tables below:

Number of atoms (M) and electrons (N) in 51 test molecules, their G3 total electronic energies (hartree) and the energy corrections (hartree) calculated with two different basis sets

| | Species | M | N | $E_{tot.el.}$(G3) | $E_{corr}$ 6-311+G(2d,p) | $E_{corr}$ 6-31G(d) |
|---|---|---|---|---|---|---|
| 1 | C4H6 (methylallene) | 10 | 30 | -155.9081 | -0.9629 | -1.0098 |
| 2 | C5H8 (isoprene) | 13 | 38 | -195.2311 | -1.2166 | -1.2755 |
| 3 | C5H10 (cyclopentane) | 15 | 40 | -196.4769 | -1.2611 | -1.3140 |
| 4 | C5H12 (n-pentane) | 17 | 42 | -197.6891 | -1.2990 | -1.3566 |
| 5 | C5H12 (neopentane) | 17 | 42 | -197.6963 | -1.3062 | -1.3632 |
| 6 | C6H8 (1,3-cyclohexadiene) | 14 | 44 | -233.3246 | -1.4309 | -1.4950 |
| 7 | C6H8 (1,4-cyclohexadiene) | 14 | 44 | -233.3242 | -1.4271 | -1.4923 |
| 8 | C6H12 (cyclohexane) | 18 | 48 | -235.7858 | -1.5153 | -1.5786 |
| 9 | C6H14 (n-hexane) | 20 | 50 | -236.9874 | -1.5525 | -1.6203 |
| 10 | C6H14 (3-methylpentane) | 20 | 50 | -236.9884 | -1.5573 | -1.6250 |
| 11 | C6H5CH3 (toluene) | 15 | 50 | -271.4507 | -1.6403 | -1.7117 |
| 12 | C7H16 (n-heptane) | 23 | 58 | -276.2857 | -1.8057 | -1.8841 |
| 13 | C8H8 (cyclooctatetraene) | 16 | 56 | -309.4598 | -1.8553 | -1.9397 |
| 14 | C8H18 (n-octane) | 26 | 66 | -315.5840 | -2.0589 | -2.1478 |
| 15 | C10H8 (naphthalene) | 18 | 68 | -385.7282 | -2.2793 | -2.3757 |
| 16 | C10H8 (azulene) | 18 | 68 | -385.6708 | -2.2914 | -2.3901 |
| 17 | CH3COOCH3 | 11 | 40 | -268.2831 | -1.3658 | -1.4499 |
| 18 | (CH3)3COH | 15 | 42 | -233.5910 | -1.3609 | -1.4392 |
| 19 | C6H5NH2 (aniline) | 14 | 50 | -287.4877 | -1.6751 | -1.7585 |
| 20 | C6H5OH (phenol) | 13 | 50 | -307.3446 | -1.6975 | -1.7887 |
| 21 | C4H6O (divinyl ether) | 11 | 38 | -231.1038 | -1.2715 | -1.3453 |
| 22 | C4H8O (tetrahydrofuran) | 20 | 40 | -232.3577 | -1.3145 | -1.3830 |
| 23 | C5H8O (cyclopentanone) | 14 | 46 | -270.4653 | -1.5241 | -1.6014 |
| 24 | C6H4O2 (benzoquinone) | 12 | 56 | -381.2975 | -1.9604 | -2.0698 |
| 25 | CH4ON2 (urea) | 8 | 32 | -225.1869 | -1.1229 | -1.2052 |
| 26 | C4H4N2 (pyrimidine) | 10 | 42 | -264.2111 | -1.4496 | -1.5206 |
| 27 | NC-CH2-CH2-CN | 10 | 42 | -264.2017 | -1.4513 | -1.5222 |
| 28 | C4H4N2 (pyrazine) | 10 | 42 | -264.2037 | -1.4534 | -1.5241 |
| 29 | CH3COCCH (acetyl acetylene) | 9 | 36 | -229.8838 | -1.2310 | -1.3001 |
| 30 | CH3CH=CHCHO(crotonaldehyde) | 11 | 38 | -231.1390 | -1.2672 | -1.3383 |
| 31 | CH3-CO-O-CO-CH3 | 13 | 54 | -381.5785 | -1.8852 | -2.0016 |
| 32 | (CH3)2CH-CN | 12 | 38 | -211.3013 | -1.2497 | -1.3088 |
| 33 | CH3-CO-CH2-CH3 | 13 | 40 | -232.3764 | -1.3100 | -1.3809 |
| 34 | (CH3)2CH-CHO | 13 | 40 | -232.3663 | -1.3123 | -1.3833 |
| 35 | C4H8O2 (1,4-dioxane) | 14 | 48 | -307.5394 | -1.6233 | -1.7169 |
| 36 | C4H8NH (tetrahydropyrrole) | 14 | 40 | -212.5006 | -1.2939 | -1.3565 |
| 37 | CH3-CH2-CH(CH3)-NO2 | 16 | 56 | -362.8203 | -1.9421 | -2.0547 |
| 38 | CH3-CH2-O-CH2-CH3 | 15 | 42 | -233.5692 | -1.3533 | -1.4260 |
| 39 | CH3-CH(OCH3)2 (.acetal) | 16 | 50 | -308.7526 | -1.6655 | -1.7617 |
| 40 | (CH3)3C-NH2 (t-butylamine) | 16 | 42 | -213.7294 | -1.3396 | -1.4088 |
| 41 | -CH=CH-N(CH3)-CH=CH- | 13 | 44 | -249.3768 | -1.4706 | -1.5399 |
| 42 | C5H10O (tetrahydropyran) | 16 | 48 | -271.6639 | -1.5695 | -1.6478 |
| 43 | CH3-CH2-CO-CH2-CH3 | 16 | 48 | -271.6753 | -1.5635 | -1.6443 |
| 44 | CH3-C(=O)-O-CH(CH3)2 | 17 | 56 | -346.8901 | -1.8740 | -1.9790 |
| 45 | C5H10NH (piperidine) | 17 | 48 | -251.8073 | -1.5487 | -1.6211 |
| 46 | (CH3)3C-O-CH3 | 18 | 50 | -272.8712 | -1.6146 | -1.6974 |
| 47 | C6H4F2(1,3-difluorobenzene) | 12 | 52 | -430.5462 | -2.0171 | -2.1432 |
| 48 | C6H4F2(1,4-difluorobenzene) | 12 | 52 | -430.5452 | -2.0175 | -2.1436 |
| 49 | C6H5F (fluorobenzene) | 12 | 44 | -331.3479 | -1.7012 | -1.7949 |
| 50 | (CH3)2CH-O-CH(CH3)2 | 21 | 52 | -312.1751 | -1.8651 | -1.9583 |
| 51 | C2F6 | 8 | 66 | -675.0219 | -2.4306 | -2.6426 |

The origin of the visible differences above between energy corrections (i.e. $|E_{corr}(HF/6\text{-}31G(d))| >> |E_{corr}(HF/6\text{-}311\text{+}G(2d,p))|$) is obviously the basis set error (and not the correlation effect). It is compensated by the differences in the RECEP-3 and REBECEP correlation energy parameters. (The molecules of the G3-3 subset containing second-row atoms like Si, P, S, Cl were excluded because of the problems with basis sets, atomic enthalpies of formation, relativistic effects, and the relatively poor MP2 molecular geometries. Also, for Na-Ar the 6-311G(d) basis set is defined in a non-uniform manner across the row.)

Experimental enthalpies of formation ($\Delta H_f^0(298K)$), deviations of G3, B3LYP and REBECEP methods from experimental enthalpies of formation in case of the 50 selected molecules from the G3/99 test set pluss urea (#25) in kcal/mol:

| | Species | experimental | | expt.-theory | | expt.-REBECEP[NPA],a | |
|---|---|---|---|---|---|---|---|
| | | $\Delta H_f^0$(298K) | Error(±) | G3 | B3LYP | 6-311+G(2d,p) | 6-31G(d) |
| 1 | C4H6 (methylallene) | 38.8 | 0.1 | 0.2 | 0.0 | 1.7 | 0.1 |
| 2 | C5H8 (isoprene) | 18.0 | 0.3 | -0.2 | -4.9 | -0.1 | -2.9 |
| 3 | C5H10 (cyclopentane) | -18.3 | 0.2 | -0.5 | -10.0 | -3.7 | -0.5 |
| 4 | C5H12 (n-pentane) | -35.1 | 0.2 | 0.3 | -7.1 | -1.5 | -1.1 |
| 5 | C5H12 (neopentane) | -40.2 | 0.2 | 0.5 | -10.5 | -5.8 | -4.4 |
| 6 | C6H8 (1,3-cyclohexadiene) | 25.4 | 0.2 | -0.9 | -9.3 | -2.5 | -2.0 |
| 7 | C6H8 (1,4-cyclohexadiene) | 25.0 | 0.1 | -1.4 | -9.7 | -0.6 | -0.7 |
| 8 | C6H12 (cyclohexane) | -29.5 | 0.2 | -0.2 | -13.4 | -5.6 | -1.9 |
| 9 | C6H14 (n-hexane) | -39.9 | 0.2 | 0.6 | -9.3 | -3.3 | -1.6 |
| 10 | C6H14 (3-methylpentane) | -41.1 | 0.2 | 0.2 | -11.7 | -6.4 | -4.5 |
| 11 | C6H5CH3 (toluene) | 12.0 | 0.1 | -0.9 | -7.6 | xx | 0.4 |
| 12 | C7H16 (n-heptane) | -44.9 | 0.3 | 0.8 | -11.7 | -4.4 | -2.3 |
| 13 | C8H8 (cyclooctatetraene) | 70.7 | 0.4 | -1.4 | -11.7 | -3.9 | -5.2 |
| 14 | C8H18 (n-octane) | -49.9 | 0.3 | 0.9 | -14.0 | -5.7 | -3.0 |
| 15 | C10H8 (naphthalene) | 35.9 | 0.4 | 0.5 | -11.7b | xx | 0.2 |
| 16 | C10H8 (azulene) | 69.1 | 0.8 | -1.6 | -11.7b | xx | -10.7 |
| 17 | CH3COOCH3 | -98.4 | 0.4 | 0.7 | -2.9 | -1.5 | 1.1 |
| 18 | (CH3)3COH | -74.7 | 0.2 | 0.8 | -9.0 | -4.6 | -3.4 |
| 19 | C6H5NH2 (aniline) | 20.8 | 0.2 | -1.3 | -2.7 | xx | -1.5 |
| 20 | C6H5OH (phenol) | -23.0 | 0.2 | -1.6 | -7.1b | xx | -1.1 |
| 21 | C4H6O (divinyl ether) | -3.3 | 0.2 | -0.2 | -1.2 | -1.0 | -3.6 |
| 22 | C4H8O (tetrahydrofuran) | -44.0 | 0.2 | -0.2 | -7.6 | -2.1 | 1.9 |
| 23 | C5H8O (cyclopentanone) | -45.9 | 0.4 | 0.7 | -8.2 | -1.9 | 2.0 |
| 24 | C6H4O2 (benzoquinone) | -29.4 | 0.8 | -1.1 | -8.6b | xx | -0.9 |
| 25 | CH4ON2 (urea) | -56.3 | 0.3 | -1.3 | xx | -1.8 | 1.0 |
| 26 | C4H4N2 (pyrimidine) | 46.8 | 0.3 | 1.7 | 5.3 | 2.7 | 4.7 |
| 27 | NC-CH2-CH2-CN | 50.1 | 0.2 | -0.2 | -2.1 | 1.9 | 6.5 |
| 28 | C4H4N2 (pyrazine) | 46.9 | 0.3 | -2.7 | 1.4 | 0.6 | 3.5 |
| 29 | CH3-CO-C-CH | 15.6 | 0.2 | -2.5 | -5.9 | -2.8 | -3.6 |
| 30 | CH3-CH=CH-CHO (crotonaldehyde) | -24.0 | 0.3 | 0.8 | -1.0 | 2.3 | 1.3 |
| 31 | CH3-CO-O-CO-CH3 | -136.8 | 0.4 | 2.1 | -4.0 | -4.1 | 0.3 |
| 32 | (CH3)2CH-CN | 5.6 | 0.3 | -1.1 | -5.4 | -0.4 | 1.2 |
| 33 | CH3-CO-CH2-CH3 | -57.1 | 0.2 | 0.3 | -4.5 | -0.5 | 1.1 |
| 34 | (CH3)2CH-CHO | -51.6 | 0.2 | -0.6 | -6.4 | -2.7 | -2.1 |
| 35 | C4H8O2 (1,4-dioxane) | -75.5 | 0.2 | 0.9 | -7.8 | -1.3 | 3.7 |
| 36 | C4H8NH (tetrahydropyrrole) | -0.8 | 0.2 | -0.7 | -5.3 | -3.2 | -0.8 |
| 37 | CH3-CH2-CH(CH3)-NO2 | -39.1 | 0.4 | 1.1 | -5.2 | -4.8 | -1.2 |
| 38 | CH3-CH2-O-CH2-CH3 | -60.3 | 0.2 | 0.8 | -4.4 | -0.8 | 1.3 |
| 39 | CH3-CH(OCH3)2 | -93.1 | 0.2 | 1.7 | -6.5 | -4.4 | -1.4 |
| 40 | (CH3)3C-NH2 (t-butylamine) | -28.9 | 0.2 | -0.1 | -6.5 | -6.4 | -5.4 |
| 41 | -CH=CH-N(CH3)-CH=CH- | 24.6 | 0.1 | -0.8 | -2.8 | -7.7 | -4.4 |
| 42 | C5H10O (tetrahydropyran) | -15.2 | 0.2 | 0.3 | -10.8 | -4.0 | 0.9 |
| 43 | CH3-CH2-CO-CH2-CH3 | -61.6 | 0.2 | 1.1 | -6.4 | -1.3 | 1.3 |
| 44 | CH3-CO-O-CH(CH3)2 | -115.1 | 0.2 | 1.3 | -8.3 | -5.2 | -0.5 |
| 45 | C5H10NH (piperidine) | -11.3 | 0.1 | -0.9 | -9.2 | -5.7 | -2.0 |
| 46 | (CH3)3C-O-CH3 | -67.8 | 0.3 | 1.4 | -10.2 | -6.9 | -5.0 |
| 47 | C6H4F2 (1,3-difluorobenzene) | -73.9 | 0.2 | 0.4 | -4.7b | xx | -1.0 |
| 48 | C6H4F2 (1,4-difluorobenzene) | -73.3 | 0.2 | 0.4 | -4.8b | xx | -0.4 |
| 49 | C6H5F (fluorobenzene) | -27.7 | 0.3 | -0.4 | -5.1b | xx | 0.7 |
| 50 | (CH3)2CH-O-CH(CH3)2 | -76.3 | 0.4 | 1.6 | -11.6 | -6.7 | -3.2 |
| 51 | C2F6 | -321.3 | 0.8 | 2.8 | -7.4 | 0.5 | 2.3 |

#27: butanedinitrile, #29: acetyl acetylene, #31:acetic anhydride, #37:nitro-s-butane,
#39: acetaldehyde dimethyl acetal, 41:N-methylpyrrole.

a.: NPA or Mulliken charges (not shown) are used in the REBECEP calculation.
The REBECEP parameters were fitted to reproduce the G3 or the experimental total energy.

b.: Natural bond orbital cannot handle linearly dependent basis sets.

B3LYP= B3LYP/6-311+G(3df,2p).

Owing to linear dependency problems, the NPA analysis with the 6-311+G(2d,p) basis set was not feasible for some molecules, see xx in column REBECEP[6-311+G(2d,p)], while all NPA charges were available for the smaller 6-31G(d) basis set.

The experimental errors are rather small for the 51 molecules (<±0.4 kcal/mol, only three <±0.8 kcal/mol (azulene, benzoqinone, $C_2F_6$)). The G3 values approximate the experimental values quite well, however, the B3LYP values show considerable error: $\Delta H_f^0$(298K,B3LYP) >> $\Delta H_f^0$(298K,expt) by average deviation -6.82 kcal/mol, while for G3, REBECEP[NPA, 6-311+G(2d,p) or 6-31G(d)] it is only 0.04 (nearly perfect), -2.7 or -1.0, resp.. Histogram of G3, B3LYP, and REBECEP deviations (expt. minus theory) for the test set of 51 molecules, each vertical bar represents the frequency of a given deviation in a 1 kcal/mol range:

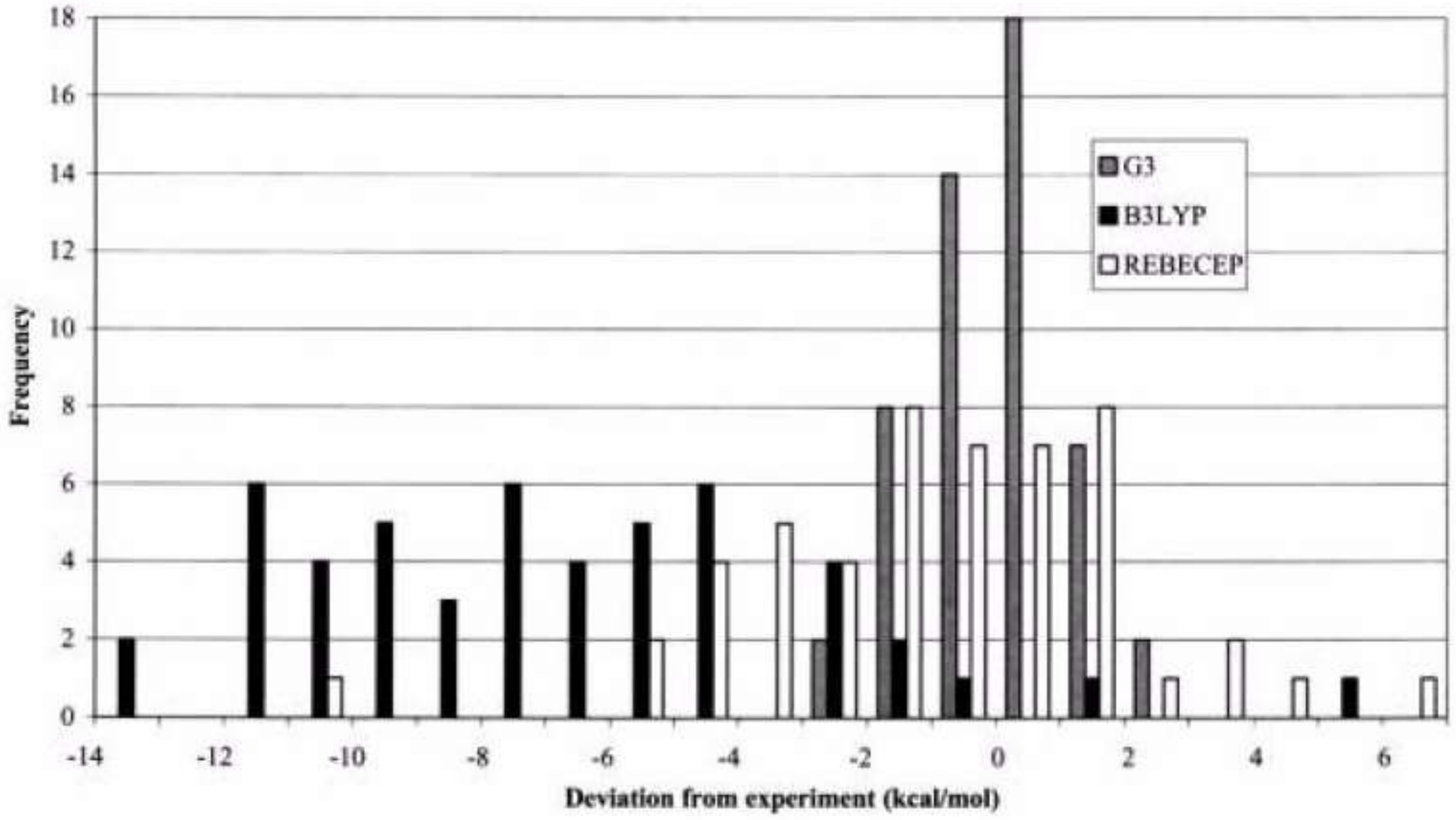

This histogram clearly shows the problem with the B3LYP method of which the distribution of errors is not a Gaussian-like, but nearly evenly distributed between -4 and -11 kcal/mol. The distribution of the REBECEP deviations is considerably more focused than the distribution of the B3LYP deviations. The REBECEP is closer to the ideal Gaussian distribution, however, it shows a composed structure (e.g. several overlapping Gaussians). The G3 method shows a relatively narrow and nearly ideal Gaussian distribution.

In agreement with literature, the above analysis suggests that the density functional theory (here B3LYP) errors accumulate in large molecules, while analyzing the REBECEP errors with respect to the number of electron pairs, it was found that no correlation in general, however, homologous series show quasi-linear accumulation of errors.

The REBECEP[6-31G(d)] with the small basis set can provide results of comparable quality to the results of the RECEP-3[6-311+G(2d,p)] performed with the considerably larger basis set. This can be attributed to the consistent behavior of 6-31G(d). The HF-SCF basis set extension energy error can be effectively approximated from partial charges. The implicit atomic partitioning of the basis set error can be done explicitly and the infinite basis set energies may be predicted from atomic parameters and charges.

**Thesis-11-Theory: Correlation energy calculation with Mulliken matrix**

Good density functional quality (B3LYP/6-31G*) ground state total electronic energies have been approximated using single point Hartree–Fock-self consistent field (HFSCF/6-31G*) total energies and its Mulliken matrix as certain density functional developed by myself. This is a development of my rapid estimation of basis set error and correlation energy from partial charges (REBECEP) method (density functional) based solely on using partial charges (e.g. Mulliken) only, published earlier. The development: (1) A larger set of atoms (H, C, N, O, F, Si, P and S) was considered as building blocks for closed shell, neutral, ground state molecules at their equilibrium geometry; (2) Geometries near equilibrium geometry were also considered; (3) A larger set, containing 115 molecules, was used to fit REBECEP parameters; (4) Most importantly, electrons belonging to chemical bonds (between atom pairs) were also considered (Mulliken matrix) in addition to the atoms (Mulliken charges), using more REBECEP parameters to fit and yielding a more flexible algorithm. With these parameters a rather accurate closed shell ground state electronic total energy can be obtained from a small basis set HFSCF calculation in the vicinity of optimal geometry. The 3.3 kcal/mol root mean square deviation of REBECEP has been improved to 1.5 kcal/mol when using Mulliken matrix instead of Mulliken charges.

= = = = = =

Jó sűrűség funkcionál (B3LYP/6-31G*) minőségű alap állapotú teljes elektronikus energia közelítését dolgoztam ki egypontos (adott geometriára vonatkozó) Hartree–Fock ön-konzisztens (HF-SCF/6-31G*) teljes energia és annak Mulliken mátrixának felhasználásával, ami egy saját fejlesztésű sűrűség funkcionál. Ez egy továbbfejlesztése a “rapid estimation of basis set error and correlation energy from partial charges (REBECEP)” előzőleg publikált módszeremnek (sűrűség funkcionál), mely egyedül a parciális töltésekre (pl. Mulliken) volt alapozva csak. A fejlesztés: (1) Egy nagyobb atom halmazt (H, C, N, O, F, Si, P és S) használtam, mint építőköveket a zárt héjú, semleges, alap állapotú molekulákra az egyensúlyi geometriájuknál; (2) Egyensúlyi geometriák közelében levő geometriákat is felhasználtam; (3) Egy nagyobb halmazt, mely 115 molekulát tartalmazott, használtam a REBECEP paraméterek illesztésére; (4) Mint legfontosabb, a kémiai kötésben részt vevő elektronokat (atom párok közt) szintén figyelembe vettem (Mulliken mátrix) az atomokkal (Mulliken töltések) egyetemben, ami több REBECEP paraméter illesztését igényelte, de flexibilisebb algoritmust eredményezett. Ezekkel a paraméterekkel egy pontosabb zárt héjú alap állapotú elektronikus teljes energia kapható egy kis bázis készletű HF-SCF számításból az optimális geometriára vagy annak környezetére. A 3.3 kcal/mol REBECEP átlagos négyzetes hiba 1.5 kcal/mol-ra javul, ha a Mulliken mátrixot használjuk a Mulliken töltések helyett.

**Sandor Kristyan: Theoretical Chemistry Accounts, 115 (2006) 298-307**

= = = = = =

Representative equations/tables/figures:

The REBECEP (rapid estimation of basis set error and correlation energy from partial charges) formula to calculate correlation energy and basis set error for ground state covalent neutral molecules in the vicinity of stationary points (here only geometry minimums) is

$$E_{corr}(\text{REBECEP, method, charge def., basis set}) \equiv$$
$$\Sigma_{A=1,\dots,M}\, E_{corr}(N_A, Z_A, \text{method, charge def., basis set}),$$

where the "method" is e.g. G2, G3 or B3LYP/6-31G*, etc. what REBECEP reproduce (aproximates) with using only the (much) faster and (much) less disc space consuming HF-SCF/basis calculation, the "charge def." is e.g. Mulliken or NPA (natural population analysis) partial cahrges, etc., M= # of atoms in the molecule, and $N_A$= electrons w/r to partial charge around atom with atomic charge $Z_A$.

The $E_{corr}(N_A, Z_A$ method, charge def., basis set) atomic energy terms are to be interpolated linearly as follows: $E_{corr}(N_A, Z_A$, method, charge def., basis set) =

$$(N_A - N_1)E_{fitpar}(N_2, Z_A, \text{method, charge def.,basis set})$$
$$+(N_2 - N_A)E_{fitpar}(N_1, Z_A, \text{method, charge def., basis set}),$$

where $N_1$ and $N_2$ are integer numbers of electrons, with $N_1 \leq N_A \leq N_2 = N_1 + 1$, and $N_A$ is the electron content around atom A based on the chosen partial charge it.

$E_{fitpar}(N_1$ or $N_2$, $Z_A$, method, charge def., basis set) are the so called REBECEP atomic parameters that transform the partial charge into energy correction (correlation and basis set). Important to notice, that these are close to the CI correlation energy of atoms ($Z_A$, $N_1$ or $N_2$).

The development of "atomic charge" to "Mulliken matrix" version of REBECEP is that the latter uses the Mulliken matrix elements between atom pairs as charges on "dummy atoms" in addition to the atomic Mulliken caharges.

Input data:

An extract from Gaussian 98 output of HF-SCF single point energy calculation for Mulliken matrix and Mulliken partial charges (the latter is a "projection" of Mulliken matrix to its diagonal elelemts called "Total atomic charges") for the molecule formaldehyde is

```
> # hf/6-31G*
> Formaldehyde (H2C=O)
> SCF Done: E(RHF) = -113.863712881 A.U. after 6 cycles
> Convg = 0.8513D-04 -V/T = 2.0037
> S**2 = 0.0000
> Condensed to atoms (all electrons):
>      1        2         3         4
> 1 O  8.013325 0.522645 -0.049815 -0.049815   ← 4x4 Mulliken matrix
> 2 C  0.522645 4.600151  0.371281  0.371281
> 3 H -0.049815 0.371281  0.596456 -0.068770
> 4 H -0.049815 0.371281 -0.068770  0.596456
> Total atomic charges:
> 1
> 1 O -0.436341   ← Mulliken partial charge
> 2 C  0.134643
> 3 H  0.150849
> 4 H  0.150849
> Sum of Mulliken charges= 0.00000
> Normal termination of Gaussian 98.
> . . .
> -0.63645253 = (-114.50016541)-(-113.86371288)
   an accurate corr. energy in hartree (B3LYP-HF)/6-31G*
> Formaldehyde (H2C=O)   ---> See molecule #13 in "Test of method" table.
```

Let us rearrange the data for this neutral molecule at equilibrium geometry:

```
> 0 Molecular charge
> 16 # of atoms in the molecule, listed below
> 8 -0.01332500 Z atomic charge, partial charge
> 6  1.39984900
> 1  0.40354400
> 1  0.40354400
> 608 -0.52264500 dummy Z between O C
> 108  0.04981500 dummy Z between O H
> 108  0.04981500 dummy Z between O H
> 608 -0.52264500 dummy Z between C O
> 106 -0.37128100 dummy Z between C H
> 106 -0.37128100 dummy Z between C H
> 108  0.04981500 dummy Z between H O
> 106 -0.37128100 dummy Z between H C
> 101  0.06877000 dummy Z between H H
> 108  0.04981500 dummy Z between H O
> 106 -0.37128100 dummy Z between H C
> 101  0.06877000 dummy Z between H H
```

where e.g. 0608= 608 is a "dummy atom" atomic charge between C(Z=6) and O(Z=8) with the first two digits (06) as number is less then the last two digits (08) as number with electron exces (partial charge) –0.52264500 between them.

Fitting REBECEP parameters:

Fitted $E_{fitpar}$(N, Z) REBECEP/6-31G* atomic parameters in hartree to use in REBECEP formula for calculating correlation energy and basis set error for closed shell neutral molecules in the vicinity of optimum geometry from HF-SCF/6-31G* Mulliken charges/matrix containing atoms listed to correct HF-SCF/6-31G* total ground state electronic energy to achieve e.g. B3LYP quality, (N = # of electrons and Z = atomic charge).

The case of Mulliken charge (CHARGE): A set of 115 molecules with their Mulliken partial charges has generated the range of these 23 parameters:

| Atom | N | Z | $E_{fitpar}$(N,Z) |
|---|---|---|---|
| H | 2 | 1 | -0.04172540 |
| C | 4 | 6 | -0.17422431 |
| C | 5 | 6 | -0.20981117 |
| C | 6 | 6 | -0.23745760 |
| C | 7 | 6 | -0.25917593 |
| N | 6 | 7 | -0.33448715 |
| N | 7 | 7 | -0.30484084 |
| N | 8 | 7 | -0.32310273 |
| O | 7 | 8 | -0.46192945 |
| O | 8 | 8 | -0.36921711 |
| O | 9 | 8 | -0.37905434 |
| F | 9 | 9 | -0.40151786 |
| F | 10 | 9 | -0.42644947 |
| Si | 12 | 14 | -0.50175714 |
| Si | 13 | 14 | -0.55615027 |
| Si | 14 | 14 | -0.57734232 |
| P | 13 | 15 | -0.55840024 |
| P | 14 | 15 | -0.61222298 |
| P | 15 | 15 | -0.63072847 |
| S | 14 | 16 | -0.63446938 |
| S | 15 | 16 | -0.67763142 |
| S | 16 | 16 | -0.67971304 |
| S | 17 | 16 | -0.68886287 |

The case of Mulliken matrix (MATRIX): The Z > 16 values are fictitious values describing atom-atom pairs, so for the N> 16. For example, Z = 106 = 100 × $Z_1$ + $Z_2$ with $Z_1 \leq Z_2 \Rightarrow$ Z1 = 1 and $Z_2$ = 6 ⇒ it is a H–C atom pair, etc. A set of 115 molecules with their Mulliken matrices for partial charge has generated the range of these 106 parameters:

| Atom | N | Z | $E_{fitpar}$(N,Z) |
|---|---|---|---|
| H | 2 | 1 | -0.06622464 |
| C | 3 | 6 | -0.23074779 |
| C | 4 | 6 | -0.16366323 |
| C | 5 | 6 | -0.19885719 |
| C | 6 | 6 | -0.23999147 |
| N | 5 | 7 | -0.33387528 |
| N | 6 | 7 | -0.30265901 |
| N | 7 | 7 | -0.32320268 |
| N | 8 | 7 | -0.31547675 |
| O | 7 | 8 | -0.27245124 |
| O | 8 | 8 | -0.37441258 |
| O | 9 | 8 | -0.42794560 |
| F | 8 | 9 | -0.56146588 |
| F | 9 | 9 | -0.45917123 |
| F | 10 | 9 | -0.40105467 |
| Si | 10 | 14 | -1.37396128 |
| Si | 11 | 14 | -0.24089557 |
| Si | 12 | 14 | -0.19384219 |
| P | 11 | 15 | -0.36525803 |
| P | 12 | 15 | -0.51376891 |
| P | 13 | 15 | -0.54378488 |
| P | 14 | 15 | -0.58356695 |
| S | 12 | 16 | -0.47375909 |
| S | 13 | 16 | -0.61628751 |
| S | 14 | 16 | -0.61866633 |
| S | 15 | 16 | -0.65943013 |
| S | 16 | 16 | -0.70051863 |
| H H | 100 | 101 | 0.02466176 |
| H H | 101 | 101 | -0.00002474 |
| H C | 105 | 106 | 0.02908445 |
| H C | 106 | 106 | 0.00010452 |
| H C | 107 | 106 | -0.01891420 |
| H N | 106 | 107 | -0.00785187 |
| H N | 107 | 107 | 0.00064296 |
| H N | 108 | 107 | -0.00441181 |
| H O | 107 | 108 | 0.04957125 |
| H O | 108 | 108 | 0.00061670 |
| H O | 109 | 108 | 0.01110888 |
| H F | 108 | 109 | -0.23221507 |
| H F | 109 | 109 | 0.00317488 |
| H F | 110 | 109 | 0.06218120 |
| HSi | 113 | 114 | -0.36008531 |
| HSi | 114 | 114 | -0.17045459 |
| HSi | 115 | 114 | 0.15404294 |
| H P | 114 | 115 | 0.04703979 |
| H P | 115 | 115 | -0.00996426 |
| H P | 116 | 115 | -0.00196052 |
| H S | 115 | 116 | 0.00484868 |
| H S | 116 | 116 | 0.00099813 |
| H S | 117 | 116 | 0.00550390 |
| C C | 605 | 606 | 0.03838783 |
| C C | 606 | 606 | -0.00024197 |
| C C | 607 | 606 | -0.03750534 |
| C C | 608 | 606 | -0.04927484 |
| C N | 606 | 607 | -0.01010163 |
| C N | 607 | 607 | -0.00009908 |
| C N | 608 | 607 | -0.01659326 |
| C O | 607 | 608 | 0.05062429 |
| C O | 608 | 608 | -0.00144555 |
| C O | 609 | 608 | -0.02567958 |
| C F | 608 | 609 | -0.05229206 |
| C F | 609 | 609 | -0.01489223 |
| C F | 610 | 609 | 0.15527481 |
| CSi | 613 | 614 | 0.17134387 |
| CSi | 614 | 614 | 0.34342506 |
| CSi | 615 | 614 | 0.69324456 |
| C P | 614 | 615 | 0.03070217 |
| C P | 615 | 615 | 0.02223860 |
| C S | 615 | 616 | -0.04885000 |
| C S | 616 | 616 | 0.00237198 |
| C S | 617 | 616 | -0.01438090 |
| N N | 706 | 707 | -0.07637497 |
| N N | 707 | 707 | 0.00162535 |
| N N | 708 | 707 | 0.01900106 |
| N O | 707 | 708 | 0.02826211 |
| N O | 708 | 708 | -0.00020507 |
| N F | 708 | 709 | 0.31644475 |
| N F | 709 | 709 | 0.01304137 |
| N F | 710 | 709 | 0.12844456 |
| O O | 807 | 808 | 0.04144183 |
| O O | 808 | 808 | 0.00205831 |
| O O | 809 | 808 | -0.06846057 |
| O F | 808 | 809 | -0.33836067 |
| O F | 809 | 809 | 0.01181880 |
| OSi | 814 | 814 | -0.28877492 |
| OSi | 815 | 814 | -0.23619244 |
| O P | 814 | 815 | -0.06232572 |
| O P | 815 | 815 | -0.00480354 |
| O P | 816 | 815 | -0.01175658 |
| O S | 815 | 816 | -0.00606231 |
| O S | 816 | 816 | -0.00386373 |
| O S | 817 | 816 | -0.01576410 |
| F F | 908 | 909 | -0.44396153 |
| F F | 909 | 909 | 0.00270366 |
| FSi | 914 | 914 | -0.12826953 |
| FSi | 915 | 914 | 0.29276534 |
| F P | 915 | 915 | 0.00146196 |
| F P | 916 | 915 | 0.04812473 |
| SiSi | 1413 | 1414 | 0.58203406 |
| SiSi | 1414 | 1414 | 0.66704466 |
| SiSi | 1415 | 1414 | 1.36228331 |
| P P | 1514 | 1515 | -0.08698784 |
| P P | 1515 | 1515 | 0.02160504 |
| P P | 1516 | 1515 | 0.07410262 |
| S S | 1615 | 1616 | -0.35180004 |
| S S | 1616 | 1616 | 0.02064828 |

Test of the method:

Molecule set used in the linear fit to get the REBECEP/6-31G* parameter set are listed. The column B3LYP ≡ (B3LYP/6-31G*)−(HF-SCF/6-31G*) is the B3LYP level correlation and basis set error in hartree. The last two columns show the deviation from B3LYP total ground state electronic energies in kcal/mol belonging to

Mulliken charge [CHARGE ≡ (B3LYP/6-31G*) − (REBECEP/6-31G*/Mulliken charge)] and

Mulliken matrix [MATRIX ≡ (B3LYP/6-31G*) − (REBECEP/6-31G*/Mulliken matrix)] methods. These two methods are supposed to reproduce the B3LYP/6-31G* correlation energy and basis set error correction and the total ground state electronic energy. In the "CHARGE" case the root mean square deviation is 3.3 kcal/mol, which is improved in "MATRIX" method to 1.5 kcal/mol. (1 hartree is about 627.5 kcal/mol). Notice that the order number quasi increases with molecular weight or $\Sigma N_A=N$, so the absolute value B3LYP (as expected by rule), but not the deviation (error) of CHARGE and MATRIX values, indicating that the REBECEP is a plausible approximation.

| Molecule | B3LYP [hartree] | CHARGE [kcal/mol] | MATRIX [kcal/mol] |
|---|---|---|---|
| 1 Methane (CH4) | −0.3233 | −1.1 | 0.2 |
| 2 Ammonia (NH3) | −0.3641 | 0.5 | 0.4 |
| 3 Water (H2O) | −0.3991 | 1.4 | 0.3 |
| 4 Hydrogen fluoride (HF) | −0.4179 | 4.1 | 0.3 |
| 5 Silane (SiH4) | −0.6586 | 1.3 | 0.0 |
| 6 Phosphine (PH3) | −0.6924 | 0.7 | −0.1 |
| 7 Hydrogen sulfide (H2S) | −0.7182 | 0.4 | 0.2 |
| 8 Acetylene (C2H2) | −0.5098 | 4.6 | 2.4 |
| 9 Ethylene (C2H4) | −0.5563 | 1.6 | 1.1 |
| 10 Ethane (C2H6) | −0.6018 | −0.5 | 0.7 |
| 11 Hydrogen cyanide (HCN) | −0.5516 | 6.3 | 2.2 |
| 12 Carbon monoxide (CO) | −0.5746 | 16.6 | 2.9 |
| 13 Formaldehyde (H2CO) | −0.6365 | 3.9 | 2.2 |
| 14 Methanol (CH3OH) | −0.6802 | 1.3 | 1.1 |
| 15 Hydrazine (H2N−NH2) | −0.6884 | 0.7 | 0.6 |
| 16 Hydrogen peroxide (H2O2) | −0.7730 | −2.0 | 0.5 |
| 17 Carbon dioxide (CO2) | −0.9522 | 4.6 | 1.7 |
| 18 Silicon monoxide (SiO) | −0.9446 | −3.6 | 0.0 |
| 19 Carbon monosulfide (CS) | −0.9057 | 7.5 | 0.1 |
| 20 Disilane (H3Si-SiH3) | −1.2777 | 1.2 | 0.0 |
| . . . . . . . . | | | |
| 100 Alanine | −1.8763 | 2.5 | 2.5 |
| 101 Allenyl-CH3 | −1.0763 | −0.9 | 1.5 |
| 102 Glycine | −1.5965 | 2.9 | 1.3 |
| 103 m-Methyl-Ethyl-Benzene | −2.3877 | 0.6 | 0.4 |
| 104 o-Methyl-Ethyl-Benzene | −2.3885 | 0.4 | 0.0 |
| 105 p-Methy- Ethyl-Benzene | −2.3879 | 0.6 | 0.3 |
| 106 Trinitro-toluol (TNT) | −4.9438 | −7.2 | 1.3 |
| 107 Valine | −2.4362 | 1.8 | −1.0 |
| 108 CH3-NH-CH2-NH-CH3 | −1.5259 | 0.9 | 2.4 |
| 109 CH3-NH-NH2 | −0.9696 | −0.1 | 1.2 |
| 110 CH3-POH-CH2-POH-CH3 | −2.9003 | −0.9 | 0.5 |
| 111 CH3-POH-POH-CH3 | −2.6305 | 0.3 | −0.8 |
| 112 CH3-SiH2-CH2-SiH3 | −1.8345 | 2.9 | 0.0 |
| 113 CH3-SiH2-SiH2-CH3 | −1.8388 | 0.3 | 0.0 |
| 114 NH2-CH2-NO2 | −1.6782 | 2.6 | 0.8 |
| 115 Quinuclidine N(CH2CH2)3CH | −2.2332 | 3.1 | 0.8 |

**Thesis-12-Theory: REZEP for calculating zero point energies (ZPE) of molecules**

Using a database of HF-SCF/6-31G(d) zero-point energies (scaled by 0.8929) and atomic partial charges of 117 closed-shell, neutral molecules containing H, C, N, O, and F atoms, relationships have been developed that permit the rapid estimation of zero-point energies from atomic partial charges (REZEP). The estimated zero-point energies have been compared to scaled HF-SCF/6-31G(d), B3LYP/6-31G(2df,p) (scaled by 0.9854), and to zero-point energies estimated from molecular stoichiometry. Sixty-nine experimental zero-point energies have also been used to check the quality of the various methods. The scaled HF-SCF and B3LYP zero-point energies show 0.4, and the stoichiometric and the proposed REZEP methods show a 1.0 kcal/mol average absolute deviation from the experimental results. New parameters have been developed for the stoichiometric method that reduces the average absolute deviation from the experimental results to 0.7 kcal/mol.

= = = = = =

Felhasználva HF-SCF/6-31G(d) szintű zéró pont energiákat (skálafaktor 0.8929) és atomi parciális töltéseket 117 zárt héjú, semleges, H, C, N, O és F atomokat tartalmazó molekula esetében, egy összefüggést fejlesztettünk ki a zéró pont energia gyors közelítésére atomi parciális töltések felhasználásával (rapid estimation of zero-point energies from atomic partial charges, REZEP). Ezen zéró pont energia közelítéseket összehasonlítottuk skálázott HF-SCF/6-31G(d), B3LYP/6-31G(2df,p) (skálafaktor 0.9854), valamint az (irodalmi) "zéró pont energia közelítése molekuláris sztöchiometriából" módszerekkel. Több, nevezetesen 69, kísérleti zéró pont energiát is teszteltünk ezen különböző módszerek minősítésére. A skálázott HF-SCF és B3LYP zéró pont energiák 0.4, a sztöichiometrikus és a kifejlesztett REZEP módszerek 1.0 kcal/mol átlagos abszolút eltérést mutattak a kísérleti értékektől. (Bármely bázisfüggvény szinten, a HF-SCF vagy B3LYP energia es atomi töltés számítás adott molekuláris geometriára jóval kisebb számítási igényű frekvencia számítás nélkül, mint azzal.) Új paramétereket fejlesztettünk ki az (irodalmi) sztöichiometrikus módszerre, amelyek az átlag abszolút eltérést 0.7 kcal/mol értékre redukálják a kísérleti eredményekre vonatkozóan.

**Adrienn Ruzsinszky - Sandor Kristyan - Jozsef L. Margitfalvi - Gabor I. Csonka: Journal of Physical Chemistry A, 107(11) (2003) 1833-1839**

= = = = = =

Representative equations/tables/figures:

Compared was explicit ZPEs for 117 molecules at HF-SCF/6-31G(d) and B3LYP/6-31G(2df,p) levels (scaled by 0.8929 and 0.9854, resp.) and ZPEs calculated from the molecular stoichiometry using Politzer' linear equation (1995, using database of 61 molecular BP86/6-31G(d,p) vibrational energy). The latter is the simple

$$\text{ZPE (Politzer, kcal/mol)} = 7.06n_H + 3.66n_C + 3.41n_N + 2.76n_O + 1.90n_F + 2.49n_{Cl} - 3.97.$$

Conclusion was: Larger database allows more accurate parameters, but more importantly, the wheigting by e.g. atomic partial charges allows even further and stronger improvement.

In our rapid estimation of zero-point energy from partial charges (REZEP) procedure, the ZPE is estimated for closed- or open shell ground state covalent molecules in the vicinity of their equilibrium geometry by an inexpensive atom by atom method:

$$\text{ZPE(REZEP)} \equiv \sum E_{ZPE}(N_A, Z_A),$$

where the sum runs for all A=1,…,M atoms in the molecule, approximating an accurate (experimental or well-tested calculating) ZPE, using a particular basis set and atomic partial charge definition. The "atomic ZPE energies" are

$$E_{ZPE}(N_A, Z_A) = (N_A - N_1)\, E_{ZPEpar}(N_2, Z_A) + (N_2 - N_A)\, E_{ZPEpar}(N_1, Z_A),$$

where $N_A = (Z_A$ - partial charge on A), the noninteger "electron content" of atom A between closest integer values $N_1 \leq N_A \leq N_2 = N_1 + 1$, $Z_A$ is the nuclear charge of atom A, as well as we used atomic charges obtained from HF-SCF/6-31G(d) results at the given molecular geometry. $E_{ZPEpar}(N,Z)$ are the „atomic ZPE parameters", for hydrogen atoms we suggest using a single parameter (as in REBECEP)

$$E_{ZPEpar}(N_A, Z_A=1) = N_A\, E_{ZPEpar}(N_2=2, Z_A=1)/2 \qquad \text{for } 0 \leq N_A \leq 2.$$

The simplicity of this suggested method is manifesting: After the HF-SCF/basis level partial charge calculation it is instant.

Parameters $E_{ZPEpar}(N,Z)$ in kcal/mol were obtained from a multilinear fit for minimizing $\sum[\text{ZPE(G3)}_i - \text{ZPE(REZEP)}_i]^2$, where $\text{ZPE(G3)}_i$ was an empirically corrected (scaled by 0.8929), fitted HF-SCF/6-31G(d) level ZPE, and $\text{ZPE(REZEP)}_i$ was calculated with 6-31G(d) basis set and with several charge definitions for 117 molecules; (the two basis sets do not have to be the same!, like in REBECEP the B3LYP/6-31G(d) geom. used for charge calc.):

| Atom | $Z_A$ | $N_1$ | Mulliken | NPA | Stockholder |
|---|---|---|---|---|---|
| H | 1 | 2 | 15.08 | 14.65 | 13.45 |
| C | 6 | 4 | 2.48 | -4.40 | ... |
| C | 6 | 5 | 1.99 | -1.00 | -2.23 |
| C | 6 | 6 | 3.99 | 3.61 | 4.02 |
| C | 6 | 7 | 3.81 | 7.05 | -9.20 |
| N | 7 | 6 | 5.74 | 1.64 | -3.41 |
| N | 7 | 7 | -0.39 | 0.99 | 3.40 |
| N | 7 | 8 | 7.79 | 8.79 | 1.25 |
| N | 7 | 9 | ... | -7.31 | ... |
| O | 8 | 8 | 0.18 | 0.86 | 2.58 |
| O | 8 | 9 | 6.22 | 7.27 | 4.55 |
| F | 9 | 9 | -1.70 | 0.02 | 2.29 |
| F | 9 | 10 | 9.45 | 8.62 | 2.17 |

Our analysis has shown that using B3LYP instead of HF-SCF for ZPEs does not provide an improvement; recall the correlation calculation, where the B3LYP is definitely better than HF-SCF. For 117 molecules, the ZPE(G3) and deviations ZPE(experimental [*CCCDB,* release 7; Sept.2002; http://srdata.nist.gov/cccbdb]) - ZPE(G3) and ZPE(G3) - ZPE(REZEP/HF-SCF/6-31G(d)/partial charge) in kcal/mol (M,N = number of nuclei, electrons, resp.):

| | | | | | = = = = = deviations | = = = = = = | |
|---|---|---|---|---|---|---|---|
| no. | species | M | N | ZPE(G3) | exptl | Mulliken NPA | Stockholder |
| 1 | methane (CH4) | 5 | 10 | 26.77 | 0.34 | -2.3 -2.8 | -2.1 |
| 2 | ammonia (NH3) | 4 | 10 | 20.73 | -0.10 | -2.1 0.0 | -0.3 |
| 3 | water (H2O) | 3 | 10 | 12.87 | 0.01 | -1.1 -1.8 | -1.6 |
| 4 | hydrogenfluoride (HF) | 2 | 10 | 5.56 | 0.36 | -2.1 -2.5 | -1.8 |
| 5 | acetylene (C2H2) | 4 | 14 | 16.50 | -0.31 | -2.2 -3.5 | -1.0 |
| 6 | ethylene (H2CdCH2) | 6 | 16 | 30.68 | 0.21 | -2.0 -2.6 | -1.4 |
| 7 | ethane (H3C-CH3) | 8 | 18 | 44.68 | 0.64 | -1.2 -1.5 | -1.3 |
| 8 | hydrogencyanide (HCN) | 3 | 14 | 10.08 | -0.31 | -1.7 -2.3 | -2.3 |
| 9 | formaldehyde (H2CdO) | 4 | 16 | 16.36 | -0.22 | -3.0 -3.1 | -2.5 |
| 10 | methanol (CH3-OH) | 6 | 18 | 31.00 | 0.01 | -0.9 -1.1 | -1.2 |
| . . . | | | | | | | |
| 110 | piperidine(C5H10NH) | 17 | 48 | 95.78 | -0.22 | 2.3 2.7 | 2.0 |
| 111 | ether (CH3)3C-O-CH3 | 18 | 50 | 98.64 | 0.01 | -0.3 0.1 | 0.5 |
| 112 | 1,3-F-benzene(C6H4F2) | 12 | 58 | 50.76 | - | 0.5 0.8 | 0.9 |
| 113 | 1,4-F-benzene(C6H4F2) | 12 | 58 | 50.74 | - | 0.3 0.7 | 0.1 |
| 114 | fluorobenzene(C6H5F) | 12 | 50 | 55.57 | - | 0.4 0.4 | 0.4 |
| 115 | (CH3)2CH-O-CH(CH3)2 | 21 | 58 | 115.75 | - | -0.1 0.6 | 1.3 |
| 116 | C2F6 | 8 | 66 | 18.46 | - | 1.2 1.0 | 1.0 |
| 117 | azulene (C10H8) | 18 | 68 | 87.62 | - | -0.1 0.1 | 0.1 |

Statistical analysis of ZPE(G3) - ZPE(REZEP/HF-SCF/6-31G(d)/partial charge) from the 117 ZPEs (kcal/mol):

| | Mulliken | NPA | Stockholder |
|---|---|---|---|
| root-mean-square deviation | 1.1 | 1.3 | 1.1 |
| average deviation | -0.3 | -0.3 | -0.2 |
| average absolute deviation | 0.9 | 1.0 | 0.9 |
| largest positive deviation | 2.7 | 2.7 | 2.7 |
| largest negative deviation | -3.0 | -3.5 | -2.5 |

Because no experimental results were used in the parametrization of Politzer' equation but using BP86/6-31G(d,p) ZPEs, re-fitting to the currently used experimental ZPEs (69 of 117, containing the restricted types of atoms) improves:

$$\text{ZPE (modified Politzer, kcal/mol)} = 6.99n_H + 3.74n_C + 3.98n_N + 3.45n_O + 2.79n_F - 4.63$$

These new parameters provide considerably better rms (0.8 kcal/mol) and average absolute deviation (0.7 kcal/mol) in comparison to ZPE(REZEP), as well as ZPE(modified Politzer) is superior to ZPE(Politzer) in relation to experimental ZPEs. However, the handicap of ZPE(modified Politzer) still remains: e.g. isomers have the same ZPE(modified Politzer) while ZPE(REZEP) treats it physically much more plausibly. Consistency: For example, comparing the values of $E_{ZPEpar}$(N=2, Z=1, 6-31G(d), charge def., G3)/2 = 6.8-7.5 kcal/mol for Hydrogen compares with values 7.06 (ZPE(Politzer)) or 6.99 (ZPE(modified Politzer)).

**Thesis-13-Theory: Performance of REZEP for ZPE calculations**

In our method „Rapid Estimation of Zero point Energies from atomic Partial charges" [REZEP, J.Phys.Chem. A 107 (2003) 1833], we have used a database of scaled HF-SCF/6-31G(d) zero point energies and atomic partial charges of more than one hundred closed shell, neutral molecules containing H, C, N, O, and F atoms. Our method showed 1.0 kcal/mol average absolute deviation from the experimental results. I have demonstrated the performance of REZEP on large molecules like naphthalene (18 atoms) and α-iso-cinchonine (44 atoms>>18) against other *ab initio* and molecular mechanics calculations in view of computation time and disc space. The latter molecule is around the limit of the HF-SCF/6-31G(d) calculation used currently in standard computers with respect to correlation calculation and frequency analysis. The twin method REBECEP [Theor. Chem. Accounts, 106 (2001) 319] for correlation calculation is also commented on along with REZEP in this respect.

= = = = = =

A „Rapid Estimation of Zero point Energies from atomic Partial charges" [REZEP, J.Phys.Chem. A 107 (2003) 1833] módszerünkben több mint száz zárt héjú, semleges molekula skálázott HF-SCF/6-31G(d) zéruspont energiáit és atomi parciális töltéseit használtuk fel, mely adatbázisban a H, C, N, O, és F atomok fordultak elő. Módszerünk 1.0 kcal/mol átlagos abszolút eltérést mutatott a kísérleti eredményektől. Demonstráltam a REZEP teljesítményét nagyobb molekulákon, mint naftalin (18 atom) és α-izo-cinkonin (44 atom>>18) más *ab initio* és molekula mechanikai számításokkal szemben a számítógépes idő és felhasznált merev lemez terület tekintetében. Az utóbbi (44 atomos) molekula kb. a HF-SCF/6-31G(d) praktikus-gyakorlati számítások határán van a jelenlegi (2006) standard számítógépeken a korreláció számítás és frekvencia analízis tekintetében. A REBECEP testvér módszert [Theor. Chem. Accounts, 106 (2001) 319] a korreláció számításokra szintén demonstráltam a REZEP mellett a teljesítmény tekintetben.

**Sandor Kristyan: PANNON-8, 2006 July, Szeged, Hungary, pp.173-178,**
**Published by Hungarian Zeolite, Association, ISBN-963-06-0138-9**

= = = = = =

Representative equations/tables/figures:

| Molecule | Method | ZPE(method) [kcal/mol] | Program | CPU time | Disc-space [MB] |
|---|---|---|---|---|---|
| Naphthalene | MMFF94 | 92.44 | Spartan [36] | 0.1 sec | - |
| $C_{10}H_8$ | HF-SCF/6-31G(d) /freq | 88.61 | Gaussian98 [3] | 89 min | 47 |
| | REZEP/6-31G(d)[a] | 87.61 | Eq.3 [2] | 2 min | 20 |
| | Stoichiometric | 88.69 | Eq. 4 | - | - |
| | Stoichiometric | 89.11 | Politzer et al. [35] | - | - |
| α-iso-cinchonine | MMFF94 | 235.56 | Spartan [36] | 1 sec | - |
| $C_{19}H_{22}N_2O$ | HF-SCF/6-31G(d) /freq | 227.41 | Gaussian98 [3] | 7067 min | 735 |
| | REZEP/6-31G(d)[b] | 224.06 | Eq.3 [2] | 71 min | 69 |
| | Stoichiometric | 231.62 | Eq. 4 | - | - |
| | Stoichiometric | 230.47 | Politzer et al. [35] | - | - |

Comparison of the performance: accuracy, as well as CPU and disc space requirement of different ZPE estimates by MMFF94, HF-SCF/6-31G(d)/freq and REZEP/6-31G(d) methods. The REZEP calculation was performed using the optimized MMFF94 equilibrium geometry, as well as a.: the naphthalene was included in the multi-linear fit, b.: the α-iso-cinhonine was not included in the fitting procedure. From a and b: REZEP parameters are transferable.

**Thesis-14-Theory: Dependence of correlation energy and zero point energy on the nuclear frame and number of electrons**

For our developed twin methods, REBECEP for correlation energy and basis set error calculation and REZEP for zero point energy estimation, both via atomic partial charges, we have used the databases of G2 and G3/99 total energies, as well as our scaled HF-SCF/6-31G(d) zero point energies along with certain atomic partial charges of more than 100 closed shell neutral molecules containing H, C, N, O, and F atoms. These methods show near 1.0 kcal/mol average absolute deviation from the experimental results with much less computational time and disc space demand. To support the theory of these two semi-empirical methods, I have carried out some statistical analysis to establish a plausible physical background for the parameters involved. The dependency is very interesting in the general theory of correlation energy and zero point energy as well.

= = = = = =

Általunk fejlesztett két testvér módszer számára [REBECEP (korrelációs energia és bázis készlet hiba számítás) és REZEP (zéró pont energia közelítés)], mindkettő az atomi parciális töltések felhasználásával, felhasználtuk a G2 és G3/99 adatbázisokból a totális energiákat, valamint az általunk végzett skálázott HF-SCF/6-31G(d) zéró pont energia alapszámításokat bizonyos atomi parciális töltésekkel egyetemben több mint 100 zárt héjú semleges molekula esetében melyek H, C, N, O, és F atomokat tartalmaztak. Ezen módszerek kb. 1.0 kcal/mol átlagos abszolút eltérést mutatnak a kísérleti eredményektől, ugyanakkor sokkal kevesebb számítási időt és lemezterületet igényelnek. Hogy alátámasszam e két szemi-empirikus módszer elméletét, bizonyos statisztikai analízist végeztem hogy megalapozzak egy plauzibilis fizikai hátteret a benne foglaltatott paraméterek számára. Ez a viszonosság rendkívül érdekes a korrelációs és zéró pont energia általános elmélete szempontjából is.

**Sandor Kristyan: Journal of Molecular Structure: THEOCHEM, 712 (2004) 153-158**

= = = = = =

Representative equations/tables/figures:

The zero point energy is a positive, additive and collective internal property, approximated by the 'harmonic' formula ZPE = $(h/2)\Sigma\nu_i$. The sum runs for the 3M-6 normal frequencies (nonlinear molecule of M atoms, number 6 is from the rigid 3-3 translations and rotations), h is the Planck constant.

The correlation and frequency calculations are very expensive at the *ab initio* level in comparison to the HF-SCF routine.

For values of terms to compare in relation to accuracy (why it cannot be neglected; in kcal/mol for ZPE and $E_{corr,\ basis\ set\ error}$ of pentane (HF-SCF/6-31G*)):

$$0 - 5 \text{ kcal/mol experimetal reaction barriers} \ll \text{ZPE} = 95 \ll E_{corr,\ basis\ set\ error} \approx 950.$$

Our approximate ZPE corrected total electronic energy in the vicinity of equilibrium geometries:

$$E_0 \approx E_0(\text{HF-SCF/6-31G*}) + E_{corr}(\text{REBECEP/6-31G*}) + \text{ZPE}(\text{REZEP/6-31G*}),$$

where the REBECEP and REZEP energies are the sum of M atomic terms, e.g.:

$$\text{ZPE(REZEP)} \equiv \Sigma\, E_{ZPE}(N_A, Z_A),$$

where the sum runs for A=1,…,M atoms in the molecule, $N_A$ is the non-integer number of electrons around atom A with nuclear charge $Z_A$, associated via a simple difference between the partial charge (Mulliken, natural population analysis, ChelpG, electrostatic charge, etc.) and $Z_A$, as well as $E_{ZPE}(N_A, Z_A)$ is a weighted average of the two neighboring atomic ZPE parameters for atoms with $Z_A$ and $N_1 < N_A < N_2 = N_1+1$, where $N_1$ and $N_2$ are the closest integers to $N_A$. Analogous relation holds for $E_{corr}$(REBECEP).

We have refined Politzer's "static" equation as

$$\text{ZPE[kcal/mol]} \approx 6.99n_H + 3.74n_C + 3.98n_N + 3.45n_O + 2.79n_F - 4.63,$$

where $n_H$ is the number of H atoms in the molecule, etc. The development of REZEP estimation over this, which makes that "dynamic" is the weighting of the atomic parameters with actual partial charges in the molecule. Its necessity is manifesting, for example, in case of a simple isomerization for which the static equation yields the same ZPE, incorrectly.

REBECEP and REZEP atomic parameters depend on basis set and partial charge type chosen, but the final ground state electronic energies depend only slightly on these choices. Reasonable choice for REBECEP and REZEP (can be independent from each other):

1.: basis set level at least HF-SCF/6-31G* (with ZPEs scaled by 0.8929 to reproduce experimental results),

2.: Mulliken partial charge ("natural" side result in HF routine).

REZEP is valid in geometric minimum (where all frequencies are real), REBECEP is valid in stationary ponts (transition sates and geometric minimum, where the spin pairing effect is satisfied).

At the limit where the molecule separates into atoms in free space, the REBECEP atomic correlation parameters tend towards the atomic correlation energies and basis set errors in free space. The REZEP atomic ZPE parameters tend towards zero, because ZPE motion vanishes in the case of individual atoms in free space.

The quasi-linear dependence of correlation energy on the number of electrons in molecular systems comes from that any electron has N-1 neighbors, and a pair has about -0.04 hartree correlation energy contribution almost independently from the molecular frame

$$E_{corr} \approx a(N-1) \quad \text{with} \quad -0.045 < a[\text{hartree}] < -0.030,$$

but the molecular frame canot be neglected for accuracy, again, the isomers indicate it with their different correlation energies. Supporting plot:

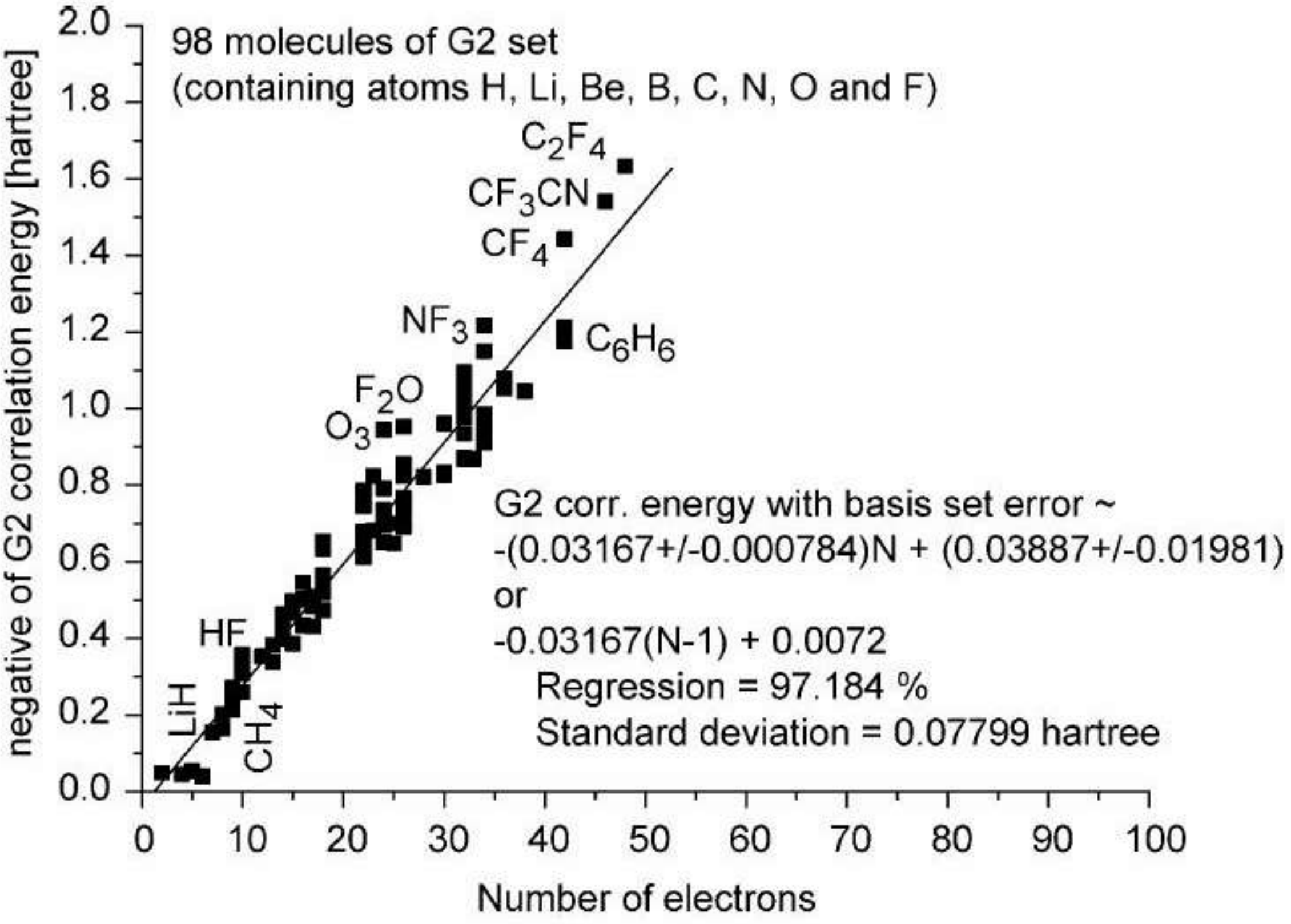

The G2 level correlation energy including basis set error (G2 - HF-SCF/6-31G*) vs. the number of electrons, N, for 98 open and closed shell molecules at their equilibrium geometry, stoichiometry is indicated for some points. (G2 is supposed to reproduce the real or CI quality $E_{total\ electr,0}$.)

The REBECEP method (useful in practice) yields considerably better results than $E_{corr} \approx a(N-1)$ (theoretically useful but not for practice) taking the partial charge, nuclear, and spin dependence of the correlation energy into account, as well as capable to approach G2 or G3 level within the chemical accuracy.

However, the plot ZPE(G3) vs. the number of electrons, N, in molecules indicates another dependence, stoichiometry is indicated for some points.

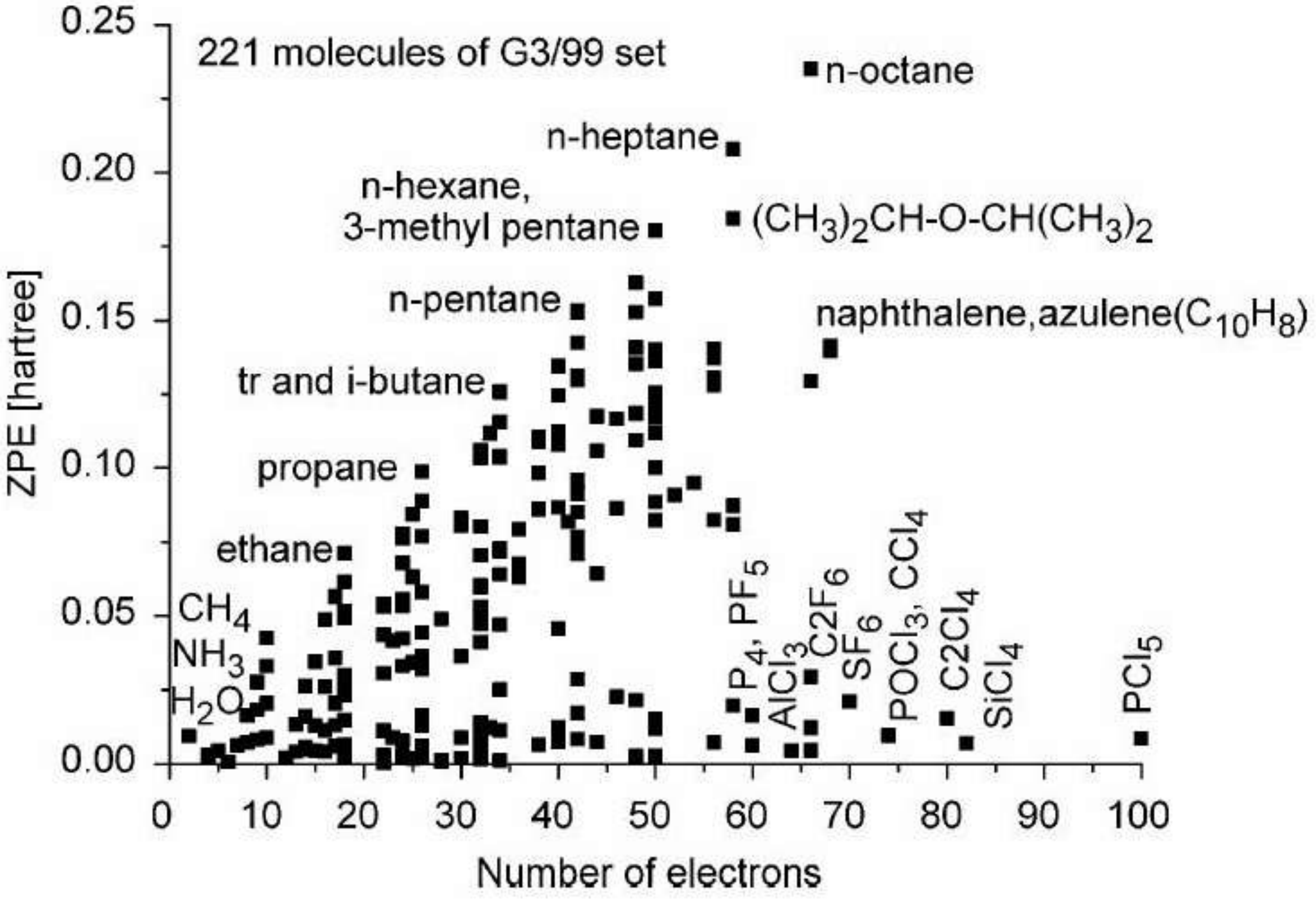

ZPE vs. N plot still shows an interesting relationship, (at least, it is manifested by the plot of $Z_A$<18 nuclei constituting neutral G3 subset molecules at their equilibrium geometry):

$$0 < ZPE \leq b(N\text{-}1) \qquad \text{with } b \approx 0.0036 \text{ hartree}$$

for the lower and upper envilopes.

Recall the classical equation for M mass points A, performing the same kind of kinetic motion in ZPE: $E_{kin} = \Sigma E_{kin,A}$, where sum runs for A=1,….M and $E_{kin,A}= ½m_A v_A^2$ for translation or $\nu= (2\pi)^{-1}(k_A/m_A)^{1/2}$ for vibration (harmonic oscillator with force constant $k_A$ connected to a large body with infinite mass). It provides the hypothesis that the ZPE vibrational motion can be described algebraically with

$$ZPE \approx E_{kin}= \Sigma h(2\pi)^{-1}(k_A/m_A)^{1/2}$$

where the sum runs for A=1,…M (number of atoms), in contrast to $E_{corr}$(REBECEP) in which the sum runs for A=1,…N (number of electrons). Indeed, the ZPE(G3) vs. $\Sigma m_A^{-1//2}$ for A=1,…M in molecules shows stronger statistical correlation:

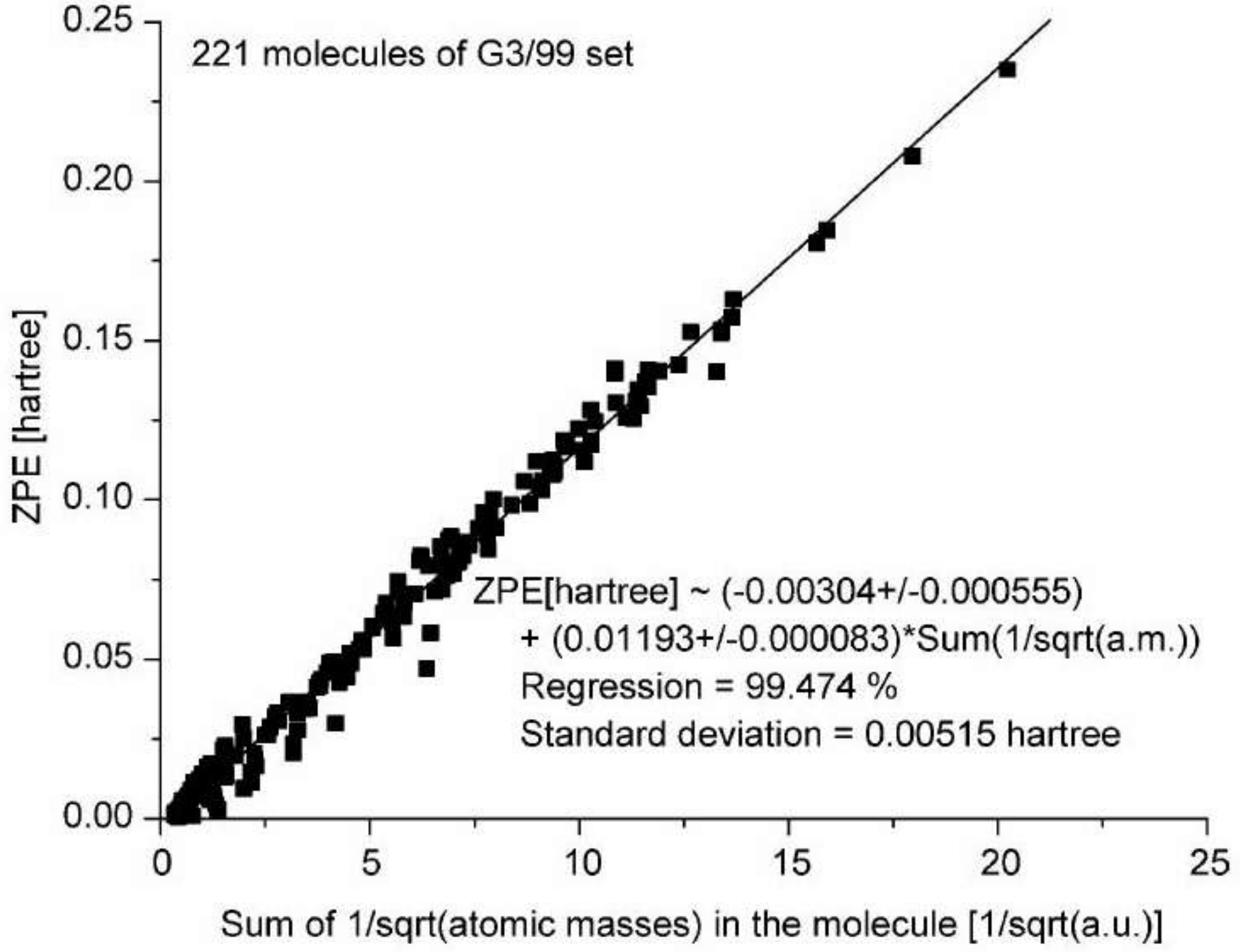

Notice, for example, the light hydrogen atom has the largest ZPE atomic parameter value as a consequence of the term $m_A^{-1/2}$ in REZEP or the 6.99 in Politzer approximation. See the upper envelope in ZPE vs. N plot: The hydrocarbons with high H content have large ZPE values, while molecules containing heavy atoms (Al, Si, P, S and Cl) without or small H content, they are less mobil, i.e. own lower ZPE values.

The classical force constant ($k_A$) that can also be associated with the atomic ZPE parameter terms relates to the partial charge, i.e. how strongly an individual atom A bonds in the molecule to the remaining part.

$E_{corr}$[hartree] $\approx$ -0.03(N-1) and ZPE[hartree] $\leq$ 0.0036(N-1) explain the magnitude difference between the two energy terms via parameters in them (0.03/0.003=10).

From the plot ZPE vs. $\Sigma m_A^{-1//2}$ one can establish an empirical rule for neutral molecules based on the strong correlation (99.474% regression and 0.00515 hartree standarddeviation) as:

$$\text{ZPE[hartree]} \approx (-0.00304\pm0.000555) + (0.01193\pm0.000083)\Sigma m_A[\text{a.u.}]^{-1//2} \approx$$
$$\approx c\, \Sigma m_A[\text{a.u.}]^{-1//2} \quad \text{with } c=0.012$$

visibly fluctuating with the nuclear frame in respect to chemical accuracy (1 kcal/mol).

The (handicap) constant -0.00304 hartree= -1.9 kcal/mol compares to the Politzer constant (-4.63 kcal/mol), although technically, it yields a ZPE value without material content (M=0).

The slope with an H atom ($m_H$=1) is {0.01193 hartree $m_H^{-1/2}$} = 7.49 kcal/mol corresponds to the 6.99 kcal/mol coefficient in modified Politzer equation, and so on.

More accurately, c should be the semi-classical atomic $c_A$= $h(2\pi)^{-1}k_A^{1/2}$ and shuld be behind the sum, depending on bond strength or partial charge in agreement with the idea of REZEP.

The

$$\text{ZPE[hartree]} = (h/2)\Sigma\nu_i \approx 0.012\Sigma m_A[\text{a.u.}]^{-1//2} \leq 0.0036(N-1)$$

approximating and limiting relationship (at least for molecules containing atomic masses $Z_A$<18) is important, because the sum of frequencies or normal vibrations (providing the ZPE value), is connected to the sum of diagonal values of the symmetric Hessian matrix. (In case of a symmetric matrix, the sum of eigenvalues is the sum of diagonal elements; the trace of symmetric real matrices is invariant to similarity transformations.) In this way these simple relationships at equilibrium molecular geometry say something about this matrix which is very difficult to obtain on *ab initio* level.

**Thesis-15-Theory: Estimating correlation energy and basis set error for Hartree–Fock SCF calculation by scaling the kinetic and repulsion energies**

For electronic ground state energy, the HF-SCF procedure minimizes the energy functional $<S|H_{\nabla}|S> + <S|H_{Rr}|S> + <S|H_{rr}|S> \; > \; <\Psi_0|H|\Psi_0> \equiv E_{electr,0}$ for a normalized single Slater determinant approximate wavefunction $\Psi_0$ (denoted as S; as well as $H_{\nabla}$, $H_{Rr}$, $H_{rr}$, H = kinetic-, nuclear–electron attraction-, electron–electron repulsion-, electronic Hamiltonian energy operators). The $<S|H|S>$ can never reach the value $E_{electr,0}$ (variation principle), causing about 1% non-negligible energy error, called correlation energy ($E_{corr}$). We could re-correct this error with scaling during the SCF subroutine by minimizing the new functional $(1 + k_c)<S|H_{\nabla}|S> + <S|H_{Rr}|S> + (1 + k_{ee})<S|H_{rr}|S>$ to estimate $E_{electr,0}$ better. The very flexible $k_c$ and $k_{ee}$, were fitted to accurate G3 electronic ground state molecular energies. They negligibly depend on atomic numbers, number of electrons and system size, as well as transferable. Numerical results and tests includig HF-SCF and B3LYP with STO-3G and 6-31G** bases sets have validated these results.

= = = = = =

Az elektronikus alapállapot energia számításához, a HF-SCF eljárás minimalizálja a $<S|H_{\nabla}|S> + <S|H_{Rr}|S> + <S|H_{rr}|S> \; > \; <\Psi_0|H|\Psi_0> \equiv E_{electr,0}$ energia functionalt egy normalizált egy Slater determinánsú közelítő hullámfüggvény segítségével a valódi (pontos) $\Psi_0$ helyett (S –el jelölve; valamint $H_{\nabla}$, $H_{Rr}$, $H_{rr}$, H = kinetikus-, mag–elektron vonzás-, elektron–elektron taszítás-, elektronikus Hamilton energia operatorok). A $<S|H|S>$ sose érheti el az $E_{electr,0}$ értékét (variációs elv), okozván egy kb. 1% nem elhanyagolható energia hibát, melynek ismert neve korrelációs energia ($E_{corr}$). E hiba javítható volt egy un. skálázással az SCF szubrutin alatt azáltal, hogy egy másik, új funkcionált, a $(1 + k_c)<S|H_{\nabla}|S> + <S|H_{Rr}|S> + (1 + k_{ee})<S|H_{rr}|S>$ kifejezést minimalizáltam az $E_{electr,0}$ becslésére. A nagyon is flexibilis $k_c$ and $k_{ee}$ paramétereket a pontos G3 elektronikus alapállapotú molekuláris energiákra illesztettem. Ezek elhanyagolhatóan függenek az atomszámtól, az elektronok számától és a rendszer méretétől, valamint transzferábilisak. Numerikus eredmények és tesztek a HF-SCF és B3LYP módszerek segítségével az STO-3G és 6-31G** bázis függvények felhasználásával igazolták eredményeimet.

<u>Sandor Kristyan: Computational and Theoretical Chemistry, 975 (2011) 20–23</u>

= = = = = =

Representative equations/tables/figures:

Minimizing the new functional

$$(1 + k_c)<S|H_{\nabla}|S> + <S|H_{Rr}|S> + (1 + k_{ee})<S|H_{rr}|S>$$

to estimate $E_{electr,0}$ better.

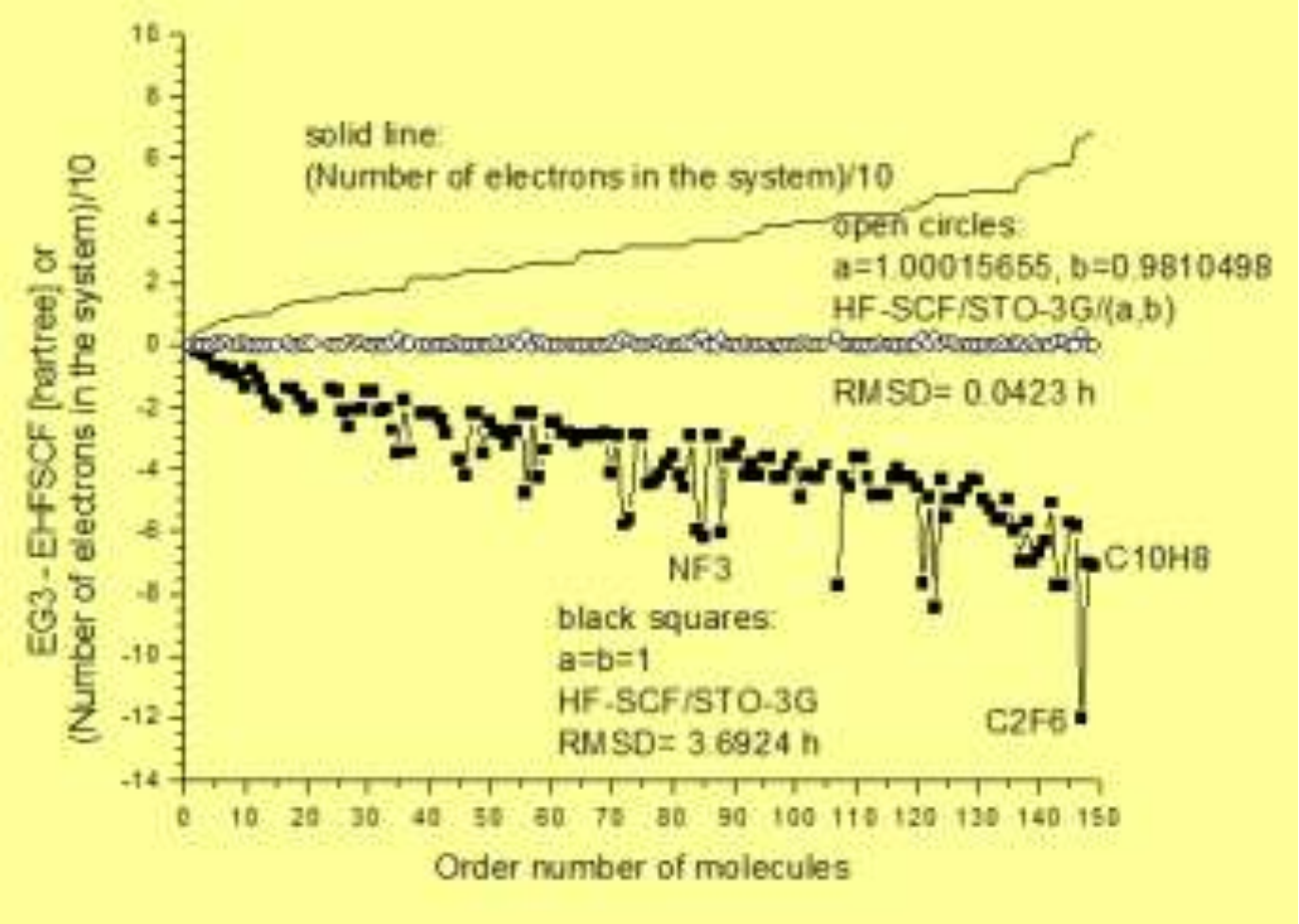

Comparing the errors of calculations on STO-3G basis level:

HF-SCF/STO-3G vs. HF-SCF/STO-3G/optimized(a,b).

Notes:

$b \equiv 1+k_c$ and $a \equiv 1+k_{ee}$

N demonstarates the system size.

Crude empirical: $E_{corr} \approx a(N-1)$.

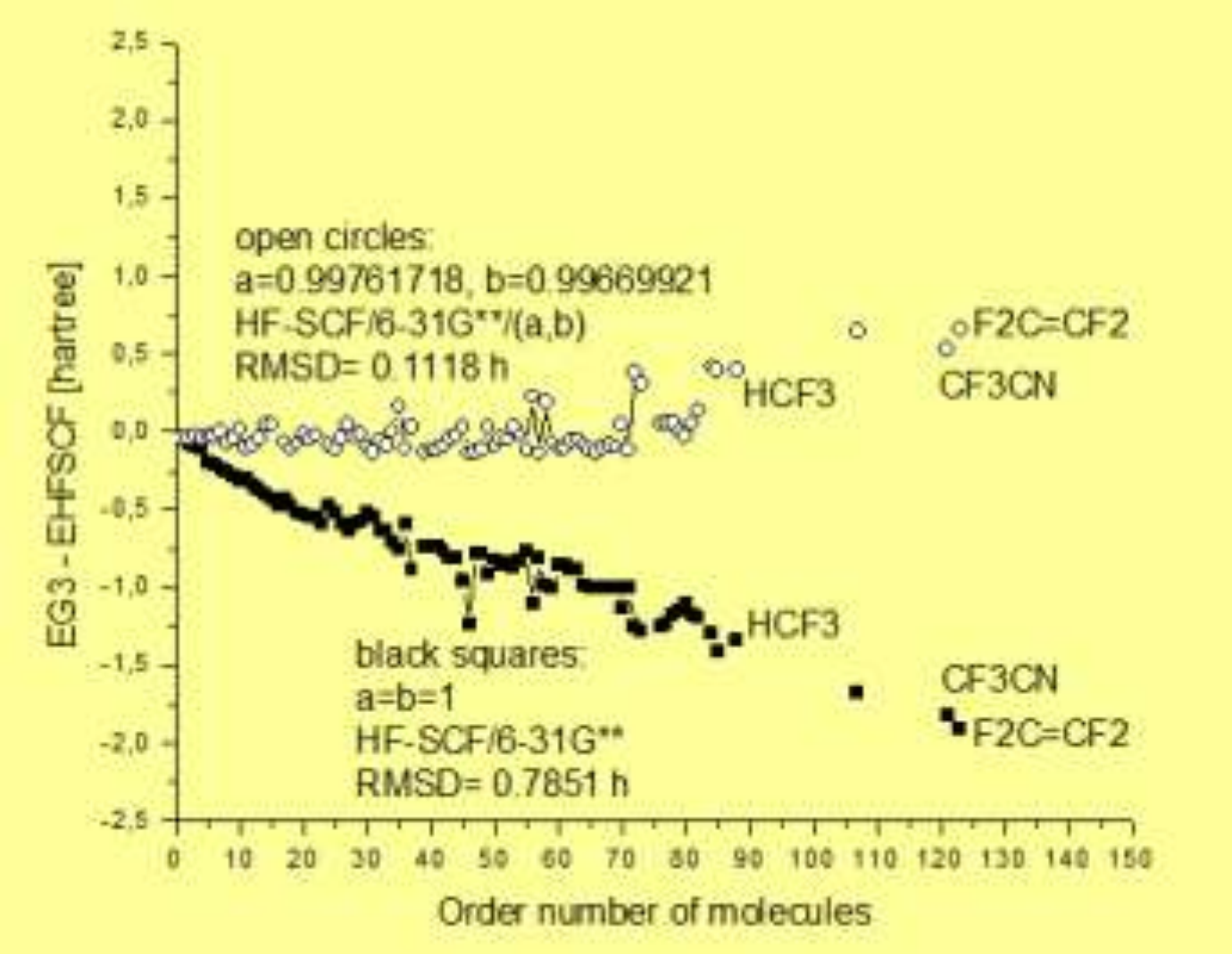

Comparing the errors of calculations on 6-31G** basis level:

HF-SCF/6-31G** vs. HF-SCF/6-31G**/optimized(a,b).

Notice the different scale in axis EG3 - EHFSCF, coming from the larger basis than STO-3G used, i.e. from lower absolute value in $E_{error}$.

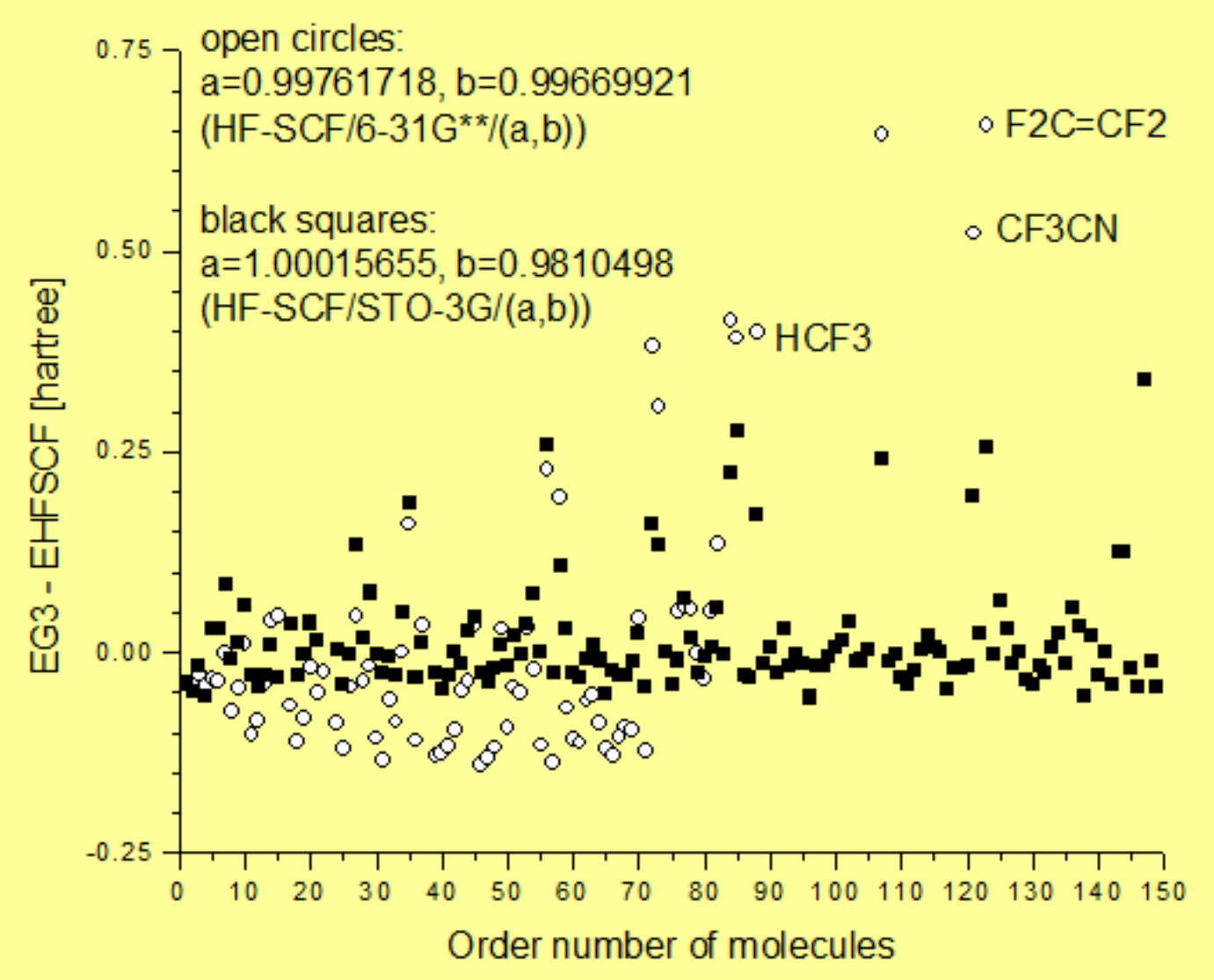

Plotting the above two figures together to compare the bases STO-3G and 6-31G** in HF-SCF/basis/optimized(a,b) method.

HF-SCF/STO-3G calculation improves better than the HF-SCF/6-31G**, although the 6-31G** is a larger basis than STO-3G. (The fit was done and analyzed for absolute energies, not for relative energies e.g. for heats of atomizations, etc..)

**Thesis-16-Theory: Variational calculation with a scaling correct moment functional**

The updated literature is listed, reviewed and summarized on the reduction of the electronic Schrödinger equation from 4N-dimensions to a nonlinear, approximate density functional of a 3 spatial dimension one-electron density for an N electron system which is tractable in practice. Using a well behaving density functional in power series with respect to density scaling within the orbital-free framework for kinetic and repulsion energy of electrons, a good density functional approximation is analyzed and solved via the Lagrange multiplier device. (The introduction of a Lagrange multiplier to ensure normalization is a new element in this part of the related, general theory.) Its relation to Hartree-Fock and Kohn-Sham formalism is also analyzed.
= = = = = =

Összegyűjtöttem és összefoglaltam a legújabb szakirodalomig az elektronikus Schrödinger egyenlet redukcióját 4N-dimenziós (spin és 3 tér) parciális differenciálegyenletről egy nem-lineáris, közelítő, csak a 3 tér dimenziós egy-elektron sűrűséget tartalmazó sűrűség funkcionál alakra az N elektron rendszerek leírásához, mely kezelhető a gyakorlatban. Felhasználva egy 'megfelelő viselkedésű' hatványsoros sűrűség funkcionált a sűrűség skálázási szabályának tekintetében az elektronok kinetikus és taszítási energiájának pálya (orbitál) mentes leírásnál, egy hasznos sűrűség funkcionál közelítést analizáltam és megoldását közöltem a Lagrange multiplikátoros módszer segítségével. (A Lagrange multiplikátoros módszer bemutatása, mely biztosítja a normalizálást, egy új dolog az ide vonatkozó elméletben.) Analizáltam a módszer kapcsolatát a Hartree-Fock and Kohn-Sham formalizmusokkal is.

**Sandor Kristyan (REVIEW ARTICLE):**
**International Journal of Quantum Chemistry 113 (2013) 1479-1492,**
= = = = = =

Representative equations/tables/figures:

$\Sigma_{j=1,\ldots n}\{(1+2/(3j))jA_j[\int\rho_0^{[1+2/(3j)]}d\mathbf{r}_1]^{j-1}\rho_0^{2/(3j)} + (1+1/(3j))jB_j[\int\rho_0^{[1+1/(3j)]}d\mathbf{r}_1]^{j-1}\rho_0^{1/(3j)}\} + v(\mathbf{r}_1) = \lambda$

Truncation n=1: $(5/3)A_1\rho_0^{2/3} + (4/3)B_1\rho_0^{1/3} + v(\mathbf{r}_1) \approx \lambda \equiv E_{electr,0} / N$

(a second order algebraic equation with $z \equiv \rho_0^{1/3}$)

$E_{electr}[\rho_0] \approx E_{electr,0,approx} \equiv \int(A_1\rho_0^{5/3} + B_1\rho_0^{4/3} + v(\mathbf{r}_1)\rho_0)d\mathbf{r}_1$

Truncation n=2: $(5/3)A_1u^4 + (8/3)A_2[\int u^8 d\mathbf{r}_1]u^2 + (4/3)B_1u^2 + (7/3)B_2[\int u^7 d\mathbf{r}_1]u + v(\mathbf{r}_1) \approx \lambda$

($u \equiv \rho_0^{1/6}$, integral-equation for $u(\mathbf{r}_1)$)

Notations: $\lambda$ = Lagrange multiplier or chemical potential $\partial E_{electr,0}/\partial N$,

$\rho_0(\mathbf{r}_1)$ = one-electron density, $\rho_0(\mathbf{r}_1)$.

**Thesis-17-Theory: Variational calculation with general density functional for ground state, recipe for SCF**

A recipe for self consistent field solution has been worked out for variational treatment of the general density functional to solve the electronic Schrödinger equation directly for ground state; reducing the N-electron problem from 4N to 3 dimensions. Using orbital-free framework, the numerical recipe originates from the linear dependence of nuclear-electron attraction functional on one-electron density ($V_{ne}[\rho_0(\mathbf{r}_1)]= -\Sigma_{A=1,\dots,M}Z_A\int\rho_0(\mathbf{r}_1)r_{A1}^{-1}d\mathbf{r}_1$) and a quadratic LCAO approximation for $\rho_0(\mathbf{r}_1)$ ($\approx (\Sigma_{k=1\dots L}d_k\ b_k(\mathbf{r}_1))^2$ or $\approx \rho_{0,HF\text{-}SCF}(\mathbf{r}_1)= 2\Sigma_{i=1\dots N/2}[\Sigma_{k=1\dots L1}c_{ik}b_k(\mathbf{r}_1)]^2$); the optimization can be done with iterative use of lin-solver.

= = = = = =

Egy ön-konzisztens eljárást dolgoztam ki általános (vagyis bármely, plauzibilis) sűrűség funkcionálnak variációs minimalizálására az elektronikus Schrödinger egyenlet megoldásához, hogy direkt módon megkapjuk egy molekula rendszer alap állapotának energiáját. Ezzel az N-elektron problémát 4N-ről 3 dimenziósra lehet redukálni. Pálya (orbitál) mentes tárgyalás mellett e numerikus módszer azért lehetséges, mert a mag-elektron taszítás funkcionálja lineárisan függ az egy-elektron sűrűségtől ($V_{ne}[\rho_0(\mathbf{r}_1)]= -\Sigma_{A=1,\dots,M}Z_A\int\rho_0(\mathbf{r}_1)r_{A1}^{-1}d\mathbf{r}_1$), valamint az egy-elektron sűrűség közelítése négyzetesen függhet a bázis függvények LCAO leírásától ($\rho_0(\mathbf{r}_1) \approx (\Sigma_{k=1\dots L}d_k\ b_k(\mathbf{r}_1))^2$ or $\approx \rho_{0,HF\text{-}SCF}(\mathbf{r}_1)= 2\Sigma_{i=1\dots N/2}[\Sigma_{k=1\dots L1}c_{ik}b_k(\mathbf{r}_1)]^2$). Az optimáláshoz csak a lineáris egyenletrendszer iteratív alkalmazása szükséges.

<u>**Sandor Kristyan: Journal of Theoretical and Applied Physics 2013, 7:61**</u>

= = = = = =

Representative equations/tables/figures:

Key property for the algorithm is that the absolute values of the three main energy terms are in about the same magnitude in an electronic molecular system, (recall e.g. the virial theorem for stationary systems as $(V_{ee}+ V_{ne}+ V_{nn})/T = -2$), for example:

H(Z=N=1) atom: $E_{electr,0}= T+ V_{ee}+ V_{ne}= 0.5 +0 -1 = -0.5$ hartree $\Rightarrow$ <u>0.5 : 0 : 1</u>,

while the much larger Ar(Z=N=18) atom has similar ratio among the three terms:

$E_{electr,0}= (0.527544 +0.264456 -1.319544)\times10^3= -527.544$ hartree $\Rightarrow$ <u>0.53 : 0.26 : 1.32</u>,

as well as the abs($V_{ne}$) term is the largest among the three, furthermore

$|\partial T(N)/\partial\rho_0|,\ |\partial V_{ee}(N)/\partial\rho_0| < |\int(v(\mathbf{r}_1)-\lambda)\ \rho_{0i}d\mathbf{r}_1|$.

The "Lagrange's method of undetermined multiplier" for the 2nd Hohenberg-Kohn theorem minimizes the functional $L^* = T[\rho_0(\mathbf{r}_1)] + V_{ee}[\rho_0(\mathbf{r}_1)] + V_{ne}[\rho_0(\mathbf{r}_1)] - \lambda(\int\rho_0(\mathbf{r}_1)d\mathbf{r}_1-N)$ with e.g. $\rho_0(\mathbf{r}_1)\approx (\Sigma_{k=1\dots L}d_k\ b_k(\mathbf{r}_1))^2$, where $b_k$ are the atomic basis func. (AO) and $d_k$ are the LCAO coeff., as well as for minimum $\partial L^*/\partial d_i = \partial L^*/\partial\lambda = 0$, yielding the quasi-linear equation system to solve iteratively (i=1,…,L, and left hand side is from particular DFT func. used):

$$\Sigma_{k=1\dots L}\ d_k^{iter\ m+1}\int(v(\mathbf{r}_1)-\lambda^{iter\ m})b_i(\mathbf{r}_1)b_k(\mathbf{r}_1)d\mathbf{r}_1 = -(1/2)\partial T[\rho_0^{iter\ m}(\mathbf{r}_1)]/\partial d_i -(1/2)\partial V_{ee}[\rho_0^{iter\ m}(\mathbf{r}_1)]/\partial d_i$$

The flow diagram of the SCF algorithm:

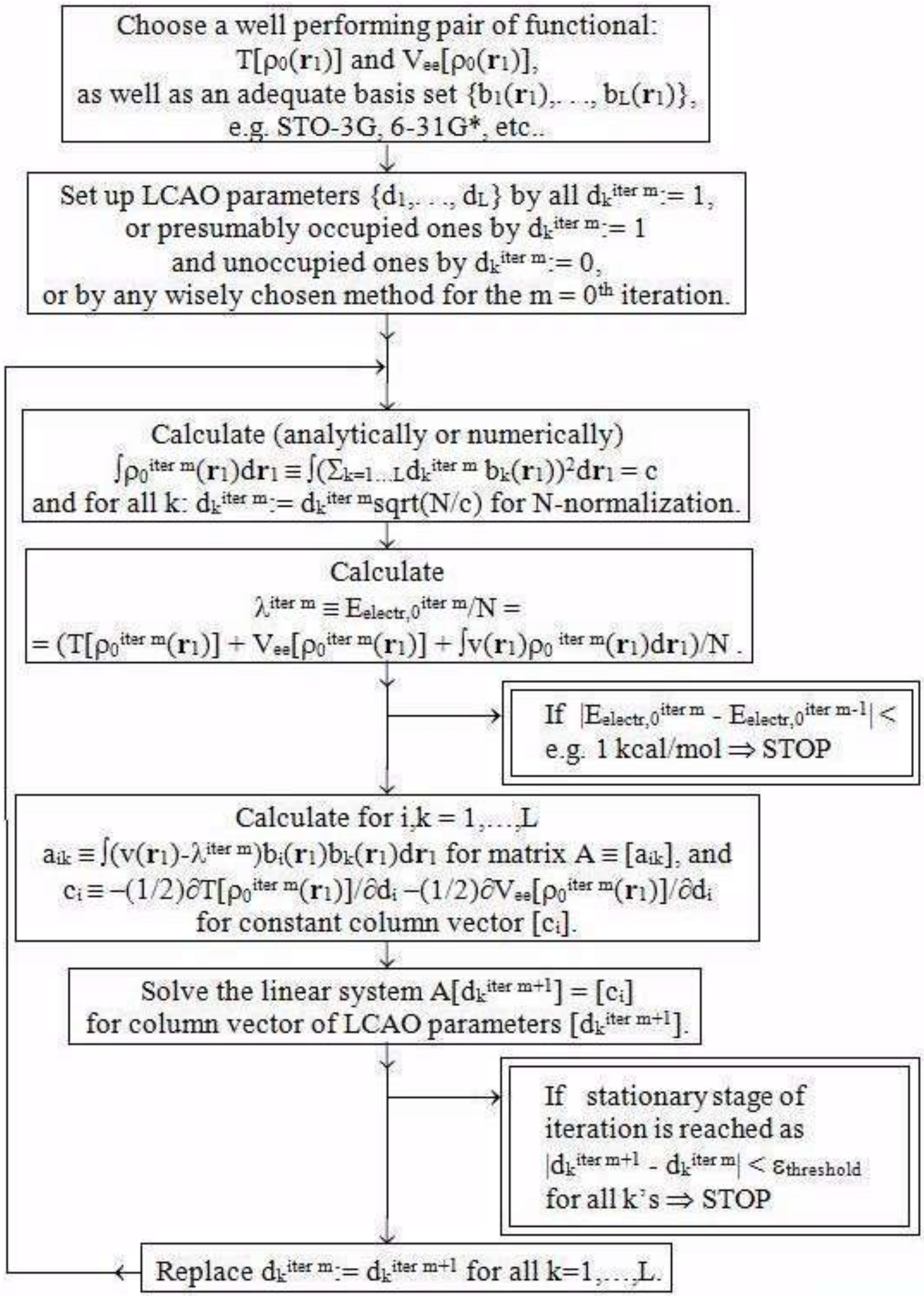

**Thesis-18-Theory: Compact one-electron density approximation for ground state electronic energy of molecular systems with Thomas–Fermi kinetic- and Parr electron–electron repulsion DFT energy functionals**

The reduction of the electronic Schrödinger equation (SE) or its calculating algorithm from 4N-dimensions to a (nonlinear, approximate) density functional of three spatial dimension one-electron density for an N-electron system was accomplished in a way of first approximation, and found to be tractable in the practice for electronic structure calculation. Using the Thomas-Fermi kinetic energy ($\sim\int\rho^{5/3}d\mathbf{r}_1$) and Parr electron–electron repulsion energy ($\sim\int\rho^{4/3}d\mathbf{r}_1$) main-term first approximation functionals, a compact one-electron density approximation for calculating ground state electronic energy from the $2^{nd}$ Hohenberg–Kohn theorem was set up. Its two parameters have been fitted to neutral and ionic atoms, which are transferable to molecules when one uses it for estimating ground-state electronic energy. The convergence is proportional to the number of nuclei (M) needing low disc space usage and numerical integration. Its properties are discussed and compared with known *ab initio* methods, and for energy differences (particularly for atomic ionization potentials) it is comparable or sometimes gives better result than those. It does not reach the chemical accuracy for total electronic energy, but beside its amusing simplicity, it is interesting in theoretical point of view, and can serve as generator function for more accurate one-electron density models.

= = = = = =

Az elektronikus Schrödinger egyenletet (SE), illetve annak számítási algoritmusát redukáltam 4N-dimenzióról egy (nem-lineáris, közelítő) három tér dimenziós egy-elektronos sűrűség funkcionállá az N-elektronos rendszerek leírására mint egy első közelítést, és azt találtam, hogy a gyakorlatban alkalmas elektron szerkezeti számításokra. Felhasználva a Thomas-Fermi kinetikus energia ($\sim\int\rho^{5/3}d\mathbf{r}_1$) és a Parr elektron–elektron taszítási energia ($\sim\int\rho^{4/3}d\mathbf{r}_1$) első közelítéses fő-tag-funkcionálokat, egy kompakt egy-elektron sűrűség közelítést dolgoztam ki az alap állapotú elektronikus energia számítására a második Hohenberg–Kohn tétel segítségével. A modell két pataméterét semleges és ionos állapotú atomok energiáira illesztettem, ami transzferálható molekulákra, pontosabban azok alap állapotú elektronikus energiáinak közelítésére. A konvergencia arányos a magok számával (M), alacsony lemez területet igénnyel, továbbá numerikus integrálás szükségeltetik. Tulajdonságait feltérképeztem és diszkutáltam, valamint összehasonlítottam ismert *ab initio* módszerekkel, és energia különbségek tekintetében (speciálisan atomi ionizációs potenciálokra) azt találtam, hogy összehasonlíthatók velük, sőt néha jobb eredményeket adnak. Nem éri el a kémiai pontosságot a totális elektronikus energia értékekre, de amellett, hogy meglepően egyszerű, érdekes elméleti szempontból, valamint generátor függvényként szolgálhat pontosabb egy-elektronos sürüségi modellekben.

<u>**Sandor Kristyan: Journal Computational Chemistry, 30 (2009) 1445–1453**</u>

= = = = = =

Representative equations/tables/figures:

With the mentioned approximate algebraic terms, the electronic SE in free space (primarily for ground states $\rho_0$) is crudely

$$c_{10}c_1\rho^{5/3} + P(\mathbf{r}_1)\rho + c_{20}c_2\rho^{4/3} \approx \rho E_{electr},$$

where $c_1 \equiv Nc_F$, $P(\mathbf{r}_1) \equiv -N\Sigma_{A=1,\ldots,M}Z_AR_{Ai}^{-1}$ and $c_2 \equiv N2^{-1/3}(N-1)^{2/3}$. With division of $\rho_0$ and using $z \equiv \rho_0^{1/3} \geq 0$ it reduces to a 2nd order algebraic equation:

$$c_{10}c_1z^2 + c_{20}c_2z + (P(\mathbf{r}_1) - E_{electr,0}) \approx 0.$$

With $A \equiv c_{10}c_1$, $B \equiv c_{20}c_2$, $P(\mathbf{r}_1)$, $\mathrm{discr}(\mathbf{r}_1) \equiv B^2 - 4A(P(\mathbf{r}_1) - E_k)$, its solution is

$$\rho_{0,k}(\mathbf{r}_1) = C[(+(\mathrm{discr}(\mathbf{r}_1))^{1/2} - B)/(2A)]^3 \quad \text{with} \quad C \equiv N/[\int[(+(\mathrm{discr}(\mathbf{q}_1))^{1/2} - B)/(2A)]^3 d\mathbf{q}_1]$$

for any nuclear configuration (index k refers to the $k^{th}$ approximation). C fixes the normalization to N (# of electrons) required for 2nd HK theorem when searching the minimum with respect to $E_k$ for the approximate functional of the ground state electronic energy:

$$E_{electr,0}(\text{approx.}, E_k) = (1/N)\int(c_{10}c_1\rho_{0,k}^{5/3} + P(\mathbf{r}_1)\rho_{0,k} + c_{20}c_2\rho_{0,k}^{4/3})d\mathbf{r}_1.$$

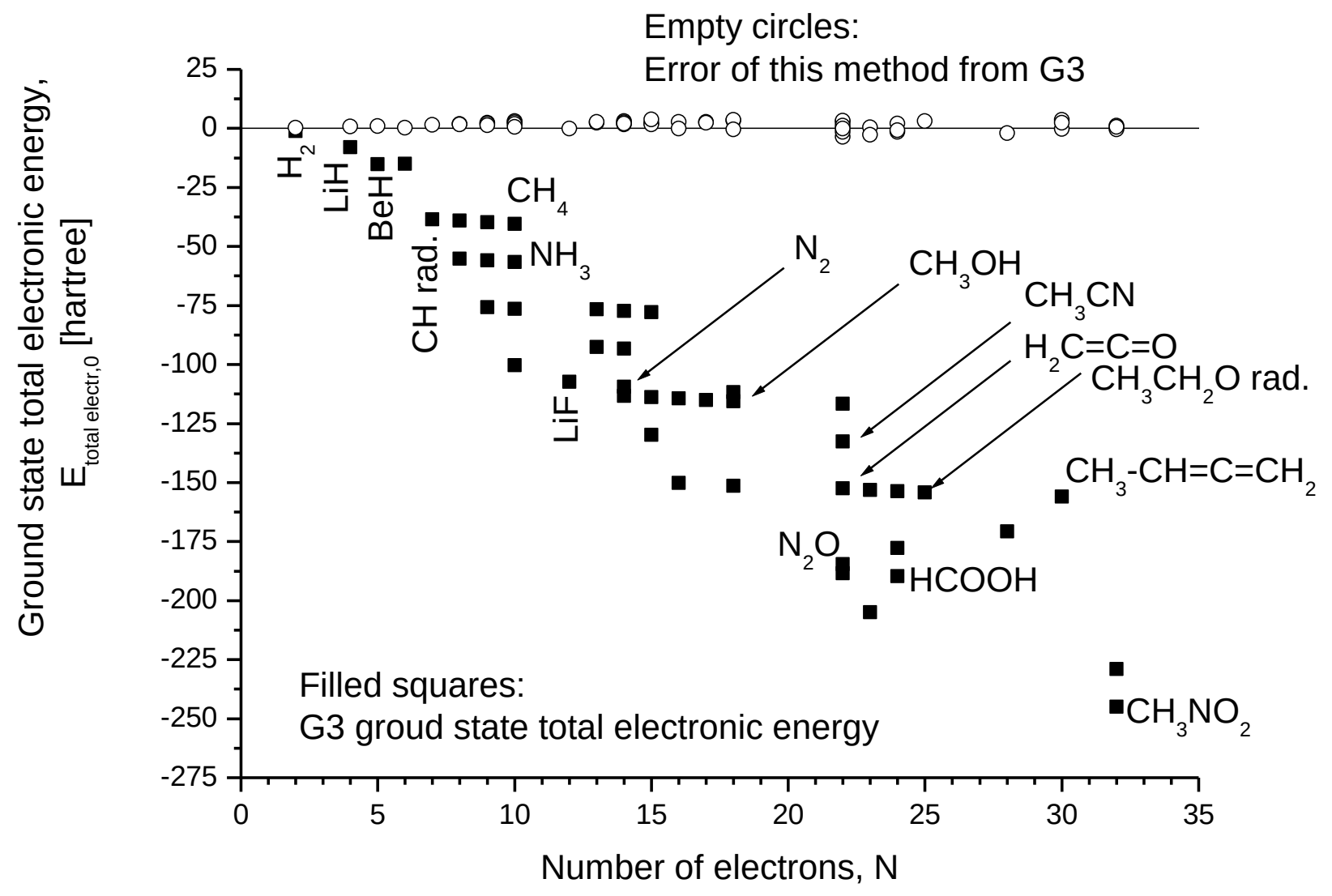

Fifty representative molecules from the G3 molecule set (some are marked) to test the transferability (to molecules) of optimized c-parameters fitted for the extremum ($\partial E_{electr,0}(\text{approx.}, E_k)/\partial E_k = 0$) to reproduce the ground state electronic energy, $E_{electr,0}(\mathrm{CI})$, of 43 atomic ions with N=3,…,Z+1 and Z=3-10 as possible.

If the integrand is simply extended with the Weizsacker correction, $c_{100}N|\nabla_1\rho_{0,k}|^2/\rho_{0,k}$, the fitting procedure for optimum yields the three dimensional parameter space point (c10, c20, c100) = (1.3546492283, 0.8311630382, 0.0924664675) for the 43 atomic ions.

**Thesis-19-Theory: Multi-electron densities from Hohenberg–Kohn theorems to variational principle**

The properties of multi-electron densities have been analyzed along with their behavior with respect to the two Hohenberg–Kohn theorems. This analysis was continued with the form of density functionals and density differential and/or integral operators on different levels of dimensions between the variational principle (4N-dimension) and Hohenberg– Kohn theorems (3-dimension). The trend in ionization potentials has been commented upon. The exact density functional operator of H-like atoms and one-electron systems has been formalized (cannot be found in the literature yet) with the two-electron systems, not only as simple ‘‘forever prototypes”, but as a certain projection of one-electron density formalism of N≥1 electron systems to N= 1 and 2. The review part of this work is focusing primarily on functional analytical properties.

= = = = = =

A több-elektron sűrűség tulajdonságait és viselkedését analizáltam a két Hohenberg–Kohn tétel vonatkozásában. Ezen analízis tartalmazza a sűrűség funkcionálok formáját és a sűrűségi differenciál és/vagy integrál operátorokat a dimenziók különböző szintjein a variációs elv (4N-dimenzió) és Hohenberg– Kohn tételek (3-dimenzió) között. Az ionizációs potenciálok trendje e vonatkozásban szintén kommentálódott. A H-szerű atomok és egy-elektron rendszerek egzakt sűrűség funkcionál operátorát algebrai formába öntöttem (nem található meg az irodalomban) a két-elektronos rendszerekével együtt, nem csak mint egyszerű ‘‘örök prototípusok”, hanem mint egy bizonyos projekciói az N≥1 elektron rendszerek egy-elektron sűrűség formalizmusának az N= 1 és 2 esetekre. A szemle része e munkának első sorban a funkcionál analitikai tulajdonságokra összpontosít.

**Sandor Kristyan (REVIEW ARTICLE):**
**Journal of Molecular Structure: THEOCHEM 858 (2008) 1–11**

= = = = = =

Representative equations/tables/figures:

Define the multi-electron densities $b_1,\ldots,b_N$ (for any $k^{th}$ excited state k=0,1,2,…) as $|\Psi(\mathbf{x}_1,\mathbf{x}_2,\mathbf{x}_3,\ldots,\mathbf{x}_N)|^2 = \Psi^* \Psi$, $b_N(\mathbf{r}_1,\mathbf{r}_2, \mathbf{r}_3\ldots, \mathbf{r}_{N-1},\mathbf{r}_N) = \int\Psi^* \Psi\, ds_1 ds_2\, ds_3\ldots ds_{N-1} ds_N$, . . . $b_2(\mathbf{r}_1, \mathbf{r}_2) = \int\Psi^* \Psi\, ds_1 ds_2 d\mathbf{x}_3\ldots d\mathbf{x}_{N-1} d\mathbf{x}_N$, $\rho(\mathbf{r}_1) \equiv b_1(\mathbf{r}_1) = \int\Psi^* \Psi\, ds_1 d\mathbf{x}_2 d\mathbf{x}_3\ldots d\mathbf{x}_{N-1} d\mathbf{x}_N$.

For example, the second-order reduced density matrix is defined in the literature as $n_2(\mathbf{x}_1,\mathbf{x}_2;\mathbf{x}'_1,\mathbf{x}'_2) = (N(N-1)/2) \int\Psi^*(\mathbf{x}'_1,\mathbf{x}'_2,\mathbf{x}_3,\ldots,\mathbf{x}_N)\Psi(\mathbf{x}_1,\mathbf{x}_2,\mathbf{x}_3,\ldots,\mathbf{x}_N)d\mathbf{x}_3\ldots d\mathbf{x}_N$ and from this, the diagonal of the spin-independent second-order density matrix is $n_2(\mathbf{r}_1,\mathbf{r}_2) = \int n_2(\mathbf{x}_1,\mathbf{x}_2;\mathbf{x}_1,\mathbf{x}_2)ds_1 ds_2 = (N(N-1)/2)b_2(\mathbf{r}_1, \mathbf{r}_2)$. The property $\int b_{i+1} d\mathbf{r}_{i+1} = b_i$, i=1,2,3,…,N-1 holds.

Define the finite series of N+2 functions: $\{b_{N+2}\equiv\Psi(\mathbf{x}_1,\mathbf{x}_2,\mathbf{x}_3,\ldots,\mathbf{x}_N)$, $b_{N+1}\equiv|\Psi(\mathbf{x}_1,\mathbf{x}_2,\mathbf{x}_3,\ldots,\mathbf{x}_N)|^2$, $b_N(\mathbf{r}_1,\mathbf{r}_2, \mathbf{r}_3\ldots, \mathbf{r}_{N-1},\mathbf{r}_N)$, $b_{N-1}$, $b_{N-2}$, …, $b_3$, $b_2(\mathbf{r}_1, \mathbf{r}_2)$, $\rho(\mathbf{r}_1) \equiv b_1(\mathbf{r}_1)\}$.

Recall: The 2nd Hohenberg-Kohn (HK) theorem states the energy variation principle (VP) for a trial one-electron density $\rho_{trial}(\mathbf{r}_1)$: For any $\rho_{trial}(\mathbf{r}_1)$ which $\rho_{trial}(\mathbf{r}_1) \geq 0$ at any $\mathbf{r}_1$ real space point and normalized to N ($\int\rho(\mathbf{r}_1)d\mathbf{r}_1 = N$) the

$$E_{electr,0} \leq E_{v,1}[\rho_{trial}(\mathbf{r}_1)] \quad \text{or} \quad E_{v,1}[b_{1,trial}]$$

is hold and the equality is hold for the ground state $\rho_0 \equiv b_{1,0}$. (The index v in $E_{v,1}$ indicates that the Hamiltonian is characterized by the nuclear frame that defines the molecular system by knowing how many electrons it has.) The analogy to the VP for wave functions is obvious:

$$E_{electr,0} \leq E_{v,N+2}[b_{N+2,trial}] .$$

Define: Use the notation t, $v_{en}$, and $v_{ee}$ for $\langle\Psi|\Psi\rangle=N$ normalized wave functions as

$$E_{v,i}[b_i] = \langle\Psi|H_\nabla\Psi\rangle+\langle\Psi|H_{Rr}\Psi\rangle+\langle\Psi|H_{rr}\Psi\rangle \equiv \int t[b_i]+\int v_{en}[b_i]+\int v_{ee}[b_i] \equiv T+V_{en}+V_{ee}$$

where the three terms are obviously the kinetic-, electron-nuclear attraction-, and electron-electron repulsion energy terms, respectively.

Statement: On N+2 level, the kinetic functional is $T= F_{kin}[b_{N+2}\equiv\Psi]= \langle\Psi|H_\nabla\Psi\rangle$, but serious difficulty starts even from the next lower level, N+1, for $T= F_{kin}[b_{N+1}\equiv|\Psi|^2]$, i.e. finding the analytical form of $F_{kin}$ as a functional of $b_{N+1}$, generally for cases i=1,2,…,N+1. For i=1, the weak Thomas –Fermi approximation, $t_1[b_1 \equiv \rho] \sim \rho^{5/3}$, holds, while for i=N+2 the exact $t_{N+2}[b_{N+2}] = -(1/2)b_{N+2}^*\nabla_1^2 b_{N+2}$ holds (*= complex conjugate).

Statement: In contrary to kinetic operators, all cases i=1,2,…, N+1,N+2 yield easy and simple nuclear-electron attraction functional:

$$v_{en}[b_i] = -\Sigma_{A=1,\ldots,M} Z_A\, b_i\, R_{A1}^{-1} = b_i\, v(\mathbf{r}_1).$$

Statement: For cases i=2,…,N+1,N+2, the electron-electron repulsion functional is also simple:

$$v_{ee}[b_i] = ((N-1)/2)\, b_i\, r_{12}^{-1},$$

but problematic for i=1. Technically,

$$v_{ee}[b_{N+2}] = ((N-1)/2)\, b_{N+1}\, r_{12}^{-1}.$$

Generalization of 1st HK theorem: The external potential $v(\mathbf{r}_i) \equiv -\Sigma_{A=1,\ldots,M} Z_A R_{Ai}^{-1}$ is determined, within a trivial additive constant, by the multi-electron density $b_i$ for $i=1,2,3,\ldots,N+2$. (The 1st HK theorem states it for i=1.)

Generalization of 2nd HK theorem: VP holds for all $i=1,2,3,\ldots,N+1$ as $E_{electr,0} \le E_{v,i}[b_{i,trial}]$, if the trial i-electron density $b_{i,trial}$, is $b_{i,trial} \ge 0$ at any $(\mathbf{r}_1,\mathbf{r}_2, \mathbf{r}_3\ldots,\mathbf{r}_i)$ 3i-dimensional real space point and normalized to N ($\int b_{i,trial}(\mathbf{r}_1,\mathbf{r}_2, \mathbf{r}_3\ldots,\mathbf{r}_i)d\mathbf{r}_1 d\mathbf{r}_2 d\mathbf{r}_3\ldots d\mathbf{r}_i = N$) and $E_{v,i}[b_{i,trial}]$ is the energy functional. (The 2nd HK theorem states it for i=1, and the VP states it for i=N+2.)

Integro-differential forms: Integrating both sides of non - relativistic, spinless, fixed nuclear coordinate electronic Schrödinger equation for a molecular system containing M atoms and N electrons with nuclear configuration $\{\mathbf{R}_A, Z_A\}_{A=1,\ldots,M}$ in free space for all spin-orbit variables $\mathbf{x}_i = (\mathbf{r}_i,s_i)$ except $\mathbf{r}_1$ after multiplying by the complex conjugate of the same $k^{th}$ exited state wave function from left, one yields for $\rho(\mathbf{r}_1)$ that

$$D[\rho] \equiv D_\nabla[\rho] + D_{Rr}[\rho] + D_{rr}[\rho] = \rho E_{electr},$$

here D is the density operator corresponding to the Hamiltonian for ground or excited states, the $(1/N)\int D[\rho]d\mathbf{r}_1 = E_{electr}$ holds. Using the anti-symmetric property of $\Psi$ for ground and excited states, the kinetic term is

$$D_\nabla[\rho] = -(1/2)\Sigma_{i=1,\ldots,N}\int\Psi^*\nabla_i^2\Psi ds_1 d\mathbf{x}_2\ldots d\mathbf{x}_N =$$
$$= -(1/2)\int\Psi^*\nabla_1^2\Psi ds_1 d\mathbf{x}_2\ldots d\mathbf{x}_N -((N-1)/2)\int\Psi^*\nabla_2^2\Psi ds_1 d\mathbf{x}_2\ldots d\mathbf{x}_N$$

with $\int D_\nabla[\rho]d\mathbf{r}_1 = -(N/2)\int\Psi^*\nabla_1^2\Psi d\mathbf{x}_1 d\mathbf{x}_2\ldots d\mathbf{x}_N$, the nuclear-electron attraction term is

$$D_{Rr}[\rho] = \rho(\mathbf{r}_1)v(\mathbf{r}_1) + (N-1)\int b_2(\mathbf{r}_1,\mathbf{r}_2)v(\mathbf{r}_2)d\mathbf{r}_2$$

with $\int D_{Rr}[\rho]d\mathbf{r}_1 = N\int\rho(\mathbf{r}_1)v(\mathbf{r}_1)d\mathbf{r}_1$, and the electron-electron repulsion term is

$$D_{rr}[\rho] = (N-1)\int b_2(\mathbf{r}_1,\mathbf{r}_2)r_{12}^{-1}d\mathbf{r}_2 + [N(N-1)/2 - (N-1)]\int b_3(\mathbf{r}_1,\mathbf{r}_2,\mathbf{r}_3)r_{23}^{-1}d\mathbf{r}_2 d\mathbf{r}_3$$

with $\int D_{rr}[\rho]d\mathbf{r}_1 = (N(N-1)/2)\int\Psi^*\Psi r_{12}^{-1}d\mathbf{x}_1 d\mathbf{x}_2\ldots d\mathbf{x}_N = (N(N-1)/2)\int b_2(\mathbf{r}_1,\mathbf{r}_2)r_{12}^{-1}d\mathbf{r}_1 d\mathbf{r}_2$ ,as well as for i=1 case and ground state $V_{ee} \equiv (1/N)\int D_{rr}[\rho_0]d\mathbf{r}_1 \approx \int[2^{-1/3} (N-1)^{2/3} \rho_0^{4/3} + \text{correction}]d\mathbf{r}_1$ which is not a very accurate approximation if the correction is neglected, or the more plausible Coulomb approximation in which the correction is the famous $E_{xc}[\rho_0]$ exchange correlation – these corrections have functional approximation difficulties for practice, but always N-representable. For i=2, $V_{ee} \equiv (1/N)\int D_{rr}[b_2]d\mathbf{r}_1 d\mathbf{r}_2 = ((N-1)/2) \int b_2\, r_{12}^{-1}d\mathbf{r}_1 d\mathbf{r}_2$, which is exact, but has N-representablility (existence of $\Psi$ for this $b_2$) problem.

Side result is that $\int\nabla_1^2\rho(\mathbf{r}_1)d\mathbf{r}_1 = 0$ for ground and excited states, as well as it holds not only for physical one-electron densities, but for HF-SCF approximate ground states as well.

For H-like atoms and one-electron molecules for ground and excited states the exact

$$D[N=1,\rho(\mathbf{r}_1)] \equiv -(1/4)\nabla_1^2\rho(\mathbf{r}_1) + (1/8)\rho(\mathbf{r}_1)^{-1}|\nabla_1 \rho(\mathbf{r}_1)|^2 +\rho(\mathbf{r}_1)v(\mathbf{r}_1) = E_{electr}\, \rho(\mathbf{r}_1)$$

holds, while for ground state the exact

$$E_{electr,0} \le \int [(1/8)\rho_{0,trial}(\mathbf{r}_1)^{-1}|\nabla_1\rho_{0,trial}(\mathbf{r}_1)|^2 +\rho_{0,trial}(\mathbf{r}_1)v(\mathbf{r}_1)]d\mathbf{r}_1$$

holds for the HK energy functional.

For ground- and excited states as well as HF-SCF ground state one-electron density

$$\int\nabla_1^2\rho(\mathbf{r}_1)d\mathbf{r}_1 = 0$$

holds, useful e.g. in correlation calculation.

**Thesis-20-Theory: Electron-electron repulsion energy participation in non-relativistic electronic Schrödinger equation via the coupling strength parameter in Hartree-Fock theory**

No detailed analysis has been published yet about the ratio or participation of electron-electron repulsion energy ($V_{ee}$) in total electronic energy – aside from virial theorem and the very detailed and well-known algorithm $V_{ee}$ is calculated during the standard HF-SCF and post-HF-SCF routines. Using particular modification of the SCF part in Gaussian package the ground and excited state solutions of the electronic Hamiltonian $H_\nabla+H_{ne}+aH_{ee}$ via the coupling strength parameter "a" have been analyzed. Technically, this modification was essentially a modification of a single line in the SCF algorithm, wherein the operator $r_{ij}^{-1}$ was overwritten as $r_{ij}^{-1} \rightarrow a*r_{ij}^{-1}$, with making "a" as input. The most important findings are that the repulsion energy $V_{ee}(a)$ is a quasi-linear function of "a", as well as the statement and analysis of the extended 1st Hohenberg-Kohn theorem ($\Psi_0(a=1) \Leftrightarrow H_{ne} \Leftrightarrow Y_0(a=0)$) and its consequences in relation to "a". The latter allows an algebraic transfer from the simpler solution of case a=0 (where the single Slater determinant is accurate form) to the realistic wanted case a=1. More, the case a=0 generates orto-normalized set of Slater determinant, which can be used as basis set for CI calculations (the 1st approximation and diagonal elements of TNRS-CI matrix are $E_{electr,k} \approx E_{electr,k}(TNRS) \equiv e_{electr,k} + (N(N-1)/2)\langle Y_k|r_{12}^{-1}|Y_k\rangle$ for ground and excited ($k \geq 0$) states with off-diagonal elements $\langle Y_{k'}|H_{ee}|Y_k\rangle$), as well as the emblematic theorems Hund's rule and virial/ Møller-Plesset/ Hohenberg-Kohn/ Koopmans/ Brillouin theorems and configuration interactions formalism in relation to "a" have been generalized.

= = = = = =

Hiányzik az irodalomból a teljes elektronikus energiában az elektron-elektron taszítás ($V_{ee}$) arányának és részvételének részletes analízise – leszámítva a viriál tételt és a részletesen ismert algoritmust, ahogy $V_{ee}$ értékét számolják a standard HF-SCF és poszt-HF-SCF algoritmusokban. A Gaussian program csomag SCF részének bizonyos módosításával, az elektronikus Hamilton operátor $H_\nabla+H_{ne}+aH_{ee}$ alap és gerjesztett állapotú megoldásait analizáltam a csatolási állandó, "a" függvényében. Lényegében, ez a módosítás egy sor volt az SCF algoritmusban, miszerint az $r_{ij}^{-1}$ operátor lett átírva mint $r_{ij}^{-1} \rightarrow a*r_{ij}^{-1}$, ahol "a" egy bemeneti értékként funkcionált. A legfontosabb felismeréseim a taszítási energia, $V_{ee}(a)$ qvázi-lineáris függése "a" értékétől, valamint az állítása és analízise a első Hohenberg-Kohn tétel ($\Psi_0(a=1) \Leftrightarrow H_{ne} \Leftrightarrow Y_0(a=0)$) kiterjesztésének és következményinek "a" tekintetében. Az utóbbi egy algebrai transzformációt biztosít az egyszerű megoldás a=0 esetéből (ahol az egy Slater determináns pontos forma) a valós és keresett a=1 esethez. Továbbá, az a=0 eset orto-normalizált Slater determináns függvények halmazát generálja, mely használható mint bázis készlet CI számítások számára (az első közelítés és egyben diagonális eleme a TNRS-CI mátrixnak $E_{electr,k} \approx E_{electr,k}(TNRS) \equiv e_{electr,k} + (N(N-1)/2)\langle Y_k|r_{12}^{-1}|Y_k\rangle$ alap és gerjesztett ($k \geq 0$) állapotokra az off-diagonális $\langle Y_{k'}|H_{ee}|Y_k\rangle$ elemekkel), valamint az emblematikus tételek mint Hund's szabály és viriál/ Møller-Plesset/ Hohenberg-Kohn/ Koopmans/ Brillouin tételek és konfigurációs kölcsönhatások formalizmusa az "a" értékének tekintetében általánosításra kerültek.

**Sandor Kristyan (#108): 16th International Conference on Density Functional Theory and its Application, CELEBRATING THE 50TH ANNIV. KOHN-SHAM THEORY, August 31 - September 4, 2015, Debrecen University, Debrecen, Hungary,**
**http://dft2015.unideb.hu/home**

**Sandor Kristyan (# 0018): 8th Molecular Quantum Mechanics 2016, Celebration of the Swedish School, An international Conference in Honour of Per E M Siegbahn and in memory of P.-O. Löwdin and B.O. Roos, June 26 - July 1, Uppsala University, Uppsala, Sweden,**
**http://www-conference.slu.se/mqm2016/**

**Sandor Kristyan (W44): Theory and Applications of Computational Chemistry 2016, August 28 - September 2, University of Washington in Seattle, USA,**
**http://www.tacc2016.org/**

= = = = = =

Representative equations/tables/figures:

The electronic Hamiltonian can be extended with coupling strength parameter (a) as

$$H(a)y_k(a) \equiv (H_\nabla + H_{ne} + aH_{ee})y_k(a) = enrg_{electr,k}(a)y_k(a)$$

of which only a=1 makes physical sense (for which the familiar notation is $\{enrg_{electr,k}, y_k\} = \{E_{electr,k}, \Psi_k\}$), and the simplest mathematical case a=0 is

$$(H_\nabla + H_{ne})Y_k = e_{electr,k} Y_k \ ,$$

in which no electron-electron interaction at all, the totally non-interacting reference system (TNRS). For eigenvalue/eigenfunction set $\{e_{electr,k}, Y_k\}$ the single Slater determinant is correct form, while for cases a≠0 it is not. The HF-SCF/basis/a approximation is achieved with using single determinats at stationary points as $\Psi_0 \approx S_0$, generally as $y_0(a) \approx s_0(a)$, and the necessity of exchange-correlation energy comes up (which is non-negligible ≈1% at a=1). The mathematical conection between cases (k',a=0) vs. (k,a=1) is

$$E_{electr,k} = e_{electr,k'} + (N(N-1)/2) <\Psi_k|r_{12}^{-1}|Y_{k'}>/<\Psi_k|Y_{k'}>$$

of the particular interest is the ground state k=k'=0

$$E_{electr,0} = e_{electr,0} + <\Psi_0|H_{ee}|Y_0>/<\Psi_0|Y_0>,$$

with further relations

$$e_{electr,0} << (e_{electr,0} + <\Psi_0|H_{ee}|\Psi_0>) \leq$$
$$\leq E_{electr,0} = (e_{electr,0} + <\Psi_0|H_{ee}|Y_0>/<\Psi_0|Y_0>) \leq$$
$$\leq (e_{electr,0} + <Y_0|H_{ee}|Y_0>)$$

as well as

$$<\Psi_0|H_{ee}|\Psi_0>) \leq <\Psi_0|H_{ee}|Y_0>/<\Psi_0|Y_0> \leq <Y_0|H_{ee}|Y_0>.$$

The derivative w/r to coupling strength parameter is

$$\partial enrg_{electr,0}(a)/\partial a = (N(N-1)/2) <y_0(a)|r_{12}^{-1}|y_0(a)>,$$
$$\partial^2 enrg_{electr,0}(a)/\partial a^2 = N(N-1) <y_0(a)|r_{12}^{-1}| \partial y_0(a)/\partial a>,$$

with the important statement:

$$\partial enrg_{electr,0}(a)/\partial a \text{ is nearly constant.}$$

Example: The HF-SCF/STO-3G/a=1 calculation for Ne atom yields the LCAO coefficients:

```
1CLOSED SHELL SCF, NUCLEAR REPULSION ENERGY IS 0.000000000 HARTREES
0CONVERGENCE ON DENSITY MATRIX REQUIRED TO EXIT IS  1.0000D-05
0  CYCLE      ELECTRONIC  ENERGY            TOTAL  ENERGY        CONVERGENCE
EXTRAPOLATION
    1       -126.604525025      -126.604525025
    2       -126.604525025      -126.604525025   1.81460D-16
0AT TERMINATION TOTAL ENERGY IS      -126.604525  HARTREES
1MOLECULAR ORBITALS                       5 OCCUPIED MO
                             1          2          3          4          5
    EIGENVALUES---   -32.21252   -1.70610   -0.54305   -0.54305   -0.54305

  1  1  NE    1S       0.99501   -0.26941    0.00000    0.00000    0.00000
  2  1  NE    2S       0.01978    1.03065    0.00000    0.00000    0.00000
  3  1  NE    2PX      0.00000    0.00000    0.00000    1.00000    0.00000
  4  1  NE    2PY      0.00000    0.00000    0.00000    0.00000   -1.00000
  5  1  NE    2PZ      0.00000    0.00000   -1.00000    0.00000    0.00000
```

while the HF-SCF/STO-3G/a=0 (no electron-electron interaction, i.e., TNRS):

```
1CLOSED SHELL SCF, NUCLEAR REPULSION ENERGY IS 0.000000000 HARTREES
0CONVERGENCE ON DENSITY MATRIX REQUIRED TO EXIT IS  1.0000D-05
0  CYCLE      ELECTRONIC  ENERGY            TOTAL  ENERGY        CONVERGENCE
EXTRAPOLATION
    1       -182.113502106      -182.113502106
    2       -182.113502106      -182.113502106   0.00000D+00
0AT TERMINATION TOTAL ENERGY IS      -182.113502  HARTREES
1MOLECULAR ORBITALS                       5 OCCUPIED MO
                             1          2          3          4          5
    EIGENVALUES---   -49.42500  -10.95959  -10.22405  -10.22405  -10.22405

  1  1  NE    1S       1.00094    0.24650    0.00000    0.00000    0.00000
  2  1  NE    2S      -0.00389   -1.03083    0.00000    0.00000    0.00000
  3  1  NE    2PX      0.00000    0.00000    1.00000    0.00000    0.00000
  4  1  NE    2PY      0.00000    0.00000    0.00000   -1.00000    0.00000
  5  1  NE    2PZ      0.00000    0.00000    0.00000    0.00000    1.00000
```

i.e. the LCAO coefficients are very close to each other between cases a=0 and 1.

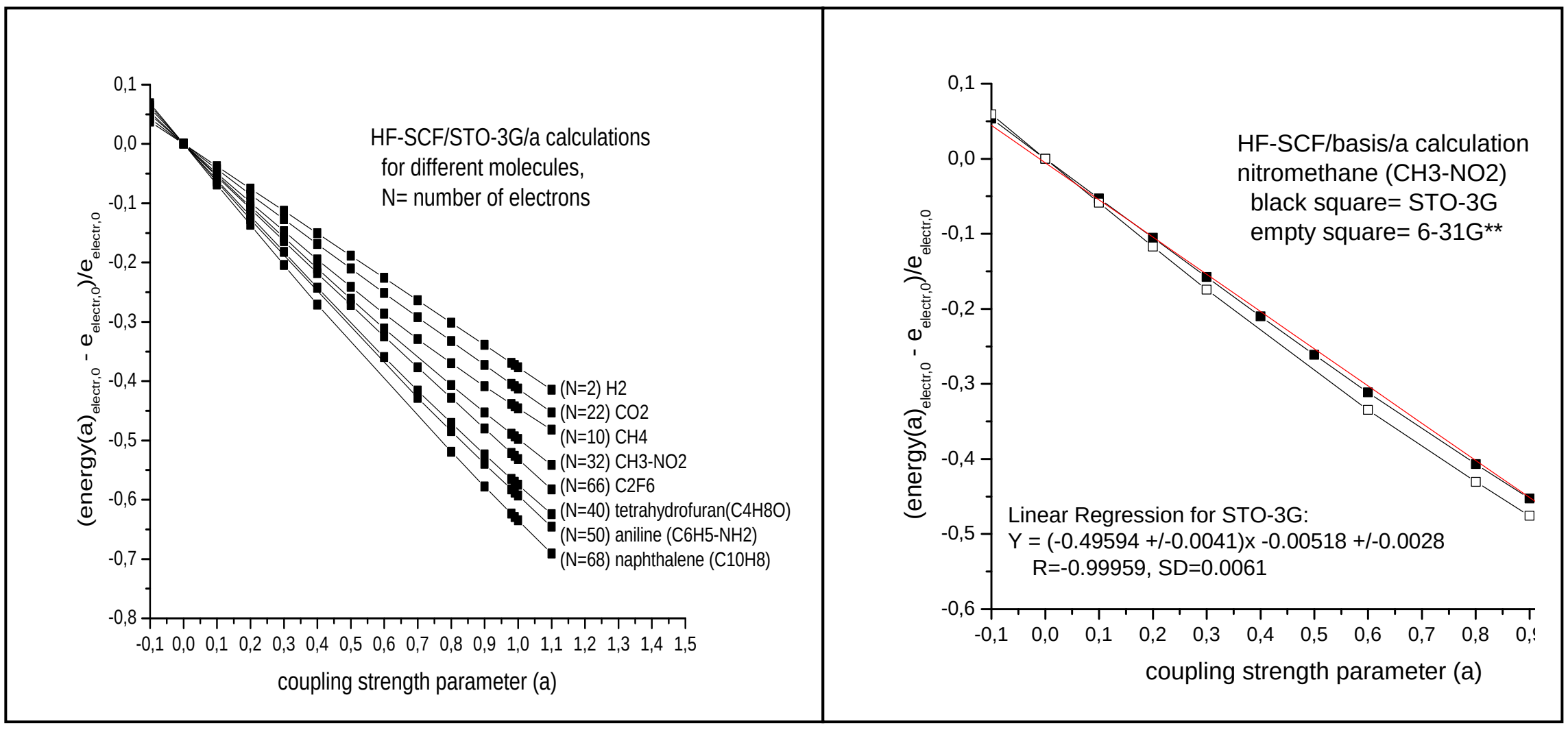

Generalization of the 1st HK theorem ($\rho_0 \Rightarrow \{N, Z_A, R_A\} \Rightarrow H \Rightarrow \Psi_0 \Rightarrow E_{electr,0}$ and all other properties, while the opposite way such as $\Psi_0 \Rightarrow \rho_0$, $E_{electr,0}$ and all other properties is obvious) provides that $\Psi_0 \Leftrightarrow H \Leftrightarrow H_\nabla+H_{ne} \Leftrightarrow Y_0$, i.e., $\Psi_0 \Leftrightarrow Y_0$, which is more generally $\rho_0(\mathbf{r}_1,a) \Leftrightarrow \rho_0(\mathbf{r}_1,a=0)$ or $\rho_0(\mathbf{r}_1,a=1)$. The inclusion of the coupling strength parameter "a" makes it more general, even for two different values of "a" as

$$\rho_0(\mathbf{r}_1,a_1) \quad \text{or} \quad y_0(a_1) \Leftrightarrow \rho_0(\mathbf{r}_1,a_2) \quad \text{or} \quad y_0(a_2).$$

In practice, the most important for DFT establishment is

$$\rho_0(\mathbf{r}_1,a=0) \text{ from } H_\nabla+H_{ne} \Leftrightarrow E_{electr,0} \text{ from } H_\nabla+H_{ne}+H_{ee}.$$

If an r-symmetric w is good enough, $wY_0$ may approach $\Psi_0$ more efficiently than $S_0$. More generally, and extending with coupling strength parameter 'a' and supposing basis set limit, the equality can (hypothetically) hold with r-symmetric w in such a way that $y_0(a)= w(\mathbf{r}_1,\mathbf{r}_2,\ldots\mathbf{r}_N,a)Y_0$, that is, how $y_0(a\neq 0)$, particularly $y_0(a=1)=\Psi_0$ and $y_0(a=0)=Y_0$ connect via w. A DFT correspondent or alternative of w, notated as $w_{DFT}(\mathbf{r}_1)$ also exists provided by the generalized 1st HK, and acting as a functional link between the real and TNRS one-electron densities as $\rho_0(\mathbf{r}_1,a=1)= [w_{DFT}(\mathbf{r}_1)]^2\rho_0(\mathbf{r}_1,a=0)$ with normalization $N= \int\rho_0(\mathbf{r}_1,a=1)d\mathbf{r}_1= \int[w_{DFT}(\mathbf{r}_1)]^2\rho_0(\mathbf{r}_1,a=0)d\mathbf{r}_1$. The w should be a well behaved (aside from $<w|w>=\infty$) r-symmetric function, but $wY_0$ must definitely be a well behaved x-anti-symmetric function with normalization constraint $<wY_0|wY_0>=1$, and the variation equation for w is $enrg_{electr,0}(a)= e_{electr,0} \; -(N/2)<wY_0|Y_0\nabla_1{}^2 w> -N<wY_0|\nabla_1 Y_0\nabla_1 w> + a<wY_0|H_{ee}|wY_0>$, of particular interest is

$$E_{electr,0}(a)= e_{electr,0} \; -(N/2)<wY_0|Y_0\nabla_1{}^2 w> -N<wY_0|\nabla_1 Y_0\nabla_1 w> + <wY_0|H_{ee}|wY_0>,$$

the a=1 case, in which $(e_{electr,0},Y_0)$ is pre-calculated. As a particular example, consider a non-relativistic atom ($1\leq Z\leq 18$, M=a=1) with N=2 electrons: $Y_0(a=0)$ contains $\phi_i(1s)= f_i=2Z^{3/2}\exp(-Z|\mathbf{r}_i|)$ with $\varepsilon_i=-Z^2/2$ in a.u. for i=1,2 (no basis set error), yielding the exact

$$-(1/2)(\nabla_1{}^2+\nabla_2{}^2)w +Z(\nabla_1|\mathbf{r}_1|\nabla_1 w+\nabla_2|\mathbf{r}_2|\nabla_2 w) +w/r_{12}= (E_{electr,0}+Z^2)w.$$

Notice that <u>not an x-anti-symmetric $S_0$, but an r-symetric w must be sought</u>! To have a feeling about the general w: For a=1 and N=2 the core spatial equation for w is the $[-(1/2)(\nabla_1{}^2+\nabla_2{}^2) + r_{12}{}^{-1}]z= \varepsilon z$ eigenvalue equation, and its analytical solution with the smallest $\varepsilon$ value is $z(\mathbf{r}_1,\mathbf{r}_2)= \exp(r_{12}/2)$ with $\varepsilon=-0.25$, representing the necessity of a correlation calculation in the anti-symmetrised approximation $\Psi_0\equiv(\alpha_1\beta_2 - \alpha_2\beta_1)\exp(r_{12}/2) \approx S_0\equiv(\alpha_1\beta_2 - \alpha_2\beta_1)p(\mathbf{r}_1)p(\mathbf{r}_2)$.

For ground (k=0) and excited (k>0) states, using the nuclear frame generated orto-normalized Slater determinant basis set $\{Y_k\}$ from TNRS (a=0) for different levels of CI calculation, the diagonal elements (k'=k):

$$<Y_k|H_\nabla+H_{ne}+aH_{ee}|Y_k>= e_{electr,k} \; + \; a(N(N-1)/2)<Y_k|r_{12}{}^{-1}|Y_k>,$$

making the link between case a=0 and 1 for ground (k=0) and excited (k>0) states as:

$$E_{electr,k} \approx E_{electr,k}(TNRS)\equiv e_{electr,k} + (N(N-1)/2)<Y_k|r_{12}{}^{-1}|Y_k>.$$

The off-diagonal elements ($k'\neq k$):

$$<Y_{k'}|H_\nabla+H_{ne}+aH_{ee}|Y_k> = a(N(N-1)/2)<Y_{k'}|r_{12}{}^{-1}|Y_k> \qquad \text{if } k'\neq k,$$

i.e. the off diagonal elements contain the electron-electron interaction only. (If the known/regular HF-SCF/basis/a=1 determinant basis set $\{S_k\}$ is used (generally $\{s_k(a)\}$), the molecular orbital energies ($\varepsilon_i$) show up in the off-diagonal elements.)

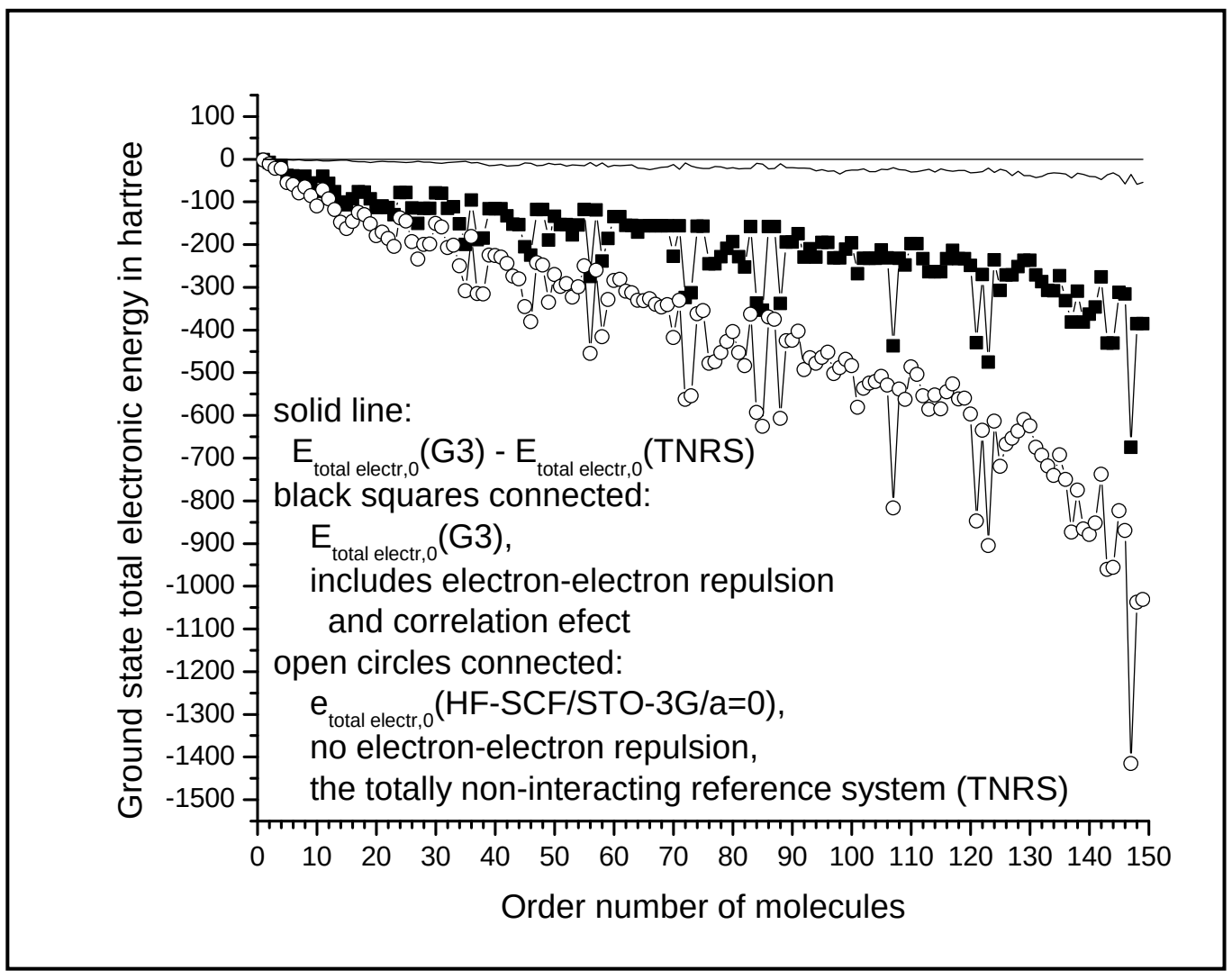

Extension of Brillouin's theorem w/r to coupling sterngth parameter for HF-SCF/basis/a (which approximates $y_0(a)$ by single determinant $s_0(a)$) is formally the same for a=1 vs. $a \neq 1$:

$$\langle s_0(a)|H_\nabla + H_{ne} + aH_{ee}|s_{0,b}{}^r(a)\rangle = 0.$$

An important consequence of this is that, the $\{Y_0, \{Y_{0,b}{}^r\}\}$ truncated basis set from a=0 (using the minimal, singly-excitated ones) can already be used as a basis set to estimate $\Psi_0(a=1)$ better than the (1,1) diagonal element ($E_{electr,0}$(TNRS)), even to estimate $\Psi_1$ also by the eigenvectors of the Hamiltonian matrix (TNRS-CI). This means that, it can provide the large part of correlation energy, and the doubly excited determinants do not have to be used to save computer time and disc space unless one needs more accurate results or higher excited states.

**Thesis-21-Theory: Analytic evaluation of Coulomb integrals for one, two and three-electron operators,** $R_{C1}^{-n}R_{D1}^{-m}$, $R_{C1}^{-n}r_{12}^{-m}$ **and** $r_{12}^{-n}r_{13}^{-m}$ **with** n, m=0,1,2

In the thesis, where R stands for nucleus-electron and r for electron-electron distances, the (n,m)=(0,0) case is trivial, the (n,m)=(1,0) and (0,1) cases are well known, fundamental milestone in integration and widely used in computation chemistry, as well as based on Laplace transformation with integrand $\exp(-a^2t^2)$. The rest of the cases are new and need the other Laplace transformation with integrand $\exp(-a^2t)$ also, as well as the necessity of a two dimensional version of Boys function comes up in case. These analytic expressions (up to Gaussian function integrand, $\exp(\pm w^2)$) are useful for manipulation with higher moments of inter-electronic distances ($r_{ij}^n$), for example in correlation calculations.
= = = = = =

A tézisben, ahol R jelöli a mag-elektron és r az elektron-elektron távolságot, az (n,m)=(0,0) eset triviális, az (n,m)=(1,0) és (0,1) esetek jól ismertek, alapvető mérföldkövek az integrálásban és széles körben alkalmazzák a számításos kémiában, valamint a Laplace transzformációra vannak alapozva az $\exp(-a^2t^2)$ integrandus segítségével. A többi eset azonban új, és egy másik Laplace transzformáció szükségeltetik hozzá az $\exp(-a^2t)$ integrandussal, továbbá a két dimenziós Boys függvény szükségessége is előjön. Ezen analitikai kifejezések (a Gauss függvény integrandus, $\exp(\pm w^2)$, erejéig) hasznosak a magasabb elektron-elektron távolság momentumokkal ($r_{ij}^n$) való manipulálásban pl. a korrelációs energiák számításához.

<u>**Sandor Kristyan: Periodica Poltechnica, Chemical Engineering, In publication in the 2017 memorial edition for György Oláh.**</u>
= = = = = =

Representative equations/tables/figures:
Briefly:

The $\int_{(R3)} \exp(-pR_{P1}^2)R_{C1}^{-n}\, d\mathbf{r}_1 = (2\pi/p)F_0(v)$ if n=1 & $(2\pi^{3/2}/p^{1/2})e^{-v}F_0(-v)$ if n=2, with $v \equiv pR_{CP}^2$ and Boys function ($F_0$), generate <u>analytic expressions for Coulomb integrals</u> with higher distance moment for n, m=0, 1, 2, i.e. for

$$\int\rho(1)R_{C1}^{-n}R_{D1}^{-m}d\mathbf{r}_1,$$
$$\int\rho(1)\rho(2)R_{C1}^{-n}r_{12}^{-m}d\mathbf{r}_1d\mathbf{r}_2,$$
$$\int\rho(1)\rho(2)\rho(3)r_{12}^{-n}r_{13}^{-m}d\mathbf{r}_1d\mathbf{r}_2d\mathbf{r}_3\,.$$

In more detail:

Expressions below have been derived not for one-electron density (ρ), but for primitive Gaussians, $G_{Ai}$, because the real or any physically realistic model $\rho(1)\geq 0$ can be well approximated as linear combination of a well chosen set $\{G_{A1}\}$. The exact theory says that the Coulomb interaction energy is represented by the two-electron energy operator $r_{12}^{-1}$, and using GTO functions in HF and post-HF theories, which is

$$G_{Ai}(a,nx,ny,nz) \equiv (x_i-R_{Ax})^{nx}(y_i-R_{Ay})^{ny}(z_i-R_{Az})^{nz}\exp(-a|\mathbf{r}_i-\mathbf{R}_A|^2)$$

with a>0 and nx, ny, nz ≥0 benefiting its important property such as $G_{Ai}(a,nx,ny,nz)G_{Bi}(b,mx,my,mz)$ is also (a sum of) GTO, the Coulomb interaction energy for molecular systems is expressed finally with the linear combination of the famous integral

$$\int G_{A1}G_{B2}\, r_{12}^{-1}d\mathbf{r}_1d\mathbf{r}_2.$$

(We use double letters for polarization powers i.e., nx, ny and nz to avoid "indice in indice", nx=0,1,2,… are the s, p, d-like orbitals, etc..) In view of the general and extreme power of series expansion in numerical calculations, however, the

$$\int G_{A1}G_{B2}\, r_{12}^{-2}d\mathbf{r}_1d\mathbf{r}_2 \quad \text{as well as} \quad \int G_{A1}G_{B2}G_{C3}r_{12}^{-n}\, r_{13}^{-m}d\mathbf{r}_1d\mathbf{r}_2d\mathbf{r}_3$$

with n,m=1,2 important terms have come up in computation chemistry, what we can call higher moments with respect to inter-electronic distances $r_{ij}$, though their analytical evaluations have not been provided yet. Particularly, in correlation calculations the

$$\{ \int[\rho(\mathbf{r}_1)]^p[\rho(\mathbf{r}_2)]^q r_{12}^{-1}d\mathbf{r}_1 d\mathbf{r}_2 \}^t, \quad \{ \int\rho^p R_{C1}^{-n}d\mathbf{r}_1 \}^t$$

and similar terms can come up (among the many other models). For analytic evaluation, the idea comes from the Laplace transformation for n= 1 and 2, respectively, as

$$R_{C1}^{-1} = \pi^{-1/2} \int_{(-\infty,\infty)} \exp(-R_{C1}^2t^2)dt \text{ (well known) and,}$$

$$R_{C1}^{-2} = \int_{(-\infty,0)} \exp(R_{C1}^2t)dt = \int_{(0,\infty)} \exp(-R_{C1}^2t)dt \text{ (new trick).}$$

**One-electron spherical Coulomb integral for $R_{C1}^{-2}$:**

$$V_{P,C}^{(n)} \equiv \int_{(R3)} \exp(-p R_{P1}^2) R_{C1}^{-n} d\mathbf{r}_1$$

$$V_{P,C}^{(2)} = (2\pi^{3/2}/p^{1/2}) \int_{(0,1)} \exp(p R_{CP}^2 (w^2-1))dw = (2\pi^{3/2}/p^{1/2})e^{-v}F_0(-v)$$

$$0 < \exp(-v) < [p^{1/2}/(2\pi^{3/2})] V_{P,C}^{(2)} < 1$$

$$V_{P,C}^{(1)} = (2\pi/p) \int_{(0,1)} \exp(-p R_{CP}^2 w^2)dw = (2\pi/p)F_0(v), \text{ (known)}$$

$$0 < \exp(-v) < [p/(2\pi)] V_{P,C}^{(1)} < 1$$

$$V_{P,C}^{(2)}(R_{CP}=0)/ V_{P,C}^{(1)}(R_{CP}=0)= (2\pi^{3/2}/p^{1/2})/(2\pi/p)= (\pi p)^{1/2}$$

$v\equiv pR_{CP}^2$ as well as $\lim V_{P,C}^{(n=1 \text{ or } 2)}=0$ if $R_{CP}\to\infty$. The integral type $\int\exp(-w^2)dw$ frequently comes up in physics, but also the $\int\exp(w^2)dw$, here.

**One-electron non-spherical Coulomb integral for $R_{C1}^{-2}$:**

$${}^{full}V_{P,C}^{(2)} \equiv \int_{(R3)} G_{P1}(p,nx1,ny1,nz1)\, R_{C1}^{-2}\, d\mathbf{r}_1$$

$$\Sigma_1 \equiv \Sigma_{i1=0}^{nx1}\Sigma_{j1=0}^{ny1}\Sigma_{k1=0}^{nz1} \binom{nx1}{i1}\binom{ny1}{j1}\binom{nz1}{k1} \quad \text{for even i1, j1, k1 only}$$

$$n1 \equiv nx1+ny1+nz1)$$

$$m1 \equiv i1+j1+k1$$

$$\Gamma_1\equiv \Gamma((i1+1)/2)\, \Gamma((j1+1)/2)\, \Gamma((k1+1)/2)$$

$$D \equiv (x_P-x_C)^{nx1-i1}(y_P-y_C)^{ny1-j1} (z_P-z_C)^{nz1-k1}$$

$${}^{full}V_{P,C}^{(2)}= 2\Sigma_1\Gamma_1 D\, p^{-(m1+1)/2} \int_{(0,1)} (w^2-1)^{n1-m1} w^{m1} \exp(p R_{CP}^2(w^2-1))\, dw$$

$${}^{full}V_{P,C}^{(1)}= 2p^{-1}\pi^{-1/2}\Sigma_1\Gamma_1 D\, p^{-m1/2}\int_{(0,1)}(-w^2)^{n1-m1} (1-w^2)^{m1/2} \exp(-p R_{CP}^2 w^2)\, dw.$$

**One-electron spherical Coulomb integral for $R_{C1}^{-n}R_{D1}^{-m}$ with n, m=1,2:**

$$V_{P,CD}^{(n,m)}\equiv \int_{(R3)}\exp(-pR_{P1}^2)R_{C1}^{-n}R_{D1}^{-m}d\mathbf{r}_1$$

$$V_{P,CD}^{(1,2)}= \pi\int_{t=(-\infty,\infty)}\int_{u=(0,\infty)}g^{-3/2}\exp(-f/g)dudt$$

$$g\equiv p + t^2 +u,\ f\equiv p\, t^2 R_{PC}^2 +p\, u\, R_{PD}^2 +u\, t^2 R_{CD}^2.$$

The algorithm is straightforward for other cases of (n,m).

**Two-electron spherical Coulomb integral for $r_{12}^{-2}$, the (n,m)=(2,0) or (0,2) case:**

$$V_{PQ}^{(n)}\equiv \int_{(R6)} \exp(-p R_{P1}^2) \exp(-q R_{Q2}^2) r_{12}^{-n} d\mathbf{r}_1 d\mathbf{r}_2$$

$$V_{P,C}^{(2)} = \int_{(R3)} \exp(-p R_{P1}^2)r_{12}^{-2} d\mathbf{r}_1 = (2\pi^{3/2}/p^{1/2}) \int_{(0,1)} \exp(p R_{P2}^2 (w^2-1))dw$$

$$V_{P,C}^{(1)} = \int_{(R3)} \exp(-p R_{P1}^2)r_{12}^{-1} d\mathbf{r}_1 = (2\pi/p) \int_{(0,1)} \exp(-p R_{P2}^2 w^2)dw, \text{ (known)}$$

with $v\equiv pqR_{PQ}^2/(p+q)$

$$V_{PQ}^{(2)} = 2\pi^3(pq)^{-1/2}(p+q)^{-1}\int_{(0,1)} \exp(v(w^2-1))dw = (2\pi^3(pq)^{-1/2}(p+q)^{-1})e^{-v}F_0(-v)$$

$$0 < \exp(-v) < [(pq)^{1/2}(p+q)/(2\pi^3)]V_{PQ}^{(2)} < 1$$

$$V_{PQ}^{(1)} = (2\pi^{5/2}/(pq)) \int_{(0,c)} \exp(-pqR_{PQ}^2 w^2)dw, \text{ (known)}$$

with $c\equiv(p+q)^{-1/2}$

$$0 < \exp(-v) < [pq(p+q)^{1/2}/(2\pi^{5/2})]V_{PQ}^{(1)} < 1$$

$$V_{PQ}^{(2)}(R_{PQ}=0)/ V_{PQ}^{(1)}(R_{PQ}=0)= (2\pi^3(pq)^{-1/2}(p+q)^{-1}/(2c\pi^{5/2}/(pq))= (\pi pq/(p+q))^{1/2}$$

as well as $\lim V_{PQ}^{(n=1 \text{ or } 2)}=0$ if $R_{PQ}\to\infty$.

**Two-electron spherical Coulomb integral for the mixed term** $R_{C1}^{-n}r_{12}^{-m}$ **with n, m=1,2:**

$$\int_{(R6)}\exp(-pR_{P1}^2)\exp(-qR_{Q2}^2)R_{C1}^{-1}r_{12}^{-1}d\mathbf{r}_1d\mathbf{r}_2=(2\pi^2/q)\int_{u=(0,1)}\int_{t=(-\infty,\infty)} g^{-3/2}\exp(-f/g)\,dtdu$$

$$f\equiv pqR_{PQ}^2u^2+pR_{PC}^2t^2+qR_{QC}^2u^2t^2\ ,\ g\equiv p+qu^2+t^2,\text{ or}$$

$$\int_{(R6)}\exp(-pR_{P1}^2)\exp(-qR_{Q2}^2)R_{C1}^{-1}r_{12}^{-1}d\mathbf{r}_1d\mathbf{r}_2=(4\pi^2/q)\int_{(0,1)}F_0(gR_{WC}^2)g^{-1}\exp(-f/g)du$$

$$f\equiv pqR_{PQ}^2u^2,\ g\equiv p+qu^2\ .$$

The $R_{WC}$ depends on u as $gR_{WC}^2= (p+qu^2)|\mathbf{R}_W-\mathbf{R}_C|^2= |p\mathbf{R}_P+qu^2\mathbf{R}_Q -g\mathbf{R}_C|^2$; the algorithm is straightforward for other cases of (n,m).

**Three-electron spherical Coulomb integral for** $r_{12}^{-n}r_{13}^{-m}$ **with n,m=1,2:**

$$V_{PQS}^{(n,m)}\equiv \int_{(R9)} \exp(-p\ R_{P1}^2)\ \exp(-q\ R_{Q2}^2)\ \exp(-s\ R_{S3}^2)\ r_{12}^{-n}r_{13}^{-m}\ d\mathbf{r}_1d\mathbf{r}_2d\mathbf{r}_3$$

$$V_{PQS}^{(1,1)}= (4\pi^{7/2}/(qs))\int_{(0,1)}\int_{(0,1)} g^{-3/2}\exp(-f/g)dudt$$

$$f\equiv pqR_{PQ}^2u^2+psR_{PS}^2t^2+qsR_{QS}^2u^2t^2,\ g\equiv p+qu^2+st^2.$$

Alternatively, with inclusion of the Boys function

$$V_{PQS}^{(1,1)}= (4\pi^2/(qs))\int_{(R3)} F_0(qR_{Q1}^2)\ F_0(sR_{S1}^2)\ \exp(-pR_{P1}^2)d\mathbf{r}_1,$$

but continuing with numerical integration.

Alternatively, with embedding the Boys function

$$V_{PQS}^{(1,1)}= (4\pi^{7/2}/(qs))\ \int_{(0,1)}\ h(u)\ g^{-1}\exp(-f/g)\ du$$

$$h(u)\equiv\ \int_{(0,c)}\exp(-g\ s\ R_{VS}^2\ w^2)dw,\ c\equiv(g+s)^{-1/2},\ f\equiv pqR_{PQ}^2u^2,\ g\equiv p+qu^2.$$

The algorithm is straightforward for other cases of (n,m).

**Appendix**: The product of two Gaussians, $G_{J1}(p_J,0,0,0)$ with J=1,…,m=2 is another Gaussian centered somewhere on the line connecting the original Gaussians, but a more general expression for m>2 comes from the elementary

$$\Sigma_J\ p_J\ R_{J1}^2 = (\Sigma_J\ p_J)\ R_{W1}^2 + (\Sigma_J\Sigma_K\, p_J\ p_K\, R_{JK}^2)/(2\Sigma_J\ p_J)$$

$$\mathbf{R}_W \equiv (\Sigma_J\, p_J\ \mathbf{R}_J)/(\Sigma_J\ p_J)$$

where $\Sigma_{J\text{ or }K}\equiv \Sigma_{(J\text{ or }K=1\text{ to }m)}$ and $R_{J1}\equiv |\mathbf{R}_J-\mathbf{r}_1|$ for $\exp(\Sigma_J\ c_J)= \Pi_{(J=1\text{ to }m)}\exp(c_J)$, keeping in mind that $R_{JJ}=0$, and the m centers do not have to be collinear.

**Appendix:** Given a single power term polynomial at $\mathbf{R}_P$, we need to rearrange or shift it to a given point $\mathbf{R}_S$. For variable x, this rearrangement is $(x-x_P)^n= \Sigma_{i=0\text{ to }n}\ c_i\ (x-x_S)^i$, which can be solved systematically and immediately for $c_i$ by the consecutive equation system obtained from the 0,1,…n$^{th}$ derivative of both sides at x:= $x_S$, yielding

$$\text{POLY}(x,P,S,n) \equiv (x-x_P)^n= \Sigma_{i=0\text{ to }n}\ (^n{}_i)(x_S-x_P)^{n-i}\ (x-x_S)^i\ ,$$

where $(^n{}_i)$=n!/(i!(n-i)!). If $x_S$=0, it reduces to the simpler well known binomial formula as $(x-x_P)^n= \Sigma_{i=0\text{ to }n}\ (^n{}_i)(-x_P)^{n-i}x^i$.

**Appendix:** De-convolution of Boys functions from $F_L(v)\equiv \int_{(0,1)} \exp(-vt^2)t^{2L}dt$, L=0,1,2,… to $F_0(v)= \int_{(0,1)} \exp(-vt^2)dt$ for v>0 and v≤0 comes from the help of partial integration yielding

$$2vF_{L+1}(v)= (2L+1)F_L(v) - \exp(-v).$$

The value of L recursively goes down to zero, and the value of $F_0(v)$ is needed only at the end. The v=0 case is trivial and the v>0 is well known in the literature but, the v<0 cases are also needed for cases described in the thesis.

# 3.2 Developing theories in physical chemistry

**Thesis-22-Theory: Reformulation of Gaussian error propagation for a mixture of dependent and independent variables**

The Gaussian error propagation to estimate standard deviation for a general expression $f(x_1,\dots,x_n,z_1,\dots,z_m)$ is generalized for practice, when the measurable quantities in its argument are correlated somehow, e.g. all $z_j$ depends on some of the independent $x_i$'s. The derivation is based on the formula for total derivative of a general multivariable function for which some of its variables are not independent from the others, yielding a counterpart to the probability (see the concept of covariance) approach of this subject.

= = = = = =

Általánosítottam a Gauss féle hibaterjedést szórás becslésére egy általános formula $f(x_1,\dots,x_n,z_1,\dots,z_m)$ esetén a gyakorlat számára, amikor a mért mennyiségek a változók között valamilyen mértékben korrelálnak, pl. minden $z_j$ függ valahogyan néhány vagy az összes független $x_i$-től. A levezetés az általános többváltozós függvények totális deriváltján alapszik, amikor néhány változója nem független a többitől. Ennek eredményeképpen a valószínűségi leírás (ld. kovariancia koncepció) alternatív formuláját vezettem le.

**<u>Sandor Kristyan: Periodica Polytechnica Chemical Engineering, 58(Sup) (2014) 49-52</u>**

= = = = = =

Representative equations/tables/figures:

Consider the $f(x_1,\dots,x_n,z)$, where $x_1,\dots,x_n$ are independent variables and $z= z(x_1,\dots,x_n)$ is dependent. Using the reformulated total derivative

$$df = \Sigma_{(i=1\dots n)}(\partial f/\partial x_i)dx_i + (\partial f/\partial z)dz = \Sigma_{(i=1\dots n)}(\partial f/\partial x_i)dx_i + (\partial f/\partial z)(\Sigma_{(i=1\dots n)}(\partial z/\partial x_i)dx_i) =$$
$$= \Sigma_{(i=1\dots n)}[(\partial f/\partial x_i) + (\partial f/\partial z)(\partial z/\partial x_i)]dx_i,$$

the Gaussian error propagation in this case is

$$(\Delta f)^2 = \Sigma_{(i=1\dots n)}[(\partial f/\partial x_i) + (\partial f/\partial z)(\partial z/\partial x_i)]^2(\Delta x_i)^2.$$

More generally, if $y= f(x_1,\dots,x_n,z_1,\dots,z_m)$ with dependent variables $z_j= z_j(x_1,\dots,x_n)$ for $j= 1,\dots,m$, then the more general "analytical formula"

$$(\Delta f)^2 = \Sigma_{(i=1\dots n)}[(\partial f/\partial x_i) + \Sigma_{(j=1\dots m)}(\partial f/\partial z_j)(\partial z_j/\partial x_i)]^2(\Delta x_i)^2,$$

compare this to the well known "probability formula" standard deviation of f (denoted as $s_f$) when its variables are not independent:

$$(s_f)^2 = (\Delta f)^2 = \Sigma_{(i=1\dots n+m)}\Sigma_{(j=1\dots n+m)} (\partial f/\partial \xi_i)(\partial f/\partial \xi_j)\mathrm{cov}(\xi_i,\xi_j).$$

**Thesis-23-Theory: Role of the surface free enthalpy excess of solid chemical elements in their melting and critical temperature**

Statistical mechanical consideration has yielded that a 5-20% increase in the average number of neighbors of an atom (or particle, $n_{avrg}$) in the surface phase between 0 K and melting temperature, $T_m$, makes the solid surface "geometrically impossible" to exist at some temperature what is called the melting temperature. This phenomenon in the surface geometry results in the collapse of crystal structure in the bulk, and the formation of surface layers of liquid begins a few atoms (particles) thick. The critical temperature can also be pictured in the same way. This study has yielded expressions for the heat of melting ($\Delta H_m$), entropy of melting ($\Delta S_m$), melting temperature ($T_m$) and critical temperature ($T_c$) in relation to surface parameters in case of pure metals, as well as picturing the phase transitions (between the solid-, liquid- and critical phases) as (not bulk, but purely a) surface phenomenon.

= = = = = =

Statisztikus mechanikai vizsgálataim azt eredményezték, hogy 5-20% növekedés a felület fázisbeli atomok (részecskék) szomszédjainak átlagos számában ($n_{avrg}$) 0 K és olvadási hőmérséklet ($T_m$) között a szilárd felület létezését "geometriailag lehetetlenné" teszi egy adott hőmérsékleten, amit olvadási hőmérsékletnek nevezünk. Ez a jelenség a felület geometriájában azt eredményezi, hogy a tömb fázis kristály szerkezete összeomlik, valamint egy felületi folyadék fázisú réteg kialakulása kezdődik, mely néhány atom (részecske) vastag. A kritikus hőmérséklet hasonlóan képzelhető el. E tanulmányban egyenleteket állítottam fel az olvadáshő ($\Delta H_m$), olvadási entrópia ($\Delta S_m$), olvadási hőmérséklet ($T_m$) és kritikus hőmérséklet ($T_c$) számára felületi paraméterek tekintetében tiszta fémek esetén, valamint a fázis átmeneteket (a szilárd-, folyadék- és kritikus fázisok között) úgy kell elképzelni mint egy (nem tömb, hanem tisztán) felületi jelenséget.

= = = = = =

Representative equations/tables/figures:

To tickle funny bones: Basic physical chemistry books talk about phase transitions in detail (melting points, boiling points, critical temperatures, etc.) but the reason why these happen is not discussed… (We do not talk here phases like plasmas (ionized gases), etc..) An empirical relation for the "free enthalpy excess of the surface" of solid chemical elements is given by $\gamma$[J/mol]=$\alpha$ $\Delta H'$, where $\Delta H'$ is the internal enthalpy (heat) of atomization and $\alpha$ changes very slightly with temperature, but its change is fumdamental for phase transitions. Expression $\alpha =(z- n_{avrg})/z$, where z is the bulk effective coordination number, mainly the 1st, or by any chance 2nd, 3rd, etc. nearest neighbors, as well as $0.2<\alpha<0.3$, has served as starting point for statistical mechanical derivation yielding:

$$\partial\alpha/\partial T = -R \ln(m)/\Delta H' ,$$

where R= gas constant, m= the number of layers in the surface phase, and it is independent of effective coordination number, crystal structure, and crystal face, as well as

$$\partial\alpha/\partial T \approx -5\text{x}10^{-5} K^{-1} \text{ for metals.}$$

Temperature dependence of $\alpha$ is quasi-linear, and let $\Delta\alpha_{cr} \equiv \alpha(T_m)-\alpha(T=0K) = -RT_m\ln(m)/\Delta H'$, where $\Delta H'$ is considered to be T-independent. From eperimental data the calculation yields

$$-0.1 < \Delta\alpha_{cr} < -0.055 \text{ for metals.}$$

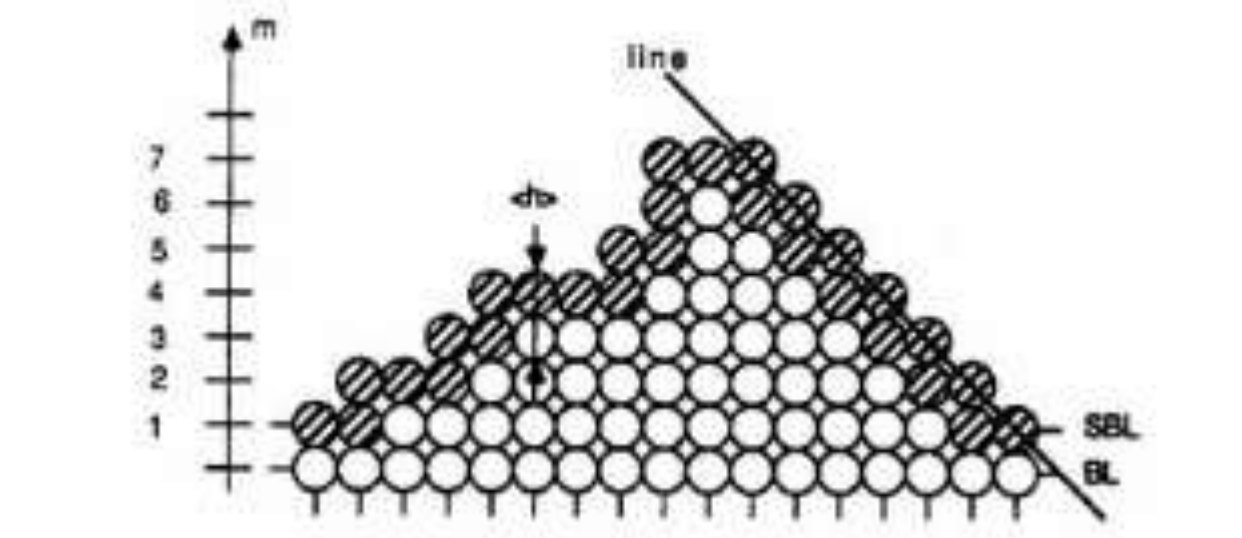

A segment of the surface of a 2D crystal of cubic p lattice with $m = 7$ and (10) face. Along the "line", the face is (11). This figure also shows the meaning of $m$ and $\langle h \rangle$. The shaded atoms are those surface atoms which have $n < z = 8$ neighbors and the others in the surface phase (in the SBL and SL) should be considered as bulk atoms ($n = z = 8$). This figure also exhibits that the value of $\gamma$ comes from the upper $\langle h \rangle \approx 2$ layers of the surface.

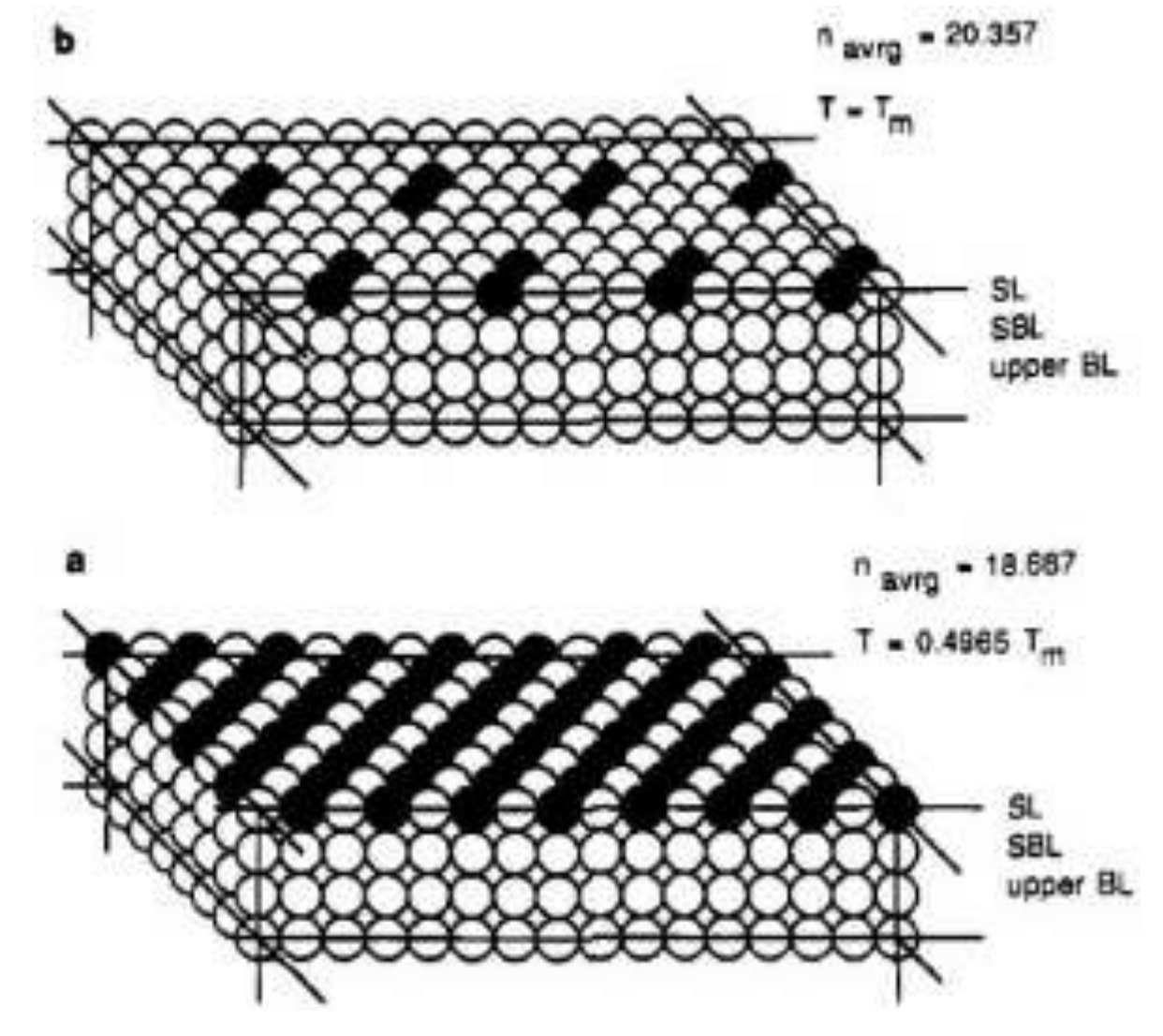

Upper figure: Characterization of surface in relation to m, depth (<h>), semi-bulk layer (SBL) and bulk layers (BL) is not trivial (see faces (10) vs. (11)).

Lower figure: Computer generated 8x16 atom size part of the surface of a 3D crystal of cubic p lattice with two surface layers (m=2) and (100) face. For the better view the blackened circles represent missing atoms in the SL, bulk atoms have $3^3-1=26$ neighbors (>20.357). One would expect rougher surface at higher temperature (T), but paradoxically that looks flatter. The $n_{avrg}$ increases ($\alpha$ decreases) in the surface with increasing T and reaches a geometrical limit it cannot increase further at a sharp, the melting temperature ($T_m$), and the crystal structure of bulk collapses via diffusing vacancies from surface, vacancies in SBL decrease $n_{avrg}$ but diffuse away into BL. It continues in liquid phase too, and a limit is reached again at critical temperature ($T_c$) when liquid surface $n_{avrg}$ equals the liquid bulk coordination, (z', and $\alpha(T_c)=0$,) i.e. "surface cannot exist" while colloidics theaches us it must...

Further derivation yields

$$T_m = |\Delta\alpha_{cr}|\Delta H'/(R\ln(m)) = \Delta n_{cr}\Delta H'/(z\,R\ln(m)),$$

where the subscript "cr" refers to the maximum increase in the number of neighbors ($n_{avrg}$) in the solid surface phase in the temperature range [0 K, $T_m$], and importantly, it cannot increase further by geometrical reason in a given crystal structure, (cr= critical, not to be confused with critical temperature). With the heat of melting, $\Delta H_m$, the liquid bulk is closer energetically to the gas phase than to the solid bulk, and instead of "solid/vacuum interface", for a "liquid/bulk interface", the $\Delta H_m = (z - z')\,\Delta H'(T_m)/z$ holds when passing from a solid to a liquid bulk, wherein all are measured quantities except the effective liquid coordination number, z', e.g. the decrease is $z-z' \approx 1.5$ in case of cubic lattice. Finally,

$$-\Delta\alpha_{cr} \approx 1 - (\rho_{liq}(T_m)/\rho_{sol}),$$

with the liquid and solid densities, and the heat of melting and entropy are

$$\Delta H_m = T_m R\ln(m) \approx -T_m(\partial\gamma/\partial T)_{avrg} \text{ (see figure below)}$$

$$\Delta S_m = \Delta H_{m,calc}/T_m = R\ln(m) \approx -(\partial\gamma/\partial T)_{avrg}.$$

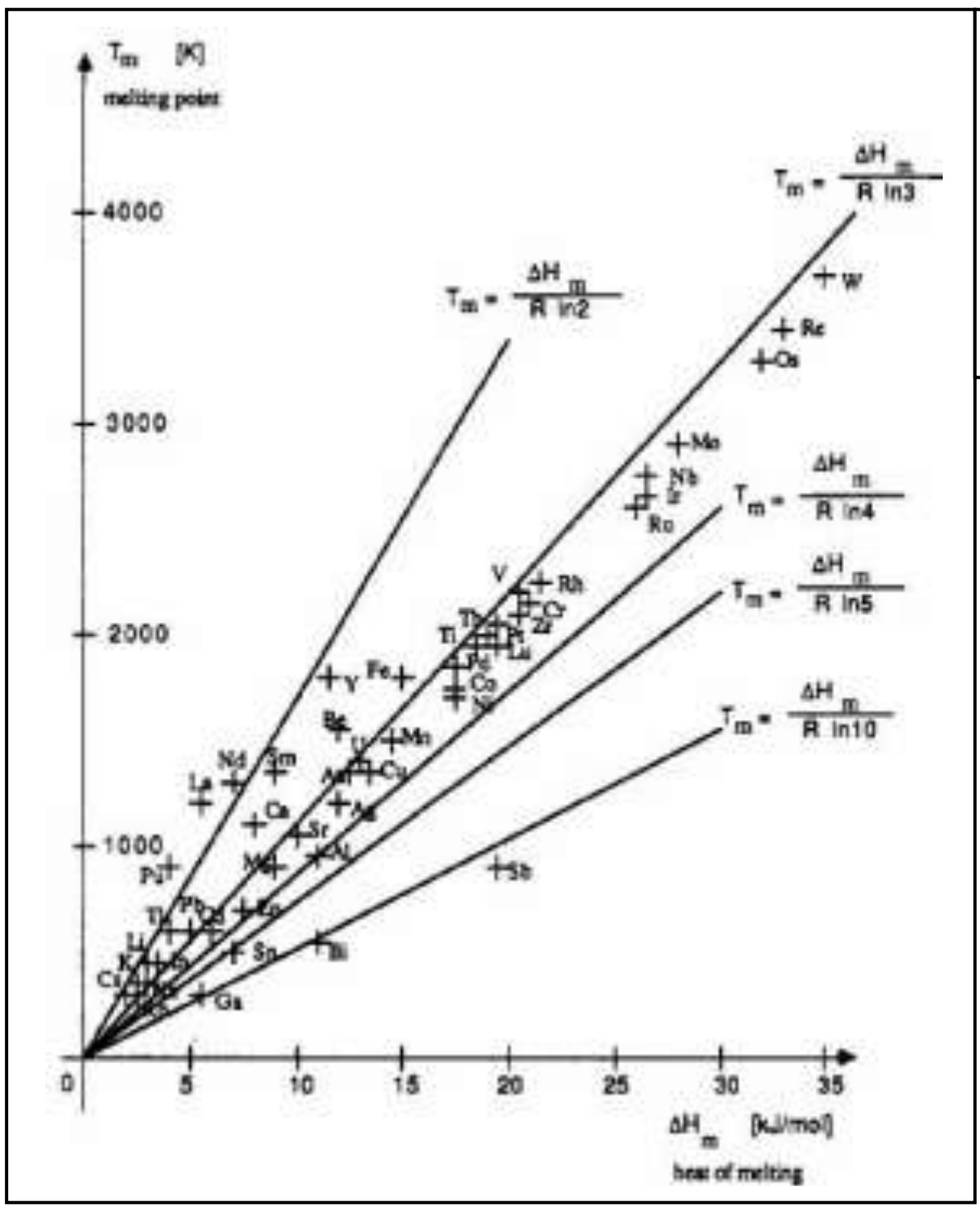

Melting temperature vs heat of melting for solid chemical elements, crosses: measured values, solid lines: theoretical curves with different values of m.

The “surface cannot exist” when the $\alpha$ of a liquid surface reaches $\alpha(T_c)=0$. The partition function is formally the same for the surface of a liquid or solid, yielding after derivation

$$T_c = \alpha(T=0K)\ \Delta H'/R\ \ln(m) = \gamma/R\ \ln(m) \approx -\gamma/(\partial\gamma/\partial T)_{avrg},$$

where $(\partial\gamma/\partial T)_{avrg}$ is the average slope for $T<T_m$. For two different metals (labeled with subscripts 1 and 2), since m, $\Delta\alpha_{cr}$ and $\Delta H'$ do not vary strongly with metals, metals and temperature, respectively, a sometimes crude, sometimes amazingly good estimation is

$$T_{c1}/T_{c2} \approx \Delta H'_1/\ \Delta H'_2 \approx T_{m1}/T_{m2}\ .$$

Furthermore, using $\alpha(T=0K)\approx 0.2$ and $|\Delta\alpha_{cr}|\approx 0.055$ common values for all metals,

$$T_c/T_m \approx \alpha(T=0K)/|\Delta\alpha_{cr}| \approx 3.6 \text{ (see figure below)},$$

$$T_c/T_m \approx \gamma/\Delta H_m\ .$$

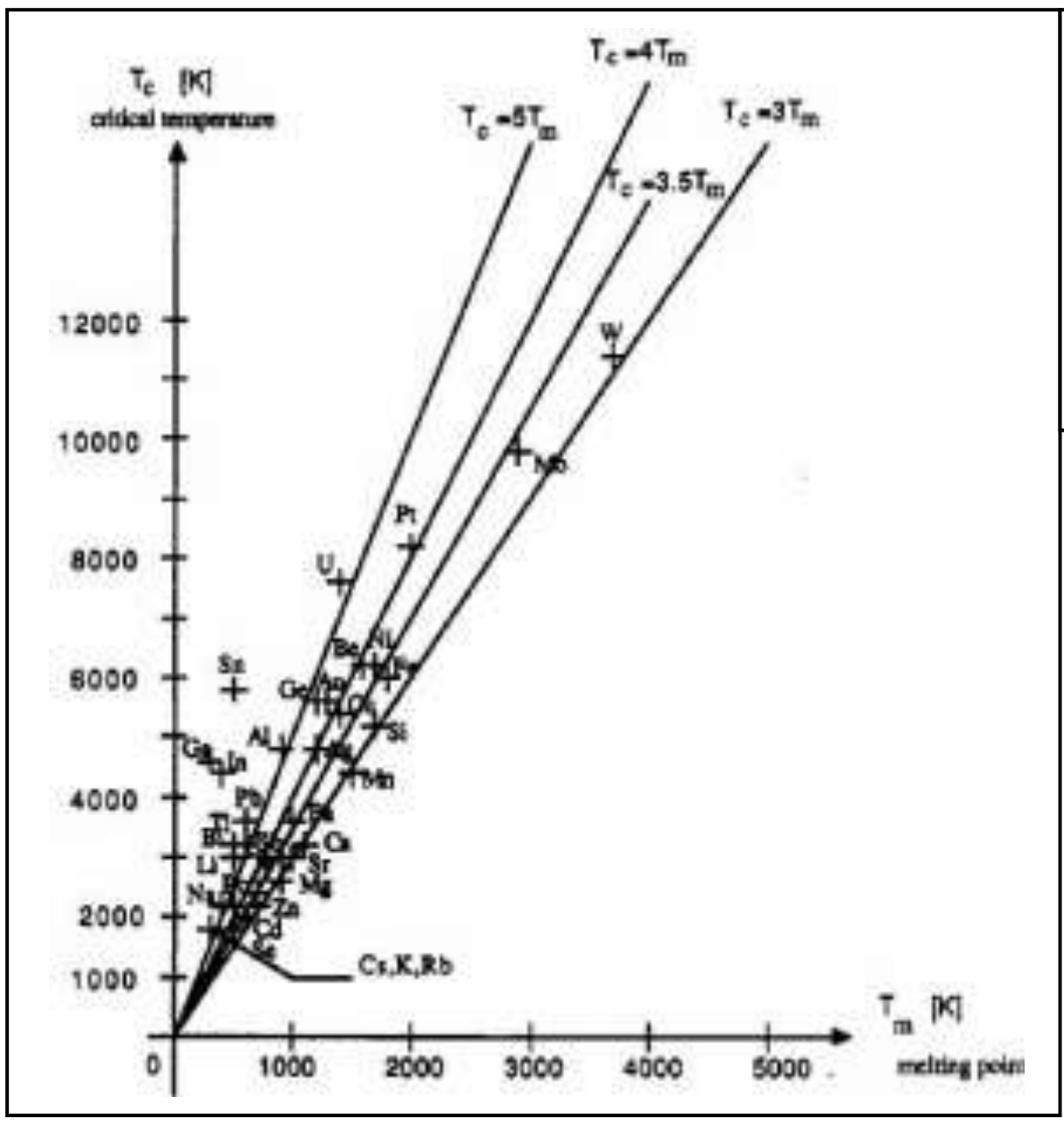

Critical temperature vs melting temperature for solid chemical elements, crosses: measured values, solid lines: theoretical curves with different slopes.

Generalization is possible from solids to compounds, in the latter molecules are located at the lattice points. After derivation, for compounds

$$\alpha(T=0K) \approx RT_c \ln(5)/\lambda ,$$

$$T_c/T_m \approx 2.5 \text{ (see figure below)},$$

$$\lambda \approx 8\, a \ln(m)/(27\, b\, \alpha(T=0K)) \approx 2a/b \text{ (see figure below)}.$$

The surface free enthalpy excesses ($\gamma$) of compounds are not easily available as they are for metals, however, the critical temperature ($T_c$), the heat of vaporization ($\lambda$), and the heat of melting ($\Delta H_m$), or van der Waals constants of gases (a, b) are easily available for compounds.

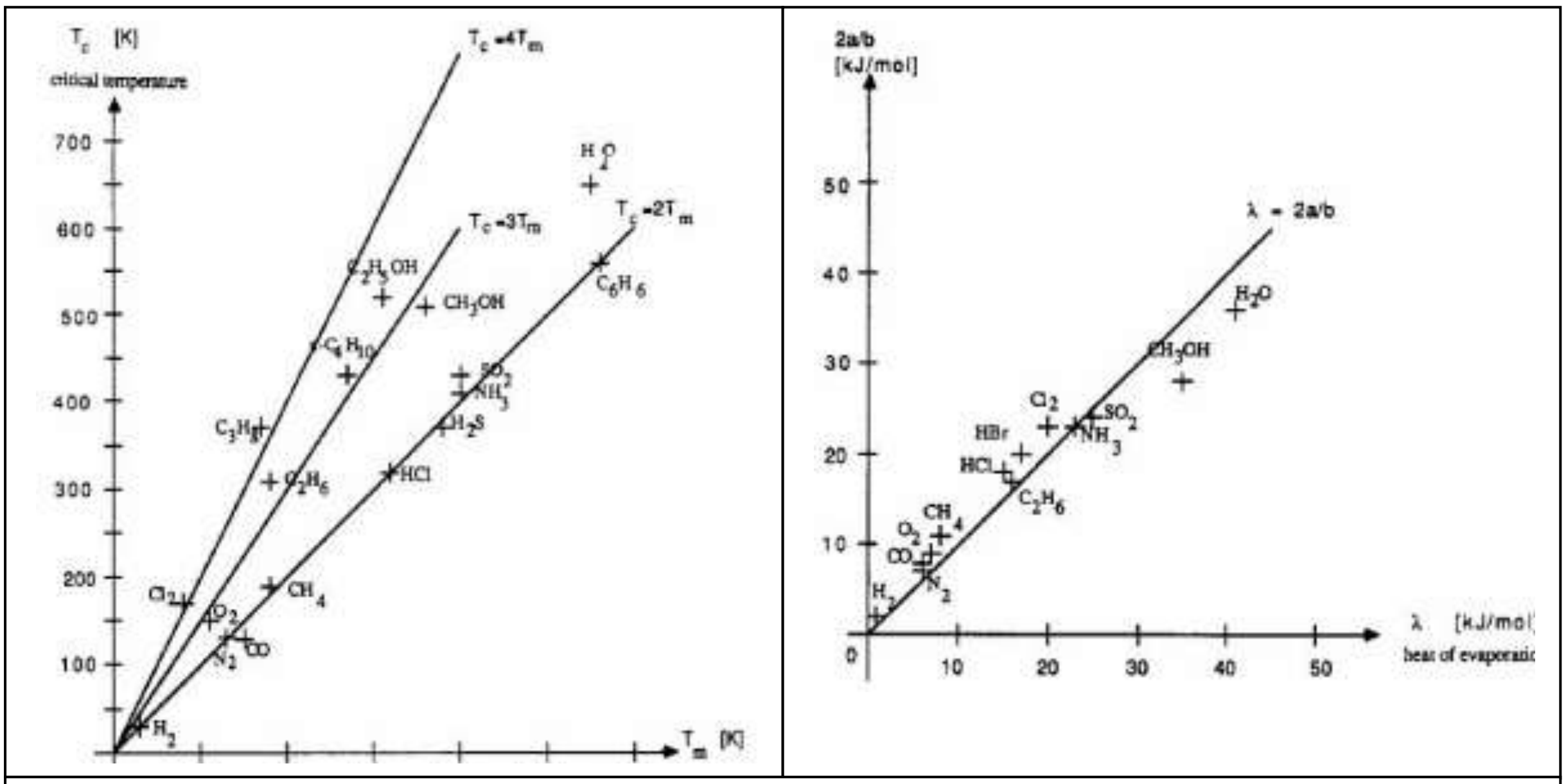

Critical temperature vs melting temperature for compounds (left), and relationship between the van der Waals constants of gases and the heat of evaporation for some compounds (right),
crosses: measured values,
solid lines (left): theoretical curves with different slopes,
solid line (right): theoretical curve, the value 2 comes from surface properties,
while the parameters (a, b, $\lambda$) are bulk properties.

In summary, one has to consider the stability of the solid surface instead of the bulk for explaining the sharp transition temperature from solid to liquid, called melting. Surface atoms (particles) tend to gain more and more neigbors with elevating temprature, (an opposit phenomenon to sublimation/evaporation, wherein the particles tend to get rid of their solid/liquid bulk (or surface) neighbors by temperature), at temperature when the (liquid) bulk (effective or average) coordination number and (average) number of neigbors of surface atoms (particles) became equal, that is the critical point.

**Thesis-24-Theory: Quaternionic treatment of the electromagnetic wave equation**

The Maxwell equations have been converted to quaternionic representation benefiting its important properties along with the short derivation and discussion of wave equation.
= = = = = =

A Maxwell egyenletek kvaternió reprezentációját mutattam be, kihasználva annak előnyeit, valamint röviden újra tárgyaltam a hullám egyenlet levezetését és diszkusszióját e tekintetben.

<u>**Sandor Kristyan – J. Szamosi: Acta Physica Hungarica, 72 (1992) 243 - 248**</u>
= = = = = =

Representative equations/tables/figures:

The quaternions are the non-commutative extention of complex numbers to four dimension as Q=a+bi+cj+dk, where a,b,c,d are real numbers, and for the symbols $i^2= j^2= k^2=$ ijk= -1, i→j→k→i cirle helps to picture ij=-ji=k, etc.; i,j,k correspond to √-1, the known complex imaginary unit, – with c=d=0 the quaternion Q reduces to a complex number. Two purely imaginary (a=0) quatenions $q_i= b_i i+c_i j+d_i k$ corresponding to 3 dim physical vectors $v_i= (b_i,c_i,d_i)$ synthesizes the dot and vector products as $q_1*q_2=Re(q_1*q_2)+Im(q_1*q_2)=-v_1v_2+v_1 \times v_2$, and $q_1*q_1= -|v_1|^2$, because by parallelity $v_1 \times v_1=0$.

The Maxwell equations (∇E=ρ/ε, ∇B=0, ∇xE=-∂B/∂t, ∇xB=μJ+με∂E/∂t) in quaternion representation are

$$\nabla*E = -\rho/\varepsilon - \partial B/\partial t,$$

$$\nabla*B = \mu J + \mu\varepsilon\partial E/\partial t,$$

and the energy density of the electrostatic and magnetostatic field is

$$w= -(1/2)\varepsilon E*E \text{ and } -(1/2)\mu^{-1}B*B, \text{ resp..}$$

The <u>associativity</u> $Q_1*Q_2*Q_3\equiv Q_1*(Q_2*Q_3)= (Q_1*Q_2)*Q_3$ for purely imaginary $q_1=q_2= \nabla\equiv e_1(\partial/\partial x)+ e_2(\partial/\partial y)+ e_3(\partial/\partial z)$ and $q_3$= E or B <u>provides the electromagnetic wave equation</u> $\mu\varepsilon\partial^2 q_3/\partial t^2= \nabla^2 q_3$ if no current (J=0) an no charge (ρ=0) as well as $c^2\mu\varepsilon=1$ for the speed of light. However, associativity is strictly hold for vectors/quaternions ($Q_i$), not necessarily for differential operators, e.g. $\nabla*(\nabla*q_3)= (\nabla*\nabla)*q_3$, but generally $q_1*(\nabla*q_2)\neq (q_1*\nabla)*q_2$, etc..

The Independent Scientific Research Institute (Box 30, CH-1211, Geneva-12, Switzerland, ISRI-05-04.25 on August 25, and Oxford, OX4 4YS, England, ISRI-05-04.26 on 6 July) has summarized the “Quaternions in mathematical physics (1): Alphabetical bibliography” in 2008 for the anniversary day of the discovery of quaternions (October 16, 1843 at Brougham Bridge on the Royal Canal, now in the Dublin suburbs), on the occasion of the bicentenary of the birth of William Rowan Hamilton (1805–2005), this work is listed in the total about 1400 (i.e. not many) related documents in that about 100 page bibliography. (As personal remark: In physics the complex numbers are fundamental, but quaternions have not provided breakthrough yet.)

**Thesis-25-Theory: Generalization of Savitzky-Golay parameters to least-square smoothing and differentiation of two-dimensional data**

The computation method to smooth and differentiate data of z=f(x,y) kind is introduced as the generalization of the Savitzky-Golay method, requiring only that the data points are equidistant in both, x and y. The smoothed data, as well as the partial derivatives can be calculated directly from data points.

= = = = = =

Az (egy-dimenziós, z=f(x)-re vonatkozó) Savitzky-Golay módszer általánosításaként kidolgoztam a z=f(x,y) típusú két-dimenziós adatpontok simításának és differenciálásának számításos eljárását, melyben az egyetlen megszorítás az adatpontok ekvidisztáns jellege az x és y változókban. A simított adatok, valamint a parciális deriváltak közvetlen számíthatók az adatpontokból.

= = = = = =

Representative equations/tables/figures:

To have a taste how it works, consider the $3^{th}$ degree 1 dim. Savitzky-Golay parameter sets

| m | smooth | df/dx | $d^2f/dx^2$ | $d^3f/dx^3$ |
|---|---|---|---|---|
| -2 | -3 | 1 | 2 | -1 |
| -1 | 12 | -8 | -1 | 2 |
| 0 | 17 | 0 | -2 | 0 |
| 1 | 12 | 8 | -1 | -2 |
| 2 | -3 | -1 | 2 | 1 |
| N | 35 | 12 | 7 | 2. |

To smooth data set $\{z_1(x_0+d), z_2(x_0+2d), z_3, z_4, z_5, z_6, z_7,\ldots,z_K(x_0+Kd)\}$ by least square (LS) fit for polynom $f(x)= a_0+ a_1x+ a_2x^2+ a_3x^3$, instead of the standard way of LS, the simple

$$z_i^{smoothed}= (-3z_{i-2}+ 12z_{i-1}+ 17z_i+ 12z_{i+1} -3z_{i+2} )/35$$

for i=3,…,K-2 from the column "smooth" of table does the same job, so for the derivatives via f(x) using the corresponding column.

The two dimension version has been worked out (extended), for example, the $2^{nd}$ degree 2 dim. smoothing parameter sets for polynom $f(x,y) = a_0+ a_1x+ a_2y+ a_3x^2+ a_4xy+ a_5y^2$ are

```
to smooth:       -13   2   7   2 -13
                   2  17  22  17   2
                   7  22  27  22   7      N= 175
                   2  17  22  17   2
                 -13   2   7   2 -13,
to calc. ∂²f/∂x∂y: -4  -2   0   2   4
                  -2  -1   0   1   2
                   0   0   0   0   0      N= 100
                   2   1   0  -1  -2
                   4   2   0  -2  -4, etc.
```

**Thesis-26-Theory: Kinetics of heterogeneous catalytic hydrogenolysis of ethane over supported catalyst (Pt, Pd, Ni), its coke formation, cleaning and reverse reaction**

This research of mine arcing over fifteen years has provided partly my D.Sc thesis (kinetics, Budapest University of Technology (Dept. of Phys. Chem.) and Institute of Isotopes of the Hung. Acad. Sci., 1980-1982) and my Ph.D thesis (poisoning, University of Texas at Arlington (Dept. of Chem.), 1983-1985) as well as a continuing research project of interest.

In the hydrogenolysis of ethane ($C_2H_6 + H_2$ –cat→ $2CH_4$), the chemisorptions of ethane and hydrogen produce a common surface species, adsorbed hydrogen, and the coverages of the two-carbon-atom surface compound and the adsorbed hydrogen are interdependent through the partial pressures of ethane and hydrogen. The kinetically slow rupture of the C-C bond can take place in an interaction theoreticaly with a free site, adsorbed hydrogen or molecular hydrogen, but analyzing the experimental results, my conclusion is that molecular hydrogen is the most probable agent in the bond splitting.

Among Pt (most expensive), Pd and Ni (less exensive), the Pt is the less and Ni is the most capable metal for coke formation (larger $C_mH_n$ quasi-twoo dimension molecular "island" formation on the surface with m>>n). In situ catalytic activation and regeneration using electrostatic field gradients has been developed as demonstrated during hydrogenolysis of ethane and ethylene on a Nickel wire catalyst, a relatively inexpensive approach to improve catalytic efficiency. Activation is observed only when the Ni wire is biased negatively with respect to the outside (e.g. Al) cylinder. This activation and, in some cases, catalytic regeneration is believed to arise from the combined effect of high field gradients and small leakage currents generated during high potential application (destroying/breaking the coke). The separate activation energies for the normal hydrogenolysis reaction (129 kJ/mol) and the self-poisoning reaction (146 kJ/mol) have been modeled and calculated.

For the reverse reaction, the kinetics of the heterogeneous catalytic decomposition (in fact, recombination) of methane ($CH_4$ –cat→ $C_2H_6 + H_2$ + coke) has also been modeled.

= = = = = =

Ez a tizenöt éven keresztül átívelő kutatásom, mely részben az egyetemi doktori disszertációmat (kinetika, Budapesti Műszaki (és Gazdaságtudományi) Egyetem (Fizikai Kémia Tanszék) és Magyar Tudományos Akadémia Izotóp Intézete, 1980-1982) és a Ph.D disszertációmat (kokszosodás, University of Texas at Arlington (Dept. of Chem.), 1983-1985) szolgáltatta, az érdeklődési körömbe vágó egyik folytatólagos kutatási projekt.

Etán hidrogenolízisében ($C_2H_6 + H_2$ –kat→ $2CH_4$), az etán és hidrogén kemiszorpciója egy közös felületi képződményt és adszorbeált hidrogént ad, továbbá a két-szénatomos felületi vegyület és az adszorbeált hidrogén borítottsága kölcsönösen függ az etán és hidrogén parciális nyomásaitól. A kinetikailag lassú C-C kötés szakadása történhet elvileg üres felületi hellyel, adszorbeált hidrogénnel vagy molekuláris hidrogénnel való kölcsönhatásban, de elemezve a kísérleti eredményeket, a következtetésem, hogy nagy valószínűség szerint a kötés szakadásának oka a harmadik eset, a molekuláris hidrogénnel való kölcsönhatás.

Pt (legdrágább), Pd és Ni (legolcsóbb) fémek közül, a Pt a legkevésbé és a Ni a leginkább alkalmas fém a kokszosodásra, mely nagyobb $C_mH_n$ kvázi-két dimenziós molekuláris “szigetek” képződése a felületen melyben m>>n. In szitu katalitikus aktiválást és regenerálást fejlesztettünk ki elektrosztatikus tér gradiens felhasználásával, melyet az etán és etilén hidrogenolízisével demonstráltunk nikkel drót katalizátoron, egy relatíve olcsó mód a katalitikus hatékonyság javítására. Aktiválást csak akkor figyelhettünk meg, amikor a nikkel drót negatív pólusként volt kapcsolva a külső (pl. Al) hengerhez (fém burkoláshoz) képest. Ez az aktiválás, és néhány esetben katalitikus regenerálás feltételezéseink szerint a nagy elektromos feszültség gradiens és egy kis elektromos áram átcsorgásának kombinált effektusából ered a nagy elektromos potenciál alkalmazásakor (mely szétrombolja/áttöri a koksz réteget). A normál hidrogenolizis (129 kJ/mol) és az ön-mérgeződő (ön-kokszosodó, 146 kJ/mol) reakciók aktiválási energiáit modelleztük és számítottuk.

A reverz reakció esetében, a metán heterogén katalitikus dekompozíciójának ($CH_4$ –kat→ $C_2H_6 + H_2$ + koksz), tulajdonképpen rekombinációjának a kinetikáját szintén modelleztem.

= = = = = =

Representative equations/tables/figures:

The mechanism of the hydrogenolysis is (*= surface site, $\theta_i$ = coverage of compound i, $p_i$= parcial pressure of gas i, $K_i$= equlibrium constant of reaction i, k= rate constant)

$$C_2H_6 + (7\text{-}m)* \rightarrow K_E \text{ (fast )} \leftarrow C_2H_m* + (6\text{-}m)H* \qquad \text{with } K_E= \theta_m \theta_H^{6-m}/(p_E\, \theta_0^{7-m})$$

$$H_2 + 2* \rightarrow K_H \text{ (fast)} \leftarrow 2H* \qquad \text{with } K_H= \theta_H^2/(p_H\, \theta_0^2)$$

$$C_2H_m* \ + \ B \text{ –k (irrev., slow)}\rightarrow \ CH_u*+CH_v* \ \text{ -(H* or } H_2\text{, fast)}\rightarrow \ CH_4(g)$$

$$\text{rate} = k\, \theta_m\, B \text{ and } B= * \text{ or } H* \text{ or } H_2(g)$$

For $CH_{u\text{ or }v}*$ the $\theta_i \approx 0 \Rightarrow \theta_0+\theta_H+\theta_m \approx 1$, using $y \equiv K_E\, p_E$, $x \equiv (K_H\, p_H)^{1/2}$, $G_6 \equiv x^{6-m}$, $G_7 \equiv x^{7-m}$ and $D \equiv y+G_6+G_7 \Rightarrow \theta_0= G_6/D$, $\theta_m= y/D$ and $\theta_H= G_7/D \Rightarrow$ the mathematically possible rates are

$$\text{rate} = k\, y\, G_6/D^2 \quad \text{if} \quad B= \theta_0$$

$$\text{rate} = k\, y\, G_7/D^2 \quad \text{if} \quad B= \theta_H$$

$$\text{rate} = k\, y\, p_H/D \quad \text{if} \quad B= p_H$$

From rate measurement experiments with Ni catalyst at 250 $^0$C and Pd catalyst at 350 $^0$C (the partial pressures of hydrogen and ethane were varied between 0.5 and 10 kPa), the calculated k, $K_E$, $K_H$ and m values can be found in J. Chem. Soc., Farad. Trans. I, 80 (1984) 1645.

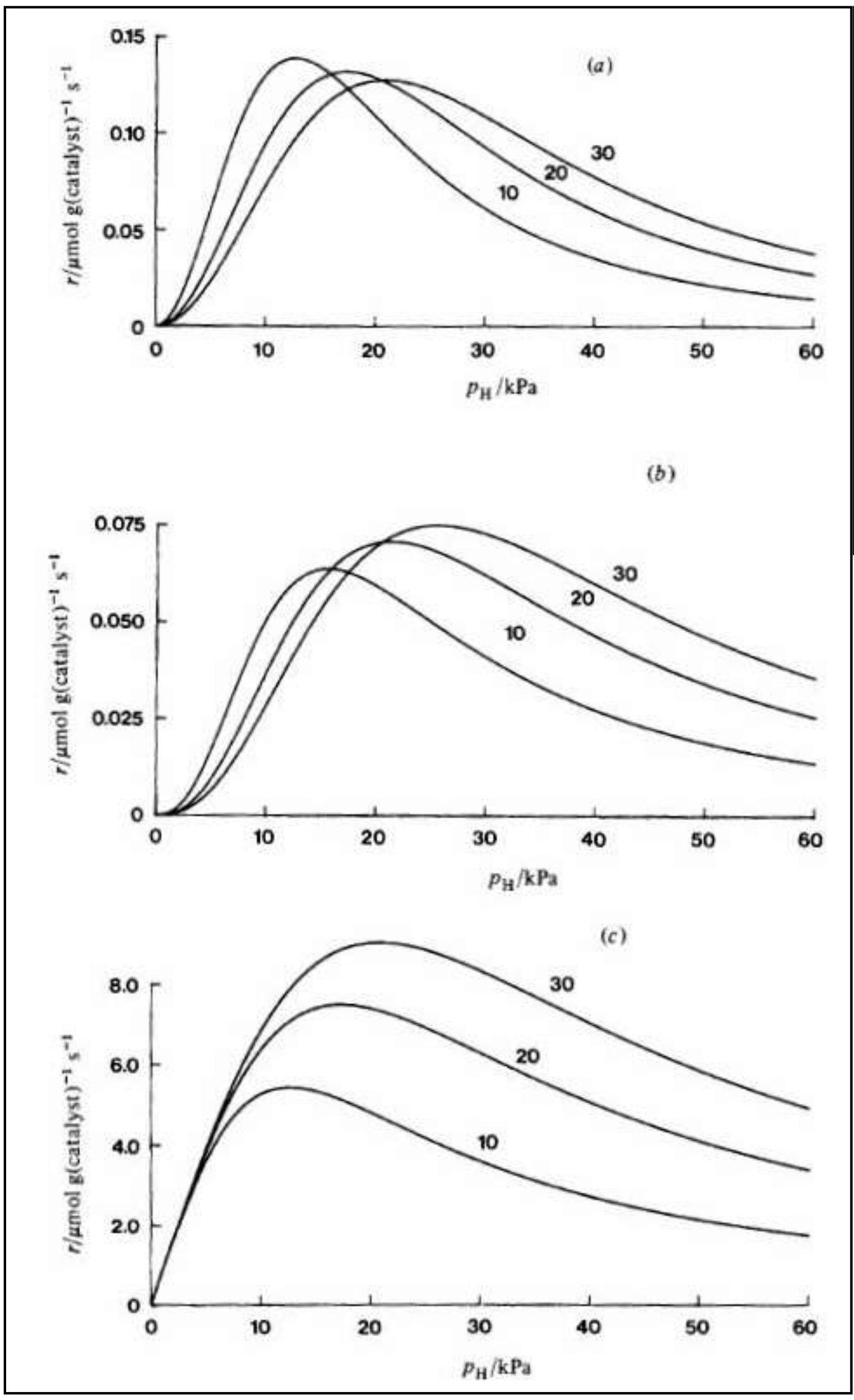

Simulated reaction rate plotted against partial pressure of hydrogen ($p_H$) at constant partial pressure of ethane ($p_E$= 10, 20 and 30 kPa), (a) B= $\theta_0$, (b) B= $\theta_H$ and (c) B= $p_H$.

The $H_2$(g) is responsible for the C-C rupture on Ni and Pd catalysts, because pattern (c) was obtained indisputably in experiments.

To investigate the mechanism of coke formation and cleaning, the reactions were carried out in a conventional glass reaction and gas handling system:

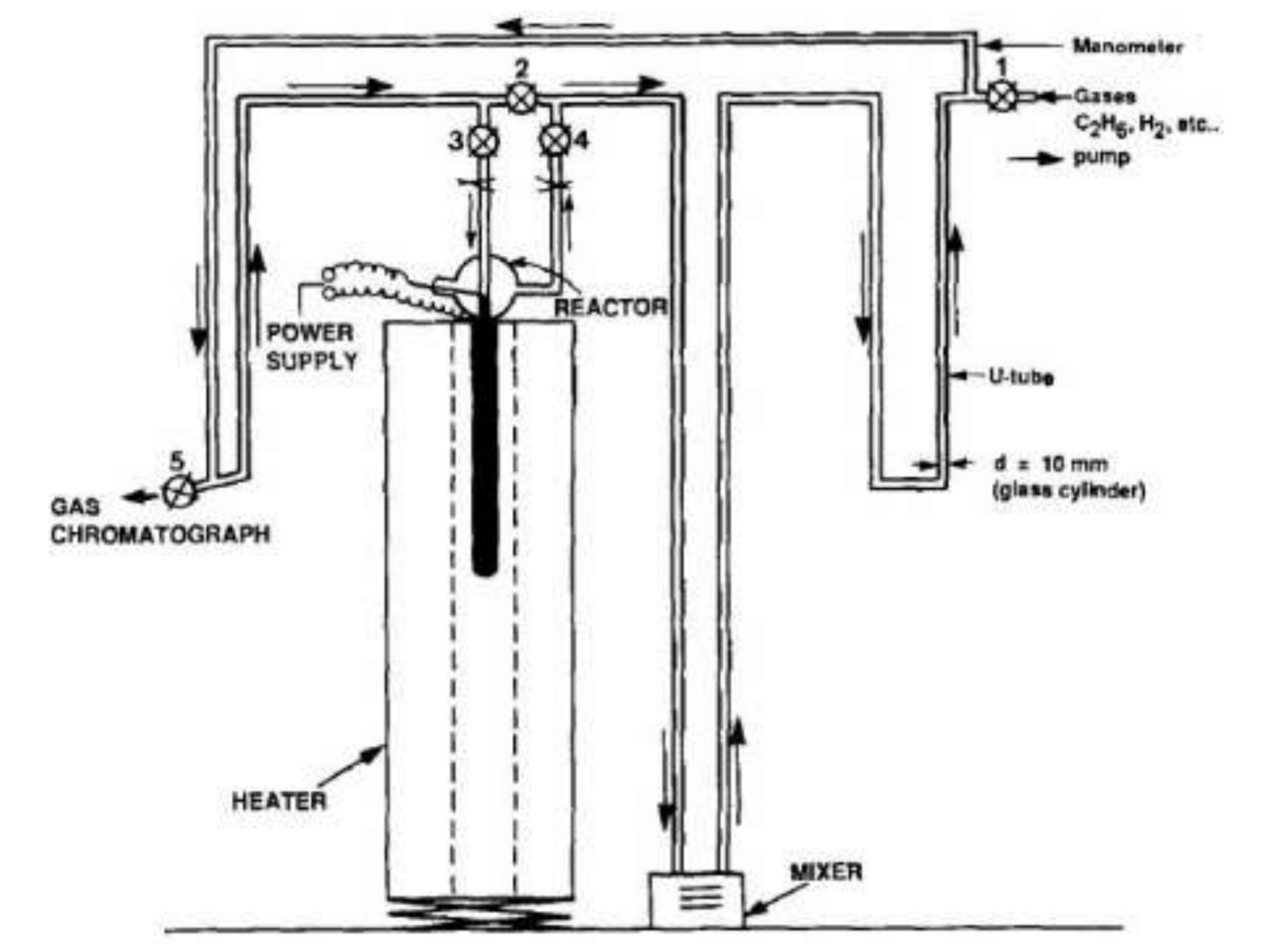

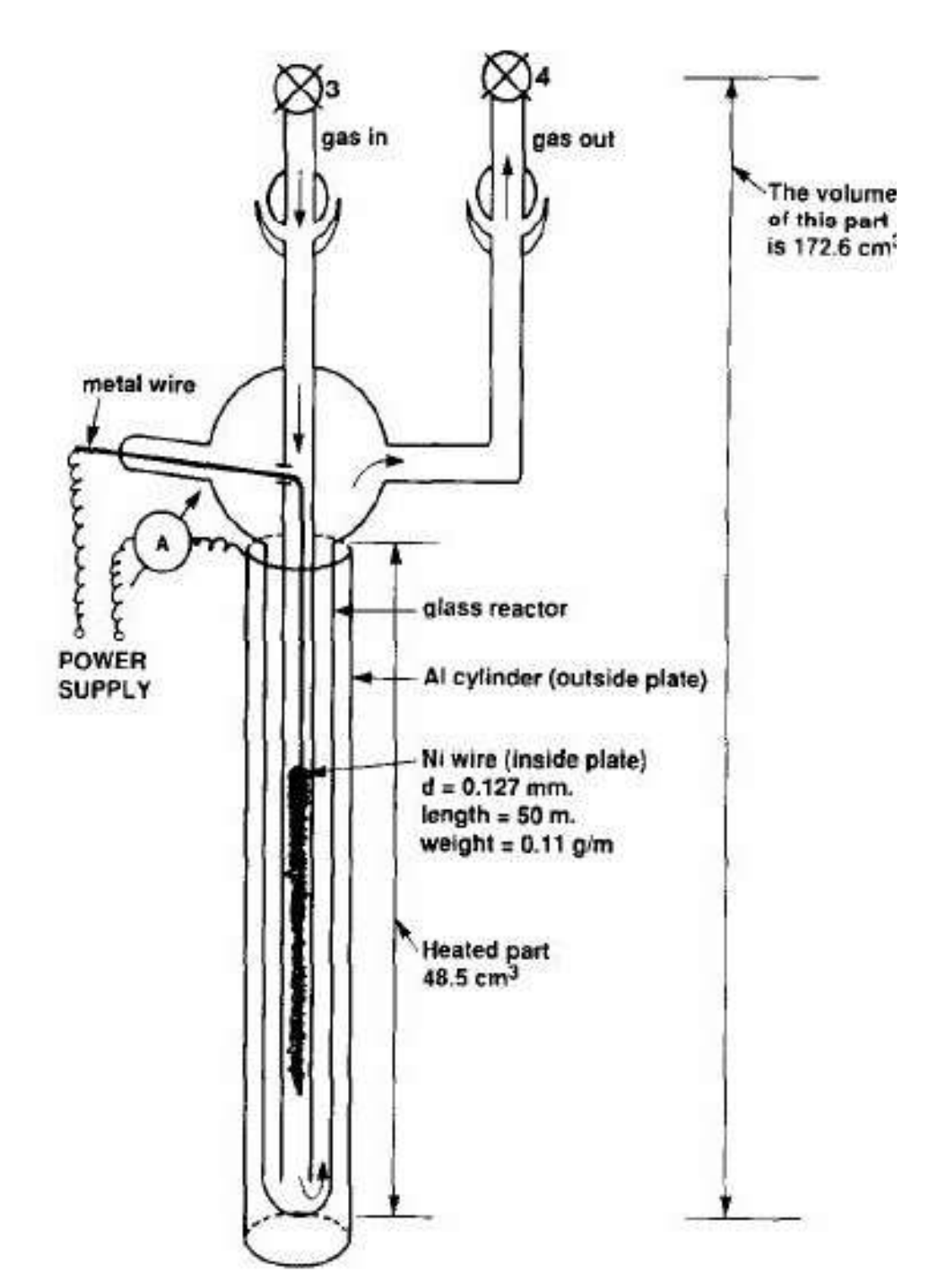

The active zone of the reactor system is plotted, showing essential dimensions, sizes and physical arrangement of electrodes in detail.

**Thesis-27-Theory: The problem of anomalous near surface diffusion: Diffusion to the surface in case of weak segregation of binary alloys**

Anomalously fast and slow diffusion, perpendicular to the surface in the near surface region of solids, have been well known in the literature (summarized, reviewed), and the possible reasons of these anomalies are discussed. A simple model has been solved to show the effect of driving force due to surface free enthalpy excess in case of weak segregation. An explicit formula has been derived for the time dependence of concentration of the segregant in the topmost layer.
= = = = = =

A szilárd testek felület közeli régiójában az anomálisan gyors és lassú felületre merőleges diffúzió jól ismert jelenség az irodalomból (összesítés és szemlézés a publikációimban), ezen anomáliák lehetséges okait elemeztem. Egy egyszerűsített modellt oldottam meg gyenge szegregáció esetére, hogy megmutassam a hajtóerő hatását, ami a felületi szabadentalpia többletnek tulajdonítható. Egy explicit formulát is levezettem a szegregátum koncentrációjának időfüggésére a legfelső rétegben.

**<u>Sandor Kristyan – J. Giber: Surface Science, 224 (1989) 476-488</u>**
**<u>J. Giber - Sandor Kristyan – P. Deak (REVIEW): Acta Phys. Hung., 65 (1989) 335-342</u>**
= = = = = =

Representative equations/tables/figures:

The segregation is weak, if the segregation number (the mol fraction in surface vs. bulk, $x^s/x_0$) is between 1 and 4.

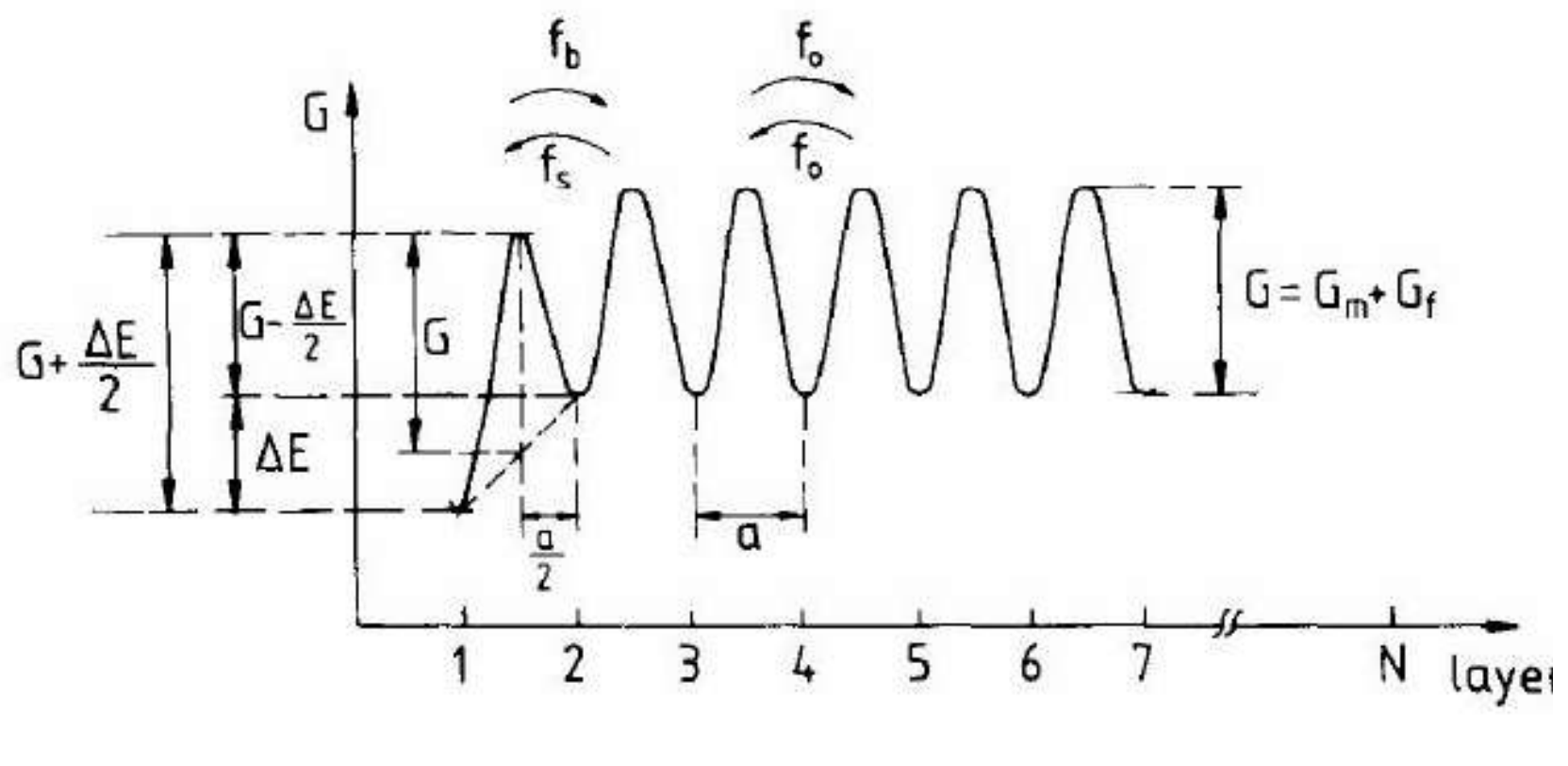

The periodic potential modified by the driving force of the segregation in the near-surface region a.: The convergence as the size (N layer bulk) of model tends to infinity,
the segregation number considered is 4.

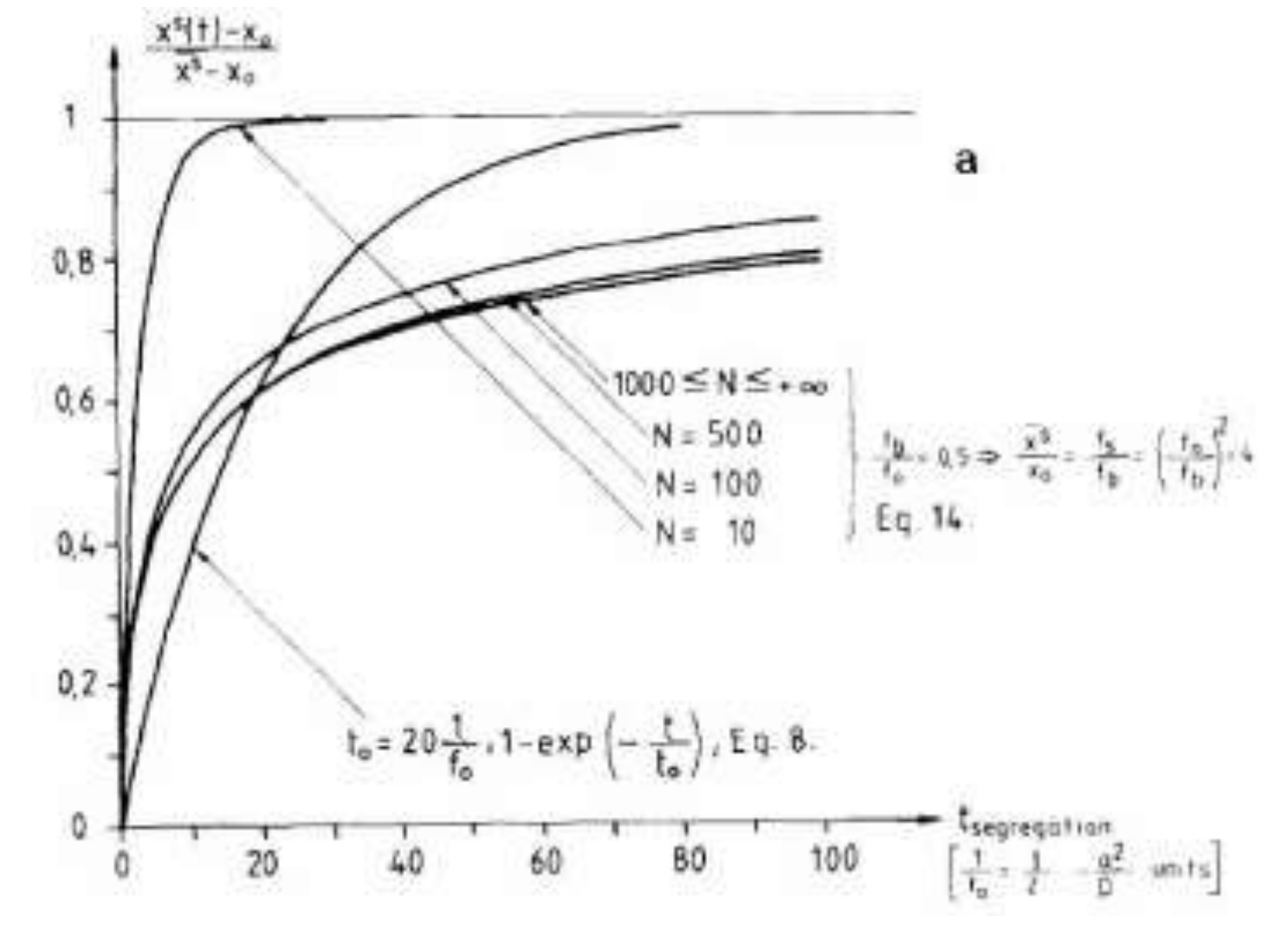

<u>⇒N=500 bulk layers can be considered as infinite bulk in this kind of calculations</u>.

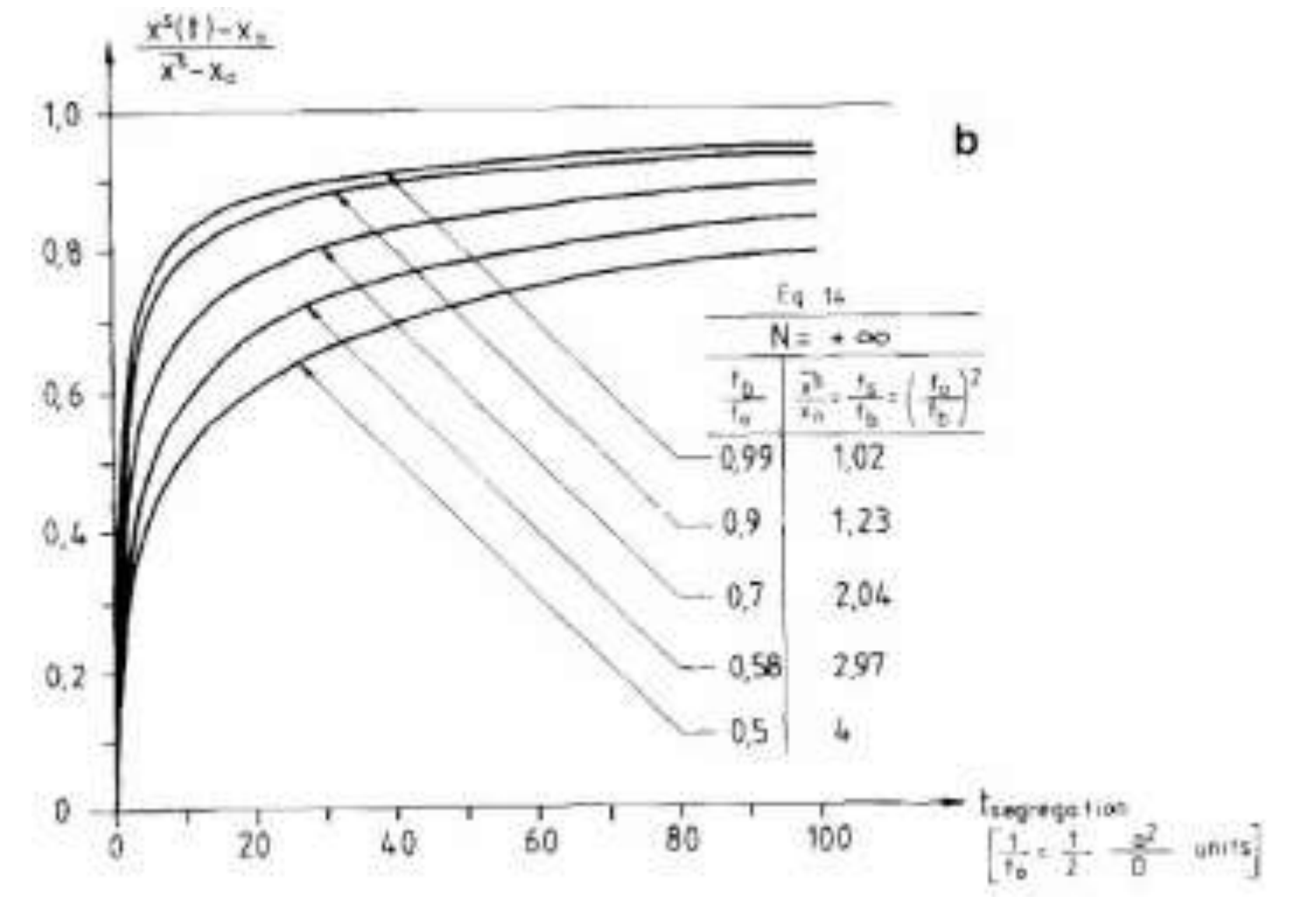

b.: Time evolution of segregant at the surface with diffrent (weak) segregation numbers

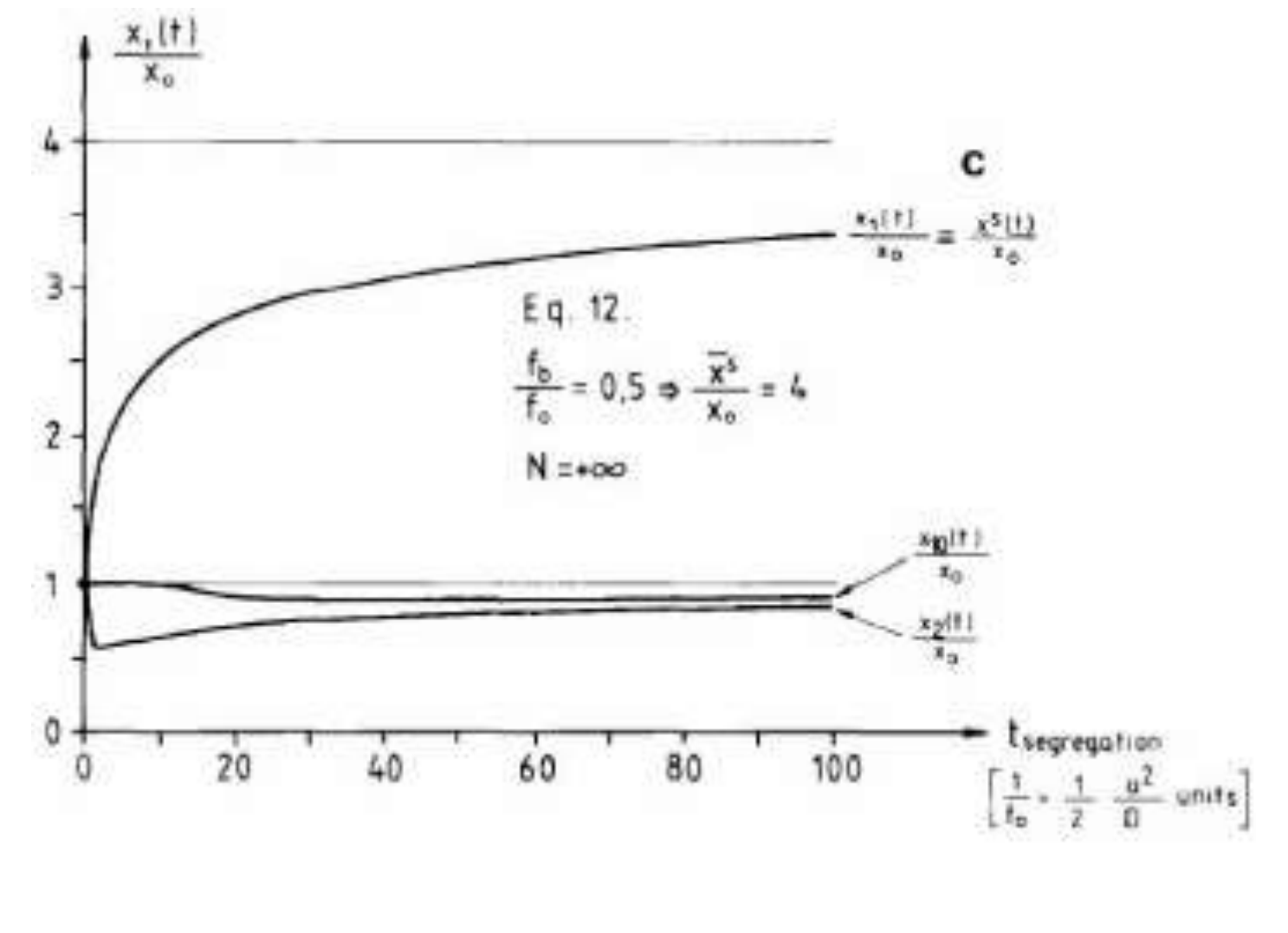

c.: Time evolution of segregant in the 1st (surface), 2nd and 10th layers with segregation number 4

As a personal remark, there was a "theoretical vacuum" in modeling the kinetics of this phenomenon. Naively, the segregant concentration in the first layer ($x^s \equiv x_1(t)$, $x^{s,eq} \equiv x_1(t=\infty)$ i.e. in the surface, and $x_0 \equiv x_\infty(t)$= constant in time, i.e. in the bulk far from surface) follows the first order kinetics pszeudo equation

$$(x^s(t) - x_0)/(x^{s,eq} - x_0) \approx 1 - \exp(-t/t_0).$$

Visibly, this looks plausible for $x_1(t)$ on figure c, but far not true for $x_i(t)$ with i=2,3,…N. However, solving the problem vith system of linear differential equations (the new feature in the papers filling up the "theoretical vacuum"), which is well known in numerical analysis along with eigensolver, but was not used in the chemical study of surface segregation of alloys, the time evolution of the segregant concentration of layer-i is

$$x^s(t) \equiv x_1(t) = a_1 + \Sigma_{i=2 \text{ to } N}\, a_i \exp(e_i t)$$

(with eigenvalues $e_1=0$ and $e_{i=2,\ldots,N} < 0$).

# 3.3 Answering chemical problems with regular computation chemistry

**Thesis-28-Application: Supporting the explanation of "specific behavior of p-aminothiophenol – silver sol system in their Ultra-Violet–Visible (UV–Visible) and Surface Enhanced Raman (SERS) spectra, as well as SERS behavior of substituted propenoic acids (used in heterogeneous catalytic asymmetric hydrogenation)" with quantum chemical calculations**

Surface Enhanced Raman Spectroscopy (SERS) behavior of silver sol (a typical SERS agent) was studied. I.: In the presence of different bifunctional thiols such as p-aminothiophenol, p-mercaptobenzoic acid, p-nitrothiophenol, p-aminothiophenol hydrochloride, and 2-mercaptoethylamine hydrochloride in diluted aqueous solution, our findings have been, 1., the p-aminothiophenol induced aggregation of citrate stabilized silver colloid originates from its electrostatic nature, 2., the azo-bridge formation cannot be the reason of the observed time dependent UV–Visible spectra, 3., certain amount of oxidized form of the probe molecule has to be present for the so-called b2-mode enhancement in the SERS spectrum of p-aminothiophenol, 4., the azo-bridge formation is responsible for the b2-mode enhancement in the SERS spectrum of p-aminothiophenol. II.: Strength and geometry of adsorption of substituted propenoic acids on silver surface were studied by SERS using silver sol. Two classes of phenylpropenoic acids studied have been distinguished, 1., the first class of propenoic acids (Series I: atropic acid, (E)-2,3-diphenylpropenoic acid, (E)-2-(2-methoxyphenyl)-3- phenylpropenoic acid, (E)-2,3-di-(4-methoxyphenyl)phenylpropenoic acid and (E)-2-(2-methoxyphenyl)-3-(4-fluorophenyl)propenoic acid) has shown strong charge transfer (CT) effect, bidentate carboxyl bonded species, the plane of the α-phenyl group is almost parallel to the silver surface, while the β-phenyl group is in tilted position depending on the type and the position of substituent(s) showing strong SERS enhanced bands, 2., the other class of propenoic acids (Series II: cinnamic acid, (E)-2-phenyl-3-(4-methoxyphenyl)propenoic acid) has shown weak electromagnetic (EM) enhancement, no significant carboxyl enhancement, the adsorbed species lie parallel to the surface. The two types of adsorption can be related to the dissociation ability of the carboxylic group, in the first case the carboxylic H dissociates, while in the second case it does not. In relation to computation chemistry, the enhanced bands strongly depend on how the molecules adsorb on the surface of Ag or Au colloids. (Interaction between the analyte and surface plasmons appearing in these metals is responsible for the enhancements in SERS spectra.) In case of composit molecules, for example, the differently substituted phenil-pyruvates or cinnamic acids, the knowledge of possible adsorption geometries is absolutely necessary to understand the SERS spectra, for which the knowledge of equilibrium geometries (as well as other properties, like partial charges, dissotiation abilities, etc.) is helpful.

= = = = = =

Az ezüst szol (egy tipikus SERS ágens) felületerősített Raman spektroszkópiai (SERS) viselkedését tanulmányoztuk. I.: Különböző bifunkcionális tiolok, mint p-aminotiofenol, p-merkaptobenzoesav, p-nitrotiofenol, p-aminotiofenol hidroklorid, és 2-merkaptoetilamin hidroklorid híg vizes oldatának jelenlétében azt találtuk, hogy, 1., a citrát stabilizált ezüst kolloidnak a p-aminotiofenol indukálta aggregációja az elektrosztatikus természetéből ered, 2., az azo-híd kialakulása nem lehet oka a megfigyelt idő függő UV–Látható spektrumoknak, 3., a próba molekula bizonyos mennyiségű oxidált formájának jelen kell lenni a p-aminotiofenol SERS spektrumában tapasztalható úgynevezett b2-mód erősítéshez, 4., az azo-híd kialakulása felelős a p-aminotiofenol SERS spektrumában tapasztalható b2-mód erősítésért. II.: A szubsztituált propénsavak adszorpciójának erősségét és geometriáját vizsgáltuk ezüst felületen SERS módszerrel ezüst szolt használva. A vizsgált fenilpropénsavak két osztályát különböztethettük meg, 1., a propénsavak egyik osztálya (Series I: atropasav (= 2-fenilakrilsav), (E)-2,3-difenilpropénsav, (E)-2-(2-metoxifenil)-3-fenilpropénsav, (E)-2,3-di-(4-metoxifenil)fenilpropénsav és (E)-2-(2-metoxifenil)-3-(4-fluorofenil)propénsav) erős töltés transzfer (charge transfer (CT)) effektust mutat, kétfogú (bidentate) karboxil mentén kötött molekulák, az α-fenil csoport síkja majdnem párhuzamos az ezüst felülettel, míg a β-fenil csoport döntött pozícióban van attól függően, hogy milyen típusú és pozíciójú a szubsztituens, ezek erős SERS erősített sávokat mutatnak, 2., a propénsavak másik osztálya (Series II: fahéjsav, (E)-2-fenil-3-(4-metoxifenil)propénsav) gyenge elektromágneses (electromagnetic (EM)) erősítést mutat, nincs szignifikáns karboxil erősítés, az adszorbeált molekulák párhuzamosan helyezkednek el a felületen. A két típusú adszorpció a karboxil csoport disszociációs képességére vezethető vissza, az első esetben a karboxilos H disszociál, míg a második esetben nem. A számítógépes elméleti kémia tekintetében, a felületerősített Raman spektrumokban az erősített sávok nagymértékben függnek attól, hogy a molekulák hogyan adszorbeálódnak az erősítést előidéző Ag vagy Au kolloid felületére. Összetett molekuláknál, mint pl. a különböző módon szubsztituált fenil-piruvátok vagy fahéjsav származékok, ahhoz, hogy megértsük a SERS spektrumokat, mindenképpen ismernünk kell a lehetséges adszorpciós geometriát, amiben segít az egyensúlyi geometria (valamint egyéb tulajdonságok, mint parciális töltések, disszociációs képességek, stb.) ismerete.

**T. Firkala, E. Tálas, J. Mihály, T. Imre, Sándor Kristyán:**
**Journal of Colloid and Interface Sciences, 410 (2013) 59–66**
**T. Firkala, E. Tálas, Sándor Kristyán, Gy. Szöllősi, E. Drotár, J. Mink, J. Mihály:**
**Journal Raman Spectroscopy, 46 (2015) 1102–1109**

= = = = = =

Representative equations/tables/figures:

Pictorial representation of Raman effects:

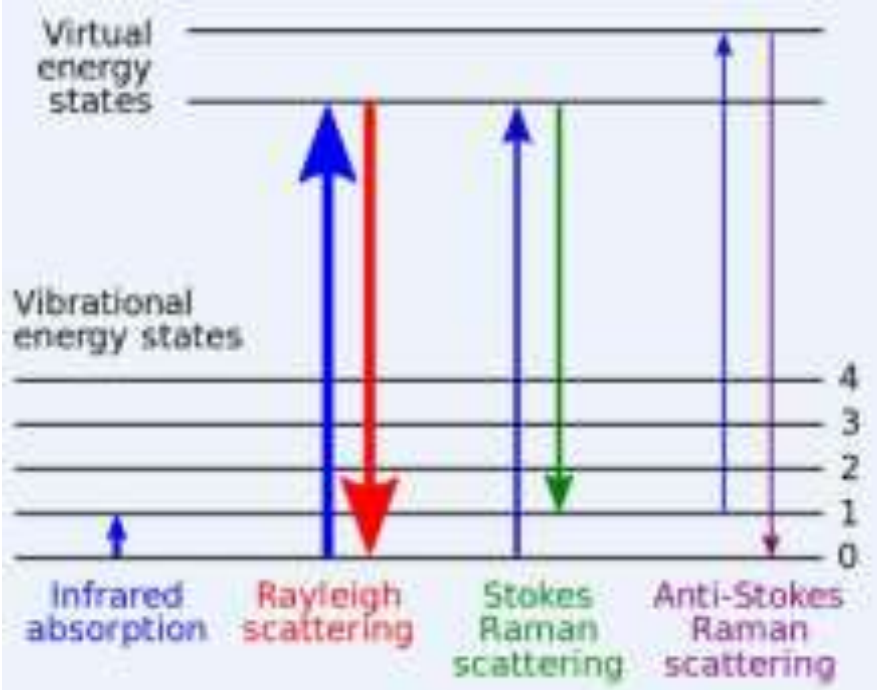

(In many quantum processes a virtual state is an intermediate state, in a multi-step process that mediates otherwise forbidden transitions. Since virtual states are not eigenfunctions of anything, normal parameters such as occupation, energy and lifetime need to be qualified. No measurement of a system will show one to be occupied, but they still have lifetimes derived from uncertainty relations. While each virtual state has an associated energy, no direct measurement of its energy is possible.)

The probability of inelastic scattering in Raman effects is very low, roughly only one photon scatters inelastically among $10^8$, however, the SERS enhancement in Raman scattering intensities can be up to 6 orders of magnitude. The complete enhancement process in surface enhanced spectroscopy (SES) particularly in SERS includes the electromagnetic mechanism (EM) effects and the chemical mechanism (CM, particularly the charge transfer (CT)):

Electromagnetic mechanism (EM): If the surface under consideration is roughened, then the electrodynamics of the irradiated surface becomes much more interesting. The key result is that surface plasmons can now be excited by electromagnetic radiation, resulting in enhanced electromagnetic fields close to the surface. (Although collective excitations of the conduction electrons known as surface plasmons exist for Ag at frequencies of roughly 3.5 eV, they cannot be excited by the electromagnetic field when irradiating a flat surface, as momentum cannot be conserved in the excitation process. If the surface under consideration is roughened, then the situation in the electrodynamics of the irradiated surface is different.) In Raman scattering, the intensity depends on the square of the incident field strength, and as a result, the intensity is enhanced relative to what it would be in the absence of the surface. The Raman emitted field may also be enhanced, though generally by a different amount than the incident field, since the frequency is different. Also the field is spatially different as it arises from an oscillating dipole located at the position of the emitting molecule. The overall enhancement associated with the incident and emitted fields is what is considered to be the EM contribution to SERS.

Chemical mechanisms (CM): It refers both to enhancements that arise from interactions between molecule and surface that require orbital overlap between the molecule and metal wavefunctions (CT), and to those interactions that do not require overlap. The most commonly considered interaction that requires overlap occurs when charge transfer between surface and adsorbate leads to the formation of excited states that serve as resonant intermediates in Raman scattering. Interactions that do not require overlap arise from electromagnetic coupling between the vibrating molecule and the metal.

In SERS phenomenon not really the "colloid" has to be emphasized, this part is rather called as "coin metals", because SERS can be observed on roughened surface of Ag, Au or Cu also, the interaction with surface plasmon is what fundamental. Generally, the metal is called SERS substrate, and for example, Ag or Au foil is roughened with electrochemical device, or one has Ag or Au nano particles adsorbed onto the surface of these foils before investigating the enhancement, (these depend on the geometry of the equipment).

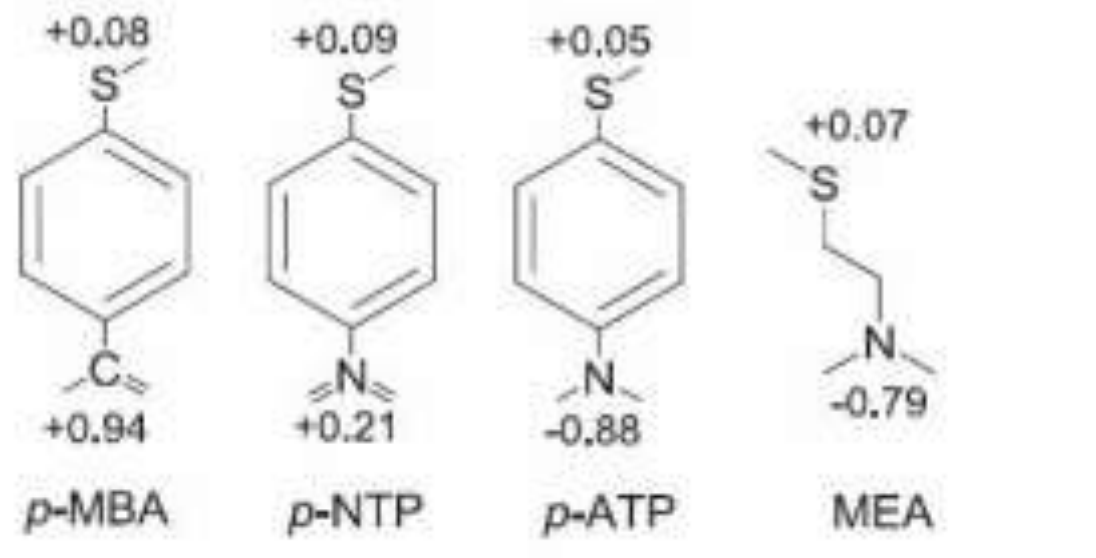

Bifunctional thiols and their partial charges on functional groups.

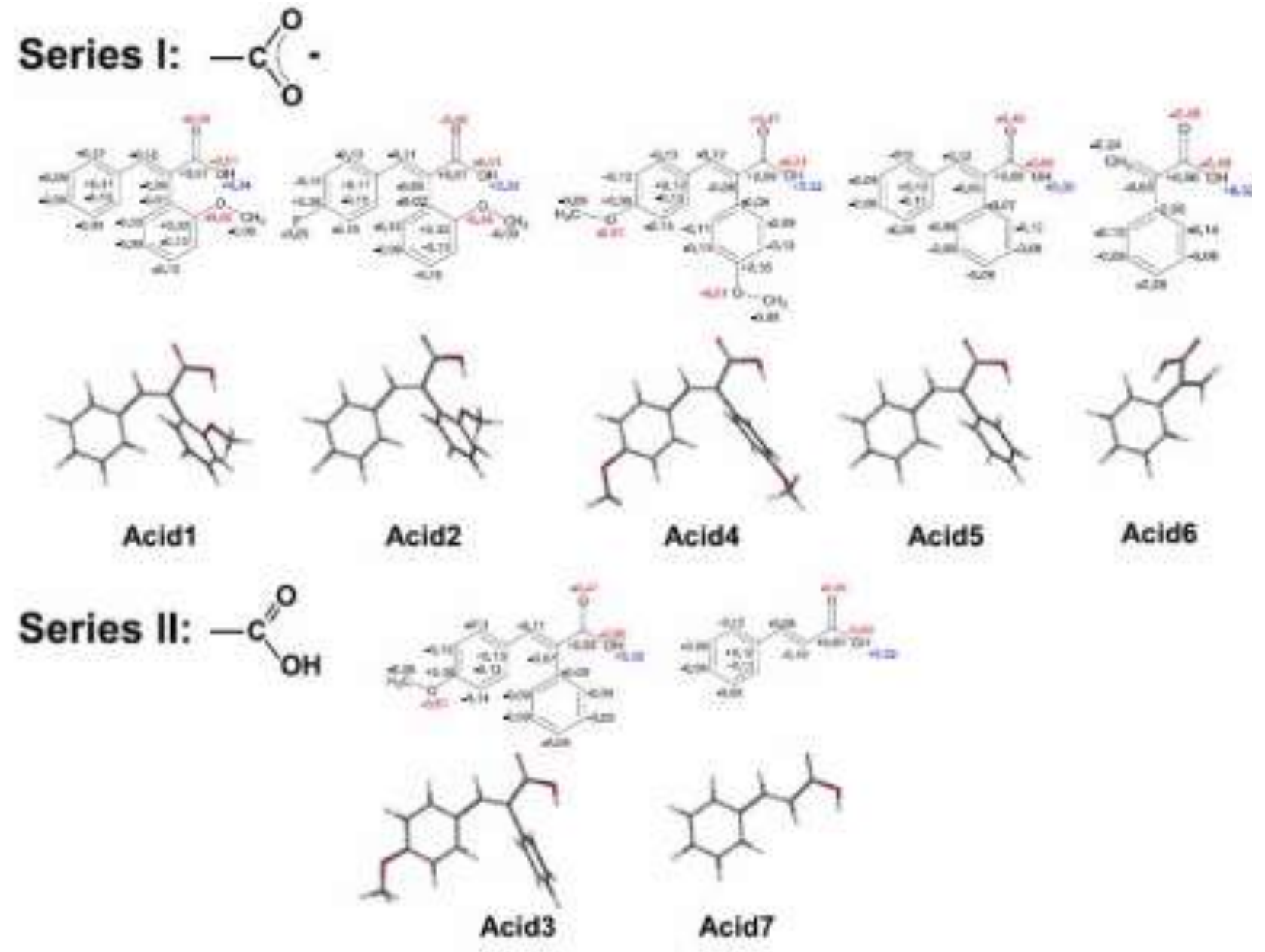

Mulliken charges and equilibrium geometries of the studied propenoic acids calculated at B3LYP/6-31 + G(d,p)[38] level of theory.

| No | Proton affinity, kcal/mol | Mulliken charges | | | |
|---|---|---|---|---|---|
| | | H8 | C7 | C8 | C8–C7 difference |
| Acid1 | −354.0 | +0.34 | −0.12 | −0.06 | 0.06 |
| Acid2 | −352.5 | +0.34 | −0.11 | −0.06 | 0.05 |
| Acid3 | −354.3 | +0.32 | −0.11 | −0.07 | 0.04 |
| Acid4 | −355.4 | +0.32 | −0.12 | −0.06 | 0.05 |
| Acid5 | −352.6 | +0.32 | −0.12 | −0.05 | 0.07 |
| Acid6 | −353.6 | +0.32 | −0.24 | +0.03 | 0.27 |
| Acid7 | −356.2 | +0.32 | −0.08 | −0.19 | −0.11 |

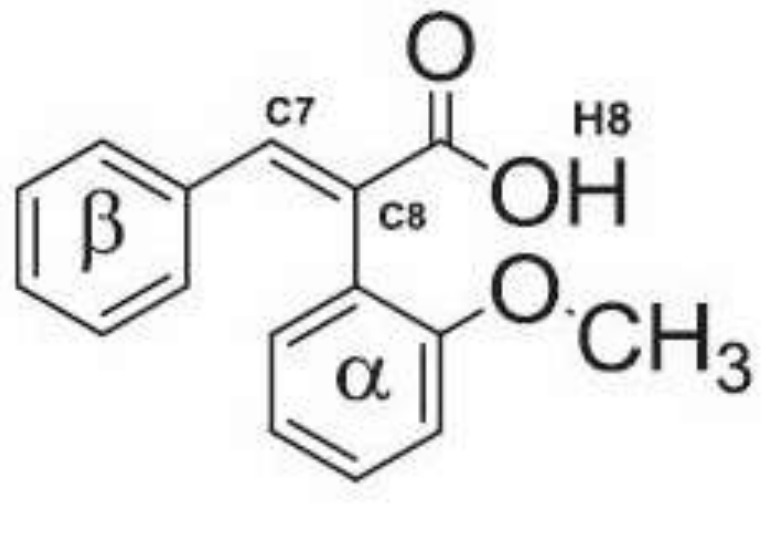

(Above: The numbering is represented on Acid1 on left.)

**Thesis-29-Application: Molecular interactions between DPPC and PMLA via modeling its measured infrared spectra**

In the problem of thermotropic and structural effects of low molecular weight poly(malic acid) (PMLA ratios related to lipid were 1 and 5 wt%) on fully hydrated multilamellar dipalmitoylphosphatidylcholine (DPPC 20 wt%) / water system, a detailed experimental investigation was done by using differential scanning calorimetry (DSC), small-angle X-ray scattering (SAXS) and transmission electron microscopy combined freeze-fracture procedure (FF-TEM). My experimentalist collages (leaded by Dr. Attila Bóta) have found that PMLA derivatives changed significantly the thermal behavior of DPPC and caused drastic loss in correlation of lamellae in the three characteristic states (e.g. in gel, rippled gel and liquid crystalline phases) along that the structural behaviors on atomic level were supported by FTIR spectroscopy. The molecular interactions between DPPC and PMLA via modeling its measured infrared spectra were simulated and their feature was interpreted. It was found that poly(malic acid) is attaching to the headgroups of the phospholipids through hydrogen bonds between the free hydroxil groups of PMLA and phosphodiester groups of DPPC.
= = = = = =

A kis molekulasúlyú poli-almasav (PMLA arány a lipid részre vonatkoztatva 1 ill. 5 wt%) termotropikus és strukturális hatásának probléma körében a teljesen hidratált multilamellaris dipalmitoil-foszfatidil-kolin (DPPC 20 wt%) / víz rendszer esetében, egy részletes kísérleti vizsgálatot végeztünk a differenciális pásztázó kaloriméter (DSC), kis szögű Röntgen diffrakció (SAXS) és a fagyasztva törés technikával kombinált transzmissziós elektron mikroszkópia (FF-TEM) felhasználásával. Kísérletekkel foglalkozó kollegáim (Dr. Bóta Attila vezetésével) azt találták, hogy a PMLA szignifikánsan megváltoztatja a DPPC termikus viselkedését és drasztikus csökkenést okoz a lamellák kölcsönhatásában a három karakterisztikus fázisban (gél, „rippled gel” és folyadék kristályos fázisban) melyet még megtámogattak a strukturális viselkedés atomi szintű vizsgálatával FTIR spektroszkópia segítségével. A DPPC és PMLA közti molekuláris kölcsönhatások modellezését végeztem, mely a mért infravörös spektrumok szimulálását és a molekuláris kölcsönhatások magyarázatát jelentette. Azt találtam, hogy a poli-almasav a foszfolipid fejcsoportjához kapcsolódik hidrogén kötések mentén a PMLA szabad hidroxil csoportjai és a DPPC foszfodiészter csoportjai között.

**Szilvia Berényi, Judith Mihály, Sandor Kristyan, Lívia Naszályi Nagy, Judit Telegdi, Attila Bóta: Biochimica et Biophysica Acta (BBA) – Biomembranes, 1828 (2013) 661-669**
= = = = = =

Representative equations/tables/figures:

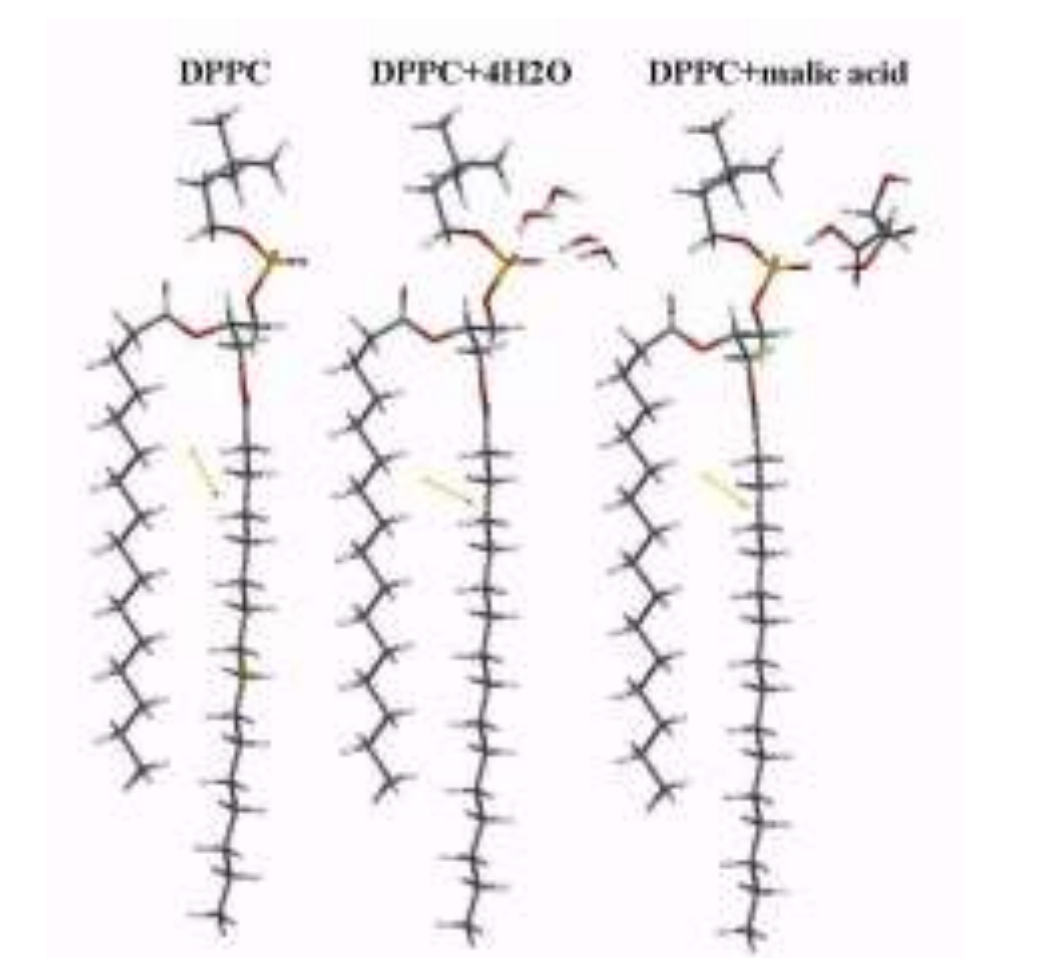

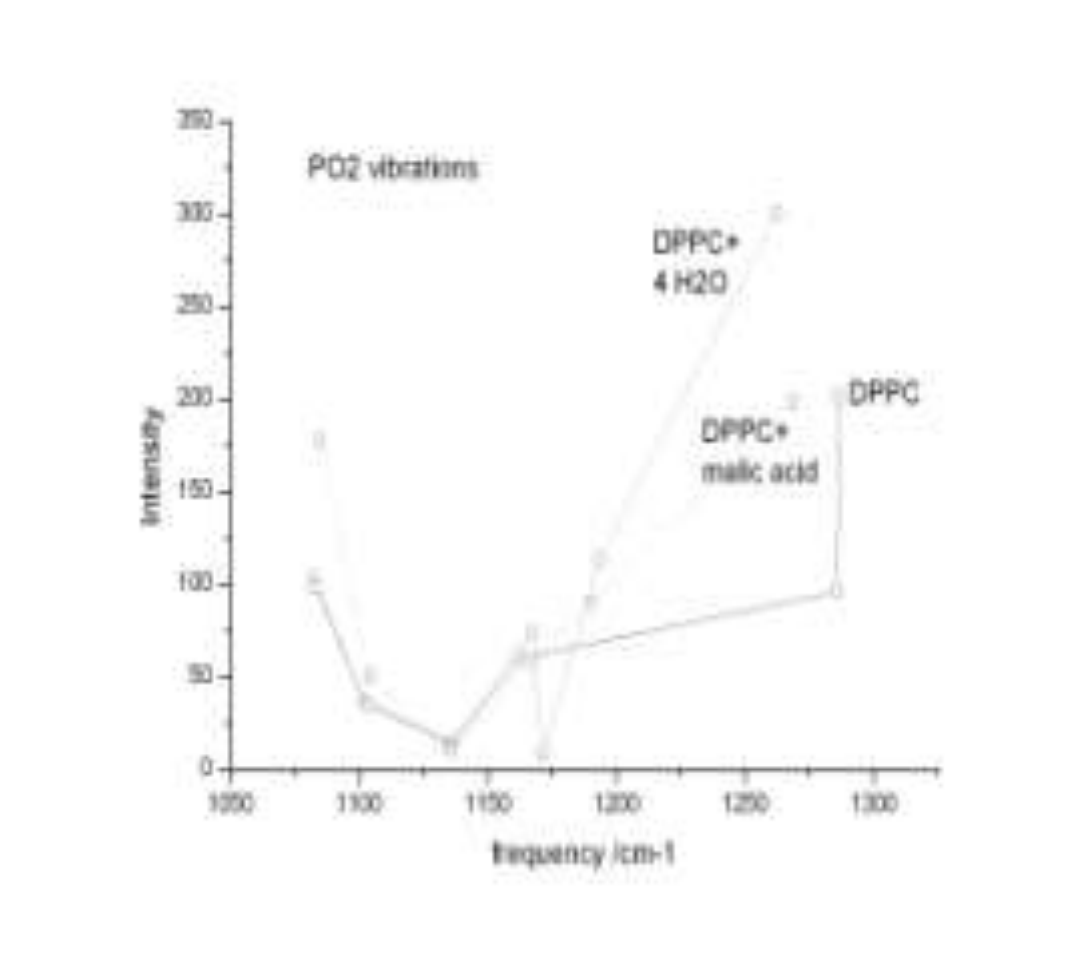

After drying the DPPC/malic-acid/water system to DPPC/malic-acid, the interesting fact is that the dehydrated $PO_2^-$ vibration shows up at about the same position as the hydrated $PO_2^-$ (1222 $cm^{-1}$), while a regular water- and malic-acid-free DPPC (dipalmitoyl-phosphatidyl-choline) system owns this peak at 1243 $cm^{-1}$. The exprmental values of $PO_2^-$ vibrations of hydrated/dehydrated DPPC at 1222/1243 (21 $cm^{-1}$ shift) were simulated providing the calculated (no scaling) values 1263/1287 (24 $cm^{-1}$ shift) and yielding the conclusion: The calculated DPPC system (dehydrated, $PO_2^-$ peak at 1287 $cm^{-1}$) separates from the DPPC+4H2O (hydrated, $PO_2^-$ peak at 1263 $cm^{-1}$) and DPPC+malic acid ($PO_2^-$ peak at 1269 $cm^{-1}$) system with about the same frequency shift. The negative electron lobes of oxygen atoms in $PO_2^-$ group in DPPC and the partially positive hydrogen atoms of two OH groups and hydrogen atom of CH in the malic acid inter-molecularly affect each other (quenching the vibration of $PO_2^-$) in the same way as the hydrogen atoms in water in the DPPC+4H2O system.

**Thesis-30-Application: Predicting the order of reactivity in catalytic hydrogenation of R–CO–X with initial partial charges**

Hydrogenation of >C=O group in various R-CO-X carbonyl compounds was compared (on 1-1 w% Re-Pt/$Al_2O_3$ catalyst; R= aliphatic, saturated and unsaturated group up to 16 carbon atoms or phenyl; X= H, R', OR', OH, $NH_2$, NR' and NR'R"). The order of reactivity (aldehydes > ketones > carboxamides > carboxylic acids > esters) was demonstrated based on catalytic experiments and computational chemistry (molecular mechanics and *ab initio* calculations). Latter was based on computing various partial charges (Mulliken, NPA, electrostatic), bond order (Löwdin) and band gap (HOMO-LUMO) on the carbonyl group as a function of R and X. Our calculations have indicated that even a simple Mulliken partial charge analysis (an instant side result of the common Hartree-Fock calculation) on the carbonyl group in gas phase can qualitatively predict its order of reactivity in the complex catalytic hydrogenation. Our calculations have also supported the experimental finding such that the effect of the chain size and length of group R on the reactivity is much smaller than that of group X in both, aliphatic and aromatic carbonyl compounds. However, the reactivity of aromatic carbonyl compounds (R = phenyl), as expected, is somewhat higher than that of aliphatic.

= = = = = =

A >C=O csoportok hidrogénezésének összehasonlítását végeztük különböző R-CO-X karbonil vegyületek esetében (1-1 w% Re-Pt/$Al_2O_3$ katalizátoron; R= alifás, telített és telítetlen csoport 16 szénatomig vagy fenil; X= H, R', OR', OH, $NH_2$, NR' és NR'R"). A reaktivitás sorrendje (aldehidek > ketonok > karboxilamidok > karbonsavak > észterek) a katalitikus kísérletek és a számításos kémia (molekuláris mechanika és *ab initio* számítások) alapján lett bemutatva, párhuzamba állítva. Az utóbbi a karbonil csoportok különböző parciális töltései (Mulliken, NPA, elektrosztatikus), kötés rendjei (Löwdin) és tiltott sávjai (HOMO-LUMO), mint az R és X funkciós csoportok függvénye, számításán alapult. Számításaink megmutatták, hogy egy egyszerű Mulliken parciális töltés analízis az erőtér mentes (gáz fázisú) karbonil csoporton (ami egy azonnali ’mellékterméke’ a közönséges Hartree-Fock számításnak) kvalitatíve meg tudja jósolni a reaktivitás sorrendjét ezekben az összetett katalitikus hdrogenezésekben. Számításaink szintén alátámasztották a kísérleti eredményeket, miszerint az R csoport hatása a lánchossz és alak szerint a reaktivitásban sokkal kisebb mint az X csoporté mindkét, alifás és aromás karbonil származékok esetén. Azonban, az aromás vegyületek reaktivitása (R = fenil), ahogy elvárható volt, valamivel magasabb mint az alifásoké.

**Sandor Kristyan, Sandor Gőbölös:**
**Computational and Theoretical Chemistry, 975 (2011) 24–30**

= = = = = =

Representative equations/tables/figures:

Experimental order of reactivity toward catalytic hydrogenation (rapid → slow):

R-CO-H > R-CO-R' >> R-CO-$NH_2$, R-CO-NHR', R-CO-NR'R" > R-COOR' ~ R-COOH.

The definition of 4 digit ID (identification) number follows the reactivity: The first two digits identify the X function group as 01 for H, 02 for –$CH_3$, 03 for –$C_2H_5$, … 11 for OH. The last two digits identify the structure R alkyl as 01 for methyl, 02 for $C_2H_5$-, 03 for $CH_2$=CH-, … 11 for n$C_{16}H_{33}$-.

Experimental activity of R-CO-X decreases downward with an increasing ID number in the table. (The effect of R is much smaller, this is the reason we identify R only in the last two digits.)

| ID number | $CH_3$-CO-X | | ID number | R-CO-H |
|---|---|---|---|---|
| 0101 | $CH_3$-CO-H (aldehydes) | | 0101 | $CH_3$-CO-H |
| 0201 | $CH_3$-CO-$CH_3$ | | 0102 | $C_2H_5$-CO-H |
| 0301 | $CH_3$-CO-$C_2H_5$ | | 0103 | $CH_2$=CH-CO-H |
| 0401 | $CH_3$-CO-O$CH_3$ | | 0104 | $CH_3$CH=CH-CO-H |
| 0501 | $CH_3$-CO-O$C_2H_5$ | | 0105 | i-propyl-CO-H |
| 0601 | $CH_3$-CO-$NH_2$ | | 0106 | i-butyl-CO-H |
| 0701 | $CH_3$-CO-NH$CH_3$ | | 0107 | $C_6H_5$-CO-H |
| 0801 | $CH_3$-CO-NH$C_2H_5$ | | 0108 | n$C_4H_9$-CO-H |
| 0901 | $CH_3$-CO-N$(CH_3)_2$ | | 0109 | n$C_8H_{17}$-CO-H |
| 1001 | $CH_3$-CO-N($CH_3$)($C_2H_5$) | | 0110 | n$C_{12}H_{25}$-CO-H |
| 1101 | $CH_3$-CO-OH | | 0111 | n$C_{16}H_{33}$-CO-H |

R-CO-X + $H_2$/catalyst ⟶ R-CH(OH)-X (1)

R-CO-X + $H_2$/catalyst ⟶ R-$CH_2$-X + $H_2O$ (2)

R-CO-X + $H_2$/catalyst ⟶ R-$CH_2$-OH + HX (3)

reaction 1: aldehydes, ketones,

reaction 2: carboxamides,

reaction 3 : carboxylic acids, esters.

Rate determining step is the attack of catalytically activated H• radical or hydride ion ($H^-$) on the partially positive carbon atom of the >C=O group. We have made a review on experimental data from literature and our measurements along with calculations, e.g.: Mulliken partial charge (HF-SCF/6-31G*) on >C=O („half electron" difference):

≈ 0.3 on C of very reactive aldehydes (ID= 0101-0111),

≈ 0.8 on C of less reactive esters, carboxamides and carboxylic acids (ID= 0401-1111), (points with the same R group are connected).

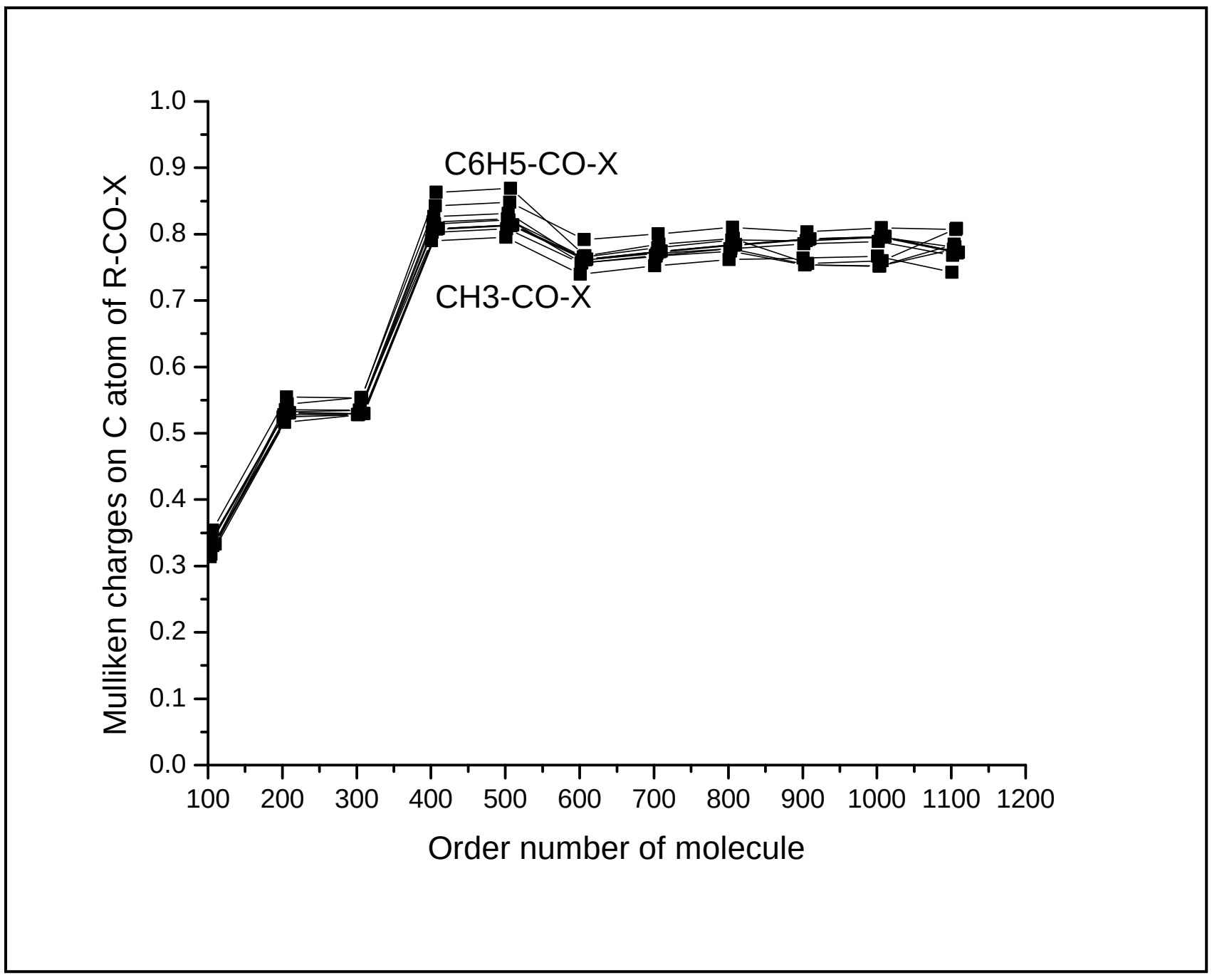

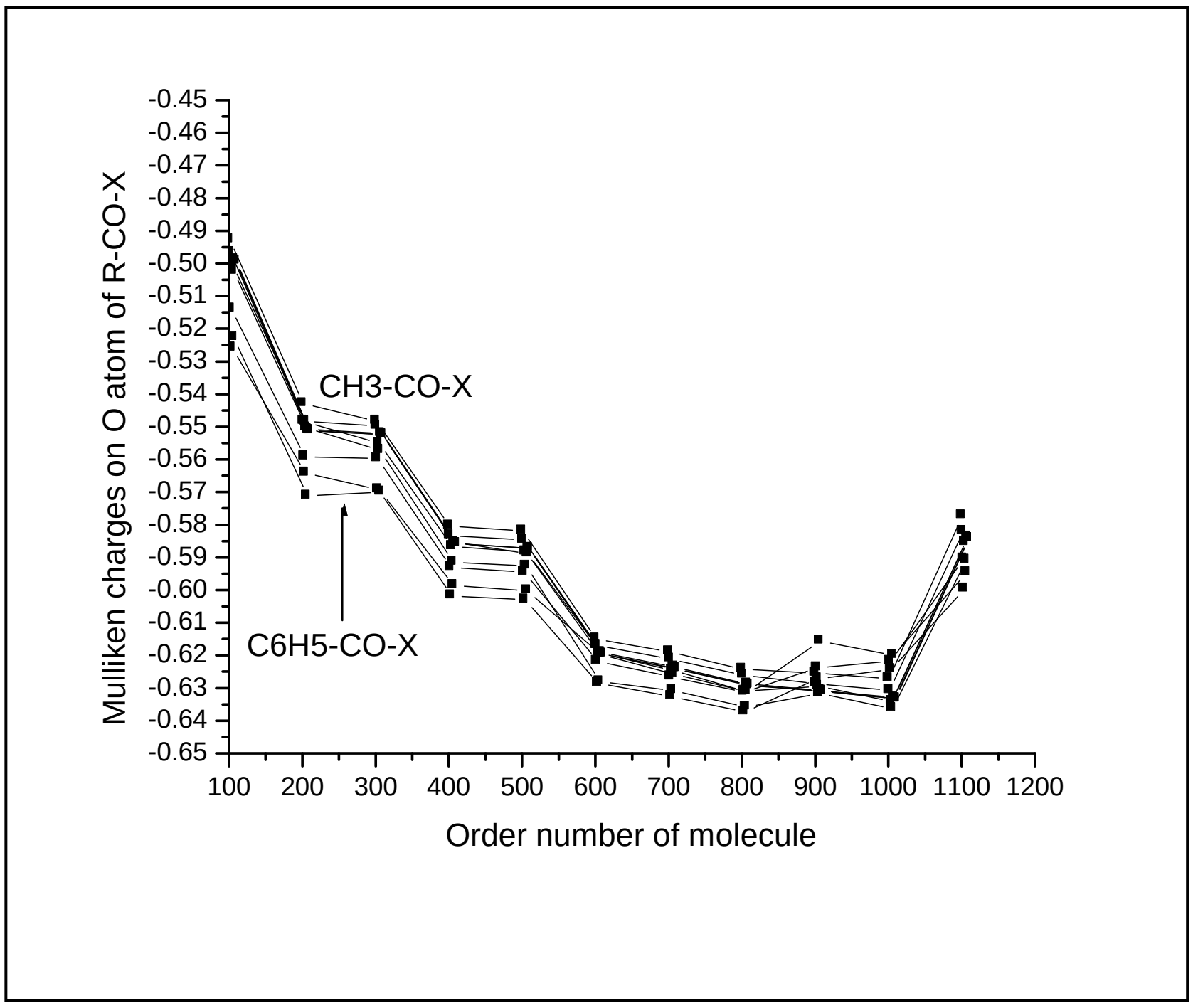

**Thesis-31-Application: Mechanisms in catalytic enantioselective hydrogenation of pyruvates come from internal rotations**

Performing detailed conformation analysis with using molecular mechanics building techniques and density functional theory methods, the main orientations of methyl pyruvate in the force field of cinchona alkaloids have been fully mapped and analyzed. Beside the known „open" and „closed" forms of cinchonidine, its „trans" conformation was also analyzed, as well as the potential surface of iso-cinchonines. In this way the origin of different mechanistic models by different authors for the hydrogenation of activated ketones on cinchona alkaloid modified platinum have been revealed: All models can be originated from the configurational properties of the internal rotation of the two-ring system in cinchona alkaloids (particularly cinchonidine, α and β-iso-cinchonine) with and without methyl pyruvate adduct formation.
= = = = = =

Részletes konformáció analízissel, felhasználva a molekula mechanika molekulaépítő módszerét és a sűrűség funkcionál elmélet módszereit, teljesen letérképeztem és analizáltam a metil piruvát fő orientációit a cinkona alkaloidok erőterében. Az ismert „open/nyílt" és „closed/zárt" cinkonidin konformációk mellett a „trans/transz" konformációt is analizáltam, valamint az izo-cinkoninok potenciál felületének esetét is. Így jutottam el a különböző szerzők által (más meggondolásból) javasolt különböző mechanisztikus modellek eddig rejtett közös eredetéhez, amiket az aktivált ketonok cinkonidin alkaloiddal módosított platina felületen történő katalitikus hidrogénezésére javasoltak: Az összes modell eredete a cinkona alkaloidokban a két-gyűrűs rendszer belső forgásának konfigurációs tulajdonságaiból származtatható (speciálisan a cinkonidin, α és β-izo-cinkonin esetében) metil piruvát addukt formálódással és anélkül.

<u>**Sandor Kristyan: Journal of Physical Chemistry C, 113 (2009) 21700–21712**</u>
= = = = = =

Representative equations/tables/figures:

The asymmetric hydrogenation of α-ketoesters and related ketones on platinum surface modified with cinchona alkaloids was described first by Orito et al. (1979):

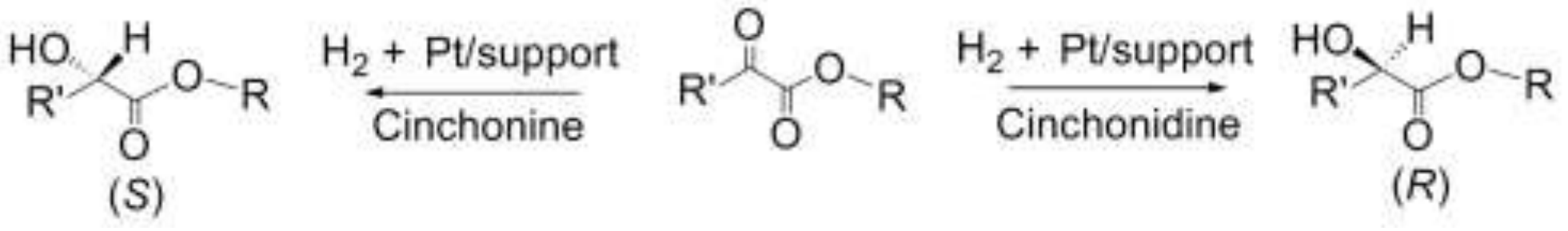

Major optical products (R or S) from the hydrogenation of methyl- or ethyl pyruvate:

| Modifier | Major product in acetic acid | Major product in toluene |
|---|---|---|
| Cinchonidine (CD) | R | R |
| Cinchonine (CN) | S | S |
| α-iso-cinchonine (αICN) | S | S |
| β-iso-cinchonine (βICN) | S | R |

Some energy relations in the complexation between CD and trans methyl-pyruvate (MP). All values are from B3LYP/6-31G* calculation and in kcal/mol, as well as in every line the lower value indicates the more stable system, while the opposit holds for values in parentheses. Every line lists an energy level diagram, wherein the column „Closed” is chosen as zero level.

| Molecule | Closed | Open | Trans |
|---|---|---|---|
| 1.: Relative total energy of CD | 0.0 | -2.33 | 3.57 |
| 2.: Relative total energy of complex {CD … MP} | 0.0 | -3.37 | 3.69 |
| 3.: Relative bound energy of complexation in {CD … MP} at 0 K, and absolute values in parentheses | 0.0 (6.71) | -0.86 (7.57) | -2.61 (9.32) |

Cinchonidine (CD) in closed form, showing the definition of the dihedral angles α, β. The α ≡ C4’C9C8N1 dihedral angle follows the rotation of quinuclidine with respect to any of the two sides of fixed quinoline, and β ≡ C10’C4’C9C8 dihedral angle follows which side of the quinoline the quinuclidine is positioned. (Particularly, the α= 60 deg value is shown, recall the minimum called “closed”.)

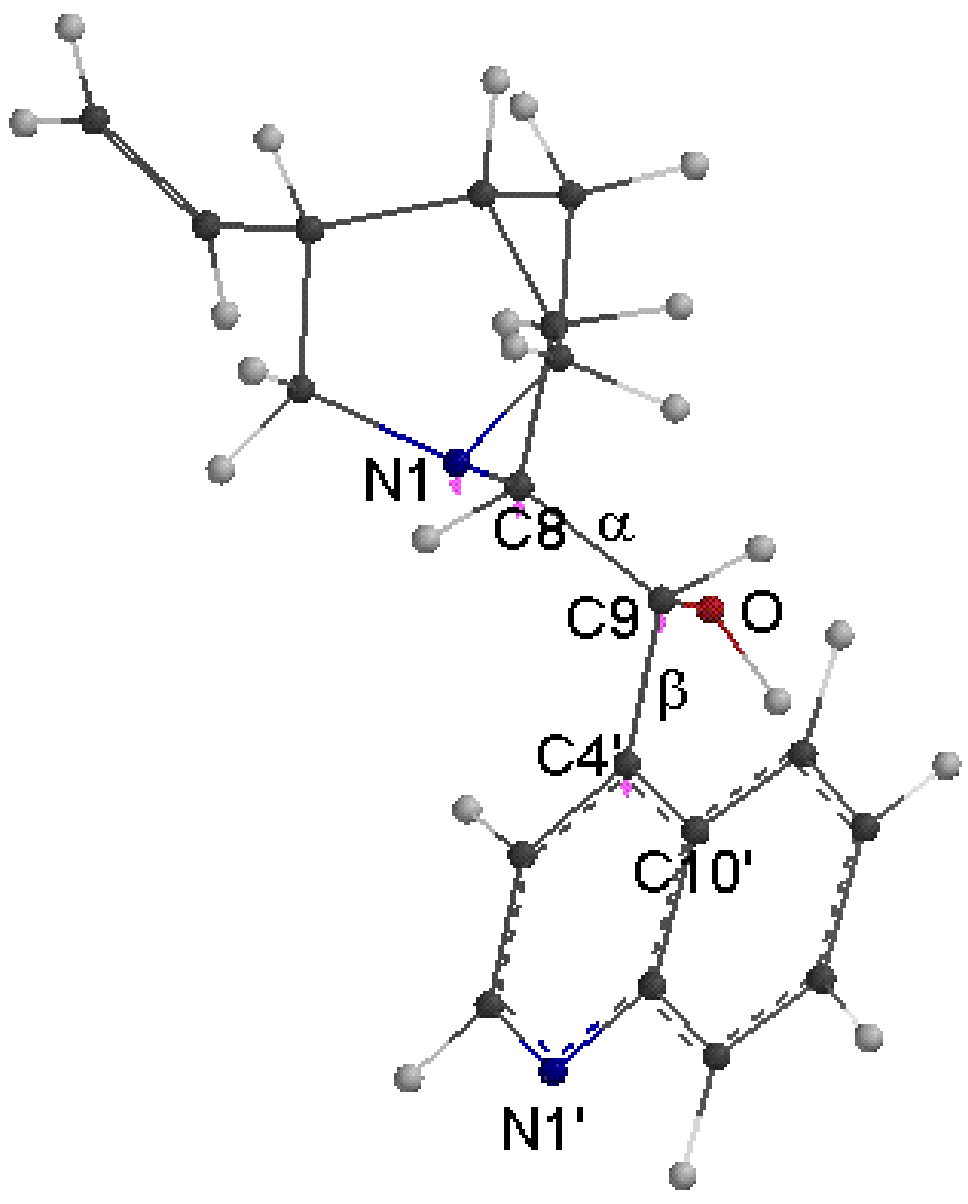

Potential energy diagram of the internal rotation of the two-ring system (quinoline and quinuclidine) relative to each other in CD on two computation levels: At all frozen $\alpha$ dihedral angles, all the other atoms were relaxed in gas phase. The reference point is chosen at $\alpha = 0$ dihedral angle where the two curves are shifted to cross and have zero values for easier comparison of the two computation levels. Minimums define the closed, open and trans configurations, marked. Ranges [0, 360] and [360, 720] (shifted by $2\pi$ for better view) for $\alpha$ correspond to trans $\beta$ and cis $\beta$ cases respectively, i.e. tells which side of the quinoline the quinuclidine ring is positioned.

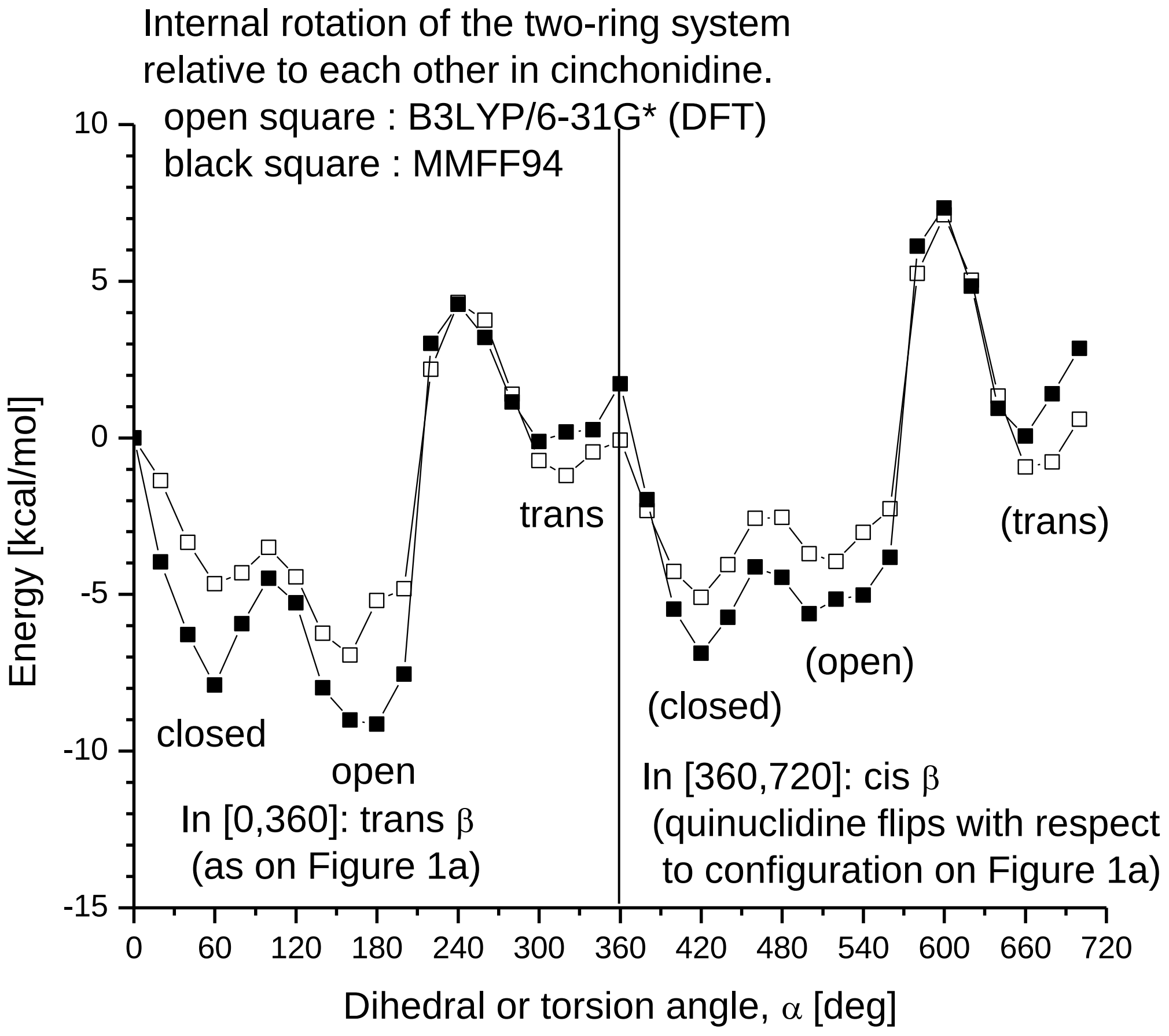

CD (particularly the trans β case) complexing with trans MP in free space. The three energetic minimums of CD alone, as a function of dihedral angle, α, determines the possible stable adducts too: These are the cases as CD captures a MP approaching it from all possible different directions and orientations. Adduct A involves the closed-, B involves the open- and C involves the trans CD, and, of course the CD gets deformed slightly by the presence of MP. (The other cases, involving cis β, cis MP, etc. combining with each other in all possible ways to form adducts are not depicted to save space, but those are essentially the same as the three adducts A, B and C here in respect to the position of the two molecules relative to each other.) The eye is perpendicular to the plain of the three coplanar quinoline rings. (In A the MP is between the eye and quinoline ring).

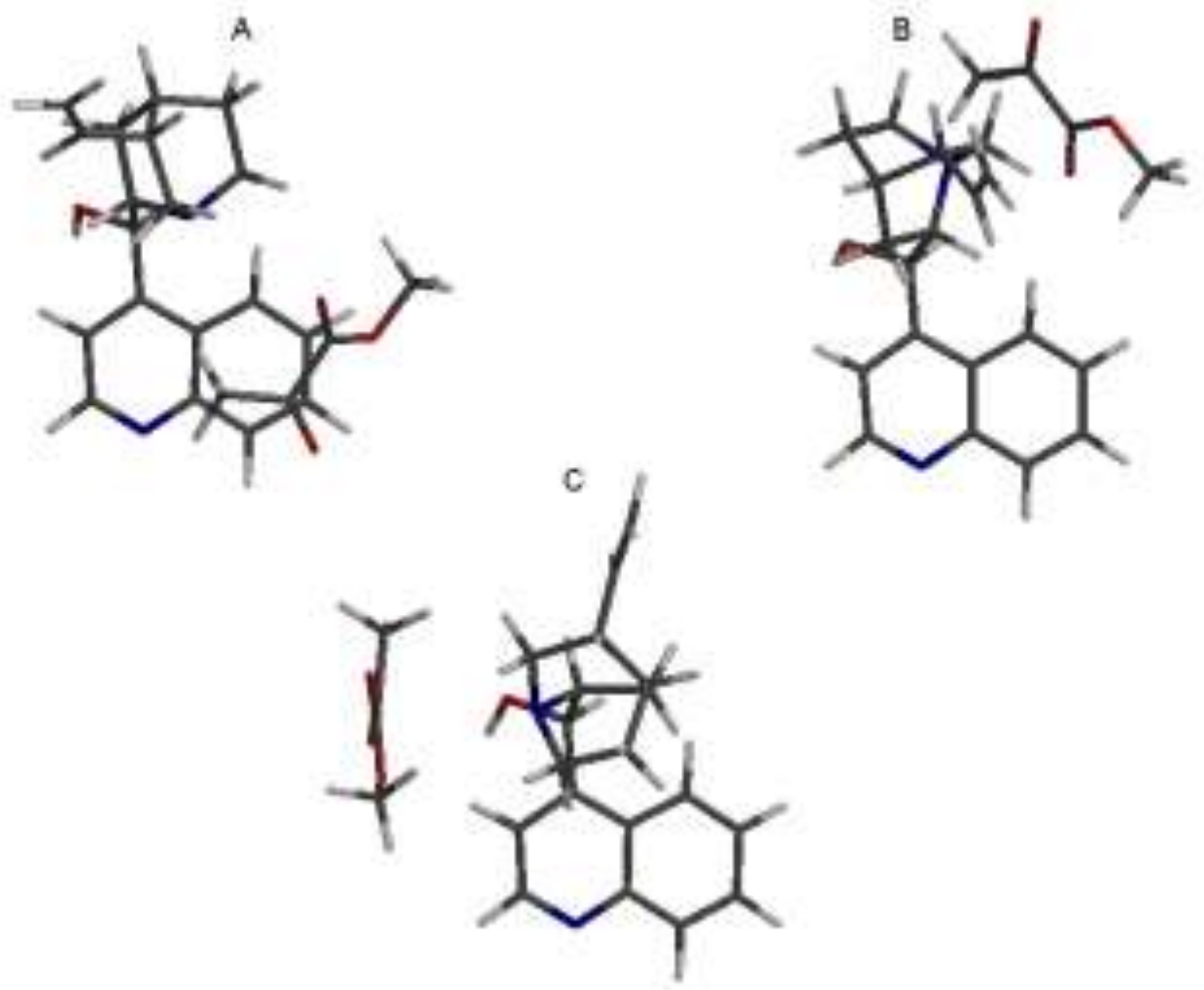

Another view: The eye sees the quinoline rings (which are closer to us than the other parts) as horizontal lines.

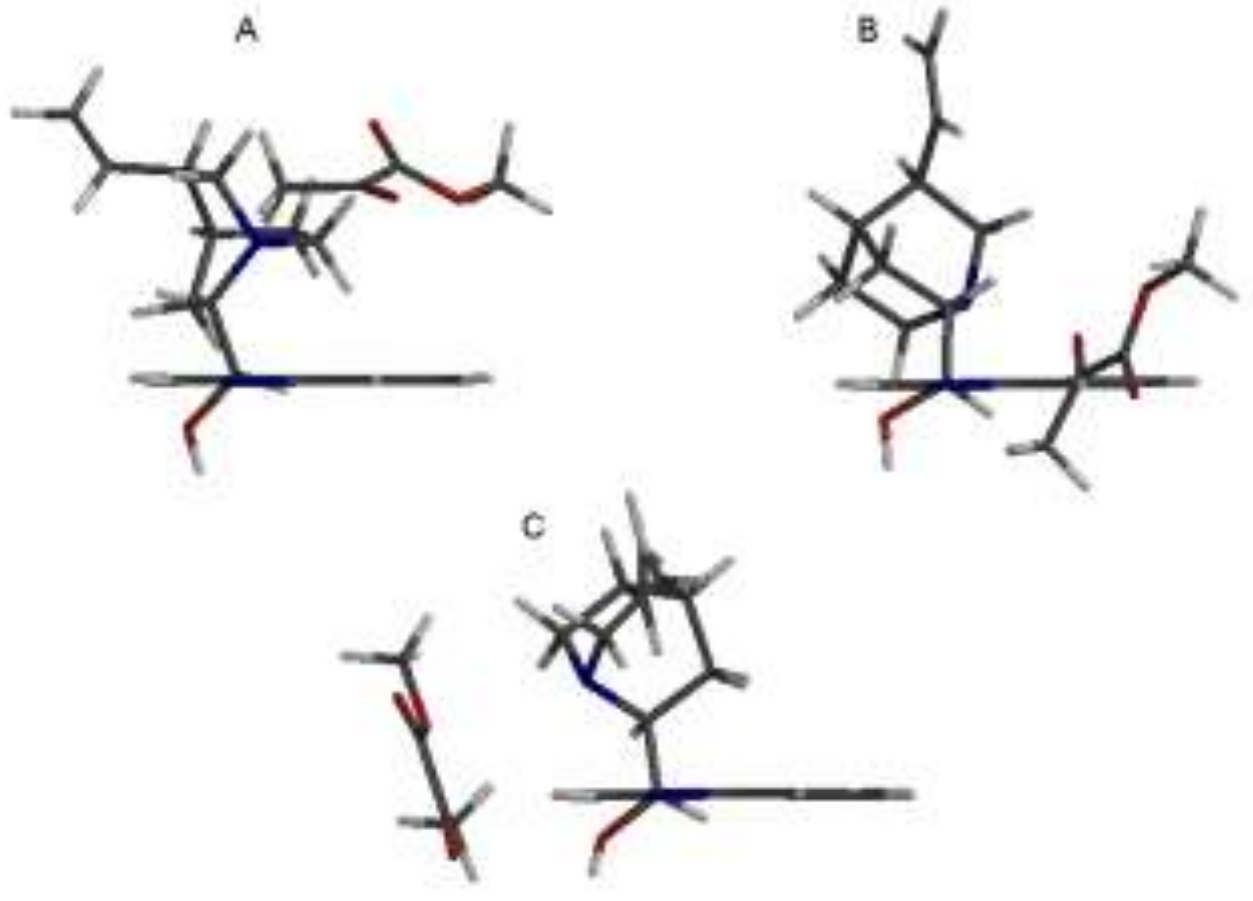

Same complexes arranged in the same order, but now the orientation (with rigid rotation) of the complexes (or adducts) are depicted as follows. The three arrangements are made as the plain of MP's lie on the same hypothetical catalytic (e.g. Pt) surface (i.e. coplanar adsorption). Supposing that the catalytically activated H atom attaches from below provided by the catalytic Pt surface (plain drawn) to the prochiral C or α-oxo atom, all these three arrangements yield the R-lactate from this trans MP. In complexes A, B and C the CD part is in very similar configuration as in the "closed" 60 deg, "open" 160 deg and "trans" 320 deg minimums of individual CD, respectively. Case A involves the closed CD configuration, compare it to model 3, case B involves the open CD configuration, compare it to model 2, (but there, the quinuclidine N is protonated), and case C configuration involves the trans CD, compare it to model 1: Model 1-2-3 proposed by Augustine, Baiker and Blaser, Margitfalvi, resp..

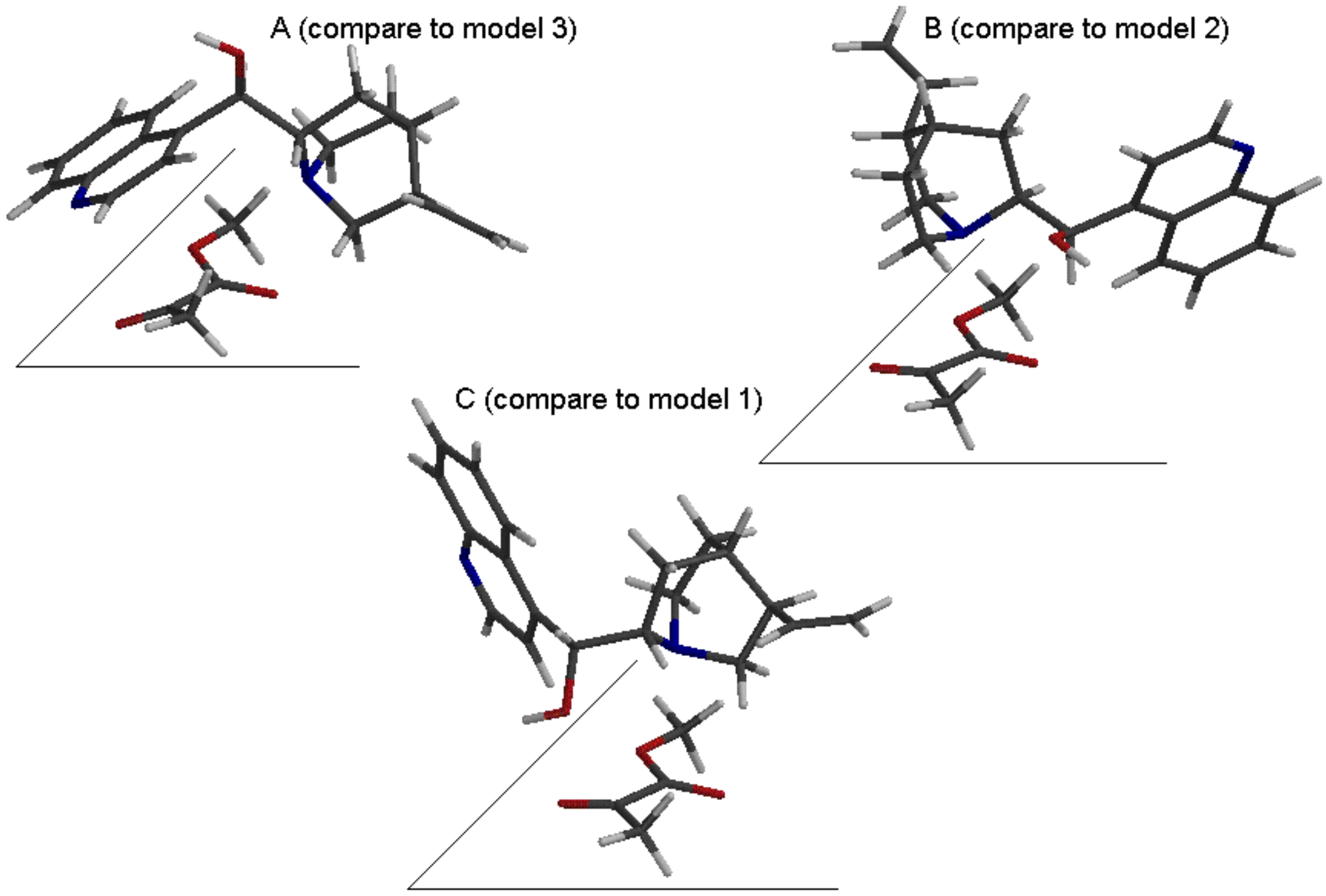

Some molecular orbitals (MO) of the complexes with same relative orientations. Upper line (1) shows LUMO, middle line (2) shows HOMO, and lower line (3) shows HOMO(-1) molecular orbitals. Left column (A) involves the closed CD complex (recall model 3), middle column (B) involves the open CD complex (recall model 2), and right column (C) involves the trans CD complex (recall model 1).

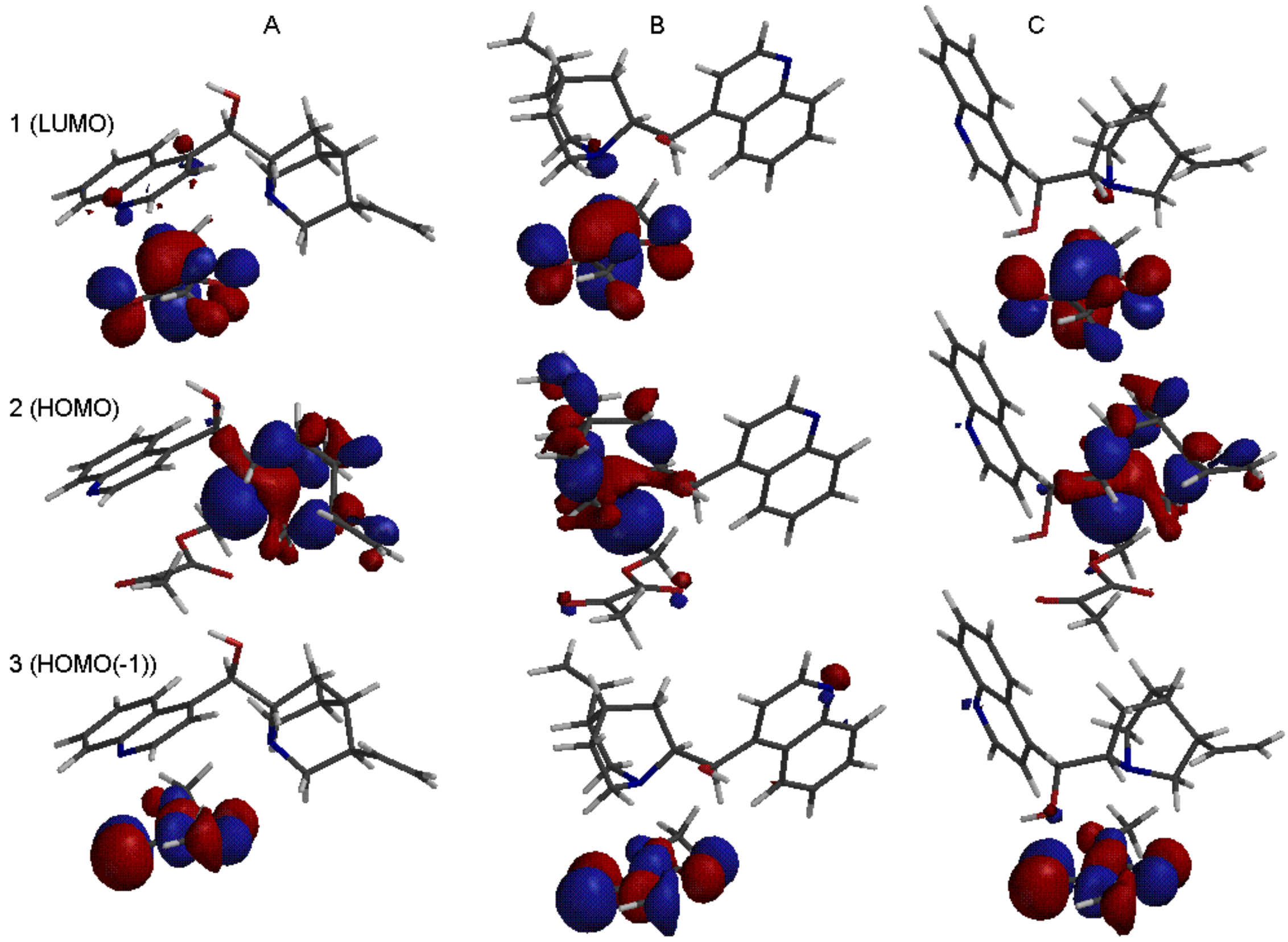

Notice that the LUMO, HOMO and HOMO(-1) orbitals belong to the adduct, as one entity or molecular system, even though we distinguish the two molecules in the adduct (with „rods" representing the chemical bonds). Though not shown, the LUMO and HOMO orbitals of individual (trans) MP have practically the same shape and localization as the LUMO (line 1) and HOMO(-1) (line 3) MO's localized on the MP part in the complexes. This can be an indication of the role of the quinuclidine part of the modifier in the enantioselective reaction, because it reveals that the individual MP topmost HOMO and LUMO surround an MO (line 2) which is localized on the quinuclidine ring when it compexates with CD. These MO's of complexes, of course, perturbate when the surface compound is in contact with a real catalytic surface.

**Thesis-32-Application: Dimer formation of cinchonidine in liquid phase: Relevance to enantioselective hydrogenation of ethyl pyruvate**

Literature data related to the possible dimer formation of cinchona alkaloids in the liquid phase have been collected and analyzed. These data have been correlated with experimental results obtained in the heterogeneous catalytic enantioselective hydrogenation of ethyl pyruvate with my experimental coworkers. In this reaction, the addition of achiral tertiary amines resulted in an increase in both, the reaction rates and enantioselectivity. The positive influence of achiral tertiary amines was attributed to the suppression of dimer formation in aprotic solvents. The results of circular dichroism spectroscopy and *ab initio* calculations provided further proof for dimer formation. Four possible cinchonidine dimer configurations were found with approximate 11–13 kcal/mol stabilization energies.

= = = = = =

A cinkona alkaloidoknak folyadék fázisban valószínűsíthető dimer formációjára vonatkozó irodalmi adatokat gyűjtöttük össze és analizáltuk. Az adatokat a kísérleti (etil piruvátok heterogén katalitikus enantioszelektív hidrogénezése) eredményekkel korreláltattuk munkatársaimmal. Ezekben a reakciókban, az akirális tercier aminok hozzáadása azt eredményezte, hogy a reakció sebesség és az enantioszelektivitás egyaránt emelkedtek. Az akirális tercier aminok pozitív hatása annak tulajdonítható következtetéseink szerint, hogy aprotikus oldószerekben a dimer formáció visszaszorul. A cirkuláris dikroizmus spektroszkópia és *ab initio* számítások eredményei további evidenciákat szolgáltattak a dimer képződésre. Négy valószínű cinkonidin dimer konfigurációt találtam, közelítőleg 11–13 kcal/mol stabilizációs energiával.

= = = = = =

Representative equations/tables/figures:

For the details of experimental results and discussion see the article.

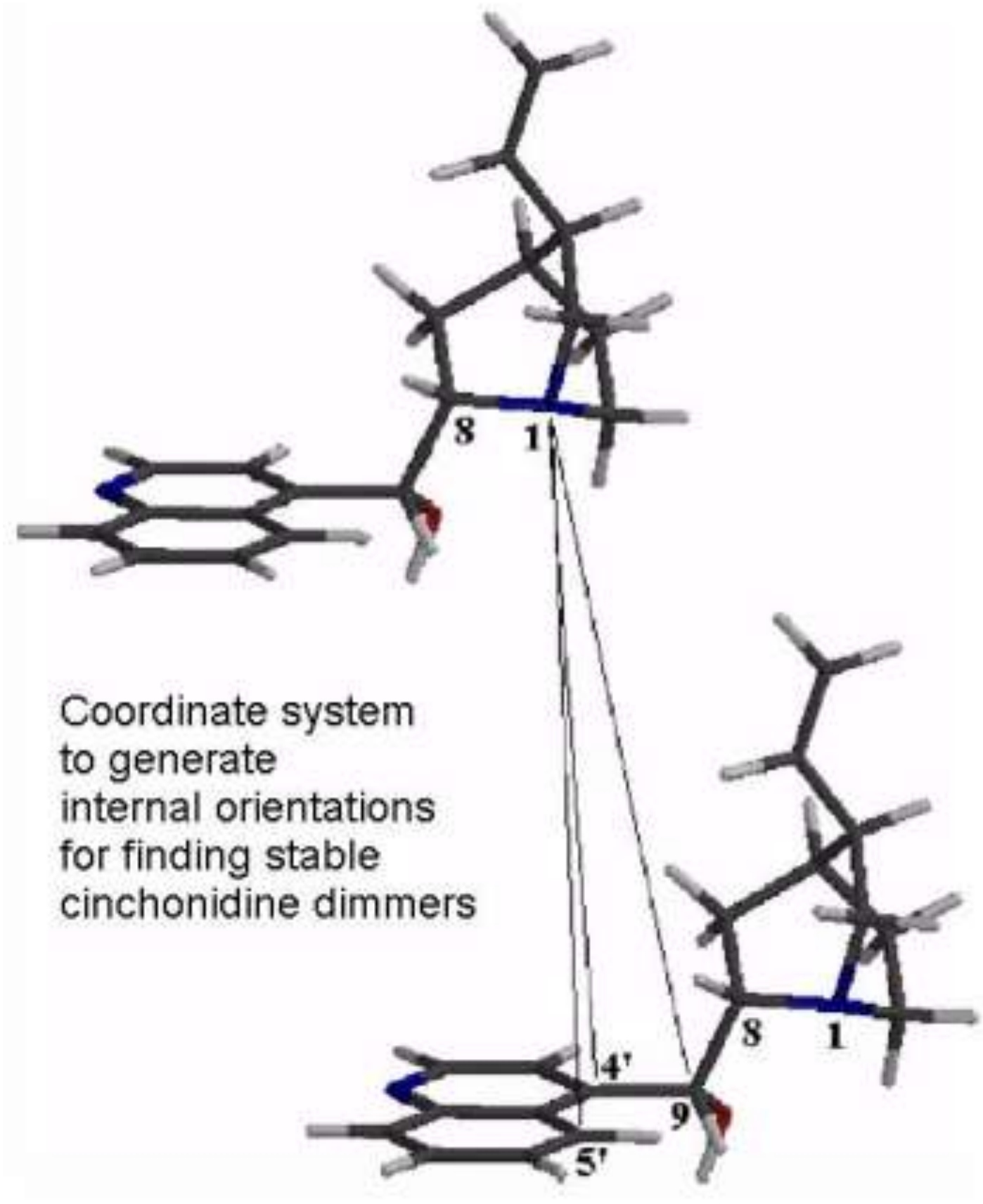

[redacted] The energetic and bond properties of cinchonidine dimers [redacted]

| Cinchonidine dimers in Figure 6 | Molecular mechanics stabilization energy (kcal/mol) | Ab initio, B3LYP/6-31G* stabilization energy (kcal/mol) | 1st main vdW bond | 2nd main vdW bond | Note: relative position of the two quinoline rings |
|---|---|---|---|---|---|
| A | −16.3 (0.0) | −12.0 (0.0) | Quinoline N···HO | Quinoline N···HO | Quasi parallel |
| B | −11.9 (4.4) | −12.8 (−0.8) | Quinuclidine N···HO | Quinoline H···quinuclidine N | Quasi-perpendicular |
| C | −13.2 (3.1) | −12.7 (−0.7) | Quinuclidine N···HO | Quinuclidine N···HO | Quasi-coplanar |
| D | −12.1 (4.2) | −10.9 (+1.1) | Quinuclidine N···HO | Quinoline N···HO | About 45° angle |

The stabilization energy is defined as the 0 K enthalpy of formation of the reaction 2 × cinchonidine→(cinchonidine)$_2$ dimer; the relative energy of the dimer is listed in parentheses with respect to dimer type A.

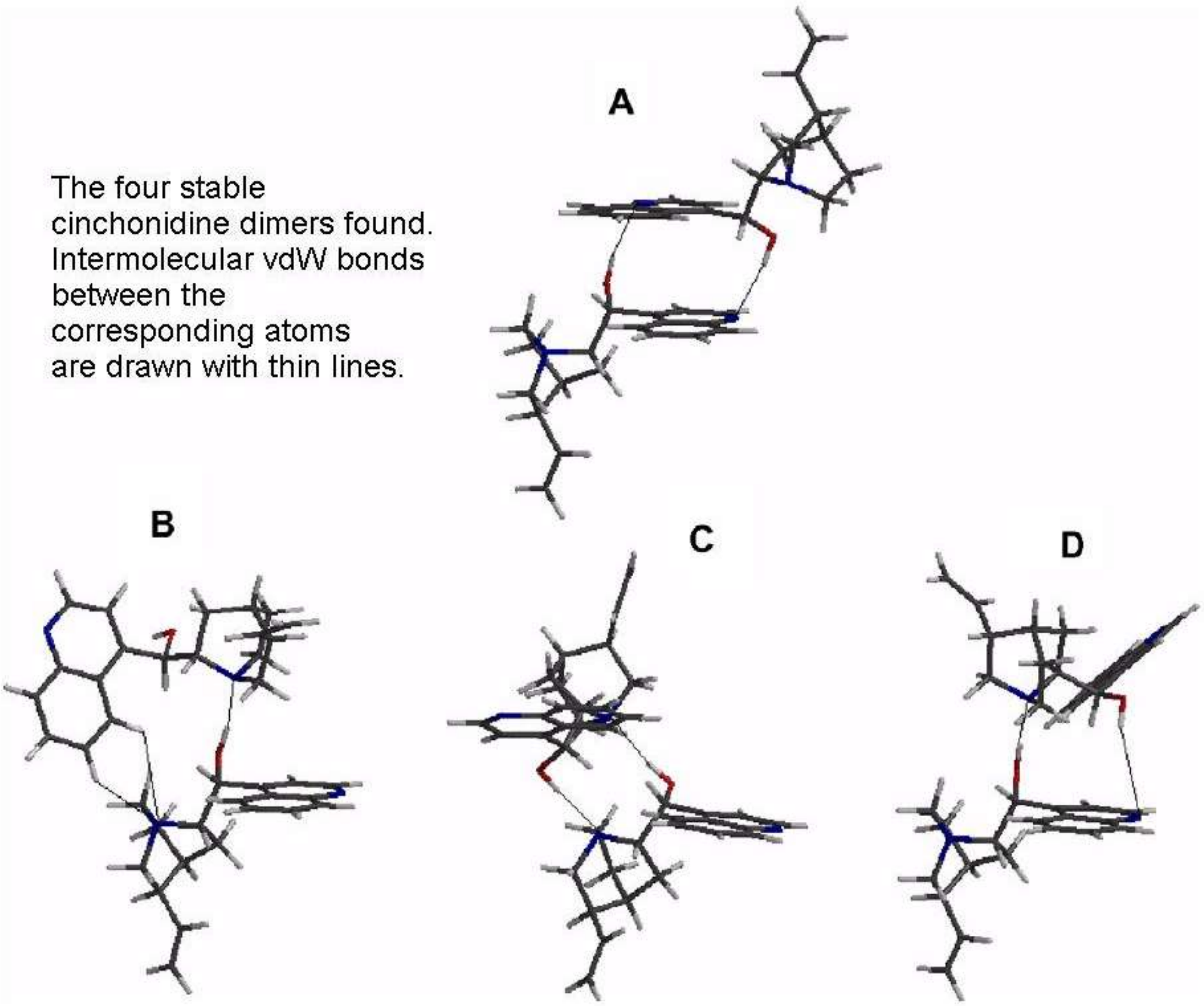

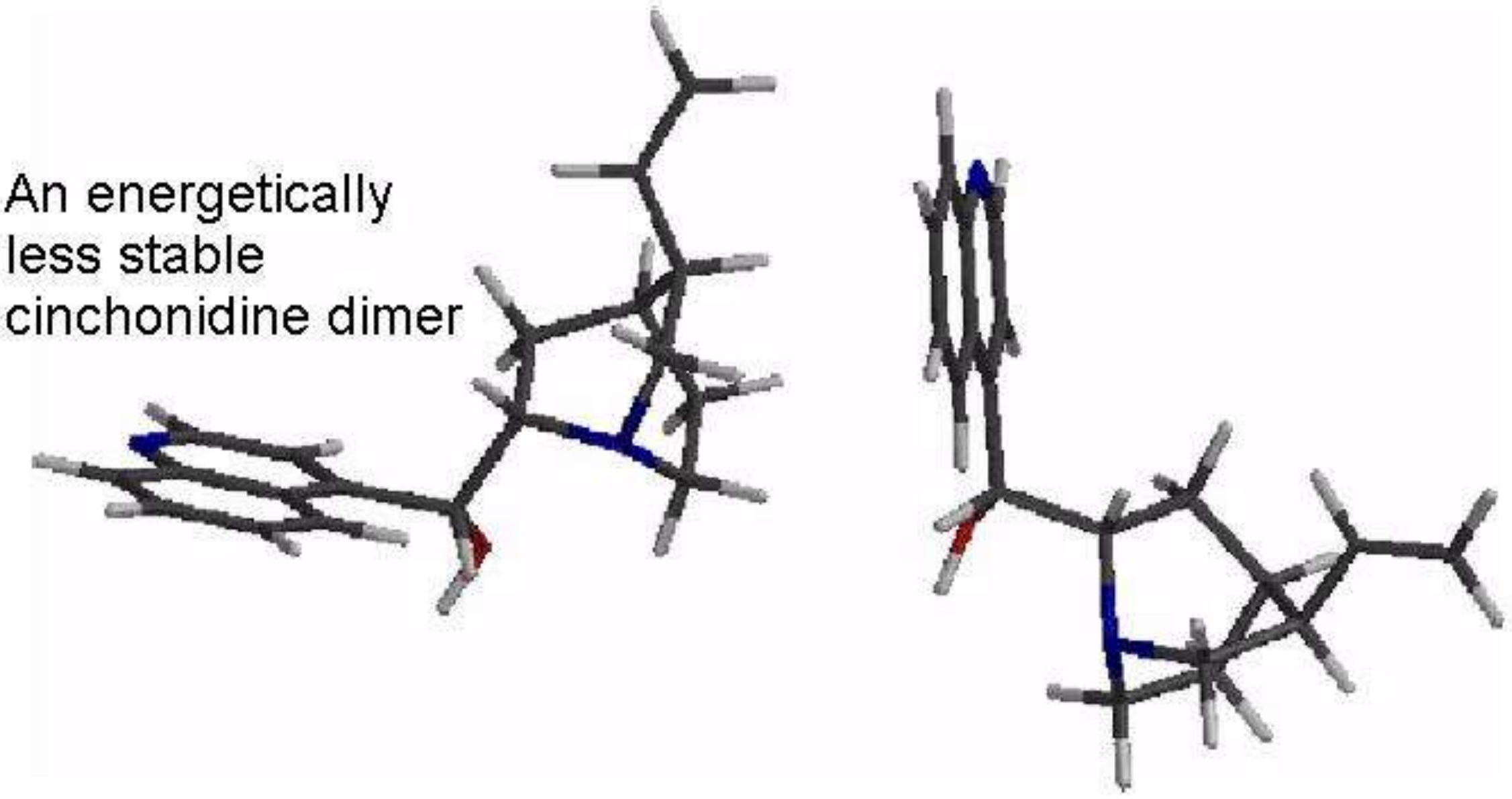
An energetically
less stable
cinchonidine dimer

**Thesis-33-Application: Modeling clusters in IR- and UV- MALDI**

A large succinic acid ($HOOC(CH_2)_2COOH$) matrix containing 7x7x7 unit cells with guest oligonucleotide AGCAGCT was modeled with molecular dynamics simulation for infrared matrix-assisted laser desorption ionization (MALDI). The laser heating of the succinic acid was simulated (missing from the literature) with λ=2940 nm infrared laser pulses and compared to ultraviolet excitation in order to elucidate the cluster formation of succinic acid in the gas phase in itself and around the analyte. At this wavelength, the laser energy is coupled into the matrix through the OH vibrations (stretch) of the carboxyl groups. The most pronounced difference observed at 1,500 K simulation is that infrared heating generates about 10–15 more succinic acid molecules bound to the analyte in noncovalent complex form than the ultraviolet mode, which generates only about 2 molecules. Energy redistribution within the matrix between the host and guest species as well as other dynamical properties was calculated. The parameter and topology data for succinic acid were optimized and ready for use in CHARMM computer code environment for simulation.

= = = = = =

Egy nagyobb, 7x7x7 elemi cellából álló, AGCAGCT oligonukleotid vendég molekulát (analit) tartalmazó borostyánkősav ($HOOC(CH_2)_2COOH$) mátrixot* modelleztem molekula dinamika (MD) segítségével az "infrared matrix-assisted laser desorption ionization (MALDI)" folyamat/módszer szimulálására. A borostyánkősav lézer fűtésének szimulálásával (λ=2940 nm infravörös lézer impulzusokkal, mely hiányzott az irodalomból), összehasonlítva a szintén szimulált ultraibolya (200–400 nm) fűtéssel (gerjesztéssel), leírtam a borostyánkősav klaszter (csoport) formálódását gáz fázisban önmagával és a vizsgálandó anyag (analit) körül. Ezen a hullámhosszon a lézer energiája a mátrixban lévő karboxil csoportok OH rezgéseit (nyújtás) gerjeszti. 1500 K szimulációs hőmérsékleten a legszembetűnőbb különbség az volt, hogy az infravörös fűtés kb. 10–15-ször több borostyánkősav molekulát hoz létre az analit körül nem-kovalens komplex kötésekkel az ultraibolya fűtéshez viszonyítva, mely utóbbi csak kb. 2 molekulával létesít ilyen kapcsolatot. A mátrixon belüli energia eloszlásra a fogadó (borostyánkősav) és vendég (analit) molekulák között, valamint egyéb dinamikai tulajdonságokra szintén végeztem számításokat. A borostyánkősav (format-tált) paraméter és topológia adatait optimalizáltam, melyek közvetlenül felhasználhatók a CHARMM számítógépes programjának futtatásánál a különféle szimulációkhoz.

*A borostyánkősav egy dikarbonsav, mely szobahőmérsékleten szilárd, színtelen és szagtalan kristályok formájában van jelen (2 molekulás elemi cella, olv.pont 185 °C, forr pont 235 °C).

**Sandor Kristyan - Akos Bencsura - Akos Vertes:**
**Theoretical Chemistry Accounts, 107 (2002) 319-325**

= = = = = =

Representative equations/tables/figures:

MALDI (volatilization can be achieved up to 200 kDa) combined with mass spectrometry (MS) is a powerful tool in the characterization of bio- and synthetic polymers. Large guest molecule (analyte) is embedded into crystals of relatively small organic molecules (host) that readily absorb the incident infrared (IR) or ultraviolet (UV) laser light: Easy evaporation of hosts help the guest to go into gas phase, while without host it would be impossible.

The simplifications or drawback of MD (i.e. no fragmentation, no ionization and/or no electronic excitation) do not allow a fair comparison with MS data, however MD is still the only method for considering large systems. The advantage of MD is that energy transfer, complexation, IR and UV excitation can be modeled. If the internal structure is not included (e.g. breathing sphere -, coarse grain moel) the system can be even larger, but features like IR excitation, or certain aspects of cluster formation cannot be considered in detail.

Choice of guest molecule: Some research groups investigate this AGCAGCT oligonucleotide because of its affinity for forming noncovalent complexes with peptides via their thymidine content, affected by the acidity of the matrix. The phosphate link between C and T contained a negative charge, while all succinic acids in the crystal generation and MD simulation were neutral molecules. Succinic acid was not in the CHARMM parameter set, so *ab initio* quantum chemical calculations (GAMESS/3-21G* basis level) was used to determine its structural parameters and ChelpG partial charge distribution. Periodic boundary conditions (toward x-y directions) were applied to model the MALDI evaporation at the succinic acid-vacuum interface, the crystal had a semi-infinite end toward the -z direction but it was also able to evaporate toward the +z spatial direction. The OH bonds of succinic acid were vibrationally excited as a model for the IR laser heating, while the UV excitation (i.e. an electronic transition followed by internal conversion) was modeled by kinetically exciting the entire molecule (because MD cannot account directly for electronic excitations). In the CHARMM environment, the Nose heating method was used, which not only provides correct canonical distribution functions, but also allows the application of selective heating for individual degrees of freedom (in contrast to the much simpler velocity scaling method for heating, which does neither). Initial structure was subjected to 300 K dynamical equilibration and obtained the structure:

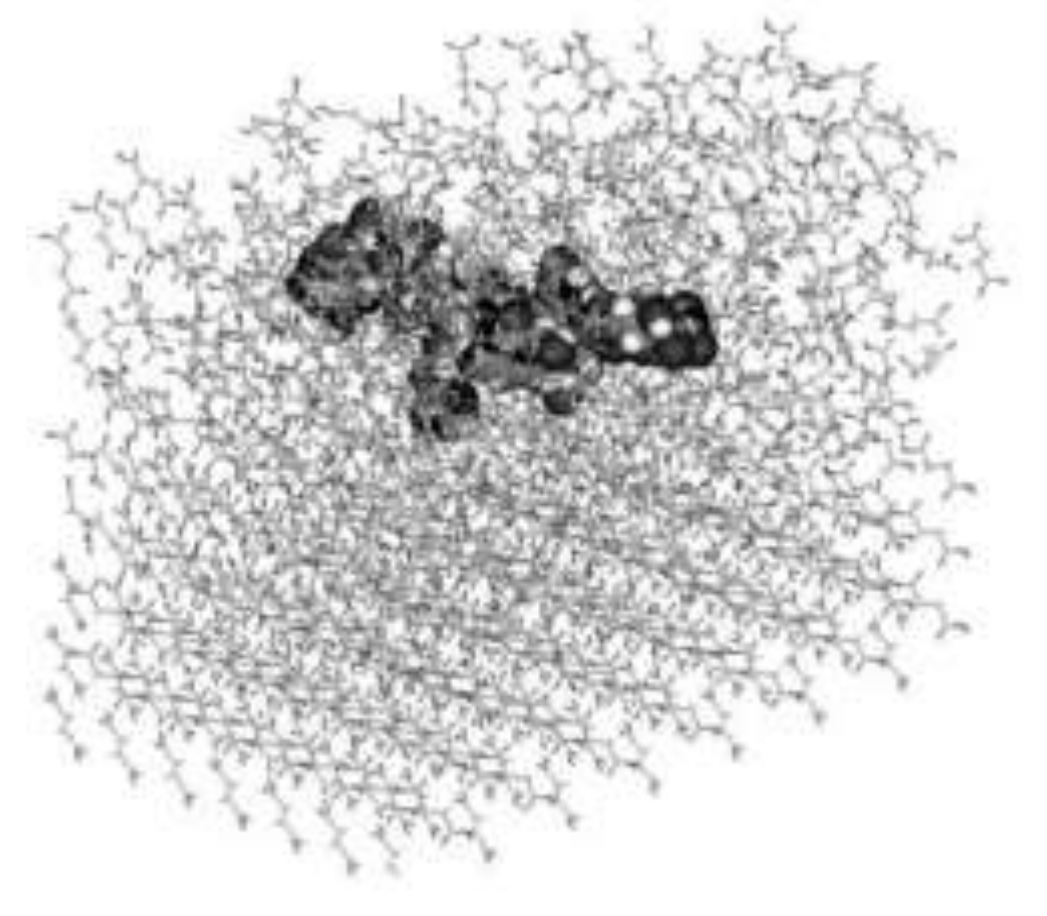

Model MALDI matrix: AGCAGCT oligonucleotide heptamer (van der Waals radii plotted) embedded in a 7x7x7 unit cell size succinic acid matrix (rod representation plotted) between the 4th and 5th layers in the +z direction, and equilibrated at 300 K with MD. This system was evaporated thereafter by model IR and UV laser heating to 1500 K. The periodic boundary condition allowed the evaporation toward the +z direction only.

Stage of evaporation of the model MALDI matrix at 1500 K after 100 ps with a slow IR laser (Nose selective model) heating for the OH groups of succinic from 300 K is plotted. The cluster formation can be seen which varies with the slow (QREF = $2x10^9$ CHARMM coupling constant) or fast (QREF= 100) IR or slow (QREF= $2x10^{10}$) UV laser heating mode. The cluster formation is analyzed in Table below.

In case of fast IR heating the temperature jumped from 300 K to 1,500 K in less than 2 ps; the two kinds of speed were supposed to model the so-called nano-laser and pico-laser excitation (heating) of MALDI.

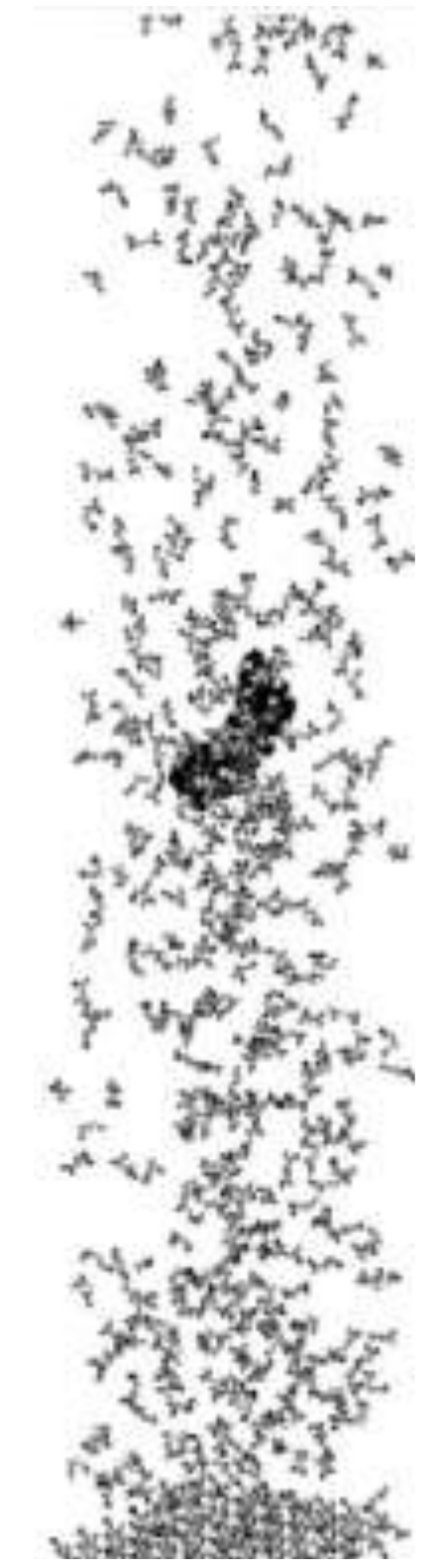

Control test: The experimental value for the enthalpy of sublimation of succinic acid, 28–29 kcal/mol, was reproduced within 1 kcal/mol, probably a fortuitous result. The calculated dynamically equilibrated succinic acid structure was compared to crystallographic data determined by X-ray diffraction at 77 K and 300 K.

Technical results for CRARMM users:

```
 1 1 SUC CB1  0.0808 -0.0664 -0.0277 0.16
 2 1 SUC CA1  0.2599 -0.0345 -0.2364 0.15
 3 1 SUC O1   0.2528  0.0775 -0.3760 0.18
 4 1 SUC OH1  0.4240 -0.1395 -0.2570 0.19
 5 1 SUC H11 -0.0253 -0.1635 -0.0924 0.23
 6 1 SUC H12  0.1816 -0.1035  0.1466 0.23
 7 1 SUC HO1  0.5355 -0.1189 -0.4025 0.21
 8 1 SUC CB2 -0.0808  0.0664  0.0277 0.16
 9 1 SUC CA2 -0.2599  0.0345  0.2364 0.15
10 1 SUC O2  -0.2528 -0.0775  0.3760 0.18
11 1 SUC OH2 -0.4240  0.1395  0.2570 0.19
12 1 SUC H21  0.0253  0.1635  0.0924 0.23
13 1 SUC H22 -0.1816  0.1035 -0.1466 0.23
14 1 SUC HO2 -0.5355  0.1189  0.4025 0.21
15 2 SUC CB1 -0.0808  0.4336  0.5277 0.16
16 2 SUC CA1 -0.2599  0.4655  0.7364 0.15
17 2 SUC O1  -0.2528  0.5775  0.8760 0.18
18 2 SUC OH1 -0.4240  0.3605  0.7570 0.19
19 2 SUC H11  0.0253  0.3365  0.5924 0.23
20 2 SUC H12 -0.1816  0.3965  0.3534 0.23
21 2 SUC HO1 -0.5355  0.3811  0.9025 0.21
22 2 SUC CB2  0.0808  0.5664  0.4723 0.16
23 2 SUC CA2  0.2599  0.5345  0.2636 0.15
24 2 SUC O2   0.2528  0.4225  0.1240 0.18
25 2 SUC OH2  0.4240  0.6395  0.2430 0.19
26 2 SUC H21 -0.0253  0.6635  0.4076 0.23
27 2 SUC H22  0.1816  0.6035  0.6466 0.23
28 2 SUC HO2  0.5355  0.6189  0.0975 0.21
```

Succinic acid unit cell parameters for CHARMM crystal generation are listed. Columns: 2nd distinguishes the two molecules (of 2x14 atoms), in 3rd the name SUC is optional, 4th identifies the atoms (first letter) and the rest describes the chemical bond connectivity necessary in MM calculation, 5th -7th contain the coordinates in ‘‘unit cell’’ unit, i.e. it must be scaled with unit cell parameters if one wants to see the distances in Angstrom, as well as the last column is the temperature factor used in the simulation.

```
   22   1

AUTO ANGLES DIHE

 RESI SUC 0.0
 GROO
 ATOM HO2  H      0.44
 ATOM OH2  OH1   -0.61
 ATOM O2   OB    -0.55
 ATOM CA2  CD     0.75
 ATOM CB2  CT2   -0.21
 ATOM H21  HA     0.09
 ATOM H22  HA     0.09
 ATOM CB1  CT2   -0.21
 ATOM H11  HA     0.09
 ATOM H12  HA     0.09
 ATOM CA1  CD     0.75
 ATOM O1   OB    -0.55
 ATOM OH1  OH1   -0.61
 ATOM HO1  H      0.44

 BOND   HO1 OH1   OH1 CA1   O1 CA1      CA1 CB1
 BOND   CB1 H11   CB1 H12   CB1 CB2
 BOND   CB2 H21   CB2 H22   CB2 CA2
 BOND   HO2 OH2   OH2 CA2   O2 CA2

IMPR  CA1 O1 CB1 OH1      CA2 O2 CB2 OH2
DONOR  HO1 OH1
DONOR  HO2 OH2
ACCE   O1 CA1
ACCE   O2 CA2
ACCE   OH1 CA1
ACCE   OH2 CA2

patch first none last none
END
```

Succinic acid topology and parameter set in CHARMM format for immediate use, showing the 3-21G* level ChelpG partial charges as well. 1st letter identifies the atom, the 2nd and 3rd letters or numbers identify the connection of chemical bond: HOOC-CH2-CH2-COOH is coded as HO2-OH2- CA2(=O2)- CB2(H21)(H22)-CB1(H11)(H12)- CA1(=O1)- OH1-HO1, ‘‘ATOM’’ and "BOND" keywords tell the MM module what those mean chemically, as well as the other keywords are CHARMM commands for the use of force field.

Chemical results:

Trajectory analysis: MALDI is a highly non-equilibrium flowing (evaporating) system, and as a consequence more than one temperature can be defined in both, in UV (entire succinic acid) and in selective IR (OH in succinic acid) excitation modes:

T(OH) > T(matrix) > T(guest)

is hold with 50-500K differences during the heating. T(matrix) has a visible jump (≈50 ps) the time when guest starts to lift off (observed in other systems, as well as in both, in UV and IR heating modes), as well as there is no experimental data on this latter fine part (jump):

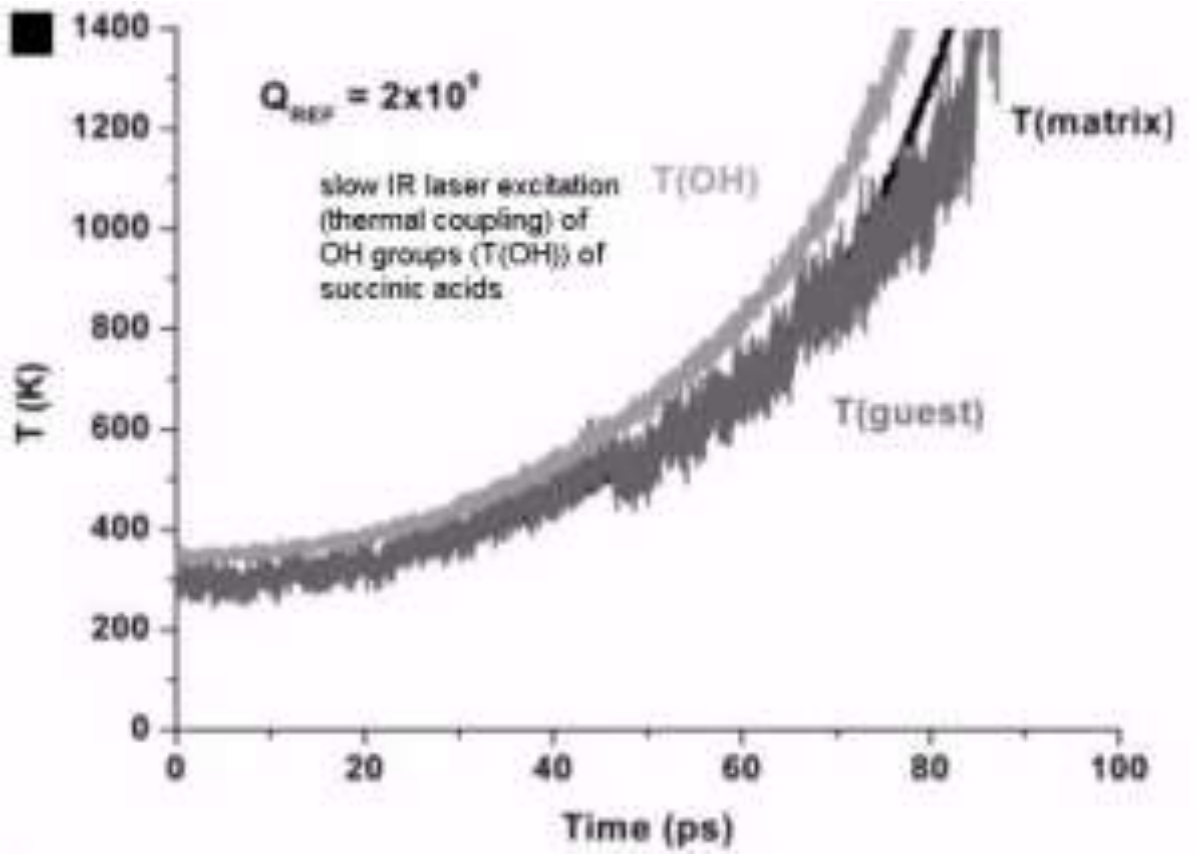

Theoretically calculated statistics of cluster formations in gas phase after 100 ps in case of different model heatings (started from 300 K and lasted to 1500 K): Total number of atoms in the system (matrix[succinic acids]+ guest[oligonucleotide, AGCAGCT]) is N(atom)=9832. Notations: i = size of cluster (# of atoms), N(i) = number of clusters with this size i,

N(cluster)= ΣN(i) = # of individual clusters (irrespective of size), SUC = succinic acid. The SUC contains 14 atoms, so multiples of number 14 in the table immediately refers to the SUC clusters or its multiples. Definition: Two atoms belong to the same cluster if the distance between them is < 3.0 Angstrom (> chem. bond, < van der Waals dist.).

Table a: ‘‘Slow’’ (QREF = 2x10$^9$) IR heating of OH in –COOH of SUC; ΣN(i)= 407

| i(cluster size) | N(i) | N(i)/N(cluster) (piecewise ratio) | i*N(i)/N(atom) (integrated volume ratio) | Note |
|---|---|---|---|---|
| 14 | 339 | 0.833 | 0.483 | SUC monomer |
| 28 | 44 | 0.108 | 0.125 | SUC dimers |
| 42 | 10 | 0.025 | 0.043 | SUC trimers |
| 56 | 8 | 0.020 | 0.046 | SUC tetramers |
| 70 | 4 | 0.010 | 0.028 | SUC pentamers |
| 452 | 1 | 0.002 | 0.046 | oligo(228 atoms) + 16 SUC |
| 2254 | 1 | 0.002 | 0.229 | 161 SUC remained on surface |

Table b: ‘‘Fast’’ (QREF = 100) IR heating of OH in –COOH of SUC; ΣN(i)= 380

| i(cluster size) | N(i) | N(i)/N(cluster) (piecewise ratio) | i*N(i)/N(atom) (integrated volume ratio) | Note |
|---|---|---|---|---|
| 14 | 334 | 0.879 | 0.476 | SUC monomer |
| 28 | 36 | 0.095 | 0.103 | SUC dimer |
| 42 | 4 | 0.011 | 0.017 | SUC trimer |
| 56 | 3 | 0.008 | 0.017 | SUC tetramer |
| 266 | 1 | 0.003 | 0.027 | SUC 19-mer |
| 382 | 1 | 0.003 | 0.039 | oligo (228 atoms) + 11 SUC |
| 3164 | 1 | 0.003 | 0.322 | 226 SUC remained on the surface |

Table c: ‘‘Slow’’(QREF = $2x10^{10}$) UV heating of SUC; ΣN(i)= 478

| i(cluster size) | N(i) | N(i)/N(cluster) (piecewise ratio) | i*N(i)/N(atom) (integrated volume ratio) | Note |
|---|---|---|---|---|
| 14 | 430 | 0.900 | 0.612 | SUC monomer |
| 28 | 38 | 0.079 | 0.108 | SUC dimers |
| 42 | 6 | 0.013 | 0.026 | SUC trimers |
| 56 | 1 | 0.002 | 0.006 | SUC tetramers |
| 70 | 1 | 0.002 | 0.007 | SUC pentamers |
| 256 | 1 | 0.002 | 0.026 | oligo(228 atoms) + 2 SUC |
| 2114 | 1 | 0.002 | 0.215 | 151 SUC remained on surface |

Table a-b-c reveal that the three distributions of clusters are quite distinguishable in these different heating modes (100 ps = after the laser pulse started).

The center of mass of the guest molecule from the surface was about 110, 450, and 250 Angstrom, resp., which can be considered as a gas phase particle. The 450 A value means that in the case of fast IR heating, the guest lifted off from the surface instantly.

The low ΣN(i)= 380 value in the “fast IR heating” case (Table b) comes from that many succinic acids remained on the surface (226, cf. Table b) – it is interesting, because in this case the temperature jump (or laser energy transfer) to 1,500 K was almost instantaneous.

The IR mode (slow or fast) produced less individual succinic acid gas phase molecules (339 or 334) than the UV laser mode (430): The UV heating, targeting the entire succinic acid molecule, shakes the matrix molecules off from the MALDI ensemble more effectively than the IR heating which targets only a particular group (here, OH) of the molecule. The 19-mer (Table b) peak is interesting, since it seems to us that the fast IR laser pulse can cut off large chunks from the solid phase MALDI matrix.

In summary these may enable us to better understand the MALDI process, as well as to predict its mass spectra theoretically.

**Thesis-34-Application: Conformational changes vs. catalytic transformation of *n*-alcanes**

Temperature induced conformational changes of various alkanes were studied. The temperature dependent molecular diameter of long chain alkanes, such as $C_8$, $C_{16}$ and $C_{24}$, have been calculated for controlling their penetration ability into zeolite channels, while the ability of n-hexane to be involved in $C_5$ and $C_6$ ring closure reactions has also been demonstrated.

= = = = = =

Tanulmányoztam több alkán hőmérséklet indukálta konformáció változását, ezen belül a hosszabb láncú, $C_8$, $C_{16}$ és $C_{24}$ alkánok hőmérséklet függő molekula átmérőit számítottam, ami kontrolálja penetrációs képességüket zeolit csatornákba, továbbá demonstráltam az n-hexán képességét a $C_5$ és $C_6$ gyűrűzárási reakciókra.

**<u>Jozsef L. Margitfalvi - Sandor Kristyan - Erno Tfirst:</u>**
**<u>Reaction Kinetics and Catalysis Letters, 74 (2001) 337-344</u>**

= = = = = =

Representative equations/tables/figures:

In the presence of heterogeneous catalysts the routes of hydrocarbon transformations depend on many factors, such as (1) the type of the catalyst in use, (2) temperature, (3) pressure, (4) presence or absence of hydrogen, (5) contact time, etc. All these parameters have been widely investigated in the last sixty years of literature. In all mechanistic studies on n-alkanes it is often assumed that these molecules have a linear form and not too much attention has been paid to the conformational changes of these molecules.

Two conformational problems are in focus here:

1.: The change of the critical diameter of $C_8$, $C_{16}$ and $C_{24}$ n-alkanes, relating to the penetration ability of long-chain hydrocarbons into the pores of zeolites.

2.: The effect of temperature on the distance between C1–C6, C1–C5 and C1–C3 carbon atoms in n-hexane, relating to the probability to have C5- and C6 type cyclization.

Computation methods:

1.: molecular dynamics (MD) at mm+ level in HyperChem/ Windows,

2.: for visualization Cerius 2 and Spartan/Silicon Graphics Octane workstation,

3.: data treatment from MD calculations was processed using Fortran/Unix.

The most stable equilibrium conformation of n-alkanes is the "stretched": All C atoms are placed in one plane and the C-C-C-C torsion angles are 180°. At higher than 0 K, however, the C atoms move relative to each other, and the chain can rotate around the C-C bonds, overcoming the energy barrier and become globular. Essentially different conformations of n-alkanes are obtained by the rotations around its n-3 bonds (rotations around the C-C bonds at the two ends will transform only the H atoms), as well as 3 local minima can be found for each rotations around the bonds (trans, gauche+, gauche-). It follows that the number of the essentially different conformations for a n-alkane containing n C atoms is in the magnitude of $3^{n-3}$.

| | |
|---|---|
| 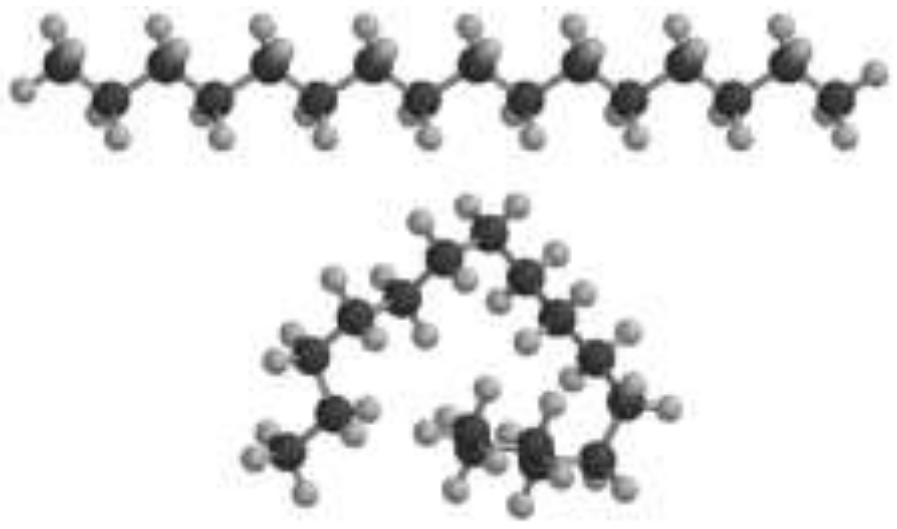 | Linear (stretched, T=0K) and globular (T>>0) structures for $C_{16}H_{34}$ molecule. Despite that some conformations transform into each other by symmetry operations and some conformations do not appear in the potential energy hypersurface because of the overlap of some atoms, the accurate number of conformations is in the magnitude of $3^{n-3}= 3^{16-3}= 3^{13}\approx 1.59x10^{6}$. |

Because of this globular structure, the area of the projection of the molecule on any plane (catalytic or zeolite pore, for example,) is always larger than the ("minimal") area of the projection of the most stable, "stretched" structure toward its longitudinal axis. (The minimal area is about the diameter of a methyl group in this case.) It is quite obvious from the figure above that the linear structure can penetrate longitudinally into a zeolite pore of about a methyl group diameter, while the globular structure cannot.

To model the penetration ability of a n-hydrocarbon chain at a real temperature into a zeolite channel with given diameter, MD simulation has to be made at that temperature, followed by analyzing many projections of the structures obtained during the simulation to get statistical quantities, which can characterize the penetration probability of the molecule into the zeolite channels. The diameter(s) of the projection on any planes around any globular (zig-zag) structure can be obtained by standard mathematical analysis, and the minimal one (d) can be singled out (taking into account that all nuclei coordinates possess a van der Waals radius sphere around). If we determine the d values of the structures obtained by MD and compute the distribution of these values, we can qualitatively characterize the ability of the penetration of the n-alkane molecules into the studied channel. In this way, the dependence of this probability on the temperature and the length of the carbon chain can be determined.

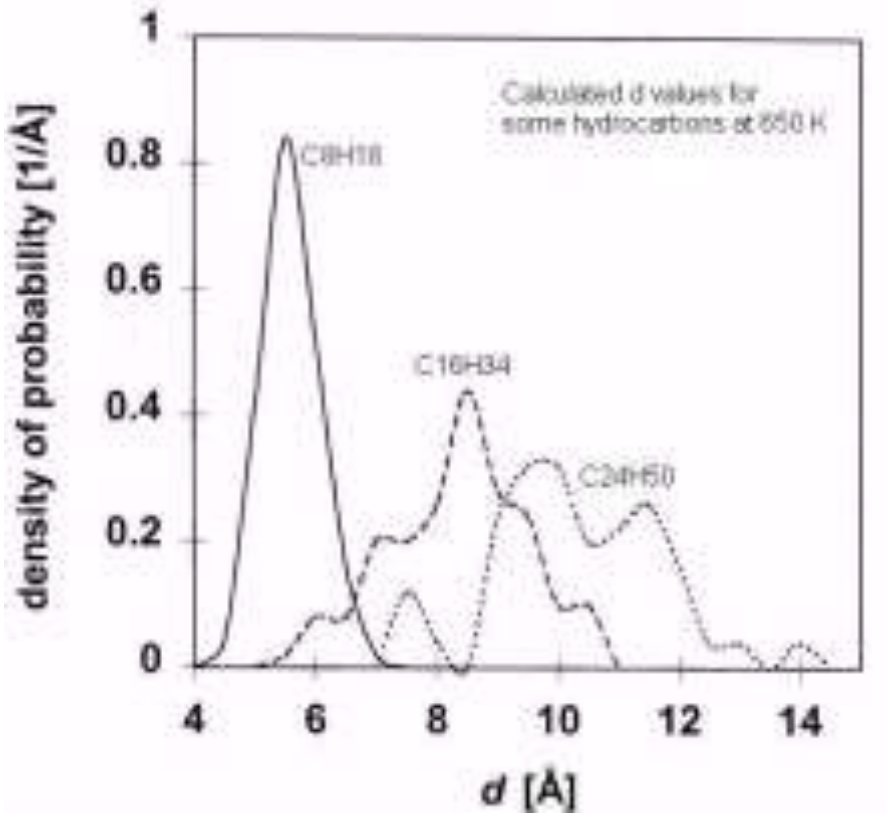

100 ps long MD simulations were performed for n-alkanes at 450-750 K with 50K steps, and the Cartesian coordinates of the generated structures were saved in every 1 ps.

Calculated d values (minima of projected diameters on all planes around) for some hydrocarbons (n=8, 16, 24) at 650 K are exhibited. (Area under the curves is 1.) As expected, mean of d increases with n; interesting is that increasing n yields more maxima, e.g. at n=24, there are two dominating diameters.

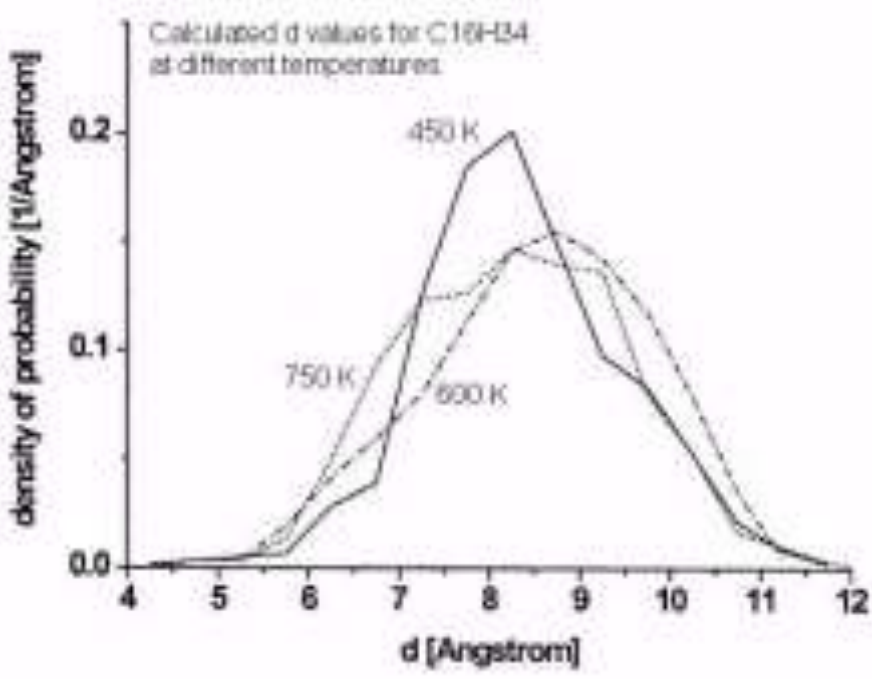

Calculated d values for $C_{16}H_{34}$ at different temperatures are exhibited. The mean of d is smaller at lower temperatures for the same carbon chain (n=16 exhibited only). However, in a broad range (450 K – 750 K) of temperature, the mean molecular diameter changes less than 1 Å. It may indicate that the strong temperature effect in catalytic and zeolite processes cannot originate from this event, but else (e.g. affinity, stherical, reaction barrier, etc.).

To model the distribution of C1-C6, C1-C5 and C1-C3 distances in n-hexane, MD calculation has to be also performed.

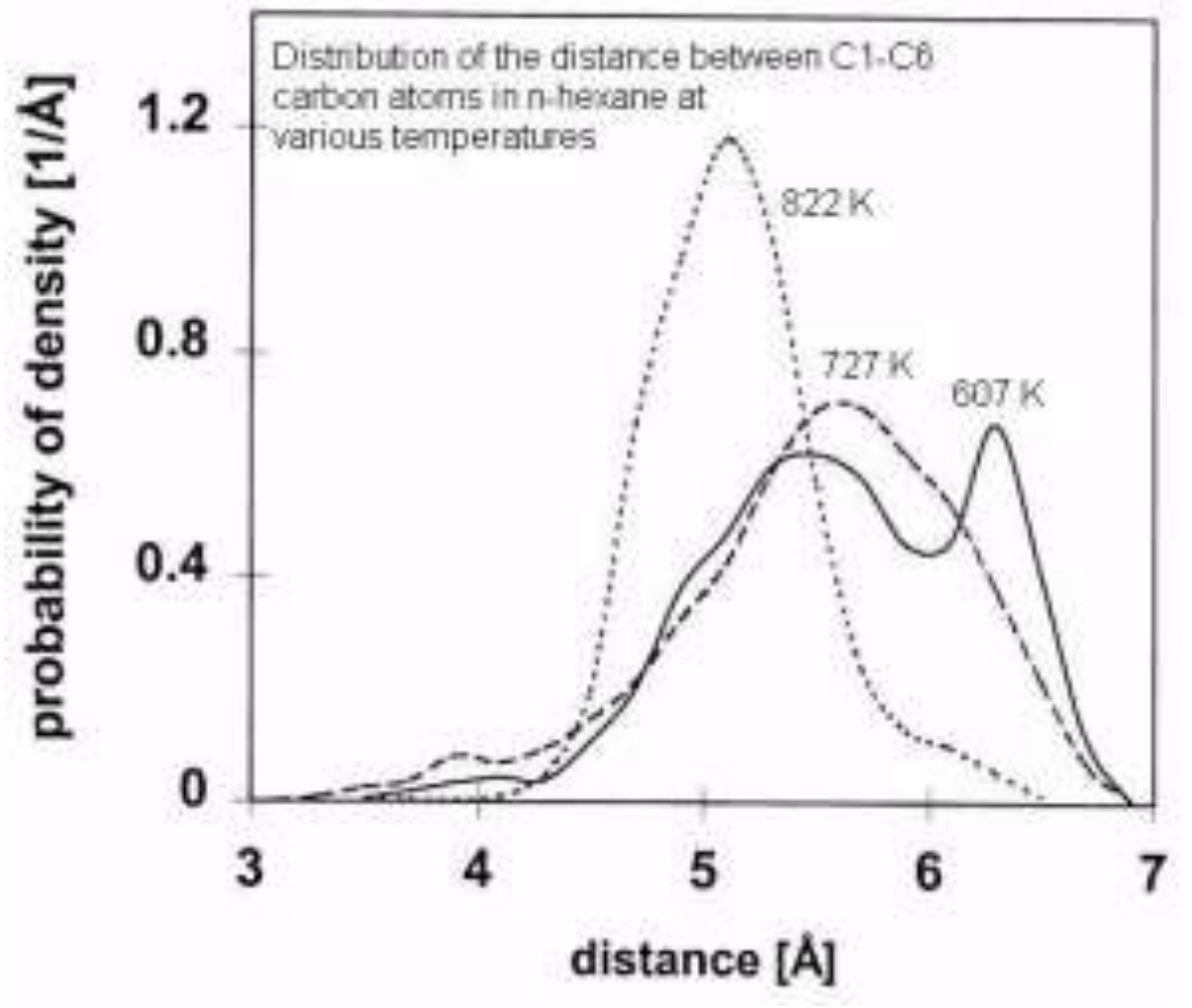

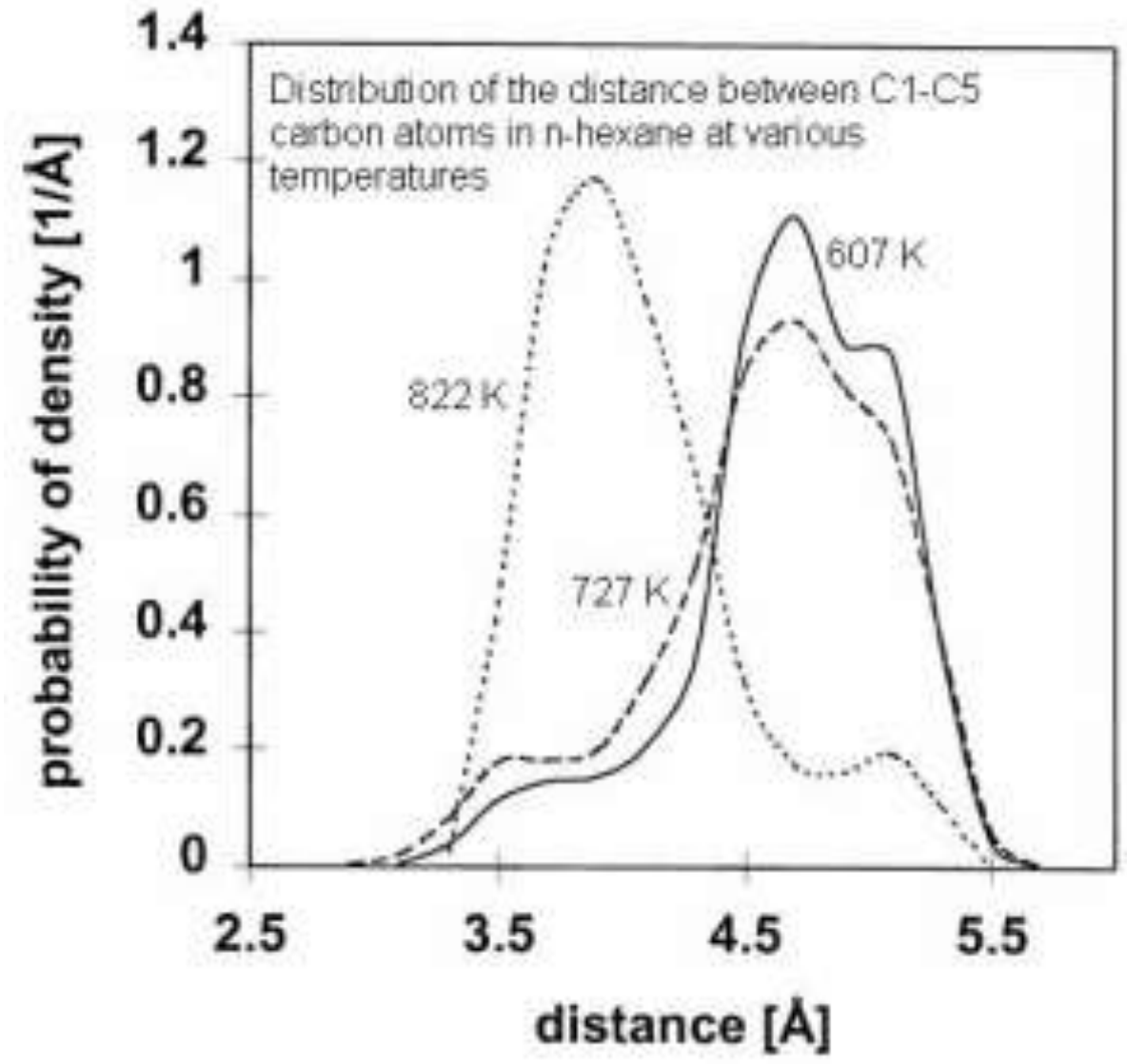

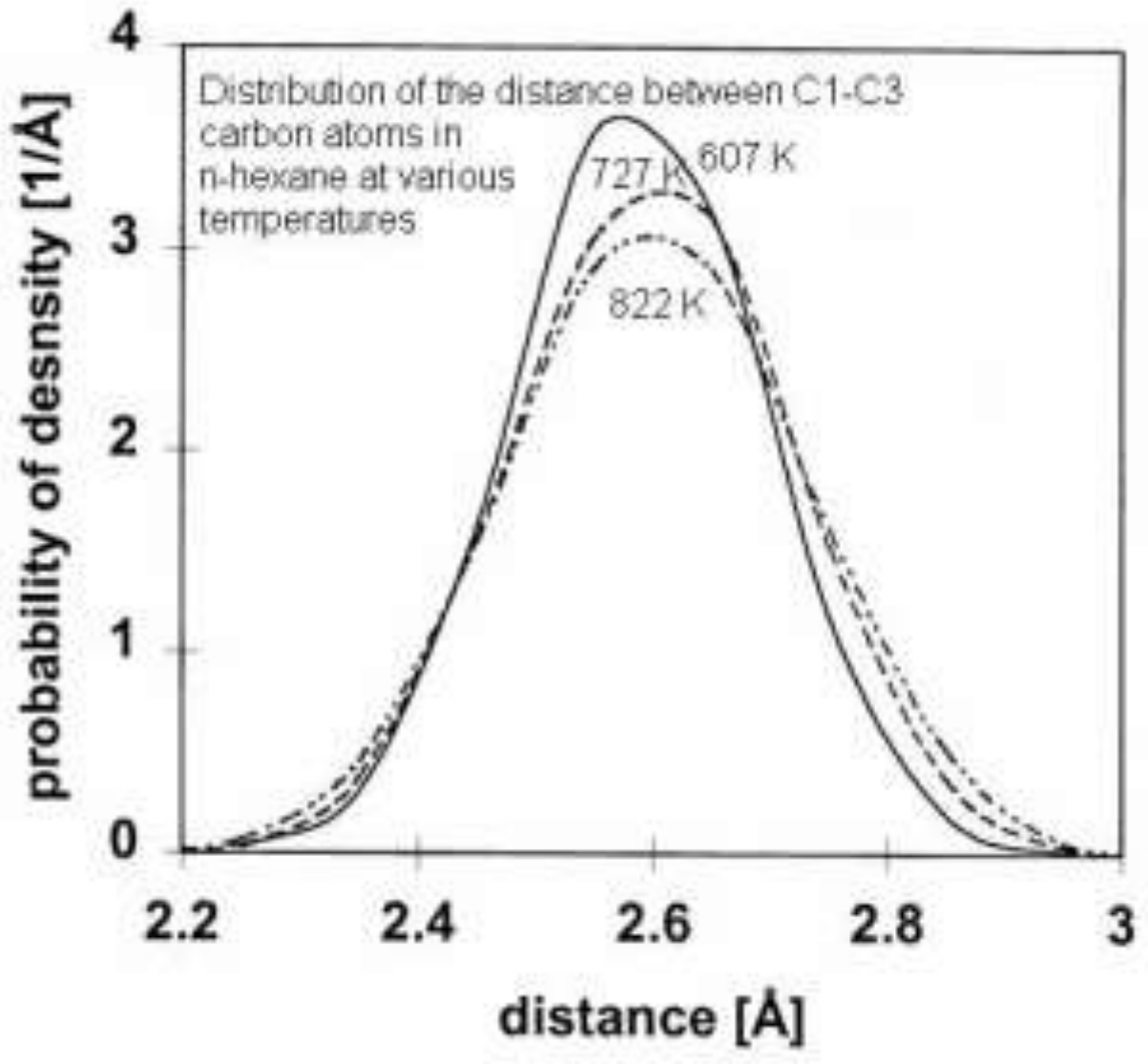

Calculated distribution function of C1-C6, C1-C5 and C1-C3 distances at temperatures 607, 727 and 822 K are shown. The distribution function for C1-C6 and C1-C5 distances implies higher probability for C5 and C6 ring closure at higher temperatures, however, as it has been expected, the C1-C3 distance is almost independent of the temperature. Of course, the complexity of the catalytic reaction of e.g. n-hexane over different catalysts is much deeper, and cannot be explain with only this kind of conformational analysis, but these conformational changes are in good accordance with experimental results obtained both, on supported and unsupported metal catalysts.

In general, investigation of isomers and/or other hydrocarbons is straightforward with this method for both, penetration ability and ring closure formation as well.

**Thesis-35-Application: Excited electronic potential energy surfaces and transition moments for the $H_3$ system**

Four electronic states of $H_3$ have been studied using a multiple-reference double-excitation configuration-interaction (MRSD-CI) method with an extensive basis set of 75 Gaussian-type atomic orbitals. A total of 1340 *ab initio* points were calculated over a wide range of $H_3$ molecular geometries. These four states include the ground state and the Rydberg 2s $^2A_1'$ and $2p_z$ $^2A_2''$ states, as well as the state that in equilateral triangular geometry is related to the ground state by a conical intersection. Electric dipole transition moments were also obtained between these states. The results show that the atomic and diatomic energetic asymptotes are accurately described. The barriers, wells, and energy differences also show good agreement compared to literature values, where available. The potential energies of the ground state and the $2p_z$ $^2A_2''$ Rydberg state display smooth and regular behavior and were fitted over the whole molecular geometries using a rotated Morse curve-cubic spline approach. The other two potential-energy surfaces reveal more complicated behaviors, such as avoided crossings, and require a different fitting procedure to obtain global fitting. The related huge literature has also been reviewed.

= = = = = =

A $H_3$ molekuláris rendszer négy elektronikai állapotát tanulmányoztam az MRSD-CI (multiple-reference double-excitation configuration-interaction) módszerrel, egy nagy bázis készlet felhasználásával, mely 75 Gauss típusú atomi pályát tartalmazott. Összesen 1340 *ab initio* pontot számoltam, letérképezve egy nagy tartományú $H_3$ molekuláris geometriát. A négy állapot tartalmazta az alap állapotot és a Rydberg 2s $^2A_1'$ és $2p_z$ $^2A_2''$ állapotokat, valamint azt az állapotot, melyben a szabályos háromszög geometriájú elrendeződések az alap állapotra vonatkoznak egy “conical intersection” tekintetében. Elektronikus dipól átmeneti momentumokat szintén számoltam ezen állapotok között. Az eredmények azt mutatták, hogy az atomi és diatomi energetikai aszimptóták pontos leírást nyertek. Az energia gátak, minimumok és különbségek szintén jó egyezést mutattak az irodalmi értékekkel, ahol hozzáférhetőek voltak. Az alap állapot potenciál felület és a $2p_z$ $^2A_2''$ Rydberg állapot sima és szabályos viselkedést mutattak, melyekre a teljes molekuláris geometriai tartományon Morse görbét fektettem spline módszerrel. A másik két potenciál felület olyan bonyolultabb viselkedésről tanúskodott mint “avoided crossings”, így ezek más illesztési procedúrát igenyeltek a globális leíráshoz. Az ide vonatkozó vaskos irodalom szinten összefoglalást nyert.

<u>**Zhengwei Peng, Sandor Kristyan, Aron Kuppermann, James S. Wright:**</u>
<u>**Physical Review A, 52 (1995) 1005-1023**</u>

= = = = = =

Representative equations/tables/figures:

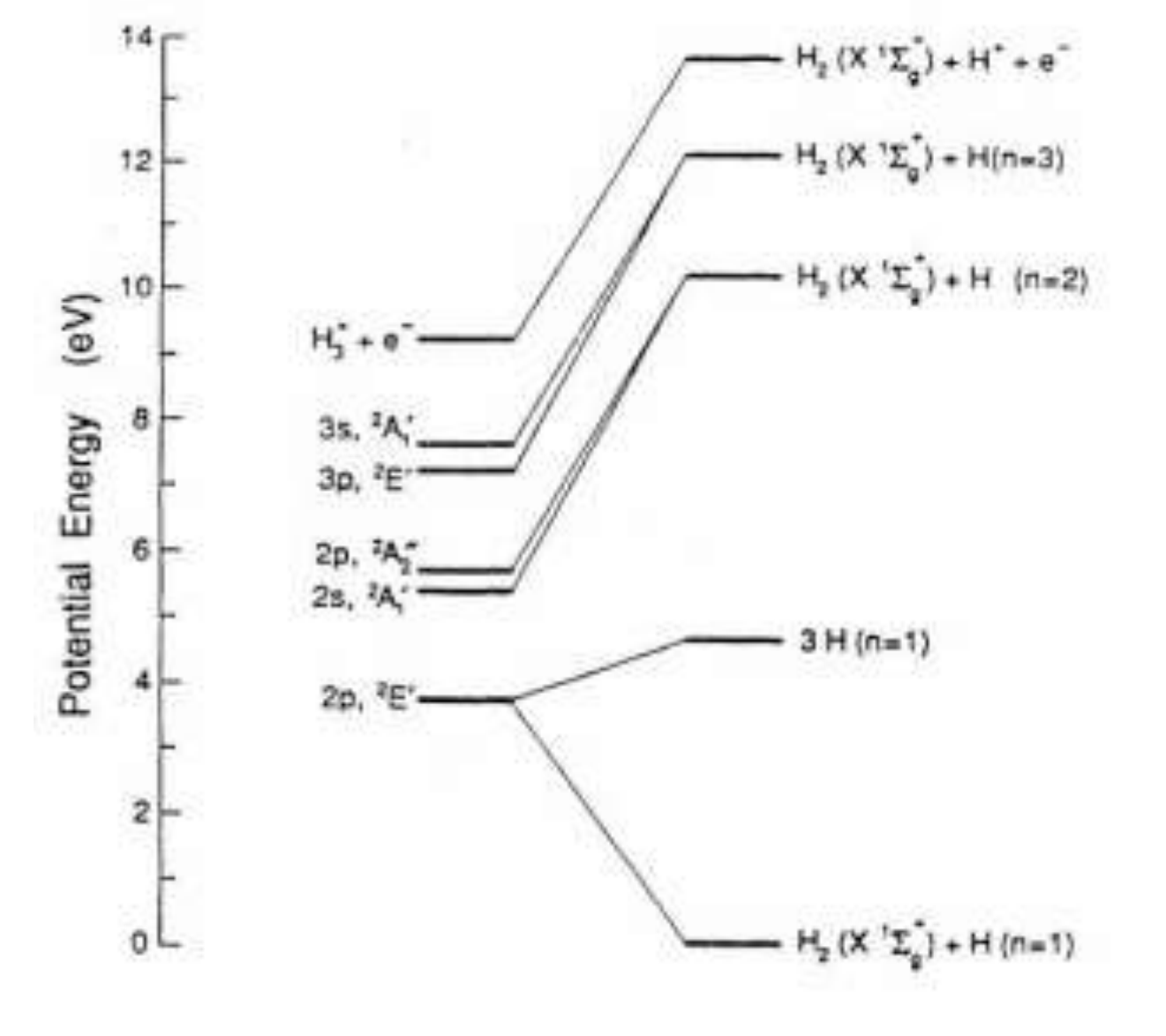

Electronic energy level and correlation diagram of $H_3$. The spacing of the $H_3$ energy levels was calculated for an equilateral triangular configuration with an internuclear distance of 1.64 bohr and referred to the energy of dissociated products by the results of a separated calculation. Since the molecular point group for an equilateral triangle is $D_{3h}$, all electronic states are labeled according to the symmetry representation of this point group, along with the labels of the united-atom limits. In this study, potential energies of the lowest four electronic states of $H_3$ have been calculated: $2p_{xy}\ ^2E'$, $2s\ ^2A_1'$, and $2p_z\ ^2A_2''$.

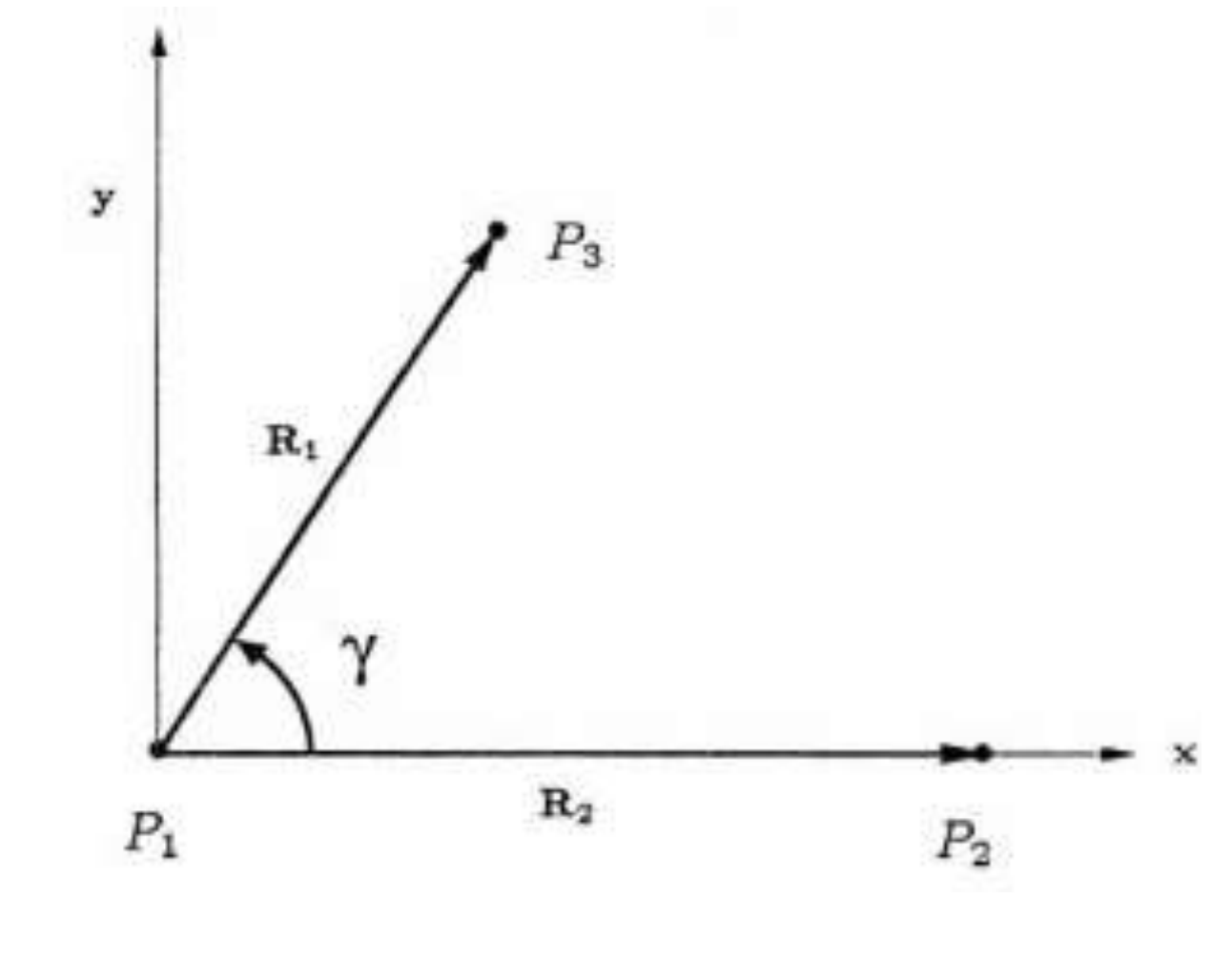

The molecule was located in the xy plane with its geometry described by two bond lengths $R_1$ and $R_2$ and one bond angle $\gamma$. All calculations were carried out using the point group $C_s$ even though at some molecular geometries the symmetry of the molecule could be higher ($D_{3h}$, etc.). In terms of the $C_s$ point group, A'-type functions are symmetric with respect to the xy plane, whereas A"-type functions are antisymmetric.

Using the symmetry notation appropriate for the equilateral triangular ($D_{3h}$) geometry, for simplicity rename the states as:

$E_1$ for the ground state $^2E'(1a'^2\,1e')$,

$E_2$ for the state degenerate with the ground one in the equilateral triangular geometry,

$E_3$ for the 2s $^2A_1'(1a'^2\,2s)$ state, and

$E_4$ for the $2p_z\ ^2A_2''\ (1a'^2\,2p_z)$ state.

The four electronic states of interest labeled as $E_i$, the i=1,2,3 are of A' symmetry, and the i=4 is A" symmetry. Although $E_1$ and $E_2$ are degenerate in the equilateral triangular geometry, such a degeneracy is lifted as soon as the triangle is distorted and this is what generates the conical intersection between the potential energy surfaces of the $E_1$ and $E_2$ states. Notation $E_i$ comes from the first letter of the word energy and is not an indication of state labels (since some of them have E symmetry only). Special attention was given to equilateral triangular ($D_{3h}$) and collinear ($C_{\infty v}$) configurations.

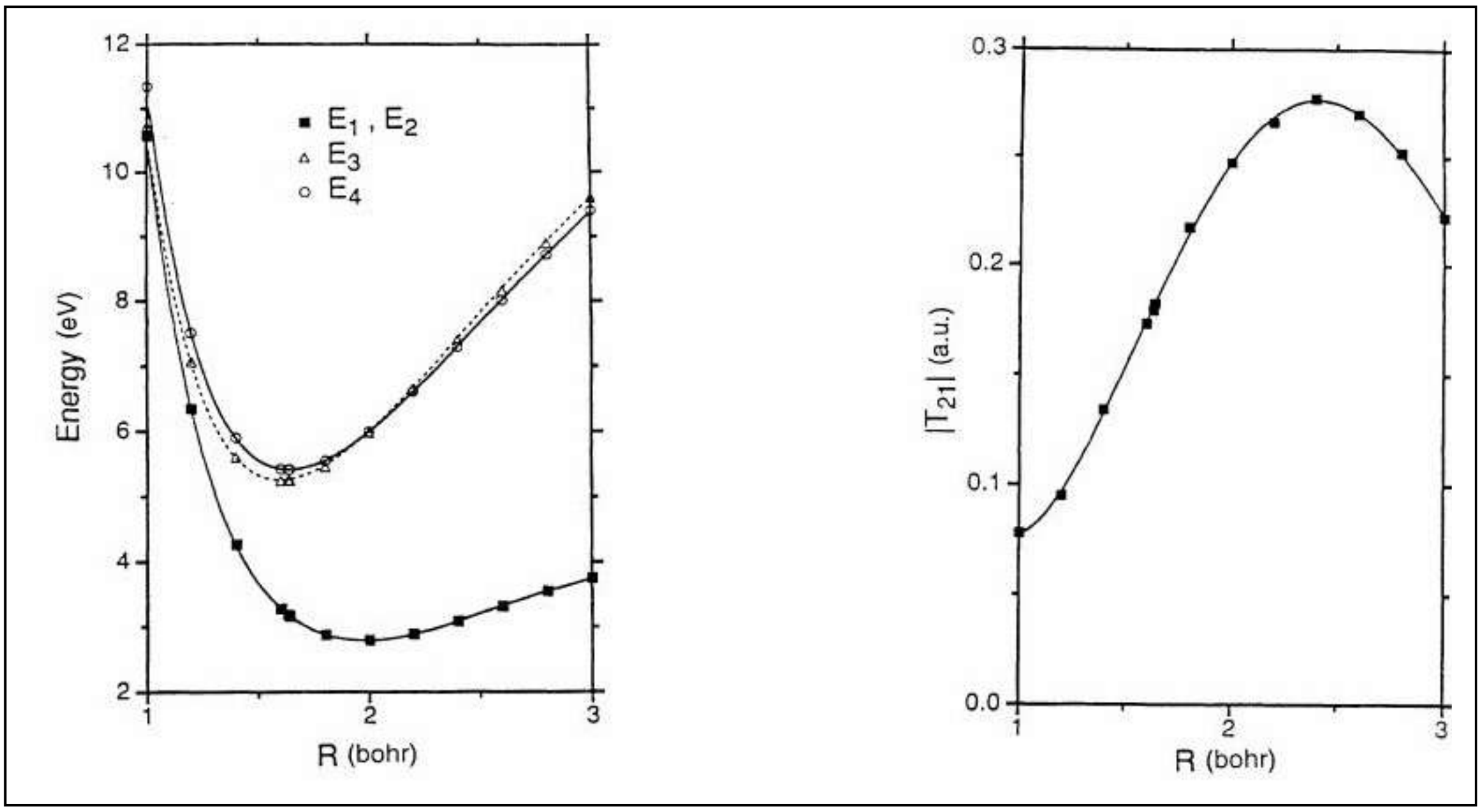

In equilateral triangle $H_3$ with side length R, the potential energy curves (left, in which nuclear configuration the $E_1$ and $E_2$ states are degenerate with each other and the energy origin is the accurate H(ls)+ $H_2(X^1\Sigma_g^+)$ value -0.5-1.174474 = -1.674474 hartree), and the magnitude of the electric dipole transition moment $\mathbf{T}_{21}$ between the $E_2$ and $E_1$ states (right) are plotted above.

Lowest conical intersection energy and its corresponding geometry ($R=R_1=R_2=R_3$) with respect to that of the separated $H+H_2$ configuration:

| | This work | LSTH | DMBE |
|---|---|---|---|
| R(bohr) | 1.973 | 1.981 | 1.973 |
| E(eV) | 2.747 | 2.756 | 2.748 |

(For the SLTH and DMBE surfaces, the accurate H + $H_2$ energy is used as the reference, for the present *ab initio* surface, the energy at the nuclear configuration with $R_1$= 1.402 bohr, $R_2$= 10 bohr, and $R_3$= $R_1$ + $R_2$ = 11.402 bohr is used instead. The difference between the second and the first of these reference energies is 0.040eV.)

Large number of further important data on the electronic potential of $H_3$ sytem, and literature review are discussed in the related article.

A typical view of a general conical intersection:

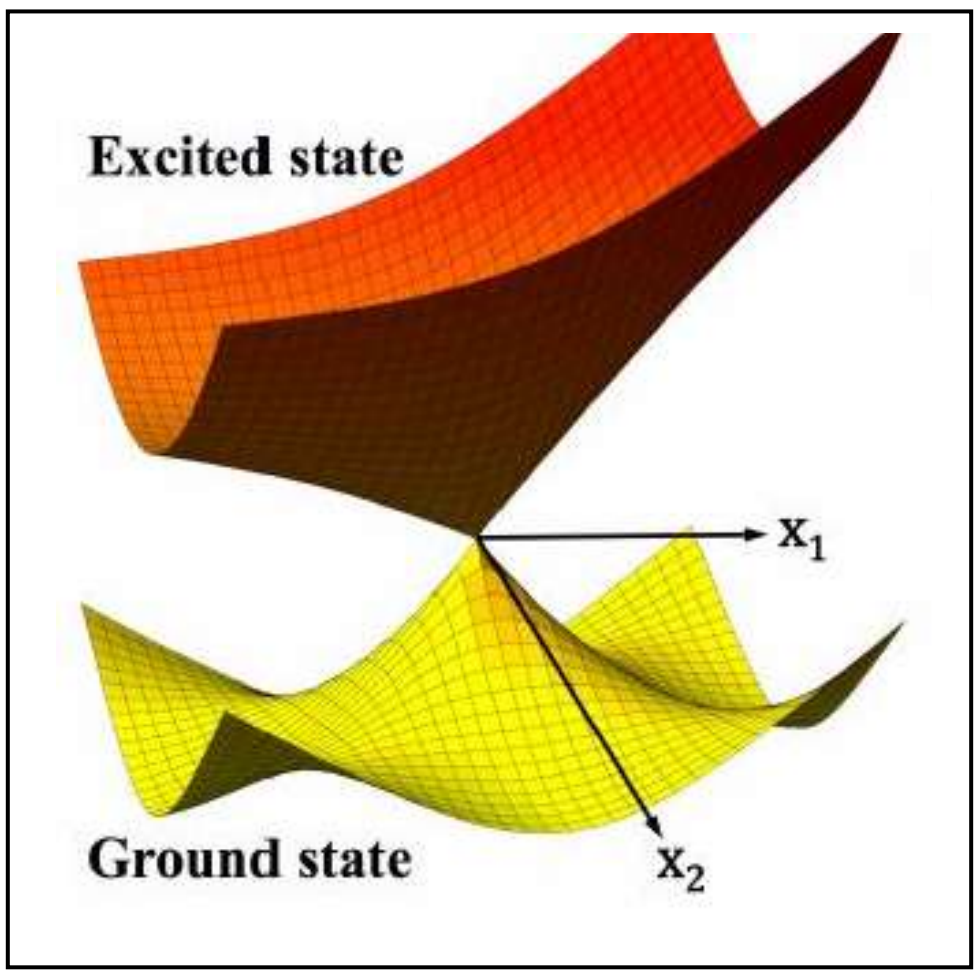

**Thesis-36-Application: Equipotential surfaces of the reaction $H+H_2$: Review on general aspects in its kinetics**

The solution of the complete Schrödinger equation for scattering and electronic structure calculations has been reviewed, focusing on the triatomic systems. The mathematical difficulties of solving the Schrödinger equation are analyzed, what can be done easily and what needs large scale computational work. An overview of basic atomic and molecular quantum mechanics is given, including how adiabatic and diabatic approximations are used in calculating scattering cross sections. Effective interaction potentials are discussed in detail, how to compute them and what they look like. Besides describing how computers have opened the door to real computation of the full spatial behavior of these multicomponent systems, some simple graphical images significantly aid the development of intuition of how the system actually behaves. The graphical representation focuses on the potential energy surfaces of the $H_3$ electronic structure.

= = = = = =

Szemle jelleggel összefoglaltam a teljes (totális) Schrödinger egyenlet megoldási problematikáját az ütközési és elektron szerkezeti számítások esetében, különös tekintettel a három atomos rendszerekre. Analizáltam a Schrödinger egyenlet megoldásának matematikai problémáit, mi az ami könnyen megoldható, és mi szükségel nagy mennyiségű számítási eljárást. Összefoglaltam az atomi és molekuláris kvantum mechanika alapjait, hogyan jön elő benne az adiabatikus és diabatikus közelítések használata az ütközési keresztmetszet számításánál. Részletesen elemeztem az effektíve kölcsönható potenciálokat, hogyan kell őket számítani és hogyan néznek ki. Leírtam, hogy a számítógépek hogyan nyitottak utat valódi számításokhoz amiben leírhatóak a térkoordináták teljes viselkedése ezen multi-komponensű rendszerek esetében, valamint hogy néhány egyszerű grafikai reprezentáció komoly segítséget nyújthat az intuíciók fejlesztésében ha meg akarjuk érteni hogyan viselkednek ezek a rendszerek. Grafikusan ábrázoltam a $H_3$ rendszer elektronikai szerkezetének energia felületeit.

**Sandor Kristyan (REVIEW ARTICLE): Computers in Physics, 8 (1994) 556-575**

= = = = = =

Representative equations/tables/figures:

The potential energy surfaces ($\Psi_{\text{total electronic},k}$, k=0,1,2…) are obtained within the adiabatic or Born–Oppenheimer approximation, in which molecular wave function the motion of nuclei ($\mathbf{R}\equiv(\mathbf{R}_1,\ldots\mathbf{R}_M)$) and electrons ($\mathbf{r}\equiv(\mathbf{r}_1,\ldots\mathbf{r}_N)$) are separated as $\Psi_{\text{total molecular}}(\mathbf{R},\mathbf{r})\approx \Psi_{\text{nuclear}}(\mathbf{R})\Psi_{\text{total electronic}}(\mathbf{r};\mathbf{R})$, where $\Psi_{\text{total electronic}}$ parametrically depends on $\mathbf{R}$ and "total" means the inclusion of nuclear-nuclear attraction energy, while in $\Psi_{\text{total molecular}}$ the "total" means nuclear and electronic motions are included simultaneously. In the neighbourhood of an avoided crossing or conical intersection (e.g. between k= 0 and 1) this assumption fails. A mathematical trick is to perform one unitary transformation (preserving norms, and thus, good for probability amplitudes) from the adiabatic representation to the so-called "diabatic" representation in which the nuclear kinetic energy operator is diagonal. For example, in the diabatic representation, the potential energy surfaces are smoother, so simple functional fit of the surface $\Psi_{\text{total electronic},k}$ (like LSTH or DMBE for $H_3$) capture much of the complexity of the original system. However, one must keep in mind that strictly diabatic states do not exist in the general case.

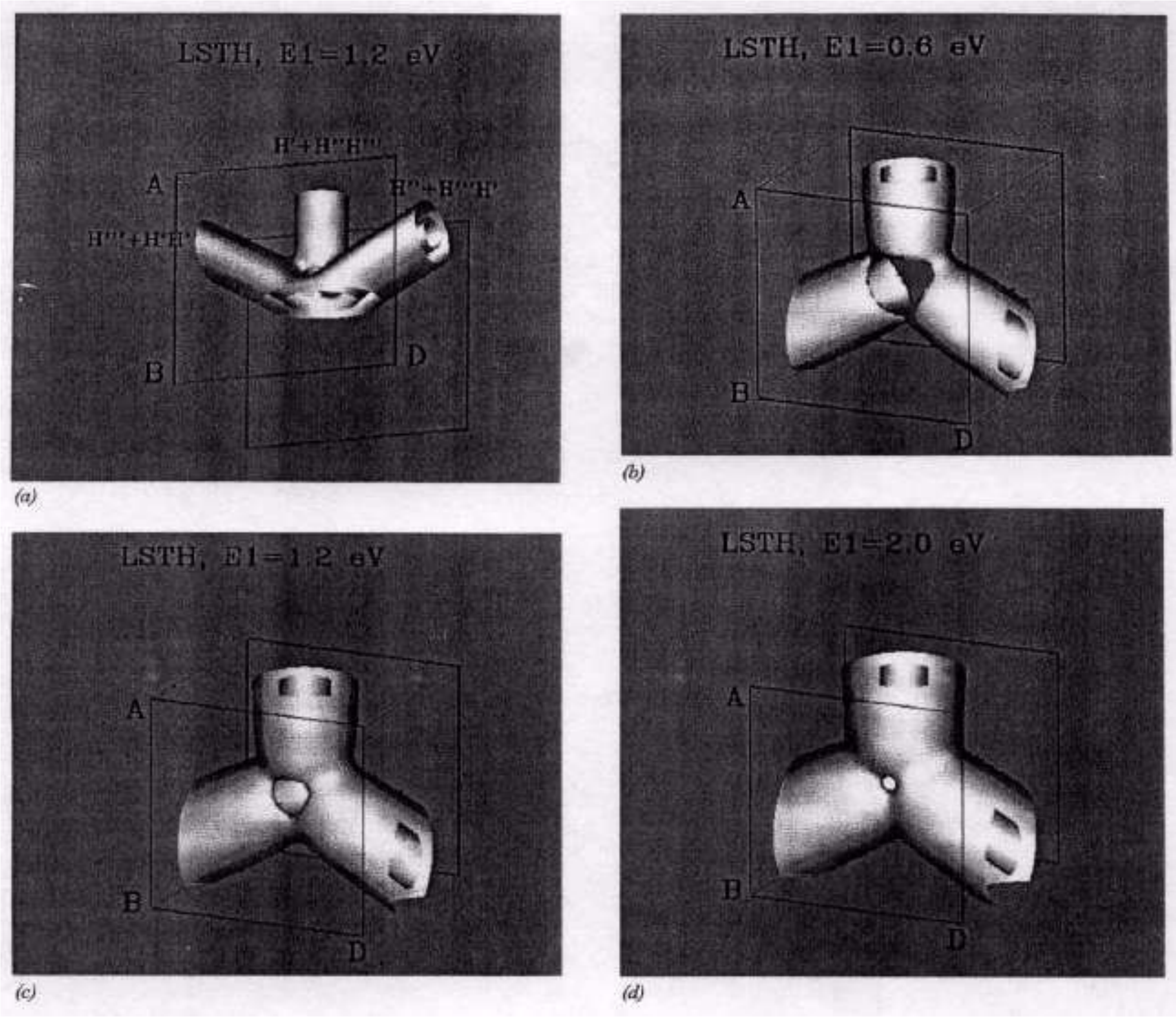

Figures show the LSTH and DMBE adiabatic equipotential surfaces of $E_1$ and $E_2$ states of the $H_3$ system in the XYZ nuclear configuration space; (Stardent AVS plots, a high-tech device in the year published at Caltech). Points A, B, C, D help the eye to orient and place the object among figures, in cube coordinates (±a, ±a, ±a) the a=8.305085 bohr with 50x50x50 mesh resolution in the calculation.

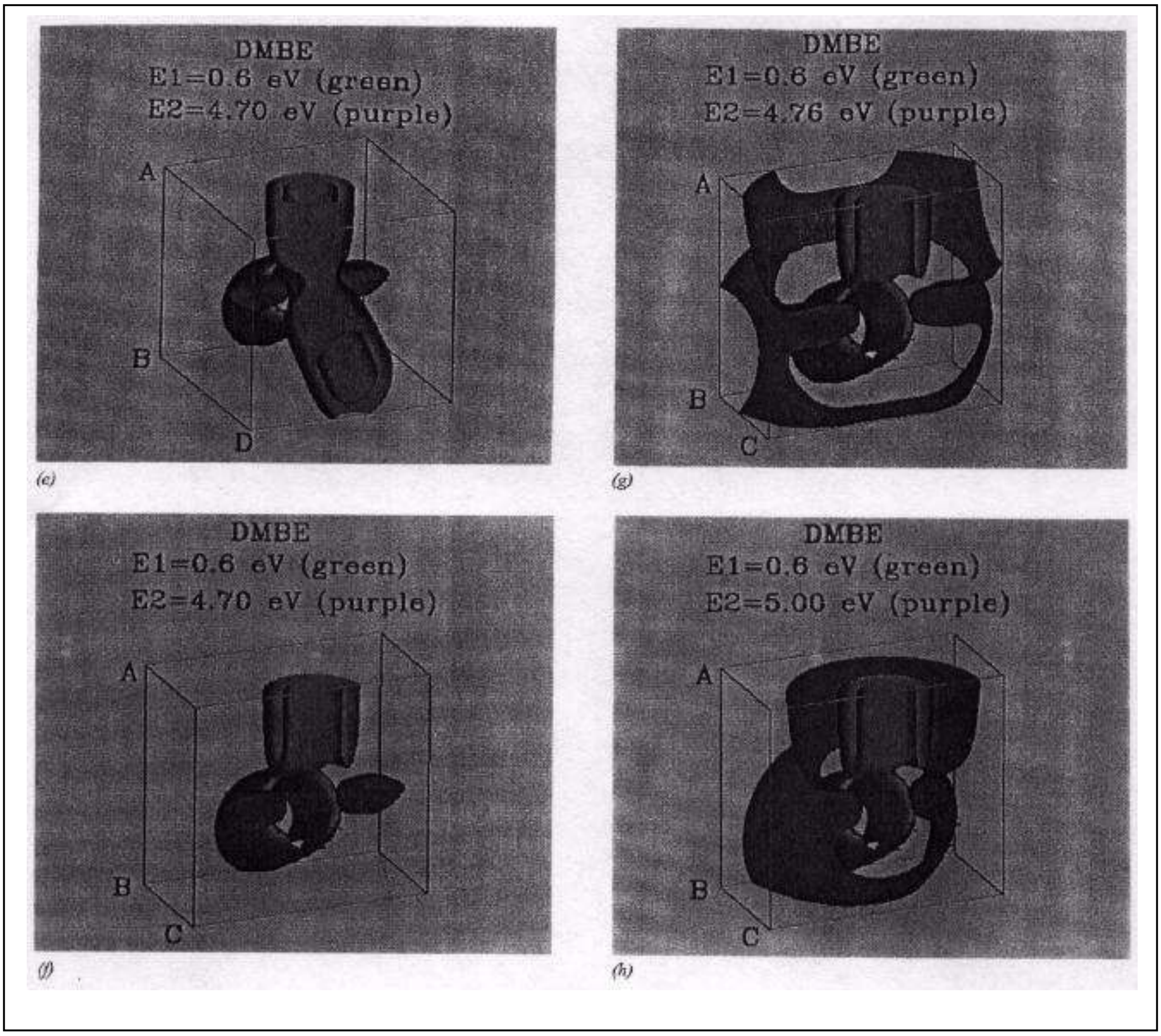

Figure a is a view from below and before the "umbrella opening", the circles at the bottom represent the same collinear nuclear configuration and the highest symmetry axis is the Z axis, a $C_{2v}$ axis, not compalible with the 3 possible outcomes of the $H+H_2$ reaction. The object is not invariant to the interchange of the three equivalent atoms. The 3 possible reaction channels with products {atom H +diatom $H_2$} are noted, the primes distinguish the different H atoms.

Figures b-h show some equipotential surfaces of $E_1$ and $E_2$ adiabatic surfaces after the "umbrella opening". The Y axis (perpendicular to plain ABD) is a $C_{3v}$ axis, the objects are invariant to the interchange of the three equivalent atoms. The line of conical intersection or line of equilateral triangle nuclear configurations of the $H_3$ system is the Y axis.

As a function of energy, the development of the hole can be followed on the equipotential surfaces of the $E_1$ state on b-d, it closes as the energy value increases, and becomes a point on the Y axis first at around 2.75 eV, as well as the relative position of equipotential surfaces of $E_1$ and $E_2$ and the shape development of the equipotential surfaces of $E_2$ can be followed on e-h.

**Thesis-37-Application: Kinetics of the HN+NO reaction**

The reaction of $^3$HN with NO producing H+$N_2O$ and $N_2$+OH has been investigated with the variational RRKM theory using existing potential energy surface data. The bimolecular constant for the loss of the reactants and those for the formation of $N_2O$ and $N_2$ have been calculated and compared with experimental results. The agreement between theory and experiment appears to be satisfactory.

= = = = = =

Létező potenciál energia felületek felhasználásával, és a variációs RRKM elmélet segítségével vizsgáltam a $^3$HN és NO gyökök reakcióját, mely H+$N_2O$ illetve OH+$N_2$ termékeket eredményezhet. Számoltam a bimolekuláris állandót a reaktánsok fogyására, és az $N_2O$ illetve $N_2$ termékek keletkezésére, valamint összehasonlítottam a kísérleti eredményekkel. Az egyezés az elmélet és kísérlet között elfogadhatónak bizonyult.

= = = = = =

Representative equations/tables/figures:

The NH radical plays a significant role in the combustion of nitramine propellants, particularly ADN (ammonium dinitramide, $[O_2N\text{-}N\text{-}NO_2]^-[NH_4]^+$), in which the reaction of $NH_{1\text{-to-}3}$+$NO_{1\text{-to-}2}$ generate major chain carriers, H and OH.

The chemical reaction system under consideration:

$$^3HN + NO \begin{cases} \rightarrow [HNNO] -(k_1)\rightarrow H + N_2O \\ \rightarrow [N_2OH] -(k_2)\rightarrow OH + N_2 \end{cases}$$

Schematic potential surface for the HNNO system:

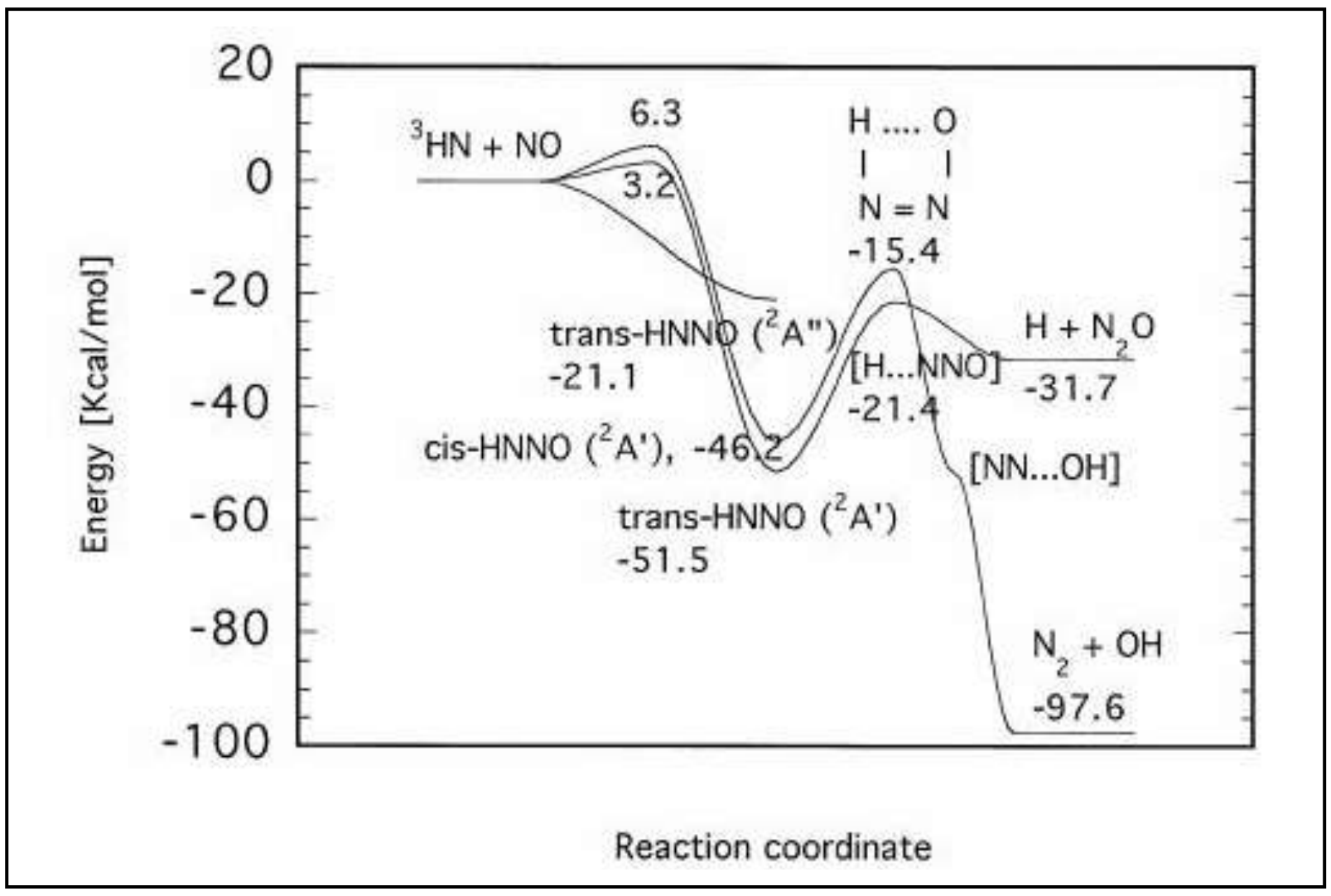

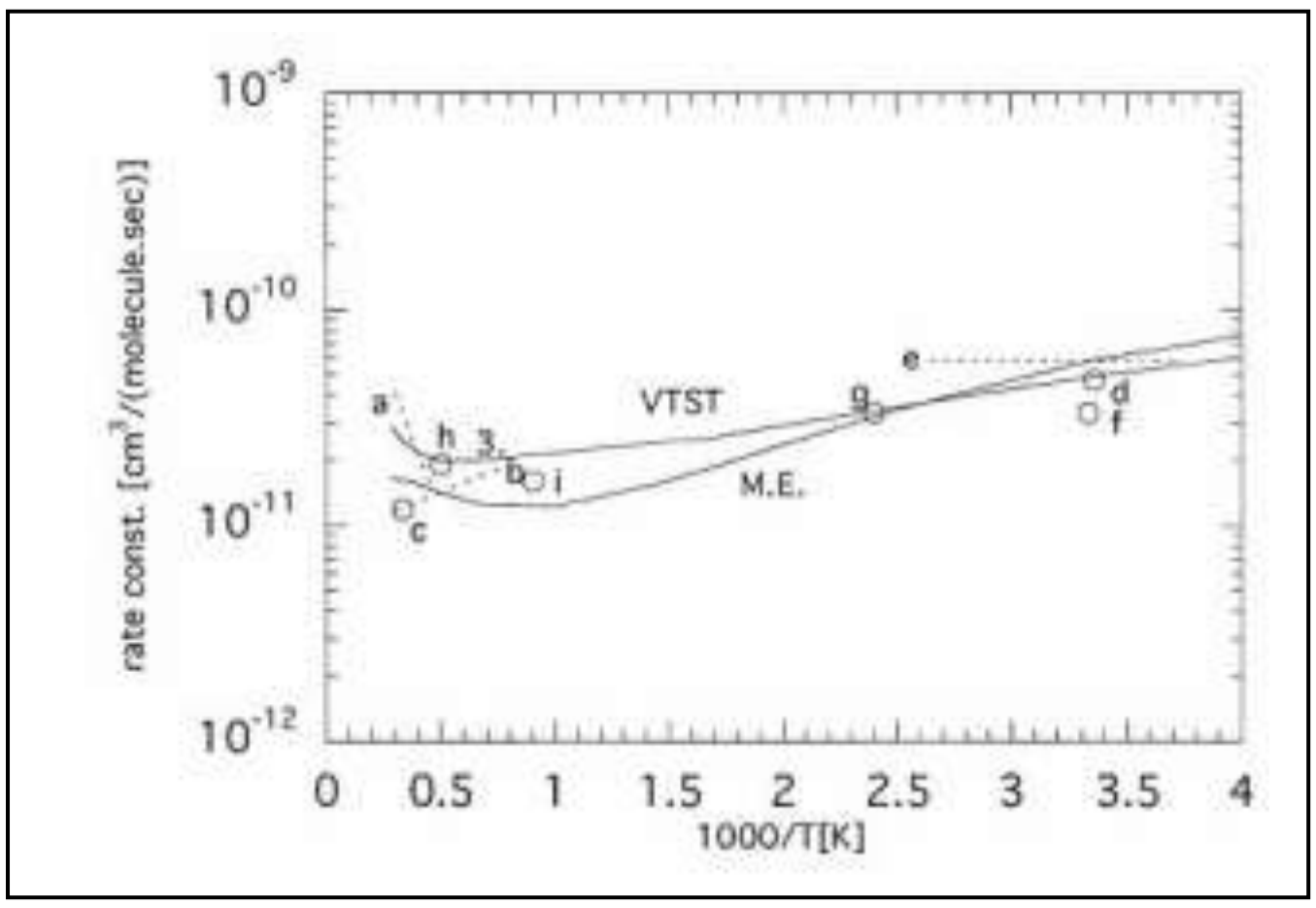

Above: High pressure limit rate for HN+NO loss. Circles and dotted lines: experiments by a. Mertens et al. (109–790 Torr), b. Miller et al., c. Yokoyama et al. (Ar, 210–870 Torr), d. Hansen et al. (298 K, 30–700 Torr), e. Harrison et al. (1 Torr), f. Cox et al., g. Gordon et al., h. Dean et al., and i. Kondo. Theoretical curves: M.E.= VRRKM microscopic rates with analytical solution of the master equation (using Beyer–Swinehart state count up to 20 kcal/mol and Whitten–Rabinowitch thereafter, 0.25 kcal/mol energy step size for the numerical integration); VTST = variational transition state (with extrapolated frequencies from HNNO ($^2$A$'$) adduct with 1.0 A$^{\circ -1}$ fitted exponential constant for vanishing frequencies).

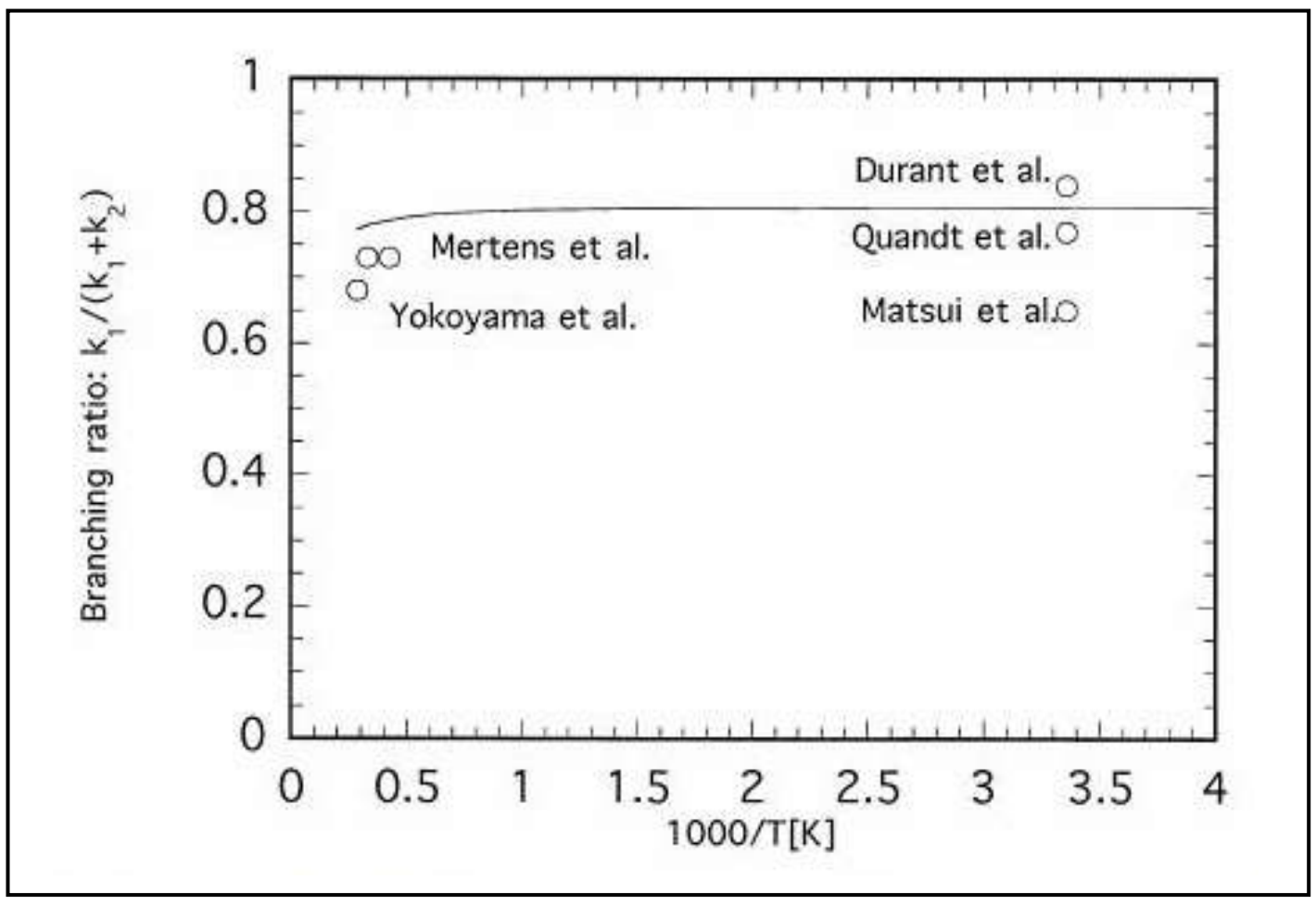

Above: Branching ratios for the two products of the reaction HN+NO (circles = experimental points; solid lines = theoretical calculation). The theoretical curve was calculated by VRRKM microscopic rates with analytical solution of the master equation. (Ar bath gas, weak collision model with 200 cm$^{-1}$, P=10–15200 Torr). Experimental points are marked with the authors.

As summary, a comparison of the collected available experimental data with the theoretical ones indicates that the VRRKM theory with the master equation can reproduce experimental data reasonably. However, these rate estimation methods and the accuracy of the potential surface calculations must be improved, and on the other hand, experimental results also suffer from large errors.

**Thesis-38-Application: Kinetics of the CN + NO reaction**

A pulsed-laser photolysis/laser-induced fluorescence technique was employed in the determination of the pressure and temperature dependence of the reaction of CN with NO in the range from 207 to 740 K and for Ar bath gas pressures ranging from 30 to 900 Torr. The variational RRKM model coupled with the one-dimensional master equation treatment has provided a reasonably satisfactory description of the available kinetic data for both, the association and dissociation processes. The association kinetic data was best fit by a collisional energy transfer parameter for the simple exponential down model gradually increasing from -35 $cm^{-1}$ at 100 K to -500 $cm^{-1}$ at 740 K. The expression reproducing the theoretical estimates for the high-pressure rate constant in the range from 207 to 740 K is $3.4x10^{-10}$exp(120/T[K]) $cm^3s^{-1}$.

= = = = = =

A CN+NO reakció nyomás és hőmérséklet függését vizsgáltuk lézerimpulzussal gerjesztett fotolízis/lézerfénnyel gerjesztett fluoreszcencia technikával a 207-740 K-es hőmérséklet és 30-900 Torr nyomású Ar fürdő (bath) gáz tartományban. A variációs RRKM modell összekapcsolva az egydimenziós mester egyenlettel elfogathatóan írta le a rendelkezésre álló kinetikai adatokat az asszociációs és disszociációs folyamatok esetében egyaránt. Az asszociációs kinetikai adatok legjobb illesztése az ütközési energia átadás (collisional energy transfer) paraméter segítségével történt a „simple exponential down model"-ben, mely fokozatosan növekszik -35 $cm^{-1}$ –től (100K) -500 $cm^{-1}$ –ig (740 K). Az elméleti közelítés a nagy nyomás tartománybeli sebességi állandóra 207 és 740 K között: $3.4x10^{-10}$exp(120/T[K]) $cm^3s^{-1}$.

**S.J.Klippenstein, D.L.Yang, T.Yu, Sandor Kristyan, M.C.Lin, S.H.Robertson: Journal of Physical Chemistry A, 102 (1998) 6973-6980**

= = = = = =

Representative equations/tables/figures:

*Ab Initio* calculations: Optimized molecular structures and vibrational frequencies were obtained for CN, NO, NCNO in the ground state singlet ($S_0$) and triplet states ($T_1$), and CNNO with various electron correlation procedures, including second-order Møller Plesset perturbation theory (MP2), configuration interaction with single and double excitations (CISD), coupled cluster techniques incorporating single and double substitutions (CCSD), and density functional theory employing the Becke3-Lee-Yang-Parr (B3LYP/6-31G*) functional. The relative energies of these same species were obtained at the G2(MP2) level of theory. The GAUSSIAN92/DFT quantum chemical software was employed in each of these evaluations. Sample analyses of the transition state separating CNNO from CO + N2 (as determined via a reaction path analysis) are also provided. This transition state is only of importance to the kinetics at temperatures higher than those considered here, and so the corresponding analyses were restricted to MP2/6-31G*, B3LYP/6-31G*, and G2(MP2) evaluations.

Kinetics:

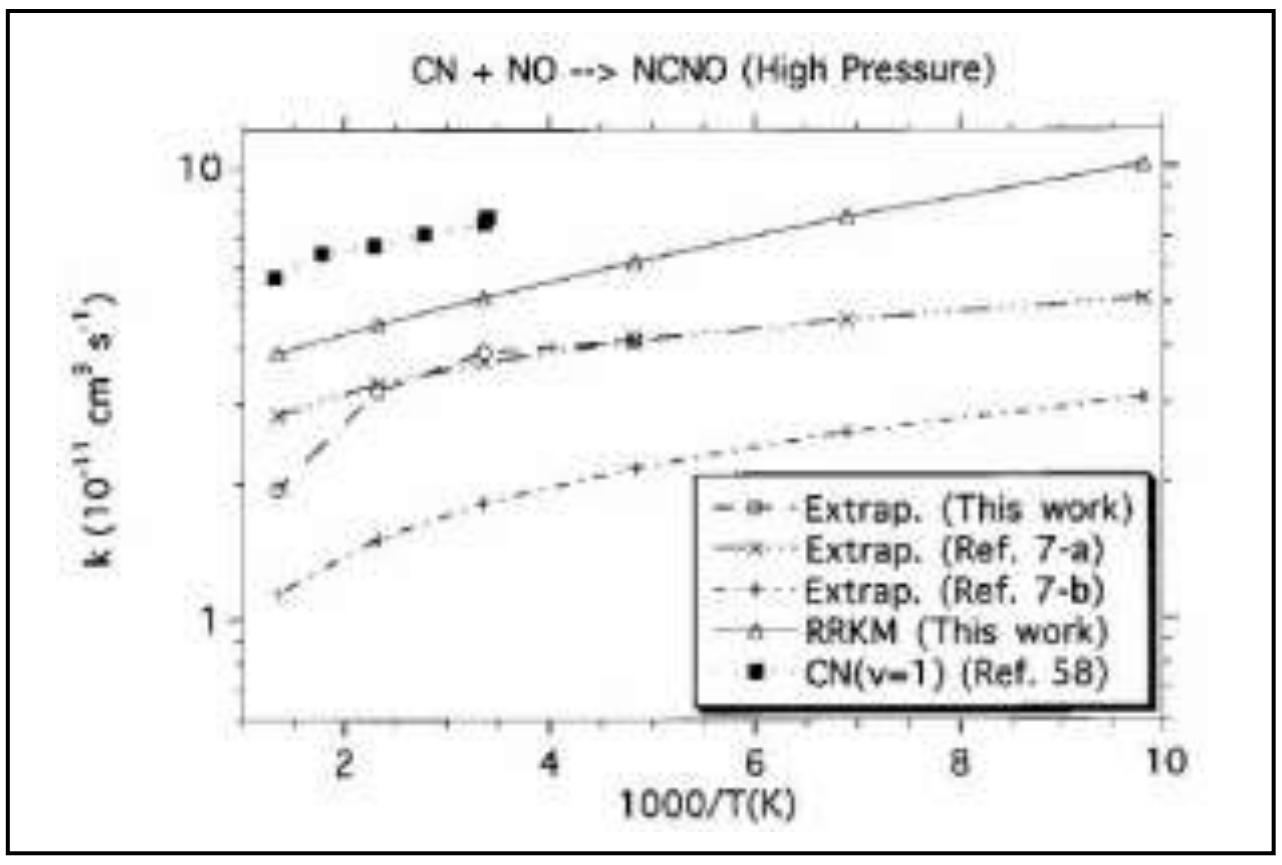

Above: Plot of the temperature dependence for the <u>high-pressure limiting thermal association rate constant</u>. The filled squares denote the CN vibrational relaxation data of Sims and Smith, the circles denote the Troe-based extrapolations of the present work, the crosses and pluses denote the corresponding extrapolations of Sims and Smith, and the triangles denote the present RRKM estimates. The lines are provided as a guide to the eye.

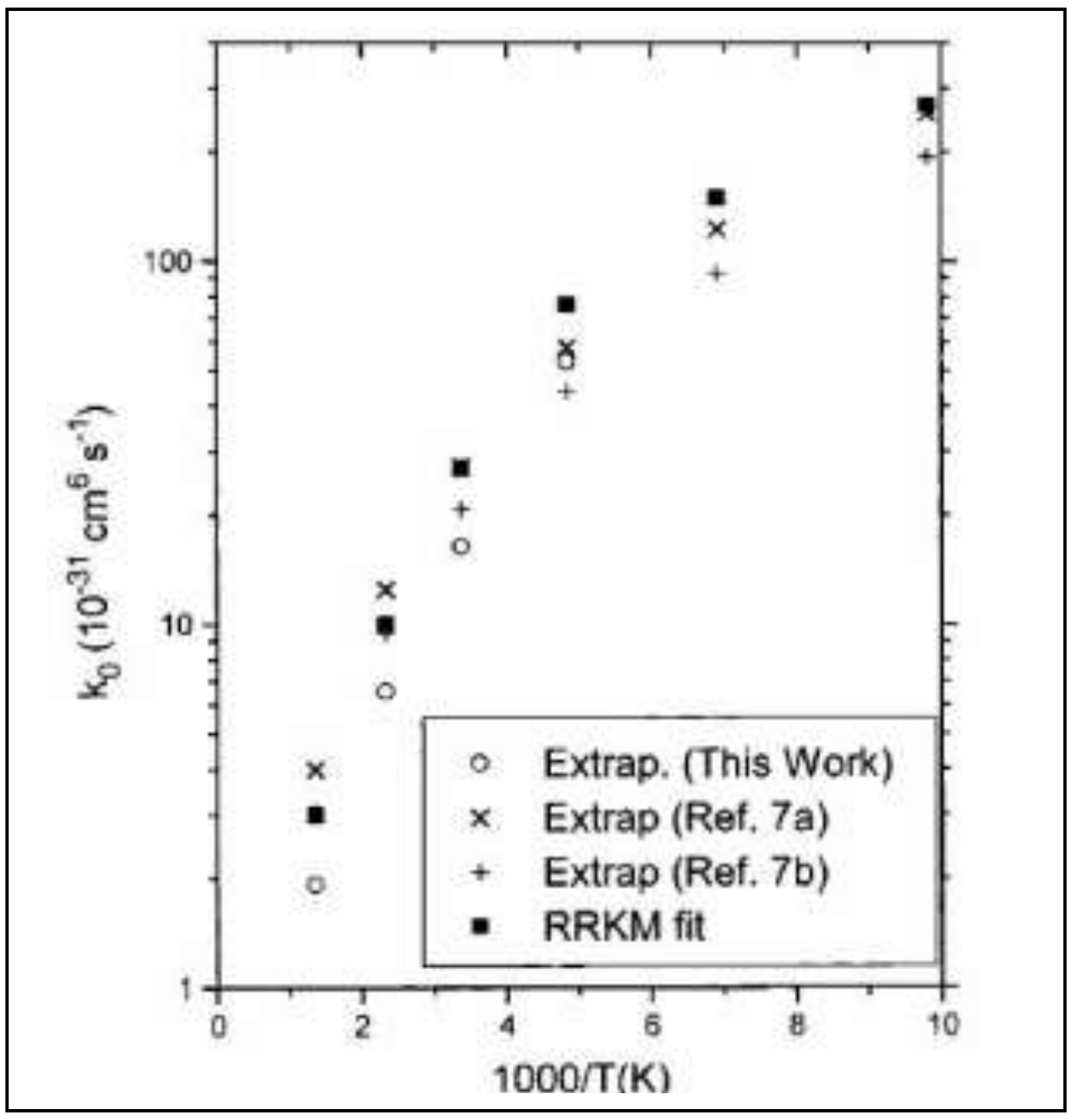

Above: Plot of the temperature dependence of the <u>low-pressure limiting thermal association rate constant</u>. The circles, pluses, and crosses are as in previous figure. The filled squares denote the present RRKM estimates employing at each temperature the „collisional energy transfer parameter" value that provides the best fit to the observed pressure dependence.

**Thesis-39-Application: Unimolecular decomposition of the phenyl radical**

The unimolecular decomposition of the $C_6H_5$ radical has been studied by *ab initio* molecular orbital and statistical-theory calculations. Three low-energy decomposition channels (including the commonly assumed *n*-$C_4H_3$+$C_2H_2$) have been identified using B3LYP/6-31G(d) level of theory, modified Gaussian-2 method, and some others to compare. RRKM calculations have been carried out for the production of *n*-$C_4H_3$+$C_2H_2$, *l*-$C_6H_4$+H, and *o*-$C_6H_4$+H with the coupled multi-channel mechanism. At T < 1500 K *o*-$C_6H_4$ is the major product of the decomposition reaction, above 1500 K the formation of *l*-$C_6H_4$ becomes competitive, however, the formation of the commonly assumed *n*-$C_4H_3$+$C_2H_2$ products was found to be least competitive. Rate constants for all three product channels have been calculated as functions of temperature and pressure for practical applications.

= = = = = =

A $C_6H_5$ gyök unimolekuláris bomlását tanulmányoztam molekuláris *ab initio* és kinetikai számításokkal. Három alacsony energiájú bomlási utat (közte az általánosan feltételezett *n*-$C_4H_3$+$C_2H_2$ termék fele vezetőt) azonosítottam, felhasználva a B3LYP/6-31G(d) szintű elméletet, a módosított Gaussian-2 módszert, és másokat összehasonlítás végett. Végeztem RRKM számításokat a *n*-$C_4H_3$+$C_2H_2$, *l*-$C_6H_4$+H, és *o*-$C_6H_4$+H termékekre mint (csatolt) több-utas mechanizmusra. T < 1500 K hőmérsékleten *o*-$C_6H_4$ a fő bomlási terméke a reakciónak, 1500 K felett a *l*-$C_6H_4$ képződése válik számottevővé, azonban úgy találtam, hogy az általánosan feltételezett *n*-$C_4H_3$+$C_2H_2$ termék képződése a legkevésbé valószínű. A háromféle termékképződés sebességi állandóit számítottam mint a hőmérséklet és nyomás függvényét praktikus alkalmazások számára.

**L.K.Madden, L.V.Moskaleva, S.Kristyan, M.C.Lin:**
**Journal of Physical Chemistry A, 101 (1997) 6790-6797**
= = = = = =

Representative equations/tables/figures:

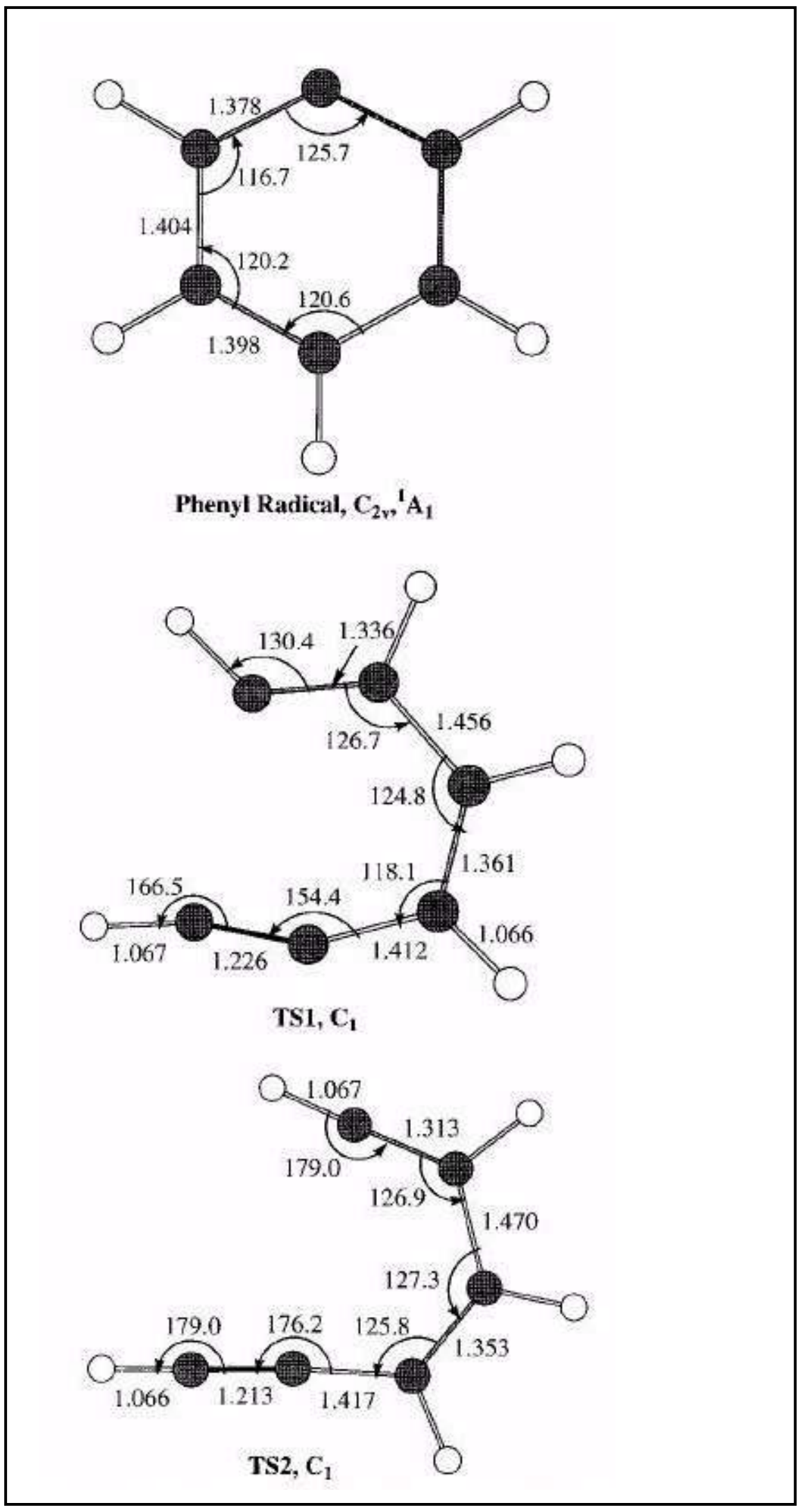

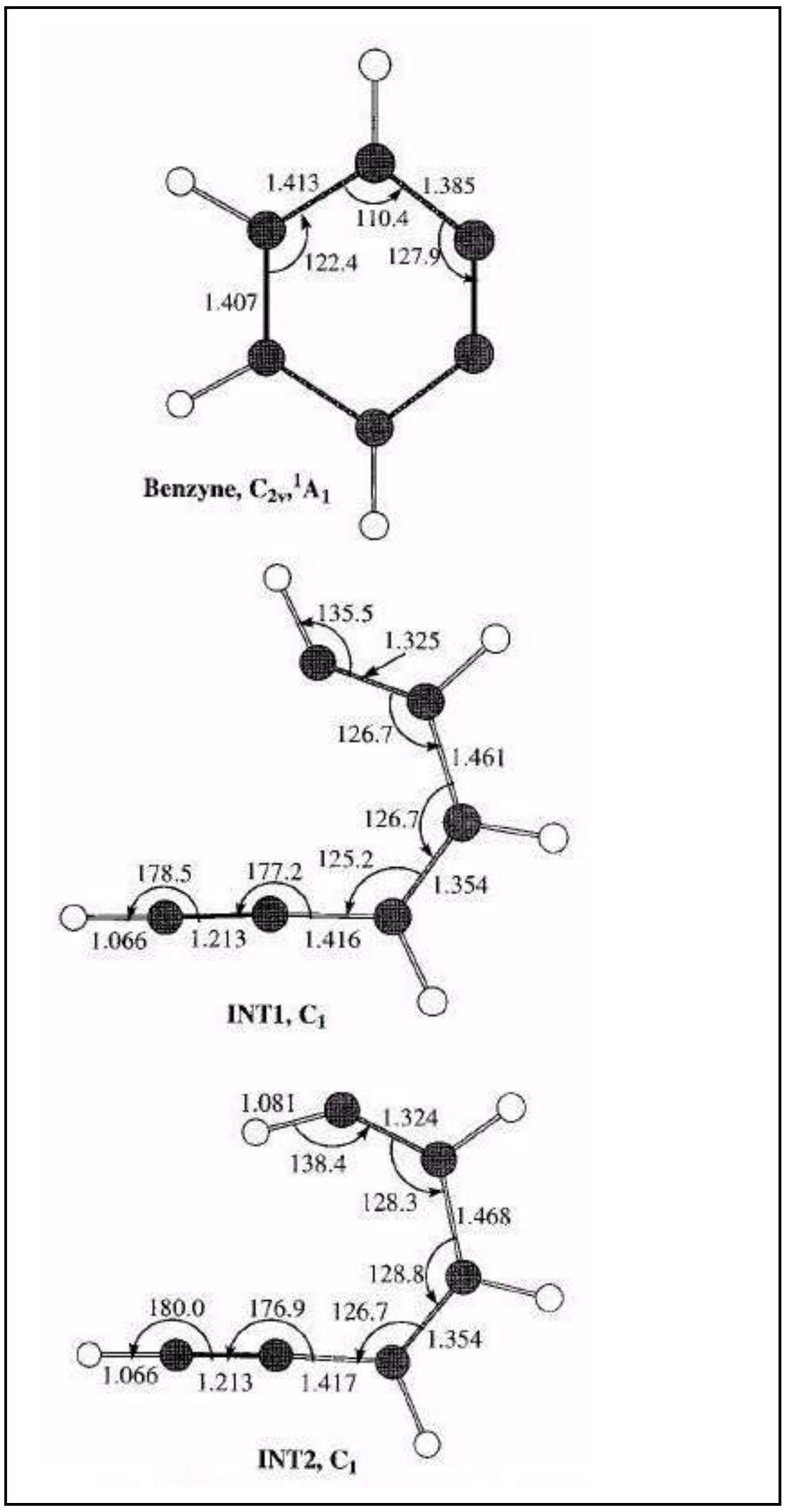

Above and below, the geometries of various species involved in the decomposition of c-$C_6H_5$ calculated at the B3LYP/6-31G(d) level of theory.

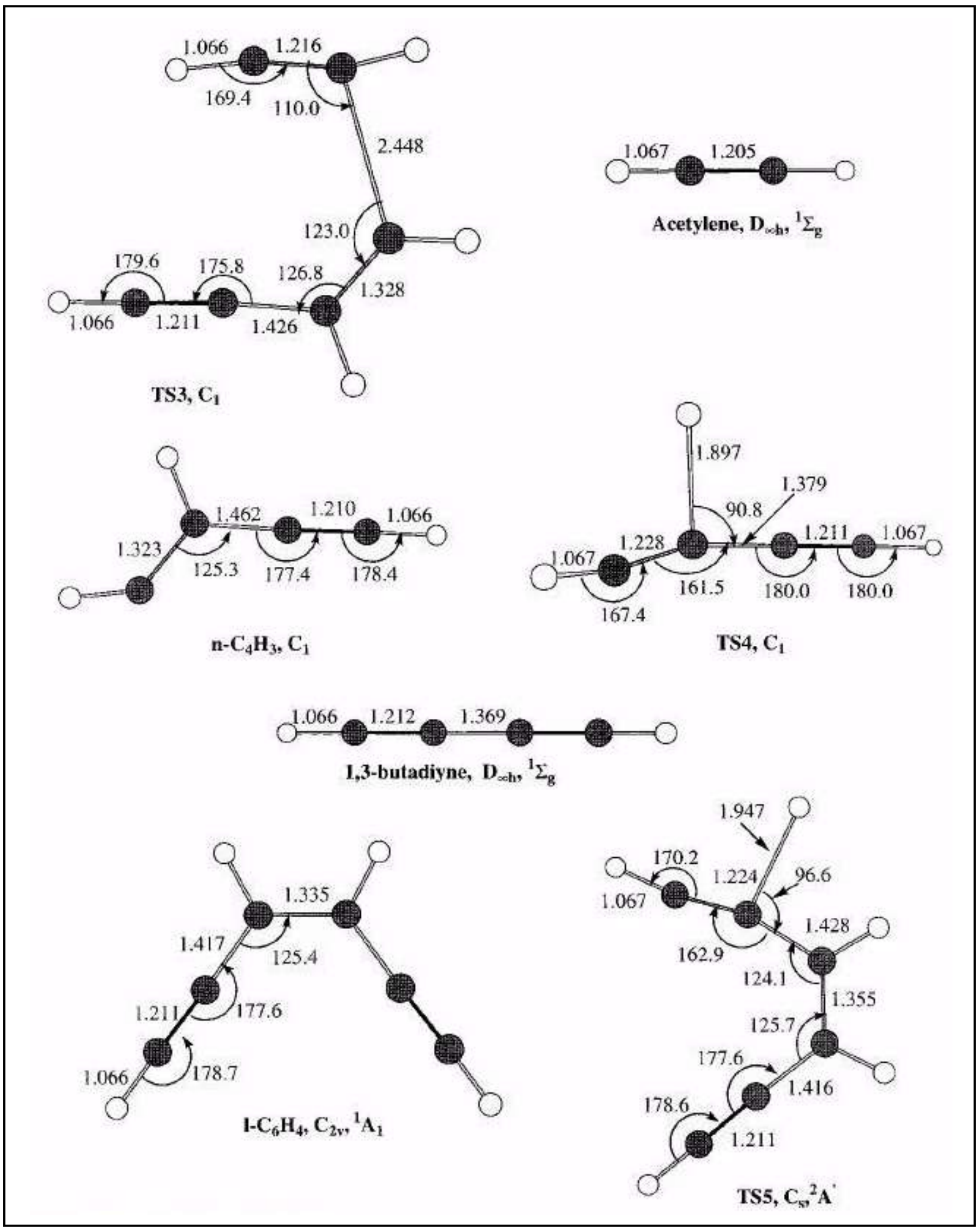

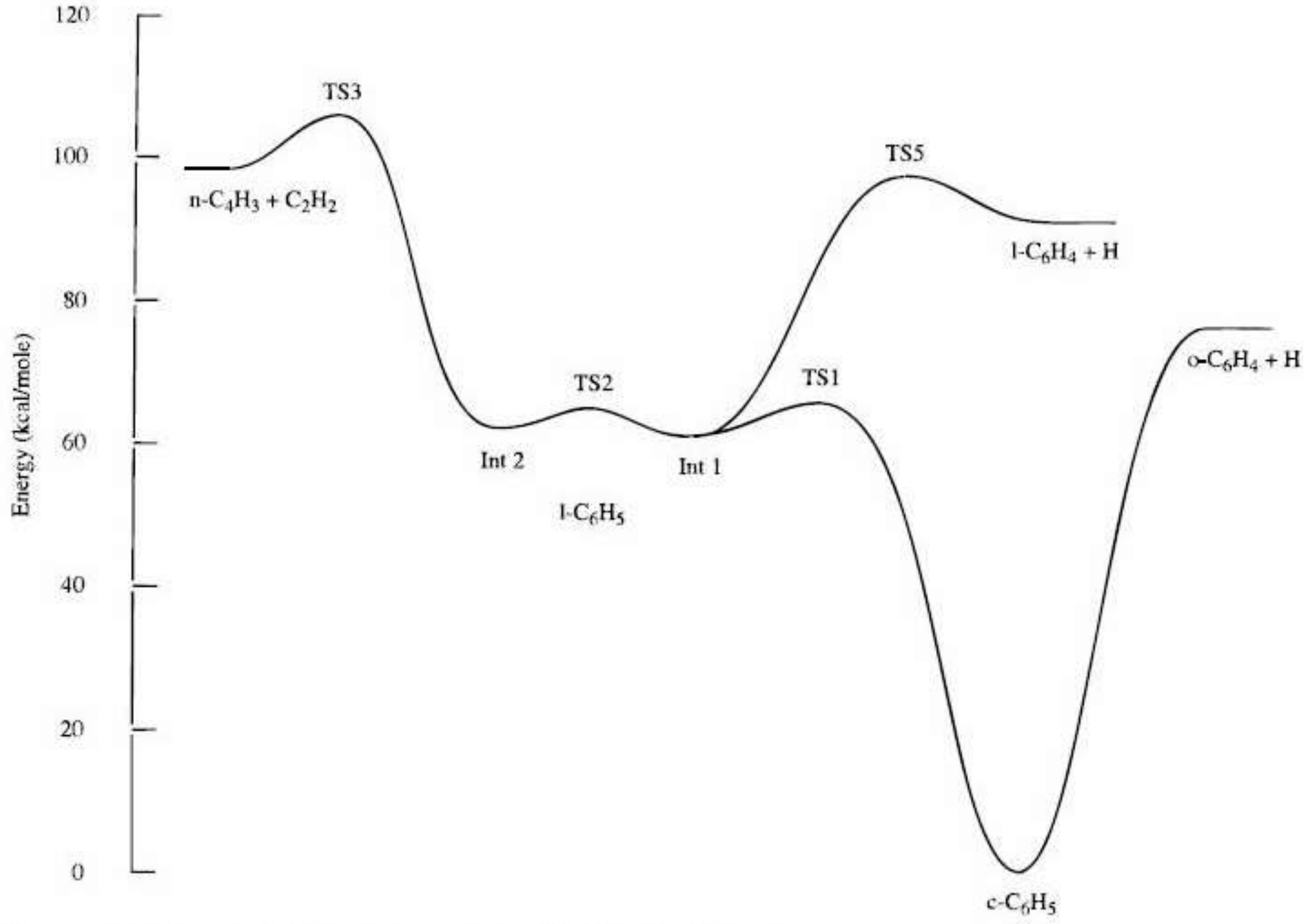

Potential energy surface of the $C_6H_5$ system at the G2M level of theory.

**Thesis-40-Application: Kinetics of the H + $HNO_3$ reaction**

The kinetics and mechanism of the H + $HNO_3$ reaction have been elucidated with *ab initio* molecular orbital and statistical-theory calculations reproducing the experimental data remarkably well. The reaction is dominated by an indirect metathetical process taking place via vibrationally excited dihydroxyl nitroxide, $ON(OH)_2$, producing OH + cis-HONO. The excited $ON(OH)_2$ also undergoes molecular elimination, yielding $H_2O$ + $NO_2$ as a minor competing reaction. The direct H abstraction reaction forming $H_2$ + $NO_3$ was found to be the least important one. At atmospheric pressure, the three rate constants have been found in units of $cm^3$/(molecule×sec) from the 300–3000 K temperature range as

$$H + HNO_3 \rightarrow H_2 + NO_3, \quad k_a = (9.24\times10^{-16})T^{1.53}e^{-8253/T},$$
$$\rightarrow OH + cis\text{-}HONO, \quad k_b = (6.35\times10^{-19})T^{2.30}e^{-3511/T},$$
$$\rightarrow H_2O + NO_2, \quad k_c = (1.01\times10^{-22})T^{3.29}e^{-3163/T}.$$

The direct mechanism $k_a$ is from conventional transition-state theory (CTST, Eyring-Polanyi) calculations, while indirect mechanisms $k_b$ and $k_c$ are from Arrhenius fits to the solution of the master equation which includes Rice–Ramsperger–Kassel–Marcus (RRKM) microscopic rate constants and tunneling corrections (intermediate $ON(OH)_2$ is long-lived).

= = = = = =

A H + $HNO_3$ reakció kinetikájának és mechanizmusának magyarázata történt molekuláris *ab initio* és kinetikai számításokkal, melyek jól leírják a kísérleti eredményeket. Egy indirekt metatézis dominálja a reakciót a vibrációsan gerjesztett dihidroxil-nitroxidon ($ON(OH)_2$) keresztül, mely OH + cis-HONO termékekhez vezet. A gerjesztett $ON(OH)_2$ kisebb mértékben szintén elreagál $H_2O$ + $NO_2$ termékekké. A direkt hidrogén elvonás $H_2$ + $NO_3$ termékekké a legkevésbé jelentős a folyamatok közül. Atmoszféra nyomáson, a három sebességi állandó [$cm^3$/(molecule×sec)] a 300–3000 K hőmérsékleti tartományban

$$H + HNO_3 \rightarrow H_2 + NO_3, \quad k_a = (9.24\times10^{-16})T^{1.53}e^{-8253/T},$$
$$\rightarrow OH + cis\text{-}HONO, \quad k_b = (6.35\times10^{-19})T^{2.30}e^{-3511/T},$$
$$\rightarrow H_2O + NO_2, \quad k_c = (1.01\times10^{-22})T^{3.29}e^{-3163/T}.$$

A direkt mechanizmus $k_a$ értékét a konvencionális átmeneti állapotok elmélete (CTST, Eyring-Polanyi) szerinti számítás szolgáltatta, míg az indirekt mechanizmus $k_b$ és $k_c$ értékeit az Arrhenius illesztés a mester egyenlet megoldására, utóbbiban felhasználva a Rice–Ramsperger–Kassel–Marcus (RRKM) mikroszkopikus sebességi állandókat es az alagút effektus korrekciót (a közbenső $ON(OH)_2$ termék hosszú életű).

**J.W.Boughton, Sandor Kristyan, M.C.Lin: Chemical Physics, 214 (1997) 219-227**

= = = = = =

Representative equations/tables/figures:

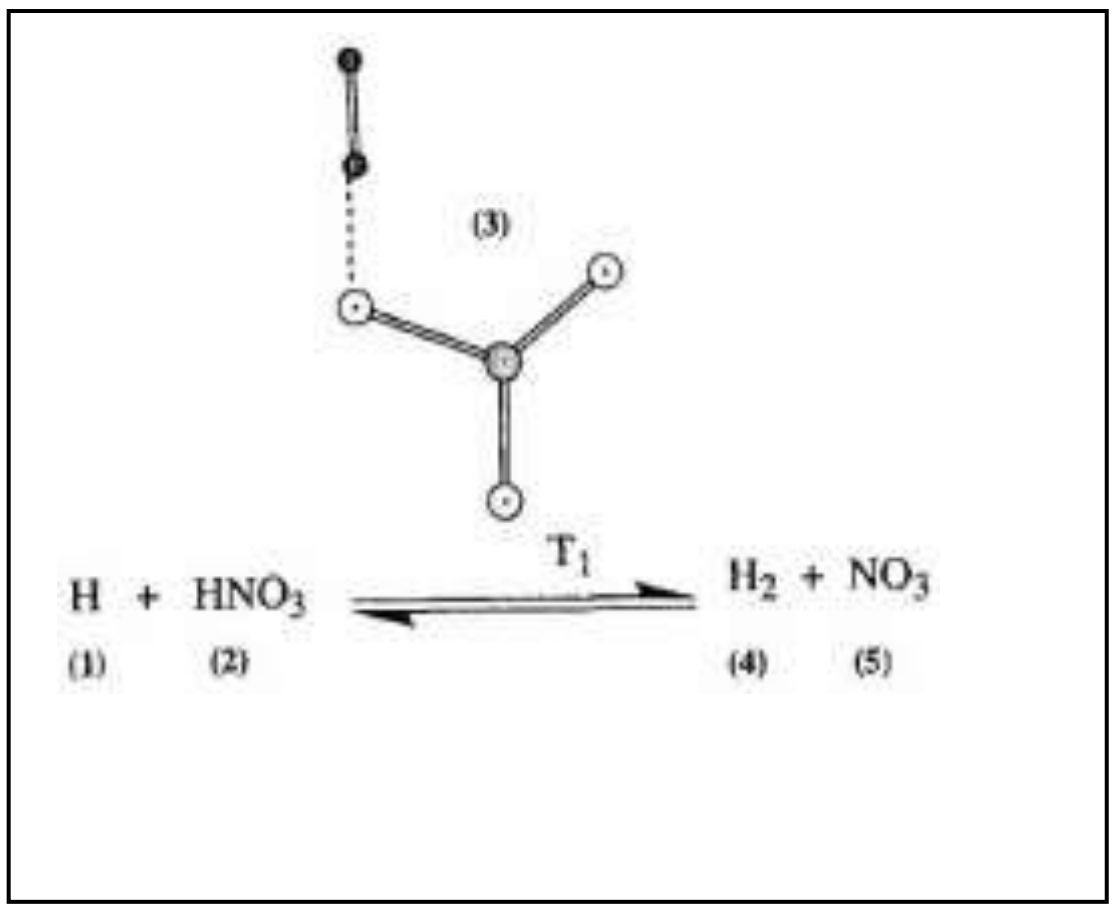

Above, scheme for reaction (a), below, for (b) and (c):

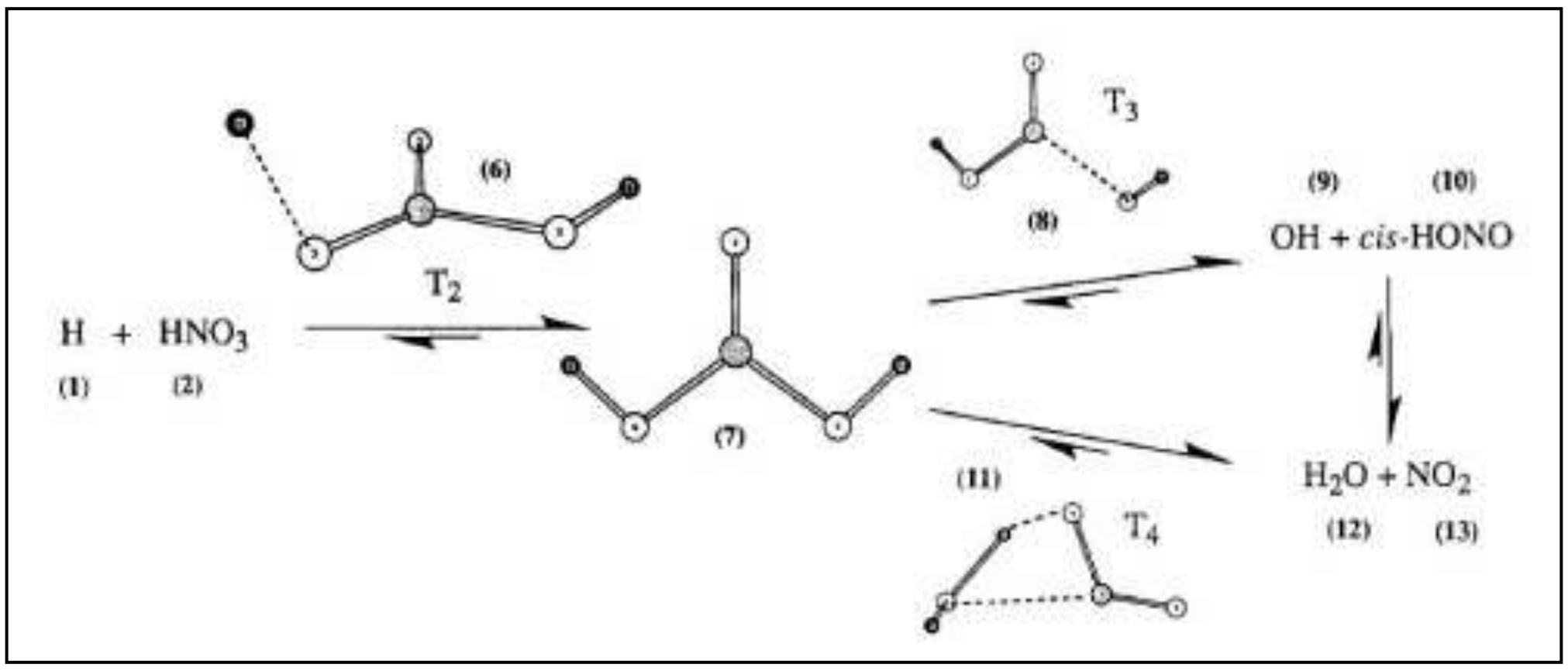

Electronic energies (hartree) at various levels of theory and relative energies (kcal/mol) at the G2M level

| Species [a] | | Becke3LYP /6-311G(d,p) | ZPE [b] (kcal/M) | Projected MP2 /6-311G(d,p) | Projected MP2 /6-311 + G(3df,2p) | RCCSD(T) /6-311G(d,p) | G2M [c]-ZPE | G2M relative |
|---|---|---|---|---|---|---|---|---|
| (1) | H | | | | | | −0.500000 | |
| (2) | $HNO_3$ | −280.967355 | 16.599 | −280.312654 | −280.491220 | −280.337065 | −280.580911 | |
| (4) | $H_2$ | −1.179572 | 6.318 | −1.160249 | −1.162725 | −1.16834 | −1.176256 | |
| (5) | $NO_3$ | −280.298542 | 6.711 | −279.662849 | −279.838761 | −279.666701 | −279.902643 | |
| (9) | OH | −75.754521 | 5.285 | −75.574339 | −75.619447 | −75.589033 | −75.650651 | |
| (10) | cis-HONO | −205.762146 | 12.697 | −205.274915 | −205.402801 | −205.301958 | −205.478804 | |
| (12) | $H_2O$ | −76.447444 | 13.374 | −76.263883 | −76.318258 | −76.276122 | −76.352257 | |
| (13) | $NO_2$ | −205.132704 | 5.527 | −204.660467 | −204.785484 | −204.672491 | −204.841218 | |
| | $H_2O + NO_2$ | −281.580148 | 18.901 | | | | −281.193475 | 0.0 |
| | OH + cis-HONO | −281.516667 | 17.982 | | | | −281.129455 | 39.254 |
| (8) | $T_3$ | −281.519255 | 20.297 | −280.837953 | −281.013492 | −280.880763 | −281.121772 | 46.390 |
| (7) | $ON(HO)_2$ | −281.533785 | 22.078 | −280.857602 | −281.038912 | −280.893459 | −281.140239 | 36.583 |
| (6) | H-($HNO_3$), $T_2$ | −281.463651 | 17.287 | −280.784588 | −280.964944 | −280.821005 | −281.066831 | 77.856 |
| | H + $HNO_3$ | −281.467165 | 16.599 | | | | −281.080911 | 68.333 |
| (3) | HH-$NO_3$, $T_1$ | −281.455030 | 14.480 | −280.778850 | −280.957383 | −280.805472 | −281.049475 | 85.940 |
| | $H_2 + NO_3$ | −281.478114 | 13.029 | | | | −281.078899 | 66.026 |

a. Geometries are optimized at the DFT Becke3LYP/6-311G(d,p) level.
b. Zero-point vibrational energies are calculated at the DFT Becke3LYP/6-311G(d,p) level.
c. Regular calculation of G2M energies.

Zero-point energy reaction coordinate diagram:

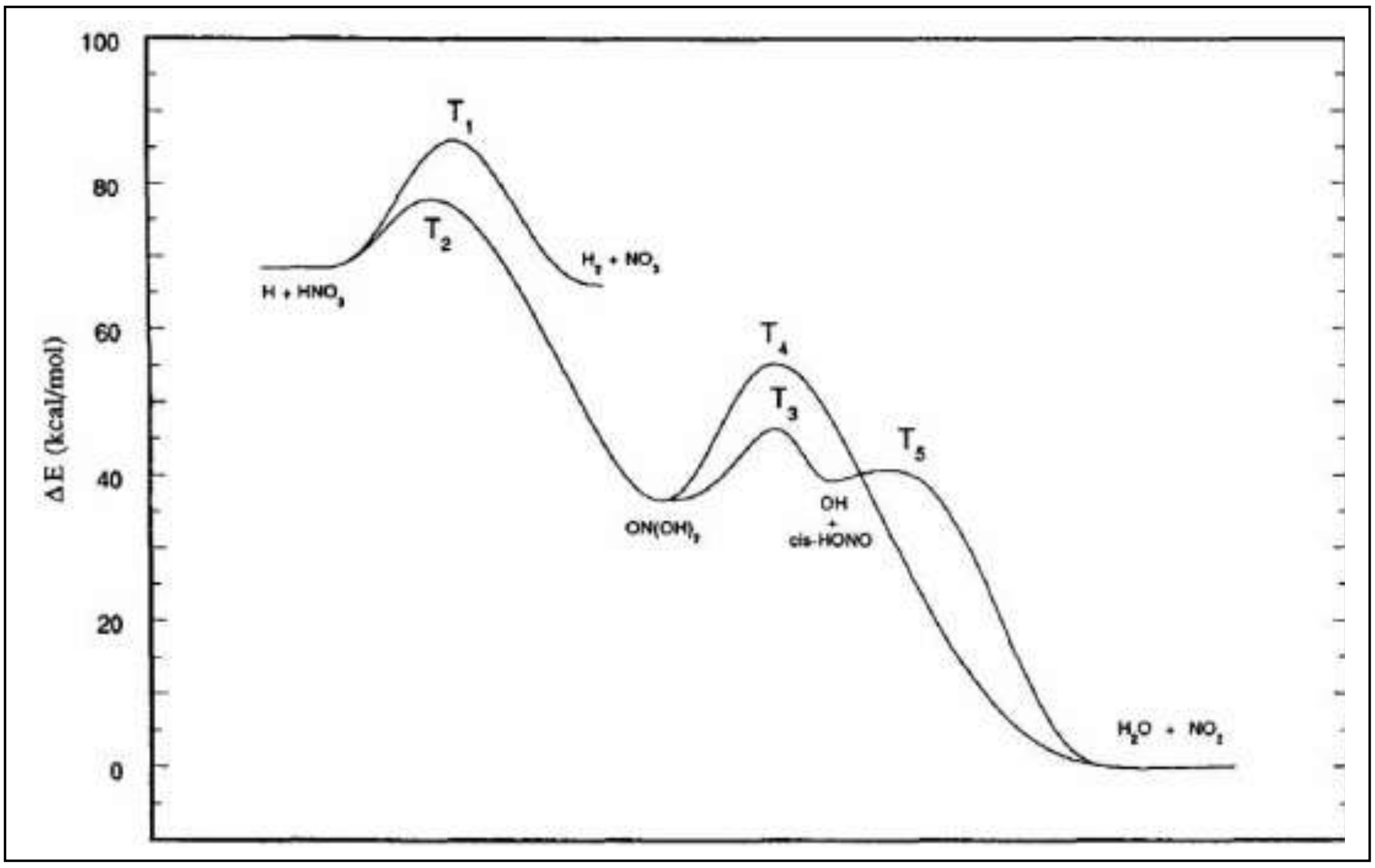

# 4. Acknowledgement

Financial and emotional support from OTKA and NKFI is kindly acknowledged:

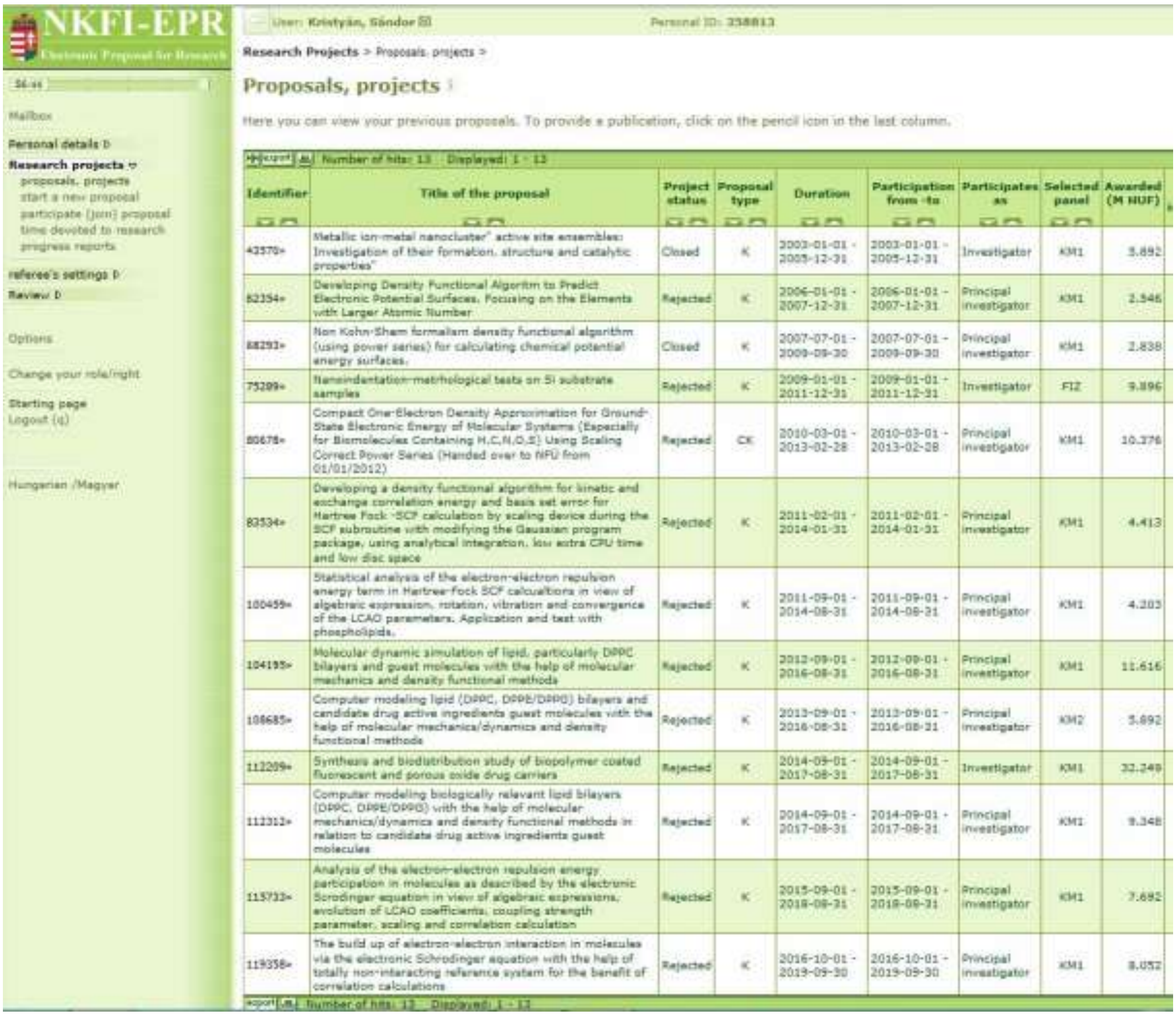

Research Projects > Proposals, projects >

## Proposals, projects

Here you can view your previous proposals. To provide a publication, click on the pencil icon in the last column.

Number of hits: 13 Displayed: 1 - 13

| Identifier | Title of the proposal | Project status | Proposal type | Duration | Participation from -to | Participates as | Selected panel | Awarded (M HUF) |
|---|---|---|---|---|---|---|---|---|
| 42570» | Metallic ion-metal nanocluster" active site ensembles: Investigation of their formation, structure and catalytic properties" | Closed | K | 2003-01-01 - 2005-12-31 | 2003-01-01 - 2005-12-31 | Investigator | KM1 | 5.892 |
| 62354» | Developing Density Functional Algoritm to Predict Electronic Potential Surfaces, Focusing on the Elements with Larger Atomic Number | Rejected | K | 2006-01-01 - 2007-12-31 | 2006-01-01 - 2007-12-31 | Principal investigator | KM1 | 2.546 |
| 68293» | Non Kohn-Sham formalism density functional algorithm (using power series) for calculating chemical potential energy surfaces. | Closed | K | 2007-07-01 - 2009-09-30 | 2007-07-01 - 2009-09-30 | Principal investigator | KM1 | 2.838 |
| 75289» | Nanoindentation-metrhological tests on Si substrate samples | Rejected | K | 2009-01-01 - 2011-12-31 | 2009-01-01 - 2011-12-31 | Investigator | FIZ | 9.896 |
| 80678» | Compact One-Electron Density Approximation for Ground-State Electronic Energy of Molecular Systems (Especially for Biomolecules Containing H,C,N,O,S) Using Scaling Correct Power Series (Handed over to NFÜ from 01/01/2012) | Rejected | CK | 2010-03-01 - 2013-02-28 | 2010-03-01 - 2013-02-28 | Principal investigator | KM1 | 10.376 |
| 83534» | Developing a density functional algorithm for kinetic and exchange correlation energy and basis set error for Hartree Fock -SCF calculation by scaling device during the SCF subroutine with modifying the Gaussian program package, using analytical integration, low extra CPU time and low disc space | Rejected | K | 2011-02-01 - 2014-01-31 | 2011-02-01 - 2014-01-31 | Principal investigator | KM1 | 4.413 |
| 100459» | Statistical analysis of the electron-electron repulsion energy term in Hartree-Fock SCF calcualtions in view of algebraic expression, rotation, vibration and convergence of the LCAO parameters. Application and test with phospholipids. | Rejected | K | 2011-09-01 - 2014-08-31 | 2011-09-01 - 2014-08-31 | Principal investigator | KM1 | 4.203 |
| 104195» | Molecular dynamic simulation of lipid, particularly DPPC bilayers and guest molecules with the help of molecular mechanics and density functional methods | Rejected | K | 2012-09-01 - 2016-08-31 | 2012-09-01 - 2016-08-31 | Principal investigator | KM1 | 11.616 |
| 108685» | Computer modeling lipid (DPPC, DPPE/DPPG) bilayers and candidate drug active ingredients guest molecules with the help of molecular mechanics/dynamics and density functional methods | Rejected | K | 2013-09-01 - 2016-08-31 | 2013-09-01 - 2016-08-31 | Principal investigator | KM2 | 5.892 |
| 112209» | Synthesis and biodistribution study of biopolymer coated fluorescent and porous oxide drug carriers | Rejected | K | 2014-09-01 - 2017-08-31 | 2014-09-01 - 2017-08-31 | Investigator | KM1 | 32.249 |
| 112312» | Computer modeling biologically relevant lipid bilayers (DPPC, DPPE/DPPG) with the help of molecular mechanics/dynamics and density functional methods in relation to candidate drug active ingredients guest molecules | Rejected | K | 2014-09-01 - 2017-08-31 | 2014-09-01 - 2017-08-31 | Principal investigator | KM1 | 9.348 |
| 115733» | Analysis of the electron-electron repulsion energy participation in molecules as described by the electronic Scrodinger equation in view of algebraic expressions, evolution of LCAO coefficients, coupling strength parameter, scaling and correlation calculation | Rejected | K | 2015-09-01 - 2018-08-31 | 2015-09-01 - 2018-08-31 | Principal investigator | KM1 | 7.692 |
| 119358» | The build up of electron-electron interaction in molecules via the electronic Schrodinger equation with the help of totally non-interacting reference system for the benefit of correlation calculations | Rejected | K | 2016-10-01 - 2019-09-30 | 2016-10-01 - 2019-09-30 | Principal investigator | KM1 | 8.052 |

Number of hits: 13 Displayed: 1 - 13